\documentclass[12pt]{suny-thesis}

\usepackage[utf8x]{inputenc}  
\usepackage{graphicx}  

\usepackage{amsmath}  
\usepackage{amsthm}
\usepackage{helvet}
\usepackage{enumerate}
\usepackage{amsfonts}
\usepackage{graphicx}
\def\be{\begin{equation}}
\def\bea{\begin{eqnarray}}
\def\ee{\end{equation}}
\def\eea{\end{eqnarray}}
\def\d{\partial}
\def\eps{\varepsilon}
\def\la{\lambda}

\def\nn{\nonumber \\}

\def\h{{1\over 2}}
\def\be{\begin{equation}}
\def\bea{\begin{eqnarray}}
\def\ee{\end{equation}}
\def\eea{\end{eqnarray}}

\newcommand{\ba}{\begin{array}}
\newcommand{\ea}{\end{array}}
\def\d{\partial}
\def\eps{\varepsilon}
\def\la{\lambda}

\def\nn{\nonumber \\}

\def\h{{1\over 2}}
\def\cH{\mathcal{H}}

\def\ch{\,\mbox{ch}}
\def\sh{\,\mbox{sh}}

\def\d{\partial}
\def\eps{\varepsilon}
\def\la{\lambda}

\def\tgt{}  
\def\nn{\nonumber \\}

\def\varkappa{\kappa}

\def\k{\kappa}

\allowdisplaybreaks  

\author{Yuri Chervonyi}
\title{Hidden symmetries in string theory}
\college{Arts \& Sciences}
\department{Physics}

\begin{document}

	\frontmatter

		\maketitle  

		\begin{abstract}  
			In this thesis we study hidden symmetries within the framework of string theory. Symmetries play a very important role in physics: they lead to drastic simplifications, which allow one to compute various physical quantities without relying on perturbative techniques. 

There are two kinds of hidden symmetries investigated in this work: the first type is associated with dynamics of quantum fields and the second type is related to integrability of strings on various backgrounds. Integrability is a remarkable property of some theories that allows one to determine all dynamical properties of the system using purely analytical methods. The goals of this thesis are twofold: extension of hidden symmetries known in General Relativity to stringy backgrounds in higher dimensions and construction of new integrable string theories. 

In the context of the first goal we study hidden symmetries of stringy backgrounds, with and without supersymmetry. For supersymmetric geometries produced by D--branes we identify the backgrounds with solvable equations for geodesics, which can potentially give rise to integrable string theories. Relaxing the requirement of supersymmetry, we also study charged black holes in higher dimensions and identify their hidden symmetries encoded in so--called Killing(-Yano) tensors. We construct the explicit form of the Killing(-Yano) tensors for the charged rotating black hole in arbitrary number of dimensions, study behavior of such tensors under string dualities, and use the analysis of hidden symmetries to explain why exact solutions for black rings (black holes with non-spherical event horizons) in more than five dimensions remain elusive. As a byproduct we identify the standard parameterization of $AdS_p \times S^q$ backgrounds with elliptic coordinates on a flat base.

The second goal of this work is construction of new integrable string theories by applying continuous deformations of known examples. We use the recent developments called (generalized) $\lambda$-deformation to construct new integrable backgrounds depending on several continuous parameters and study analytical properties of the such deformations.

		\end{abstract}
	

		\begin{acknowledgments}  
			I am exceptionally grateful to my supervisor Oleg Lunin for his guidance, willingness to answer all my questions and countless discussions though my PhD, without which this research would have been impossible. I am also thankful to the String theory group, especially Jia Tian, Chris Simmons, Daniel Robbins, and the entire Physics Department for many insightful discussions and useful meetings.

			I would also like to thank the Physics Department for awarding me with the Mr. and Mrs. Mahmood Alam Award and the Jagadish Garg Award for Outstanding Research.
		\end{acknowledgments}

		\tableofcontents

		\listoffigures

	\mainmatter


		\chapter{Outline}

		String theory unifies the two biggest challenges of fundamental physics, quantum gravity and theory of strong interactions, in context of gauge/gravity duality. A lot of insights to these problems were and still are being extracted by studying black holes. Another important tool, which has played the key role in understanding of gauge/gravity duality, is \textit{integrability}, a very special feature of some theories, which allows to compute scattering amplitudes, spectrum, and other physical quantities without relying on perturbative techniques such as Feynman diagrams. However, only few integrable theories are known so far and the goal of the integrability program is finding other, closer to the real world problems, integrable theories. 
Unfortunately, there is no general method of classifying such theories, but one approach is to start with a system, which is known to be integrable, and perform a deformation preserving it, so--called integrable deformation.

The method of integrable deformations has been proven to be a very powerful technique for generating integrable theories. It has led to expanding the list of integrable backgrounds in string theory from several discrete points (flat space and few $AdS_p\times S^q$ spaces) to families of theories specified by continuous parameters. Several families have been constructed so far. The most recent development is called the $\lambda$--deformation and it interpolates between the gauged Wess--Zumino--Witten model and the non-Abelian T--dual of the Principal Chiral Model. Interestingly, the embedding of the deformed action into string theory results in producing two prescriptions for computing the dilaton, based on bosonic cosets and supercosets respectively. While the metric obtained from both constructions is the same, fluxes turn out to be different. Another striking feature of the $\lambda$--deformed backgrounds is the absence of isometries, which makes constructing the deformed backgrounds especially interesting. In particular, in Chapter \ref{ChapterLambda} the supergravity background of the $\lambda$--deformed $AdS_3\times S^3$ supercoset is constructed and a progress towards constructing the deformation of $AdS_5\times S^5$ is reported, which, however, still remains an open problem. In Chapter \ref{ChapterGenLambda} the generalized $\lambda$--deformations are studied, which are based on solutions of the Yang--Baxter equation. Specifically, the deformation of $AdS_3\times S^3, AdS_2\times S^2$ are constructed. Furthermore, an interesting feature of the (generalized) $\lambda$--deformed coset backgrounds, which was observed before for the ordinary deformation, was proven: it was shown that the frames for both ordinary and generalized $\lambda$--deformed backgrounds are obtained from the non-deformed ones by a simple dressing procedure resulting in multiplication by a constant matrix. 

Proving integrability in string theory requires finding the complete infinite set of conserved quantities, which is a very complicated problem. In Chapter \ref{ChapterGeodesics} we look at the problem from a different perspective, focusing on integrability of point particles as a necessary condition for integrability of strings. This leads to drastic simplifications and allows us to classify all possibly integrable backgrounds produced by supersymmetric D branes. Specifically, we reduce the problem to studying separability of Hamilton--Jacobi equation and analyze backgrounds constructed from: a single type of branes, Dp branes dissolved in D(p+4) branes, rotating branes describing all microscopic states of D1--D5 black hole and finally bubbling solutions. Surprisingly, even though we consider curved backgrounds, separation of Hamilton--Jacobi equation happens only in the elliptic coordinates. As a by--product of this analysis we find that the standard parameterization of all $AdS_p\times S^q$ can be interpreted as the elliptic coordinates on the flat base.

In Chapter \ref{ChapterKillingYano} this work is complemented by studying non--supersymmetric backgrounds, in particular a broad class of \textit{black holes}. These geometries are known to possess hidden symmetries, which play an important role in understanding their dynamics. Specifically, symmetries provide a new way for studying various astrophysical processes such as the radiation produced by infalling particles, the plasma acceleration around black holes, gravitational wave production in star collisions. Moreover, they simplify studying stability of black holes and the calculations of the quasi-normal modes. Symmetries that are responsible for all these simplifications are encoded in Killing--(Yano) tensors, which can be viewed as generalizations of Killing vectors to higher ranks. A Killing tensor is associated with separability of Hamilton--Jacobi and Klein--Gordon equations, and a Killing--Yano tensor is associated with separability of the Dirac equation. Surprisingly, Killing(--Yano) tensors were not well studied in string theory and supergravity, so we investigate transformations of these objects under S and T dualities and find the unique modification of the equation for Killing--Yano tensor in the presence of fluxes. As a byproduct, we construct the modified Killing--Yano tensors for a wide class of NS--NS backgrounds including the most general rotating black hole in arbitrary dimensions and a system of fundamental strings and NS5 branes. We also find that the equations for Killing vectors and Killing tensors are not modified by the NS--NS fields.

In Chapter \ref{ChapterBlackRings} the framework of Killing tensors is used for investigating properties of higher dimensional black rings, the geometries with non--spherical event horizons. Such solutions have been found in five dimensions but their higher dimensional counterparts are still missing. In this work we show that some hidden symmetries of the black ring solutions cannot appear in higher dimensions. The analysis is performed for geometries with and without supersymmetry.

		\chapter{(Non)-Integrability of Geodesics in D-brane Backgrounds }\label{ChapterGeodesics}

		Motivated by the search for new backgrounds with integrable string theories, in this chapter we classify the D-brane geometries leading to integrable geodesics. Our analysis demonstrates that the Hamilton-Jacobi equation for massless geodesics can only separate in elliptic or spherical coordinates, and all known integrable backgrounds are covered by this separation. In particular, we identify the standard parameterization of $AdS_p \times S^q$ with elliptic coordinates on a flat base. We also find new geometries admitting separation of the Hamilton-Jacobi equation in the elliptic coordinates. Since separability of this equation is a necessary condition for integrability of strings, our analysis gives severe restrictions on the potential candidates for integrable string theories. 

\bigskip

This chapter is based on the results published in \cite{ChL}.

\section{Introduction and summary}

Over the last two decades, AdS/CFT correspondence \cite{AdSCFT} has led to great advances in our understanding of gauge theories and string theory on Ramond--Ramond backgrounds. A special role in this progress has been played by integrability, a surprising property of field theories, which allows one to compute spectrum, correlation functions, and scattering amplitudes \cite{review_AdSCFT_int} using an infinite set of conserved charges \cite{DNW}. Originally integrable structures were discovered in the ${\mathcal N}=4$ super--Yang--Mills theory, but they have been  extended to other systems\footnote{The examples include the marginal deformation of ${\mathcal N}=4$ super--Yang--Mills \cite{margin,Frolov}, the three dimensional Chern--Simmons theory \cite{Chern} and two--dimensional CFT \cite{AdS3CFT2}.}, and it is important 
to classify field theories admitting integrability. A promising approach to such classification, which is based on analyzing behavior of strings on a dual background, has led to ruling out integrability for the superconformal theory on a quiver \cite{Zayas} and for a certain deformation of ${\mathcal N}=4$ SYM \cite{StepTs}\footnote{See \cite{Zayas1} for further discussion of non-integrability and chaos in the context of AdS/CFT correspondence.}. In this chapter we will analyze integrability of strings on a large class of Ramond--Ramond backgrounds, rule out integrability for a wide range of field theories, and identify the potential candidates for integrable models. 

To put our results in perspective, let us briefly review the status of integrability in ${\mathcal N}=4$ SYM (or in string theory on $AdS_5\times S^5$). In the planar limit, the field theory can be solved by the Bethe ansatz \cite{bethe}, and the spectrum of strings on the gravity side can be found by solving the Landau--Lifshitz model \cite{TsLL}. The agreement between these two exact solutions provides a highly nontrivial check of the AdS/CFT correspondence. The methods of 
\cite{bethe,TsLL} are applicable only to the light states, whose conformal dimension obeys the relation 
\bea\label{Planar}
\Delta\ll N,
\eea
While the techniques of \cite{bethe} are not applicable when inequality (\ref{Planar}) is violated, the integrability might still persist in this case, at least for some sectors of the theory. Violation of (\ref{Planar}) implies that excitations of $AdS_5\times S^5$ might contain D--branes in addition to the fundamental strings, and generic excitations of this type are very complicated. Fortunately, some states violating the condition (\ref{Planar}) still have very simple behavior: these are the BPS states with\footnote{Here $\Delta$ is a conformal dimension of the state, and $J$ is its R charge. For simplicity we are focusing on 1/2--BPS states, but condition (\ref{NearBPS}) can be easily generalized to BPS states with lower amount of supersymmetry.} $\Delta=J$. On the gravity side of the correspondence, the BPS states are represented by supergravity modes or by D--branes, depending on the value of $J$. To have interesting dynamics, one can introduce some fundamental strings in addition to these BPS branes and to replace (\ref{Planar}) by
\bea\label{NearBPS}
\Delta-J\ll N.
\eea
As we already mentioned, the planar techniques of \cite{bethe} are not applicable to the states (\ref{NearBPS}) which violate (\ref{Planar}), and in this chapter we will use alternative methods to study integrability of such states. 

The most useful version of the AdS/CFT duality involves field theory on $R\times S^3$, then the bulk configurations satisfying  (\ref{NearBPS}) are represented by fundamental strings in the presence of giant gravitons \cite{giant}. The interactions between these objects can be very complicated \cite{BalFeng}, but additional simplifications occur for semiclassical configurations of giant gravitons with $J\sim N^2$, which can be viewed as classical geometries \cite{LLM}. In this regime of parameters, integrability of the sector (\ref{NearBPS}) reduces to integrability of strings on the bubbling geometries constructed in \cite{LLM}. If the CFT is formulated on $R^{3,1}$, the counterparts of the bubbling geometries are given by brane configurations describing the Coulomb branch of ${\mathcal N}=4$ SYM \cite{KrausLarsen}. In the latter case, one can introduce an additional deformation which connects $AdS_5\times S^5$ asymptotics to flat space and see whether integrability persists for such configurations. 

It turns out that the answer to the last question is no, and this result discovered in \cite{StepTs} was the main motivation for our investigation. As demonstrated in \cite{StepTs}, addition of one to the harmonic function describing a single stack of D3 branes destroys integrability of the closed strings on a new asymptotically-flat background. Since continuation to the flat asymptotics destroys the dual field theory, this procedure appears to be more drastic than a transition to the Coulomb branch, which corresponds to a normalizable excitations, so the latter might have a chance to remain integrable. In this chapter we focus on geometries dual to the Coulomb branch of ${\mathcal N}=4$ SYM (either on $R^{3,1}$ or on $R\times S^3$) and on similar geometries involving other D branes. A different class of theories, which involves putting D branes on singular manifolds, was explored in \cite{Zayas}, where it was demonstrated that strings are not integrable on the conifold. From the point of view of field theory, this result pertains to the vacuum of ${\mathcal N}=1$ SYM with a quiver gauge group, which is complementary to our analysis of excited states in ${\mathcal N}=4$ SYM.

To identify the backgrounds leading to integrable string theories, one has to analyze the equations of motion for the sigma model and to determine whether they admit an infinite set of conserved quantities. Instead of solving this complicated problem, we will focus on necessary conditions for integrability and demonstrate that strings are not integrable on a large class of backgrounds created by D--branes.
Integrability on a given background should persist for string of arbitrary size, and in the limit of point-like strings it 
leads to integrability of null geodesics\footnote{In this chapter we focus on Ramond--Ramond backgrounds produced by D--branes, but in the presence of the NS--NS $B$ field, pointlike strings could carry additional charges, which modify equations for the geodesics.}, which implies that the motion of a particle is characterized by 10 conserved quantities, matching the number of the degrees of freedom $x^i$. Massless geodesics can be found by solving the Hamilton--Jacobi (HJ) equation,
\bea\label{HJone}
g^{MN}\frac{\d S}{\d X^M}\frac{\d S}{\d X^N}=0,
\eea 
where $S$ is the action of a particle, and $g_{MN}$ is the background metric. The system is called integrable if the HJ equation separates 
\cite{Arnold}, i.e., if there exists a new set of coordinates $Y^M$, such that 
\bea\label{SeparIntro}
S(Y_0,\dots Y_9)=\sum_{I=0}^9 S_I(Y_I).
\eea
This also implies that the HJ equation has ten independent integrals of motion. Non--trivial examples of geometries leading to integrable geodesics  include Kerr--Neumann black hole \cite{Carter} and its generalizations to Kerr--NUT--AdS spacetimes in higher dimensions \cite{FrolovKerrNUTAdS}. To rule out integrability of geodesics on a particular background, it is sufficient to demonstrate that separation (\ref{SeparIntro}) cannot be accomplished in any set of coordinates. 

In this chapter we will analyze the motion of  massless particles in the geometries produced by stacks of parallel Dp branes and identify the distributions of branes which lead to integrable HJ equation (\ref{HJone}). Specifically, we will focus on supersymmetric configurations of D$p$--branes with flat worldvolume\footnote{In section \ref{Bubbles} we will also discuss a special class of spherical branes.}, and assume that Ramond--Ramond $(p+1)$--form sourced by the branes is the only nontrivial flux in the geometry. This implies that metric $g_{ij}$ has the form \cite{HorStr}
\bea\label{PreGeomT}
g_{ij}dx^idx^j=\frac{1}{\sqrt{H}}\eta_{\mu\nu}dx^\mu dx^\nu+
\sqrt{H}ds^2_{base},
\eea
where the first term represents $(p+1)$--dimensional Minkowski space parallel the branes, and $H$ is a harmonic function on the $(9-p)$--dimensional base space. We will further assume that the base space is flat.

For a single stack of D$p$--branes, the HJ equation separates in spherical coordinates, and this well-known case is reviewed in section 
\ref{Examples}. This section also includes another example, separation in elliptic coordinates, which plays an important role in the subsequent discussion. 

Section \ref{SecGnrGds} describes our main procedure, which is subsequently used to study geodesics on a variety of backgrounds. 
In subsection \ref{RedTo2D} we demonstrate that the HJ equation (\ref{HJone}) does not separate unless the metric on the base space has the form 
\bea\label{BaseMetrIn}
ds_{base}^2=dr_1^2+dr_2^2+r_1^2d\Omega_{d_1}^2+
r_2^2d\Omega_{d_2}^2,
\eea
and $H$ depends only on $r_1$ and $r_2$. The most general harmonic function $H$ leading to integrable HJ equation is derived in section 
\ref{SectPrcdr}, and equations (\ref{TheSolnMthree})-(\ref{TheSolnMtwo}) summarize the main result of this chapter for branes with flat worldvolume. The brane configurations giving rise to geometries 
(\ref{TheSolnMthree})--(\ref{TheSolnMtwo}) are analyzed in section \ref{Sources}. The results of section 
\ref{SecGnrGds} imply that $(Y_0,\dots,Y_{10})$ leading to separation (\ref{SeparIntro}) must reduce to the elliptic coordinates discussed in section \ref{Examples}.

Section \ref{SecKillWave} discusses physical properties of the geometries leading to integrable geodesics. We demonstrate that separability persists for the wave equation beyond the eikonal approximation, a property that have been observed earlier for various black holes \cite{Carter,cvetLars}. In section \ref{SecKill} we show that separability of the wave equation is associated with a hidden symmetry of the background, and we construct the conformal Killing tensor associated with this symmetry. In section \ref{NonintStr} we apply the techniques of  \cite{Zayas,StepTs} to demonstrate that most backgrounds with separable wave equation do not lead to integrable string theories. 

In section \ref{DpDp4} our results are generalized to D$p$--branes dissolved in D$(p+4)$--branes, the system which plays an important role in understanding the physics of black holes \cite{StrVafa}. We find that in asymptotically--flat space there are no integrable solutions apart from the spherically--symmetric distribution of branes. However, there are several separable configurations in the near--horizon limit of D$(p+4)$--branes, and the most general D$p$--D$(p+4)$ configurations leading to separable HJ equation are presented in (\ref{H1D1D5}), (\ref{H2D1D5}). 

In section \ref{Rotations} we consider another generalization by allowing the branes to rotate, i.e., by breaking the Poincare symmetry on the brane worldvolume. Although the general analysis of rotating branes is beyond the scope of this chapter, we consider the special class of rotating solutions which cover all microscopic states of the D1--D5 black hole \cite{lmPar,lmm}. Such solutions are parameterized by curves in eight--dimensional space, and our analysis demonstrates that HJ equation is not separable unless this curve is a simple circle. For such configuration the separable coordinates have been found before \cite{MultiStr}, and we will demonstrate that these coordinates reduce to a special case of the general elliptic coordinates discussed in section \ref{SectPrcdr}.

In section \ref{Bubbles} we consider a different class of rotating solutions, which describes all half--BPS states of IIB supergravity supported by the five--form field strength \cite{LLM}. We demonstrate that there are only three bubbling solutions leading to separable HJ equation for the geodesics: 
AdS$_5\times$S$^5$, pp--wave and a geometry dual to a single M2 brane. In section \ref{StdPar} we discuss the equation for geodesics on bubbling geometries in M theory, and we demonstrate that the elliptic coordinates emerging from the separation of variables in the geometries of \cite{LLM} coincide with standard parameterization of AdS$_5\times$S$^5$, AdS$_7\times$S$^4$, and AdS$_4\times$S$^7$.

\section{Examples: spherical and elliptic coordinates}\label{Examples}

We begin with discussing two known examples of brane configurations which lead to integrable equations for the geodesics. First we recall the situation for a single stack of Dp branes. In this case the metric has the form
\bea\label{DbrnMetr}
ds^2&=&H^{-1/2}\eta_{\mu\nu}dx^\mu dx^\nu+H^{1/2}(dr^2+r^2 d\Omega_{8-p}^2),
\eea
where
\bea\label{HrmFncSph}
 H=a+\frac{Q}{r^{7-p}}
\eea
and $d\Omega_{8-p}^2$ is the metric on a $(8-p)$--dimensional sphere:
\bea\label{8dSphere}
d\Omega_{8-p}^2&=&h_{ij}dy^i dy^j.
\eea
The Hamilton--Jacobi equation for a particle propagating in the geometry (\ref{DbrnMetr}) has the form
\bea
\left(\frac{\d S}{\d r}\right)^2+\frac{h^{ij}}{r^2}\frac{\d S}{\d y^i}\frac{\d S}{\d y^j}+H\eta^{\mu\nu}\frac{\d S}{\d x^\mu}\frac{\d S}{\d x^\nu}=0,
\eea
and variables in this equation separate:
\bea\label{HJact}
S=p_\mu x^\mu+S_{L}(y)+R(r).
\eea 
Here
\bea\label{RHJ}
(R')^2+\frac{L^2}{r^2}+ H p_\mu p^\mu=0,\\
\label{AngHJ}
{h^{ij}}\frac{\d S_L}{\d y^i}\frac{\d S_L}{\d y^j}=L^2.
\eea
Solution (\ref{HJact}) has $(p+1)$ integrals of motion $p_\mu$, $8-p$ independent integrals coming from $S_L$ (this is ensured by the isometries of the sphere, the explicit form of $S_L$ is given in appendix \ref{AppSphere}), and one integration constant coming from the differential equation (\ref{RHJ}). 
This implies that action (\ref{HJact}) can be written in terms of $10$ conserved quantities, so the geometry (\ref{DbrnMetr}) leads to integrable geodesics. As demonstrated in \cite{StepTs}, integrability does not persist for strings, unless $a=0$ in (\ref{HrmFncSph}). Spherical coordinates (\ref{DbrnMetr}) will play an important role in our construction since any localized distribution of D branes leads to a harmonic function which approaches (\ref{HrmFncSph}) at infinity. Thus, any set of separable coordinates must reduce to (\ref{DbrnMetr}) far away from the branes. 

\begin{figure}[h!]
\centering
\includegraphics[width=0.8\textwidth]{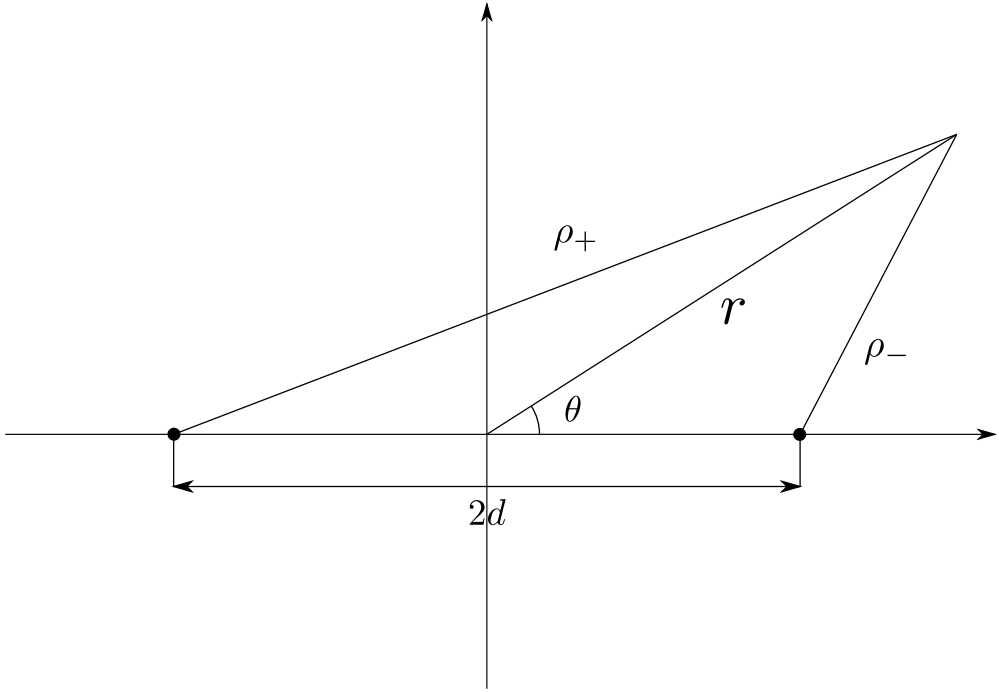} 
\caption[Elliptic coordinates]{Geometrical meaning of $\rho_+$ and $\rho_-$ appearing in the definition of the elliptic coordinates (\ref{DefEllCoord}).}
\label{Fig:EllCoor}
\end{figure}

Our second example deals with two stacks of Dp branes separated by a distance $2d$ (see figure \ref{Fig:EllCoor}). The metric produced by this configuration is given by 
\bea\label{2cntrMetr}
ds^2&=&H^{-1/2}\eta_{\mu\nu}dx^\mu dx^\nu+H^{1/2}(dr^2+r^2d\theta^2+r^2\sin^2\theta d\Omega_{7-p}^2),
\eea
where
\bea\label{EllHarm}
 H=a+\frac{Q}{\rho_+^{6-p}}+\frac{Q}{\rho_-^{6-p}},\qquad
\rho_\pm=\sqrt{r^2+d^2\pm 2rd\cos\theta}
\eea
Introducing coordinates $y_i$ on the $(7-p)$--dimensional sphere and separating variables in the action,
\bea\label{HJactEl}
S=p_\mu x^\mu+S_{L}(y)+R(r,\theta),
\eea 
we can rewrite (\ref{HJone}) as a PDE for $R(r,\theta)$:
\bea\label{EllHJ}
(\d_r R)^2+\frac{1}{r^2}(\d_\theta R)^2+\frac{L^2}{r^2\sin^2\theta}+ H p_\mu p^\mu=0
\eea
This equation describes the motion of a particle in a two--center potential, and it is well--known that for $p=5$ it can be separated further by introduction of the elliptic coordinates \cite{Landau}:
\bea\label{DefEllCoord}
\xi=\frac{\rho_++\rho_-}{2d},\qquad \eta=\frac{\rho_+-\rho_-}{2d},
\eea
In appendix \ref{SprElliptic} we review the separation procedure that leads to (\ref{HJactEl}) and write down the explicit form of $R$ (see (\ref{RActEll})).
For future reference we quote the asymptotic relation between elliptic and spherical coordinates: 
\bea\label{AsympEll}
\xi=\frac{r}{d}+\frac{d}{2r}\sin^2\theta + O\left(\frac{1}{r^3}\right),\qquad \eta=\cos\theta+\frac{d^2}{2r^2}\cos\theta\sin^2\theta
+O\left( \frac{1}{r^4} \right).
\eea

This completes our review of spherical and elliptic coordinates, and in the remaining part of this chapter we will investigate whether the separation of variables persists for more general geometries produced by Dp--branes.

\section{Geodesics in D--brane backgrounds}
\label{SecGnrGds}

We now turn to the main topic of this chapter: the analysis of geodesics in the geometry produced by D--branes \cite{HorStr}:
\bea\label{PreGeom}
g_{MN}dX^MdX^N&=&\frac{1}{\sqrt{H}}\eta_{\mu\nu}dx^\mu dx^\nu+
\sqrt{H}ds^2_{base},\\
\nabla^2_{base}H&=&0\nonumber
\eea
We will further assume that the base space is flat:
\bea\label{FlatBase}
ds_{base}^2=dx^jdx^j.
\eea
The massless geodesics in the background (\ref{PreGeom}) are integrable if and only if the HJ equation (\ref{HJone}) separates in some coordinates $(Y_0,\dots,Y_{9})$, as in (\ref{SeparIntro}).
In section \ref{RedTo2D} we will use this separation and the Laplace equation 
\bea\label{LaplAA}
\frac{\d^2 H}{\d x_j\d x_j}=0
\eea
to demonstrate that function $H$ can only depend on two of the ten coordinates $(Y_0,\dots,Y_{9})$. The further analysis presented in section \ref{SectPrcdr} demonstrates that $(Y_0,\dots,Y_{9})$ must reduce to a slight generalization of the elliptic coordinates presented in section \ref{Examples}. Although the Laplace equation (\ref{LaplAA}) is satisfied away from the branes, any nontrivial function $H$ must have sources, and the brane configurations leading to a separable HJ equation are analyzed in section \ref{Sources}.

\subsection{Reduction to two dimensions}
\label{RedTo2D}

Let us assume the separation (\ref{SeparIntro}) in the HJ equation (\ref{HJone}) and explore the consequences for the harmonic function $H$ appearing in (\ref{PreGeom}). Metric (\ref{FlatBase}) is invariant under $SO(9-p)$ transformations, but part of this rotational symmetry might be broken by the harmonic function $H$. Let $SO(d_1+1)\times SO(d_2+1)\times SO(d_k+1)$ be the maximal subgroup of $SO(9-p)$ preserved by $H$, then the metric on the base space can be written as
\bea\label{BaseMetr}
ds_{base}^2=dr_1^2+r_1^2d\Omega_{d_1}^2+\dots+
dr_k^2+r_k^2d\Omega_{d_k}^2,
\eea
and $H$ becomes a function of $(r_1,\dots,r_k)$.  Moreover, since all rotational symmetries have been isolated, we conclude that\footnote{If this relation is not satisfied for $i=1$, $j=2$, then $SO(d_1+1)\times SO(d_2+1)$ is enhanced to $SO(d_1+d_2+2)$.}
\bea\label{NoRotation}
(r_i\d_j-r_j\d_i)H\ne 0.
\eea
If the branes are localized in a compact region, then at sufficiently large values of 
$$R=\sqrt{r_1^2+\dots+r_k^2}$$ 
function $H$ satisfies the Laplace equation
\bea\label{LaplRk}
\frac{1}{r_1^{d_1}}\frac{\d}{\d r_1}\left[r_1^{d_1}\frac{\d H}{\d r_1}\right]+\dots+
\frac{1}{r_k^{d_k}}\frac{\d}{\d r_k}\left[r_k^{d_k}\frac{\d H}{\d r_k}\right]=0,
\eea
and asymptotic behavior of function $H$ is given by 
\bea\label{Hasymt}
H\sim a+\frac{Q}{(r_1^2+\dots+r_k^2)^{q}}.
\eea
Here $a$ is a parameter, which is equal to zero for the near--horizon geometries and which can be set to one for asymptotically--flat solutions. 

Our goal is to classify the backgrounds (\ref{PreGeom}) that lead to separable HJ equations for geodesics, and we begin with separating variables associated with symmetries. Poincare invariance of (\ref{PreGeom}) and rotational invariance of the base metric (\ref{BaseMetr}) allow us to write the action appearing in the HJ equation (\ref{BaseMetr}) as
\bea\label{GenAction}
S=p_\mu x^\mu+\sum_{i=1}^k S^{(d_i)}_{L_i}(\Omega_{d_i})+
{\tilde S}(r_1,\dots r_k).
\eea
Here $p_\mu$ is the momentum of the particle in $p+1$ directions longitudinal to the branes, and 
$S^{(d_i)}_{L_i}(\Omega_{d_i})$ is the part of the action that depends on coordinates $y_1,\dots,y_{d_i}$ of the sphere $\Omega_{d_i}$. The label $L_i$ represents the angular momentum of a particle along this sphere, and it is defined by relation
\bea\label{AngHJaa}
{h^{ij}}\frac{\d S_L}{\d y^i}\frac{\d S_L}{\d y^j}=L_i^2.
\eea
An explicit construction of $S^{(d)}_{L}(\Omega)$ is presented in appendix \ref{AppSphere}. 

Substitution of (\ref{GenAction}) in the HJ equation (\ref{HJone}) leads to equation for ${\tilde S}$:
\bea\label{HJaa}
\sum_{i=1}^k\left[\left(\frac{\d {\tilde S}}{\d r_i}\right)^2+\frac{L_i^2}{r_i^2}\right]+ H p_\mu p^\mu=0.
\eea
A special case of this equation with $k=1$ separates in spherical coordinates (see section \ref{Examples}), and we will now prove the equation (\ref{HJaa}) does not separate if $k>2$. The separation of (\ref{HJaa}) for $k=2$ will be discussed in section \ref{SectPrcdr}. 

Separability of equation (\ref{HJaa}) should persist for all values of angular momenta, so we begin with setting all $L_j$ to zero and $p_1=\dots=p_p=0$. The resulting equation (\ref{HJaa}) can be viewed as a HJ equation on an effective $(k+1)$--dimensional space
\bea\label{trrMetr}
ds^2=-H^{-1}dt^2+dr_1^2+\dots+dr_k^2
\eea
A general theory of separable HJ equations on curved backgrounds has a long history (see \cite{MorseFesh}), and a complete classification is presented in \cite{KalMil1, KalMil2}. In particular, this theory distinguishes between ignorable directions (which correspond to Killing vectors) and non-ignorable ones. Clearly, the time direction in (\ref{trrMetr}) is ignorable, but (\ref{trrMetr}) does not have additional Killing vectors which commute with $\d_t$. Indeed, any such vector would be a Killing vector of the $k$--dimensional flat space, so it must be a combination of translations and rotations in $r_i$. However, the asymptotic behavior of function $H$ (\ref{Hasymt}) breaks translational symmetry, and our assumption (\ref{NoRotation}) destroys the rotational Killing vectors, so if the HJ equation for the metric (\ref{trrMetr}) separates in some coordinates 
$(t,x_1,\dots x_k)$, only one of them (specifically, $t$) can correspond to an ignorable direction. Moreover, the discrete symmetry $t\rightarrow -t$ of  (\ref{trrMetr}) guarantees that $t$ does not mix with 
$(x_1,\dots x_k)$ in the metric, and such orthogonality leads to simplifications in the general analysis of \cite{KalMil1, KalMil2}.

Specifically, according to theorem 6 of \cite{KalMil2} separation of variables in (\ref{trrMetr}) implies that\footnote{The discussion of \cite{KalMil1,KalMil2} is more general: it allows mixing between ignorable and essential coordinates.} 
\begin{enumerate}[(1)]
\item{There exist $k$ independent conformal Killing tensors $A^{(a)}$ with components $(A^{(a)}_{ij},A^{(a)}_{tt})$.}
\item{Each of the one--forms $dx^l=\omega^l_i dr^i$ is a simultaneous eigenform of all $A_{(a)}^{ij}$ with eigenvalues 
$\rho_{(a,l)}$. This implies that  a projector ${P^{(l)i}}_j$ onto $dx^l$ satisfies equation
\bea\label{KMeigen}
(A_{(a)}^{ij}-\rho_{(a,l)}g^{ij}){P^{(l)j}}_k=0.
\eea
}
\item{The metric in coordinates $dx^l$ is diagonal, so projectors ${P^{(l)i}}_j$ commute with $A_{(a)}^{ij}$ and 
$g^{ij}$, and projectors with different values of $l$ project onto orthogonal subspaces. Since the number of projectors is equal to the number of coordinates, we arrive at the decomposition
\bea\label{KMdecomp}
A_{(a)}^{ij}=\sum_l^k \rho_{(a,l)}h_l P_{(l)}^{ij},\quad g^{ij}=\sum_l^k h_l P_{(l)}^{ij}
\eea }
\item{The components of $A_{(a)}$ along the Killing direction 
satisfy an overdefined system of differential equations:
\bea\label{KMProj}
\d_i\left[A_{(a)}^{tt}\right]-\sum_l^k 
\rho_{(a,l)}{P_i}^{(l)j}\d_j g^{tt}=0
\eea
}
\end{enumerate}
Notice that in \cite{KalMil2} the theorem is formulated in terms of coordinates $x_i$, so it does not use the projectors. For our purposes the covariant formulation given above is more convenient, in particular, to rule our the separation of variables, we will have to work in the original coordinates $r_i$ and demonstrate that the required Killing tensors $A_{(a)}$ do not exist. Notice that equation (\ref{KMProj}) can be rewritten in the form which does not refer to projectors (and thus to coordinates $x_i$): multiplying this equation by $g^{ij}$ and using (\ref{KMdecomp}), we find
\bea\label{kkk}
g^{ij}\d_j(A_{(a)}^{tt})-A_{(a)}^{ij}\d_j g^{tt}=0.
\eea
This relation is equivalent to $(tti)$ component of the equation for the Killing tensor $A_{(a)}$. 

The theorem quoted above implies that $A^{(a)}_{ij}$ are conformal Killing tensors on the $k$--dimensional base of (\ref{trrMetr}), and, for the flat base, all such tensors can be written as quadratic combinations of $k(k+3)(k+4)(k+5)/12$ Killing vectors \cite{Weir}\footnote{The explicit form of the conformal Killing vectors is given in appendix \ref{AppFltKil}.}:
\bea
A^{(a)}_{ij}=\sum_{m,n}b^{(a)}_{m,n} V^{(m)}_i
V^{(n)}_j
\eea
Equations (\ref{KMeigen}) and (\ref{kkk}) give severe restrictions on coefficients $b^{(a)}_{m,n}$, but fortunately the consequences of (\ref{KMeigen}) have been analyzed elsewhere. Indeed, equation (\ref{KMeigen}) does not involve $g_{tt}$, so it remains the same for $H=1$, when (\ref{trrMetr}) gives the flat space, and the corresponding HJ equation gives an eikonal approximation for the standard wave equation. It is well-known that in $3+1$ dimensions ($k=3$) the latter can only be separated in ellipsoidal coordinates and their special cases \cite{MorseFesh}, and generalization of this result to $k>3$ is presented in \cite{KalMil1, KalMil2}. This leads to the conclusion that the HJ equation in the metric (\ref{trrMetr}) with $k>2$ can only separate in ellipsoidal coordinates or in the degenerate form thereof. Before ruling out this possibility, we briefly comment on the peculiarities of the two--dimensional base. In this case the conformal group becomes infinite-dimensional, 
so the base space admits an infinite number of the conformal Killing tensors. This situation will be analyzed in section \ref{SectPrcdr}.

To summarize, we concluded that for $k>2$, the HJ equation can only be separable in some special case of ellipsoidal coordinates $(x_1,\dots,x_k)$, which are defined by \cite{Jacobi,Miller4}
\bea\label{ref317}
&&r_i^2=-\left[\prod_j(a_i^2+x_j)\right]\left[\prod_{j\ne i}\frac{1}{(a_j^2-a_i^2)}\right].
\eea
Here $(a_1,\dots,a_k)$ is the set of positive constants, which specify the ranges of variables $x_i$:
\bea
x_1\ge -a_1^2\ge x_2\ge -a_2^2\ge \dots x_k\ge -a_k^2,
\eea 
Rewriting the metric (\ref{trrMetr}) in terms of $x_i$ and substituting the result into (\ref{HJaa}), we find the HJ equation in ellipsoidal coordinates (see appendix \ref{AppEllips} for detail):
\bea\label{HJelps}
\sum_{i=1}^k\left[\frac{1}{h_i^2}\left(\frac{\d {\tilde S}}{\d x_i}\right)^2+\frac{L_i^2}{r_i^2}\right]+ H p_\mu p^\mu=0.
\eea
Here $h_i$ is defined by 
\bea
h_i^2=\frac{1}{4}\left[\prod_{j\ne i}(x_i-x_j)\right]\left[\prod_j\frac{1}{a_j^2+x_i}\right].
\eea
Function $H$ appearing in (\ref{HJelps}) must satisfy equation (\ref{LaplRk}) away from the sources, and appendix \ref{AppEllips} we demonstrate that for such functions equation 
(\ref{HJelps}) never separates in ellipsoidal coordinates. This shows that the HJ equation can only be integrable for $k=1$ (the situation considered in section \ref{Examples}) and for $k=2$, which will be analyzed in the next subsection.

\subsection{Separation of variables and elliptic coordinates}
\label{SectPrcdr}

In the last subsection we have demonstrated that the HJ equation (\ref{HJaa}) is not integrable unless the flat base has the form (\ref{BaseMetr}) with $k\le 2$ and $H$ is a function of $r_1$ and $r_2$ only. In section \ref{Examples} we have already discussed $k=1$ and this subsection is dedicated to the analysis of $k=2$. To simplify some formulas, we slightly deviate from the earlier notation and write the metric (\ref{PreGeom}) with the base (\ref{BaseMetr}) for $k=2$ as
\bea\label{SSmetr}
ds^2=\frac{1}{\sqrt{H}}\eta_{\mu\nu}dx^\mu dx^\nu+
\sqrt{H}(dr^2+r^2d\theta^2+r^2\cos^2\theta d\Omega_{m}^2+r^2\sin^2\theta d\Omega_{n}^2),
\eea
The connection to coordinates of (\ref{BaseMetr}) is obvious:
\bea
r_1=r\cos\theta,\quad r_2=r\sin\theta,
\eea
with $d_1=m, d_2=n$. The arguments presented in the last subsection ensure that $H$ appearing in (\ref{SSmetr}) can only depend on $r$ and $\theta$, i.e., the distribution of Dp branes that sources this harmonic function is invariant under $SO(m+1)\times SO(n+1)$ rotations. Notice that
\bea\label{MNasPQ}
p=7-m-n.
\eea

The Poincare and $SO(m+1)\times SO(n+1)$ symmetries of (\ref{SSmetr}) lead to a partial separation of the Hamilton--Jacobi equation (\ref{HJone}) for geodesics (this equation is a counterpart of (\ref{GenAction})): 
\bea\label{GenSeparAct}
S=p_\mu x^\mu+S^{(m)}_{L_1}(y)+S^{(n)}_{L_2}({\tilde y})+R(r,\theta),
\eea
where $S^{(m)}_{L_1}$ and $S^{(n)}_{L_2}$ satisfy differential equations
\bea\label{HJSphere}
{h^{ij}}\frac{\d S^{(m)}_{L_1}}{\d y^i}\frac{\d S^{(m)}_{L_1}}{\d y^j}=L_1^2,\qquad
{{\tilde h}^{ij}}\frac{\d S^{(n)}_{L_2}}{\d {\tilde  y}^i}\frac{\d S^{(n)}_{L_2}}{\d{\tilde y}^j}=L_2^2,
\eea
and $L_1$, $L_2$ are angular momenta on the spheres. An explicit solution of equations (\ref{HJSphere}) is presented in appendix 
\ref{AppSphere}.

Substitution of (\ref{GenSeparAct}) into (\ref{HJone}) leads to the equation for $R(r,\theta)$:
\bea\label{TheHJ}
(\d_r R)^2+\frac{1}{r^2}(\d_\theta R)^2+
\frac{L_1^2}{r^2\cos^2\theta}+
\frac{L_2^2}{r^2\sin^2\theta}+
 p_\mu p^\mu H(r,\theta)=0.
\eea
We recall that $H(r,\theta)$ is a harmonic function describing the distribution of D branes, so away from the sources it satisfies 
the Laplace equation on the base of the ten--dimensional metric (\ref{SSmetr}):
\bea\label{LaplEqn}
\frac{1}{r^{m+n+1}}\d_r(r^{m+n+1}\d_r H)+\frac{1}{r^2\sin^{n}\theta\cos^{m}\theta}
\d_\theta(\sin^{n}\theta\cos^{m}\theta\d_\theta H)=0.
\eea

Let us assume that the massless HJ equation (\ref{TheHJ}) separates in coordinates $(x_1,x_2)$. In particular, this implies that the metric in $(x_1,x_2)$ must have a form \cite{Stackel,Eisenhart}
\bea\label{StackMetr}
dr^2+r^2d\theta^2=A(x_1,x_2)\left[e^{g_1}(dx_1)^2+
e^{g_2}(dx_2)^2\right],
\eea
where $g_1(x_1,x_2)$ and $g_2(x_1,x_2)$ satisfy the St\"ackel conditions:
\bea\label{Stackel}
\d_j\d_i g_l-\d_j g_l\,\d_i g_l+\d_j g_l\,\d_i g_j+
\d_i g_l\,\d_j g_i=0,\quad i\ne j
\eea
We wrote the St\"ackel conditions (\ref{Stackel}) for an arbitrary number of coordinates to relate to our discussion in section 
\ref{RedTo2D}, but in the present case ($k=2$) equation 
(\ref{Stackel}) gives only two relations ($i=1$, $j=2$, $l=1,2$):
\bea\label{Stack2D}
\d_2\d_1 g_1+\d_2 g_1\,\d_1 g_2=0,\qquad
\d_2\d_1 g_2+\d_1 g_2\,\d_2 g_1=0
\eea
In particular, we find that
\bea
g_1-g_2=f_1(x_1)-f_2(x_2),
\eea
so by adjusting function $A$ in (\ref{StackMetr}) we can set
\bea
g_1=f_1(x_1),\qquad g_2=f_2(x_2).
\eea
We can further redefine variables, 
$x_1\rightarrow {\tilde x}_1(x_1)$, $x_2\rightarrow {\tilde x}_2(x_2)$  to set $g_1=g_2=0$, at least locally.\footnote{Notice that separation in variables $(x_1,x_2)$ implies separation in 
$({\tilde x}_1,{\tilde x}_2)$.} Introducing $x={\tilde x}_1$, $y={\tilde x}_2$, we rewrite (\ref{StackMetr}) as 
\bea\label{RThAxY}
dr^2+r^2d\theta^2=A(x,y)\left[dx^2+dy^2\right].
\eea
To summarize, we have demonstrated that in two dimensions the St\"ackel conditions (\ref{Stackel}) imply that coordinates $x_i$ separating the HJ equation are essentially the same as the conformally--Cartesian coordinates $(x,y)$ in (\ref{RThAxY})\footnote{Although any two--dimensional metric can be written as $A[(d{\tilde x}_1)^2+(d{\tilde x}_2)^2]$ in \textit{some} coordinates, a priori the HJ equation does not have to separate in $({\tilde x}_1,{\tilde x}_2)$. It is the St\"ackel conditions (\ref{Stack2D}) that guarantee that any set of separable coordinates can be rewritten in a conformally--flat form without destroying the separation.}. In higher dimensions, conditions (\ref{Stackel}) are less stringent than the requirement for coordinates $x_i$ to be conformally--Cartesian: for example, conditions (\ref{Stackel}) are satisfied by spherical coordinates that have
\bea
g_1=0,\quad g_2=2\ln  x_1,\quad g_2=2\ln (x_1\sin x_2),
\eea
but there is no change of coordinates of the form 
${\tilde x}_i(x_i)$ that allows one to write 
\bea
(dx_1)^2+x_1^2dx_2^2+[x_1\sin x_2]^2 (dx_3)^2=
A\sum (d{\tilde x}_i)^2.
\eea
Moreover, for $k>2$, a relation 
\bea\label{Jul14}
\sum_{i=1}^k (dr_i)^2=A\sum_{i=1}^k (d{\tilde x}_i)^2
\eea
implies that $A$ must be equal to constant, so if conformally--Cartesian coordinates ${\tilde x}_i$ separate the HJ equation, then 
${\tilde x}_i$ must be Cartesian\footnote{To see this, one has to evaluate the Riemann tensor for both sides of (\ref{Jul14}).}.

Returning to $k=2$, we will now find restrictions on $A(x,y)$ and $H(r,\theta)$. First we define ${\tilde R}(x,y)$ by
\bea
{\tilde R}(x,y)\equiv R(r,\theta).
\eea
Then equation (\ref{RThAxY}) can be used to rewrite (\ref{TheHJ}) in terms of ${\tilde R}$, and we find the necessary and sufficient conditions for the Hamilton--Jacobi equation (\ref{TheHJ}) to be separable:
\bea\label{XYdiff}
&&(\d_r R)^2+\frac{1}{r^2}(\d_\theta R)^2=\frac{1}{A(x,y)}\left[(\d_x {\tilde R})^2+(\d_y {\tilde R})^2\right],
\\
\label{XYalg}
&&\frac{L_1^2}{r^2\cos^2\theta}+
\frac{L_2^2}{r^2\sin^2\theta}
-M^2 H(r,\theta)=\frac{1}{A(x,y)}\left[U_1(x)+U_2(y)\right].
\eea
Here $M$ is an effective mass in $(p+1)$ dimensions defined by
\bea
M^2=-p_\mu p^\mu.
\eea

The construction of the most general harmonic function 
$H(r,\theta)$ that admits the separation of variables (\ref{XYdiff})--(\ref{XYalg}) will be performed in three steps:
\begin{enumerate}
\item{Determine the restrictions on function $A(x,y)$ imposed by equation (\ref{XYdiff}).}
\item{Use equation (\ref{XYalg}) to find $H(r,\theta)$ corresponding to a given $A(x,y)$.}
\vskip -1.5cm
\bea\label{ThreeSteps}
\eea
\item{\vskip -0.2cm Use the Laplace equation (\ref{LaplEqn}) to find further restrictions on $A(x,y)$.}
\end{enumerate}
To implement the first step, it is convenient to introduce complex variables
\bea\label{defCmplVar}
z=x+iy,\qquad w=\ln \frac{r}{l}+i\theta.
\eea  
Here $l$ is a free parameter which has dimension of length. 
Rewriting equation (\ref{XYdiff}) in terms of complex coordinates, 
\bea\label{FromZtoW}
\d_w R\d_{\bar w}R=\frac{l^2e^{w+{\bar w}}}{A}\d_z {\tilde R}\d_{\bar z}{\tilde R}.
\eea
we conclude  that\footnote{The alternative solution, $\d_z{\tilde R}=\d_{\bar z}{\tilde R}=0$, leads to ${\tilde R}=R=\mbox{const}$, which does not solve (\ref{TheHJ}).}
\bea
\frac{\d z}{\d w}\frac{\d { z}}{\d {\bar w}}=0,
\eea
so $z(w,{\bar w})$ is either holomorphic or anti--holomorphic. Without loss of generality we assume 
that
\bea\label{holomZ}
z=h(w),
\eea 
then equation (\ref{FromZtoW}) gives an expression for $A(x,y)$:
\bea
A=\frac{l^2e^{w+{\bar w}}}{|h'|^2}=\frac{r^2}{|h'|^2}
\eea

The second step amounts to rewriting equation (\ref{XYalg}) as
\bea\label{HarmRTheta}
H(r,\theta)=
\frac{1}{M^2}\left[
\frac{L_1^2}{r^2\cos^2\theta}+
\frac{L_2^2}{r^2\sin^2\theta}-
\frac{|h'|^2}{r^2}\left[
U_1\left(\frac{h+{\bar h}}{2}\right)+
U_2\left(\frac{h-{\bar h}}{2i}\right)\right]
\right].
\eea

Implementation of the third step amounts to finding expressions for $h(w)$ and $U_1(x)$, $U_2(y)$ which are consistent with Laplace equation (\ref{LaplEqn}) for function $H$. Physically interesting configurations correspond to branes distributed in a compact spacial region, so at large values of $r$ function $H$ behaves as 
\bea\label{LeadHarm}
H=a+\frac{Q}{r^{7-p}}+O(r^{p-8}).
\eea
Here $Q$ is the total brane charge, and $a$ is a parameter, which can be set to zero for asymptotically flat space, and which is equal to zero for the near--horizon geometry of branes (cf. equation (\ref{HrmFncSph})). Keeping only the two leading terms in (\ref{LeadHarm}), we conclude that the Hamilton--Jacobi equation (\ref{TheHJ}) separates in coordinates $(r,\theta)$, and the solution is  similar to the first example discussed in section \ref{Examples}. The subleading corrections in (\ref{LeadHarm}) obstruct the separation in spherical coordinates, but for some harmonic functions $H$ the Hamilton--Jacobi equation (\ref{TheHJ}) separates in a different coordinate system, as in the second example discussed in section \ref{Examples}. Notice that at large values of $r$ the new coordinate system must approach spherical coordinates since (\ref{LeadHarm}) depends only on $r$, and for elliptic coordinates such asymptotic reduction was given by equation (\ref{AsympEll}). 

In appendix \ref{AppPrcdr} we construct the most general expressions for $h(w)$, $U_1(x)$, $U_2(y)$ by starting with asymptotic relations (\ref{LeadHarm}) and
\bea\label{BoundCond}
z=w+O\left(\frac{l}{r}\right),
\eea
and writing expansions in powers of $l/r$. Notice that these asymptotics lead to the unique value of $l$ for a given configuration of branes (for example, $l=d/2$ in (\ref{ElliptCase})). Requiring that function (\ref{HarmRTheta}) satisfies the Laplace equation 
(\ref{LaplEqn}), the boundary condition (\ref{LeadHarm}), and remains regular at sufficiently large $r$, we find three possible expressions for $h$ and $U_1$, $U_2$:\footnote{To avoid unnecessary complications in (\ref{TheSolnMthree})--(\ref{TheSoln67}) we set $a=0$ in these expressions, but constant $a$ can be added to the harmonic function without destroying the separation (\ref{XYalg}). This leads to minor changes in $U_1$ and $U_2$.}
\bea
\label{TheSolnMthree}
&&\hskip -1.4cm \mbox{I}:h(w)=w,\quad
H=\frac{Q}{r^{m+n}},\quad
{U}_1(r)=-\frac{QM^2}{r^{m+n-2}},\quad 
{U}_2(\theta)=\frac{L_1^2}{\cos^2 \theta}+\frac{L_2^2}{\sin^2 \theta}.\\
&&\phantom{\frac{e^x}{r^t}}\nonumber\\
\label{TheSolnMone}
&&\hskip -1.4cm \mbox{II}:h(w)=\ln\left[\frac{1}{2}\left\{e^w+\sqrt{e^{2w}-4}\right\}\right],\quad
H=\frac{1}{(\cosh^2 x-\cos^2 y)}
\frac{\tilde Q}{\sinh^{n-1}x
\cosh^{m-1}x}.
\nonumber\\
&&\hskip -1.2cm{U}_1(x)=-\frac{{\tilde Q}(Md)^2}{\sinh^{n-1}x
\cosh^{m-1}x}-\frac{L_1^2}{\cosh^2 x}+
\frac{L_2^2}{\sinh^2 x},\quad 
{U}_2(y)=\frac{L_1^2}{\cos^2 y}+\frac{L_2^2}{\sin^2 y}.\\
&&\phantom{\frac{e^x}{r^t}}\nonumber\\
\label{TheSolnMtwo}
&&\hskip -1.4cm \mbox{III}:h(w)=\ln\left[\frac{1}{2}\left\{e^w+\sqrt{e^{2w}+4}\right\}\right],\quad
H=\frac{1}{(\cosh^2 x-\sin^2 y)}
\frac{\tilde Q}{\sinh^{m-1}x
\cosh^{n-1}x}.
\nonumber\\
&&\hskip -1.2cm{U}_1(x)=-\frac{{\tilde Q}(Md)^2}{\sinh^{m-1}x
\cosh^{n-1}x}-\frac{L_2^2}{\cosh^2 x}+
\frac{L_1^2}{\sinh^2 x},\quad 
{U}_2(y)=\frac{L_1^2}{\cos^2 y}+\frac{L_2^2}{\sin^2 y}.
\eea
~\\
The derivation of these constraints is presented in appendix \ref{AppPrcdr}. 

For six-- and seven--branes\footnote{Condition (\ref{MNasPQ}) as well as restrictions $m,n\ge 0$ imply that ansatz (\ref{SSmetr}) covers only $p\le 7$} (i.e., for $m+n<2$), harmonic function is slightly more general:
\bea\label{TheSoln67}
\mbox{I}:&&H=-\frac{1}{M^2r^2}\left[\frac{C_1}{r^{m+n-2}}+\frac{C_3}{\sin^{n-1}\theta\cos^{m-1}\theta}\right.\nonumber\\
&&\qquad\qquad\qquad\left.
+\frac{C_4}{\sin^{n-1}\theta} F\left(\frac{m-1}{2},
\frac{3-n}{2},\frac{m+1}{2};\cos^2\theta\right)\right]\nonumber
 \nonumber\\
\mbox{II}:&&H=\frac{1}{(\cosh^2 x-\cos^2 y)}\left[
\frac{\tilde Q}{\sinh^{n-1}x
\cosh^{m-1}x}+\frac{P}{\sin^{m-1}y\cos^{n-1}y}\right],\\
\mbox{III}:&&H=\frac{1}{(\cosh^2 x-\sin^2 y)}\left[
\frac{\tilde Q}{\sinh^{m-1}x
\cosh^{n-1}x}+\frac{P}{\cos^{m-1}y\sin^{n-1}y}\right].\nonumber
\eea
The terms proportional to $P$ are ruled out by the boundary condition (\ref{LeadHarm}) for $p<6$, but they are allowed for $p=6,7$. 

\bigskip

Notice that case III can be obtained from case II by interchanging the spheres $S^m$  and $S^n$, so, without loss of generality, we can focus on solutions I and II. Case I corresponds to spherical coordinates, and case II corresponds to elliptic coordinates discussed in the second example of section \ref{Examples}: as demonstrated in appendix \ref{AppPrcdr}, expression (\ref{TheSolnMone}) for $h(w)$,
\bea\label{HolomElpt}
h(w)=\ln\left[\frac{1}{2}\left\{e^w+\sqrt{e^{2w}-4}\right\}\right],
\eea
is equivalent to 
\bea\label{ElliptCase}
\cosh x=\frac{\rho_++\rho_-}{2d},\quad \cos y=\frac{\rho_+-\rho_-}{2d},\quad
\rho_\pm=\sqrt{r^2+d^2\pm 2rd\cos\theta},\quad
d=2l.
\eea
Thus $(x,y)$ are equivalent to the elliptic coordinates $(\xi,\eta)$ defined by (\ref{DefEllCoord}). For future reference, we rewrite the metric (\ref{SSmetr}) in terms of $x$ and $y$ 
(see (\ref{metrAA})):
\bea\label{SSmetrAA}
ds^2&=&\frac{1}{\sqrt{H}}\eta_{\mu\nu}dx^\mu dx^\nu+
\sqrt{H}ds_{base}^2,
\\
ds_{base}^2&=&
4l^2\left[(\cosh^2x-\cos^2y)(dx^2+dy^2)+
\sinh^2 x\sin^2 y d\Omega_{n}^2+
\cosh^2 x\cos^2 y d\Omega_{m}^2\right].\nonumber
\eea

\subsection{Properties of the brane sources}\label{Sources}

In section \ref{SectPrcdr} we have classified the geometries which lead to separable Hamilton--Jacobi equations for geodesics. The Laplace equation (\ref{LaplEqn}) played an important role in our construction, but this equation is only satisfied away from the sources. In this subsection we will analyze the solutions (\ref{TheSolnMthree})--(\ref{TheSolnMtwo}) to find the sources of the Poisson equation and to identify the corresponding distribution of branes. As already discussed in section \ref{Examples}, the spherically symmetric distribution (\ref{TheSolnMthree}) corresponds to a single stack of Dp branes. 

Since solution (\ref{TheSolnMtwo}) can be obtained from (\ref{TheSolnMone}) by interchanging $S^m$ and $S^n$, it is sufficient to discuss only (\ref{TheSolnMone}). Function $H$ defined by (\ref{TheSolnMone}) becomes singular at
\bea\label{SingOne}
x=0,\quad y=0
\eea
for all values of $(m,n)$, at\footnote{Recall that coordinate $\theta$ in the metric (\ref{SSmetr}) is bounded by $\pi$ if $m=0$ or by $\pi/2$ if $m>0$, and ranges of $y$ and $\theta$ are the same.}
\bea\label{SingTwo}
x=0,\quad y=\pi\mbox{\ \ if\ \ }m=0,
\eea
and at 
\bea\label{SingThree}
x=0\mbox{\ \ if\ \ }n>1.
\eea
The first condition (\ref{SingOne}) implies that $e^z=e^{x+iy}=1$, and since $z=h(w)$,  it can be rewritten as
\bea
e^w+\sqrt{e^{2w}-4}=2
\eea
using $h(w)$ from (\ref{TheSolnMone}). Solving this equation and recalling the definition of $w$ (\ref{defCmplVar}), we find the first singular locus, which is present for all values of $(m,n)$:
\bea\label{EndPoints}
r=2l,\quad \theta=0.
\eea

For $m=0$, we have an additional locus (\ref{SingTwo}), and repeating the steps above, we find a counterpart of (\ref{EndPoints}):
\bea\label{EndPointOne}
m=0:\quad r=2l,\quad \theta=\pi.
\eea
Equations (\ref{EndPoints}) and (\ref{EndPointOne}) describe two point--like sources, and we
have already encountered these points in the original elliptic coordinates discussed in section \ref{Examples}. 
In the remaining part of this section we will focus on $m>0$. 

For $n=0,1$, the $m$--dimensional sphere described by (\ref{EndPoints}) is the only singularity of the harmonic function, and for $n>1$ there is an additional locus given by (\ref{SingThree}). To formulate (\ref{SingThree}) in terms of $r$ and $\theta$, we first use the expression 
(\ref{TheSolnMone}) for $h(w)$, to rewrite  (\ref{SingThree}) as
\bea\label{Jun11}
x=\mbox{Re}\left\{\ln\left[\frac{1}{2}\left\{e^w+\sqrt{e^{2w}-4}\right\}\right]\right\}=0
\eea
The last relation can be rewritten as
\bea
\left|e^w+\sqrt{e^{2w}-4}\right|=2\quad
\Rightarrow\quad
\left|re^{i\theta}+\sqrt{r^2e^{2i\theta}-4l^2}\right|=2l,
\eea
so there must exist an angle $\psi$, such that
\bea
r+\sqrt{r^2-4l^2e^{-2i\theta}}=2le^{i\psi}.
\eea
Solving the last equation for $r$, we find
\bea
r=l(e^{i\psi}+e^{-2i\theta-i\psi})
\eea
The right--hand side of the last equation must be real, 
this implies that (\ref{SingThree}) is equivalent to
\bea\label{Rpsi}
n>1:\qquad\theta=0,\qquad r=2l\cos\psi
\eea
Substituting this relation into (\ref{ElliptCase}) and recalling that $d=2l$, we conclude that $\psi=y$. 

\begin{figure}
\begin{minipage}[h!]{0.4\linewidth}
\center{\includegraphics[width=1\linewidth]{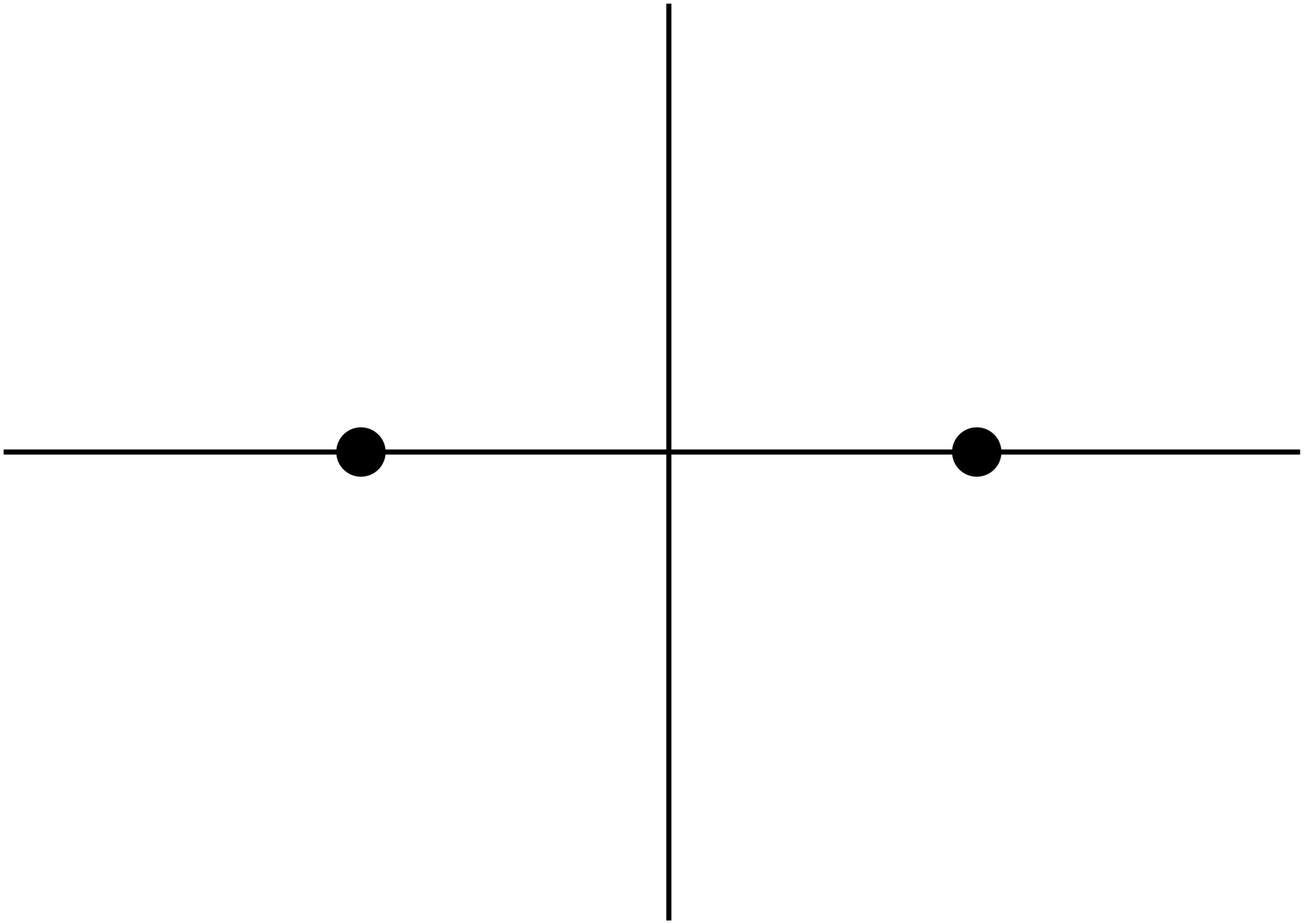}} (a) $m=0,n\le1$ \\
\end{minipage}
\hfill
\begin{minipage}[h!]{0.4\linewidth}
\center{\includegraphics[width=1\linewidth]{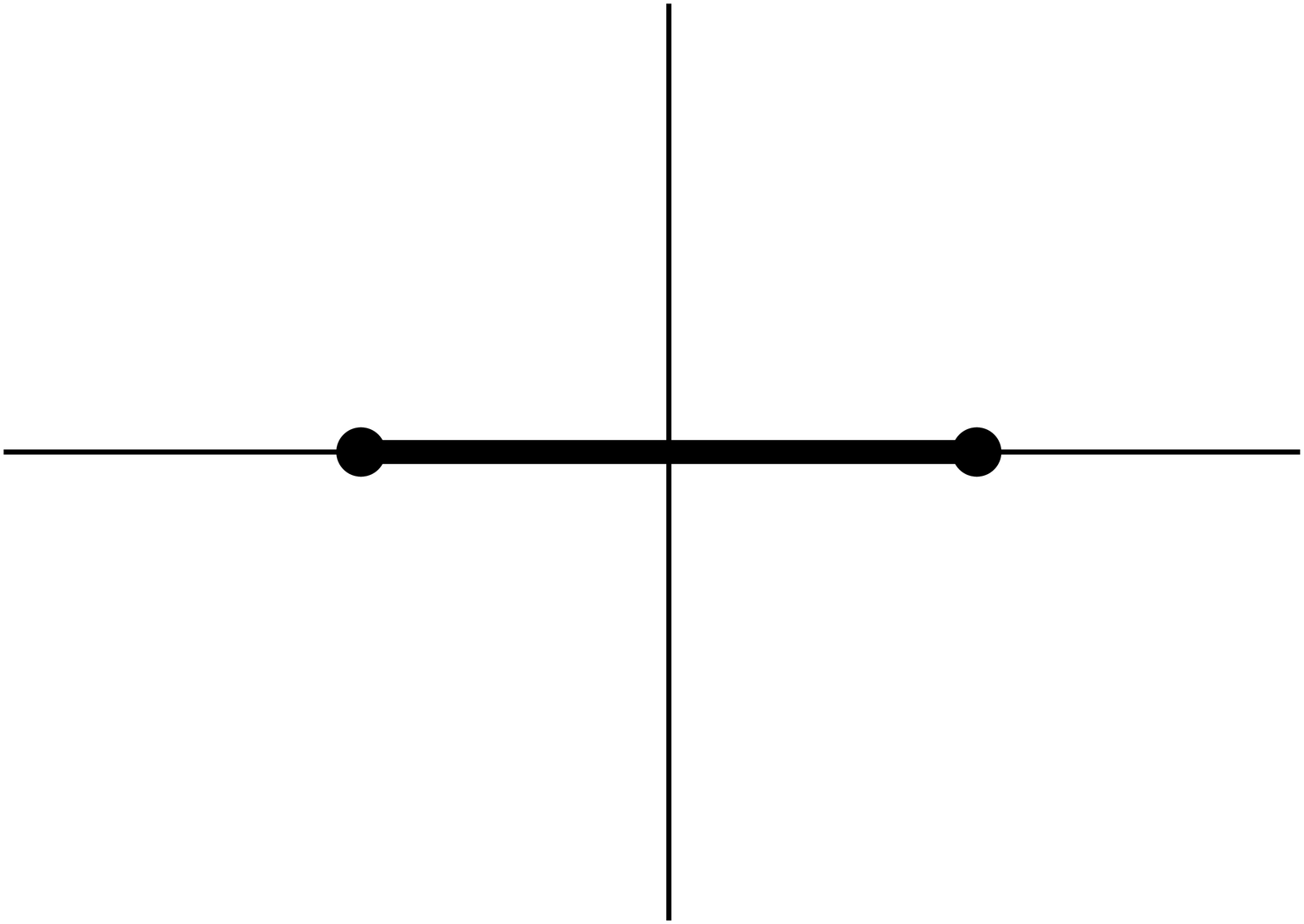}} (b) $m=0,n>1$ \\
\end{minipage}
\begin{center}
\hskip 1.2cm
\end{center}
\begin{minipage}[h!]{0.4\linewidth}
\center{\includegraphics[width=1\linewidth]{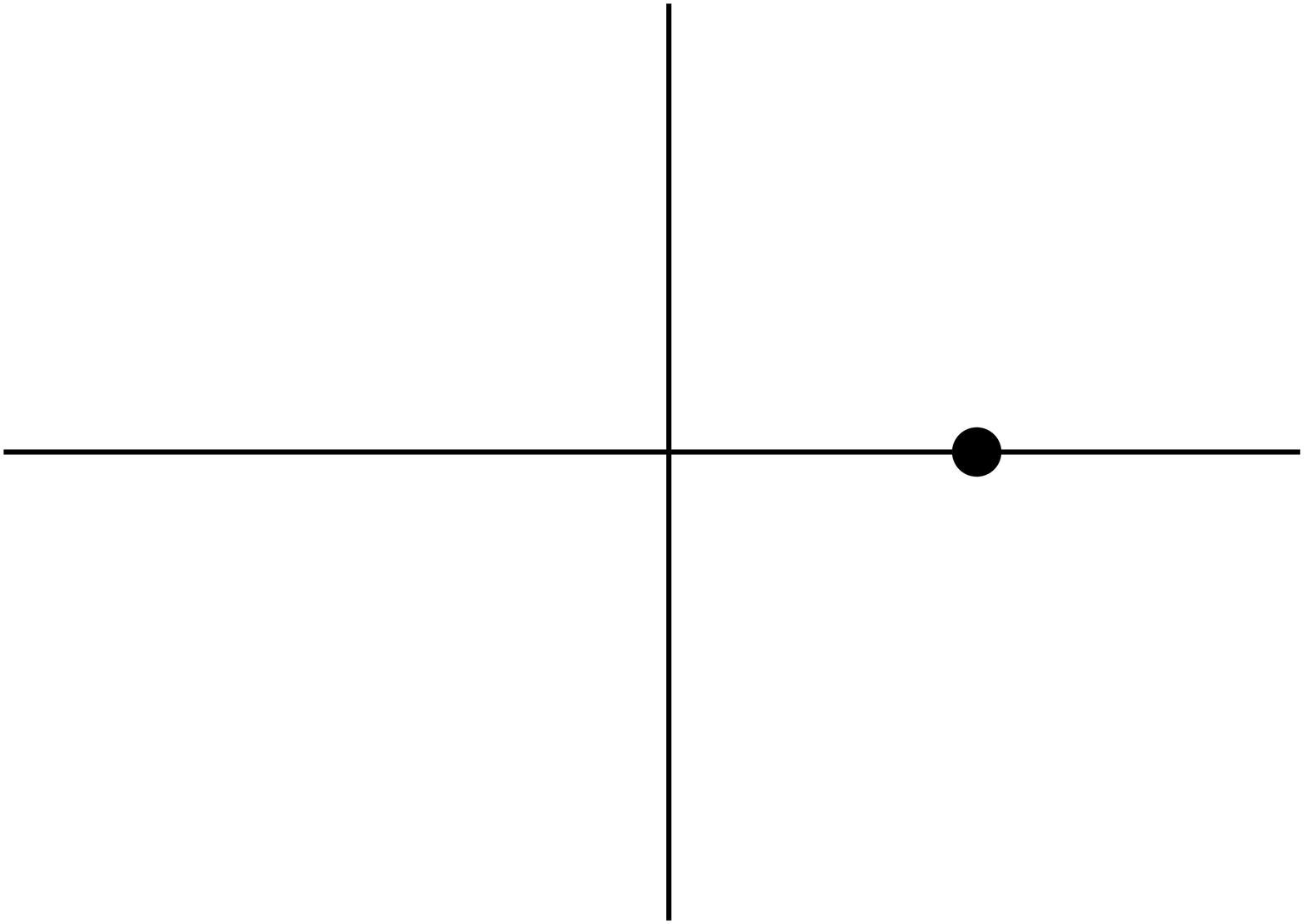}} (c) $m>0,n\le1$ \\
\end{minipage}
\hfill
\begin{minipage}[h!]{0.4\linewidth}
\center{\includegraphics[width=1\linewidth]{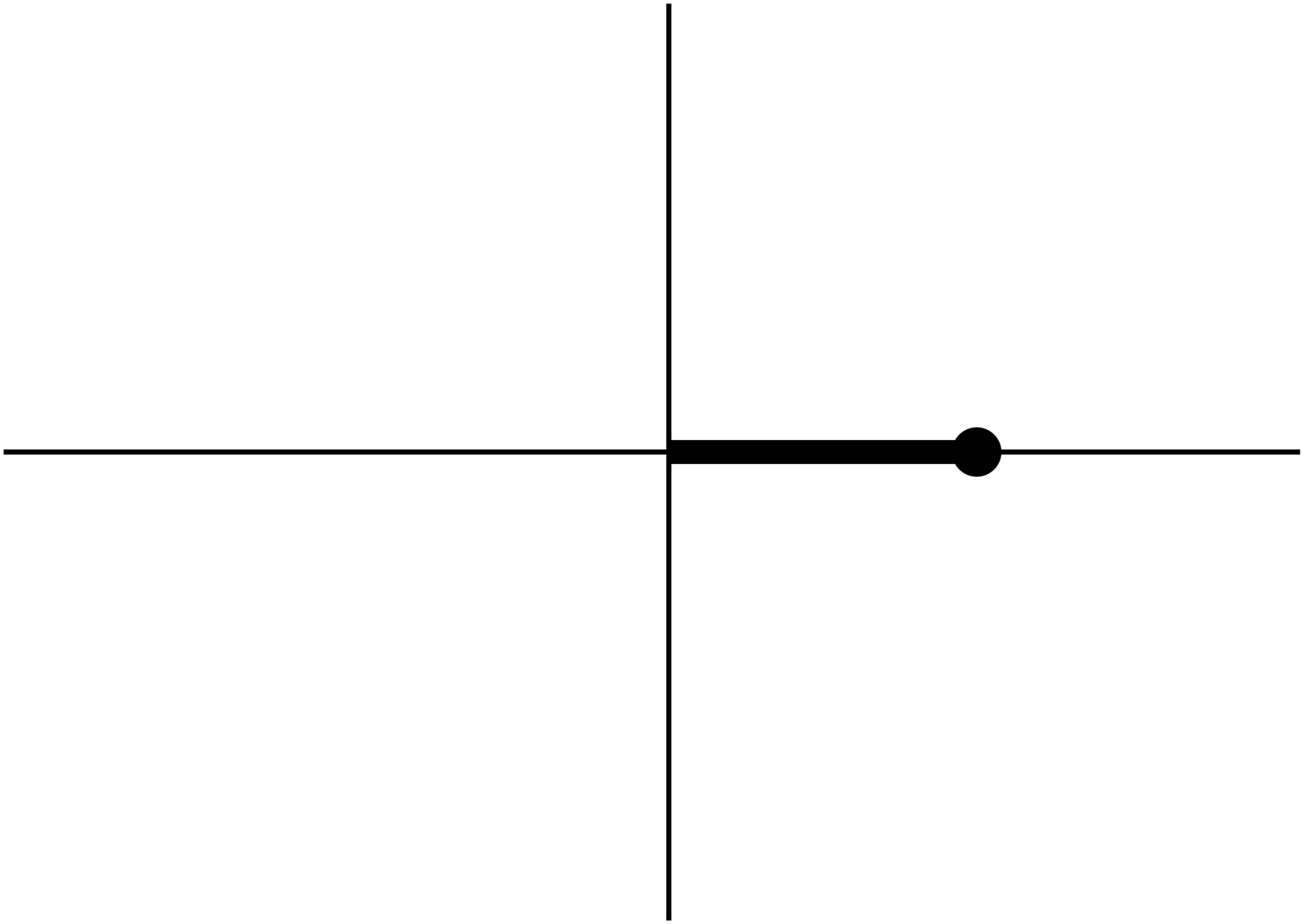}} (d) $m>0,n>1$ \\
\end{minipage}
\caption[Distribution of branes with integrable geodesics]{Distribution of branes for various values of  $(m,n)$.}
\label{Fig:Sources}
\end{figure}

To give a geometric interpretation of (\ref{Rpsi}), we recall that, for $m>0$, $y$ ranges from zero to $\pi/2$, so (\ref{Rpsi}) describes a line connecting $r=0$ with $r=2l$. The geometry (\ref{SSmetr}) has an $m$-dimensional sphere attached to every of this line, so the singular locus (\ref{Rpsi}) has a topology of $(m+1)$--dimensional disk. For $m=0$, 
$y$, and $\theta$ range from zero to $\pi$, so (\ref{Rpsi}) represents a line connecting two singular points (\ref{EndPoints}), 
(\ref{EndPointOne}). The pictorial representation of singular loci in $(r,\theta)$ plane is given in figure \ref{Fig:Sources}.

\bigskip

\noindent
We will now combine (\ref{EndPoints}) and (\ref{Rpsi}) to analyze the brane distribution for $m>0$.

\begin{enumerate}[(a)]
\item{$n=0$.\\
In this case, the $m$--dimensional sphere described by (\ref{EndPoints}) is the only singularity of the harmonic function, and in the vicinity of this singularity we find
\bea\label{Jun16}
H=\frac{{\tilde Q}x}{x^2+y^2}+regular
\eea
To give a geometric interpretation of this expression, we consider the base metric (\ref{SSmetr}) in the vicinity of singularity (\ref{EndPoints}):
\bea\label{Jul15}
ds^2_{9-p}\approx (2l)^2d\Omega_m^2+dr^2+(2l)^2d\theta^2.
\eea
It is convenient to introduce polar coordinates $(R,\Phi)$ by
\bea\label{RPhiCrd}
r=2l+R\cos\Phi,\quad 2l\theta=R\sin\Phi.
\eea
Recalling that $\theta$ varies from $-\frac{\pi}{2}$ to $\frac{\pi}{2}$ when $n=0$, we conclude that $\Phi\in[0,2\pi)$ (or $\Phi\in[-\pi,\pi)$, see below), as long as $R$ remains small. Then metric (\ref{Jul15}) takes the standard form
\bea\label{Jun16ac}
ds^2_{9-p}\approx (2l)^2d\Omega_m^2+dR^2+R^2d\Phi^2.
\eea
To write $(x,y)$ in terms of $(R,\Phi)$, we use the real form (\ref{ElliptCase}) of the holomorphic map (\ref{TheSolnMone}). In particular, for small 
$x$ and $y$ we find
\bea\label{Jun16b}
&&\hskip -1.7cm x^2+y^2\approx 2(\cosh x-\cos y)=\frac{2\rho_-}{2l}=
\frac{2}{2l}\sqrt{(r-2l)^2+4rl(1-\cos\theta)}
\approx\frac{R}{l}\\
&&\hskip -1.7cm x^2\approx 2(\cosh x-1)\approx\frac{1}{2l}\left[(r-2l)+\sqrt{(r-2l)^2+4rl(1-\cos\theta)}\right]
\approx\frac{R}{l}\cos^2\frac{\Phi}{2}.
\nonumber
\eea
Extraction of the square root from the last expression should be done carefully: positivity of the harmonic function (\ref{Jun16}) (or (\ref{TheSolnMone})) requires that $x>0$. Thus we can write
\bea\label{Jun16a}
x=\sqrt{\frac{R}{l}}\cos\frac{\Phi}{2},
\eea 
as long as $\Phi\in[-\pi,\pi)$. The range $\Phi\in[0,2\pi)$ is equivalent from the point of view of (\ref{RPhiCrd}), but it leads to a more complicated counterpart of (\ref{Jun16a}), and it will not be explored further. Substitution of (\ref{Jun16b}) and (\ref{Jun16a}) into equation (\ref{Jun16}) leads to a simple expression for the harmonic function in the vicinity of the sources:
\bea\label{Jun16ab}
H={\tilde Q}\sqrt{\frac{l}{R}}\cos\frac{\Phi}{2}+regular
\eea
We recall that $R=0$ corresponds to the $m$--dimensional sphere (\ref{EndPoints}) with radius $2l$. Notice that this expression never becomes negative since 
$\Phi\in[-\pi,\pi)$. 
}
\item{$n=1$.\\
Here again, the $m$--dimensional sphere described by (\ref{EndPoints}) is the only singularity of the harmonic function, and in the vicinity of this singularity we find
\bea\label{Jun16e}
H=\frac{A}{x^2+y^2}+regular=\frac{Al}{R}+regular
\eea
As before, we used (\ref{Jun16b}) to rewrite the harmonic function in terms of coordinates (\ref{RPhiCrd}), and in this case
\bea
ds^2_{9-p}\approx (2l)^2d\Omega_m^2+dR^2+R^2d\Phi^2+
R^2\sin^2\Phi d\phi^2,\qquad 0\le\Phi<\pi\nonumber
\eea
Formula (\ref{Jun16e}) gives a standard harmonic function in three dimensions transverse to $S^m$, so this configuration corresponds to D--branes uniformly distributed over $S^m$.
}
\item{$n>1$.\\
In this case the singularity consists of a line (\ref{Rpsi}) connecting $r=0$ and (\ref{EndPoints}) with sphere $S^m$ fibered over it. In the vicinity of $\theta=0$, the base of the metric (\ref{SSmetr}) becomes
\bea\label{Jun18}
ds^2_{9-p}\approx dr^2+r^2d\Omega_m^2 +r^2d\theta^2+
r^2\theta^2d\Omega_n^2,
\eea
and the locus (\ref{Rpsi}) is an $m+1$--dimensional ball with metric
\bea
ds^2_{sing}\approx  dr^2+r^2d\Omega_m^2,\qquad 0\le r<2l
\eea
The singularity corresponds to $x=0$ (recall (\ref{SingOne}) and (\ref{SingThree})), and the leading contribution to the harmonic function (\ref{TheSolnMone}) for small $x$ is
\bea\label{Jun18b}
H=\frac{1}{x^2+\sin^2 y}
\frac{\tilde Q}{x^{n-1}}+\dots
\eea
For small $x$, metric (\ref{SSmetrAA}) becomes
\bea\label{Jun18a}
ds^2_{base}\approx 4l^2\cos^2 y d\Omega_m^2+
4l^2(x^2+\sin^2 y)\left[dx^2+dy^2\right]+
4l^2x^2\sin^2 y d\Omega_n^2.
\eea
Away from $y=0$, we can neglect $x^2$ in comparison with $\sin^2y$, then function (\ref{Jun18b}) describes the Coulomb potential produced by D--branes uniformly distributed over $S^m$. Rewriting (\ref{Jun18a}) as
\bea\label{Jun18ab}
ds^2_{base}\approx dR^2+R^2 d\Omega_m^2+
(4l^2-R^2)\left[dx^2+x^2d\Omega_n^2\right],
\eea 
we find the charge density $\rho$:
\bea\label{DensNgr1}
H&=&\frac{\rho}{[(4l^2-R^2)x^2]^{(n-1)/2}}+\dots,\nonumber\\
\rho&=&4l^2{\tilde Q}(4l^2-R^2)^\frac{n-3}{2}=
(2l)^{n-1}{\tilde Q}\sin^{n-3}y
\eea
As expected, the charge density vanishes on the boundary (\ref{EndPoints}) of the ball, where $y=0$. 
}
\end{enumerate}
To summarize, we found that for $n\le 1$ the brane sources are localizes on the $m$--dimensional sphere, they produce a Coulomb potential (\ref{Jun16e}) for $n=1$ and a potential (\ref{Jun16ab}) with a fractional power of the radial coordinate $R$ for $n=0$. For $n>1$, the branes are located on a line connecting $r=0$ and (\ref{EndPoints}) with sphere $S^m$ fibered over it ($(R,\Omega_m)$ subspace of (\ref{Jun18ab})). These sources produce a Coulomb potential in $(n+1)$ transverse directions with charge density (\ref{DensNgr1}).

\section{Beyond geodesics: wave equation, Killing tensors, and strings}
\label{SecKillWave}

The main goal of this chapter is identification of backgrounds which can potentially lead to integrable string theories. As discussed in the introduction, integrability can be ruled out by looking at relatively simple equations for the light modes of strings (massless particles), and in the last section we demonstrated that classical equations of motion for such particles are integrable only for the harmonic functions given by (\ref{TheSolnMthree})--(\ref{TheSolnMtwo}). It is natural to ask whether separability of the HJ equation (\ref{HJone}) in the elliptic coordinates (\ref{ElliptCase}) persists at the quantum level and whether it is related to some hidden symmetry of the system. In section \ref{SecWave} we analyze integrability of the wave equation, a quantum counterpart of (\ref{HJone}), and in section \ref{SecKill} we identify the Killing tensor responsible for the separation. Finally in section \ref{NonintStr} we investigate the question whether integrability of geodesics persists for finite size strings.

\subsection{Separability of the wave equation}
\label{SecWave}

In this subsection we will analyze the wave equation which governs dynamics of a minimally--coupled massless scalar:
\bea\label{WaveEqn}
\frac{1}{\sqrt{-g}}\partial_{M}\left( g^{MN}
\sqrt{-g}\partial_{N} \Psi \right)=0.
\eea
Most D$p$--branes generate a nontrivial dilaton, so the last equation would look differently in the string and in the Einstein frames\footnote{Unlike (\ref{WaveEqn}), the HJ equation (\ref{HJone}) is invariant under conformal rescaling of the metric, so the results of section \ref{SecGnrGds} are valid in both the string and the Einstein frames.}, and here we will focus on the most interesting case of the Einstein frame:
\bea\label{EinstFrame}
g^{(E)}_{MN}dx^Mdx^N=
e^{-\Phi/2}\left[
\frac{1}{\sqrt{H}}\eta_{\mu\nu}dx^{\mu}dx^{\nu}+
\sqrt{H}ds^2_{base}\right],\quad e^{2\Phi}=H^{(3-p)/2}.
\eea
Since the HJ equation (\ref{HJone}) arises in the eikonal approximation of (\ref{WaveEqn}), the arguments presented in sections 
\ref{SecGnrGds} imply that (\ref{WaveEqn}) is not integrable unless the metric $ds^2_{base}$ has the form (\ref{SSmetrAA}) and $H$ is given by (\ref{TheSolnMone})\footnote{Solution (\ref{TheSolnMthree}) leads to a trivial separation in spherical coordinates, and solution 
(\ref{TheSolnMtwo}) reduces to (\ref{TheSolnMone}).}, although these conditions are not sufficient for integrability of (\ref{WaveEqn})\footnote{For example, as we will see below, equation (\ref{WaveEqn}) does not separate if $g_{MN}$ is a metric in the string frame, although it still reduces to (\ref{HJone}) in the eikonal approximation.}. 

To write the wave equation in the geometry (\ref{EinstFrame}), we recall the metric on the base (\ref{SSmetrAA}),
\bea\label{WWbase}
ds_{base}^2=(\cosh^2x-\cos^2y)(dx^2+dy^2)+\sinh^2x\sin^2y d\Omega_n^2+\cosh^2x\cos^2y d\Omega_m^2,
\eea
and introduce a convenient notation:
\bea
A=(\cosh^2x-\cos^2y),\quad X=\sinh^nx\cosh^mx,\quad
Y=\sin^ny\cos^my. 
\eea
Evaluating the determinant of the metric,
\bea\label{detEinst}
\sqrt{-g^{(E)}}= e^{-5\Phi/2}H^{2-p/2}X(x)Y(y)A,
\eea
and substituting (\ref{EinstFrame}), (\ref{WWbase}), (\ref{detEinst}) into (\ref{WaveEqn}), we find the wave equation:
\bea\label{WaveEinst}
&&H\partial_{\mu}\partial^{\mu}\Psi +
\frac{1}{\sinh^2x\sin^2y}\Delta_{\Omega_n}\Psi+\frac{1}{\cosh^2x\cos^2y}\Delta_{\Omega_m}\Psi\nonumber\\
&&+\frac{H^{(p-3)/2}e^{2\Phi}}{AXY}\left\{\partial_x \left[ H^{(3-p)/2}e^{-2\Phi}XY\partial_x \Psi \right] + \partial_y \left[ H^{(3-p)/2}e^{-2\Phi}XY \partial_y \Psi \right] \right\}=0
\eea
This equation separates since $H^{(3-p)/2}e^{-2\Phi}=1$ (see (\ref{EinstFrame})). Specifically, if we write\footnote{Here $Y_{k}(\Omega_m)$ and 
$Y_{l}(\Omega_n)$ are standard spherical harmonics with angular momenta $k$ and $l$. For example, $\Delta_{\Omega_n}Y_l(\Omega_n)=-l(l+n-1)Y_l(\Omega_n)$.}
\bea\label{MultSepar}
\Psi=e^{ip_\mu x^\mu}Y_{k}(\Omega_m)
Y_{l}(\Omega_n)F(x)G(y),
\eea
then (\ref{WaveEinst}) splits into two ordinary differential equations with separation constant $\Lambda$: 
\bea\label{SeprWave1}
\left[-\left(a\cosh^2 x+\frac{{\tilde Q}}{\sinh^{n-1}x\cosh^{m-1}x}\right)p_\mu p^\mu -
\frac{l(l+n-1)}{\sinh^2x}+\frac{k(m+k-1)}{\cosh^2x}\right]F\nonumber\\
+\frac{1}{X}\left[XF' \right]'=\Lambda F
\eea
\bea\label{SeprWave2}
ap_\mu p^\mu\cos^2 y -\left[
\frac{l(l+n-1)}{\sin^2y}+\frac{k(m+k-1)}{\cos^2y}\right]G+\frac{1}{Y}
\left[YG' \right]'=-\Lambda G
\eea
We used the expression for the harmonic function,
\bea\label{KGharm}
H=a+\frac{1}{(\cosh^2 x-\cos^2 y)}
\frac{{\tilde Q}}{\sinh^{n-1}x\cosh^{m-1}x}
\eea
found in section \ref{SecGnrGds} (see (\ref{TheSolnMone}), (\ref{TheSolnMtwo})). 

Let us now demonstrate that the wave equation (\ref{WaveEqn}) does not separate in the string metric unless 
$p=3$. The string--frame counterpart of (\ref{WaveEinst}) can be obtained by formally setting $e^{2\Phi}=1$ in that equation:
\bea\label{WaveStr}
&&H\partial_{\mu}\partial^{\mu}\Psi +
\frac{1}{\sinh^2x\sin^2y}\Delta_{\Omega_n}\Psi+\frac{1}{\cosh^2x\cos^2y}\Delta_{\Omega_m}\Psi\nonumber\\
&&+\frac{H^{(p-3)/2}}{AXY}\left\{\partial_x \left[ H^{(3-p)/2}XY\partial_x \Psi \right] + \partial_y \left[ H^{(3-p)/2}XY \partial_y \Psi \right] \right\}=0
\eea
Clearly, the multiplicative separation (\ref{MultSepar}) does not work for this equation unless $p=3$. Since coordinates $(x,y)$ are uniquely fixed by the discussion of the HJ equation (which is an eikonal limit of (\ref{WaveStr})) presented in section \ref{SecGnrGds}, to rule out the separation, it is sufficient to show that a substitution of
\bea\label{PsdMultSepar}
\Psi=e^{ip_\mu x^\mu}Y_{k}(\Omega_m)
Y_{l}(\Omega_n)F(x)G(y)P(x,y)
\eea
into (\ref{WaveStr}) does not lead to separate equations for $F$ and $G$ for any \textit{fixed} function $P(x,y)$. To demonstrate this, we perform such substitution and rewrite the result as
\bea\label{ToSeparate}
&&-A\left[Hp_{\mu}p^{\mu}+
\frac{l(l+n-1)}{\sinh^2x\sin^2y}+\frac{k(k+m-1)}{\cosh^2x\cos^2y}\right]\nonumber\\
&&\qquad+\frac{H^{(p-3)/2}}{P}\left\{
\frac{1}{X}\partial_x \left[ H^{(3-p)/2}X\partial_x P \right]+
\frac{1}{Y}\partial_y \left[ H^{(3-p)/2}Y\partial_y P \right]\right\}\nonumber\\
&&\qquad+\frac{1}{F}\left\{F'' + 
F'\partial_x \ln\left[ H^{(3-p)/2}X P^2 \right] \right\}\\
&&\qquad
+\frac{1}{G}\left\{G'' + 
G'\partial_y \ln\left[ H^{(3-p)/2}Y P^2 \right] \right\}=0\nonumber
\eea
Since $H$ and $P$ are fixed functions, the first two lines of (\ref{ToSeparate}) remain the same for all $F$ and $G$, so equation (\ref{ToSeparate}) does not separate unless its third line is only a function of $x$ and the forth line is only a function of $y$. This implies that
\bea\label{tmpWaveSep}
\d_x\d_y\ln\left[ H^{(3-p)/2}P^2 \right]=0\quad\Rightarrow
\quad P=H^{(p-3)/4}P_1(x)P_2(y).
\eea
Functions $P_1$ and $P_2$ can be absorbed into $F$ and $G$ (recall (\ref{PsdMultSepar})), so we set $P_1=P_2=1$. Direct substitution into (\ref{ToSeparate}) shows that the third line of that equation obstructs separation unless $p=3$.

To summarize, we have demonstrated that while the HJ equation is integrable in the elliptic coordinates (\ref{ElliptCase}), its quantum version may or may not be separable depending on the frame. In particular, the equation for the minimally--coupled massless scalar separates in the Einstein, but not in the string frame.

\subsection{Killing tensor}
\label{SecKill}

Separation the Hamilton--Jacobi and Klein--Gordon equations implies an existence of nontrivial conserved charges which are associated with symmetries of the background. In the simplest case, such symmetries are encoded in the Killing vectors, which correspond to invariance of the metric under reparametrization
\bea
x^M\rightarrow x^M+V^M(x),
\eea
where $V^M(x)$ satisfies the equation 
\bea
V_{M;N}+V_{N;M}=0.
\eea
This symmetry guarantees a conservation of the charge
\bea
Q_V=V^M\frac{\d S}{\d x^M},
\eea
and momenta $p_\mu$ appearing in (\ref{GenAction}) were examples of such charges. Although not every separation of variables can be associated with Killing vector (separation between $x$ and $y$ coordinates found in sections \ref{SecGnrGds} and \ref{SecWave} is our prime example), the general theory developed in 
\cite{Carter,KalMil1,KalMil2,Miller3} guarantees that any such separation is related to a symmetry of the background, which is encoded by a (conformal) Killing tensor. In this subsection we will discuss the conformal Killing tensor associated with separation (\ref{TheSolnMone}) and (\ref{SeprWave1})--(\ref{SeprWave2}).

The conformal Killing tensor of rank two satisfies equation \cite{Penrose}
\bea\label{CKTdef}
\nabla_{(M}K_{NL)}=\frac{1}{2}W_{(M}g_{NL)},
\eea
which is solved in appendix \ref{AppKill}. Here we will deduce the same solution by using the separation of variables found in section \ref{SecGnrGds}. A conformal Killing tensor $K_{MN}$ always implies that 
\bea\label{IntMtnKil}
I=K^{MN}\d_M S\d_N S
\eea
is an integral of motion of the massless HJ equation\footnote{A Killing tensor $K_{MN}$, which has $V_M=0$ in (\ref{CKTdef}),  implies that (\ref{IntMtnKil}) is an integral of motion of the massive HJ equation.}, so $K^{MN}$ can be extracted from the know separation. Going to the eikonal approximation in (\ref{WaveEinst}) with harmonic function (\ref{KGharm}), we find
\bea\label{WaveKill}
&&\left[a\cosh^2x+\frac{{\tilde Q}}{\sinh^{n-1}x\cosh^{m-1}x}\right]
\d_{\mu}S\d^{\mu}S+
\frac{1}{\sinh^2x}h^{ij}\d_i S\d_j S\nonumber\\
&&\qquad\qquad-
\frac{1}{\cosh^2x}{\tilde h}^{ij}{\tilde \d}_i S{\tilde\d}_j S
+\d_x S\d_x S\\
&&=-a\d_{\mu}S\d^{\mu}S-\frac{1}{\sin^2y}h^{ij}\d_i S\d_j S-\frac{1}{\cos^2y}{\tilde h}^{ij}{\tilde \d}_i S{\tilde\d}_j S
-\d_y S\d_y S\nonumber
\eea
For separable solutions, both sides of this equation must be constant, and identifying this constant with $-I$ in (\ref{IntMtnKil}), we find
\bea\label{Jul25}
K^{MN}p_Mp_N=ap_\mu p^\mu+\frac{1}{\sin^2y}h^{ij}p_ip_j+\frac{1}{\cos^2y}{\tilde h}^{ij}
{\tilde  p}_i {\tilde p}_j +p_yp_y
\eea
In appendix \ref{AppKill} this expression is derived in a geometrical way by solving the equation (\ref{CKTdef}) for the Killing tensor. 

\subsection{Non--integrability of strings}\label{NonintStr}

In section \ref{SecGnrGds} we have classified all supersymmetric configurations of Dp--branes that lead to integrable equations for null geodesics. Specifically, we demonstrated that a metric (\ref{PreGeom})--(\ref{FlatBase}) leads to a separable HJ equation (\ref{HJone}) if and only if it has the form 
\bea\label{IntGeodMetr}
ds^2=\frac{1}{\sqrt{H}}\eta_{\mu\nu}dx^{\mu}dx^{\nu}+\sqrt{H}(dr^2+r^2d\theta^2+r^2\cos^2\theta d\Omega_m^2 +r^2\sin^2\theta d\Omega_n^2)
\eea
with the harmonic function $H$ from (\ref{TheSolnMthree})--(\ref{TheSolnMtwo}). In this subsection we will investigate whether integrability of geodesics extends to strings with finite size. Our discussion will follow the logic presented in \cite{StepTs}, and to compare our results with ones from that paper we rewrite the metric in terms of a new function $f=H^{1/4}$ so the metric (\ref{IntGeodMetr}) becomes:
\bea\label{MetricStrInt}
ds^2=\frac{1}{f^2}\eta_{\mu\nu}dx^{\mu}dx^{\nu}+f^2(dr^2+r^2d\theta^2+r^2\cos^2\theta d\Omega_m^2+r^2\sin^2\theta d\Omega_n^2).
\eea
To demonstrate integrability of strings on a particular background, one has to find an infinite set of integrals of motions, and this ambitious problem has only been solved for very few geometries \cite{RoibPolch,Frolov}. However, to \textit{rule out} integrability of sigma model on a given background, it is sufficient to start with a particular solution and look at linear perturbations around it. If such linearized problem  has no integrals of motion, one concludes that the original system is not integrable. This approach has been used in \cite{Zayas,StepTs} to rule out integrability of strings on the conifold and on the asymptotically--flat geometry produced by a  single stack of Dp--branes. The analysis presented in this subsection is complimentary to  \cite{StepTs}: we still focus on the near--horizon limit (where strings are known to be integrable for a single stack), but allow a nontrivial distribution of sources. 

The equation for linear perturbations around a given solution of a dynamical system is known as Normal Variational Equation (NVE) \cite{ziglinMor}, and to determine whether NVE is integrable, one can use the Kovacic algorithm\footnote{The Kovacic algorithm is implemented in Maple and one can use the function kovacicsols to check integrability of particular physical systems.}\cite{Kovacic_alg}.
Thus to demonstrate that the string theory on a particular background is not integrable one needs to perform the following steps:
\begin{enumerate}
\item write down the equations of motion
\item compute the variational equations
\item choose a particular solution and consider the normal equations (NVE)
\item algebrize NVE (rewrite equations as differential equations with rational coefficients) and transform them to normal form to make NVE be suitable for using the Kovacic algorithm
\item apply the Kovacic algorithm to the obtained NVE, if it fails the system is non--integrable.
\end{enumerate}
Now we apply this method to check integrability of strings in the background (\ref{MetricStrInt}). We begin with looking at the Polyakov action \cite{PolAction, Polyakov}
\begin{equation}
S=-\frac{1}{4 \pi \alpha^{'}} \int d\sigma d\tau G_{MN}(X) \partial_a X^{M} \partial^{a} X^{N}.
\end{equation}
supplemented by the Virasoro constraints
\bea\label{Virasoroo}\label{Virasoro11}
G_{MN}\dot{X^M} X^{'N}=0,\\
\label{Virasoro21}
G_{MN}(\dot{X^M} \dot{X^N}+X^{'M} X^{'N})=0.
\eea

For a specific string ansatz on 2--sphere,
\begin{equation}\label{StringAnsatz}
x^{0}=t(\tau),\quad r=r(\tau), \quad \phi=\phi(\sigma), \quad \theta=\theta(\tau),
\end{equation}
the system has an effective Lagrangian density
\begin{equation}
{\mathcal L}=-f^{-2}\dot{t}^2+f^2\dot{r}^2+f^2r^2(-\sin^2\theta\phi^{'2}+\dot{\theta}^2).
\end{equation}
Equations of motion for cyclic variables $t,\phi$ lead to two integrals of motion ($E$ and $\nu$), and combining this with Virasoro constraint (\ref{Virasoro21}) we find
\begin{eqnarray}
\dot{t}&=&Ef^2,\nonumber\\
\phi'&=&\nu=const,\\
E^2&=&\dot{r}^2+r^2\dot{\theta}^2+\nu^2r^2\sin\theta^2.
\nonumber
\end{eqnarray}
These equations can be derived from the effective Lagrangian
\begin{equation}
L=f^2(\dot{r}^2+r^2\dot{\theta}^2-r^2\sin^2\theta\nu^2-E^2).
\end{equation}
Expanding around a particular solution,
\bea
 r=\frac{E}{\nu}\sin{\nu\tau},\quad \theta=\frac{\pi}{2},
\eea
we find the following NVE for $\eta\equiv \delta\theta$ (see appendix \ref{NonIntStrsDBr} for detail):
\begin{equation}\label{NVERes}
\eta^{''}+\left[ \frac{\ddot{r}}{\dot{r}^2}+2\left(\frac{f'_r}{f}+\frac{1}{r}\right) \right] \eta' - \left[ \frac{f^{''}_{\theta\theta}}{f} \frac{1}{r^2} -3\left(\frac{f'_{\theta}}{f}\right)^2\frac{1}{r^2}-\left(\frac{f'_{\theta}}{f}\right)^2\frac{\nu^2}{\dot{r}^2}-\frac{f^{''}_{\theta\theta}}{f}\frac{\nu^2}{\dot{r}^2}+\frac{\nu^2}{\dot{r}^2} \right]\eta=0.
\end{equation}

To proceed we need to choose a particular configuration corresponding to the specific function $f$. In section \ref{SecGnrGds} we have demonstrated that equations for geodesics are integrable only if function $H$ is given by 
(\ref{TheSolnMthree})--(\ref{TheSoln67}), and here we consider (\ref{TheSolnMone})  ignoring the $P$--term in (\ref{TheSoln67})\footnote{We will discuss NVE associated with this term in appendix \ref{AppNintGen}.}:
\bea\label{HGen}
H&=&f^4=\frac{\tilde{Q}}{\cosh^2x-\cos^2y}\sinh^{1-n}x\cosh^{1-m}x\nonumber\\&&=\frac{d^2\tilde{Q}}{\rho_+\rho_-}\left[ \frac{\rho_++\rho_-}{2d} \right]^{1-m}\left[ \left(\frac{\rho_++\rho_-}{2d}\right)^2-1\right]^{\frac{1-n}{2}},\\
\rho_\pm&=&\sqrt{r^2+d^2\pm 2rd\cos\theta}\nonumber.
\eea
Here we used the map (\ref{ElliptCase}) between the coordinates $(x,y)$ and $(\rho_{+},\rho_{-})$.
After carrying out all calculations one obtains the following NVE
\bea\label{NVEGen}
\eta^{''}&+&\frac{U}{D}\eta=0,\nonumber\\
U&=&E^4 \left[-d^4(n-1)^2-2 d^2 r^2 \left((m-3) n-5 m+n^2+2\right)-r^4
   (m+n-4) (m+n)\right] \nonumber\\
   &&+2 E^2 r^2 \nu^2 \Big[d^4 ((n-2) n+5)+2 d^2 r^2 (m (n-5)+(n-4) n+9)\nonumber\\
   &&+r^4 \left(m^2+2 m (n-3)+(n-6) n+4\right)\Big]-r^4 \nu^4 \Big[d^4 (n-3) (n+1)\\&&+2 d^2
   r^2 \left((m-5) n-5 m+n^2+4\right)+r^4 \left(m^2+2 m (n-4)+(n-8) n-4\right)\Big],\nonumber\\
D&=&16 r^2 \left(d^2+r^2\right)^2 [E^2-(r \nu)^2]^2.\nonumber
\eea
Application of the Kovacic algorithm to (\ref{NVEGen}) shows that the system is not integrable unless $d=0$, $m+n=4$ (this corresponds to AdS$_5\times$S$^5$). Detailed description of the method used in this section and complete calculations are presented in appendices \ref{NVEReview}--\ref{AppNintGen}.

To summarize, we have demonstrated that integrability of geodesics discussed in section \ref{SecGnrGds} does not persist for classical strings, and AdS$_5\times$S$^5$ is the only static background produced by a single type of D--branes placed on a flat base that leads to an integrable string theory\footnote{Analysis presented in this subsection does not rule out integrability on backgrounds containing NS--NS fluxes in addition to D--branes or on geometries produced by several types of branes, such as D$p$--D$(p+4)$ system discussed in the next section.}.

\section{Geodesics in static D$p$--D$(p+4)$ backgrounds}\label{DpDp4}

In the sections \ref{Examples}-\ref{SecGnrGds} we analyzed geodesics in the backgrounds produced by a single type of D branes. However, some of the most successful applications of string theory to black hole physics \cite{StrVafa} and to study of strongly coupled gauge theories \cite{HanWitt} involve intersecting branes, and in this section our analysis will be extended to a particular class of brane intersections. Specifically, we will extend the results of sections \ref{SecGnrGds} to 1/4--BPS configurations involving D$p$ and D$(p+4)$ branes. The geometries produced by such ``branes inside branes" continue to play an important role in understanding the physics of black holes, and a progress in understanding of the infall problem and Hawking radiation requires a detailed analysis of geodesics and waves on the backgrounds produced by D$p$--D$(p+4)$ systems. In this section we will continue to explore static configurations, and a large class of stationary solutions produced by D1--D5 branes will be analyzed in the next section. 

Let us consider massless geodesics in the geometry produced by D$p$ and D$(p+4)$ branes \cite{TsBranes}:
\bea\label{D1D5mGn}
ds^2=\frac{1}{\sqrt{H_1H_2}}\eta_{\mu\nu}dx^\mu dx^\nu+
\sqrt{H_1H_2}ds^2_{base}+\sqrt{\frac{H_1}{H_2}}dz_4^2.
\eea
The first and the second terms describe the spaces 
parallel/transverse to the entire D$p$--D$(p+4)$ system, and the four--dimensional torus represented by $dz_4^2$ is wrapped by D$(p+4)$ branes. We assume that D$p$ branes are smeared over the torus\footnote{Some localized solutions are also known \cite{CherkHash}, but we will not discuss them here.}. Metric (\ref{D1D5mGn}) contains two harmonic functions, $H_1$ and $H_2$, which are sourced by D$p$ and D$(p+4)$ branes. Away from the sources, these functions satisfy the Laplace equation on the $(5-p)$--dimensional flat base with metric $ds^2_{base}$.

Let us assume that geometry (\ref{D1D5mGn}) leads to a separable Hamilton--Jacobi equation. Then arguments presented in section \ref{SectPrcdr} imply that $H_2$ can only depend on two coordinates, $(r_1,r_2)$, where metric $ds^2_{base}$ has the form (\ref{BaseMetr}). To see this, we separate the Killing directions in the action
\bea
S=p_\mu x^\mu+q_i z^i+S_{base},
\eea
and rewrite the HJ equation (\ref{HJone}) as
\bea\label{HJD1D5w}
(\nabla S_{base})^2+p_\mu p^\mu H_1H_2+q_iq_iH_2=0.
\eea
Here we define
\bea
M^2=-p^\mu p_\mu,\qquad N^2=q_iq_i
\eea
If a given distribution of branes corresponds to integrable geodesics, then equation (\ref{HJD1D5}) should be separable for all allowed values of $M$ and 
$N$. Since equation (\ref{HJD1D5}) is analytic in these parameters, separability must persist even in the unphysical region where $M=0$ and $N$ is 
arbitrary\footnote{For asymptotically--flat solutions, $H_1$ and $H_5$ go to one at infinity, so $M>N$.}. In this region, equation (\ref{HJD1D5w}) reduces to  
(\ref{TheHJ}) with $H\rightarrow H_2$, $p_\mu p^\mu\rightarrow N^2$,  then arguments presented in section \ref{RedTo2D} reduce the problem to 
$H_2(r,\theta)$ with base space
\bea\label{Jul16}
ds_{base}^2=dr^2+r^2d\theta^2+r^2\cos^2\theta d\Omega_m^2+
r^2\sin^2\theta d\Omega_n^2,
\eea
and the analysis of section \ref{SectPrcdr} leads to three possible solutions ((\ref{TheSolnMthree}), (\ref{TheSolnMone}), (\ref{TheSolnMtwo})) for $(x,y)$ and $H_2$. 

Let us now demonstrate that $H_1$ can only be a function of $r$ and $\theta$. Indeed, separation for $M=0$ implies that  
\bea\label{GenSeparActw}
S_{base}={\tilde S}(y,{\tilde y})+R(r,\theta),
\eea
where $y_i$ are coordinates on $S^m$, and ${\tilde y}_j$ are coordinates on $S^n$. Substituting (\ref{Jul16}), (\ref{GenSeparActw}) into (\ref{HJD1D5w}) and differentiating the result with respect to $y_k$, we find
\bea
\frac{\d}{\d y_k}\left[
\frac{1}{r^2\cos^2\theta}h^{ij}
\left(\frac{\d{\tilde S}}{\d y_i}\right)
\left(\frac{\d{\tilde S}}{\d y_j}\right)+
\frac{1}{r^2\sin^2\theta}{\tilde h}^{ij}
\left(\frac{\d{\tilde S}}{\d {\tilde y}_i}\right)
\left(\frac{\d{\tilde S}}{\d {\tilde y}_j}\right)-H_1H_2M^2\right]=0.
\eea
Rewriting this relation as 
\bea
\frac{\d H_1}{\d y_k}=\frac{1}{M^2H_2}
\frac{\d}{\d y_k}
\left[\frac{1}{r^2\cos^2\theta}h^{ij}
\left(\frac{\d{\tilde S}}{\d y_i}\right)
\left(\frac{\d{\tilde S}}{\d y_j}\right)+
\frac{1}{r^2\sin^2\theta}{\tilde h}^{ij}
\left(\frac{\d{\tilde S}}{\d {\tilde y}_i}\right)
\left(\frac{\d{\tilde S}}{\d {\tilde y}_j}\right)\right],\nonumber
\eea
we conclude that $H_1$ develops unphysical singularities at 
$\theta=0,\frac{\pi}{2}$ for arbitrarily large $r$ unless 
$(\d H_1/\d y_k)=0$. Similar argument demonstrates that 
$(\d H_1/\d {\tilde y}_k)=0$, so $H_1$ can only depend on 
$(r,\theta)$.

To summarize, separability of the HJ equation (\ref{HJD1D5w}) requires the functions $H_1$ and $H_2$ to depend only on 
$(r,\theta)$, then the action has the form (\ref{GenSeparAct}),
\bea\label{GenSeparAbb}
S=p_\mu x^\mu+q_i z^i+S^{(m)}_{L_1}(y)+S^{(n)}_{L_2}({\tilde y})+R(r,\theta),
\eea
where $S^{(m)}_{L_1}$ and $S^{(n)}_{L_2}$ satisfy equations (\ref{HJSphere}). This results in the HJ equation
\bea\label{HJD1D5}
(\d_r R)^2+\frac{1}{r^2}(\d_\theta R)^2+
\frac{L_1^2}{r^2\cos^2\theta}+
\frac{L_2^2}{r^2\sin^2\theta}-H_1H_2M^2+N^2H_2=0,
\eea
and our analysis of separation at $M=0$ leads to three possible solutions ((\ref{TheSolnMthree}), (\ref{TheSolnMone}), (\ref{TheSolnMtwo})) for $(x,y)$ and $H_2$. As already discussed in section \ref{SectPrcdr}, solutions (\ref{TheSolnMone}) and  
(\ref{TheSolnMtwo}) are related by interchange of two spheres, so without loss of generality, we will focus on (\ref{TheSolnMthree}) and (\ref{TheSolnMone}).

Spherical coordinates (\ref{TheSolnMthree}) lead to separable equation (\ref{HJD1D5}) if an only if $H_1$ and $H_2$ do not depend on $\theta$, and since these functions have to be harmonic we find
\bea
H_1=a+\frac{Q_p}{r^{m+n}},\qquad H_2=a+\frac{Q_{p+4}}{r^{m+n}}.
\eea 
Here $a=1$ for asymptotically--flat space, and $a=0$ for the near--horizon solution. The corresponding metric 
(\ref{D1D5mGn}) gives the geometry produced by a single stack of D$p$--D$(p+4)$ branes \cite{HorStr}. 

Separation in the elliptic coordinates (\ref{TheSolnMone}) leads to 
\bea\label{H2D1D5}
&&\qquad\qquad H_2=a+\frac{1}{(\cosh^2 x-\cos^2 y)}
\frac{A}{\sinh^{n-1}x
\cosh^{m-1}x},\\
\label{ElCrdD1D5}
&&\cosh x=\frac{\rho_++\rho_-}{2d},\quad \cos y=\frac{\rho_+-\rho_-}{2d},\quad
\rho_\pm=\sqrt{r^2+d^2\pm 2rd\cos\theta},
\eea
and now we will determine the corresponding function $H_1$. Equation (\ref{HJD1D5}) separates in coordinates (\ref{ElCrdD1D5}) 
if and only if
\bea
N^2H_2-M^2H_1H_2=\frac{1}{(\cosh^2 x-\cos^2 y)}\left[V_1(x)+V_2(y)\right].
\eea
Since $H_2$ is already given by (\ref{H2D1D5}), the last relation implies that\footnote{Notice that $M\ne 0$ due to equation (\ref{HJD1D5}).}
\bea\label{H1H2z}
H_1H_2=\frac{1}{(\cosh^2 x-\cos^2 y)}\left[{\tilde V}_1(x)+{\tilde V}_2(y)\right].
\eea
First we set $a=0$ in (\ref{H2D1D5}), then equation (\ref{H1H2z}) becomes
\bea\label{H1H2y}
H_1=\frac{1}{\tilde Q}{\sinh^{n-1}x
\cosh^{m-1}x}\left[{\tilde V}_1(x)+{\tilde V}_2(y)\right].
\eea
To determine ${\tilde V}_1(x)$ and ${\tilde V}_2(y)$, we recall that, away from the sources, function $H_1$ must satisfy the Laplace equation (\ref{LaplInXY}):
\bea\label{LaplInD1D5}
\frac{1}{\sinh^{n}x\cosh^{m}x}\frac{\d}{\d x}\left[
\sinh^{n}x\cosh^{m}x\frac{\d H_1}{\d x}
\right]+
\frac{1}{\sin^{n}y\cos^{m}y}\frac{\d}{\d y}\left[
\sin^{n}y\cos^{m}y\frac{\d H_1}{\d y}
\right]=0
\eea
Substituting (\ref{H1H2y}) into (\ref{LaplInD1D5}) and performing straightforward algebraic manipulations, we find the most general solutions for $H_1$: 
\bea\label{H1D1D5}
\mbox{D0-D4:}&&(m,n)=(3,0)\quad H_1=C_1+C_2\left\{\arctan\left[\tanh\frac{x}{2}\right]+\frac{\sinh x}{2\cosh^2x}\right\}\nonumber\\
&&(m,n)=(2,1)\quad H_1=C_1+\frac{C_2}{\cosh x}\left\{1+\cosh x\ln\left[\tanh\frac{x}{2}\right]\right\}\nonumber \\
&&(m,n)=(1,2)\quad H_1=C_1+\nonumber\\&&\frac{C_2}{2\sqrt{2}\sinh x}\left\{ -1 +2\sinh x\left( \mathrm{arctanh}\left[\tanh\frac{x}{2}\right] + 2\Pi\left[ -1; -\arcsin\left[\tanh\frac{x}{2}\right] |1 \right] \right) \right\} \nonumber\\
&&(m,n)=(0,3)\quad H_1=C_1+C_2\left\{\coth x+\ln\left[\tanh\frac{x}{2}\right]\right\}\nonumber\\
\mbox{D1-D5:}&&(m,n)=(2,0)\quad H_1=C_1+C_2\tanh x\nonumber\\
&&(m,n)=(1,1)\quad C_1+C_2\ln[\tanh x]+C_3\ln[\tan y]+C_4\ln[\sin(2y)\sinh(2x)]\nonumber\\
&&(m,n)=(0,2)\quad H_1=C_1+C_2\coth x\nonumber\\
\mbox{D2-D6:}&&(m,n)=(1,0)\quad H_1=C_1+C_2\arctan\left[\tanh\frac{x}{2}\right]\nonumber\\
&&(m,n)=(0,1)\quad H_1=C_1+C_2\ln\left[\tanh\frac{x}{2}\right]\nonumber\\
\mbox{D3-D7:}&&(m,n)=(0,0)\quad H_1=C_1+C_2 x
\eea
Here $\Pi[n;\phi|m]$ is the incomplete elliptic integral.

So far we have assumed that $a=0$ in $H_2$. The case $a=1$, ${\tilde Q}=0$ corresponds to D$p$ branes only, so it is covered by discussion in section \ref{SecGnrGds}. Solutions with nonzero $a$ and ${\tilde Q}$ can be analyzed by looking at formal perturbation theory in $a$, and it turns out that $H_1$ must be constant for such solutions.

\section{Geodesics in D1--D5 microstates}\label{Rotations}

In the last three sections we have analyzed geodesics in a variety of static backgrounds produced by D--branes. In general, supersymmetric geometries are guaranteed to have a time--like (or light--like) Killing vector, so they must be stationary, but not necessarily static.  In particular, an interest in stationary geometries produced by the D1--D5 branes has been generated by the fuzzball proposal for resolving the black hole information paradox 
\cite{lmPar,fuzz}. According to this picture, microscopic states accounting for the entropy of a black hole have nontrivial structure that extents to the location of the na\"ive horizon, and the black hole geometry emerges as a course graining over such structures. Although the vast majority of fuzzballs is expected to be quantum, some fraction of microscopic states should be describable by classical geometries, and study of this subset has led to important insights into qualitative properties of generic microstates \cite{3charge}.

The fuzzball program has been particularly successful in identifying the microscopic states corresponding to D1--D5 black hole, where all microstate geometries have been constructed in \cite{lmPar,lmm}. Moreover, a strong support for the fuzzball picture came from analyzing the properties of these metrics \cite{tube}, and success of this study was based on separability of the wave equation on a special classes of metrics. In this chapter we have been  focusing on separability of the HJ equation as necessary condition for integrability of strings, but such separability also implies separability of the wave equation. This provides an additional motivation for studying the HJ equation for microscopic states in the D1--D5 system.

In this section we will mostly focus on the HJ equation for particles propagating on metrics constructed in \cite{lmPar}, and extension to the geometries for the remaining microstates of D1--D5 black holes \cite{lmm} will be discussed in the end\footnote{Note that the first studies of geodesics in the five-dimensnioal version of D1--D5--p system, the BMPV black hole, was performed in \cite{Gibbons:1999uv, Herdeiro:2000ap}, where it was shown that the integrability is related to the existence of a Stackel-Killing tensor}. The solutions of \cite{lmPar},
\bea\label{GenD1D5J}
ds^2&=&\frac{1}{\sqrt{H_1H_2}}\left[-(dt-A_idx^i)^2+(du+B_idx^i)^2\right]+
\sqrt{H_1H_2}dx_idx_i+\sqrt{\frac{H_1}{H_2}}dz_4^2,
\nonumber\\
&&d(\star_x dH_1)=d(\star_x dH_2)=0,\qquad dB=-\star_x dA,
\eea
generalize the static metric (\ref{D1D5mGn}) with $p=1$ by allowing the branes to vibrate on the four dimensional base, which is transverse to D1 and D5, and geometries of \cite{lmm} account for fluctuations on the torus. While the metric (\ref{GenD1D5J}), supplemented by the appropriate matter fields given in \cite{lmPar}, always gives a supersymmetric solution of supergravity away from the sources, the bound states of D1 and D5 branes, which are responsible for the entropy of a black hole, have additional relations between $H_1$, $H_2$ and $A$. Such bound states are uniquely specified by a closed contour ${F}_i(v)$ in four non--compact directions, and the harmonic functions are given by \cite{lmPar}\footnote{Relations (\ref{MicroHarm}) contain a constant parameter $\alpha$. Solutions with $\alpha=1$ correspond to asymptotically--flat geometries, and metrics with $\alpha=0$ asymptote to AdS$_3\times$S$^3$.}
\bea\label{MicroHarm}
H_1=\alpha+\frac{Q_5}{L}\int_0^L\frac{|\dot{\mathbf F}|^2dv}{|{\mathbf x}-{\mathbf F}|^2},\quad
H_2=\alpha+\frac{Q_5}{L}\int_0^L\frac{dv}{|{\mathbf x}-{\mathbf F}|^2},\quad
A_i=-\frac{Q_5}{L}\int_0^L\frac{\dot{ F}_idv}{|{\mathbf x}-{\mathbf F}|^2}.
\eea 
Remarkably, the resulting metric (\ref{GenD1D5J}) is completely smooth and horizon--free in spite of an apparent coordinate singularity at 
the location of the contour \cite{lmm}. To avoid unnecessary complications, we will focus on a special case 
$|{\dot {\mathbf F}}|=1$ (which leads to $H_1=H_2\equiv H$), although our results hold for arbitrary ${\dot {\mathbf F}}$. 

Applying the arguments presented in section \ref{RedTo2D} to metric (\ref{GenD1D5J}), we conclude that this geometry must preserve $U(1)\times U(1)$ symmetry of the base space, i.e., the profile ${F}_i(v)$ must be invariant under shifts of $\phi$ and $\psi$ in\footnote{This is a counterpart of (\ref{BaseMetr}) with $d_1=d_2=1$ and the base in (\ref{SSmetr}) with $m=n=1$.}
\bea
dx_idx_i=dr^2+r^2d\theta^2+r^2\cos^2\theta d\psi^2+
r^2\sin^2\theta d\phi^2.
\eea
This implies that the singular curve, $x_i=F_i(v)$ can only contain concentric circles with radii $r=(R_1,R_2,\dots,R_n)$ in the $\theta=\frac{\pi}{2}$ plane and concentric circles with radii $r=({\tilde R}_1,{\tilde R}_2,\dots,{\tilde R}_l)$ in 
the $\theta=0$ plane. First we focus on circles in the $\theta=\frac{\pi}{2}$ plane and demonstrate that separability of the HJ equation implies that $n=1$. Then we will show that $n=1$ also implies that $l=0$. 

\begin{enumerate}[1.]
\item{\textbf{Circles in the $(x_1,x_2)$ plane.}

For a single circular contour $(r=R_1,\theta=\frac{\pi}{2})$, the integrals (\ref{MicroHarm}) have been evaluated in \cite{MultiStr}, and superposition of these results gives the harmonic functions for several circles:
\bea\label{ExplsD1D5J}
H&=&\alpha+\sum_{i=1}^n\frac{Q_i}{f_i},\quad A=\sum_{i=1}^n \frac{2Q_iR_i s_i}{f_i}\frac{r^2\sin^2\theta}{r^2+R_i^2+f_i}d\phi,\nonumber\\
f_i&=&\sqrt{(r^2+R_i^2)^2-4r^2R_i^2\sin^2\theta}.
\eea
Here $Q_i$ is a five--brane charge of a circle with radius $R_i$, and $s_i$ is a sign that specifies the direction for going around this circle.  

Let us assume the the HJ equation (\ref{HJone}) separates in some coordinates. Metric (\ref{GenD1D5J}) has eight Killing directions ($t,u,\phi,\psi$ and the torus), they can be separated in the action:
\bea
S=p_tt+p_u u+J_\phi\phi+J_\psi\psi+q_iz_i+{\tilde S}(r,\theta),
\eea
Our assumption of integrability amounts to further separation of ${\tilde S}(r,\theta)$ in some coordinates $(x,y)$. 
In particular, for $J_\psi=J_\phi=p_u=0$, the HJ equation (\ref{HJone}) in the metric (\ref{GenD1D5J}) can be written as
\bea\label{HJD1D5J}
(\d_r {\tilde S})^2+\frac{1}{r^2}(\d_\theta {\tilde S})^2-p_t^2\left[H^2-\frac{(A_\phi)^2}{r^2\sin^2\theta}\right]=0.
\eea
This equation looks very similar to (\ref{TheHJ}), but in practice it is easier to analyze: we don't need to impose the Laplace equation (as we did for (\ref{TheHJ})), since the explicit forms of $H$ and $A$ are known (see (\ref{ExplsD1D5J})). The discussion of section 
\ref{SectPrcdr} implies that (\ref{HJD1D5J}) should be viewed as a relation between $h(w)$ (which is defined by (\ref{defCmplVar}) and (\ref{holomZ})) and $(R_1,\dots,R_n)$. In particular, substitution of the perturbative expansion (\ref{AppPert}) for $h(w)$ leads to an infinite set of constraints on $(R_1,\dots,R_n)$. To write these constraints, we introduce a convenient notation: 
\bea
D_k=\sum_{j=1}^n Q_j (s_jR_j)^k.
\eea
Then separation in $(k+2)$-rd order of perturbation theory gives a constraint
\bea\label{ConstrD1D5J}
(D_0)^{k-1} D_{k}=(D_1)^{k} \quad k\ge1.
\eea
Already the first nontrivial relation ($k=2$) implies that $R_1=R_2=\dots=R_n$, so it is impossible to have more than one circle (see below). We conclude that the HJ equation does not separate on the background (\ref{GenD1D5J}), (\ref{ExplsD1D5J}) unless $n=1$. The remaining constraints (\ref{ConstrD1D5J}) for $k>2$ are automatically satisfied for this case. 

We will now prove that equation (\ref{ConstrD1D5J}) with $k=2$ implies that $R_1=\dots=R_n$, and the readers who are not interested argument can go directly to part 2. Let us order the radii by $R_1\ge R_2\ge\dots\ge R_n$ and define a function 
\bea
G(R_1,\dots R_n)\equiv D_0D_2-(D_1)^2.
\eea
Then the derivative
\bea
\frac{\d G}{\d R_1}=2Q_1(D_0 R_1-s_1D_1)=
2Q_1\sum_{j=1}^n Q_j(R_1-s_1s_jR_j)
\eea
is positive unless $R_1=\dots=R_n$ and $s_1=\dots=s_n$, so function $G$ reaches its minimal value when all radii are equal, and it is this value, 
\bea
G_{min}=G(R_1,\dots R_1)=\left[\sum Q_j\right]
\left[\sum Q_j R_1^2\right]-\left[\sum Q_j s_jR_1\right]^2=0,
\eea
that gives (\ref{ConstrD1D5J}) for $k=2$. We conclude the equation $G=0$ implies that $R_1=\dots=R_n$.}
\item{\textbf{Circles in orthogonal planes.}

Having established that separation requires to have at most one circle in $\displaystyle\theta=\frac{\pi}{2}$ plane, we conclude that the same must be true about 
$\theta=0$ plane, but in principle it is possible to have one circle in each of the two planes. In this case we find
\bea\label{MuntiString}
&&H=\alpha+\sum_{i=1}^2\frac{Q_i}{f_i},\quad A=\frac{2Q_1R_1r^2\sin^2\theta}{f_1(r^2+R_1^2+f_1)}d\phi+
\frac{2Q_2R_2 r^2\cos^2\theta}{f_2(r^2+R_2^2+f_2)}d\psi,\nonumber\\
&&f_1=\sqrt{(r^2+R_1^2)^2-4r^2R_1^2\sin^2\theta},\quad f_2=\sqrt{(r^2+R_2^2)^2-4r^2R_2^2\cos^2\theta}
\eea
and for $J_\psi=J_\phi=p_y=0$ the HJ equation becomes
\bea
(\d_r {\tilde S})^2+\frac{1}{r^2}(\d_\theta {\tilde S})^2-p_t^2\left[H^2-\frac{(A_\phi)^2}{r^2\sin^2\theta}-\frac{(A_\psi)^2}{r^2\cos^2\theta}\right]=0
\eea
Fifth order of perturbation theory gives a relation $Q_1Q_2=0$, which implies that there is no separation in the geometry produced by two orthogonal circles. 
}
\item{\textbf{Separable coordinates.}

The perturbative procedure implemented in part 1 also gives the expression for $h(w)$  in terms of $D_k$ for the configurations satisfying (\ref{ConstrD1D5J}):
\bea\label{Jul17}
h(w)&=&\ln\left[\frac{1}{2}\left( e^{w}+\sqrt{e^{2w}+\frac{D_1}{lD_0}} \right) \right].
\eea
We have already encountered this holomorphic function in (\ref{TheSolnMone})--(\ref{TheSolnMtwo}) (depending on the sign of $D_1$), and demonstrated that it corresponds to elliptic coordinates (\ref{ElliptCase}) or (\ref{RepCase3}). For completeness we present the expression for $H^2$ that clearly demonstrates the separation of variables in (\ref{HJD1D5J}):
\bea
H^2&=&\frac{1}{A}\left[  2\alpha D_0+\frac{16e^{2x}D_0^6l^4}{(4e^{2x}D_0^2l^2+D_1^2)^2} +\alpha^2\left( e^{2x}+\frac{e^{-2x}D_1^4}{16l^4D_0^4} \right) + \frac{\alpha^2}{2} \left( \frac{D_1}{lD_0} \right)^2 \cos2y  \right],
\nonumber\\
&&A=l^2(\sinh^2 x+\cos^2y).
\eea
}
\item{\textbf{Separation with non--vanishing angular momenta.}

So far we have demonstrated that the HJ equation with $J_\psi=J_\phi=p_u=0$ separates only for microstate whose harmonic 
functions are given by (\ref{ExplsD1D5J}) with $k=1$, and separation takes place in the elliptic coordinates (\ref{Jul17}), 
(\ref{ElliptCase}). A direct check demonstrates that this separation persists for all values of momenta, when the relevant HJ equation is 
\bea\label{D1D5JhjGen}
(\d_r S)^2+\frac{1}{r^2}(\d_\theta S)^2+\frac{(B_\psi p_u-J_\psi)^2}{r^2\cos^2\theta}+
\frac{(J_\phi + A_\phi p_t)^2}{r^2\sin^2\theta}+ H^2(p_u^2-p_t^2)=0
\eea
and\footnote{To compare with \cite{MultiStr}, we replaced $R_1$ in (\ref{ExplsD1D5J}) by $a$ and $f_1$ by $f$.}
\bea\label{D1D5Jaa}
H&=&\alpha+\frac{Q}{f},\quad A_\phi=\frac{2Qa}{f}\frac{r^2\sin^2\theta}{r^2+a^2+f},\quad
B_\psi=\frac{2Qa}{f}\frac{r^2\cos^2\theta}{r^2+a^2+f}\\
f&=&\sqrt{(r^2+a^2)^2-4r^2a^2\sin^2\theta}.\nonumber
\eea
}
\end{enumerate}
To summarize, we have demonstrated that the HJ equation (\ref{HJone}) separates for the stationary D1--D5 geometry 
(\ref{GenD1D5J})--(\ref{MicroHarm}) with $|{\dot{\mathbf F}}|=1$ if and only if the string profile is circular, i.e., the harmonic functions 
are given by (\ref{D1D5Jaa}). 
\bigskip

Notice that integrability of (\ref{D1D5JhjGen}) follows from separability of the wave equation in the background (\ref{GenD1D5J}), 
(\ref{D1D5Jaa}), which has been discovered long time ago \cite{cvetLars,tube}. Let us clarify the relation between variables used in these papers and the elliptic coordinates  (\ref{ElliptCase}), (\ref{Jul17}). 

To separate the wave equation in the metric (\ref{GenD1D5J}) with harmonic functions (\ref{D1D5Jaa}), one can use the coordinates $(r',\theta')$ which appear naturally if the D1--D5 solution is viewed as an extremal limit of a black hole \cite{cvet,cvetLars,tube}:
\bea\label{D1D5Jprime}
ds^2&=&-\frac{1}{H}\left(dt-\frac{aQ}{f}\sin^2\theta' d\phi\right)^2+\frac{1}{H}\left(du+\frac{aQ}{f}\cos^2\theta' d\psi\right)^2+dz_idz_i
\nonumber\\
&&+Hf\left[\frac{(dr')^2}{(r')^2+a^2}+(d\theta')^2\right]+H\left[(r'\cos\theta')^2d\psi^2+
(r^2+a^2)\sin^2\theta'd\phi^2\right],\nonumber\\
f&=&(r')^2+a^2\cos^2\theta',\qquad H=\alpha+\frac{Q}{f}.
\eea
The relation between $(r,\theta)$ and $(r',\theta')$ was found in \cite{MultiStr}\footnote{This comes from interchanging $(r,\theta)$ with 
$(r',\theta')$ in formula (4.7) of \cite{MultiStr} and (\ref{D1D5Jprime}) is obtained from setting $Q_1=Q_5$ in equation (5.12) in that paper.}
\bea\label{RThTrnls}
r=\sqrt{(r')^2+a^2\sin^2\theta'},\qquad \cos\theta=\frac{r'\cos\theta'}{\sqrt{(r')^2+a^2\sin^2\theta'}}.
\eea
Looking at the $(r',\theta')$ sector of the metric:
\bea
ds^2=(Q+f)\left[\frac{(dr')^2}{(r')^2+a^2}+(d\theta')^2\right]
\eea
we arrive at natural ``conformally--Cartesian" coordinates $(x,y)$:
\bea\label{RYprime}
r'=a\sinh x,\qquad \theta'=y.
\eea
Substituting this into (\ref{RThTrnls}) we find the expressions for $(r,\theta)$ in terms of $(x,y)$:
\bea
r=a\sqrt{\sinh^2 x+\sin^2y},\quad \cos\theta=\frac{\sinh x\cos y}{\sqrt{\sinh^2 x+\sin^2y}},
\eea
Using the definitions (\ref{defCmplVar}), we conclude that 
\bea
w&\equiv&\ln r+i\theta=\frac{1}{2}\ln\left[\sinh^2 x+\sin^2y\right]+i\arccos\left[\frac{\sinh x\cos y}{\sqrt{\sinh^2 x+\sin^2y}}\right]\nonumber\\
&=&\ln(\sinh z).
\eea
is a holomorphic function of $z=x+iy$, as expected from the general analysis presented in section \ref{SectPrcdr}. Inverting the last expression, we recover the relation (\ref{HolomElpt}). 
\bea
z=\ln\left[\frac{1}{2}\left(e^{w}+ \sqrt{e^{2{w}}+1}\right)\right].
\eea
Thus we conclude that coordinates (\ref{RYprime}) used in \cite{cvet,tube} are essentially the elliptic coordinates up to a minor redefinition of $r'$. We will come back to this feature in section \ref{StdPar}.

\section{Geodesics in bubbling geometries}\label{Bubbles}

Given integrability of sigma model on $AdS_5\times S^5$, it is natural to look at deformations of this 
background which might preserve integrable structures. In particular, reference \cite{StepTs} demonstrated that deformation of $AdS_5\times S^5$ to asymptotically-flat geometry by adding one to the harmonic function destroys integrability of sigma model, although the Hamilton--Jacobi equation for geodesics remains separable. The extension from $AdS_5\times S^5$ to flat geometry is only possible if one choses flat metric on the worldvolume of D3 branes\footnote{This corresponds to field theory living on $R^{3,1}$, which is dual to the Poincare patch of $AdS_5$}, and in section \ref{SecGnrGds} we analyzed several classes of geometries produced by flat Dp branes. From the point of view of AdS/CFT correspondence, it is equally interesting to look at field theories on $R\times S^3$, which are dual to geometries produced by spherical D3 branes. The most symmetric geometry of this type is a direct product of global $AdS_5$ and a five dimensional sphere, but less symmetric examples are also known \cite{LLM,quarter}. In this section we will apply the techniques developed in section \ref{SecGnrGds} to identify the most general geometries of \cite{LLM} with separable geodesics. 
 
We begin with recalling the metrics of $1/2$--BPS geometries constructed in \cite{LLM}\footnote{These metrics are supported by the five--form field strength, and expression for $F_5$ can be found in \cite{LLM}. Notice that our notation in (\ref{GenBubble}) slightly differs from one in \cite{LLM}: we  replaced $y$ of \cite{LLM} by $Y$ to avoid the confusion with coordinate $y$ introduced in section \ref{SectPrcdr}.}:
\bea\label{GenBubbleM}
ds^2&=&-h^{-2}(dt+V_i dx^i)^2+h^2(dY^2+dx_1^2+dx_2^2)+Ye^G d\Omega_3^2+Ye^{-G} d{\tilde\Omega}_3^2,\\
h^{-2}&=&2Y\cosh G,\qquad 
YdV=\star_3 dz,\qquad
z=\frac{1}{2}\tanh G\nonumber
\eea
The solutions are parameterized by one function
$z(x_1,x_2,Y)$ that satisfies the Laplace equation,
\bea\label{bbLplM}
\d_i\d_i z+Y\d_Y\left(Y^{-1}\d_Y z\right)=0,
\eea
and obeys the boundary conditions 
\bea\label{DropletM}
z(Y=0)=\pm \frac{1}{2}.
\eea
The regions with $z=\frac{1}{2}$ form droplets in $(x_1,x_2)$ plane, and any configuration of droplets leads to the unique regular geometry. Solutions (\ref{GenBubbleM}) are dual to half--BPS states in $N=4$ super--Yang--Mills theory, and droplets in  $(x_1,x_2)$ plane correspond to eigenvalues of the matrix model describing such states 
\cite{beren} (see \cite{LLM} for detail).

\begin{figure}[]
\begin{minipage}[h!]{0.4\linewidth}
\center{\includegraphics[width=1\linewidth]{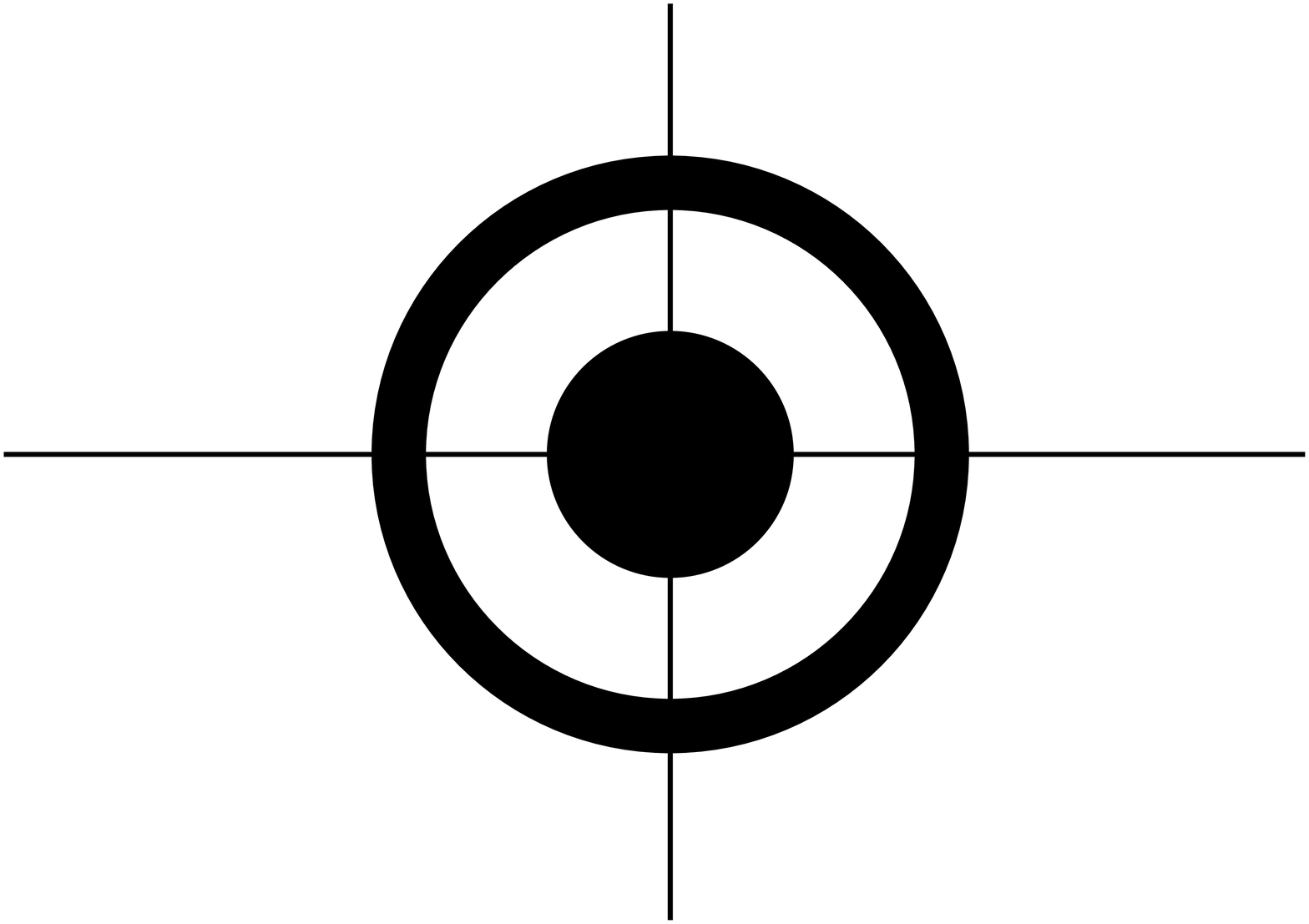}} (a) \\
\end{minipage}
\hfill
\begin{minipage}[h!]{0.25\linewidth}
\center{\includegraphics[width=1\linewidth]{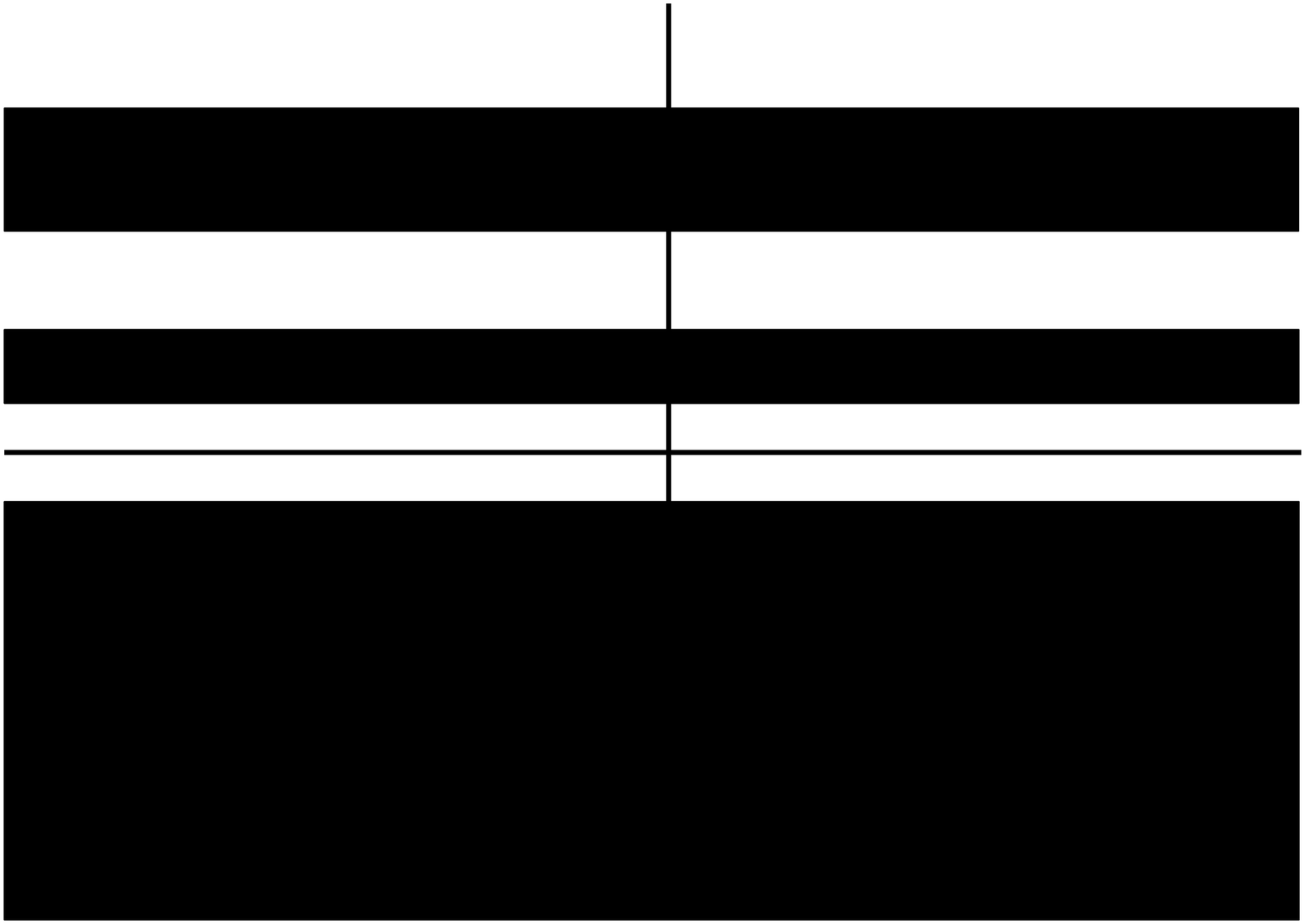}} (b) \\
\end{minipage}
\hfill
\begin{minipage}[h!]{0.25\linewidth}
\center{\includegraphics[width=1\linewidth]{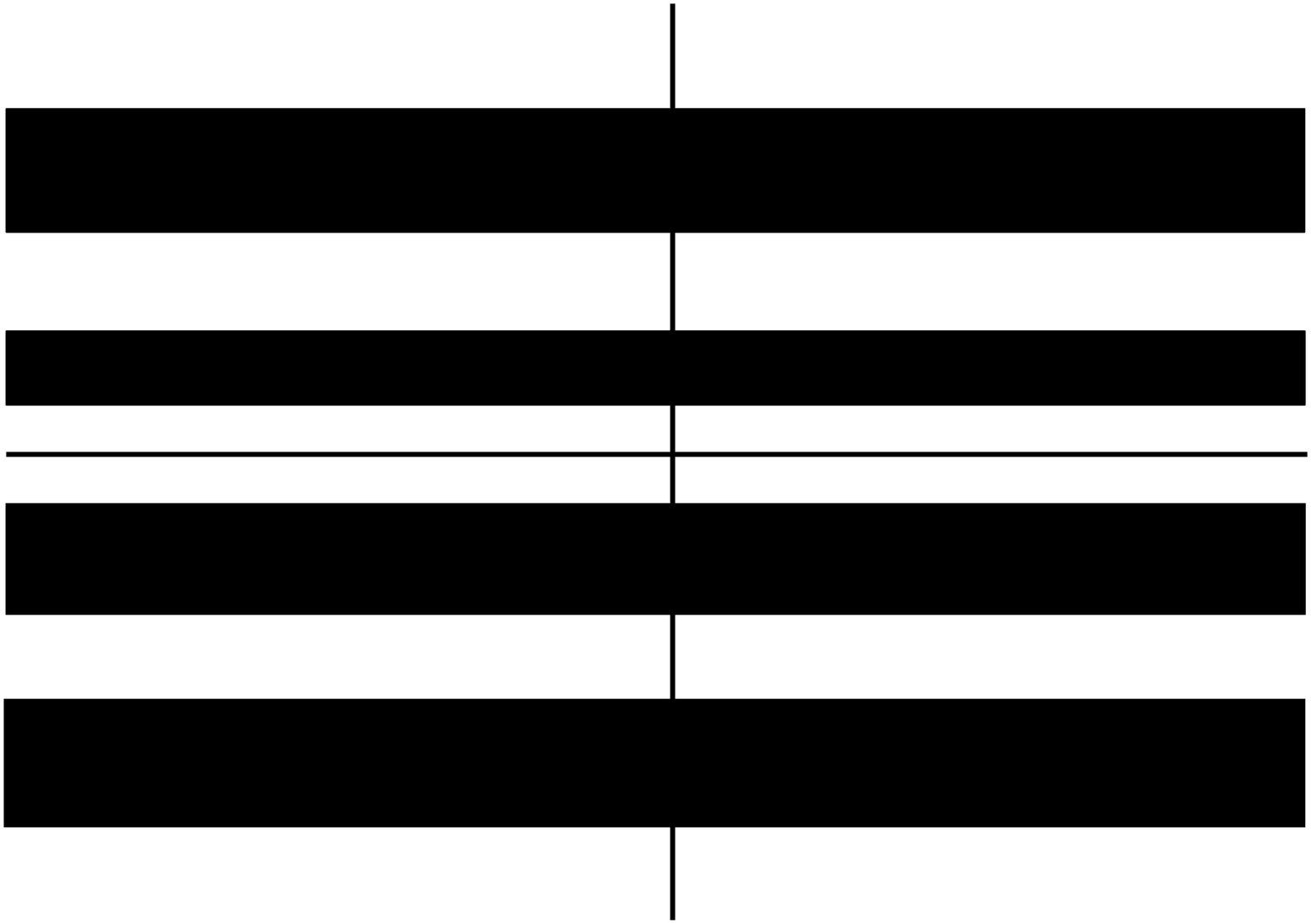}} (c) \\
\end{minipage}
\caption[Boundary conditions for geometries with rotational and translational isometries]{Boundary conditions in the $Y=0$ plane corresponding to geometries with rotational and translational isometries, (\ref{RingRed}) and (\ref{StripRed}).}
\label{Fig:Bubbles}
\end{figure}

\begin{figure}
\centering
\begin{minipage}[h!]{0.4\linewidth}
\center{\includegraphics[width=1\linewidth]{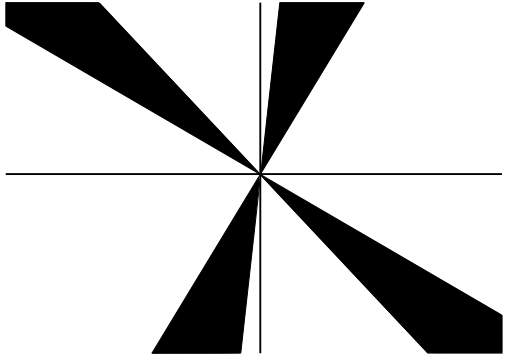}}
\end{minipage}
\caption[Boundary conditions for geometries with $\d_r z=0$.]{Boundary conditions in the $Y=0$ plane corresponding to geometries with $\d_r z=0$.}
\label{Fig:Wedges}
\end{figure}

A generic distribution of droplets in $(x_1,x_2)$ plane leads to solution (\ref{GenBubbleM}), which has a nontrivial dependence upon three coordinates 
$(x_1,x_2,Y)$. Repeating the arguments presented in section \ref{RedTo2D}, one can show that geodesics can only be integrable if at least one of these coordinates corresponds to a Killing direction. Such configurations can be obtained by performing a dimensional reduction of (\ref{GenBubbleM}) along one of the directions in $(x_1,x_2)$ plane. Only two such reductions are possible \footnote{There are also counterparts of (\ref{RingRed}) with $\d_r z=0$, which correspond to wedges in the $(x_1,x_2)$ plane ( figure \ref{Fig:Wedges}). However, such configurations lead to singular geometries, see \cite{Jabb} for further discussion. The counterpart of (\ref{StripRed}) with $\d_2 z=0$ is related to (\ref{StripRed}) by rotation.}:
\bea\label{RingRed}
\d_\phi z=0:&& x_1+ix_2=r\cos\theta e^{i\phi},\quad
Y=r\sin\theta,\quad 0\le\theta<\pi,\\
\label{StripRed}
\d_{1} z=0:&& x_2=r\cos\theta,\quad
Y=r\sin\theta,\quad 0\le\theta<\pi.
\eea
Reduction (\ref{RingRed}) corresponds to concentric rings in the $(x_1,x_2)$--plane (see figure \ref{Fig:Bubbles}(a)), and it describes excitations of $AdS_5\times S^5$. Reduction (\ref{StripRed}) corresponds to parallel strips in $(x_1,x_2)$--plane (see figure \ref{Fig:Bubbles}(b,c)), which can describe either excitations of the pp--wave or states of Yang--Mills theory on $S^1\times R$. The three cases depicted in figure \ref{Fig:Bubbles} are analyzed in the appendix \ref{AppBbl}, and here we just summarize the results. 

\begin{enumerate}[(a)]
\item{\textbf{Geometries with AdS$_5\times$S$^5$ asymptotics.}\\
The boundary conditions depicted in figure \ref{Fig:Bubbles}(a) lead to geometries (\ref{GenBubbleM}), which are invariant under rotations in $(x_1,x_2)$ plane, and such solutions are conveniently formulated in terms of coordinates introduced in 
(\ref{RingRed}):
\bea
ds^2=-h^{-2}(dt+V_\phi d\phi)^2+h^2(dr^2+r^2d\theta^2+r^2\cos^2\theta d\phi^2)+
Ye^G d\Omega_3^2+Ye^{-G} d{\tilde\Omega}_3^2.
\nonumber
\eea
The complete solution of the Laplace equation (\ref{bbLplM}) and expression for $V_\phi$ for this case were found in \cite{LLM}:
\bea\label{zVa}
{z}=\frac{1}{2}+\frac{1}{2}\sum_{i=1}^n(-1)^{i+1}\left[\frac{r^2-R_i^2}{\sqrt{(r^2+R_i^2)^2-4R_i^2r^2\cos^2\theta}}-1\right],
\nonumber \\
V_\phi=-\frac{1}{2}\sum_{i=1}^n(-1)^{i+1}\left[ \frac{r^2+r_i^2}{\sqrt{(r^2+R_i^2)^2-4R_i^2r^2\cos^2\theta}} - 1 \right],
\eea 
Summation in (\ref{zVa}) is performed over $n$ circles with radii $R_i$, and following conventions of \cite{LLM} we will take $R_1$ to be the radius of the largest circle. For example, a disk corresponds to one circle, a ring to two circles, and so on. 

The HJ equation for the solutions specified by (\ref{zVa}) is analyzed in the appendix \ref{BubblesCircles}, where it is demonstrated that integrability leads to an infinite set of relations between radii $R_i$. Specifically, the expressions defined by 
\bea\label{RiRingA}
D_{p}\equiv\sum_{j=1}^n (-1)^{j+1} (R_{j})^p
\eea
must satisfy the relations (\ref{ReqnRing}):
\bea\label{ReqnRingA}
(D_2)^{k-1}D_{2(k+1)}=(D_4)^k.
\eea
As demonstrated in appendix \ref{BubblesCircles}, this requirement implies that $n<2$ in (\ref{zVa}), so variables separate only for flat space ($n=0$) and for AdS$_5\times$S$^5$ ($n=1$) (figure \ref{Fig:BubblesInt}(a)). Moreover, construction presented in the appendix \ref{BubblesCircles} gives the unique set of separable coordinates (\ref{BblElptCrdA}) for AdS$_5\times$S$^5$
\bea\label{BblElptCrdM}
\cosh x=\frac{\rho_++\rho_-}{2d},\quad \cos y=\frac{\rho_+-\rho_-}{2d},\quad
\rho_\pm=\sqrt{r^2+R_1^2\pm 2rR_1\cos\theta},
\eea
and their relation with standard parameterization of this manifold will be discussed in section \ref{StdPar}. 

}
\item{
\textbf{Geometries with pp--wave asymptotics}\\
We will now discuss the geometries with translational $U(1)$ symmetry (\ref{StripRed}), which correspond to parallel strips in the $(x_1,x_2)$ plane (see figure \ref{Fig:Bubbles}(b,c)). It is convenient to distinguish two possibilities: $z$ can either approach different values $x_2\rightarrow \pm\infty$ (as in figure \ref{Fig:Bubbles}(b) or approach the same value on both sides (as in figure \ref{Fig:Bubbles}(c)). Here we will focus on the first option, which corresponds to geometries with plane wave asymptotics, and the second case will be discussed in part (c).

Pp--wave can be obtained as a limit of $AdS_5\times S^5$ geometry by taking the five--form flux to infinity \cite{BMN}. This limit has a clear representation in terms of boundary conditions in $(x_1,x_2)$ plane: taking the radius of a disk (figure \ref{Fig:BubblesInt}(a)) to infinity, we recover a half-filled plane corresponding to the pp--wave (see figure \ref{Fig:BubblesInt}(c)). Taking a similar limit for a system of concentric circles (figure \ref{Fig:Bubbles}(a)), we find excitations of pp--wave geometry by 
a system of parallel strips (see figure \ref{Fig:Bubbles}(b)). Since strings are integrable on the pp--wave geometry 
\cite{BMN}, it is natural to ask whether such integrability persists for the deformations represented in figure \ref{Fig:Bubbles}(b). We will now rule out integrability on the deformed backgrounds by demonstrating that even equations for massless geodesics are not integrable. 

Solutions of the Laplace equation (\ref{bbLplM}) corresponding to the boundary conditions depicted in figure \ref{Fig:Bubbles}(b) were found in \cite{LLM}, and their explicit form is given by (\ref{zVpwave}). Such solutions are parameterized by the strip boundaries, and the black strip number $i$ is located at $d_{2i-1}<x_2<d_{2i}$. In appendix \ref{StripsPlane} we use the techniques developed in section \ref{SecGnrGds} to demonstrate that the HJ equation can only be separable when $n=0$ in (\ref{zVpwave}), i.e., when the solution represents an unperturbed pp--wave:
\bea
ds^2&=&-2dt dx_1-(x^2+y^2)dt^2+dx^2+x^2d\Omega^2+dy^2+y^2d{\tilde\Omega}^2,\\
&&r_1=x,\quad r_2=y.\nonumber
\eea
Interestingly, some special solutions also separate for the pp--wave with an additional strip (see figure \ref{Fig:BubblesInt}(d)). Specifically, the geodesics which do not move on the spheres and along $x_2$ direction (i.e., geogesics with $p=0$, $L_1=L_2=0$ in (\ref{F26})) separate in coordinates $(x,y)$ defined by (\ref{hOneStripHalfPlane}):
\bea
x+iy&=&w+\ln\left[\frac{1}{2}\left(\sqrt{1-\frac{d_0}{le^w}}+1\right)\right] +\ln\left[\frac{1}{2}\left(\sqrt{1-\frac{d_2}{le^w}}+1\right)\right],\\
w&=&\ln\frac{r}{l}+i\theta.\nonumber
\eea

\begin{figure}[]
\begin{minipage}[h!]{0.2\linewidth}
\center{\includegraphics[width=1\linewidth]{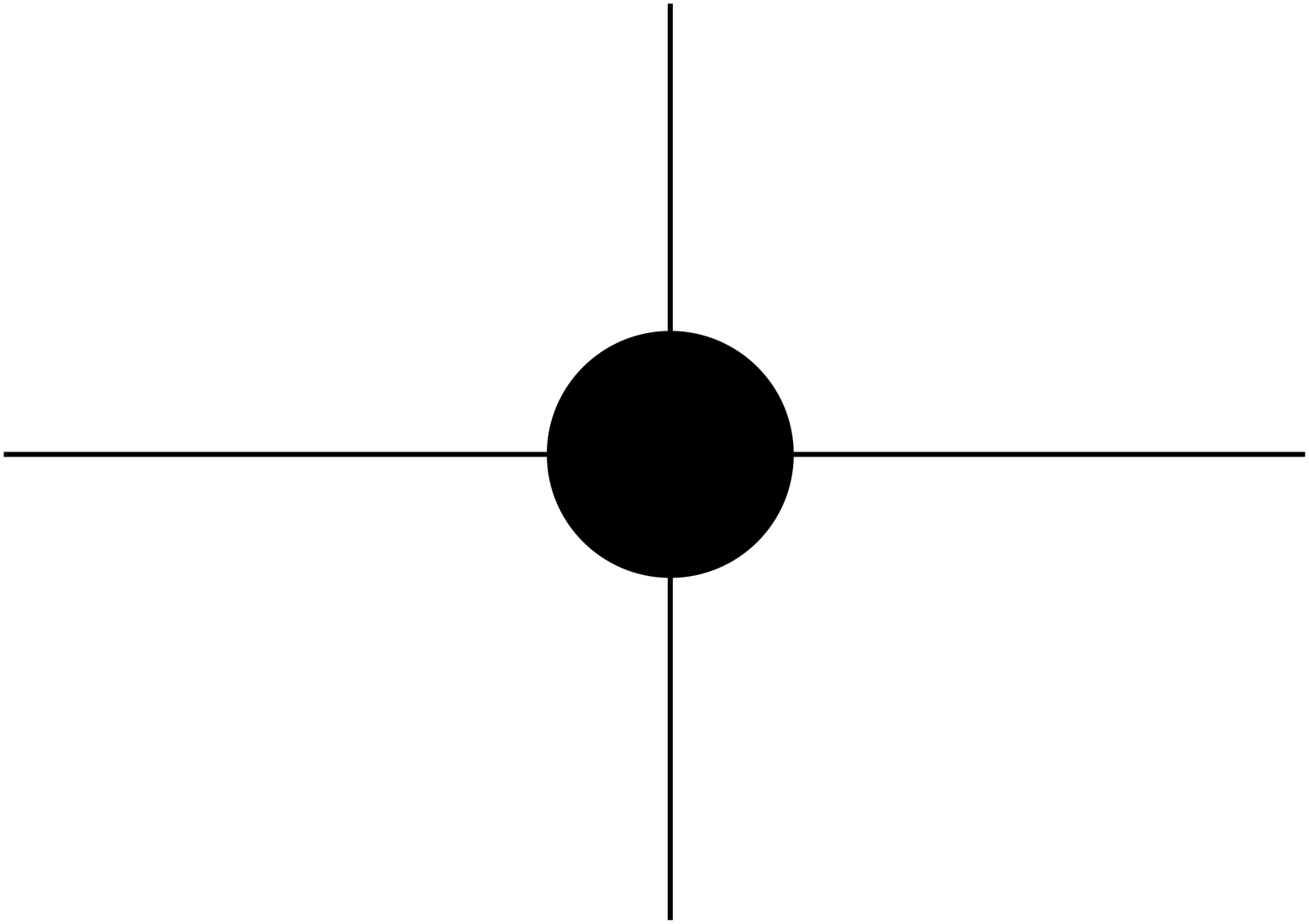}} (a) \\
\end{minipage}
\hfill
\begin{minipage}[h!]{0.2\linewidth}
\center{\includegraphics[width=1\linewidth]{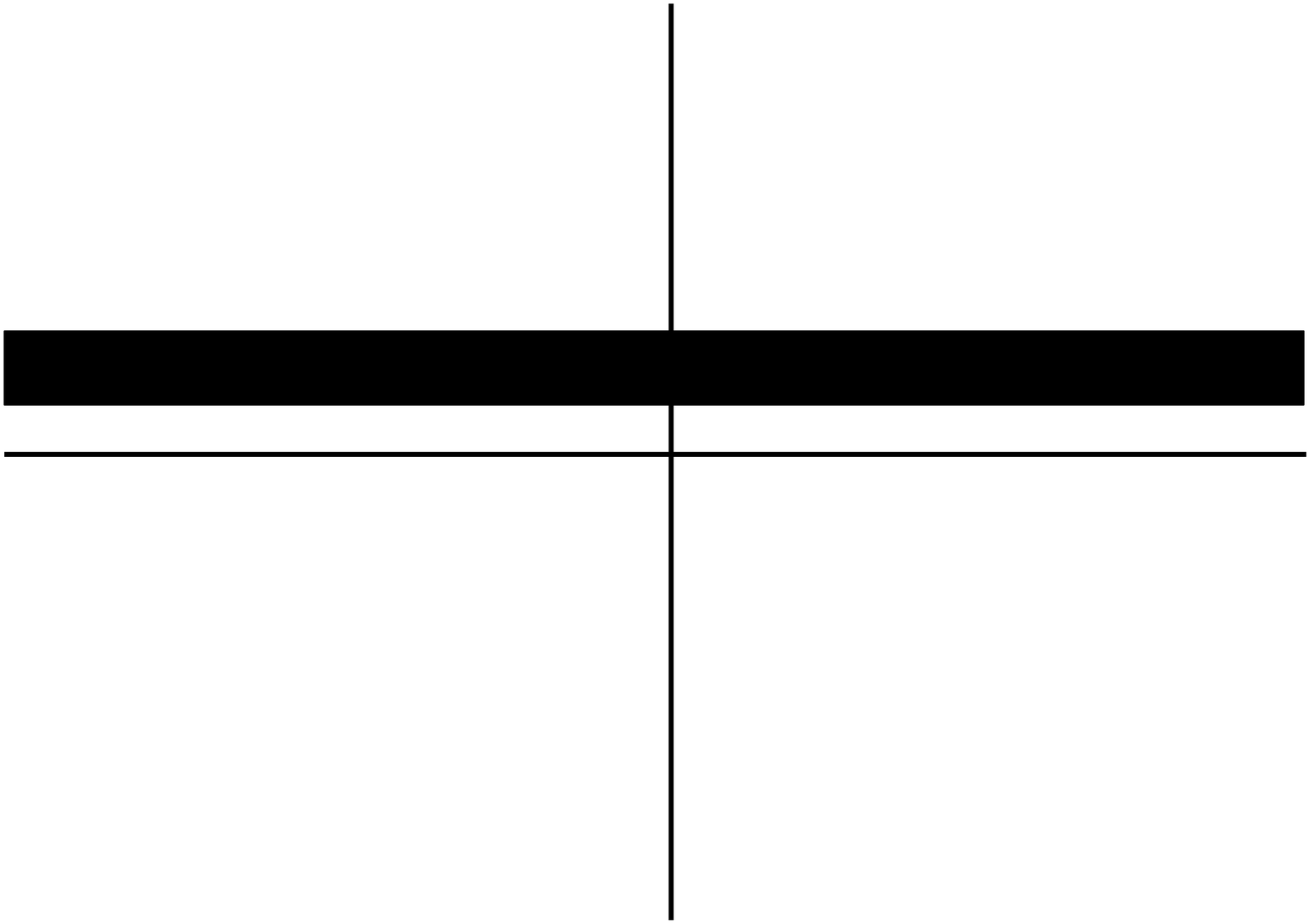}} (b) \\
\end{minipage}
\hfill
\begin{minipage}[h!]{0.2\linewidth}
\center{\includegraphics[width=1\linewidth]{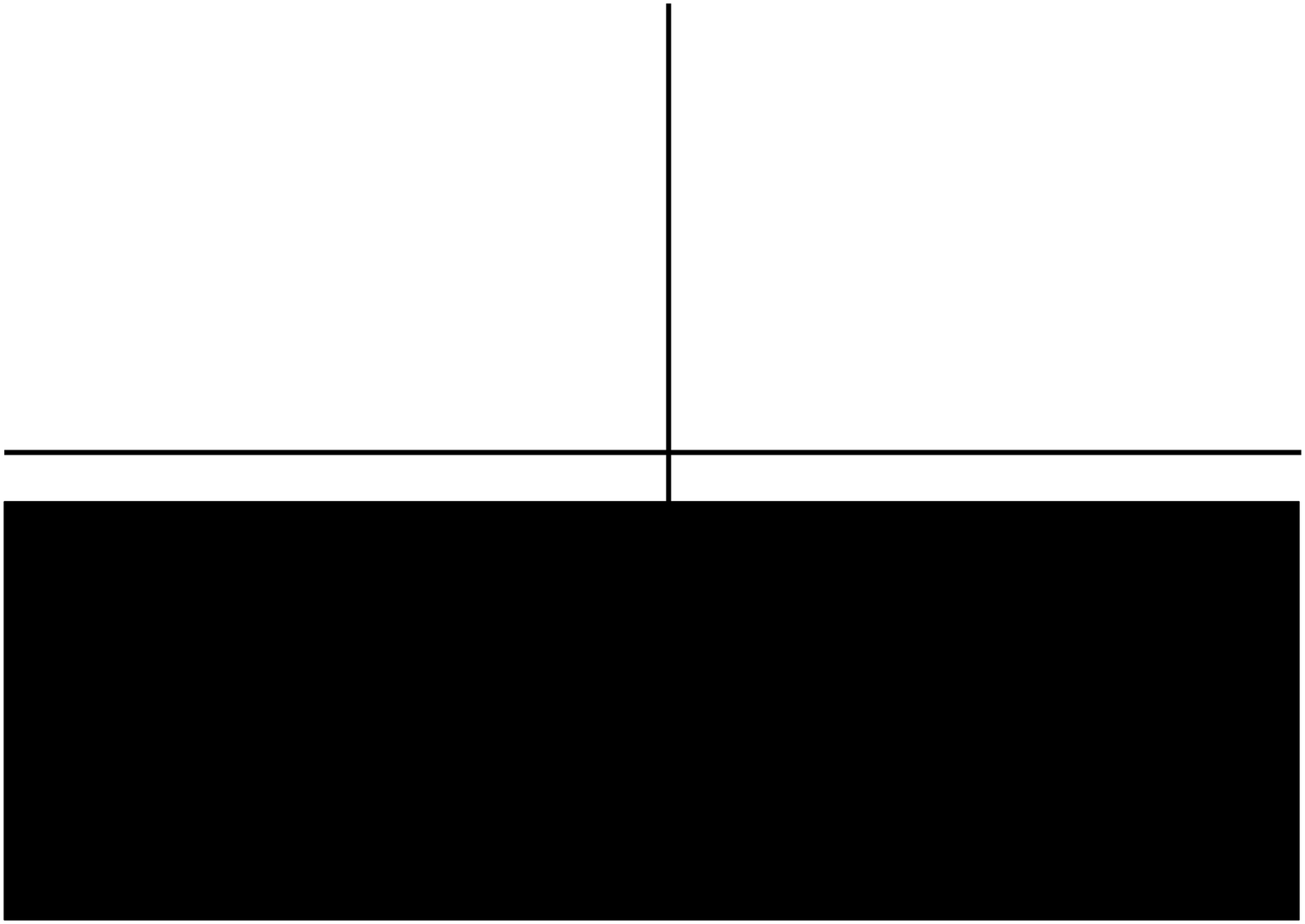}} (c) \\
\end{minipage}
\hfill
\begin{minipage}[h!]{0.2\linewidth}
\center{\includegraphics[width=1\linewidth]{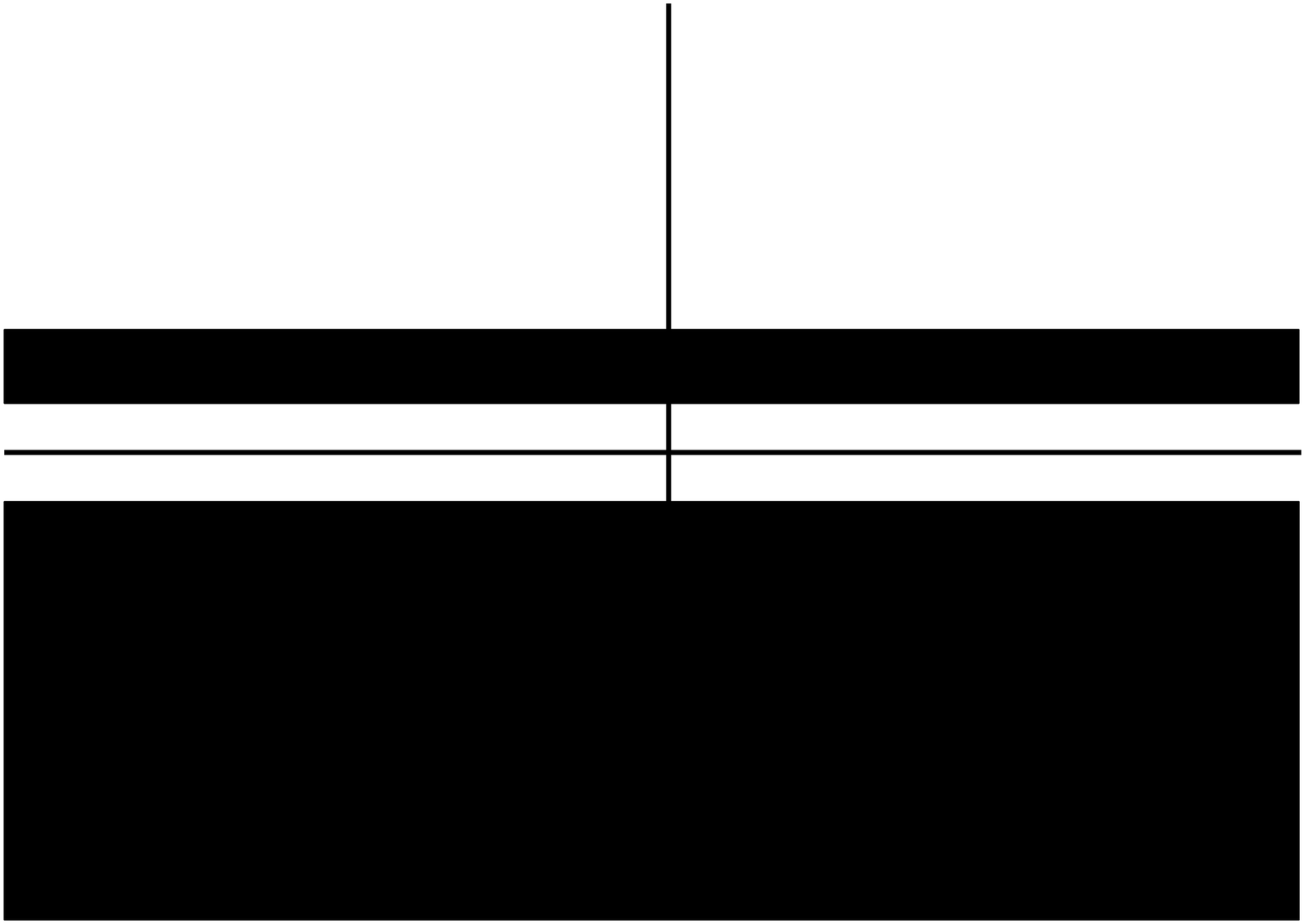}} (d) \\
\end{minipage}
\caption[Boundary conditions for geometries with integrable geodesics]{Boundary conditions in the $Y=0$ plane corresponding to geometries with integrable geodesics.}
\label{Fig:BubblesInt}
\end{figure}

} 

\item{\textbf{Geometries dual to SYM on a circle}\\
Finally we consider configuration depicted in figure \ref{Fig:Bubbles}(c). As discussed in \cite{LLM}, these configurations are dual to Yang--Mills theory on $S^3\times S^1\times R$, and since we are only keeping zero modes on the sphere, the solutions (\ref{GenBubbleM}) correspond to BPS states in two--dimensional gauge theory on a circle.   

The solution of the Laplace equations corresponding to figure \ref{Fig:Bubbles}(c) is given by (\ref{zVstrip}), and in appendix \ref{SecStrips}, it is shown that only a single strip (figure \ref{Fig:BubblesInt}(b)) leads to a separable HJ equation. For completeness we present the expression for the natural coordinates:
\bea
x+iy&=&2\ln\left[ \frac{1}{2}
\left(\sqrt{e^w-(d_2/l)}+e^{w/2}\right)\right]\\
w&=&\ln\frac{r}{l}+i\theta.\nonumber
\eea
}
\end{enumerate}
To summarize, we have demonstrated that from the infinite family of the 1/2--BPS geometries constructed in \cite{LLM}, only AdS$_5\times$S$^5$, pp--wave, and a single M2 brane give rise to integrable geodesics (figure \ref{Fig:BubblesInt}). This implies that the short strings can only be integrable on these backgrounds. Notice, however, that our results do not extend to equations of motion for D3 branes (which are expected to be integrable, as least for 1/2--BPS objects) since the HJ equation (\ref{HJone}) did not take into account coupling to the RR field. We also found that the separable coordinates $(x,y)$ given by (\ref{BblElptCrdM}) are the same elliptic coordinates (or their limits) as the one encountered in section \ref{SectPrcdr}, and in the next section we will discuss the relation between $(x,y)$ and the standard parameterization of AdS$_5\times$S$^5$.

\section{Elliptic coordinates and standard parameterization of AdS$_p\times$S$^q$}\label{StdPar}

In the last section we have demonstrated that the HJ equation for geodesics on 1/2--BPS geometries of \cite{LLM} separates only for 
AdS$_5\times$S$^5$ and for its pp--wave limit. Moreover, this separation happens in the elliptic coordinates (\ref{BblElptCrdM}). On the other hand, equations for supergravity fields on AdS$_5\times$S$^5$ are usually analyzed in the standard parameterization\footnote{We denoted an azimuthal direction on $S^5$ by $\chi$ to avoid confusion with coordinate $\theta$ introduced in (\ref{RingRed}).}:
\bea\label{StndAdS5}
ds^2=L^2\left[-\cosh^2\rho dt^2+
d\rho^2+\sinh^2\rho d\Omega_3^2+
d\chi^2+\cos^2\chi d{\tilde\phi}^2+\sin^2\chi d{\tilde\Omega}^2\right]
\eea
and the detailed study of \cite{Nvnh} uses the explicit  $SO(4,2)\times SO(6)$ symmetry of this metric to separate the resulting equations and to find the mass spectrum of supergravity modes on AdS$_5\times$S$^5$. This suggests a close relation between the elliptic coordinates and (\ref{StndAdS5}), which will be clarified in this subsection. We will also discuss the elliptic coordinates for AdS$_7\times$S$^4$, AdS$_4\times$S$^7$, and AdS$_3\times$S$^3$.

\bigskip
\noindent
\textbf{Standard parameterization of AdS$_5\times$S$^5$.}

To relate the standard parameterization (\ref{StndAdS5}) with elliptic coordinates (\ref{BblElptCrdM}), we first recall the map between (\ref{StndAdS5}) and variables used in (\ref{GenBubbleM}) (see \cite{LLM}):
\bea
x_1+ix_2=L^2\cosh\rho\cos\chi e^{i\phi},\quad Y=L^2\sinh\rho\sin\chi, \quad 
\phi={\tilde\phi}-t.
\eea
Comparing this with (\ref{RingRed}), we relate the standard coordinates (\ref{StndAdS5}) with $(r,\theta)$,
\bea\label{Jul20}
L^2\sinh\rho\sin\chi=r\sin\theta,\qquad
L^2\cosh\rho\cos\chi=r\cos\theta,
\eea
and substitution into (\ref{BblElptCrdM}) gives
\bea\label{XYellAdS5}
\rho_\pm=R_1\left[\cosh\rho\pm\cos\chi\right]
\quad\Rightarrow\quad
x=\rho,\quad y=\chi.
\eea
We conclude that the standard coordinates (\ref{StndAdS5}) on 
AdS$_5\times$S$^5$ can be viewed as elliptic coordinates on the base of the LLM geometries (\ref{GenBubbleM}) in IIB supergravity. 

\bigskip
\noindent
\textbf{1/2--BPS geometries in M-theory.}

Let us now turn to the LLM geometries in M-theory \cite{LLM}:
\bea\label{bblMth}
ds_{11}^2&=&-4e^{2\lambda}(1+Y^2e^{-6\lambda})(dt+V_i dx^i)^2+\frac{e^{-4\lambda}}{1+Y^2e^{-6\lambda}}\left[ dY^2+e^D (dx_1^2+dx_2^2) \right]\nonumber\\
&+&4e^{2\lambda}d\Omega_5^2+Y^2e^{-4\lambda}d\tilde{\Omega}_2^2\nonumber\\
e^{-6\lambda}&=&\frac{\d_Y D}{Y(1-Y\d_Y D)},\qquad
V_i=\frac{1}{2}\epsilon_{ij}\d_jD.\nonumber
\eea
Metric (\ref{bblMth}) is parameterized by one function $D$ satisfying the Toda equation,
\bea\label{Toda}
(\d_1^2+\d_2^2)D+\d_Y^2 e^D=0,
\eea 
on a three--dimensional base,
\bea\label{TodaBase}
ds_{base}^2=dY^2+e^D\left[dx_1^2+dx_2^2\right],
\eea
and some known boundary conditions in the $Y=0$ plane. Although (\ref{Toda}) is much more complicated than the Laplace equation (\ref{bbLplM}), we expect that repetition of the arguments presented in section \ref{Bubbles} ensures that function $D$ can only depend on two rather than three variables, and in this case (\ref{Toda}) can be rewritten as a Laplace equation via a nonlocal change of variables \cite{Ward}. Specifically, for the rotationally--invariant case, it is convenient to rewrite the metric on the base as\footnote{We introduced an obvious notation: $x_1+ix_2=Re^{i\phi}$, $X=\ln R$.}
\bea\label{RotToTrnsl}
ds_{base}^2=dY^2+e^D\left[dR^2+R^2d\phi^2\right]=
dY^2+e^{\tilde D}\left[dX^2+d\phi^2\right]
\eea
Function ${\tilde D}=D+2\ln R$ defined above satisfied the same Toda equation (\ref{Toda}) as $D$, and in terms of $(X,Y)$ this equation can be rewritten as
\bea\label{TrnslToda}
\d_X^2{\tilde D}+\d_Y^2 e^{\tilde D}=0
\eea
The Toda equation with translational invariance along $x_1$ can also be written as (\ref{RotToTrnsl})--(\ref{TrnslToda}) after a replacement $(x_1,x_2,D)\rightarrow (\phi,X,{\tilde D})$. 
A nonlocal change of coordinates \cite{Ward,LLM},
\bea\label{Ward}
e^{\tilde D}=\zeta^2,\quad Y=\zeta\d_\zeta V,\quad
X=\d_\eta V,
\eea
maps the nonlinear Toda equation (\ref{TrnslToda}) into the Laplace equation for function $V$:
\bea
\zeta^{-1}\d_\zeta(\zeta\d_\zeta V)+\d_\eta^2 V=0.
\eea
Unfortunately, the boundary conditions for $V$ are rather complicated \cite{LLM} (see \cite{LinMald} for a detailed discussion), and simple expressions for $D$ and $V_\phi$ similar to (\ref{zVa}) are not known. Nevertheless, the results presented in sections \ref{Rotations} and \ref{Bubbles} strongly suggest that the HJ equation on the geometries (\ref{bblMth}) would only separate on the most symmetric backgrounds: AdS$_7\times$S$^4$, AdS$_4\times$S$^7$, and the pp wave. Let us discuss the relation between the standard parameterizations of these backgrounds and the elliptic coordinates.

Since elliptic coordinates are only defined in flat space, we begin with rewriting the $(X,Y)$ sector of (\ref{RotToTrnsl}) in a conformally--flat form. It turns out that this is accomplished by going to coordinates $(\xi,\eta)$ defined by (\ref{Ward}), and metric (\ref{RotToTrnsl}) becomes
\bea\label{Oct8}
ds_{base}^2=\zeta^2\left\{\left[(\d_\zeta\d_\eta V)^2+(\d^2_\eta V)^2\right]\left(d\zeta^2+d\eta^2\right)+
d\phi^2\right\}
\eea
After introducing the standard polar parameterization $(r, \theta)$,
\bea\label{DefZetaEta}
\zeta=r\sin\theta,\quad \eta=r\cos\theta,
\eea
one can define the elliptic coordinates by (\ref{ElliptCase}).  We will now compare these coordinates with the standard parameterization of AdS$_7\times$S$^4$ and AdS$_4\times$S$^7$.

\bigskip
\noindent
\textbf{Standard parameterization of AdS$_7\times$S$^4$.}

Solution corresponding to AdS$_7\times$S$^4$,
\bea\label{AdS7S4}
ds^2=4L^2\left[d\rho^2+\sinh^2\rho d\Omega_5^2-\cosh^2\rho dt^2\right]+L^2\left[d\chi^2+\sin^2\chi d\Omega_2^2+\cos^2\chi d{\tilde\phi}^2\right],
\eea 
is given by equation (3.15) in \cite{LLM} with a replacement 
$r\rightarrow 2\sinh\rho$:
\bea
x_1+ix_2=\cosh^2\rho\cos\chi e^{i\phi},\qquad
Y=\frac{1}{L^3}\sinh^2\rho\sin\chi,\quad 
e^D=\frac{1}{L^{6}}\tanh^2\rho.
\eea
Substitution into (\ref{Ward}) gives the expression for $\zeta$ and equations for $V$: 
\bea\label{WardA}
\zeta=\frac{1}{2L^3}\sinh(2\rho)\cos\chi,\quad 
\d_\zeta V=\tanh\rho\tan\chi,\quad
\d_\eta V=\ln\left[\cosh^2\rho\cos\chi\right].
\eea
Coordinate $\eta$ can be determined using the relation $d\eta=\star_2 d\zeta$ (recall (\ref{Oct8}) and $(\rho,\chi)$ sector of (\ref{AdS7S4})):
\bea\label{etaA}
\eta=\frac{1}{2L^3}\cosh(2\rho)\sin\chi.
\eea
For completeness we also write the expression for $V$, which comes from integrating the differential equations in (\ref{WardA}), although it will not play any role in our discussion:
\bea
V=\frac{1}{2L^3}\sin\chi\left[\cosh(2\rho)\ln(\cos\chi\cosh^2\rho)-1\right]+
\frac{1}{2L^3}\ln\left[\tan\frac{\chi+\frac{\pi}{2}}{2}\right].\nonumber
\eea

To deduce the elliptic coordinates, we begin with finding the counterpart of equations (\ref{Jul20}) by rewriting the left--hand sides of (\ref{DefZetaEta}) in terms of $(\rho,\chi)$:
\bea\label{Jul20a}
\frac{1}{2L^3}\sinh(2\rho)\cos\chi=r\sin\theta,\qquad
\frac{1}{2L^3}\cosh(2\rho)\sin\chi=r\cos\theta.
\eea
Substitution of these relations into the definition (\ref{ElliptCase}) with $R_1=1/(2L^3)$, leads to identification of the elliptic coordinates 
$(x,y)$ with $(\rho,\chi)$ (compare with equation (\ref{XYellAdS5})):
\bea\label{XYellAdS7}
x=2\rho,\quad y=\frac{\pi}{2}-\chi.
\eea
This implies that the standard parameterization of AdS$_7\times$S$^4$ has a simple geometrical meaning: coordinates $(\rho,\chi)$ coincide with elliptic coordinates on the two--dimensional space spanned by $(\eta,\zeta)$.

\bigskip
\noindent
\textbf{Standard parameterization of AdS$_4\times$S$^7$.}

Solution corresponding to AdS$_4\times$S$^7$,
\bea
ds^2&=&L^2\left[d\rho^2+\sinh^2\rho d\Omega_2^2-\cosh^2\rho dt^2\right]+4L^2\left[d\chi^2+\sin^2\chi d\Omega_5^2+\cos^2\chi d{\tilde\phi}^2\right],\nonumber
\eea 
is given by equation (3.16) in \cite{LLM} with a replacement 
$r\rightarrow 2\sinh\rho$:
\bea
x_1+ix_2=\sqrt{\cosh\rho}
\cos\chi e^{i\phi},\qquad
Y=\frac{1}{L^3}\sinh\rho\sin^2\chi,\quad 
e^D=\frac{4}{L^{6}}\cosh\rho\sin^2\chi.
\eea
Substitution into (\ref{Ward}) gives the expression for $\zeta$ and $\eta$:
\bea\label{WardB}
\zeta=\frac{1}{L^3}\cosh\rho\sin(2\chi),\quad
\eta=-\frac{1}{L^3}\sinh\rho\cos(2\chi).
\eea
For completeness we also give the equations for $V$ and their solution:
\bea
&&\qquad\qquad\d_\zeta V=\frac{1}{2}\tanh\rho\tan\chi,\quad
\d_\eta V=\ln\left[
\sqrt{\cosh\rho}\cos\chi\right]\nonumber\\
&&V=\frac{1}{2L^3}\left[\sinh\rho-2\arctan\left[\tanh\frac{\rho}{2}\right]-2\cos(2\chi)\sinh\rho\ln\left[\sqrt{\cosh\rho}\cos\chi\right]\right].
\nonumber
\eea

Combining (\ref{WardB}), and analog of (\ref{DefZetaEta})\footnote{We redefined angle $\theta$ in (\ref{DefZetaEta}). Alternatively, one can keep (\ref{DefZetaEta}) and shift $\theta$ in (\ref{ElliptCase}).},
\bea
\zeta=r\cos\theta,\quad \eta=-r\sin\theta.
\eea
and (\ref{ElliptCase}) with $R_1=1/(L^3)$, we identify the elliptic coordinates $(x,y)$ with $(\rho,\chi)$ (compare with equations (\ref{XYellAdS5}) and (\ref{XYellAdS7})):
\bea\label{XYellAdS4}
x=\rho,\quad y=\frac{\pi}{2}-2\chi.
\eea

\bigskip
\noindent
\textbf{Standard parameterization of AdS$_3\times$S$^3$.}

As our final example of elliptic coordinates, we consider AdS$_3\times$S$^3$ in global parameterization, which can be obtained by taking the near horizon limit ($H\rightarrow Q/f$) in (\ref{D1D5Jprime}) \cite{BalMM}:
\bea\label{MetrAdS3}
ds^2&=&Q\left[-((r')^2+a^2)\frac{dt^2}{Q^2}+\frac{(dr')^2}{(r')^2+a^2}+\frac{(r'du)^2}{Q^2}+
(d\theta')^2+\sin^2\theta' d{\tilde\phi}^2+\cos^2\theta' d{\tilde\psi}^2\right],\nonumber\\
&&{\tilde\phi}=\phi+\frac{a}{Q}t,\qquad {\tilde\psi}=\psi+\frac{a}{Q}u.
\eea
Rewriting the metric in terms of the elliptic coordinates defined by 
(\ref{RYprime}):
\bea\label{AdS3Elps}
ds^2=Q\left[-\cosh^2 x\frac{(adt)^2}{Q^2}+dx^2+\sinh^2 x\frac{(adu)^2}{Q^2}+
dy^2+\sin^2y d{\tilde\phi}^2+\cos^2y d{\tilde\psi}^2\right],\nonumber\\
\eea
we conclude that these coordinates give the standard parameterization of AdS$_3\times$S$^3$.

\bigskip
\noindent
\textbf{Pp--wave limits of AdS$_p\times$S$^q$.}

We conclude this section by commenting on the pp--wave limits of AdS$_p\times$S$^q$. 

The pp--wave limit of AdS$_5\times$S$^5$,
\bea\label{AdS5pp}
ds^2=-2dt dx_1-(r_1^2+r_2^2)dt^2+dr_1^2+r_1^2d\Omega^2+dr_2^2+r_2^2d{\tilde\Omega}^2,
\eea
is obtained by taking
\bea
z=\frac{1}{2}\frac{x_2}{\sqrt{x_2^2+Y^2}},\quad V=\frac{1}{2}\frac{1}{\sqrt{x_2^2+Y^2}}dx_1
\eea 
in (\ref{GenBubbleM}) and setting
\bea
Y=r_1r_2,\quad x_2=\frac{1}{2}(r_1^2-r_2^2).
\eea
This leads to a very simple relation for the polar coordinates defined by (\ref{StripRed})
\bea\label{PolarA5S5}
r_1=\sqrt{2}r\cos\frac{\theta}{2},\quad r_2=\sqrt{2}r\sin\frac{\theta}{2}.
\eea
Equation for geodesics in the geometry (\ref{AdS5pp}) separates in variables $(r_1,r_2)$, which can be obtained from the elliptic coordinates (\ref{Jul20})--(\ref{XYellAdS5}) by taking the pp--wave limit: 
\bea\label{PpAdS5}
L\rightarrow \infty,\qquad \mbox{fixed}\quad r_1=Lx,\quad r_2=Ly.
\eea
We conclude that in the pp--wave limit, the elliptic coordinate degenerate into the radii of the three--spheres. 

\bigskip

The pp--wave limit of AdS$_p\times$S$^q$ in M theory,
\bea\label{PPM}
ds^2=-2dtdx_1-(r_2^2+r_5^2)dt^2+dr_2^2+r_2^2d\Omega_2^2+
dr_5^2+r_5^2d{\Omega}_5^2
\eea
is given by equation (3.14) of \cite{LLM}:
\bea
Y=\frac{r_5^2 r_2}{4},\quad 
x_2=\frac{r_5^2}{4}-\frac{r_2^2}{2},\quad 
e^D=\frac{r_5^2}{4}
\eea
This translates into
\bea
\zeta=\frac{r_5}{2},\quad \eta=\frac{r_2}{2},\quad
V=\frac{r_2r_5^2}{8}-\frac{r_2^3}{12}=\zeta^2\eta-\frac{2}{3}\eta^3
\eea
via (\ref{Ward})\footnote{Since we are dealing with translational rather than rotational symmetry, ${\tilde D}=D$ and $X=x_2$ in (\ref{Ward})}. These expressions can be obtained from (\ref{WardA})--(\ref{etaA}) or from (\ref{WardB}) by taking the large--$L$ limits, and in both cases we arrive at a counterpart of (\ref{PpAdS5}):
\bea
L\rightarrow \infty,\qquad \mbox{fixed}\quad r_2=Lx,\quad r_5=L\left(\frac{\pi}{2}-y\right).
\eea
As in the case of the type IIB pp--waves, we conclude that the elliptic coordinates degenerate into the radii of the spheres. 

\bigskip

The pp--wave limit of the AdS$_3\times$S$^3$ geometry (\ref{AdS3Elps}) is obtained by 
writing
\bea
y=\frac{\hat y}{\sqrt{Q}},\quad x=\frac{\hat x}{\sqrt{Q}},\quad
u=\frac{Q\hat u}{a},\quad t=\frac{Q}{2a}\left(
{\hat x}^++\frac{{\hat x}^-}{Q}\right),\quad
{\tilde\psi}=\frac{1}{2}\left({\hat x}^+-\frac{{\hat x}^-}{Q}\right)
\eea
and sending $Q$ to infinity, while keeping all hatted variables fixed. This results in the metric
\bea
ds^2=-d{\hat x}^+d{\hat x}^--\frac{1}{4}({\hat x}^2+{\hat y}^2)(d{\hat x}^+)^2
+d{\hat x}^2+{\hat x}^2 d{\hat u}^2+
d{\hat y}^2+{\hat y}^2 d{\tilde\phi}^2.\nonumber
\eea
Once again, elliptic coordinates degenerate to the radii of the one--spheres. 

\bigskip

To summarize, we have demonstrated that in all examples of  AdS$_p\times$S$^q$, where the HJ equation separates between the sphere and AdS in global coordinates, such separation emerges as a particular case of integrability in elliptic coordinates, and standard parameterization of AdS$_p\times$S$^q$ coincides with elliptic coordinates on the relevant flat base. In the pp--wave limits of AdS$_p\times$S$^q$, the elliptic coordinates reduce to the radii of the appropriate spheres.

\section{Discussion}

Integrability of geodesics and Klein--Gordon equation has led to numerous insights into physics of black holes. While the black hole solutions are few and far between, the large classes of supersymmetric geometries are known, and in this chapter we have classified such solutions with integrable geodesics. This integrability is demonstrated to imply that the HJ equation must separate in the elliptic coordinates. 
For branes with flat worldvolumes, such separation, that extends the known result for the spherical coordinates, can only occur for special distributions of sources, which are analyzed in section \ref{SecGnrGds}. For the curved supersymmetric branes, the elliptic coordinates can only be introduced in the most symmetric cases, and as demonstrated in section \ref{Bubbles}, all these situations reduce to AdS$_p\times$S$^q$ or their pp--wave limits. 

Our results rule out integrability of $\mathcal{N}=4$ SYM beyond the large N limit. Specifically, we proved that the excitations of strings around heavy supersymmetric states ($\Delta\sim N^2$) are not integrable. While this is consistent with general expectations, it is somewhat surprising that none of the 1/2--BPS geometries give rise to integrable sectors. It would be interesting to extend this result to states with fewer supersymmetries. 

Our results also have unfortunate consequences for the technical progress in the fuzzball program. While a large number of geometries corresponding to microscopic states of black holes have been constructed in the last decade, the detailed calculations of the absorption/emission rates have only been performed for the simplest cases. Such calculations are based on solving the Klein--Gordon equation, and as we demonstrated in section \ref{Rotations}, this equation, as well as the Hamilton--Jacobi equation for geodesics, cannot separate beyond the known cases. Our results do not imply that a study of geodesics on a particular background is hopeless. A lot of useful information can be extracted by performing numerical integration of the equations of motion and by studying some special configurations rather than generic geodesics.

		\chapter{Killing--Yano tensors in string theory}\label{ChapterKillingYano}

		In the previous chapter we studied motion of massless geodesics to classify all potentially integrable backgrounds produced by D branes. It was mentioned that separability of the equations of motion for massless geodesics is encoded in so-called Killing tensors, so in this chapter we extend our analysis and study these objects in details. In particular, we construct the Killing(-Yano) tensors for a large class of charged black holes in higher dimensions and study general properties of such tensors, in particular, their behavior under string dualities. Killing(-Yano) tensors encode the symmetries beyond isometries, which lead to insights into dynamics of particles and fields on a given geometry by providing a set of conserved quantities. By analyzing the eigenvalues of Killing tensors, we provide a prescription for constructing several conserved quantities starting from a single object, and we demonstrate that Killing tensors in higher dimensions are always associated with ellipsoidal coordinates. We also determine the transformations of Killing(-Yano) tensors under string dualities, and find the unique modification of the Killing-Yano equation consistent with these symmetries. These results are used to construct the explicit form of Killing(-Yano) tensors for the Myers-Perry black hole in arbitrary number of dimensions and for its charged version.

\bigskip

This chapter is based on the results published in \cite{KYTinSt} and has the following organization. 

In sections \ref{SectionKTKYT} and \ref{SecKYTranks} we review some well-known properties of Killing(--Yano) tensors, and in section \ref{SecSeparKT} we rewrite them in a slightly unusual form which becomes crucial for the subsequent discussion. Usually one uses a Killing tensor (KT) to produce a conserved quantity which leads to separation of the Hamilton--Jacobi and Klein--Gordon equations, and only one such quantity can be constructed from a given Killing tensor. In section  \ref{SecSeparKT} we argue that if one looks further and studies \textit{eigenvalues} of a Killing tensor, then a single KT can lead to a \textit{family} of conserved quantities since the detailed analysis of eigenvalues allows one to construct a family of Killing tensors from a single representative using an \textit{algebraic} procedure (i.e., without solving differential equations). As a bi--product of this analysis we also demonstrate that separation caused by nontrivial Killing tensors in any number of dimensions can only happen in (degenerate) ellipsoidal coordinates, this generalizes the earlier result of \cite{ChL} to non--supersymmetric geometries. In section \ref{SecKYTranks} we also show that eigenvectors of Killing tensors lead to simple expressions for  Killing--Yano tensors when the latter exist.

After developing this general technology in section \ref{SecMyersPerry} we use it to write Killing--Yano and Killing tensors for the Myers--Perry black holes \cite{MyersPerry} in arbitrary number of dimensions with arbitrary number of rotations. In section \ref{SecMPF1} this construction is extended to charged solutions built from Myers--Perry geometries by application of the solution--generating dualities, and relatively simple explicit expressions for the Killing(--Yano) tensors are derived.

The general effects of string dualities on Killing(--Yano) tensors are discussed in section \ref{SecKillingsDualities}, where it is demonstrated that Killing vectors (KV) and Killing tensors survive under dualities if certain conditions on the Kalb--Ramond field are satisfied. The resulting transformations for the KV and KT are derived\footnote{For Killing vectors, a very nice interpretation of the transformation law in terms of the Double Field Theory \cite{DFT}  is discussed in section \ref{SecKVDuality}, but unfortunately a natural embedding of KT and KYT in this formalism is still missing.}. For the Killing--Yano tensors (KYT) the situation is rather different: while dualities generically destroy the standard KYT, they preserve the modified version of the KYT equation, which is derived in section \ref{SectionModifiedKYT}. We also demonstrate that such duality--invariant modification is unique and derive the transformation laws for the Killing--Yano tensor.  Several examples of the modified KY tensors are discussed in section \ref{SecExamplesF1NS5}. 

While studying massless particles, one encounters Conformal Killing(--Yano) tensors (CKT and CKYT), interestingly their behavior under string dualities has some unusual aspects. The conformal objects are discussed throughout the chapter along with their standard counterparts. 
Some technical details are presented in appendices.

\section{Introduction and summary}\label{SectionIntro}

Symmetries of dynamical equations have always played very important role in string theory. Conformal symmetry of the worldsheet led to Polyakov's reformulation of the theory \cite{Polyakov}, making it amenable to quantization, and provided powerful tools for performing calculations \cite{BPZ}. Study of string dualities \cite{duality} led to great insights into dynamics of string theory at strong coupling and to formulation of the gauge/gravity duality \cite{AdSCFT}. More recently discovery of hidden symmetries of equations for a classical string led to the discovery of integrability \cite{RoibPolch,bethe}, which stimulated a great progress in understanding of string dynamics and gauge/gravity duality (see \cite{review_AdSCFT_int} for the review and list of references). To gain additional insights into properties of quantum gravity and strong interactions it is very important to look for new examples of integrable string backgrounds. Since at low energies strings behave as point--like particles, integrable structures must give rise to hidden symmetries of supergravity, which will be investigated in this chapter. 

Integrability of classical strings on certain backgrounds is guaranteed by an infinite number of conserved quantities which can be extracted from reformulating the dynamical equations as a linear Lax pair \cite{LaxPair}. Unfortunately, there is no algorithmic procedure for constructing such pairs, and they have to be guessed. 
Interestingly, there exists a procedure for demonstrating that a particular background does not have a Lax pair, and it has been applied in \cite{Zayas,StepTs} to rule out several promising candidates, such as strings on a conifold and on asymptotically--flat geometry produced by D3 branes. Unfortunately, this procedure for ruling out integrability is rather complicated, and it has to be applied on a case--by--case basis, so in \cite{ChL} we used a different approach based on the study of geodesics. Since at low energies strings behave as point particles, integrability must survive as a hidden symmetry of such objects, and this gives a very coarse necessary condition for integrability, which can be tested for large classes of backgrounds. Interestingly, this condition was sufficient for ruling out integrability on all known supersymmetric  geometries produced by D--branes, with an exception of AdS$_p\times$S$^q$ and a couple of other examples \cite{ChL}. Of course, to analyze the integrability of geodesics one has to start with explicit solutions, and the nontrivial integrable deformations of  AdS$_p\times$S$^q$ \cite{margin,Eta} had to be constructed using special techniques rather than obtained as members of known families\footnote{Analysis of \cite{ChL} focused only on geometries supported by the Ramond--Ramond fluxes, which allowed us to analyze very large families. The `isolated points' discussed \cite{margin,Eta} contained mixed fluxes, and they would have survived the analysis of \cite{ChL} had it been performed. Integrability of strings on the beta--deformed backgrounds \cite{margin} has been discussed in \cite{BetaIntegr,Frolov}.}. This chapter is a continuation of the  program initiated in \cite{ChL}: it extends the earlier results to geometries without supersymmetry, and, more importantly, it uncovers the hidden symmetries underlying integrability of geodesics. In spite of this continuity, this chapter does not require familiarity with \cite{ChL}.

Study of geodesics has a long history in general relativity, and the most powerful methods are based on the analysis of the Hamilton--Jacobi (HJ) equation. It is well-know that such equation separates if the background contains cyclic (ignorable) directions, but sometimes separation happens even between non--cyclic coordinates. The simplest example of such `accidental separation' comes from the three--dimensional flat space in spherical coordinates: the polar angle $\theta$ separates in the HJ equation, although the metric depends on this coordinate. In this case the separation can be attributed to the SU(2) symmetries of the sphere, but similar argument cannot be applied to the Kerr black hole, which has only U(1)$\times$U(1) isometry, although the $\theta$ coordinate still separates.  The technical aspects of this separation will be reviewed in section \ref{SecSeparKT}, and here we just recall that the separation is associated with a hidden symmetry encoded in the Killing tensor (KT) \cite{Carter,Penrose}. The same tensor also leads to separation of the Klein--Gordon equation even beyond the eikonal approximation. The Kerr metric also gives rise to separable Dirac equation, this is guaranteed by an additional symmetry encoded in the Killing--Yano tensor (KYT) \cite{Yano}.  Over the last four decades Killing(--Yano) tensors have been found for other geometries both in general relativity \cite{GRyano} and in string theory \cite{KYdual,StrYano}, and in this chapter we will construct KYT for a large class geometries in arbitrary numbers of dimensions, which contains most of the known examples as special cases. 

Killing(--Yano) tensors encode all continuous symmetries of solutions in general relativity, but string theory also has discrete symmetries associated with dualities, which can be promoted to a continuous group of 
{solution-generating transformations} in supergravity. This leads to a very natural question: what happens  with Killing(--Yano) tensors under action by this group? Answering this question is one of the main goals of this chapter. A slightly different question was answered in the article \cite{KYdual}, which identified the subset of duality transformation leaving the Killing--Yano tensor invariant. As we will see, in general both Killing and Killing--Yano tensors are changed by the dualities, even the 
\textit{equation for the KYT is modified}. However, for the special cases discussed in  \cite{KYdual} our results agree with that paper. In this chapter we focus on dualities in the NS--NS sector since our preliminary study of the Ramond--Ramond backgrounds indicates that T duality applied to such geometries may change the rank of the KYT and even produce Killing--Yano tensors of mixed rank. A very brief discussion of this point is given in section \ref{SectionModifiedKYT}. 


\section{Killing(--Yano) tensors in higher dimensions}
\label{SectionKTKYTGeneral}
\subsection{Killing tensors and Killing--Yano tensors}
\label{SectionKTKYT}

Symmetries play very important role in physics, and symmetries of geometries are encoded in Killing vectors and Killing tensors. In this section we will review some well--known properties of these objects and establish the notation which will be used in the rest of the chapter. 

We begin with recalling that  the Killing vector (KV) is defined as a vector field $V$ which leaves the metric invariant. In other words, the Lie derivative of the metric along $V$ must vanish:
\bea\label{KVDefA}
{\mathcal L}_V g_{MN}=0,
\eea
Relation (\ref{KVDefA}) can be rewritten as 
\bea
{\mathcal L}_V g_{MN}=V^P\d_P g_{MN}+\d_M V^P g_{PN}+\d_N V^P g_{MP}=
\nabla_M V_N+\nabla_N V_M=0,
\eea
and it implies that the metric does not change under an infinitesimal transformation
\bea 
x'^M=x^M+\epsilon V^M.
\eea
Since Killing vectors encode symmetries, they are always associated with conserved quantities. Specifically, the expression
\bea
I=V_M\frac{dx^M}{ds}
\eea 
is conserved along any geodesic.

The correspondence between Killing vectors and integrals of motion is not one--to-one: some conserved quantities are not associated with KV. However, it was shown by Penrose and Walker \cite{Penrose} that any integral of motion that depends on momentum comes either from a Killing vector or from a rank--two Killing tensor as
\bea
I=K_{MN}\frac{dx^M}{ds}\frac{dx^N}{ds},
\eea 
where $K_{MN}$ satisfies a linear equation
\bea\label{KTeqnDef}
\nabla_{M}K_{NP}+\nabla_{N}K_{MP}+\nabla_{P}K_{MN}=0.
\eea
To determine whether the integrals of motion survive in quantum theory as well, one should analyze 
separability of the Klein--Gordon equation, and as shown in \cite{Miller3}, the relevant conserved quantity must be associated with eigenvalues of the differential operator 
\bea
{\hat K}\equiv \frac{1}{\sqrt{-g}}\d_M\left[\sqrt{-g}K^{MN}\d_N\right]+k(x)
\eea
with some function $k(x)$. As demonstrated in \cite{Moon,Miller3}, operator ${\hat K}$ commutes with $\nabla_M\nabla^M$ if and only if $K^{MN}$ satisfies equation (\ref{KTeqnDef}) and one more condition which will not be discussed here.

In general, presence of the Killing tensor does not imply separability of the Dirac equation, this requires existence of an anti--symmetric Killing--Yano tensor (KYT) $Y_{MN}$ which satisfies the defining equation
\cite{Yano}
\bea\label{KYTdef}
\nabla_M Y_{NP}+\nabla_N Y_{MP}=0.
\eea
This equation can be generalized to tensors of arbitrary rank as \cite{Tachibana} 
\bea\label{DefKYT}
\nabla_{(M}Y_{N)P_1\dots P_{k-1}}=0,\quad Y_{P_1\dots P_k}=Y_{[P_1\dots P_k]}.
\eea
In four dimensions KYT of rank $k>2$ can be dualized into vectors and scalars, but in string theory one encounters interesting solutions of (\ref{DefKYT}), which will be discussed  throughout this chapter. It is also possible to define Killing tensors of rank $k>2$ as solutions of the equation \cite{Penrose}
\bea
\nabla_{(M_1}K_{M_2\dots M_{k+1})}=0,
\eea
but such objects will not play any role in our discussion.

Any KYT gives rise to a Killing tensor of rank two via the relation
\bea\label{KTfromKYTodrin}
{K}_{MN}={Y}_M{}^{A_1\dots A_{k-1}}{Y}_{NA_1\dots A_{k-1}}.
\eea
This equation has a simple interpretation: separability of the Dirac equation implies one for the Klein--Gordon equation in the same coordinates. In section \ref{SecKTandHJ} we will present a detailed analysis of Killing tensors and outline a procedure for ``extracting the square root'' from them which allows one to construct the Killing--Yano tensors, if they exist.

So far we discussed the integrals of motion for massive particles, but some additional symmetries might arise in the massless case. For example, while the metric
\bea
ds^2=dr^2+r^2d\phi^2
\eea
is not invariant under rescaling of $r$ coordinate, massless particles are not sensitive to such rescaling, so while 
\bea
{\mathcal V}=r\d_r
\eea
is not a Killing vector, it does lead to conserved quantities for \textit{massless} particles. Such \textit{conformal Killing vectors} (CKV) satisfy equation
\bea
\nabla_M {\mathcal V}_N+\nabla_N {\mathcal V}_M=vg_{MN},
\eea
where $v$ is an arbitrary functions of all coordinates. If $v$ is a constant, then the corresponding CKV is called homothetic \cite{Homothetic}, and such vectors will play an important role in the analysis presented in section \ref{SubsectionCKV}.

Recall that the conformal Killing(--Yano) tensors (CKT and CKYT) are defined as solutions of equations \eqref{CKTdef}
\bea\label{CKTdefKYT}
&&\nabla_{(M_1}{\mathcal K}_{M_2...M_{k+1})}=W_{(M_1...M_{k-1}}g_{M_{k}M_{k+1})},\\
&&\nabla_{(M_1}{\mathcal Y}_{M_2)...M_{k+1}}=g_{M_1M_2}Z_{M_3...M_{k+1}}+
\sum_{i=3}^{k+1}(-1)^ig_{M_i(M_1}Z_{M_2)...{M_{i-1}}M_{i+1}...M_{k+1}}.\nonumber
\eea
with coordinate--dependent tensors $W$ and $Z$. Notice that under rescaling of the metric, CKV, CKT and CKYT transform in a simple way\footnote{The relevant transformations are derived in Appendix \ref{AppConfResc}.}, so they survive S duality and transition from the string to the Einstein frame. Ordinary Killing vectors have the same feature, as long as we impose a reasonable restriction on the dilaton:
\bea
{\mathcal L}_V e^{2\Phi}=V^M\d_M e^{2\Phi}=0.
\eea
On the other hand, the ordinary KT and KYT are usually destroyed by coordinate--dependent rescaling of the metric, so they exist only in one frame. Conformal transformations of the KT and KYT  are discussed in Appendix \ref{AppConfResc}.

We will mostly focus on rank--2 KT and CKT, and they can be constructed by squaring KYT or CKYT:
\bea\label{KTfromKYT}
{\mathcal K}_{MN}={\mathcal Y}_M{}^{A_1\dots A_{k-1}}{\mathcal Y}_{NA_1\dots A_{k-1}}, \quad 
W_{M}=2{\mathcal Y}_{MA_1\dots A_{k-1}}Z^{A_1\dots A_{k-1}}.
\eea
For rank-1 and rank--2 (C)KYT this construction is well-known, and direct computation shows that it works for all $k$. 

Conformal Killing tensors ${\mathcal K}_{MN}$ with $W_M=-\nabla_M \phi$ have a special property: they can be extended to the standard KT $K_{MN}$ by
\be\label{KTfromCKT}
K_{MN}={\mathcal K}_{MV}+\phi g_{mn}.
\ee
To see this one can take a covariant derivative of \eqref{KTfromCKT} and symmetrize the result:
\bea
\nabla_{(M}K_{NP)}=\nabla_{(M} {\mathcal K}_{NP)}+\nabla_{(M} \phi g_{NP)}=0.
\eea
This construction will be illustrated in section \ref{SecKYTranks} by comparing KT and CKT for  rotating black holes.

\subsection{Killing tensors and the Hamilton--Jacobi equation}
\label{SecKTandHJ}
\label{SecSeparKT}

Solutions of the equation for the KT,
\bea\label{KTeqn4}
\nabla_P K_{MN}+\nabla_M K_{NP}+\nabla_N K_{PM}=0
\eea
form a linear space, in particular, a `trivial subspace' is spanned by combinations of the metric and Killing vectors,
\bea\label{Ktriv}
K^{triv}_{MN}=e_0 g_{MN}+\sum_{i,j} e_{ij}V^{(i)}_MV^{(j)}_N,
\eea
with constant coefficients $e_0$, $e_{ij}$. In this subsection we will establish a one--to--one correspondence between \textit{nontrivial} Killing tensors and separation of variables in the Hamilton--Jacobi equation \eqref{HJone}
\bea\label{HJoneKYT}
g^{MN}\d_M S\d_N S+\mu^2=0.
\eea

\subsubsection{Killing tensors from the Hamilton--Jacobi equation}

There are several notions of separability for equation (\ref{HJoneKYT}), and we focus on the standard one by assuming that 
\bea\label{SeparSimple}
S=S(x_1,\dots x_k)+S(x_{k+1}\dots x_n).
\eea
This assumption can be generalized to R--separability as
\bea\label{SeparCompl}
S=S(x_1,\dots x_k)+S(x_{k+1}\dots x_n)+S_0(x_1\dots x_n),
\eea
where $S_0(x_1\dots x_n)$ is a \textit{known} function of its arguments\footnote{The counterpart of (\ref{SeparCompl}) for the Schr{\"o}dinger equation is $$\Psi=X(x_1\dots x_k)Y(x_{k+1}\dots x_n)\Psi_0(x_1\dots x_n)$$ with \textit{known} function $\Psi_0$. For non-trivial $\Psi_0$ this is known as R--separation \cite{Miller3}.} \cite{MorseFesh}. However, this generalization will not play any role in our discussion.

Equation (\ref{HJoneKYT}) separates as (\ref{SeparSimple}) if and only if three conditions are satisfied:
\begin{enumerate}[(a)]
\item{Coordinates $x^M$ can be divided into cyclic coordinates $z$ and two other groups, which will be denoted by $x$ and $y$. The metric does not depend on coordinates $z$.}
\item{There exists a separation function $f$, such that
\bea\label{gEigenTens}
 g^{MN}=\frac{1}{f}\left(X^{MN}+Y^{MN} \right),\quad
 \d_x Y^{MN}=\d_y X^{MN}=0,\ X^{y^i M}=0,\ Y^{x^iM}=0.
\eea
}
\item{Function $f$ can be decomposed as
\bea\label{Eqn6Other}
f=f_x-f_y,\qquad \d_y f_x=0,\quad \d_x f_y=0,\quad \d_zf_x=\d_z f_y=0.
\eea
}
\end{enumerate} 
Conditions (a)--(c) allow us to rewrite equation (\ref{HJoneKYT}) as
\bea
X^{MN}\d_M S\d_N S+\mu^2 f_x=-Y^{MN}\d_M S\d_N S+\mu^2 f_y,
\eea
where the left--hand side depends only on $x$, and the right--hand side depends only on $y$. This implies that 
\bea\label{SeparInt}
I\equiv\left[X^{MN}-f_xg^{MN}\right]\d_M S\d_N S
\eea
must be an integral of motion, and as such it must be associated with a Killing tensor:
\bea\label{IntKT}
I=K^{MN}\d_M S\d_N S.
\eea
We conclude that separation of variables (a)--(c) is associated with Killing tensor
\bea\label{KeigenTens}
K^{MN}=X^{MN}-\frac{f_x}{f}\left( X^{MN}+Y^{MN} \right)=-\frac{f_yX^{MN}+f_xY^{MN}}{f}.
\eea
If condition (c) is not satisfied, then equation (\ref{HJoneKYT}) separates only for $\mu=0$, and the associated \textit{conformal} Killing tensor is 
\bea
K^{MN}=X^{MN}.
\eea

After reviewing the standard procedure for extracting the Killing tensor from separation of variables \cite{Carter, Penrose}, we discuss the inverse problem: recovery of separation from a given Killing tensor.


\subsubsection{Separation of variables from Killing tensor}
\label{SecSeparKTsub}

Every Killing tensor gives rise to an integral of motion via (\ref{IntKT}), and such constant must be associated with separation of variables as in (\ref{SeparInt}). While the separation functions $(f_x,f_y)$ and the corresponding tensors $(X^{MN},Y^{MN})$ are encoded in the Killing tensor, extracting them requires further analysis, and as we will demonstrate,
 this analysis may lead to an entire family of the Killing tensors which can be constructed \textit{algebraically} from one representative. Schematically our results can be represented as
\bea\label{diagram}
\begin{array}{c}
\mbox{Eigenvalues}\\ \mbox{of KT}
\end{array}
\quad\Rightarrow\quad \mbox{separation} \quad\Rightarrow\quad 
\begin{array}{c}
m\mbox{--parameter}\\
\mbox{family of KTs}
\end{array}\, \quad\Leftrightarrow\quad
\begin{array}{c}
m\,\mbox{conserved}\\\mbox{charges}
\end{array}
\eea

To justify the usefulness of eigenvalues we recall equations (\ref{gEigenTens}) and (\ref{KeigenTens}):
\bea
g^{MN}=\frac{1}{f}\left(X^{MN}+Y^{MN} \right),\qquad
K^{MN}=-\frac{f_yX^{MN}+f_xY^{MN}}{f}
\eea
and consider an eigenvalue problem:
\bea\label{Keigen}
K^{MN}Z_N=\Lambda g^{MN}Z_N.
\eea
Assuming that metric has at least one non--cyclic direction\footnote{This assumption is violated only for flat space in Cartesian coordinates.} $x$ and that there is at least one component $K^{xN}\ne 0$, the $M=x$ component of (\ref{Keigen}) becomes
\bea
-\frac{f_y}{f}X^{xN}Z_N=\Lambda \frac{1}{f}\,X^{xN}Z_N\quad\Rightarrow\quad
\Lambda=-f_y.
\eea
In other words, some eigenvalues of the Killing tensor give the separation functions, and corresponding eigenvectors can be used to recover the relevant tensors $(X^{MN},Y^{MN})$. The cyclic coordinates complicate this construction, so they should be ignored to recover the separation function and added back in the end. Specifically, we propose the following procedure for extracting the separation function from the Killing tensor:
\begin{enumerate}[(1)]
\item Find the eigenvalues and eigenvectors of the KT:
\bea\label{KTanGmain}
K_{MN}=\sum_a \Lambda_a e^{(a)}_M e^{(a)}_N,\qquad 
g_{MN}=\sum_a  e^{(a)}_M e^{(a)}_N.
\eea
Notice that some eigenvalues may vanish of be degenerate.
\item Build the projectors\footnote{To avoid cumbersome formulas, we focus on non--degenerate eigenvalues. In general the left hand side of (\ref{ProjLam}) should refer to an eigenvalue $\Lambda$ and the right--hand side should contain summation over all $a$ with $\Lambda_a=\Lambda$. Since degeneracy clutters notation without introducing new effects, we use (\ref{ProjLam}).} 
\bea\label{ProjLam}
P^{(a)}_{MN}=e^{(a)}_M e^{(a)}_N.\nonumber
\eea
Projector $P$ will be called cyclic if
\bea\label{CyclProj}
\sum_N {[P^{(a)}]_M}^N\d_N \Lambda_b=0 \quad\mbox{for all}\quad (a,b).
\eea
If all projectors are cyclic, the Killing tensor can be built from Killing vectors and the metric.
\item Remove all directions associated with cyclic projectors and construct the reduced metric and Killing tensor:
\bea\label{KandGred}
K^{red}_{MN}=\left[{\sum_a} \Lambda_a e^{(a)}_M e^{(a)}_N\right]_{red},\qquad 
g^{red}_{MN}=\left[{\sum_a}  e^{(a)}_M e^{(a)}_N\right]_{red}.\nonumber
\eea
Non--cyclic components of equation (\ref{KTeqn4}) imply that $K^{red}_{MN}$ is a Killing tensor for  
$g^{red}_{MN}$. Nontrivial $K^{red}_{MN}$ and $g^{red}_{MN}$ imply that Killing tensor cannot be constructed from the Killing vectors and the metric. 
\item Separation of variables implies that 
\bea
\sum_M e^{(a)}_M dx^M=\sqrt{g_a}dx^a,\qquad \d_j\d_k\ln g_m=0\
\mbox{for different}\ (i,j,k).
\eea
Then analysis of the Killing equations shows that \textit{generically} the reduced metric and Killing tensor must have the form
\bea\label{KillLaMain}
&&ds^2_{red}=\sum_k g_k (dx_k)^2,\quad 
K_{red}=\sum_k \Lambda_k g_k (dx_k)^2,\nn
&&g_k=h_k(x_k)\prod_{j\ne k} [x_k-x_j],\quad \Lambda_j=\d_j\Lambda,
\eea
where $\Lambda(x_1\dots x_n)$ is a linear polynomial in every $(x_1\dots x_n)$ symmetric under interchange of every pair of arguments. 
\item
Separation of variables in the reduced metric is accomplished by multiplying the reduced HJ equation by
\bea\label{RhoKmain}
\rho_k=\prod_{j\ne k} [x_k-x_j].
\eea
Then the reduced HJ equation can be written as
\bea\label{SeparHJmain}
\frac{1}{h_k}(\d_k S)^2=\sum_{p=0}^{n-1} (x_k)^p I^{(k)}_p(x_1\dots x_{k-1},x_{k+1}\dots x_n),
\eea
which implies that all $I^{(k)}_p$ must be constant\footnote{Integrals of motion $I^{(k)}_p$ are closely related to the separation constants which arise from breaking the HJ equation into pieces using St{\"a}ckel determinant. A detailed discussion of the St{\"a}ckel's method can be found in chapter 5 of \cite{MorseFesh}.}. This construction separates variable $x_k$, and other coordinates can be separated in the same fashion 
\item
After coordinates $(x_1\dots x_n)$ have been constructed, cyclic directions can be added back, and upon multiplication by (\ref{RhoKmain}) the complete $d$--dimensional HJ equation takes the form (\ref{SeparHJmain}). This follows from the fact that $K$ from (\ref{KTanGmain}) was a Killing tensor for the $d$--dimensional metric.
\item
A given Killing tensor corresponds to a particular function $\Lambda$ in (\ref{KillLaMain}), and a family of Killing tensors for the reduced metric can be constructed by keeping the same coordinates and introducing an arbitrary polynomial $\Lambda$.
\end{enumerate}
Steps (1)--(7) outline our construction, and the details and justification are presented in the Appendix \ref{AppEllipsKYT}. A different approach to separation functions and Killing tensors was developed in \cite{KalMil1}, and our results are consistent with theirs. 

Expressions (\ref{KillLaMain}) generalize Jacobi's ellipsoidal coordinates \cite{Jacobi} to curved space, and we derived them assuming that the dependence on $(x_1\dots x_n)$ is generic. Specifically we assumed that $g_1$ depends on all $n$ coordinates. It is also possible to have some degenerate cases where some $x_j$ does not appears in $g_1$, but such solutions can be obtained by taking some singular limits of the ellipsoidal coordinates. In the appendix \ref{AllFlatEllips} we review such singular limits for the ellipsoidal coordinates in flat three--dimensional space.

\bigskip

To summarize, in this subsection we clarified the relation between Killing tensors and separation of variables.  
It is well--known that separation of variables leads to a Killing tensor, which is associated with a conserved quantity \cite{Carter, Penrose}, but in higher dimensions, where the metric can depend on three or more variables and may admit more than one nontrivial Killing tensor, the correspondence is more interesting. As illustrated in the diagram (\ref{diagram}), a single separation of variables may give rise to a family of Killing tensors, and the entire family can be constructed from a single member by studying its eigenvalues. In section \ref{SecMyersPerry} our construction will be applied to an important example of the Myers--Perry black hole, and in section \ref{SecMPF1} it will be extended to the charged version of that solution. But first we discuss the additional symmetry structures which appear when the geometry admits a Killing--Yano tensor.

\subsection{Killing--Yano tensors of various ranks}
\label{SecKYTranks}

While Killing--Yano tensors (KYT) of rank two are well-known from general relativity in four dimensions, the objects with higher rank are less familiar, so in this subsection we will present several examples of such Killing--Yano tensors and discuss their relation to Killing tensors.

Recall that the Killing--Yano tensors are defined as solutions of equation (\ref{DefKYT})
\bea\label{DefKYT1}
\nabla_{(M}Y_{N)P_1\dots P_{k-1}}=0,\quad Y_{P_1\dots P_k}=Y_{[P_1\dots P_k]}.
\eea
As reviewed in section \ref{SectionKTKYT}, any Killing--Yano tensor leads to a Killing tensor via (\ref{KTfromKYTodrin}). For example, any $d$--dimensional space admits a trivial KYT of rank $d$, which is defined as a volume form, and it squares to the metric. Nontrivial KYT may square to the metric as well, as illustrated by our first example: a space that has a factorized form
\bea
ds^2=g_{mn}(x)dx^mdx^n+h_{\mu\nu}(y)dy^\mu dy^\nu,
\eea
where two subspaces have the same dimensionality $n$. Then volume forms on $x$ and $y$ spaces give rise to a family of Killing--Yano tensors:
\bea\label{AdSKYT}
&&Y=c_1 \mbox{Vol}_{g}+c_2 \mbox{Vol}_{h}\quad\Rightarrow \nn
&&K_{MN}dX^MdX^N=(n-1)!\left[c_1^2 g_{mn}(x)dx^mdx^n+c_2^2 h_{\mu\nu}(y)dy^\mu dy^\nu\right].
\eea
It is clear that a non--trivial KY tensor can square to the metric as long as $c_1^2=c_2^2$. For generic values of constants $c_1$ and $c_2$ Killing tensor has two distinct eigenvalues, and each of them has 
degeneracy $n$.

A large class of geometries admitting Killing--Yano tensors comes from rotating black 
holes\footnote{Another interesting class of geometries admitting Killing--Yano tensors comes from putting 
D--branes on singular points of Calabi--Yau manifolds. Killing--Yano tensors for Sasaki-Einstein manifolds appearing in this construction have been recently constructed in \cite{BabVis}.}, and in the next section we will construct the KYTs for black holes with arbitrary number of rotations. Before performing this general analysis we review the situation for the well--known example of the Kerr black hole \cite{Kerr} and extract important lessons from it. The non--trivial Killing tensor for the Kerr geometry was constructed by Carter \cite{Carter}, and we begin with rewriting the metric in convenient frames defined as eigenvectors of that KT:
\bea\label{Kerr4D}
ds^2&=&-e_t^2+e_r^2+e_\theta^2+e_\phi^2,\nonumber\\
e_{t}&=&\frac{\sqrt{\Delta}}{\rho}(dt-as_\theta^2 d\phi),\quad 
e_{\phi}=\frac{s_\theta}{\rho}\left[(r^2+a^2)d\phi-adt\right],\quad
e_{r}=\frac{\rho}{\sqrt{\Delta}}dr,\quad e_{\mathbf\theta}=\rho d\theta,\nonumber\\
\Delta&=&r^2+a^2-2mr,\quad \rho^2=r^2+a^2c_\theta^2,\quad
c_\theta=\cos\theta,\quad s_\theta=\sin\theta.
\eea
Then expressions for the Killing and Killing--Yano tensors become very compact:
\bea\label{Kerr4DKT}
K&=&r^2\left[e_\phi^2+e_\theta^2\right]+(ac_\theta)^2\left[e_t^2-e_r^2\right],\quad
Y=r e_\theta\wedge e_\phi+(ac_\theta) e_r\wedge e_t.
\eea
We observe that the eigenvalues of $K$ ($r^2$ and $-(ac_\theta)^2$) appear in pairs, and $Y$ is constructed from these eigenvalues and corresponding eigenvectors in a simple way. As we will see in the next section, this double degeneracy persists in all even dimensions. Notice that the separating function defined in the previous subsection is equal to the difference of eigenvalues, and in the present case equation (\ref{Eqn6Other}) becomes
\bea
f_x=r^2,\quad f_y=-(ac_\theta)^2,\quad f=r^2+(ac_\theta)^2.
\eea

In odd dimensions the situation is different\footnote{Since the number of eigenvalues is odd, the double degeneracy is not possible. To avoid unnecessary complications, we write (\ref{5DKerrNtr}) for one rotation, more general case will be discussed in the next section.}, and to get some insights, we look at a rotating black hole in five dimensions \cite{MyersPerry}. Solving equations for the Killing--Yano tensor, constructing the corresponding KT, and defining the frames as its eigenvalues, we find
\bea\label{5DKerrNtr}
ds^2&=&-e_t^2+e_r^2+e_\theta^2+e_\phi^2+e_\psi^2,\nonumber\\
K&=&r^2\left[e_\phi^2+e_\theta^2\right]+(ac_\theta)^2\left[e_t^2-e_r^2\right]+
[r^2-(ac_\theta)^2]e_\psi^2,\\
Y&=&\left[r e_\theta\wedge e_\phi+(ac_\theta) e_r\wedge e_t\right]\wedge e_\psi.\nonumber
\eea
The frames are defined by
\bea\label{5DKerrNtrFrm}
e_{t}&=&\frac{\sqrt{\Delta}}{\rho}(dt-as_\theta^2 d\phi),\quad 
e_{\phi}=\frac{s_\theta}{\rho}\left[(r^2+a^2)d\phi-adt\right],\nonumber\\
e_{r}&=&\frac{\rho}{\sqrt{\Delta}}dr,\quad e_{\mathbf\theta}=\rho d\theta,\quad 
e_\psi=rc_\theta d\psi,\\
\Delta&=&r^2+a^2-M,\quad \rho^2=r^2+a^2c_\theta^2.\nonumber
\eea
Notice that eigenvalues of $K$ come in two pairs and one special value corresponding to $e_\psi$. In the next section we will demonstrate that this pattern persists in all odd dimensions with arbitrary number of rotations. As expected from (\ref{Eqn6Other}), the separating function $f$ is equal to the difference of two non--cyclic eigenvalues 
\bea
f_x=r^2,\quad f_y=-(ac_\theta)^2,\quad f=r^2+(ac_\theta)^2,
\eea
but now the Killing tensor has an additional eigenvector $e_\psi$ associated with cyclic coordinates, and the corresponding eigenvalue is 
\bea
\Lambda_\psi=f_x+f_y=r^2-(ac_\theta)^2.
\eea
Analysis of section \ref{SecKTandHJ} did not put any restrictions on cyclic eigenvectors and eigenvalues. 

In addition to the standard KYT, rotating black holes may admit a conformal KYT, which satisfies  equations (\ref{CKTdefKYT}) and gives rise to a conformal KT (CKT) via (\ref{KTfromKYT}). In particular, the CKYT and CKT for the Kerr metric \eqref{Kerr4D} are 
\bea\label{CKYTKerr4d}
\mathcal{Y}&=&re_r\wedge e_t-(ac_\theta)e_\theta\wedge e_\phi,\quad 
Z=dt-\frac{2mr}{\rho\sqrt{\Delta}}e_t,\nn
\mathcal{K}&=&r^2[e_t^2-e_r^2]+(ac_\theta)^2[e_\theta^2+e_\phi^2],\quad 
W=-d[r^2-a^2c_\theta^2],
\eea
and for the rotating black hole in five dimensions (\ref{5DKerrNtrFrm}) they are given by
\bea\label{CKYTrank2Kerr5d}
\label{CKYTKerr5d}
\mathcal{Y}&=&r e_r\wedge e_t-(ac_\theta)e_\theta\wedge e_\phi,\quad
Z=dt-\frac{M}{\rho\sqrt{\Delta}}e_t,\nn
\mathcal{K}&=&r^2[e_t^2-e_r^2]+(ac_\theta)^2[e_\theta^2+e_\phi^2],\quad
W=-d[r^2-a^2c_\theta^2].
\eea
Notice that vectors $W$ appearing in (\ref{CKYTKerr4d}) and (\ref{CKYTKerr5d}) are written as gradients of scalar functions, which means that they give rise to standard Killing tensors via (\ref{KTfromCKT}). Direct calculations show that application of 
(\ref{KTfromCKT}) to (\ref{CKYTKerr4d}) and (\ref{CKYTKerr5d}) leads to the Killing tensors given in (\ref{Kerr4DKT}) and (\ref{5DKerrNtr}). Conformal KYT 
(\ref{CKYTKerr4d}) and (\ref{CKYTKerr5d}) will play an important role in the general analysis presented in section \ref{SecKillingsDualities}. 

\section{Example: Killing--Yano tensors for the Myers--Perry black hole}
\label{SecMyersPerry}

In this section we construct a family of Killing--(Yano) tensors for the Myers--Perry black hole using the techniques introduced in section \ref{SecSeparKT}. The cases of odd and even dimensions have to be treated differently, so we begin with MP solution in even dimensions ($d=2n+2$) \cite{MyersPerry}:
\bea\label{MPeven}
ds^2&=&-dt^2+\frac{mr}{FR}\Big(dt+\sum_{i=1}^n a_i\mu_i^2 d\phi_i\Big)^2+\frac{FR dr^2}{R-mr}
+\sum_{i=1}^n(r^2+a_i^2)\Big(d\mu_i^2+\mu_i^2 d\phi_i^2\Big)\nn
&&+r^2d\alpha^2.
\eea
Here variables $(\mu_i,\alpha)$ are subject to constraint
\bea
\alpha^2+\sum_{i=1}^n\mu_i^2=1,
\eea
and functions $F$, $R$ are defined by 
\bea\label{MPdefFR}
F=1-\sum_{k=1}^n\frac{a_k^2\mu_k^2}{r^2+a_k^2},\quad R=\prod_{k=1}^n (r^2+a_k^2).
\eea

To find the KYT for the geometry (\ref{MPeven}) we observe that the square of the KYT gives a KT with some components along non--cyclic coordinates, so following the general procedure outlined in section 
\ref{SecSeparKT}, we begin with looking at the non--cyclic part of the metric:
\bea\label{MPevNC}
ds_{NC}^2=\frac{FR dr^2}{R-mr}+\sum_{i=1}^n(r^2+a_i^2)d\mu_i^2+r^2\Big(d[1-\sum_{i=1}^n\mu_i^2]^{1/2}\Big)^2.
\eea
As demonstrated in section \ref{SecSeparKTsub}, in the appropriate frames the Killing tensor and geometry (\ref{MPevNC}) must have the form\footnote{In this section we have to distinguish between $e^a=e^a_M dx^M$ and 
$e_a=e_a^M\d_M$, so the frame indices are written in the appropriate places. In the rest of the chapter we abuse notation and write $e_a=e^a_M dx^M$ to simplify formulas.}
\bea\label{ay2}
K_{mn}dx^m dx^n=\sum_m \Lambda_m (e^m)^2,\quad ds_{NC}^2=\sum_m (e^m)^2,
\eea
where 
\bea\label{ay2a}
e^m=h_m(x_m)\left[\prod_{k\ne m}[x_m-x_k]\right]dx^m,\quad 
\Lambda_m=\d_m \Lambda(x_0\dots x_{n}),
\eea
and $\Lambda$ is a symmetric polynomial linear in every argument. To determine the new coordinates $(x_1\dots x_{n+1})$ in terms of $(r,\mu_1\dots\mu_n)$ we begin with $m=0$ case when metric (\ref{MPevNC}) becomes flat and the relation between $(x_0\dots x_{n+1})$ and $(r,\mu_1\dots\mu_n)$ is given in terms of well--known ellipsoidal coordinates \cite{MorseFesh}:
\bea\label{EplsdEven}
x_0=r^2,\quad (a_i\mu_i)^2=\frac{1}{c_i^2}\prod_{k=1}^n (a_i^2+x_k),\quad
c_i^2=\prod_{k\ne i}(a_i^2-a_k^2).
\eea
Note that here the variables are arranged in the following order
\bea
r^2>0>x_1>-a_1^2>x_2>-a_2^2>\cdots>x_n>-a_n^2.
\eea
It turns out that mass does not spoil this relation, and in terms of $(x_0\dots x_{n})$ metric (\ref{MPevNC}) takes the form (\ref{ay2})--(\ref{ay2a}):
\bea\label{FrameX}
e^r=\frac{dr}{\sqrt{R-mr}}\sqrt{\prod_{k}(r^2-x_k)}
,\qquad e^{x_i}=\frac{1}{2}dx_i\sqrt{\frac{(r^2-x_i)}{-x_i H_i}\prod_{k\ne i}(x_i-x_k)}.
\eea
From now on Latin indices take values $(1\dots n)$, and we also define convenient quantities $d_i$, $H_i$ and rewrite $FR$ in terms of the new coordinates:
\bea\label{MPnotation}
d_i=\prod_{k\ne i}(x_i-x_k),\quad H_i=\prod_k(x_i+a_k^2),\quad
FR=\prod_k(r^2-x_k).
\eea

So far we have ignored the cyclic coordinates since components of the Killing tensor in these directions contain an ambiguity of adding an arbitrary combination of Killing vectors:
\bea\label{AddKV}
K^{MN}\rightarrow K^{MN}+\sum_{a,b}c_{ab}V^M_a V^N_b,\quad
V_0=\d_t,\quad V_i=\d_{\phi_i}.
\eea
Once the proper non--cyclic coordinates $(x_0\dots x_{n})$ are found, we can determine the remaining components of the Killing tensor by studying the separation of variables associated with it. Specifically, we 
look at the Hamilton--Jacobi equation associated with (\ref{MPeven}) and write it in coordinates $(x_0\dots x_{n})$:
\bea\label{qq1}
\sum_i\frac{4H_i (-x_i)}{(r^2-x_i)d_i}(\d_i S)^2+\frac{R-mr}{FR}(\d_r S)^2+g^{ab}\d_a S\d_b S=-\mu^2.
\eea
To separate $r$ coordinate, we have to multiply the last relation by 
\bea\label{RhoR}
\rho_r=R F=\prod_k (r^2-x_k)
\eea 
and introduce integrals of motion $I_k$ as coefficients in front of various powers of $r$. Then we will find
\bea\label{IkIntgr}
(R-mr)(\d_r S)^2=\sum_{k=0}^n I_k r^{2k}.
\eea
Notice that one Killing tensor leads to several integrals of motion, while the standard prescription \cite{Carter, Penrose} allows us to construct only one:
\bea
I=K^{MN}\d_M S\d_N S.
\eea
The `extra' conserved quantities came as the result of our analysis of eigenvalues: the coordinates $(r^2,x_1\dots x_n)$ define a \textit{family} of the Killing tensors parameterized by the polynomial $\Lambda$,
and the coordinates can be extracted from any special solution. Then starting with any member of the family and analyzing its eigenvalues, we can recover other Killing tensors by changing coefficients in $\Lambda$, as summarized by (\ref{diagram}). 

Extraction of the explicit expressions for $I_k$ is straightforward, but we will be interested in a different aspect of (\ref{IkIntgr}). To extend the relations (\ref{ay2}) beyond non--cyclic variables, we should identify the relevant cyclic frames, in particular, they should form pairs with $e_r$ and $e_{x_i}$\footnote{This follows  from the existence of the Killing--Yano tensor, as discussed below.}. To extract the partner of $e_r$, we set $(r^2-x_i)\rightarrow 0$ in (\ref{IkIntgr})\footnote{This is a very formal manipulation: although we set  $(r^2-x_i)\rightarrow 0$ for all $i$, we assume that $x_i-x_j\ne 0$. The goal of this operation is to remove all $x$--dependent terms from (\ref{IkIntgr}). We also recall that (\ref{IkIntgr}) comes from multiplying  (\ref{qq1}) by (\ref{RhoR}).}, then the right--hand side coming from (\ref{qq1}) contains only one frame:
\bea
e_t\propto \d_t-\sum_i \frac{a_i}{r^2+a_i^2}\d_{\phi_i}
\eea
Raising the index and normalizing this frame, we find
\bea\label{FrameT}
e^t=\sqrt{\frac{R-mr}{FR}}\left[dt+\sum_i\frac{G_i}{a_ic_i^2}d\phi_i\right],\quad
G_i\equiv \prod_k(x_k+a^2_i)
\eea
To extract the remaining frames, we write a counterpart of (\ref{IkIntgr}) by multiplying (\ref{qq1}) by 
\bea\label{RhoI}
\rho_i=(r^2-x_i)\prod_{k\ne i}(x_i-x_k)=(r^2-x_i)d_i.
\eea
This gives
\bea
\sum_i4H_i x_i(\d_i S)^2=\sum_{k=0}^n I^{(i)}_k (x_i)^{2k}.\nonumber
\eea
As before, we formally replace $(r^2-x_i)$ and $(x_j-x_i)$ by zero to extract 
\bea\label{FrameI}
e_i\propto \d_t-\sum_k\frac{a_k}{a_k^2+x_i}\d_{\phi_i}\quad\Rightarrow\quad
e^i=\sqrt{\frac{H_i}{d_i(r^2-x_i)}}\left[dt+\sum_k\frac{G_k(r^2+a_k^2)}{a_kc_k^2(x_i+a_k^2)}d\phi_k\right].
\eea
For future reference we summarize the frames and notation associated with Myers--Perry black hole in even dimensions\footnote{See (\ref{MPdefFR}), (\ref{EplsdEven}), (\ref{FrameX}), (\ref{MPnotation}), (\ref{FrameT}), (\ref{FrameI}).} 
\bea\label{AllFramesMP}
e^t&=&\sqrt{\frac{R-mr}{FR}}\left[dt+\sum_k\frac{G_k}{a_kc_k^2}d\phi_k\right],\quad
e^r=\sqrt{\frac{FR}{{R-mr}}}dr,
\nn
e^i&=&\sqrt{\frac{H_i}{d_i(r^2-x_i)}}\left[dt+\sum_k\frac{G_k(r^2+a_k^2)}{a_kc_k^2(x_i+a_k^2)}d\phi_k\right],\quad
e^{x_i}=\sqrt{-\frac{(r^2-x_i)d_i}{4 x_i H_i}}dx_i\nn
e_t&=&-\sqrt{\frac{R^2}{FR(R-mr)}}\left[\d_t-
\sum_k\frac{a_k}{r^2+a_k^2}\d_{\phi_k}\right],\quad 
e_r=\sqrt{\frac{R-mr}{FR}}\d_r,\nn
e_i&=&-\sqrt{\frac{H_i}{d_i(r^2-x_i)}}\left[\d_t-\sum_k\frac{a_k}{x_i+a_k^2}\d_{\phi_k}
\right],\quad e_{x_i}=\sqrt{-\frac{4x_iH_i}{d_i(r^2-x_i)}}\d_{x_i}\\
d_i&=&\prod_{k\ne i}(x_i-x_k),\quad H_i=\prod_k(x_i+a_k^2),\quad
G_i= \prod_k(x_k+a^2_i),
\nn
R&=&\prod_{k} (r^2+a_k^2),\quad
FR=\prod_k(r^2-x_k),\quad c_i^2=\prod_{k\ne i}(a_i^2-a_k^2).\nonumber
\eea
In terms of frames (\ref{AllFramesMP}) the metric and the Killing tensor become \eqref{SymmKT}
\bea\label{SymmKTKYT}
&&ds^2=-(e^t)^2+(e^r)^2+\sum_k [(e^{x_k})^2+(e^k)^2],\nn
&&K_{MN}dx^Mdx^N=\Lambda_r[-(e^t)^2+(e^r)^2]+\sum_k \Lambda_k [(e^{x_k})^2+(e^k)^2].
\eea
Here $\Lambda_r$ and $\Lambda_k$ are symmetric polynomials, as guaranteed by the general construction of section \ref{SecSeparKT}. The most general KT is obtained by adding Killing vectors (see (\ref{AddKV})) and the metric to the last expression, and this leads to modification of eigenvalues. We are primarily interested in KT that comes from squaring a Killing--Yano tensor, this requires a double degeneracy in the eigenvalues, so (\ref{SymmKTKYT}) is the most natural choice. 

The simplest KYT  is the volume form,
\bea\label{KYTnn}
Y^{(2n)}=e^t\wedge e^r\wedge \prod_k \left[e^{x_k}\wedge e^k\right],
\eea
and its square gives a trivial KT with $\Lambda_r=\Lambda_k=1$ in (\ref{SymmKTKYT}). 
Experience with KYT for the Kerr metric suggests that there is also a KYT of rank $2(n-1)$ and it should have the form
\bea\label{Ydown2}
Y^{(2n-2)}=\la_r \prod_k \left[e^{x_k}\wedge e^k\right]+
e^t\wedge e^r\left[\sum_i \la_i \prod_{k\ne i} \left[e^{x_k}\wedge e^k\right]\right].
\eea
In the four--dimensional Kerr metric we had
\bea\label{KYTLaKerr}
\la_r=\sqrt{r^2},\quad \la_1=\sqrt{-x_1},\quad \Lambda_r=x_1,\quad \Lambda_1=r^2,
\eea
and generalization to higher dimensions is straightforward\footnote{The sign difference between \eqref{KYTLaKerr} and \eqref{LaDwn2} is explained by different conventions for Kerr BH (where we use $ \sqrt{a^2}=a$) and Myers-Perry BH (where $\sqrt{a_k^2}=a_k$) and the relation $a_1=-a$.}:
\bea\label{LaDwn2}
\lambda_r=\sqrt{r^2},\quad \lambda_k=-\sqrt{-x_k},\quad \Lambda_r=\sum_k x_k,\quad
\Lambda_i=r^2+\sum_{k\ne i}x_k.
\eea
Direct calculation shows that (\ref{Ydown2}) with (\ref{LaDwn2}) solves the equation for the KYT. A clear pattern appears: 
\begin{quote}
\textit{To construct a KYT of rank $2(n-k)$ one should start with (\ref{KYTnn}) and symmetrically remove $k$ pairs using the rule
\bea\label{ReplaceKYT}
e^t\wedge e^r\rightarrow \sqrt{r^2},\quad e^{x_i}\wedge e^i\rightarrow -\sqrt{-x_i}.
\eea
Then the square of this KYT is the KT (\ref{SymmKTKYT}) with}
\bea\label{ReplaceKYT1}
\Lambda_r=\d_{x_0}\Lambda,\quad \Lambda_i=\d_i\Lambda,\quad \Lambda=x_0x_1\dots x_k+perm,\quad x_0=r^2.
\eea 
\end{quote}
For example, for $k=2$ this procedure gives
\bea
Y^{(2n-4)}&=&\la_r \sum_j \la_j\prod_{k\ne j} \left[e^{x_k}\wedge e^k\right]+
e^t\wedge e^r\left[\sum_{j<m} \la_j\la_m \wedge \prod_{k\ne j,m} \left[e^{x_k}\wedge e^k\right]\right],
\\
\lambda_r&=&\sqrt{r^2},\quad \lambda_k=-\sqrt{-x_k},\quad \Lambda_r=\sum_{k<m} x_kx_m,\quad
\Lambda_i=r^2\sum_k x_k+\sum_{j<k}x_jx_k.\nonumber
\eea
Rather than proving the procedure (\ref{ReplaceKYT}) we connect it to a very nice discussion of \cite{Kub1,Kub2,KubThes,Kub3}, where it was shown that a family of KYT can be constructed starting from 
\bea\label{Kub1}
h=\sum_i a_i\mu_id\mu_i\wedge\Big[a_i dt+(r^2+a_i^2)d\phi_i\Big]+rdr\wedge \Big[dt+\sum_i a_i\mu_i^2d\phi_i\Big]
\eea
by applying an operation
\bea\label{Kub2}
Y^{2(n-k)}=\star\left[\wedge h^k\right].
\eea
While our equations (\ref{ReplaceKYT}), (\ref{ReplaceKYT1}) give simpler expressions for the KYT and KT due to the use of convenient frames, they reduce to the construction (\ref{Kub1})--(\ref{Kub2}) once (\ref{Kub1}) is rewritten in the frames  \eqref{AllFramesMP}:
\bea\label{HInFrame}
h=re^r\wedge e^t+\sum_i\sqrt{-x_i}e^{x_i}\wedge e^i.
\eea
Construction (\ref{Kub2})--(\ref{HInFrame})  is proven in Appendix \ref{AppPCKYT}, and here we just outline the steps:
\begin{enumerate}
\item Expression (\ref{HInFrame}) gives a conformal Killing--Yano tensor (CKYT) for the Myers--Perry black hole, and the two--form $h$ is closed.
\item The product ${\mathcal Y}=[\wedge h^k]$ has the same properties as $h$ (i.e., it is a closed CKYT).
\item A Hodge dual of any closed CKYT is a KYT.
\end{enumerate}
Justifications of these statements are scattered throughout the literature 
\cite{Carig,Kub1,KubThes}, and Appendix \ref{AppPCKYT} provides streamlined derivations. 
Construction (\ref{Kub2})--(\ref{HInFrame}) of the KYT will be extended to a charged black hole in section \ref{SecMPF1}.

\bigskip

We conclude this section by a brief discussion of the Myers--Perry black hole in odd dimensions. Instead of starting with (\ref{MPeven}) one should begin with
\bea\label{MPodd}
ds^2=-dt^2+\frac{mr^2}{FR}\Big(dt+\sum_{i=1}^n a_i\mu_i^2 d\phi_i\Big)^2+\frac{FR dr^2}{R-mr^2}
+\sum_{i=1}^n(r^2+a_i^2)\Big(d\mu_i^2+\mu_i^2 d\phi_i^2\Big),
\eea
then repetition of the previous analysis leads to the counterpart of (\ref{AllFramesMP}):
\bea\label{AllFramesMPOdd}
e^t&=&
\sqrt{\frac{R-mr^2}{FR}}\left[dt+\sum_k\frac{a_kG_k}{c_k^2}d\phi_k\right],\quad
e^r=\sqrt{\frac{FR}{R-mr^2}}dr,
\nn
e^i&=&\sqrt{-\frac{H_i}{x_id_i(r^2-x_i)}}\left[dt+
\sum_k\frac{G_ka_k(r^2+a_k^2)}{c_k^2(x_i+a_k^2)}d\phi_k\right],\quad
e^{x_i}=\sqrt{\frac{d_i(r^2-x_i)}{4H_i}}dx_i,
\nonumber\\
e_t&=&-\sqrt{\frac{R^2}{FR(R-mr^2)}}\left[ \d_t
-\sum_k\frac{a_k}{r^2+a_k^2}\d_{\phi_k}\right],\quad e_r=\sqrt{\frac{R-mr^2}{FR}}\d_r,\\
e_i&=&-\sqrt{-\frac{H_i}{x_id_i(r^2-x_i)}}\left[\d_t-\sum_k\frac{a_k}{x_i+a_k^2}\d_{\phi_k}
\right],\quad e_{x_i}=\sqrt{\frac{4H_i}{d_i(r^2-x_i)}}\d_{x_i},\nonumber
\eea
and to one more frame that was not present in the even--dimensional case:
\bea
e^\psi=\sqrt{\frac{\prod a_i^2}{r^2\prod (-x_k)}}\left[dt +
\sum_k \frac{G_k(r^2+a_k^2)}{c^2_k a_k} d\phi_k \right],~ 
e_\psi=-\sqrt{\frac{\prod a_i^2}{r^2\prod (-x_k)}}\left[\d_t-\sum_k\frac{1}{a_k}\d_{\phi_k}
\right].
\eea
Notice that one of the relations (\ref{EplsdEven}) between Myers--Perry and ellipsoidal coordinates is modified\footnote{Notice that in contrast to the even-dimensional case, where $\mu_i$ were not constrained, now there is a relation $\sum \mu_i^2=1$, and, as a consequence, there only $n-1$ coordinates $x_i$.}:
\bea\label{EplsdOdd}
\mu_i^2=\frac{1}{c_i^2}\prod_{k=1}^{n-1} (a_i^2+x_k).
\eea
This leads to a new expression for
\bea\label{FROdd}
FR=r^2\prod_k(r^2-x_k)
\eea
and we still have the remaining relations
\bea\label{OtherOdd}
d_i&=&\prod_{k\ne i}(x_i-x_k),\quad H_i=\prod_k^n(x_i+a_k^2),\quad
G_i= \prod_k^{n-1}(x_k+a^2_i),
\nn
R&=&\prod_{k}^n (r^2+a_k^2),\quad c_i^2=\prod_{k\ne i}(a_i^2-a_k^2).
\eea
Note a very special form of the relative coefficients in frames $e_a$: they depend only on $r$ in 
$e_t$, only on $x_i$ in $e_i$, and they are constant in $e_\psi$.

The Killing--Yano tensors are still given by construction (\ref{Kub2}) with
\bea
h=re^r\wedge e^t+\sum_i\sqrt{-x_i}e^{x_i}\wedge e^i.
\eea
The separation factors are 
\bea
\rho_r=r^2\prod_j^{n-1}(r^2-x_j),\quad \rho_i=x_i(r^2-x_i)\prod_{k\ne i}[x_i-x_k].
\eea
This reduces to (\ref{RhoR}), (\ref{RhoI}) if we introduce $x_{n}\equiv 0$.


\section{Killing(--Yano) tensors and string dualities}
\label{SecKillingsDualities}

In this section we will analyze transformations of various tensors under string dualities. Specifically, we will focus on T dualities along $U(1)$ isometries and assume that Killing--(Yano) tensors do not depend on coordinates parameterizing the isometries. We will also consider larger classes of U duality transformations. Our results are summarized below:
\begin{itemize}
\item Generically, the Killing vectors depending on the direction of T duality are destroyed (as we will show in section \ref{SubsectionzKV}), and Killing vectors with trivial dependence on the duality direction survive the duality, as long as original fluxes respect the symmetry associated with Killing vectors (see section \ref{SubsectionKV}).
\item Conformal Killing vectors are destroyed by the T duality with an exception of the homothetic CKV. The latter acquire nontrivial dependence upon the duality direction in 
the dual geometry (see section \ref{SubsectionCKV}).
\item KT equation remains the same, but there are constraints on the $B$ field and the dilaton (\ref{Hcond2fields}), \eqref{ConstrBKT},  (see section \ref{SecKTdual}).
\item Extension of T duality to the CKT is possible only for special solutions, and some examples are presented in Appendix \ref{AppCKT}.
\item KYT equation is modified by terms containing the Kalb--Ramond field \eqref{KYTBfield}, and there is an additional constraint \eqref{RestBKYTGuess}  (or, more generally, (\ref{ConstrBcompBdy})) on this field (see section \ref{SectionModifiedKYT}). 
\item Extension of T duality to CKYT is possible only for special solutions.
\end{itemize}
We will now discuss all theses properties in detail.


\subsection{Killing vectors and T duality}
\label{SecKVDuality}

In this subsection we will analyze the transformations of the Killing vectors under combinations of T dualities and reparametrizations. The most natural formalism for such study is provided by the Double Field Theory (DFT) \cite{DFT}, which is reviewed in Appendix \ref{AppDFT}, and a very simple interpretation of our results in terms of this approach is presented in the end of section \ref{SubsectionKV}.

We will begin with a pure metric 
\be\label{DimRedKV}
ds^2=e^C[dz+A_idx^i]^2+\hat g_{ij}dx^idx^j, \quad B_{MN}=0
\ee
that admits two Killing vectors, $Z=\d_z$ and $V=V^M\d_M$, and study the transformation of vector $V$ under T duality along $z$ direction. We will look at three situations and the results are summarized as follows:
\begin{enumerate}[(a)]
\item The $z$--independent vectors $V$ (i.e., vectors commuting with $Z$) have counterparts after T duality, and the transformation law is derived in section \ref{SubsectionKV}.
\item The $z$--dependent vectors $V$ (i.e., vectors with $[V,Z]\ne 0$) may be destroyed by the duality transformation, and in general the numbers of such vectors before and after T duality do not match. Some examples are discussed in section \ref{SubsectionzKV}.
\item Conformal Killing Vectors of the original geometry are destroyed by T duality unless one introduces $z$--dependence in the dual frame. This construction is discussed in section \ref{SubsectionCKV}.
\end{enumerate}
In case (a) we will find an additional constraint on the Kalb--Ramond field after duality:
\bea\label{HfieldCond}
H_{MNP}V^P=\nabla_M W_N-\nabla_N W_M,\quad\mbox{with arbitrary}\quad W_N,
\eea 
and we will demonstrate that any geometry that has a Killing vector $V$ satisfying (\ref{HfieldCond}) can be dualized in a direction commuting with $V$ without destroying the Killing vector. We will also show that condition (\ref{HfieldCond}) arises naturally from the equation for a Killing vector in DFT.  

\subsubsection{Killing vectors commuting with T duality direction}
\label{SubsectionKV}

Let us first assume that geometry (\ref{DimRedKV}) solves Einstein's equations without $B$ field, and that it admits a Killing vector $V$:
\bea\label{Apr5}
\nabla_M V_N+\nabla_N V_M=0
\eea
which commutes with $Z=\d_z$. In Appendix \ref{AppDimRed1} we perform dimensional reduction of this equation in geometry (\ref{DimRedKV}) before and after T duality in $z$ direction. Using tildes to denote the quantities after T duality, we find various components of (\ref{Apr5}) and its dual counterpart:
\bea\label{KVAllEqs}
\begin{array}{|r|c|c|}
\hline
&\nabla_M V_N+\nabla_M V_N=0&\Big.\nabla_M\tilde V_N+\nabla_N \tilde V_M=0\\
\hline
zz:&\d_rC V^r=0&\Big.\d_rC \tilde V^r=0\\
mz:&F_{mr}V^r=\d_m(e^{-C}V_z)&\d_m(e^{C}\tilde V_z)=0\\
mn:&\hat\nabla^mV^n+\hat\nabla^nV^m=0&\Big.\hat\nabla^m\tilde V^n+\hat\nabla^n\tilde V^m=0\\
\hline
\end{array}
\eea
Here $\hat\nabla$ denotes the covariant derivative corresponding to metric ${\hat g}_{ij}$. 

Comparison of two columns on (\ref{KVAllEqs}) leads to the transformation law
\bea\label{May15}
\tilde V^r=V^r, \quad \tilde V^z\equiv e^C\tilde V_z=\mbox{const}.
\eea
Relation (\ref{May15}) ensures that the Killing equations after T duality are satisfied, but the $(mz)$ component of the original equation imposes a constraint on the new $B$ field:
\bea\label{BConstrComp}
{\tilde B}_{mz}=A_m\quad \Rightarrow\quad {\tilde H}_{zmp}V^p=\d_m(e^{-C}V_z).
\eea
Notice that this is the only relation in the dual frame that contains the original $V_z$. 

The implications of the constraint (\ref{BConstrComp}) are analyzed in Appendix \ref{AppDRKV}, where it is shown that a pair of relations
\bea\label{BConstrCovar}
&&\nabla_M V_N+\nabla_M V_N=0,\nonumber\\
&&H_{MNP}V^P=\nabla_M W_N-\nabla_N W_M
\eea
is preserved by T duality as long as one imposes the the transformation 
\bea\label{May5a}\label{KVtransf}
&&{\tilde V}^a=V^a,\quad {\tilde W}_z= -e^{-C}V_z,\quad
{\tilde V}_z= -e^{-C}W_z,\nonumber\\
&&{\tilde W}_n=W_n-{\tilde A}_n e^{-C}V_z-A_n W_z+\d_n f,
\eea
with arbitrary function $f$. Although we motivated (\ref{BConstrCovar}) by starting with a pure metric, the map (\ref{KVtransf}) leaves (\ref{BConstrCovar}) invariant for arbitrary configurations of the $B$ field before and after the duality.

The system (\ref{BConstrCovar}) is the unique extension of the equation for Killing vector consistent with T duality, and in Appendix \ref{AppDFT} we show that (\ref{BConstrCovar}) can be written as a single equation for a Killing vector on an extended space used in the Double Field Theory (DFT). 
Specifically, if the metric and the $B$ field are combined in a single matrix (\ref{DefDFT})\footnote{In equations (\ref{DefDFTmain})--(\ref{LieDFTmain}) and in Appendix 
\ref{AppDFT} we deviate from the notation used throughout this chapter and denote the spacetime indices by lower--case letters, while reserving the capital ones to label the ``double space". This notation is standard in the DFT literature.}
\bea\label{DefDFTmain}
\cH_{IJ}=\begin{pmatrix}g^{ij}& -g^{ik}B_{kj}\\B_{ik}g^{kj}&g_{ij}-B_{ik}g^{kl}B_{lj} \end{pmatrix},
\eea
then equations (\ref{BConstrCovar}) appear as different components of a single equation for $\xi^P$:
\be\label{LieDFTmain}
{L}_{\xi}\mathcal{H}_{MN}\equiv\xi^P\d_P\cH_{MN}+(\d_M\xi^P-\d^P\xi_M)\cH_{PN}+(\d_N\xi^P-\d^P\xi_N)\cH_{MP}=0
\ee
Here $\xi^I=(\tilde\lambda_i,\lambda^i)$ is the generalized gauge parameter, where $\tilde\lambda_i$ corresponds to the gauge transformation of the Kalb--Ramond field $B_{ij}$ and  $\lambda^i$ generates diffeomorphisms. Equation (\ref{LieDFTmain}), which involves the generalized Lie derivative in double space $L_\xi$, implies that the system (\ref{BConstrCovar}) is covariant under combinations of diffeomorphisms and T--dualities.


\subsubsection{Killing vectors with $z$ dependence}
\label{SubsectionzKV}

In the previous subsection we assumed that components of the Killing vector $V$ did not depend on the direction of T duality\footnote{In covariant form this condition is written as $[Z,V]=0$.} and demonstrated that components of the Killing vector transform in a simple way (\ref{KVtransf}). Here we will use several examples to argue that situation for the $z$--dependent Killing vectors is rather different: even the number of such vectors can be changed by application of T duality. 

We begin with the simplest example of a pure metric
\be
ds^2=f(dz^2+dy^2)+g_{mn}dx^m dx^n
\ee
which admits a Killing vector corresponding to rotations in the $(y,z)$ plane:
\be\label{OrigKV}
V=y\d_z-z\d_y.
\ee
Performing the T duality along $z$ direction and  
solving equations for the Killing vector in the dual configuration,
\bea
ds^2=\frac{dz^2}{f}+f dy^2+g_{mn}dx^m dx^n,
\eea
we find that there are only two KVs with nontrivial $(y,z)$ components:
\bea
V=c_1\d_y+c_2\d_z
\eea
unless $f=const$, where there is also a counterpart of (\ref{OrigKV}):
\bea
V=f^2y\d_z-z\d_y.
\eea
We conclude that the $z$--dependent Killing vector (\ref{OrigKV}) disappears unless $f$ is equal to constant. 

The same phenomenon can be seen in a more interesting geometry produced by smeared fundamental strings \cite{FundString,HorStr}:
\bea\label{Apr5a}
ds^2&=&H^{-1}(dz^2-dt^2)+dr^2+r^2d\Omega_p^2+\sum_{k=1}^{7-p} dx_kdx_k, \nn
B&=&(H^{-1}-1)dt\wedge dz, \quad e^{2\Phi}=H^{-1}, \quad H=1+\frac{Q}{r^{p-1}}.
\eea
The most general Killing vector with $(z,t)$ components has the form
\bea
V=c_1\d_t+c_2\d_z+c_3(t \d_z+z \d_t).
\eea
T duality along $z$ direction leads to a metric produced by a plane wave, which has only two independent Killing vectors with components in $(t,z)$ directions:
\bea
V=c_1\d_t+c_2\d_z.
\eea
Once again, $z$--dependent Killing vector disappears after T duality.
In section \ref{SectionModifiedKYT} we will encounter a similar situation with Killing--Yano tensors (KYT): at first sight they seem to be destroyed by T duality. To cure this problem we will modify the equation for KYT by adding an extra term containing the Kalb--Ramond field. This solution would not work in the present case: since the geometry dual to (\ref{Apr5a}) does not contain matter fields, the original equation (\ref{Apr5}) is the unique relation consistent with invariance under diffeomorphisms. 

To summarize, we conclude that $z$--dependent Killing vectors can appear and disappear under T dualities, so they don't have well--defined transformation properties. We expect the situation to be at least as bad for the Killing(--Yano) tensors, so in sections \ref{SecKTdual} and \ref{SectionModifiedKYT} we will focus only on $z$--independent objects. However, 
$z$--dependence can lead to very interesting effects for conformal Killing vectors, which will be discussed now.

\subsubsection{Conformal Killing Vectors and T duality.}
\label{SubsectionCKV}

Conformal Killing vectors (CKV) do not leave the metric invariant, but rather they lead to rescalings by a conformal factor. Such vectors satisfy differential equation
\be
\nabla_M {\mathcal V}_N+\nabla_N {\mathcal V}_M=g_{MN}v,
\ee
with some function $v$. Dimensional reduction of this equation gives the counterpart of 
(\ref{KVAllEqs})\footnote{Recall that we are starting with a pure metric, so there are no $g_{zm}$ components after duality. Reductions (\ref{CKVAllEqs}) and (\ref{CKVAllEqsZ}) follow directly from Appendix \ref{AppDimRed1}.} :
\bea\label{CKVAllEqs}
\begin{array}{|c|c|c|}
\hline
&\nabla_M {\mathcal V}_N+\nabla_N {\mathcal V}_M=g_{MN}v
&\Big.\tilde\nabla_M \tilde {\mathcal V}_N+\tilde\nabla_N\tilde {\mathcal V}_M=\tilde g_{MN}\tilde v\\
\hline
zz&\h\d_re^C {\mathcal V}^r=e^Cv&\Big.\h\d_re^{-C} \tilde {\mathcal V}^r=e^{-C}\tilde v\\
mz&F_{mr}{\mathcal V}^r=\d_m(e^{-C}{\mathcal V}_z)&\d_m(e^{C}\tilde {\mathcal V}_z)=0\\
mn&\tilde\nabla^m{\mathcal V}^n+\tilde\nabla^n{\mathcal V}^m=g^{mn}v&
\Big.\hat\nabla^m\tilde {\mathcal V}^n+\hat\nabla^n\tilde {\mathcal V}^m=g^{mn}\tilde v\\
\hline
\end{array}
\eea
Imposing the relation ${\mathcal V}^n=\tilde{\mathcal V}^n$, we conclude that $v={\tilde v}$, then $(zz)$ components lead to contradiction unless $C$ is a constant or $v$ is equal to zero. To cure this problem, we allow $z$ dependence in the conformal Killing tensor after duality and replace (\ref{CKVAllEqs}) by\footnote{Notice that introduction of $z$ dependence after duality puts the initial and final system on a different footing. Similar situation is encountered in the non--Abelian T duality \cite{NabTdual}, but there an analog of $z$--dependence is introduced for the dynamical fields, while here we are looking at the Killing vectors.}
\bea\label{CKVAllEqsZ}
\begin{array}{|c|c|c|}
\hline
&\nabla_M {\mathcal V}_N+\nabla_N {\mathcal V}_M=g_{MN}v
&\Big.\tilde\nabla_M \tilde {\mathcal V}_N+\tilde\nabla_N\tilde {\mathcal V}_M=\tilde g_{MN}\tilde v\\
\hline
zz&\h\d_re^C {\mathcal V}^r=e^Cv&
\Big.\d_z{\tilde{\mathcal V}}_z+\h\d_re^{-C} \tilde {\mathcal V}^r=e^{-C}\tilde v\\
mz&F_{mr}{\mathcal V}^r=\d_m(e^{-C}{\mathcal V}_z)&\d_m(e^{C}\tilde {\mathcal V}_z)+\d_z {\tilde{\mathcal V}}_m=0\\
mn&\hat\nabla^m{\mathcal V}^n+\hat\nabla^n{\mathcal V}^m=g^{mn}v&
\Big.\hat\nabla^m\tilde {\mathcal V}^n+\hat\nabla^n\tilde {\mathcal V}^m=g^{mn}\tilde v\\
\hline
\end{array}
\eea
Once again setting
\bea
\tilde {\mathcal V}^n={\mathcal V}^n, \quad \tilde v=v,
\eea
we find a system of equations for ${\tilde{\mathcal V}}^z$:
\bea
 {\mathcal V}^r\d_r C=2v,\quad \d_z {\tilde{\mathcal V}}^z=2 \tilde v,\quad \d_m{\tilde {\mathcal V}}^z=0
\eea
since the original CKV ${\mathcal V}$  does not depend of $z$. 
Integrability conditions for the last two equations imply that ${\tilde v}$ must be constant, so the CKV ${\mathcal V}$ must be homothetic. A simple example of a homothetic KV comes from rescaling of the flat space by a constant factor:
\bea
ds^2=\eta_{MN}dx^Mdx^N,\quad 
{\mathcal V}_M dx^M=\eta_{MN}x^N dx^M,\quad
v=1.
\eea

To summarize, for every homothetic CKV we find the complete set of transformations,
\be\label{SolCKVzDep}
{\tilde {\mathcal V}}^m={\mathcal V}^m,\quad {\tilde v}=v=\mbox{const},\quad 
{\tilde {\mathcal V}}^z=2zv+const
\ee
that produces a CKV after T duality. Non--homothetic conformal Killing Vectors are destroyed by T duality.

\subsection{Killing tensors in the NS sector}
\label{SecKTdual}

In this subsection we study the behavior of Killing tensors (KT) under $O(d,d)$ transformations, which include boosts, T dualities and rotations, and then extend the construction to the full NS sector by incorporating transformations involving S dualities. 

As discussed in section \ref{SecSeparKT} equation \eqref{KTeqn4} has reducible solution spanned by combinations of the metric and Killing vectors,
\bea\label{KtrivA}
K^{triv}_{MN}=e_0 g_{MN}+\sum_{ij} e_{ij}V^{(i)}_MV^{(j)}_N,
\eea
with constant coefficients $e_0$, $e_{ij}$. In section \ref{SecKVDuality} we showed that Killing vectors are preserved by the $O(d,d)$ transformations if conditions (\ref{BConstrCovar}) are satisfied. This implies that the expression (\ref{KtrivA}) for the ``trivial Killing tensor'' holds for the entire $O(d,d)$ orbit. Here we will focus on non--trivial Killing tensors, which can be either destroyed or modified by T duality, and we identify a subset of $O(d,d)$ transformations which do not lead to destruction of a nontrivial KT. The non--trivial Killing tensors can be found either by solving equation (\ref{KTeqn4}) or by separating the Hamilton--Jacobi equation \cite{Carter}, and the second approach is more convenient for the study of T duality. The relationship between Killing tensors and separation of the massive Hamilton--Jacobi equation has been reviewed in section \ref{SecKTandHJ}, and in this subsection these results will be extended to charged solutions. An alternative approach based on dimensional reduction of KT equation is discussed in Appendix \ref{AppDRKT4}. 

In subsection \ref{SecKTOdd} we focus on the $O(d,d)$ orbit which generates fundamental strings from pure metric, and in subsection \ref{SecKTBeyondOdd} these results are extended to general F1--NS5 solutions. As we will see, existence of KT imposes certain restrictions on the Kalb--Ramond field, and they are discussed in subsection \ref{SecCondsBfield}. Finally in subsection \ref{SecCondsBfieldDR} we use an alternative method (dimensional reduction) to derive the covariant form of the constraint on the $B$ field.

\subsubsection{Killing tensors and $O(d,d)$ transformations}
\label{KTOdd}
\label{SecKTOdd}

We begin with a pure metric that solves source--free Einstein equations in $D$ dimensions, admits a Killing tensor, and has $d$ cyclic directions $\phi^a$. Such geometry can be written in a reduced form:
\bea\label{TorusMetr}
ds^2=G_{ab}(d\phi^a+V^a_m dx^m)(d\phi^b+V^b_n dx^n)+h_{mn}dx^m dx^n.
\eea
This metric has an obvious $GL(d)$ symmetry that rotates cyclic directions into each other, but in supergravity this symmetry is enhanced to $O(d,d)$, which acts on the metric and on the Kalb--Ramond $B$ field \cite{Odd,GenMetric}. This symmetry is extended further to $O(D,D)$ via the Double Field Theory (DFT) formalism \cite{DFT}, which is reviewed in Appendix \ref{AppDFT}.

Specifically, a $2D\times 2D$ matrix written in $D\times D$ blocks
\bea
M=\left[
\begin{array}{cc}
G^{-1} & -G^{-1}B \\
BG^{-1} & G-BG^{-1}G
\end{array}
 \right]
\eea
is transformed under a global $O(D,D)$ as 
\bea\label{Mrot}
M\rightarrow \Omega M\Omega^T,
\eea
where
\bea
\Omega \eta\Omega^T=\eta, \qquad \eta=\left[
\begin{array}{cc}
0 & 1 \\
1 & 0
\end{array}
 \right].
\eea
Here $\eta$ is a metric for a group $O(D,D)$.

Since we are starting with a pure metric, the initial matrix $M$ is given 
by\footnote{Note that $g^{ab}$, $q^{am}$ and $h^{mn}$ are the components of $D\times D$ matrix $G^{-1}$.}
\bea\label{ModuliPureMetr}
M=\left[
\begin{array}{cc|cc}
g^{ab}&q^{am}&0&0\\
q^{ma}&h^{mn}&0&0\\
\hline
0&0&G_{ab}&G_{am}\\
0&0&G_{ma}&G_{mn}\\
\end{array}
\right].
\eea
Parameterizing the $O(d,d)$ rotations by $d\times d$ matrices $A,C,D,E$ as
\bea\label{OddTrans}
\Omega=\left[
\begin{array}{cc|cc}
A&0&E&0\\
0&I_{D-d}&0&0\\
\hline
C&0&D&0\\
0&0&0&I_{D-d}\\
\end{array}
\right],\quad 
\left[
\begin{array}{cc}
A^T&C^T\\
E^T&D^T
\end{array}\right]
\left[\begin{array}{cc}
0&I_d\\
I_d&0
\end{array}\right]
\left[
\begin{array}{cc}
A&E\\
C&D
\end{array}\right]=
\left[\begin{array}{cc}
0&I_d\\
I_d&0
\end{array}\right]
\eea
we find the transformed metric with upper indices
\bea\label{RotateMDFT}
\Omega M \Omega^T=
\left[
\begin{array}{c|c}
\begin{array}{cc}
AgA^T+EGE^T&A q\\
qA^T&h
\end{array}&
\bullet\\
\hline
\bullet&\bullet
\end{array}
\right]
\eea
Here and below $G$ denotes a $d\times d$ matrix with components $G_{ab}$.
The survival of the Killing tensor under transformation with arbitrary $A$ and $B$ implies that the following four quantities must separate:
\bea\label{SepCondPureMetr}
fg^{ab},\quad fq^{am},\quad fh^{mn},\quad fG_{ab}.
\eea
The first three conditions are satisfied before the $O(d,d)$ transformation since metric (\ref{TorusMetr}) had a Killing tensor. Separation in the dual frame requires  $fG_{ab}$ to separate with \textit{the same} function $f$. 
Combining this with results of section \ref{SecKTandHJ} we arrive at the following conclusion:
\begin{enumerate}[(1)]
\item{Every KT is associated with a unique function $f$, which can be determined from the HJ equation or from eigenvalues, and with corresponding variables $(x,y)$.}
\item{T dualities and rotations in a sector spanned by cyclic coordinates $\phi_a$ do not spoil separation of variables for a given KT if and only if
\bea\label{DefTranslDir}
\d_x\d_y[f G_{ab}]=0.
\eea
}
\end{enumerate}
So far we have separated coordinates into cyclic and non--cyclic, but equation (\ref{DefTranslDir}) suggests a more refined distinction: among cyclic coordinates $\phi^a$ we identify the subsector where  (\ref{DefTranslDir}) holds and call the corresponding cyclic directions translational, and the remaining directions will be called rotational\footnote{Strictly speaking one should define coordinates are rotational and translational with respect to a particular Killing tensor: the same cyclic coordinate might by translational for one KT and rotational for another. Since we are dealing with one tensor at a time referring to a direction as simply translational should not cause confusions.}. A simple example demonstrates the origin of these names: in the metric 
\bea
ds^2=dr^2+r^2 d\theta^2+r^2\sin^2\theta (d\phi^1)^2+(d\phi^2)^2
\eea
coordinate $\phi^2$ would be called translational and coordinate $\phi^1$ would be called rotational since in this case $x=r$, $y=\theta$, and $f=r^2$. For many aspects of our discussion rotational coordinates appear on the same footing as non--cyclic ones. 

Once we have demonstrated that the Killing tensor is not destroyed by the $O(d,d)$ transformations as long as expressions (\ref{SepCondPureMetr}) separate, we can ask about transformation laws for this tensor. Recall that Killing vectors with upper components were unaffected by the $O(d,d)$ transformations, but Killing tensor has a more interesting behavior. The third expression in (\ref{SepCondPureMetr}) indicates that the separation function cannot be affected by the $O(d,d)$ transformations since $h^{mn}$ is invariant under them. This implies simple relations for the Killing tensors before and after T duality\footnote{For simplicity we are focusing on Killing tensor which separates two non--cyclic coordinates $x$ and $y$. Generalization to ore coordinates is straightforward, but the notation becomes cumbersome.
}:
\bea\label{KillTilde}
K^{MN}&=&X^{MN}-f_xg^{MN}, \quad {\tilde K}^{MN}={\tilde X}^{MN}-f_x{\tilde g}^{MN}.
\eea
We use tildes to denote the expressions after T duality. As discussed in section \ref{SecSeparKT}, separation in the original metric implies that
\bea
g^{MN}=\frac{1}{f}\left[X^{MN}+Y^{MN}\right],\nonumber
\eea
and the last condition in (\ref{SepCondPureMetr}) leads to an additional relation
\bea
G_{ab}=\frac{1}{f}\left[{\hat X}_{ab}+{\hat Y}_{ab}\right],
\eea
where $X,{\hat X}$ are functions of $x$ and $Y,{\hat Y}$ are functions of $y$. Then transformation (\ref{RotateMDFT}),
\bea
{\tilde g}^{ab}=\left[AgA^T+EGE^T\right]^{ab},\quad
{\tilde g}^{am}=A_{ab}g^{bm},\quad {\tilde g}^{mn}=g^{mn}\nonumber
\eea
gives
\bea\label{XforF1}
{\tilde X}^{ab}=\left[AXA^T+E{\hat X}E^T\right]^{ab},\quad 
{\tilde X}^{am}=A_{ab}X^{bm},\quad {\tilde X}^{mn}=X^{mn}.
\eea
Along with (\ref{KillTilde}) this completely determines the transformation of the Killing tensor under the action of $O(d,d)$.


To summarize, we have demonstrated that transformation (\ref{RotateMDFT}) preserve the Killing tensor as long as all directions $\phi^a$ in (\ref{TorusMetr}) are chosen to be translational, and all cyclic rotational directions are absorbed in $h_{mn}$. Notice, however, that some components on the Killing tensor are modified according to (\ref{KillTilde}), (\ref{XforF1}). Transformations (\ref{RotateMDFT}) allow 
one to generate a large class of charged solutions of supergravity starting from a simple neutral ``seed'', and this technique has been used to generate large classes of charged black holes in \cite{Sen,Cvetic}. 
One can also start with a ``seed'' which already contains a nontrivial Kalb--Ramond field, and the generalization of our analysis is straightforward. 

Suppose that metric (\ref{TorusMetr}) is supported by the $B$ field and the dilaton which are invariant under translations in $\phi$ directions:
\bea
\d_{\phi^a}e^{2\Phi}=0,\quad {\mathcal L}_{\phi^a} B=0.
\eea
Then application of the rotation \eqref{Mrot} with $\Omega$ given by \eqref{OddTrans} to the initial moduli matrix\footnote{Note that $Q$ is the full inverse metric, for example $Q^{aM}B_{Mb}=g^{as}B_{cb}+q^{as}B_{sb}$.}
\bea
M=\left[
\begin{array}{cc|cc}
g^{ab}&q^{am}&-Q^{aM}B_{Mb}&-Q^{aM}B_{Mm}\\
q^{nb}&h^{nm}&-Q^{nM}B_{Mb}&-Q^{nM}B_{Mm}\\
\hline
B_{aM}Q^{Mb}&B_{aM}Q^{Mm}&G_{ab}-B_{aM}Q^{MN}B_{Nb}&G_{am}-B_{aM}Q^{MN}B_{Nm}\\
B_{nM}Q^{Mb}&B_{nM}Q^{Mm}&G_{nb}-B_{nM}Q^{MN}B_{Nb}&G_{nm}-B_{nM}h^{MN}B_{Nm}\\
\end{array}
\right]
\eea
gives\footnote{Recall that indices of rotational matrices appearing in (\ref{OddTrans}) go only over specific subsets $A_{as},E_{am},C_{ma},D_{mn}$, so for example $(AQ)_a{}^M=A_{ab}Q^{bM}$.}
\bea
\Omega M \Omega^T=\left[
\begin{array}{c|c}
\begin{array}{cc}
AgA^T-AQBE^T + EBQA^T+E(G-BQB)E^T & Aq+EBQ\\
hA^T-QBE^T&h
\end{array}&
\bullet\\
\hline
\bullet&\bullet
\end{array}
\right].\nonumber
\eea
The new metric admits a Killing tensor if and only if the following combinations of the original quantities separate:
\bea\label{SepCondB}
\boxed{
fg^{ab},\quad fB_{aM}g^{Mb},\quad fB_{aM}g^{Mm},\quad  f(g_{ab}-B_{aM}g^{MN}B_{Nb}),
\quad fg^{am},\quad   fg^{mn}.}
\eea

In spite of the appearances, conditions (\ref{SepCondB}) are invariant under gauge transformations of the $B$ field. We will demonstrate this for the most interesting case where $B_{aM}$ has both legs in the cyclic directions (one of them translational and the other one is either translational or rotational). Indeed, separability of the second and third expressions in coordinates $(x,y)$ implies that
\bea\label{Feb7a}
\d_x\d_y(fg^{NM}B_{Mb})=0,
\eea
next recalling that that $\d_x\d_y(fg^{NM})=0$, the last condition can be rewritten in the gauge--invariant form:
\bea\label{Hcond1}
\d_y(f g^{NM})H_{xMb}+\d_x (f g^{NM}) H_{y Mb}+f g^{NM}\d_xH_{y Mb}=0.
\eea
Similarly, separability of the fourth expression in (\ref{SepCondB}) can be rewritten as
\bea\label{Hcond}
\d_x\d_y (fg_{ab})-fg^{MN}H_{y aM}H_{xNb}-fg^{MN}H_{xaM}H_{y Nb}=0.
\eea
By construction, constraints on the $B$ field for any point on an $O(d,d)$ trajectory passing through a pure metric are just separability conditions for the initial metric \eqref{SepCondPureMetr}.

\subsubsection{Conditions on the $B$ field from dimensional reduction}
\label{SecCondsBfieldDR}

So far we have been studying transformation of Killing tensors under $O(d,d)$ rotations using separation of HJ equation. Now we will use an alternative approach based on dimensional reduction to derive the unique covariant form of the constraint on the $B$ field, and the result is given by \eqref{ConstrBKT}.

Let us start with a standard Killing tensor equation
\bea\label{KTEq}
\nabla_M K_{NP}+\nabla_N K_{MP}+\nabla_P K_{MN}=0,
\eea
and perform dimensional reduction of the metric along $z$ direction:
\bea
ds^2=e^C[dz+A_i dx^i]^2+\hat{g}_{ij} dx^i dx^j.
\eea
The details of such reduction are given in Appendix \ref{AppDRKT4}, in particular $mnp$ components of the Killing tensor equation \eqref{KTEq}
\bea
\hat{\nabla}^m K^{np}+\hat{\nabla}^n K^{mp}+\hat{\nabla}^p K^{mn}=0
\eea
transform under T duality into 
\bea
\hat{\nabla}^m \tilde{K}^{np}+\hat{\nabla}^n \tilde{K}^{mp}+\hat{\nabla}^p \tilde{K}^{mn}=0.
\eea
We conclude that the KT equation is not modified by the $B$ field, in contrast to Killing-Yano tensor case, which will be discussed in section \ref{SectionModifiedKYT}. Next we look at the $mnz$ components
\bea\label{mnzKT}
{\hat g}^{ma}\left[\hat\nabla_a(e^{-C}K^n{}_{z})+F_{ba}K^{nb}\right]+(m\leftrightarrow n)=0.
\eea
Under T duality along $z$ direction $F_{mn}$ transforms into $H_{mnz}$ ($H=dB$), so we conclude that T dual counterpart of \eqref{mnzKT} should give an equation involving the $B$ field. As demonstrated in Appendix \ref{AppDRKT4}, the only covariant form of such equation is
\bea\label{ConstrBKT}
{\tilde H}_{AMP}\tilde K_N{}^A+{\tilde H}_{ANP}\tilde K_M{}^A=e^{C/2}{\tilde\nabla}_M[e^{-C/2}{\tilde W}_{NP}]+e^{C/2}{\tilde\nabla}_N[e^{-C/2}{\tilde W}_{MP}].
\eea
Recall that we had a similar expression as a constraint on the $B$ field for a Killing vector \eqref{BConstrCovar}.

Notice that the equation \eqref{mnzKT} has an interesting interpretation in terms of Lie derivatives. As
 shown in Appendix \ref{AppDRKT4} for the KT constructed from squaring a Killing vector as 
 $K^{mn}=V^m V^n$, equation \eqref{mnzKT} reduces to a combination of Lie derivatives of $A_m$ (recall that $F_{mn}=\d_m A_n-\d_n A_m$) along the Killing vector $V^m$ 
\bea
lhs=V^n \mathcal{L}_V A^m+V^m \mathcal{L}_V A^n.
\eea

To summarize we have used dimensional reduction to demonstrate that requirement of covariance of Killing tensor under T duality leads to the unique constraint on the $B$ field \eqref{ConstrBKT} similar to the equation on the $B$ field satisfied by Killing vectors. We will now discuss the behavior of Killing tensors under the U--duality group that extends $O(d,d)$ transformations, and demonstrate that covariance under such dualities leads to additional constraints on the Kalb--Ramond field.



\subsubsection{Extension beyond $O(d,d)$}
\label{SecKTBeyondOdd}

\begin{figure}
    \centering
    \includegraphics[width=1\textwidth]{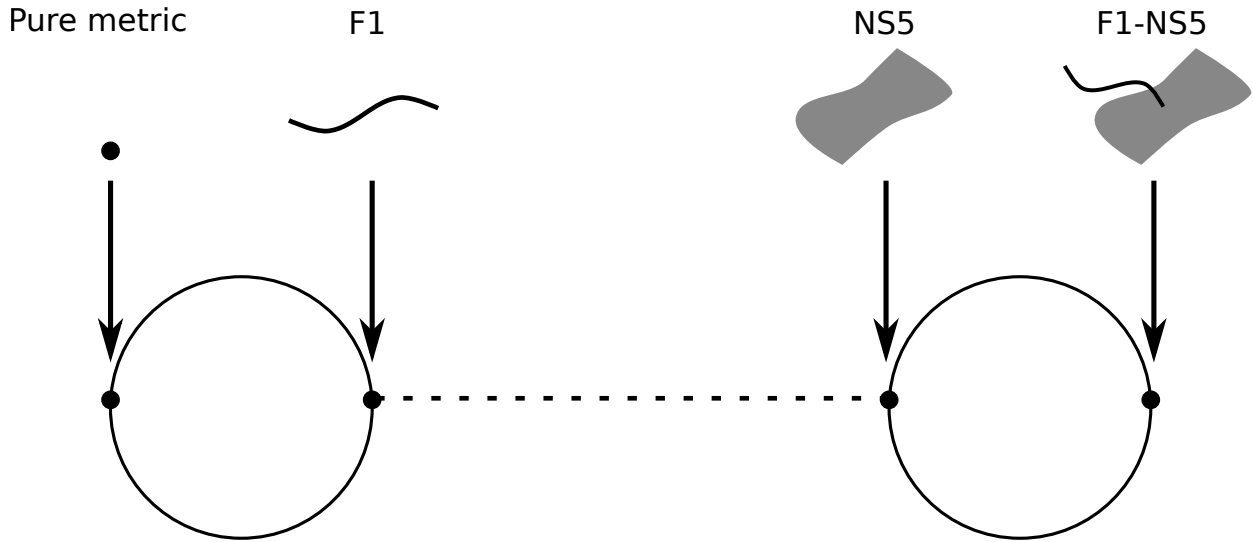}
    \caption[Pictorial representation of the duality chain (\ref{DualChainShort})]{Pictorial representation of the duality chain (\ref{DualChainShort}).
    Applying 
    $O(d,d)$ transformations (the left solid circle) to a pure metric, one produces solutions 
    of the `F1 type', then the `bridge'  (dashed line) discussed in section 
    \ref{SecKTBeyondOdd} connects the F1 geometry with a pure NS5. Additional 
    $O(d,d)$ transformations, represented by the solid circle on the right, produce the general F1--NS5 solution.}
    \label{fig:F1NS5}
\end{figure}

In this chapter we are studying the symmetries of the NS sector of string theory\footnote{Solutions for the Ramond--Ramond fluxes are also interesting, but our construction of the modified Killing--Yano tensors discussed in section \ref{SectionModifiedKYT} needs further generalization to include such geometries.}, and so far we have only discussed the geometries related to pure metric by $O(d,d)$ transformations. Inclusion of S duality allows one to produce more general NS--NS backgrounds, and in this subsection our construction is extended to such geometries. 

In the context of black hole physics $O(d,d)$ transformation are often used to generate solutions with electric $B$ field\footnote{The most notable exceptions from this rule are gravity duals of non--commutative field theories \cite{nonCom}, beta--deformation of pure geometry \cite{margin}, and generation of NS5--brane from KK monopole. From our perspective, all these operations give the solution of type `F1'.}, so we will call them `F1 geometries', even if they do not describe fundamental strings. To generate NS5 branes from black holes one has to use a specific combination of T and S dualities, and we will denote the resulting geometry by `NS5', even though it can contain more general fluxes.  This chain of dualities is shown in Figure \ref{fig:F1NS5}.

To generate the `NS5 geometry' we begin with a ten--dimensional metric reduced on $T^p\times T^4$:
\bea\label{Jan13a}
ds_P^2=H_{\alpha\beta}[dy^\alpha+Y^\alpha][dy^\beta+Y^\beta]
+G_{ab}(dz^a+A^a)(dz^b+A^b)+h_{mn}dx^mdx^n
\eea
To generate a magnetic NS flux, we perform the following dualities \cite{MultiStr,lmPar}\footnote{A detailed discussion of this duality map will be presented in the next section, where a more involved chain (\ref{DualChain}) will be used to add charges to various black holes.}: 
\bea\label{DualChainShort}
P\stackrel{T_{y}}{\longrightarrow}F\stackrel{S}{\longrightarrow}
D1\stackrel{T_z}{\longrightarrow}D5\stackrel{S}{\longrightarrow}NS5.
\eea
Notice that various labels just indicate the type of flux (i.e., F1 is an electric $B$--field, D5 is a magnetic $C^{(2)}$ and so on) rather than presence of branes. 

T dualities along $y$ directions produces F1 solution, and subsequent S duality gives
\bea
ds_{D1}^2&=&\sqrt{\mbox{det}\,H}\left[{\tilde H}_{\alpha\beta}dy_\alpha dy_\beta+G_{ab}(dz^a+A^a)(dz^b+A^b)+h_{mn}dx^mdx^n\right],\nonumber\\
e^{2\Phi}&=&{\mbox{det}\,H},\quad C^{(2)}=dy_\alpha\wedge Y^\alpha ,\quad {\tilde H}_{\alpha\beta}=[H^{-1}]_{\alpha\beta}.\nonumber
\eea
The outcome of four T dualities along $z$ directions depends on the presence of $z^a$ in $Y^\alpha$. If $Y^\alpha$ has no legs along $z$ directions, then T dualities produce a six--form, which can be dualized back to $C^{(2)}$. Any leg pointing in $z$ direction leads to $C^{(4)}$, and this RR flux can\rq{}t be removed by S duality. Thus to end up with NS system we require $Y$ to point only in the non--compact directions. Then T dualities along $z$ directions give
\bea
ds_{D5}^2&=&\sqrt{\mbox{det}\,H}\left[{\tilde H}_{\alpha\beta}dy_\alpha dy_\beta+h_{mn}dx^mdx^n\right]+\frac{1}{\sqrt{\mbox{det}\,H}}{\tilde G}_{ab}dz_adz_b,\nonumber\\
e^{2\Phi}&=&\frac{1}{{\mbox{det}}\,H {\mbox{det}}\,G},\quad C^{(2)}=dy_\alpha\wedge Y^\alpha\wedge\prod (dz^a+A^a),\quad
B=dz_a\wedge A^a,\nonumber
\eea
where ${\tilde G}$ is the inverse matrix of $G$. To avoid the RR fields after S duality, we must require $A^a=0$, this leads to the final result:
\bea\label{NS5metr}
ds_{NS5}^2&=&\sqrt{\mbox{det}\,G}\left[\mbox{det}\,H{\tilde H}_{\alpha\beta}dy_\alpha dy_\beta+{\mbox{det}\,H}h_{mn}dx^mdx^n+{\tilde G}_{ab}dz_adz_b\right],\nonumber\\
e^{2\Phi}&=&{{\mbox{det}}\,H {\mbox{det}}\,G},\quad 
B=dy_\alpha\wedge Y^\alpha\wedge\prod (dz^a+A^a).
\eea

Separation of the Hamilton--Jacobi equation in the geometry (\ref{Jan13a}) implies (among other things) the separation of
\bea
f h^{mn}\d_m S\d_n S,\quad f H^{\alpha\beta},
\eea
and for the geometry (\ref{NS5metr}) we need 
\bea
\frac{\tilde f}{{\mbox{det}\,H}\sqrt{{\mbox{det}~}G}}h^{mn}\d_m S\d_n S,
\quad
\frac{\tilde f}{{{\mbox{det}\,H}\sqrt{{\mbox{det}~}G}}}{\tilde H}^{\alpha\beta},\qquad 
\frac{\tilde f}{\sqrt{{\mbox{det}~}G}}{\tilde G}^{ab}
\eea
to separate for some function ${\tilde f}$. Setting
\bea\label{TildeNSf}
{\tilde f}={f}{{\mbox{det}\,H}}{\sqrt{{\mbox{det}~}G}}
\eea
we must require 
\bea\label{SeparCondNS5}
\d_x\d_y[{f}h^{mn}\d_m S\d_n S]=0,
\quad
\d_x\d_y[f{\tilde H}^{\alpha\beta}]=0,\ \d_x\d_y{\tilde f}=0,\ 
\d_x\d_y[f{{\mbox{det}\,H}}{\tilde G}^{ab}]=0.
\eea
The first condition is automatic, the second one is similar to the requirement for T duality (recall that 
${\tilde H}=H^{-1}$), and the last two relations are new. As before, the old and the new Killing tensors are expressed as (\ref{KillTilde})
\bea\label{KillTildeA}
K^{MN}=X^{MN}-f_xg^{MN}, \quad {\tilde K}^{MN}={\tilde X}^{MN}-{\tilde f}_x{\tilde g}^{MN},
\eea
although now tildes refer to the NS5 system. Repeating the steps which led to (\ref{XforF1}), we find
\bea\label{XforNS5}
{\tilde X}^{\alpha\beta}=\left[f H^{\alpha\beta}\right]_x,\
{\tilde X}^{ab}=\left[f{{\mbox{det}\,H}}{\tilde G}^{ab}\right]_x,\
{\tilde X}^{mn}=X^{mn},\quad {\tilde f}_x=[{f}{{\mbox{det}\,H}}{\sqrt{{\mbox{det}~}G}}]_x.
\eea
Equations (\ref{TildeNSf}), (\ref{KillTildeA}), (\ref{XforNS5}) give the Killing tensor ${\tilde K}$ in terms of the the original metric, in particular, we observe that the expression for ${\tilde K}$  in terms of $K$ is rather complicated. This reinforces the principle introduced in section \ref{SecKTandHJ}: to study the Killing tensors and their transformations under dualities, it is convenient to begin with finding the eigenvectors and eigenvalues of the tensors since the map (\ref{XforNS5}) between $X$ and ${\tilde X}$ is relatively simple. 
Several explicit examples of Killing tensors for F1--NS5 systems are presented in Appendix \ref{App5D6DandKT}.

\subsubsection{Conditions on the $B$ field from separation of variables}
\label{SecCondsBfield}

Equation \eqref{SeparCondNS5} gives the separability condition for the NS5 metric, and now we present the constraints on the $B$ field. In Section \ref{SecKTOdd} such restrictions were found by requiring separability of 
the metrics on any $O(d,d)$ orbit which starts from a pure metric, and now we impose the same requirement on the $O(d,d)$ orbit staring from an NS5 solution\footnote{The first orbit generates fundamental strings and momentum, and the second one generates F1--NS5--P system}. We will find that separability of F1--NS5--P geometries is guaranteed by (\ref{SeparCondNS5}) and constraints (\ref{Hcond2f}), (\ref{Hcond2fields}), (\ref{Hcond2sep}) on the Kalb--Ramond field of the original F1 system.

We start with constraints \eqref{Hcond1} and \eqref{Hcond} derived for the F1 orbit
\bea\label{Hconds}
&&\d_y(f g^{mM})H_{xMb}+\d_x (f g^{mM}) H_{y Mb}+f g^{mM}\d_xH_{y Mb}=0,\nn
&&\d_x\d_y (fg_{ab})-fg^{MN}H_{y aM}H_{xNb}-fg^{MN}H_{xaM}H_{y Nb}=0,
\eea
and require them to hold for NS5 solutions as well. Then using the relation between metrics for F1 and NS5 (\ref{NS5metr}),
\bea
g_{MN}^{NS5}=F g_{MN}^{F1}, \quad f^{NS5}=F f^{F1},\quad 
F\equiv \sqrt{\mbox{det}G}\,\mbox{det}H=e^{-2\Phi_{F1}}
\eea
and electric--magnetic duality transformation, we can rewrite (\ref{Hconds}) in terms of the metric and the $B$ field for F1. The detailed calculations presented in the Appendix \ref{AppCondBNS5} give
\bea\label{Hcond2fa}
&&\d_x\d_y[g_{ab}fF]+\frac{f}{F}g_{ab}\Big[\d_x\ln F \d_y\ln F+ \frac{1}{2}H_{xMN}H_y{}^{MN}\Big]=0
\eea
and
\bea\label{Hcond2f}
&&\d_y(f g^{mM})H_{xMb}+\d_x (f g^{mM}) H_{y Mb}+f g^{mM}\d_xH_{y Mb}=0,\\
&&\d_y(f g^{mM})\tilde{H}_{xMb}+\d_x (f g^{mM}) \tilde{H}_{y Mb}+\frac{1}{F}f g^{mM}\d_x(F\tilde{H}_{y Mb})=0,\nonumber
\eea
where $\tilde{H}=\star_6H^{(F1)}$ is the Hodge dual dual of $H^{(F1)}$ with respect to the metric $h_{mn}$.

Interestingly, in all examples we have considered, two terms in equation (\ref{Hcond2fa}) vanish separately, and perhaps such `coincidence' is guaranteed by equations of motion of supergravity for the NS5 brane, but we have not investigated this further. Vanishing of the first term in equation (\ref{Hcond2fa}) implies separation of a very interesting duality--invariant quantity
\bea
g^{(F1)}_{ab}f^{(F1)}e^{-2\Phi_{F1}}=g^{(NS5)}_{ab}f^{(NS5)}e^{-2\Phi_{NS5}}.
\eea
Then vanishing of the second term in \eqref{Hcond2fa} implies a relation in the F1 frame:
\bea\label{Hcond2fields}
\d_x\Phi \d_y\Phi+\frac{1}{8} H_{xMN}H_y{}^{MN}=0.
\eea

To summarize, the separability of the F1--NS5--P geometries obtained form the F1 system is guaranteed by equation (\ref{SeparCondNS5}), conditions (\ref{Hcond2f}), (\ref{Hcond2fields}) on the $B$ field of the original F1 system, and 
\bea\label{Hcond2sep}
\d_x\d_y[g_{ab}fe^{-2\Phi}]=0.
\eea

\subsection{T duality and the modified Killing--Yano equation}
\label{SectionModifiedKYT}

In this subsection we investigate the behavior of (conformal) Killing--Yano tensors under T dualities. We will show that generically T duality destroys Killing--Yano tensors, but there is a unique modification of the KYT equation which is invariant under T duality. For the geometries without Kalb--Ramond field, this modified Killing--Yano (mKY) equation reduces to the standard one (\ref{DefKYT}), but in general it also  contains contributions from the $B$ field. To motivate the mKYT equation, we apply T duality to  a pure metric. This leads to the unique modification of KYT equation in the dual frame, and we will demonstrate that such modification remains invariant under any combination of diffeomorphisms and T dualities.
 
Let us start with a standard equation for the Killing--Yano tensor (\ref{KYTdef})  
\bea\label{KYTdefAA}
\nabla_M Y_{NP}+\nabla_N Y_{MP}=0
\eea
and perform a dimensional reduction of the metric along $z$ direction:
\be\label{DimRedSetup2}
ds^2=e^C[dz+A_idx^i]^2+\hat g_{ij}dx^idx^j.
\ee
In the first step of our analysis we also assume that geometry (\ref{DimRedSetup2}) has a trivial Kalb--Ramond field.
The details of the reduction are given in Appendix \ref{AppDimRed1}, in particular, the $(mnp)$ component of the KY equation can be read off from (\ref{DimRedL}) by setting $L=Y$:
\bea\label{mnpComp}
\nabla^mY^{np}+\h F^{mp}Y^n{}_z+(m\leftrightarrow n)=0,
\eea
where $F=dA$ is the field strength associated with graviphoton. 
We will now look for the modification of the KYT equation in the dual frame that satisfies five requirements:
\begin{enumerate}[(1)]
\item The equation should be linear in the dual Killing-Yano tensor ${\tilde Y}$.
\item Its $(mnp)$ component must reproduce (\ref{mnpComp}) and other components must be consistent with dimensional reduction of (\ref{KYTdefAA}).
\item The equation must be invariant under gauge transformations of the $B$ field.
\item The new terms to be at most linear in  $B$ field since equations (\ref{mnpComp}) are linear in $F_{ab}$.  This implies that the modified KY equation should be linear in $H_{MNP}$.
\item The square of the modified KYT should give a Killing tensor in the dual frame. 
\end{enumerate}
As demonstrated in the in Appendix \ref{AppBeqKYT}, there exists a unique modification of equation (\ref{KYTdefAA}) which satisfies all these requirements, and it reads\footnote{The only alternative corresponds to changing the sign of $H$ in (\ref{KYTBfield}) and sign of $\tilde Y^n{}_z$ in (\ref{KYTtransYcomp}).}
\be\label{KYTBfield}
\boxed{
{\tilde\nabla}_M {\tilde Y}_{NP}+
{\tilde\nabla}_N {\tilde Y}_{MP}+
\frac{1}{2}\tilde H_{MPA}{\tilde g}^{AB}{\tilde Y}_{NB}+
\frac{1}{2}\tilde H_{NPA}{\tilde g}^{AB}
{\tilde Y}_{MB}=0.}
\ee
Moreover, the Kalb--Ramond field in the dual frame satisfies a constraint
\bea\label{RestBKYTGuess}
{{\tilde H}_{Q[MN}{\tilde Y}_{P]}^{\ \ Q}+\d_{[P} C\,{\tilde Y}_{MN]}=
-\d_{[P} {\tilde W}_{MN]}}
\eea
with some antisymmetric tensor ${\tilde W}_{MN}$.
Under the T duality the components of the mKYT transform as
\be\label{KYTtransYcomp}
\tilde Y^{mn}=Y^{mn}, \quad \tilde Y^n{}_z=e^{-C}Y^n{}_z.
\ee 
The counterpart of the constraint (\ref{RestBKYTGuess}) in the original metric (\ref{DimRedSetup2}) is 
\be\label{PureMetrConst}
dC\wedge dY=0.
\ee
Notice that (\ref{KYTBfield}) can be interpreted as a standard KYT equation with connection modified by torsion \cite{Strominger}
\bea
\Gamma_{MN}^P\rightarrow \Gamma_{MN}^P-\frac{1}{2}H^P{}_{MN}.
\eea
In Appendix \ref{AppComplexStructure} we discuss transformation of K{\"a}hler structure under T duality and demonstrate that a counterpart of the transformation 
(\ref{KYTtransYcomp}) maps the K{\"a}hler form into complex structure satisfying the Strominger's system for manifolds with torsion \cite{Strominger}.

Although equation (\ref{KYTBfield}) was derived by applying T duality to a pure metric, the result is invariant under any combination of T dualities and diffeomorphisms. In Appendix 
\ref{AppBeqKYT} we demonstrate that T duality maps any solution $Y_{MN}$ of (\ref{KYTBfield}) in an arbitrary geometry (\ref{DimRedSetup2})  supported by the $B$ field into a solution 
${\tilde Y}_{MN}$ of the same equation in the dual frame. The transformation (\ref{KYTtransYcomp}) between tensors can be viewed as an extension of Buscher's rules to Killing--Yano tensors.
The constraint (\ref{RestBKYTGuess}) is generalized as
\bea\label{ConstrBcompBdy}
&&
g^{ms}\d_sC {\hat Y}^n{}_{z}-g^{ns}\d_sC{\hat Y}^m{}_z+
g^{ns}G_{sr}Y^{rm}-g^{ms}G_{sr}Y^{rn}=0,\\
&&\quad G_{mn}\equiv e^{C/2}F_{mn}-e^{-C/2}{\tilde F}_{mn},\quad 
{\hat Y}_z{}^s\equiv e^{-C/2}Y_z{}^s\,,
\nonumber
\eea
where $F_{mn}$ and ${\tilde F}_{mn}$ are graviphotons in the original and dual frames. 
Notice that ${\hat Y}_z{}^s$ remains invariant under T duality, and $G_{mn}$ 
changes sign.

To summarize, we have demonstrated that the requirement of covariance under T duality leads to the unique equation (\ref{KYTBfield}) for the KYT, and the original equation (\ref{KYTdefAA}) is transformed into the system (\ref{KYTBfield})--(\ref{RestBKYTGuess}). In other words, unlike the KV and KT equations which are unaffected by the Kalb--Ramond field, the equation for the Killing--Yano tensor is modified, which is not very surprising since fermions interact with the $B$ field. In all three cases (KV, KT, mKYT) the Kalb--Ramond field satisfies additional constraints in the dual frame (see (\ref{BConstrCovar}), 
(\ref{ConstrBKT}), (\ref{RestBKYTGuess})). 

Although Ramond--Ramond fluxes appeared in the intermediate stages of the duality chain (\ref{DualChainShort}), neither the initial nor the final point contained such fields. Unfortunately an extension of our analysis to Ramond--Ramond backgrounds leads to certain complications, which we now discuss. Starting with a pure metric and performing a T duality, we find the new mKYT from (\ref{KYTtransYcomp}):
\be
\tilde Y^{mn}=Y^{mn}, \quad \tilde Y^n{}_z=e^{-C}Y^n{}_z.
\ee 
Since the mKYT equation is written in the string frame, S duality induces a conformal rescaling of such metric, so generically the modified Killing--Yano tensor is destroyed by such operation. To save it we have two option for the equation after the duality:
\begin{enumerate}[(a)]
\item Postulate that in the presence of the Ramond--Ramond fluxes, the covariant derivatives appearing in the mKYT should be computed using 
$g'_{MN}=e^{-\Phi}g_{MN}$ rather that $g_{MN}$, and $H_3$ should be replaced by $F_3$. While consistent with S duality, this prescription  does not reduce to the standard KYT in the NS--NS backgrounds with non--trivial dilaton, so it should be abandoned. 
\item Postulate that the modified KYT equation survives S duality only if the constraint 
\bea\label{ConstrDilat}
g^{AB}\d_B\Phi Y_{AM}=0 
\eea
is satisfied. Then the discussion presented in the Appendix \ref{AppCKYT} implies that the mKYT transforms according to (\ref{eqnA8})
\bea
Y'_{NP}=e^{-3\Phi/2}{\tilde Y}_{NP},
\eea
where ${\tilde Y}_{NP}$ satisfies equation (\ref{KYTBfield}) before S duality, and $\Phi$ is the dilaton for the NS system. 
\end{enumerate}
Although option (b) is not ruled out, the constraint (\ref{ConstrDilat}) is rather restrictive. Moreover, even assuming that this constraint is satisfied, and equation (\ref{KYTBfield}) does hold for the type IIB theory with replacement $H_3\rightarrow F_3$, an additional T duality to type IIA supergravity leads to rather unusual structures. By applying the dimensional reduction and T duality to Ramond--Ramond background, we found that the KY equation in the dual frame mixes tensors of different ranks. For example, starting with mKYT $Y_{MN}$ one produces an equation that mixes $Y^{(1)}_M$ and $Y^{(3)}_{MNP}$. This is not surprising since something similar happens for components of $F_3$, but KYT become rather complicated. While it would be very interesting to study the properties of such objects with mixed ranks and perhaps embed them in the democratic formalism \cite{Democratic}, this direction will not be pursued here.

\bigskip

Finally we comment on behavior of conformal Killing(--Yano) tensors. As demonstrated in section \ref{SubsectionCKV}, T duality introduces $z$--dependence in conformal Killing vectors, so such dependence should be allowed in  CKT as well.  Dimensional reduction for a relatively simple case $A_m=0$ is performed in Appendix \ref{AppCKT}, where we demonstrate that generically CKTs are destroyed by T duality. 
However, the CKT does survive the duality if two additional conditions (\ref{CKTConstr1}) and (\ref{CKTConstr2}) are satisfied. The same conclusion holds for a conformal mKYT: it survives T duality only in very special cases. 


\section{Examples of the modified KYT for F1--NS5 system}
\label{SecExamplesF1NS5}

In this section we present several examples of the modified Killing--Yano tensors introduced in section \ref{SectionModifiedKYT}. As we saw in section \ref{SecMyersPerry}, the ordinary Killing--Yano tensors exist for a large class of black holes described by the Myers--Perry solutions, and these geometries automatically solve our modified equation since they do not have a Kalb--Ramond field. However, string theory provides a very nice generating technique that allows one to start with a known solution of general relativity and construct black holes with various charges by applying string dualities \cite{Sen,Cvetic,cvet}. In this chapter we are focusing only on the NS--NS sector of string theory, so we will use the special cases of the general techniques introduced in \cite{Sen,Cvetic,cvet} to produce black holes with fundamental string and NS5--brane charges\footnote{The geometries containing D--branes are also interesting, but the full theory of modified Yano--Killing tensors for such solutions has not been developed yet. In particular, as we mentioned in section \ref{SectionModifiedKYT}, some D--brane backgrounds would contain Yano--Killing tensors of mixed ranks, and we hope to return to a detailed study of such objects in the future.}. For such special cases, it is convenient to specify the duality transformations more explicitly. 

We will start with a rotating black hole in $d<10$ dimensions and boost it in one of the $10-d$ direction. Then application of T duality along that direction produces a non--extremal fundamental string. To arrive 
at an NS5--brane (and more generally at a combination of strings and NS5--branes), one has to apply a more sophisticated procedure introduced in \cite{MultiStr,lmPar}:
\begin{enumerate}[1.]
\item Start with a rotating Myers--Perry black hole with mass $m$ in $d<6$ dimensions, perform a trivial embedding into the ten--dimensional type IIA supergravity, and identify a five--dimensional torus $T^4\times S^1$ orthogonal to the black hole. 
\item Perform a boost by $\alpha$ along $S^1$ direction\footnote{Following \cite{MultiStr,lmPar}, we will call the corresponding coordinate $y$ and parameterize the boost by $\alpha$, where 
$\tanh\alpha\equiv v/c$.} and T--dualize along $S^1$. This produces a black fundamental string wrapping one of the compact directions. 
\item Perform an S duality followed by four T dualities along $T^4$ and another S duality. The resulting metric describes a non--extremal rotating NS5 brane. 
\item Perform another boost by $\beta$ in the $S^1$ direction followed by T duality. This gives a non--extremal F1--NS5 system with mass $m$ and charges
\bea
Q_1=m\sinh^2\beta,\qquad Q_5=m\sinh^2\alpha.
\eea
\end{enumerate}
For future reference we summarize the duality chain using a simple diagram:
\bea\label{DualChain}
\mbox{BH}\ \rightarrow\
\mbox{P}_\alpha\ \rightarrow\
\mbox{F1}_\alpha\ \rightarrow\ \mbox{D1}_\alpha\ \rightarrow\ \mbox{D5}_\alpha
\ \rightarrow\ \mbox{NS5}_\alpha\ \rightarrow\ 
\left(\begin{array}{c}
\mbox{NS5}_\alpha\\ P_\beta
\end{array}\right)\ \rightarrow 
\left(\begin{array}{c}
\mbox{NS5}_\alpha\\ F1_\beta
\end{array}\right).
\eea
In this section we use $y$ to denote the $S^1$ direction. Notice that if we are adding only the F1 charge, the duality chain stops after the first two steps, and four--dimensional torus is not needed. Thus such charge can be added to the Myers--Perry black hole in $d<10$ dimensions\footnote{This construction also works for the embedding of the $d$--dimensional Myers--Perry black hole to the bosonic string as long as $d<26$.}, and we derive the explicit expression for the corresponding mKYT in section \ref{SecMPF1}. On the other hand, addition of the NS5 charge needs $T^4\times S^1$, so it only works for black holes with $d<6$. Since we are interested in asymptotically--flat geometries, the BTZ black hole \cite{BTZ} will not appear in the discussion, so $d$ can take only two values ($d=4,5$). These cases are discussed in sections \ref{SecEx5d} and \ref{SecEx6d}. Our results are summarized in table \ref{Table1}.

\begin{table}
\centering
    \begin{tabular}{ | l | c | c | c | c |}
    \hline
    {}		&  \multicolumn{2}{|c|}{4D} & \multicolumn{2}{|c|}{5D} \\ \hline
    {}		& extremal 	 & non--extremal& extremal & non-extremal  \\ \hline
    F1      & 	M		 & 			M	& M 	   & M 		       \\ \hline
    NS5     & 	M		 & 			--	& M 	   & M 		       \\ \hline
 F1--NS5($Q_1=Q_5$)	& 	-- & 		--	& C,M	   & M	           \\ \hline
 F1--NS5($Q_1\ne Q_5$)& 	--		 & 	--	& M 	   & M 		       \\ 
    \hline
    \end{tabular}
    \caption{Summary of the results for the F1--NS5 system constructed from four-- and five--dimensional black holes using the procedure (\ref{DualChain}). Here M denotes the modified KYT and C correspond to the conformal KYT.}
    \label{Table1}
\end{table}


\subsection{Charged Myers--Perry black hole}

\label{SecMPF1}

In our first example we add charges to the Myers--Perry black hole discussed in section 
\ref{SecMyersPerry} by applying the duality chain (\ref{DualChain}) and discuss the modified Killing--Yano tensor for the resulting solution. The transition from F1 to NS5 in 
(\ref{DualChain}) involves the electric--magnetic duality, which depends on the dimension of the black hole, so it is convenient to study individual black holes separately, and we will do that in sections \ref{SecEx5d}, \ref{SecEx6d}. In this section we will focus the first two 
\textit{algebraic} steps in the duality chain (\ref{DualChain}) to generate a rotating black hole with F1 charge. 

As demonstrated in Appendix \ref{AppChargedMP}, the charged Myers--Perry black hole admits a family of modified Killing--Yano tensors, which generalizes 
(\ref{AllFramesMP})--(\ref{HInFrame}): the tensors are still given by 
(\ref{Kub2}), (\ref{HInFrame})\footnote{In this subsection we have to distinguish between $e^a=e^a_M dx^M$ and 
$e_a=e_a^M\d_M$, so the frame indices are written in the appropriate places. In the rest of the chapter we abuse notation and write $e_a=e^a_M dx^M$ to simplify formulas.}
\bea\label{ChMPHYT}
Y^{(2n-2p)}=\star\left[\wedge h^p\right],\quad
h=re^r\wedge e^t+\sum_k\sqrt{-x_k}e^{x_k}\wedge e^k\,,
\eea
but the frames are modified 
\bea\label{ChMPframe}
e^r&=&\sqrt{\frac{FR}{R-mr}}dr,\quad e^{x_i}=\sqrt{-\frac{(r^2-x_i)d_i}{4x_iH_i}}dx_i,\nn
e^t&=&\frac{1}{h_1}
\sqrt{\frac{R-mr}{FR}}\left[\ch_\alpha dt+\sh_\alpha dy+\sum_k\frac{G_k}{a_kc_k^2}d\phi_k\right],\nn
e^y&=&\frac{1}{h_1}\left[\sh_\alpha dt+\ch_\alpha dy-
\frac{mr\sh_{\alpha}}{FR}dt
-\frac{mr\sh_{\alpha}\ch_\alpha}{FR}\sum_k\frac{d\phi_k}{a_kc_k^2}\right],\\
e^i&=&\frac{1}{h_1}\sqrt{\frac{H_i}{d_i(r^2-x_i)}}\left[\ch_\alpha dt+\sh_\alpha dy+
\sum_k\frac{G_k(r^2+a_k^2)}{c_k^2a_k(x_i+a_k^2)}\left\{1+
\frac{mr\sh^2_\alpha (r^2-x_i)}{FR(r^2+a_k^2)}
\right\}d\phi_i\right].\nonumber
\eea
The expressions for $c_i,d_i,H_i,G_i,(FR)$ are still given by (\ref{AllFramesMP}), and 
\bea\label{h1even}
h_1=1+\frac{m\sh_\alpha^2}{FR}.
\eea
Expressions for the inverse frames exhibit a clear separation between non--cyclic coordinates $(r,x_i)$:
\bea\label{ChMPframeDwn}
e_r&=&\sqrt{\frac{R-mr}{FR}}\d_r,\quad 
e_{x_i}=\sqrt{-\frac{4x_iH_i}{d_i(r^2-x_i)}}\d_{x_i},\quad
e_y=\ch_\alpha\d_y-\sh_\alpha\d_t,
\nn
e_t&=&-\sqrt{\frac{R^2}{FR(R-mr)}}\left[\ch_\alpha \d_t-\frac{\sh_\alpha}{R}(R-mr)\d_y
-\sum_k\frac{a_k}{r^2+a_k^2}\d_{\phi_k}\right],\\
e_i&=&-\sqrt{\frac{H_i}{d_i(r^2-x_i)}}\left[\ch_\alpha\d_t-\sh_\alpha \d_y-\sum_k\frac{a_k}{x_i+a_k^2}\d_{\phi_k}
\right].\nonumber
\eea
For the odd dimensions we find 
\bea\label{ChMPframeOdd}
e^r&=&\sqrt{\frac{FR}{R-mr^2}}dr,\quad e^{x_i}=\sqrt{\frac{d_i(r^2-x_i)}{4H_i}}dx_i,\nn
e^t&=&\frac{1}{h_1}
\sqrt{\frac{R-mr^2}{FR}}\left[\ch_\alpha dt+\sh_\alpha dy+\sum_k\frac{a_kG_k}{c_k^2}d\phi_k\right],\nn
e^y&=&\frac{1}{h_1}\left[\sh_\alpha dt+\ch_\alpha dy-
\frac{m\sh_{\alpha}}{FR}dt
-\frac{m\sh_{\alpha}\ch_\alpha}{FR}\sum_k\frac{a_kd\phi_k}{c_k^2}\right],\\
e^i&=&\frac{1}{h_1}\sqrt{-\frac{H_i}{x_id_i(r^2-x_i)}}\left[\ch_\alpha dt+\sh_\alpha dy+
\sum_k\frac{G_ka_k(r^2+a_k^2)}{c_k^2(x_i+a_k^2)}\left\{1+
\frac{m\sh^2_\alpha (r^2-x_i)}{FR(r^2+a_k^2)}
\right\}d\phi_k\right],\nonumber\\
e^\psi&=&\frac{1}{h_1}\sqrt{\frac{\prod a_i^2}{r^2\prod x_k}}\left[\ch_\alpha dt +
\sh_\alpha dy+\sum \frac{G_k(r^2+a_k^2)}{c^2_k a_k}\left\{1-
\frac{a^2_k m\sh_\alpha^2}{FR(r^2+a_k^2)} \right\}d\phi_k \right].\nonumber
\eea
and
\bea\label{ChMPframeDwnOdd}
e_r&=&\sqrt{\frac{R-mr^2}{FR}}\d_r,\quad 
e_{x_i}=\sqrt{\frac{4H_i}{d_i(r^2-x_i)}}\d_{x_i},\quad
e_y=\ch_\alpha\d_y-\sh_\alpha\d_t,
\nn
e_t&=&-\sqrt{\frac{R^2}{FR(R-mr^2)}}\left[\ch_\alpha \d_t-\frac{\sh_\alpha}{R}(R-mr^2)\d_y
-\sum_k\frac{a_k}{r^2+a_k^2}\d_{\phi_k}\right],\\
e_i&=&-\sqrt{-\frac{H_i}{x_id_i(r^2-x_i)}}\left[\ch_\alpha\d_t-\sh_\alpha \d_y-\sum_k\frac{a_k}{x_i+a_k^2}\d_{\phi_k}
\right],\nonumber\\
e_\psi&=&-\sqrt{\frac{\prod a_i^2}{r^2\prod x_k}}\left[\ch_\alpha\d_t-\sh_\alpha \d_y-\sum_k\frac{1}{a_k}\d_{\phi_k}
\right].\nonumber
\eea
The expressions for $c_i,d_i,H_i,G_i,(FR)$ are still given by (\ref{FROdd}), (\ref{OtherOdd}), and $h_1$ is given by 
(\ref{h1even}).

\subsection{F1--NS5 system from the Kerr black hole.}
\label{SubsectionExamples5D}
\label{SecEx5d}

Application of the duality chain (\ref{DualChain}) to the Kerr black hole (\ref{Kerr4D}) gives a rotating F1--NS5 system, and the complete solution is presented in Appendix \ref{App4DKerr} (see equation (\ref{NonExtr5D})). Explicit calculations show that the modified Killing--Yano equation (\ref{KYTBfield}) does not have nontrivial solutions\footnote{As shown in table \ref{Table1}, extremal F1--NS5 and non--extremal NS5 also don't admit  mKYT.}, so in this subsection we will focus on two special cases when the mKYT exists: the non--extremal fundamental string and the extremal NS5 brane. In the first case the existence of solution is guaranteed by the general construction presented in section \ref{SectionModifiedKYT} as long as condition (\ref{PureMetrConst}) is satisfied, and in the second case the mKYT comes from solving the Killing equations.

Application of the first two steps in the duality sequence (\ref{DualChain}) to Kerr geometry (\ref{Kerr4D}) leads to the system which we called F1$_\alpha$, and the corresponding geometry describes a non--extremal fundamental string with charge $Q_1=2m\sh_\alpha^2$:
\bea\label{KerrString}
ds^2&=&\frac{dy^2}{h_\alpha}+\frac{\rho^2}{\Delta}dr^2+\rho^2 d\theta^2-\left[\frac{\Delta}{\rho^2}+
\frac{4(mr \sh_\alpha)^2}{\rho^4h}\right](\ch_\alpha dt-as_\theta^2 d\phi)^2\nonumber\\&&
\qquad+\frac{s_\theta^2}{\rho^2}\left[(r^2+a^2)d\phi-
a\ch_\alpha dt\right]^2+(\sh_\alpha dt)^2\\
B_2&=&\frac{2mr \sh_\alpha}{\rho^2h}\left[\ch_\alpha dt-as_\theta^2 d\phi\right]\wedge dy,\qquad e^{2\Phi}=\frac{1}{h_\alpha}.
\nonumber
\eea
Here we defined
\bea
&&\rho^2=r^2+a^2c^2_\theta, \quad \Delta=r^2+a^2-2mr, \quad 
h_\alpha=1+\frac{2mr\sh_\alpha^2}{\rho^2}.
\nonumber
\eea
Transformation (\ref{KYTtransYcomp}) leads to the modified Killing--Yano tensor for (\ref{KerrString})
\bea\label{mKYTF15D}
Y&=&\frac{1}{h_\alpha}\left\{rs_\theta d\theta\wedge\left[(r^2+a^2)d\phi-a\ch_\alpha dt\right]+
ac_\theta dr\wedge \left[\ch_\alpha dt-as_\theta^2 d\phi\right]\right\}
\nonumber\\
&&+\frac{\sh_\alpha}{h_\alpha}(ac_\theta dr-ars_\theta d\theta)\wedge dy+
\frac{h_\alpha-1}{h_\alpha}r^3s_\theta d\theta\wedge d\phi.
\eea
To compare it with \eqref{Kerr4DKT}, we construct the Killing tensor $K_{MN}=-Y_{MA}{Y^A}_{N}$, define the frames as eigenvectors of this tensor, and rewrite the answer in terms of them: 
\bea\label{aa12}
ds^2&=&-e_t^2+e_y^2+e_r^2+e_\theta^2+e_\phi^2,\nn 
Y&=&r e_\theta\wedge e_\phi+a c_\theta e_r\wedge e_t,\quad
K=r^2[e_\theta^2+e_\phi^2]+(ac_\theta)^2[e_t^2-e_r^2]\nn
e_t&=&\frac{\sqrt{\Delta}}{\rho h_\alpha}\left(\sh_\alpha dy+\ch_\alpha dt-as^2_\theta d\phi\right),\quad e_r=\frac{\rho}{\sqrt{\Delta}}dr,\quad 
e_\theta=\rho d\theta,\\
e_y&=&\frac{1}{h_\alpha}\left[\ch_\alpha dy+\sh_\alpha
\left(1-\frac{2mr}{\rho^2}\right)dt+
\frac{amr s^2_\theta \sh_{2\alpha}}{\rho^2}d\phi \right],\nn
e_\phi&=&\frac{s_\theta}{\rho h_\alpha}\left(-a\sh_\alpha dy-a\ch_\alpha dt+(r^2+a^2+2mr\sh^2_\alpha)d\phi \right). \nonumber
\eea
Notice that eigenvalues of the Killing tensor and mKYT do not depend on the boost parameter $\alpha$.

The duality sequence (\ref{DualChain}) involves D-branes supported by Ramond--Ramond flux, and the analysis presented in section \ref{SectionModifiedKYT} does not apply to T duality performed in such systems. It would be interesting to generalize our discussion of mKYT to the geometries with Ramond--Ramond fields, but such analysis goes beyond the scope of this work. Instead we applied the duality chain (\ref{DualChain}) to the Kerr black hole and solved the mKYT equations for the resulting F1--NS5 geometry. We found that the mKYT does not exist in the system involving NS5 branes unless one takes an extremal limit and sets the F1 charge to zero:
\bea
m\rightarrow 0,\ Q_1\rightarrow 0,\quad \mbox{fixed}\quad Q_5=2m\sinh^2\alpha.
\eea
The resulting geometry,
\bea\label{KerrNS54d}
ds^2&=&-dt^2+h\left[\frac{\rho^2}{r^2+a^2}dr^2+\rho^2 d\theta^2+(r^2+a^2) s_\theta^2d\phi^2+dy^2\right],\nonumber\\
B_2&=&\frac{Q_5(r^2+a^2) c_\theta}{\rho^2} d\phi \wedge dy,\quad e^{2\Phi}=h, \quad h=1+\frac{Q_5 r}{r^2+a^2},
\eea
admits the unique mKYT
\bea\label{mKYTNS55D}
Y&=&hdy\wedge(rs_\theta d\theta-c_\theta dr)+hd\phi\wedge [rs_\theta^2 dr+(r^2+a^2)s_\theta c_\theta d\theta]\nonumber\\
&=&hd\left[rc_\theta dy-\frac{1}{2}(r^2+a^2) s_\theta^2 d\phi\right]
\eea
which was found by the direct calculation. Introducing convenient frames, we can rewrite this KYT and its square as
\bea\label{KYT4DNS5Extr}
Y&=&e_r\wedge e_y-e_\theta \wedge e_\phi, \qquad K\equiv -Y_{MA}Y^A{}_N dx^Mdx^N=e_r^2+e_y^2+e^2_\theta +e^2_\phi,\nn
e_t&=&dt,\quad 
e_r=\sqrt{\frac{\rho^2+Qr}{r^2+a^2}}dr,\quad e_\theta=\sqrt{\rho^2+Qr},\\
e_{y}&=&\frac{1}{\rho^2}\sqrt{(r^2+a^2)(\rho^2+Qr)}\left[\cos\theta dy+r\sin^2\theta d\phi \right],\nn
e_{\phi}&=&
\frac{\sin\theta}{\rho^2}\sqrt{(\rho^2+Qr)}\left[r dy-(r^2+a^2)\cos\theta d\phi \right].\nonumber
\eea
Notice that square of the KYT gives the spacial part of the metric, which can be viewed as a linear combination of two `trivial' Killing tensors: one coming form the metric and one built from the square of the Killing vector $\d_t$.

An additional T duality along $y$ direction in (\ref{KerrNS54d}) produces a metric of the extremal KK monopole, and application of (\ref{KYTtransYcomp}) to (\ref{mKYTNS55D}) gives the standard KYT for the monopole:
\be
Y=dr\wedge[(Q+r\sin^2\theta)  d\phi +\cos\theta dy]+d\theta\wedge[\cos\theta\sin\theta(a^2+r^2) d\phi-r\sin\theta dy].
\ee
In the frames we find
\bea
Y&=&e_r\wedge e_y-e_\theta \wedge e_\phi,\qquad K=e_r^2+e_y^2+e^2_\theta +e^2_\phi,\nn
e_t&=&dt,\quad e_r=\sqrt{\frac{\rho^2+Qr}{r^2+a^2}}dr,\quad e_\theta=\sqrt{\rho^2+Qr},\nn
e_{y}&=&\sqrt{\frac{r^2+a^2}{\rho^2+Qr}}\left[\cos\theta dy+(Q+r\sin^2\theta)d\phi\right],\\
e_{\phi}&=&\frac{\sin\theta}{\sqrt{\rho^2+Qr}}\left[r dy -\cos\theta(r^2+a^2)d\phi \right].\nonumber
\eea
Once again, the KYT squares to a `trivial'' Killing tensor.


\subsection{F1--NS5 system from the five--dimensional black hole.}
\label{SubsectionExamples6D}
\label{SecEx6d}

Application of the duality chain (\ref{DualChain}) to the five--dimensional black hole gives another example of the rotating F1--NS5 system, the complete geometry was found in 
\cite{cvet,3charge}, and it is given by equation \eqref{NonExtr6D}. This subsection  discusses the modified Killing--Yano tensor for this solution.

Recalling that even the neutral five--dimensional black hole had the KYT of rank three rather than two (see section \ref{SecKYTranks}), we should look at the obvious extension of  \eqref{KYTBfield} to such objects\footnote{The discussion presented in Appendix \ref{AppBeqKYT} trivially extends to KY tensors of arbitrary rank.}:
\bea\label{KYTBfieldrank3}
{\nabla}_M { Y}_{NPQ}+
{\nabla}_N {Y}_{MPQ}+
\frac{1}{2} H_{MPA}{ g}^{AB}{Y}_{NB Q}+
\frac{1}{2} H_{MQA}{ g}^{AB}{ Y}_{NPB}\nonumber\\
+\frac{1}{2} H_{NPA}{ g}^{AB}{Y}_{MB Q}+
\frac{1}{2} H_{NQA}{ g}^{AB}{Y}_{MPB}=0.
\eea
The general construction of section \ref{SectionModifiedKYT} guarantees existence of the mKYT for $\alpha=0$ (as long as constraint (\ref{PureMetrConst}) is satisfied), but the generation of the NS5 branes goes through Ramond--Ramond fluxes, which can potentially destroy the modified KYT. Remarkably, the tensor survives, and solution of \eqref{KYTBfieldrank3} for the geometry \eqref{NonExtr6D} is 
\bea\label{mKYT6DNonExtr}
Z^{-1}Y&=&-ad[r^2\cos^2\theta] dt d\psi -a\mu_A\mu_B d[(r^2+a^2-M)\sin^2\theta]d\phi dy
\\ 
&&+
a\mu_A d[(r^2+a^2-M)\sin^2\theta]  dt  d\phi-a\mu_Bd[r^2\cos^2\theta] dy d\psi 
+\sigma d[\sin^2\theta] d\phi  d\psi \nonumber
\eea
with
\bea
Z&=&\frac{r^2+A^2+a^2c_\theta^2}{r^2+B^2+a^2c_\theta^2},\quad
\sigma=\frac{(a^2 - M)A^2 B^2 - [a^2 +  A^2 + B^2 ] M r^2 - M r^4}{\sqrt{A^2+M}\sqrt{B^2+M}},
\nn
\mu_A&=&\frac{A}{\sqrt{M+A^2}},\quad \mu_B=\frac{B}{\sqrt{M+B^2}},\quad
s_\theta=\sin\theta,\quad c_\theta=\cos\theta,\\
A&=&\sqrt{M}\sinh\alpha,\qquad B=\sqrt{M}\sinh\beta.\nonumber
\eea
Although expression (\ref{mKYT6DNonExtr}) is already relatively simple, we also rewrite it in frames to connect to the general analysis presented in section \ref{SecKYTranks}. Constructing the Killing tensor $K_{MN}=-Y_{MA}{Y^A}_{N}$ and defining the frames as its eigenvectors, we find 
\bea\label{5DmKYT}
ds^2&=&-e_t^2+e_y^2+e_r^2+e_\theta^2+e_\phi^2+e_\psi^2,\nn
 Y&=&\left(a\sqrt{2 A^2 + 2Mc^2_{\theta}}\,e_r\wedge e_t+\sqrt{2M(A^2+r^2)-2a^2A^2}e_\theta\wedge e_\phi\right)\wedge e_\psi,\nn
 K&=&-\frac{1}{2}Y_{MA}{Y^A}_{N}=[M(A^2+r^2)-a^2A^2][e^2_\theta+e^2_\phi]+
 a^2 (A^2 + M c_\theta^2)[e^2_t-e^2_r]\nn
 &&+
[a^2(A^2+M c_\theta^2)-(M(A^2+r^2)-a^2A^2)]e_\psi^2,
 \eea
 where the frames are given by 
 \bea
e_t&=&\frac{r}{\rho^2H_1} \sqrt{\frac{M\Delta\rho^2H_5}{M(A^2+r^2)-a^2A^2}}
\left[\ch_\beta dt+\sh_\beta dy-a(\sh_\alpha c_\theta^2d\psi-\ch_\alpha s_\theta^2 d\phi)\right],\nn
e_y&=&\frac{1}{2\rho^2 H_1}\Bigg[
2\sh_\beta\left(\rho^2-M\right)dt+2\rho^2\ch_\beta dy+aM\sh_{2\beta}(
\sh_\alpha c_\theta^2d\psi-\ch_\alpha s_\theta^2 d\phi) \Bigg], \nn
e_r&=&\sqrt{\frac{A^2+\rho^2}{\Delta}}dr,\quad e_\theta=\sqrt{A^2+\rho^2}d\theta,\\
e_\phi&=&\frac{s_\theta c_\theta}{\rho H_1}\sqrt{\frac{MH_5}{A^2+Mc^2_{\theta}}}\Bigg[a\ch_\beta dt+a\sh_\beta dy+
 (B^2+r^2)(\sh_\alpha d\psi + \ch_\alpha d\phi)+a^2\ch_\alpha d\phi \Bigg],\nonumber\\
e_\psi&=&\frac{1}{\rho^2H_1\sqrt{A^2+Mc^2_{\theta}}\sqrt{M(A^2+r^2)-a^2A^2}}
\Bigg[
\nn
&&[r^2+(a^2-M)c_\theta^2]\left(\frac{1}{2}aM\sh_{2\alpha}
(\ch_\beta dt+\sh_\beta dy)+
Ma^2\sh_\alpha s_\theta^2 d\phi
\right)
\nn
&&+\left[M(r^2+A^2)(r^2+B^2) + MA^2 c_\theta^2 r^2-A^2 B^2 a^2 \right]
(\sh_\alpha s_\theta^2 d\phi-\ch_\alpha c_\theta^2 d\psi)\Bigg].\nonumber
\eea
For $\alpha=0$ we find
\bea
e_t&=&\frac{1}{\rho^2H_1} \sqrt{{\Delta\rho^2}}
\left[\ch_\beta dt+\sh_\beta dy+as_\theta^2 d\phi\right],\nn
e_y&=&\frac{1}{2\rho^2 H_1}\Bigg[
2\sh_\beta\left(\rho^2-M\right)dt+2\rho^2\ch_\beta dy-aM\sh_{2\beta} s_\theta^2 d\phi \Bigg], \nn
e_r&=&\sqrt{\frac{\rho^2}{\Delta}}dr,\quad e_\theta=\sqrt{\rho^2}d\theta,\quad
e_\psi=r\cos\theta d\psi,\\
e_\phi&=&\frac{s_\theta }{\rho H_1}\Bigg(a\ch_\beta dt+a\sh_\beta dy+
(a^2+M\sh^2_\beta+r^2)d\phi \Bigg).\nonumber
\eea
This is the special case of (\ref{ChMPframeOdd}) for $n=1$ and one rotation parameter. 
Finally we give the expression for the mKYT (\ref{5DmKYT}) in the extremal limit ($M=0$ with fixed $A,B$):
\bea
Z^{-1}Y
&=&-\frac{1}{2}d\left[(dt+dy)\left\{(r^2c_\theta^2)d\psi-
(r^2+a^2)s_\theta^2 d\phi\right\}+ABa\cos^2\theta d\phi d\psi\right].
\eea


\subsection{Conformal Killing--Yano tensors}

We conclude this section with discussing the CKYT for rotating F1--NS5 systems. 
Explicit calculations show that the geometry obtained by application of (\ref{DualChain}) to the Kerr solution (\ref{Kerr4D}) does not have CKYT. On the other hand 
F1--NS5 system constructed from the five--dimensional black hole (\ref{5DKerrNtrFrm}) does admit a CKYT if and only if $Q_1=Q_5$. In this case the metric has the form
\bea
ds^2&=&-e_t^2+e_\phi^2+e_y^2+e_\psi^2+e_r^2+e_\theta^2,\nonumber\\
e_{t}&=&\frac{\sqrt{\Delta}}{\rho}(dt-as_\theta^2 d\phi),\quad 
e_{\phi}=\frac{s_\theta}{\rho}\left[(r^2+a^2+Q)d\phi-adt\right],\\
e_{r}&=&\frac{\rho}{\sqrt{\Delta}}dr,\quad e_{\mathbf\theta}=\rho d\theta,\quad 
e_\psi=\frac{c_\theta}{\rho}[(Q+r^2)d\psi-a dy],\quad e_y=\frac{r}{\rho}[dy+ac_\theta^2 d\psi],
\nonumber\\
\Delta&=&r^2+a^2-M,\quad \rho^2=r^2+a^2c_\theta^2+Q,\nonumber
\eea
and the corresponding CKYT and CKT are given by
\bea\label{CKYT_D1D5_NonExtr}
{\mathcal Y}&=&\rho (e_r\wedge e_t\wedge e_y+e_\theta \wedge e_\phi \wedge e_\psi),\quad Z=\frac{1}{\rho^2}(ac_\theta e_\psi-re_y)\wedge (\sqrt{\Delta}e_t+as_\theta e_\phi),\nn
{\mathcal K}&=&\rho^2[e_t^2-e_r^2-e_y^2+e_\theta^2+e_\psi^2+e_\phi^2], \quad W=-d[r^2-a^2c_\theta^2].
\eea
Since $W$ is a total derivative, the general prescription (\ref{KTfromCKT}) can be used to construct a standard Killing tensor
\bea
K=-[2(ac_\theta)^2+Q][-e_t^2+e_r^2+e_y^2]+[2r^2+Q][e_\theta^2+e_\psi^2+e_\phi^2].
\eea
Conformal Killing tensors for four-- and five--dimensional black holes discussed in this section were constructed in \cite{cvetLars} via separation of variables.

\section{Discussion}\label{SectionDiscussion}

In this chapter we analyzed hidden symmetries of stringy geometries and their behavior under string dualities. In particular, we demonstrated that in the presence of the Kalb--Ramond field the equation for the Killing--Yano tensor is modified as (\ref{KYTBfield}), and this is the unique modification consistent with string dualities. The transformations laws for the Killing vectors, tensors, and Killing--Yano tensors are given by \eqref{KVtransf}, (\ref{KillTilde})--(\ref{XforF1}), (\ref{KYTtransYcomp}). We have also demonstrated that nontrivial Killing tensors in arbitrary number of dimensions are always associated with ellipsoidal coordinates, and we used this observation to construct the (modified) Killing(--Yano) tensors for the Myers--Perry black hole (\eqref{AllFramesMP}, \eqref{Kub2}, \eqref{HInFrame}), its charged version \eqref{ChMPHYT}--\eqref{ChMPframe}, and for several examples of F1--NS5 geometries (\eqref{KYT4DNS5Extr}, \eqref{mKYT6DNonExtr}--\eqref{5DmKYT}). 

This work has several implications. First and foremost, the modified equation for the Killing--Yano tensor  (\ref{KYTBfield}) provides a new powerful tool for studying symmetries of stringy geometries, which can extend the successful applications of the standard Killing--Yano tensors to physics of black holes \cite{cvetLars, KYTbh, tube}. Also, the understanding of hidden symmetries developed in this chapter can be used to extend the `no-go theorems' for integrability \cite{ChL} to backgrounds without supersymmetry. Finally, the explicit Killing--Yano tensors for the Myers--Perry black hole and its charged version constructed in sections \ref{SecMyersPerry} and \ref{SecMPF1} generalize most of the previously known examples and provide the largest known class of KYT.

		\chapter{Towards higher dimensional black rings}\label{ChapterBlackRings}

		In the previous chapter we studied objects underlying separability of equations of motion for massive and massless particles, Killing-(Yano) tensors, and now we apply the developed framework to explore properties of black rings, black holes with events horizons of the non-spherical topology. Despite all attempts, exact solutions for black rings in more than five dimensions remain elusive. In this chapter we clarify some of the reasons for that, in particular we show that a peculiar symmetry of the five--dimensional black ring - separability of the base - cannot occur in dimensions higher than five. We also construct supersymmetric solutions that have symmetries of 5D supersymmetric black ring and show that they do not have regular horizons.

\bigskip

This chapter is based on the results published in \cite{Chervonyi:2015uua} and has the following structure.

In Section \ref{SecNeutralBR} we try to generalize a common symmetry of five--dimensional black holes and black rings to higher dimensions and show that solutions with such symmetries do not exist. In Section \ref{SecSUSYBR} we show that symmetries of 5D supersymmetric black ring do not survive in higher dimensions as well.

\section{Introduction}

\noindent

Stationary, asymptotically flat four--dimensional black holes must have only spherical event horizons \cite{Hawking}. On the other hand, black holes in higher dimensions are less restricted and allowed to possess horizons of various topologies. The first example of a black hole with a non--spherical horizon was found by Emparan and Reall \cite{BRwick}, who used Kaluza--Klein C--metric of \cite{BRorigin} to construct five--dimensional black ring with the horizon of topology $S^1\times S^2$.
Alternative methods of constructing five--dimensional black rings are based on the generalized Weyl ansatz \cite{BRWeyl} or the inverse scattering method \cite{BelinskiZaharkov}. The latter approach was used to construct the black ring with two rotations \cite{BRinvsc} and several other configurations of black rings and black holes in five dimensions \cite{InvScatRings}. Unfortunately both the Weyl and the inverse scattering approaches rely on the presence of $D-2$ commuting Killing vectors, so they cannot be used to construct black rings  in $D>5$. In the absence of methods for finding the exact solutions with non--spherical topology in higher dimensions several approximate techniques have been developed, for example, the matching asymptotic expansions \cite{MatchingAsymp} and the blackfold effective theory \cite{Blackfolds}. Along with numerical \cite{NumReviews,NumNewTopologies} and approximate \cite{ApproxNewTopologies} methods, the blackfolds have been used to shows existence of solutions with non--spherical topologies such as helical black strings/rings, non--uniform black cylinder and several other possibilities \cite{NewTopologies,NewTopologies2}.
However, despite all recent results the exact solutions with non--spherical horizon topology are known only in  $D=5$. Approximate and numerical higher--dimensional solutions were constructed in \cite{ApproxhighBR, AppHighRingsOthers} and \cite{NumHighRings} respectively.

Another interesting direction towards finding black rings in higher dimensions is based on using supersymmetry. The supersymmetric five--dimensional black ring was constructed in \cite{SUSYring} and extended to a larger class of solutions in \cite{SUSYringTubes,SUSYrings}. However, analogously to the neutral case, SUSY black rings in higher dimensions ($D>5$) are still unknown, moreover there is even less progress in this direction. To summarize, the exact solutions for higher dimensional black rings remain elusive, and in this chapter we clarify some of the reasons for that.

\section{Separability of the neutral black ring}
\label{SecNeutralBR}

The neutral  five--dimensional black ring was constructed in \cite{BRwick,BRWeyl} and reviewed in Appendix \ref{AppMPandBR}.
We  begin by noting that the black ring metric \eqref{AppBRmetric} has a structure of a $t$--fiber over the four--dimensional base, which is conformally separable:
\bea\label{BRbase}
ds_{base}^2=\frac{R^2 F(x)F^2(y)}{(x-y)^2}\left[-\Big(\frac{G(y)d\phi^2}{F^2(y)}+\frac{dy^2}{F(y)G(y)}\Big)+\Big(\frac{G(x)d\psi^2}{F^2(x)}+\frac{dx^2}{F(x)G(x)}\Big)\right].
\eea
Moreover it is separable in two different coordinate systems as we will show in a moment. In order to do that we recall that separability of the massless Hamilton--Jacobi equation
\bea\label{MasslessHJ}
g^{MN}\d_M S\d_N S=0
\eea
can be encoded in the conformal Killing tensor $\mathcal{K}$ of rank two\footnote{Killing tensor generalizes the well--known notion of Killing vector.}, which satisfies the following equation \cite{Tachibana}
\bea\label{CKTeq}
\nabla_{(M} \mathcal{K}_{NP)}=W_{(M}g_{NP)},
\eea
where $W_M$ is the associated vector. 
To find conformal Killing tensors one can solve the general equation \eqref{CKTeq} or extract it from the metric via \eqref{MasslessHJ} if it is written in the separable coordinates. Usually such tensors are used to construct the conserved quantities through
\bea
I=\mathcal{K}^{MN}\d_M S \d_N S,
\eea
or to extract the separable coordinates. Procedures of constructing Killing tensors from the metric, extracting separable coordinates from the tensors are described in Section 2 of \cite{KYTinSt} and here we outline the results.

The massless Hamilton--Jacobi equation \eqref{MasslessHJ} separates if there exists a function $f$, such that
\bea
g^{MN}=\frac{1}{f}\left(X^{MN}+Y^{MN} \right), \quad \d_x Y^{MN}=\d_y X^{MN}=0.
\eea
Then the conformal Killing tensor can be read off as
\bea\label{CKTfromSep}
\mathcal{K}^{MN}=X^{MN}.
\eea
Applying this procedure to the base of the black ring \eqref{BRbase} we get
\bea
(\mathcal{K}^{MN})^{(x)}\d_N\d_N=\frac{F^2(x)}{G(x)}\d_\psi^2+F(x)G(x)\d_x^2.
\eea
Here the index ${}^{(x)}$ indicates that this tensor was read off from the $x$--dependent part of the metric\footnote{Note that one always can read off another tensor from the $y$--dependent part of the metric, but these tensors will not be independent.}. Lowering the indices gives
\bea\label{CKTring1}
(\mathcal{K}_{MN})^{(x)}dx^Mdx^N=\left[\frac{R^2 F(x)F^2(y)}{(x-y)^2}\right]^2\left[\frac{G(x)d\psi^2}{F^2(x)}+\frac{dx^2}{F(x)G(x)}\right].
\eea
The solution \eqref{AppBRmetric},\eqref{BRbase} contains both the black ring and the black hole with one rotation (for details see Appendix \ref{AppMPandBR}), which allows us to make an assumption that higher dimensional black rings are in the same class of solutions as the Myers--Perry black holes with one rotation \cite{MyersPerry}. So we can study higher--dimensional neutral black holes and based on the results make conclusions about the corresponding rings.

The Myers--Perry black holes are reviewed in Appendix \ref{AppMPandBR}, and here we start with writing the tensor \eqref{CKTring1} in the standard Myers--Perry coordinates for the static case. Substituting the map \eqref{AppStaticMap} into \eqref{CKTring1} we get
\bea\label{CKT2}
(\mathcal{K}_{MN})^{(x)}dx^M dx^N=\frac{r^6 \cos^2\theta}{2R^2}d\psi^2-r^4\cos^4\theta d\psi^2+\frac{r^6 \cos^2\theta}{R^2}\left[d\ln\frac{\cos\theta}{r}\right]^2,\quad m=2R^2.
\eea

We have found the conformal Killing tensor associated with separability of the base in the ring--like coordinates, and now we want to find all separable coordinate systems for static neutral black holes followed by analysis of the cases with one rotation. Solving the conformal Killing tensor equation \eqref{CKTeq} on the base of the Tangherlini solution \cite{Tangherlini} with $D\ge 5$
\bea\label{SchwBase}
ds_{base}^2=\frac{dr^2}{1-\frac{m}{r^{D-3}}}+r^2 (d\theta^2+\sin^2\theta d\phi^2 +\cos^2\theta d\Omega_{D-4}^2),
\eea
we obtain the following non-trivial conformal Killing tensors\footnote{Note that here we do not write expression for the associated vectors $W^{(i)}$ entering right--hand side of the conformal Killing tensor equation \eqref{CKTeq} because they can be easily recovered from the corresponding conformal Killing tensors $\mathcal{K}^{(i)}$.}:
\bea\label{CKTsSchw}
\mathcal{K}_{MN}^{(1)}dx^Mdx^N&=&r^4{(p+q)^{4/(D-3)}}\cos^2\theta \Big[\left(d\ln \cos\theta-
\frac{2}{D-3}d\ln[p+q]\right)^2+d\Omega^2_{D-4}\Big]\nn
\nn
\mathcal{K}_{MN}^{(2)}dx^Mdx^N&=&r^4{(q-p)^{4/(D-3)}}\cos^2 \theta \Big[\left(d\ln \cos\theta+
\frac{2}{D-3}d\ln[p+q]\right)^2+d\Omega^2_{D-4}\Big],\nn
\mathcal{K}_{MN}^{(3)}dx^Mdx^N&=&\frac{r^{D-1}}{r^{D-3}-m}dr^2,\quad q=r^{(D-3)/2},\quad p=\sqrt{q^2-m}.
\eea
Now we identify the separable coordinate systems associated with these conformal Killing tensors. 

Using the prescription \eqref{CKTfromSep} we extract the obvious tensor from \eqref{SchwBase}
\bea
(\mathcal{K}^{MN})^{(r)}\d_M\d_N=\frac{r^{D-3}-m}{r^{D-5}}(\d_r)^2\quad \Rightarrow \quad \mathcal{K}^{(r)}=\mathcal{K}^{(3)}.
\eea
Here the index ${}^{(r)}$ indicates that this tensor is associated with the $r$ coordinate. We conclude that $\mathcal{K}^{(3)}$ is associated with separability in the standard Myers--Perry coordinates.

Next by comparing the expressions \eqref{CKT2} and \eqref{CKTsSchw} we find that the tensor responsible for separation in the ring--like coordinates is 
\bea
\mathcal{K}^{(x)}=\frac{1}{2m}\left(\mathcal{K}^{(1)}+\mathcal{K}^{(2)}\right)+\mathcal{K}^{(3)}.
\eea
Thus we have associated two independent conformal Killing tensors with separable coordinate systems, namely the ring--like and standard Myers--Perry coordinates, but the complete solution of the conformal Killing equation \eqref{CKTeq} gives three independent tensors, so it is natural to ask about the meaning of the third one. To answer this question we recall the procedure of extracting coordinates from Killing tensors described in Section 2 of \cite{KYTinSt}. It was shown that the separable coordinates can be obtained from the eigenvectors of corresponding tensors. For the tensors \eqref{CKTsSchw} we find
\bea\label{CoordsTable}
\begin{array}{|r|c|c|c|}
\hline
      & \mathcal{K}^{(1)} & \mathcal{K}^{(2)} & \mathcal{K}^{(3)} \\ \hline
    \hat{x} & \displaystyle\frac{\sin\theta}{\sqrt{q+p}} & \sqrt{q+p}\sin\theta & q \\ \hline
    \hat{y} & -\displaystyle\frac{\cos\theta}{\sqrt{q+p}} & -\sqrt{q+p}\cos\theta & \theta \\ \hline
\end{array}
\qquad q=r^{(D-3)/2},\quad p=\sqrt{q^2-m}.
\eea
To summarize we see that the expressions for conformal Killing tensors \eqref{CKTsSchw} have universal character in all dimensions. It means that the bases of \textit{neutral static black holes have the same symmetries in any dimension} and five dimensions is not an exception. Now we will check this statement in the rotating case. Direct solving the conformal Killing tensor equation and treating rotation as a perturbative parameter shows that introduction of rotation decreases the number of conformal Killing tensors down to one in dimensions higher than five. In five dimensions turning on rotation destroys one of the tensors, and the survivors are $\mathcal{K}^{(3)}$ and $\mathcal{K}^{(1)}+\mathcal{K}^{(2)}$ corresponding to separation in the standard Myers--Perry and ring--like coordinates respectively. In dimensions higher than five only one tensor responsible for separation in the Myers--Perry coordinates survives. We conclude that if higher dimensional black rings are in the same class of solution as black holes they would not have the separable bases unlike their five--dimensional counterpart.


\section{Supersymmetric black rings}
\label{SecSUSYBR}

In the previous section we have been focusing on the neutral black rings and now we switch to the analysis of their supersymmetric counterparts. Consider the five dimensional SUSY black ring constructed in \cite{SUSYring} and recall that it was found 
by utilizing a very special feature of the neutral black ring - separability of the fiber. The solution was written in the form
\bea\label{SUSYBR5D}
ds^2=-f^2(dt+\omega)^2+f^{-1}h_{mn}dx^mdx^n,\quad m,n=1,..,4,
\eea
and the fiber one--form $\omega$ was assumed to satisfy
\bea
\omega=\omega_\phi d\phi+\omega_\psi d\psi,\quad \d_x\omega_{\psi}=\frac{y^2-1}{1-x^2}\d_y\omega_{\phi}.
\eea
Furthermore 5D SUSY ring was embedded into M theory in \cite{SUSYringTubes}, which is a good starting point for constructing higher dimensional SUSY rings. Following \cite{SUSYringTubes} the solution reads
\bea\label{11sol}
ds_\mathit{11}^2 &=&  ds^2_\mathit{5} + X^1 \left( dz_1^2 +
dz_2^2 \right) + X^2 \left( dz_3^2 + dz_4^2 \right) +
X^3 \left( dz_5^2 + dz_6^2 \right),\nonumber \\
\mathcal{A} &=& A^1 \wedge dz_1 \wedge dz_2 + A^2 \wedge dz_3 \wedge dz_4 +
A^3 \wedge dz_5 \wedge dz_6.
\eea
Here $\mathcal{A}$ is the three-form potential with four-form field
strength $\mathcal{G}=d\mathcal{A}$. The solution is specified by three scalars $X^i$, and three one-forms
$A^i$, which are defined on a five-dimensional
spacetime with metric $ds^2_\mathit{5}$:
\bea
\label{5dsol}
ds_\mathit{5}^2 &=& -(H_1 H_2 H_3)^{-2/3} (dt + \omega)^2 +
(H_1 H_2 H_3)^{1/3} d\mathbf{x}_\mathit{4}^2,\nonumber \\
A^i &=& H_i^{-1}(dt +\omega) +
 \frac{q_i R^2}{r^2+R^2\cos^2\theta} (\sin^2\theta d\psi -\cos^2\theta d\phi ),\\
X^i &=& H_i^{-1} (H_1 H_2 H_3)^{1/3},\nn
d\mathbf{x}_\mathit{4}^2&=&(r^2+R^2\cos^2\theta)\left(\frac{dr^2}{r^2+R^2}+d\theta^2 \right)+(r^2+R^2)\sin^2\theta d\psi^2+r^2 \cos^2\theta d\phi^2.\nonumber
\eea
Here $H_i$ are harmonic functions on the flat four--dimensional base $d\mathbf{x}_4^2$. 

Now we want to write the prototype of 7D SUSY black ring while keeping the symmetries of \eqref{11sol}, such as the flat base and several $Z_2$ symmetries. First we extend the flat base to six dimensions effectively absorbing $z_5,z_6$ into $d\mathbf{x}_4^2$ and focus on symmetries associated with the rest of $z_i$. Recalling the equations of motion in 11D SUGRA
\bea\label{EOM11D}
R_{MN}-\frac{1}{12}\left(\mathcal{G}_{MABC}\mathcal{G}_N{}^{ABC}-\frac{1}{12}g_{MN}\mathcal{G}^2 \right)=0,\nn
d\star\mathcal{G}+\frac{1}{2}\mathcal{G}\wedge \mathcal{G}=0
\eea 
we note that even though, for example, $(z_1,z_2)\to -(z_1,z_2)$ is not a symmetry of the field strength $\mathcal{G}$, it is a symmetry of equations of motion \eqref{EOM11D}. The rest of the symmetries together with their restrictions on the metric and three--form are collected in the following table\footnote{Here $z_i$ are denoted as $i$. For example, $g_{z_1z_1}=g_{11}$, etc.}:
\bea\label{SymmsTable}
\begin{array}{|l|l|}
\hline
     \mbox{Symmetry} & \mbox{Prohibited~expressions} \\ \hline
     (z_1,z_2)\to -(z_1, z_2) & g_{13}, g_{14}, g_{23}, g_{24}\\ 
      & \mathcal{A}_{13}, \mathcal{A}_{14}, \mathcal{A}_{23}, \mathcal{A}_{24}\\ \hline
     z_1\leftrightarrow z_2, z_3\leftrightarrow z_4& g_{11}\ne g_{22}, g_{33}\ne g_{44}\\ \hline
     (z_1,z_3)\to -(z_1, z_3) & g_{12}, g_{34}\\ \hline
\end{array}
\eea
The first line constrains the three form $\mathcal{A}$ and partially fixes the metric.
The rest constrains the metric resulting in the following ansatz
\bea\label{11Dansatz}
ds_{11}^2&=&-\hat{H}_1(dt+\omega)^2+\hat{H}_2 d\mathbf{x}_6^2+\hat{H}_3(dz_1^2+dz_2^2)+\hat{H}_4(dz_3^2+dz_4^2),\nn
\mathcal{A}&=&A^1\wedge dz_1\wedge dz_2+A^2\wedge dz_3\wedge dz_4+\mathcal{C},
\eea
where $\hat{H}_{i}, i=1,..,4$ are unknown functions, $d\mathbf{x}_6^2$ is the flat six--dimensional space, $A^{1,2}, \mathcal{C}$ are respectively one-- and three--forms on the seven--dimensional base. Performing the dimensional reduction along one of $z_i$ followed by three T dualities along remaining $z_i$ and finally S duality gives the solution in IIB SUGRA:
\bea
ds_{10}^2&=&-\tilde{H}_1^2(dt+\omega)^2+\tilde{H}_2^2(dz_2+fdt+\alpha)^2+\tilde{H}_3^2 d\mathbf{x}_6^2+\tilde{H}_4^2(dz_{3}^2+dz_4^2),\nn
B&=&\beta_1dt+\beta_2dz_2+gdtdz_2+\tilde{\omega}_2, \quad e^{2\Phi}=\tilde{H}_5.
\eea
Here one--forms  $\omega,\alpha$ are defined on the flat six--dimensional base. Solving the Killing spinor equations for this ansatz reveals that the most general solution is governed by the chiral null model \cite{HorStr}:
\bea
ds^2&=&\frac{2}{H}dz_2(dt+\omega)+{\mathcal F}dz_2^2+d\mathbf{x}_6^2+(dz_3^2+dz_4^2),
\quad e^{-2\phi}=H,\nn
d\star_6(dH)&=&0,\quad d\star_6(d\omega)=0,\quad d\star_6(d[H {\mathcal F} ])=0.
\eea
This family of solutions does not admit a horizon with a non--zero area, but does give rise to a stretched horizon \cite{Sen} and upon dualization to D1--D5 frame may lead to completely regular geometries \cite{lmPar,lmm}.

We conclude that the natural extension 
of 5D SUSY black ring to higher dimensions by keeping its symmetries (the flat base and several $Z_2$ symmetries) leads to the chiral null model, which means the absence of 
a horizon. So in order to produce the finite ring--like horizons in higher dimensions 
one must consider the non--flat bases.


\section{Discussion}

In the first part of this work we have shown that unlike neutral static black holes, which have the same symmetries in all dimensions, rotating higher dimensional black holes/rings do not. In particular, we found that separability of the base in more than one coordinate system is a special feature of the low--dimensional rotating black holes ($D\le 5$), which makes the 5D black ring metric to have such a simple structure. Our results show that if higher dimensional black rings are described by the same class of solutions as the black holes (as it occurs in 5D), then their metric will not have a simple and symmetric structure.

In the second part of this work we have shown that generalization of
5D supersymmetric black ring to higher dimensions while keeping its symmetries (in particular, the flat base) results in vanishing horizon. It would be interesting to extend this analysis to the non--flat bases.

		\chapter{Supergravity background of the $\lambda$--deformed $AdS_3\times S^3$ supercoset}\label{ChapterLambda}

		Three previous chapters were devoted to studying motion of point-like particles on various stringy backgrounds. Now we turn our attention to exploring dynamics of strings and study a powerful approach of generating new integrable theories, the method of integrable deformations. In particular, we focus on the recent development called $\lambda$-deformation and in this chapter construct the solution of type IIB supergravity describing the integrable lambda-deformation of the $AdS_3 \times S^3$ supercoset. While the geometry corresponding to the deformation of the bosonic coset has been found in the past, our background is more natural for studying superstrings, and several interesting features distinguish our solution from its bosonic counterpart. We also report progress towards constructing the lambda-deformation of the $AdS_5 \times S^5$ supercoset.

\bigskip

This chapter is based on the results published in \cite{ChLLambda} and has the following organization. In section \ref{SecReview} we review the procedure for constructing the $\la$--deformation, which will be used in the rest of the chapter. In section \ref{SecAdS3} we use this procedure to construct the metric and the dilaton for the deformed AdS$_3\times$S$^3$, but unfortunately construction of Ramond--Ramond fluxes requires a separate analysis. In section \ref{SecAdS3RR} we determine these fluxes by solving supergravity equations, and in sections \ref{SecAdS3SpecCases}--\ref{SecAltParamAdS3} we find some interesting connections between the new background and solutions which exist in the literature. Section \ref{SectAdS5} reports progress towards constructing the $\la$--deformation for super--coset describing strings on AdS$_5\times$S$^5$. Specifically, we determine the metric and the dilaton, but unfortunately we were not able to compute the Ramond--Ramond fluxes. The $\la$--deformation of AdS$_2\times$S$^2$ constructed in \cite{BTW} is reviewed in Appendix \ref{AppAdS2}, and its comparison with higher dimensional cases is performed throughout the chapter.

\section{Introduction}
 
Integrability is a remarkable property, which has led to a very impressive progress in understanding of string theory over the last two decades (see \cite{review_AdSCFT_int} for review). While initially integrability was discovered for isolated models, such as strings on AdS$_p\times$S$^q$ \cite{RoibPolch, bethe, IntCases, IntAdS2, AdS3CFT2}, later larger classes of integrable backgrounds have been constructed by introducing deformations parameterized by continuous variables. The first example of such family, known as beta deformation \cite{BetaIntegr, Frolov}, has been found long time ago \cite{margin}, but recently two new powerful tools for constructing 
integrable string theories have emerged. One of them originated from studies of the Yang--Baxter sigma models \cite{BosonicYangBaxter,Qdeform,Yoshida}, and it culminated in construction of new integrable string theories, which became known as $\eta$--deformations \cite{Eta,Delduc1,OtherEta,MoreEta,MoreEta2}. The second approach originated from the desire to relate two classes of solvable sigma models, the Wess--Zumino--Witten \cite{WZW} and the Principal Chiral \cite{PCM} models, and it culminated in the discovery of a one--parameter family of integrable conformal field theories, which has WZW and PCM as its endpoints \cite{SftsGr,HMS,ST}\footnote{See \cite{PreST} for earlier work in this direction.}. This connection becomes especially interesting when the PCM point represents a string theory on AdS$_p\times$S$^q$ space, and the corresponding families, which became known as $\la$--deformations, have been subjects of recent investigations \cite{SfetsosAdS5,HT,Holl2,BTW}. A close connection between the $\eta$ and $\la$ deformations has been demonstrated in \cite{HT}. In this chapter we study the $\la$--deformation for AdS$_3\times$S$^3$ and AdS$_5\times$S$^5$.

While the metrics for the $\la$--deformation of AdS$_p\times$S$^q$ have been constructed in \cite{ST,SfetsosAdS5}, the issue of the fluxes supporting these geometries has not been fully resolved. Although the metric for the deformation can be uniquely constructed starting from the corresponding coset,  there are two distinct prescriptions for the dilaton: one is based on a bosonic coset \cite{ST}, and the other one uses its supersymmetric version \cite{HMS}. In the first case the deformations for all AdS$_p\times$S$^q$ have been constructed in a series of papers \cite{ST,SfetsosAdS5}, while in the second case, which is more natural for describing superstrings, only the result for AdS$_2\times$S$^2$ is known \cite{BTW}. In this chapter we construct the geometry describing the $\la$--deformed AdS$_3\times$S$^3$ supercoset and report progress towards finding the deformed AdS$_5\times$S$^5$ solution.

\section{Brief review of the $\la$--deformation}
\label{SecReview}

We begin with reviewing the procedure for constructing the NS--NS fields for the $\la$--deformed cosets.  Such deformation belongs to a general class of two--dimensional integrable systems with equations of motion in the form \cite{SftsGr}
\bea\label{EOMcrnt}
&&\d_\mu I^\mu=0,\nn
&&\d_\mu I_\nu-\d_\nu I_\mu+[I_\mu, I_\nu]=0,
\eea
where currents $I_\mu$ take values in a semi--simple Lie algebra. Integrability of this system can be demonstrated by writing it as a zero--curvature condition for a linear problem\footnote{We denote that spectral parameter by $\Lambda$ instead of the conventional $\la$ to avoid confusion with a variable governing the deformation.}:
\bea
&&\mathcal{D}_\mu\Psi=0,\qquad \mathcal{D}_\mu(\Lambda)=\d_\mu+\frac{\Lambda^2}{\Lambda^2-1}I_\mu+\frac{\Lambda}{\Lambda^2-1}\epsilon_{\mu\rho}I^\rho,\nn
&&[\mathcal{D}_\mu(\Lambda),\mathcal{D}_\nu(\Lambda)]=0\,.\quad
\eea
Two well--known examples of the integrable systems described by equations (\ref{EOMcrnt}) are the  Principal Chiral Model (PCM) \cite{PCM} and the Wess-Zumino-Witten model \cite{WZW} for a group $G$:
\bea\label{PCM}
S_{PCM}(\tilde{g})&=&-\frac{\kappa^2}{\pi}\int\mbox{Tr}(\tilde{g}^{-1}\d_+\tilde{g} \tilde{g}^{-1} \d_-\tilde{g}),\quad I_\mu=\tilde{g}^{-1}\d_\mu \tilde{g},\\ 
\label{WZW}
S_{WZW}(g)&=&-\frac{k}{2\pi}\int\mbox{Tr}\left( g^{-1}\d_+g g^{-1}\d_- g \right)+\frac{ik}{6\pi}\int_B \mbox{Tr}(g^{-1}d g)^3,\quad I_\mu=g^{-1}\d_\mu g,
\eea
and the $\la$--deformation interpolates between these systems. This deformation utilizes two important symmetries of (\ref{PCM}) and (\ref{WZW}): the global $G_L\times G_R$ symmetry of the PCM and the  $G_{L,cur}\times G_{R,cur}$ symmetry of the current algebra of the WZW.

\bigskip
\noindent
$\la$--\textbf{deformation for groups}

Let us review the construction introduced in \cite{SftsGr}, which allows one to interpolate between the systems (\ref{PCM}) and (\ref{WZW}) while preserving integrability. To find such $\la$ deformation, one adds the PCM and WZW models (\ref{PCM}), (\ref{WZW}) for the same group $G$ and gauges the $G_L\times G_{diag,cur}$ subgroup of global symmetries. This is accomplished by modifying the derivative in the PCM as 
\bea
\d_\pm\tilde{g}\to D_\pm\tilde{g}=\d_\pm\tilde{g}-A_\pm\tilde{g},
\eea
and by gauging the resulting WZW model. Integrating out the gauge fields $A_\pm$, one arrives at the final action \cite{SftsGr}\footnote{We follow the conventions of \cite{HT,BTW}, and the deformation parameter $\la$ used in \cite{SftsGr,ST} is equal to $\la^2_{our}$.}
\bea\label{ActDeform}
S(g)&=&S_{WZW}(g)+\frac{1}{\pi}\frac{k^2}{k+\kappa^2}\int J_+^a(1-\lambda^2 D)^{-1}_{ab}J^b_-,\quad
\lambda^2=\frac{k}{k+\kappa^2},\quad 0\le\lambda\le 1,\\
 D_{ab}&=&\mbox{Tr}(t_a g^{-1} t_b g),\quad J_\pm^a=-i\mbox{Tr}(t^a \d_\pm g g^{-1}):\quad J_+^a=R_\mu^a\d_+ X^\mu,\quad J_-^a=L_\mu^a\d_- X^\mu\,.\nonumber
\eea
Deformation (\ref{ActDeform}) interpolates between the PCM ($\lambda=1$) and the WZW model ($\lambda=0$) while preserving integrability \cite{SftsGr}.
 
 To extract the gravitational background describing the deformation, one rewrites (\ref{ActDeform}) as
\bea
S(g)=S_{WZW}(g)+\frac{k^2}{\pi}\int (R^T  M^{-1} L)_{\mu\nu} \d_+ X^\mu \d_- X^\nu,\qquad
M=(k+\kappa^2)(1-\lambda^2 D),
\eea
and compares the result with the action of the sigma model
\bea
S=\frac{1}{2}\int (G+B)_{\mu\nu}\d_+ X^\mu \d_- X^\nu\,.
\eea
This leads to the metric and to the Kalb-Ramond field:
\bea\label{GeneralMetric}
ds^2&=&\frac{k}{2\pi}L^T L+\frac{k^2}{2\pi}L^T(DM^{-1}+[M^{-1}]^TD^T)L\\
B&=&\frac{1}{1-\lambda^4}\left( B_0+\frac{\lambda^2}{2}L^T\left[(D^T-\lambda^2)^{-1}-(D-\lambda^2)^{-1} \right]\wedge L\right),\nonumber
\eea
where $B_0$ is a Kalb-Ramond field of an undeformed WZW model with the field strength
\bea
H_0=-\frac{1}{6}f_{abc}L^a\wedge L^b\wedge L^c.
\eea
Recalling the definition of $M$ and the relation $D^TD=1$, one can rewrite the metric in terms of convenient frames:
\bea
ds^2&=e^a e^a,\quad e^a=\sqrt{k(1-\lambda^4)}(D-\lambda^2)^{-1}_{ab}L^b.
\eea
Expressions for $D$ and $L$ are given in (\ref{ActDeform}).

\bigskip
\noindent
\textbf{Dilaton for the $\la$--deformation}

Although extraction of the metric and the Kalb--Ramond field for the lambda deformation is rather straightforward, the procedure for calculating the dilaton is controversial. The original proposal of \cite{SftsGr} suggested the expression
\bea\label{BosDilatonGr}
e^{-2\Phi_B}=e^{-2\Phi_0} k^{\mbox{dim} G}\mbox{det} (\lambda^{-2}-D),
\eea
which can be written as 
\bea\label{BosDilaton1Gr}
e^{2\Phi_B}=\frac{1}{\mbox{det}[(Ad_f-\lambda^{-2})|_{\hat{f}}]},
\eea
where the determinant is taken in the algebra. In \cite{HMS} it was argued that for supergroups and supercosets an alternative expression is more appropriate:
\bea\label{SuDilatonGr}
e^{2\Phi}=\frac{1}{\mbox{sdet}[(Ad_f-\lambda^{-2})|_{\hat{f}}]},
\eea
Here the superdeterminant is computed in the full superalgebra $\hat{f}$. The difference between (\ref{BosDilaton1Gr}) and (\ref{SuDilatonGr}) originates from difference in the gauge fields which have been integrated out.

Recalling that an element of a superalgebra can be written as 
\bea\label{SuCosetDef} 
{\mathcal M}=\left[\begin{array}{c|c}
A&B\\
\hline
C&D
\end{array}
\right],
\eea
where $(A,D)$ are even and $(B,C)$ are odd blocks \cite{ArutFrolov,Beisert}, the expression (\ref{SuDilatonGr}) becomes
\bea\label{SuDilaton1}
e^{2\Phi}=\frac{\mbox{det}[(Ad_f-\lambda^{-2})|_{\hat{f}_1\oplus \hat{f}_3}]}{\mbox{det}[(Ad_f-\lambda^{-2})|_{\hat{f}_0\oplus \hat{f}_2}]}.
\eea
Here $\hat{f}_0$ and  $\hat{f}_2$ refer to the  even subspaces $A$ and $D$, while $\hat{f}_1$ and  $\hat{f}_3$ 
refer to the odd subspaces $B$ and $C$. In this chapter we will refer to (\ref{BosDilaton1Gr}) (which is equal to the denominator of (\ref{SuDilaton1})) as the \textit{bosonic prescription}, and the numerator of (\ref{SuDilaton1}) would be called the \textit{fermionic contribution} to the dilaton.

\bigskip
\noindent
$\la$--\textbf{deformation for cosets}

The extension of the $\la$--deformation to cosets $G/H$ is presented in \cite{ST}. Separating the generators $T^A$ of $G$ into $T^a$ corresponding to $H \subset G$ and $T^\alpha$ corresponding to the coset $G/H$, one finds the metric
\bea\label{MainCoset}
ds^2&=&e^\alpha e^\alpha,\quad e^\alpha=-\sqrt{\frac{k(1-\lambda^4)}{2\lambda^4}}(M^{-1})^{\alpha B}L^B,\nn
M_{AB}&=&\begin{bmatrix} (D-1)_{ab} & D_{a \beta}\\ D_{\alpha b} & (D-\lambda^{-2}1)_{\alpha \beta} \end{bmatrix}, \quad D_{AB}=\mbox{Tr}(T_A g^{-1} T_B g),\\
L^A&=&-i\mbox{Tr}(g^{-1}dg T^A),\quad \mbox{Tr}(T_A T_B)=\delta_{AB}.\nonumber
\eea
The expression for the dilaton is given by the generalizations of (\ref{BosDilaton1Gr}) and (\ref{SuDilaton1}) \cite{ST,HMS,HT}:
\bea\label{BosDilaton1}
e^{2\Phi_B}&=&\frac{1}{\mbox{det}[(Ad_f-1-(\lambda^{-2}-1)P_\lambda)]},\\
\label{SuDilaton}
e^{2\Phi}&=&\frac{\mbox{det}[(Ad_f-1-(\lambda^{-2}-1)P_\lambda)|_{\hat{f}_1\oplus \hat{f}_3}]}{\mbox{det}[(Ad_f-1-(\lambda^{-2}-1)P_\lambda)|_{\hat{f}_0\oplus \hat{f}_2}]}.
\eea
Here $P_\la$ is a projector which separates the generators of $H$ and the coset $G/H$, and it has the 
form \cite{HMS}
\bea\label{PlaDef}
P_\la=P_2+\frac{\la}{\la+1}[P_1-\la P_3],\quad P_1+P_3=1.
\eea
Here $P_2$ is the projector in the bosonic sector, which can be written as
\bea
P_2=\begin{bmatrix} 0_{ab} & 0_{a \beta}\\ 0_{\alpha b} & 1_{\alpha \beta} \end{bmatrix}.
\eea
The action of fermionic projectors $P_1$ and $P_3$ is evaluated on a case--by--case basis, and we will address this question in the sections \ref{SecAdS3} and \ref{SectAdS5}. 
Notice that $P_2$ has already appeared in the matrix $M$ defined in (\ref{MainCoset}):
\bea
M_{AB}=D_{AB}-1-(\la^{-2}-1)P_2=Ad_f-1-(\lambda^{-2}-1)P_2.
\eea

We conclude this discussion with reviewing a very interesting observation made in \cite{SfetsosAdS5}: factorization of the $\la$--dependence in the determinant of $M_{AB}$. This technical simplification becomes especially useful in the AdS$_5\times$S$^5$ case, where one has to deal with large matrices. Following \cite{SfetsosAdS5}, we write $M_{AB}$ as a product of two block--triangular matrices: 
\bea\label{TriangM}
M=\left[\begin{array}{cc}
{\mathbf A}&0\\
{\mathbf C}&{\mathbf I}
\end{array}\right]\left[\begin{array}{cc}
{\mathbf I}&{\mathbf A}^{-1}{\mathbf B}\\
0&{\mathbf P}
\end{array}\right]\,.
\eea
As demonstrated in \cite{SfetsosAdS5}, matrix ${\mathbf P}$ has eigenvalues $\la^{-2}\pm 1$, so the coordinate dependence of the bosonic dilaton (\ref{BosDilaton1}) comes from $\mbox{det}\,{\mathbf A}$. We find that direct evaluation of the determinant of $M$ is easier than construction of ${\mathbf P}$, but our final results confirm that the coordinate dependence of $\mbox{det}\, M$ is inherited from $\mbox{det}\, {\mathbf A}$. 


\section{Deformation of AdS$_3\times$S$^3$}
\label{SecAdS3}

Let us apply the procedure reviewed in the last section to  AdS$_3\times$S$^3$. The bosonic part of the sigma model is described by a product of two cosets
\bea\label{BosCosetAdS3}
\frac{SU(2)\times SU(2)}{SU(2)_{diag}}\times \frac{SU(1,1)\times SU(1,1)}{SU(1,1)_{diag}},
\eea
and the full string theory is described by a super--coset \cite{AdS3PSU}\footnote{Various aspects of integrability of string on this background are further discussed in \cite{AdS3Int}.}
\bea\label{SuCosetAdS3}
\frac{PSU(1,1|2)^2}{SU(1,1)\times SU(2)}\,.
\eea
In section \ref{SecAdS3metr} we construct the metric and the bosonic contribution to the dilaton for the cosets (\ref{BosCosetAdS3}), (\ref{SuCosetAdS3}). While this will give the full answer for (\ref{BosCosetAdS3}), the dilaton for the supercoset (\ref{SuCosetAdS3}) also receives a fermionic contribution, which will be evaluated in section \ref{SecAdS3ferm}. In section \ref{SecAdS3RR} we construct the Ramond--Ramond fluxes supporting the $\la$--deformed supercoset (\ref{SuCosetAdS3}), and properties of the new geometries are discussed in sections \ref{SecAdS3SpecCases} and  \ref{SecAltParamAdS3}.

\subsection{Metric and the bosonic dilaton}
\label{SecAdS3metr}

The metric is constructed using the bosonic coset (\ref{BosCosetAdS3}), then $S^3$ and $AdS_3$ decouple, and they can be studied separately. We begin with analyzing the sphere, and deformation of $AdS_3$ can be found by performing an analytic continuation. 

\noindent
\textbf{Deformation of the sphere.}

To describe the coset $\frac{SU(2)_l\times SU(2)_r}{SU(2)_{diag}}$, we use the algebraic parameterization introduced in \cite{ST}:
\bea\label{GandGprime}
g_l=\left[\begin{array}{cc}
\alpha_0+i\alpha_3&\alpha_2+i\alpha_1\\
-\alpha_2+i\alpha_1&\alpha_0-i\alpha_3
\end{array}\right],\qquad
g_r=\left[\begin{array}{cc}
\beta_0+i\beta_3&\beta_2+i\beta_1\\
-\beta_2+i\beta_1&\beta_0-i\beta_3
\end{array}\right]\,,
\eea
where variables $\alpha_k$, $\beta_k$ are subject to the determinant constraints
\bea\label{DetConstrt}
\sum (\alpha_k)^2=1,\qquad \sum (\beta_k)^2=1.
\eea
Gauging of the diagonal part of $SU(2)_l\times SU(2)_r$ makes the description (\ref{GandGprime}) redundant, and to remove the unphysical degrees of freedom we impose a convenient gauge, which was also used in \cite{ST}. Acting on $g_l$ as $g_l\rightarrow h^{-1}g_l h$, we can set $\alpha_2=\alpha_3=0$, then the remaining $U(1)$ transformations $h=\exp[ix \sigma_1]$ can be used to set $\beta_3=0$:
\bea\label{GaugeFixedS3}
\alpha_2=\alpha_3=\beta_3=0.
\eea
Following \cite{ST} we introduce a convenient coordinate $\gamma$ and solve the constraints (\ref{DetConstrt}) to express all remaining components of  $g_1$ and $g_2$  in terms of $(\alpha_0,\beta_0,\gamma)$:
\bea\label{OurVariablesAdS3}
 \beta_1\equiv \frac{\gamma}{\sqrt{1-\alpha_0^2}},\qquad\alpha_1=\sqrt{1-\alpha_0^2},\quad \beta_2=\sqrt{1-\beta_0^2-\frac{\gamma^2}{{1-\alpha_0^2}}}\,.
\eea
To simplify notation, we will drop the subscripts of ${\alpha_0}$ and $\beta_0$. 

The elements of $SU(2)_l\times SU(2)_r$ can be represented as block--diagonal $4\times 4$ matrices:
\bea\label{CosetElementS3}
g=\left(\begin{array}{cc}
g_l&0\\ 0&g_r
\end{array}\right),\qquad g^\dagger g=I,
\eea
then the generators corresponding to the subgroup $H$ and to the coset $G/H$ can be written in terms of the Pauli matrices\footnote{Recall that construction (\ref{MainCoset}) is based on normalized generators, and factor $1/2$ in \eqref{GeneratorsS3} ensures that $\mbox{Tr}(T_AT_B)=\delta_{AB}$.}.
\bea\label{GeneratorsS3}
H=SU(2)_{diag}:&&T_a=\frac{1}{2}\left[\begin{array}{cc}
\sigma_a &0\\ 0&\sigma_a
\end{array}\right],\quad a=1,2,3;\nn
G/H=\frac{SU(2)_l\times SU(2)_r}{SU(2)_{diag}}:&&T_\alpha=\frac{1}{2}\left[\begin{array}{cc}
\sigma_{\alpha-3} &0\\ 0&-\sigma_{\alpha-3}
\end{array}\right],\quad \alpha=4,5,6.
\eea
Substitution of \eqref{CosetElementS3}--\eqref{GeneratorsS3}, where $g_l,g_r$ are given by \eqref{GandGprime}, \eqref{GaugeFixedS3}, into the defining relations (\ref{MainCoset})\footnote{Recall  the ranges of indices in \eqref{MainCoset}: $a=\{1,2,3\}$, $\alpha=\{3,5,6\}$, $B=\{1,...,6\}$.} leads to the metric \cite{ST}
\bea\label{S3metr}
ds^2&=&\frac{k}{2(1-\lambda^4)\Lambda}\Delta_{\mu\nu} dx^\nu dx^\nu ,\qquad 
\Lambda=(1-\alpha^2)(1-\beta^2)-\gamma^2,\nn
\Delta_{\alpha\alpha}&=&4(1+\lambda^2)^2-\beta^2(3+\lambda^2)(1+3\lambda^2),\qquad
\Delta_{\alpha\gamma}=-\beta(1-\lambda^2)^2\\
\Delta_{\beta\beta}&=&4(1+\lambda^2)^2-\alpha^2(3+\lambda^2)(1+3\lambda^2),\qquad
\Delta_{\beta\gamma}=-\alpha(1-\lambda^2)^2
\nn
\Delta_{\gamma\gamma}&=&(1-\lambda^2)^2,\qquad
\Delta_{\alpha\beta}=\alpha\beta(1-\lambda^2)^2+4\gamma (1+\lambda^2)^2.\nonumber
\eea

\noindent
\textbf{Deformation of AdS$_3$.}

The deformation of the AdS$_3$ is constructed by performing an analytic continuation of \eqref{S3metr}. 
The defining relation for $g\in SU(1,1)_l\times SU(1,1)_r$ is 
\bea\label{gAdSFull}
g=\left[\begin{array}{cc}
g_l&0\\ 0&g_r
\end{array}\right],\quad
g^{\dagger}\Sigma_4g=\Sigma_4,\quad 
\Sigma_4=\begin{bmatrix}
1&0&0&0\\
0&-1&0&0\\
0&0&1&0\\
0&0&0&-1
 \end{bmatrix},
\eea
and it can be enforced by starting with an element of $SU(2)_l\times SU(2)_r$, renaming the coordinates as 
\bea\label{AnalContAdS3}
\alpha\to \tilde{\alpha},\quad \beta\to \tilde{\beta},\quad \gamma\to \tilde{\gamma},\quad k\rightarrow -k,
\eea
and changing their range from 
\bea\label{Range1}
0<\alpha^2<1,\quad 0<\beta^2<1,\quad \gamma^2<(1-\alpha^2)(1-\beta^2)
\eea
to
\bea\label{Range2}
1<\tilde{\alpha}^2,\quad 1<\tilde{\beta}^2,\quad \tilde{\gamma}^2<(\tilde{\alpha}^2-1)(\tilde{\beta}^2-1).
\eea
To view this transition as a proper analytic continuation, one can introduce alternative coordinates $(a,b,\gamma)$ as
\bea
a^2=1-\alpha^2,\quad b^2=1-\beta^2.
\eea
Then transition from (\ref{Range1}) to (\ref{Range2}) amounts to a continuation from real to imaginary $(a,b)$. This changes the signature from $(+++)$ to $(--+)$, and by changing the sign of $k$ we recover $(++-)$. 

Analytic continuation (\ref{AnalContAdS3}) along with the replacement $k\to -k$ gives the metric for the $\lambda$--deformed $AdS_3$
\bea\label{AdS3metr}
d\tilde{s}^2&=&\frac{k}{2(1-\lambda^4)\tilde{\Lambda}}\tilde{\Delta}_{\mu\nu} dx^\nu dx^\nu ,\qquad \tilde{\Lambda}=(\tilde{\alpha}^2-1)(\tilde{\beta}^2-1)-\tilde{\gamma}^2,\nn
\tilde{\Delta}_{\tilde{\alpha}\tilde{\alpha}}&=&-4(1+\lambda^2)^2+\tilde{\beta}^2(3+\lambda^2)(1+3\lambda^2),\qquad
\tilde{\Delta}_{\tilde{\alpha}\tilde{\gamma}}=\tilde{\beta}(1-\lambda^2)^2\\
\tilde{\Delta}_{\tilde{\beta}\tilde{\beta}}&=&-4(1+\lambda^2)^2+\tilde{\alpha}^2(3+\lambda^2)(1+3\lambda^2),\qquad
\tilde{\Delta}_{\tilde{\beta}\tilde{\gamma}}=\tilde{\alpha}(1-\lambda^2)^2
\nn
\tilde{\Delta}_{\tilde{\gamma}\tilde{\gamma}}&=&-(1-\lambda^2)^2,\qquad
\tilde{\Delta}_{\tilde{\alpha}\tilde{\beta}}=-\tilde{\alpha}\tilde{\beta}(1-\lambda^2)^2-4\tilde{\gamma} (1+\lambda^2)^2.\nonumber
\eea

\noindent
\textbf{Dilaton and RR fields for the bosonic coset.}

The deformation of AdS$_3\times$S$^3$ constructed in \cite{ST} is described by the metric $\{$(\ref{S3metr}), (\ref{AdS3metr})$\}$ and the dilaton corresponding to the \textit{bosonic} prescription (\ref{BosDilaton1}):
\bea
e^{-2\Phi_B}=\frac{2\Lambda (1-\lambda^2)^2(1+\lambda^2)}{\lambda^6}\,
\frac{2\tilde{\Lambda}(1-\lambda^2)^2(1+\lambda^2)}{\lambda^6}=e^{-2\Phi_0}\Lambda \tilde{\Lambda},
\eea
This article also listed the corresponding Ramond--Ramond fields:
\bea\label{C2Bos}
C_2=
\frac{4k\lambda}{1-\lambda^2}
\left[\tilde{\beta} \beta d\tilde{\alpha}\wedge d\alpha+2\tilde{\beta} \alpha d\tilde{\alpha}\wedge d\beta-\tilde{\beta}d\tilde{\alpha}\wedge d\gamma+\tilde{\alpha}\alpha d\tilde{\beta}\wedge d\beta-\alpha d\tilde{\gamma}\wedge d\beta\right].
\eea
However, as argued in \cite{HMS,HT,BTW}, the dilaton (\ref{SuDilaton}) for the supercoset is more natural for describing superstrings, and in the next subsection we will find the appropriate expression and construct the corresponding Ramond--Ramond fluxes.

\subsection{Fermionic contribution to the dilaton}
\label{SecAdS3ferm}

In this subsection we will construct the dilaton for the supercoset (\ref{SuCosetAdS3}) using the prescription (\ref{SuDilaton}). Before focusing on (\ref{SuCosetAdS3}), we will outline the procedure for applying (\ref{SuDilaton}) to a supermatrix (\ref{SuCosetDef}) constructed from extending an algebra of the bosonic coset $(G_1/H_1)\times (G_2/H_2)$. 

A supersymmetric extension of an algebra ${\mathfrak g}_1\times {\mathfrak g}_2$ has the form
\bea\label{SuGenerator}
{\mathcal M}=\left[\begin{array}{cc}
{\tgt{g}}_{1}&{\tgt{f}}_{12}\\ {\tgt{f}}_{21}&{\tgt{g}}_{2}
\end{array}\right],\qquad {\tgt{g}}_{1}\in {\mathfrak g}_1,\quad {\tgt{g}}_{2}\in {\mathfrak g}_2,
\eea
and to find the supercoset, we should fix the gauge corresponding to subalgebras ${\mathfrak h}_1$, 
${\mathfrak h}_2$ and evaluate the relevant projectors  $P_\lambda$. This can be done in five steps:
\begin{enumerate}
\item Find an automorphism $J_1$ of algebra ${\mathfrak g}_1$ which leaves invariant only the elements of 
${\mathfrak h}_1$. In other words, $g\in {\mathfrak g}_1$ satisfies the condition
\bea\label{JprojBos}
J_1^{-1} {g}J_1={g}
\eea
if and only if ${g}\in {\mathfrak h}_1$. Automorphism $J_2$ in ${\mathfrak g}_2$ is defined in a similar way. 
\item
Construct an automorphism of the super-algebra as
\bea
{\mathcal P}=\left[\begin{array}{cc}
J_1&0\\ 0&J_2
\end{array}\right],
\eea
and project out the elements ${\mathcal M}$ which are left invariant under such automorphism\footnote{Sometimes this condition requires a modification: as we will see in section \ref{SectAdS5}, in the case of AdS$_5\times$S$^5$ it must be replaced by ${\mathcal P}^{-1}{\mathcal M}{\mathcal P}={\mathcal M}^T$.}: 
\bea\label{SuProj}
{\mathcal P}^{-1}{\mathcal M}{\mathcal P}={\mathcal M}\,.
\eea
For bosonic generators this reduces to (\ref{JprojBos}) and its counterpart for ${\mathfrak g}_2$, while the projections for the  fermionic matrices are
\bea\label{FermProj}
J_1^{-1}{\tgt{f}}_{12}J_2={\tgt{f}}_{12},\qquad J_2^{-1}{\tgt{f}}_{21}J_1={\tgt{f}}_{21}.
\eea
\item Construct the projector $P_2$ acting on bosonic generators by requiring that $[1-P_2]$ kills the same elements as (\ref{JprojBos}) and its counterpart with $J_2$. Such $P_2$ projects on the bosonic part of the supercoset.
\item Construct projector $P_3$ acting on fermionic generators by requiring that $P_3$ keeps the same elements as (\ref{FermProj}). The fermionic projector complementary to $P_3$ is $P_1=1-P_3$.
\item Construct the projector $P_\la$ using the definition (\ref{PlaDef}). Substitution of this expression into (\ref{SuDilaton}) or (\ref{BosDilaton1}) and evaluation  of the resulting determinant gives the dilaton for the (super)coset. 
\end{enumerate}
To apply this procedure to the $AdS_3 \times S^3$ coset (\ref{SuCosetAdS3}), we observe that 
${\mathfrak g}_1$ represents the algebra of (\ref{CosetElementS3}),
\bea
g\in {\mathfrak g}_1:\quad g=\left(\begin{array}{cc}
g_l&0\\ 0&g_r
\end{array}\right),\quad g_l\in {\mathfrak{su}}(2),\ g_r\in {\mathfrak{su}}(2),
\eea
while the elements of 
${\mathfrak h}_1={\mathfrak{su}}(2)_{diag}$ have the form
\bea
\left[\begin{array}{cc}
g&0\\ 0&g
\end{array}\right],\qquad g\in {\mathfrak{su}}(2).
\eea
This leads to two options for the automorphism $J_1$:
\bea
J_1=\pm\left[\begin{array}{cc}
0&1_{2\times 2}\\ 1_{2\times 2}&0
\end{array}\right]\,.
\eea
Expression for $J_2$ is constructed in a similar way, and putting these results together, we find two options for the automorphism ${\mathcal P}$:
\bea\label{CalPforAdS3}
{\mathcal P}=\left[\begin{array}{cccc}
0&1_{2\times 2}&0&0\\ 
1_{2\times 2}&0&0&0\\
0&0&0&1_{2\times 2}\\ 
0&0&1_{2\times 2}&0
\end{array}\right]\quad\mbox{or}\quad
{\mathcal P}=\left[\begin{array}{cccc}
0&1_{2\times 2}&0&0\\ 
1_{2\times 2}&0&0&0\\
0&0&0&-1_{2\times 2}\\ 
0&0&-1_{2\times 2}&0
\end{array}\right]\,.
\eea
The fermionic generators of $PSU(1,1|2)\times PSU(1,1|2)$ appearing in (\ref{SuGenerator}) obey the relation 
\bea
f_{12}=-i\Sigma_4 (f_{21})^\dagger
\eea
with $\Sigma_4$ given in (\ref{gAdSFull}), and projection (\ref{SuProj}) leads to further constraints. It is convenient to decouple $f_{12}$ and $f_{21}$ by working with holomorphic and anti--holomorphic coordinates. Relations (\ref{FermProj}) isolate $4+4$ components of $f_{12}$ and $f_{21}$ killed by $P_1$, while $P_3$ kills the complementary $4+4$ components\footnote{Recall that even though $f_{12}$ and $f_{21}$ are represented by $4\times 4$ matrices, each of these objects has only $8$ nonzero components. The details are discussed in the Appendix \ref{AppPSU}, here we just refer to the explicit form of the $\mathfrak{psu}(1,1|2)\times\mathfrak{psu}(1,1|2)$ matrix (\ref{MAdS3Split}), which clearly exhibits the non--vanishing elements.}. Extraction of $(P_1,P_2,P_3)$, construction of $P_\la$ via (\ref{PlaDef}), and evaluation of superdeterminant  (\ref{SuDilaton}) gives the same dilaton for both choices (\ref{CalPforAdS3}):
\bea\label{SuDilatonAdS3}
e^{\Phi}&=&Qe^{\Phi_B},\quad e^{\Phi_B}=\frac{1}{\sqrt{[(1-\alpha^2)(1-\beta^2)-\gamma^2][(\tilde{\alpha}^2-1)(\tilde{\beta}^2-1)-\tilde{\gamma}^2]}},\nn
Q
&=&(1-\lambda^2)^4\Big[\gamma+\tilde{\gamma}-\frac{4\la(1+\la^2)}{(1-\la^2)^2}(\alpha{\tilde\beta}+\tilde{\alpha}\beta) +
\frac{\la^4+6\la^2+1}{(\la^2-1)^2}(\alpha\beta+\tilde{\alpha}\tilde{\beta} )\Big]^2\,.
\eea

We conclude this section by analyzing the symmetries of the metric $\{$(\ref{S3metr}), (\ref{AdS3metr})$\}$ and the dilaton (\ref{SuDilatonAdS3}), which will be used for constructing the Ramond--Ramond fluxes. First, it is clear that neither the metric nor the dilaton has continuous symmetries, but all NS--NS fluxes are invariant under several discrete transformations:
\bea\label{DicreteSymm}
S_1:&&\alpha\leftrightarrow \beta,\qquad {\tilde\alpha}\leftrightarrow{\tilde\beta}\,;\nn
S_2:&&\alpha\leftrightarrow {\tilde\alpha},\qquad {\beta}\leftrightarrow{\tilde\beta},\quad\gamma\leftrightarrow{\tilde\beta},\quad
k\leftrightarrow(-k)\,.
\eea
These symmetries will be used in the next section to select a natural solution for the RR field $C_2$.

\subsection{Ramond--Ramond fluxes}
\label{SecAdS3RR}

Although the Ramond--Ramond fluxes for the lambda--deformed backgrounds can be extracted from the fermionic part of the sigma model, such problem is notoriously complicated \cite{BTW}. When similar deformation were analyzed in the past, the RR fluxes were obtained by solving supergravity equations \cite{Eta,OtherEta,BTW}, and in this section we will follow the same route. We will demonstrate that under very weak assumptions, supergravity gives the unique expression for all fluxes. 

Since the undeformed AdS$_3\times$S$^3$ geometry is supported by the Ramond--Ramond three-form, we assume that the situation will remain the same after the deformation, so the relevant part of action for the type IIB supergravity reads \cite{SchwWest}
\bea
S=\int d^6 x\sqrt{-g}\left[ e^{2\Phi}(R+4(\d \Phi)^2) - \frac{1}{12}F_{mnp}F^{mnp} \right].
\eea
This leads to the equations of motion
\bea\label{EOM1}
&&\nabla^2 e^{-2\Phi}=0,\\
\label{EOM2}
&&\nabla_m F^{mnk}=0, \\
\label{EOM3}
&&e^{-2\Phi}(R_{mn}+2\nabla_m\nabla_n \Phi)=\frac{1}{4}\left( F_{mpq}F_n{}^{pq}-\frac{1}{6}g_{mn}F_{spq}F^{spq} \right)
\eea
and the first one is solved by metric (\ref{S3metr}), (\ref{AdS3metr}) and the dilaton (\ref{SuDilatonAdS3}).

To construct an expression for $C_2$, we observe that the left--hand side of the Einstein's equation (\ref{EOM3}) has the structure
\bea
\frac{P}{Q^2},
\eea
where  $Q$ is given by (\ref{SuDilatonAdS3}), and $P$ is a polynomial in $(\alpha,\beta,\gamma,{\tilde\alpha},{\tilde\beta},{\tilde\gamma})$. This suggests a natural ansatz for $C_2$:
\bea
C_2=\frac{1}{Q}{\tilde C}_{\mu\nu}dx^\mu\wedge dx^\nu,
\eea
where all ${\tilde C}_{\mu\nu}$ are polynomials of degree two\footnote{The degree comes from counting powers in the left--hand side of the Einstein's equations.} in 
$(\alpha,\beta,\gamma,{\tilde\alpha},{\tilde\beta},{\tilde\gamma})$. This ansatz leaves 
\bea
\frac{6\times 5}{2}\times \left[1+6+6+\frac{6\times 5}{2}\right]=420
\eea
undetermined coefficients. We then found the most general solution for ${\tilde C}_{\mu\nu}$ following these steps:
\begin{enumerate}
\item Solving equations (\ref{EOM2})--(\ref{EOM3}) for $\la=0$, when the metric and the dilaton are relatively simple, we reduced the number of undetermined coefficients to 43.
\item Solving equations (\ref{EOM2})--(\ref{EOM3})  in the first order in $\la$, we reduced the number of undetermined coefficients in the \textit{zeroth} order to 42.
\item Eliminating the gauge freedom, we demonstrated that the solution at the zeroth order in $\la$ is \textit{unique} up to a gauge transformation.
\end{enumerate}
Once uniqueness of the solution for $\la=0$ is demonstrated, we can choose a convenient gauge which respects the discrete symmetries (\ref{DicreteSymm}):
\bea
&&C_{\alpha{\tilde\alpha}}=\frac{k}{Q}\left[2-(\beta^2+{\tilde\beta}^2)\right],\quad
C_{\beta{\tilde\beta}}=-\frac{k}{Q}\left[2-(\alpha^2+{\tilde\alpha}^2)\right],\quad\\
\quad
&&C_{\alpha{\tilde\beta}}=-C_{\beta{\tilde\alpha}}=k\frac{{\tilde\gamma}-\gamma}{Q},\quad
C_{\alpha{\tilde\gamma}}=-\frac{k{\tilde\beta}}{Q},\quad
C_{\beta{\tilde\gamma}}=\frac{k{\tilde\alpha}}{Q},\quad
C_{\gamma{\tilde\alpha}}=-\frac{k{\beta}}{Q},\quad C_{\gamma{\tilde\beta}}=\frac{k{\alpha}}{Q}\,.
\nonumber
\eea
This solution is odd under $S_1$ and $S_2$. The uniqueness of the solution in the zeroth order in $\la$ guarantees 
that, up to a gauge transformation, there is a unique gauge potential $C_2$, at least in the perturbative expansion in powers of $\la$. Making a guess consistent with symmetries (\ref{DicreteSymm}), we arrive at the final solution 
\bea\label{C2forAdS3}
&&C_{\alpha{\tilde\alpha}}=\frac{\hat k}{Q}\left[2+c_1\beta{\tilde\beta}-c_3(\beta^2+{\tilde\beta}^2)\right],\quad
\quad
C_{\alpha{\tilde\beta}}=-C_{\beta{\tilde\alpha}}=\frac{\hat k}{Q}({\tilde\gamma}-\gamma),\nn
&&C_{\alpha{\tilde\gamma}}=\frac{\hat k}{Q}[c_2\beta-{\tilde\beta}]\quad
C_{\beta{\tilde\beta}}=-\frac{\hat k}{Q}\left[2+c_1\alpha{\tilde\alpha}-c_3(\alpha^2+{\tilde\alpha}^2)\right],\quad
C_{\beta{\tilde\gamma}}=-\frac{\hat k}{Q}[c_2\alpha-{\tilde\alpha}],\nn
&&C_{\gamma{\tilde\alpha}}=-\frac{\hat k}{Q}[{\beta}-c_2{\tilde\beta}],\qquad 
C_{\gamma{\tilde\beta}}=\frac{\hat k}{Q}[{\alpha}-c_2{\tilde\alpha}]
\\
&&c_1=2c_2c_3,\qquad c_2=\frac{2\la}{1+\la^2},\qquad c_3=\frac{\la^4+6\la^2+1}{(\la^2-1)^2},\quad {\hat k}=\frac{k(1+\lambda^2)}{1-\lambda^2}\,.\nonumber
\eea
Notice that, unlike the solution (\ref{C2Bos}) with the ``bosonic dilaton'', the field (\ref{C2forAdS3}) has a  complicated lambda dependence, and the situation is similar in the AdS$_2\times$S$^2$ case, which is reviewed in the Appendix \ref{AppAdS2}. In particular, while the field  (\ref{C2Bos}) vanishes at the WZW point ($\la=0$), our solution for the supercoset (\ref{C2forAdS3}) goes to a nontrivial limit, and, as we will see in section \ref{SectAdS5} and in the Appendix \ref{AppAdS2}, the same phenomenon persists for AdS$_2\times$S$^2$ and AdS$_5\times$S$^5$.

\bigskip

To summarize, the $\la$--deformed version of AdS$_3\times$S$^3$ is described by the metric (\ref{S3metr}), (\ref{AdS3metr}), the dilaton (\ref{SuDilatonAdS3}), and the Ramond--Ramond two--form (\ref{C2forAdS3}). In the next subsection we will analyze some special cases of this geometry.

\subsection{Special cases}
\label{SecAdS3SpecCases}

The solution (\ref{S3metr}), (\ref{AdS3metr}), (\ref{SuDilatonAdS3}), (\ref{C2forAdS3}) simplifies in several special cases, and we will briefly discuss these interesting limits.

The gauged WZW model is obtained by setting $\la=0$: 
\bea\label{WZWsss}
ds^2&=&\frac{k}{2\Lambda}\left[4(1-\beta^2)d\alpha^2+4(1-\alpha^2)d\beta^2+8\gamma d\alpha d\beta+(d\gamma-\beta d\alpha-\alpha d\beta)^2\right]+\nn
&&+\frac{k}{2\Lambda}\left[4({\tilde\beta}^2-1)d{\tilde\alpha}^2+4({\tilde\alpha}^2-1)d{\tilde\beta}^2-
8{\tilde\gamma} d{\tilde\alpha} d{\tilde\beta}+(d{\tilde\gamma}-{\tilde\beta} d{\tilde\alpha}-
{\tilde\alpha} d{\tilde\beta})^2\right]\,,\nn
\qquad 
\Lambda&=&(1-\alpha^2)(1-\beta^2)-\gamma^2,\qquad 
\tilde{\Lambda}=(\tilde{\alpha}^2-1)(\tilde{\beta}^2-1)-\tilde{\gamma}^2\,,\\
e^{\Phi}&=&\frac{Q}{\sqrt{\Lambda{\tilde\Lambda}}}e^{\Phi_B},\quad
Q
=\Big[\gamma+\tilde{\gamma}+\alpha\beta+\tilde{\alpha}\tilde{\beta} \Big]^2\,,\nn
C_2&=&\frac{k}{Q}\left[({\tilde\alpha}d\beta-{\tilde\beta}d\alpha)\wedge (d{\tilde\gamma}+{\tilde\alpha}d{\tilde\beta}+{\tilde\beta}d{\tilde\alpha})-
({\alpha}d{\tilde\beta}-{\beta}d{\tilde\alpha})\wedge (d{\gamma}+{\alpha}d{\beta}+{\beta}d{\alpha})
\right.\nn
&&\left.+({\tilde\gamma}-{\gamma}+{\tilde\alpha}\tilde\beta-\alpha{\beta})(d\alpha\wedge d{\tilde\beta}-d\beta\wedge d{\tilde\alpha})+2(d\alpha\wedge d{\tilde\alpha}-d\beta\wedge d{\tilde\beta})
\right]\,.\nonumber
\eea
This should be contrasted with bosonic gWZW, which has the dilaton
\bea
e^{\Phi}&=&\frac{1}{\sqrt{\Lambda{\tilde\Lambda}}}
\eea
and vanishing $C_2$ (see (\ref{C2Bos})). A similar contrast is encountered in the AdS$_2\times$S$^2$ and AdS$_5\times$S$^5$ cases, which discussed in section \ref{SectAdS5} and in the Appendix \ref{AppAdS2}.

Note that the metric (\ref{WZWsss}) for the $SO(4)/SO(3)$ gWZW model has been discussed in 
\cite{Fradkin,MoreEta}, where the element of the coset was defined as
\bea\label{Fradkin}
g&=&g_1(\varphi)g_2(\theta)g_3(2t)g_2(\theta)g_1(\varphi),\\
g_k(\alpha)&=&\exp(\alpha T_{k,k+1}),\quad (T_{k,k+1})_i^j=\delta_{k,i}\delta_{k+1}^j-\delta_{k+1,i}\delta_k^j,\quad k=1,2,3.\nonumber
\eea
The coordinates used in (\ref{WZWsss}) are related to the Euler angles (\ref{Fradkin}) as
\bea
\alpha&=&\cos\varphi\cos t\cos\theta+\sin\varphi\sin t,\nn
\beta&=&\cos\varphi\cos t\cos\theta-\sin\varphi\sin t,\\
\gamma&=&-\cos^2\varphi\sin^2 t+\cos^2 t(\cos^2\theta\sin^2\varphi+\sin^2\theta).\nonumber
\eea

\bigskip

Another interesting limit is obtained by setting $\la=1$. However, this limit should be approached with a great care since denominators contain $(\la^2-1)$. We will follow the procedure discussed in \cite{HT} adopting it to our coordinates. To arrive at a sensible limit, we rescale the coordinates on the sphere as 
\bea
\alpha\propto\frac{1}{\eps},\quad \beta\propto \frac{1}{\eps},\quad \gamma\propto\frac{1}{\eps^2}
\eea
and send $\eps$ to zero. This gives the metric of the $\eta$--deformed $S^3$ \cite{OtherEta}, and to see this, we introduce the standard coordinates 
$(r,\phi,\varphi)$ by 
\bea\label{KappaLimit}
\alpha&=&\frac{1}{\eps}\, \frac{e^{i(\varphi+\phi)}r}{(1+\lambda^2)\sqrt{2(1-r^2)}},\quad 
\beta=\frac{1}{\eps}\, \frac{e^{i(\varphi-\phi)}r}{(1+\lambda^2)\sqrt{2(1-r^2)}},\nn 
\gamma&=&\frac{1}{\eps^2}\, \frac{e^{2i\varphi}(2(1+\lambda^2)^2-(1-\lambda^2)^2r^2)}{2(1-\lambda^4)^2(1-r^2)},\quad \eps\to 0.
\eea
Performing a similar change of variables on AdS$_3$ along with an analytic continuation 
\bea
\phi\to\psi,\quad \varphi\to t,\quad r\to i\rho,\quad k\to-k,
\eea
and sending $\eps$ to zero, we arrive at the metric and the dilaton
\bea
ds^2&=&\frac{h}{2}\Big(\frac{1}{1-\varkappa^2 r^2}\left[(1-r^2)d\varphi^2+\frac{dr^2}{1-r^2} \right]+r^2d\phi^2\nn
&&+\frac{1}{1+\varkappa^2 \rho^2}\left[-(1+\rho^2)dt^2 + \frac{d\rho^2}{1+\rho^2}\right]+\rho^2d\psi^2\Big),\\
e^{\Phi}&=&(1+{\tilde\la}^2)^4\frac{\left[2(1-{\tilde\la}^2)S^2\cos(\varphi-t)-4{\tilde\la}\rho rS\cos(\phi-\psi)\right]^2}{
S^2\sqrt{(1-{\tilde\la}^2)^2+(1+{\tilde\la}^2)^2\rho^2}\sqrt{(1-{\tilde\la}^2)^2-(1+{\tilde\la}^2)^2r^2}},\nn
S&\equiv& \sqrt{(1+\rho^2)(1-r^2)},\quad \varkappa=\frac{1+{\tilde\la}^2}{1-{\tilde\la}^2},\quad 
h=\frac{(1-{\tilde\la}^2)^2}{k(1+{\tilde\la}^2)},\quad {\tilde\la}=i\la\,.
\nonumber
\eea 
This geometry describes the $\eta$--deformed AdS$_3\times$S$^3$ \cite{OtherEta}, and similar relations between $\la$-- and $\eta$--deformations have been explored in \cite{HT}.


\subsection{Alternative parameterizations}
\label{SecAltParamAdS3}

In subsections \ref{SecAdS3metr}--\ref{SecAdS3RR} we derived the full supergravity solutions corresponding to the $\la$--deformed supercoset, but the metric for this geometry has already appeared in the literature \cite{ST,HT}. We used the parameterization of \cite{ST}, and in this subsection we will discuss the relation with the coordinates used in \cite{HT} and discuss one more parameterization which becomes useful for comparing AdS$_3\times$S$^3$ and AdS$_5\times$S$^5$ solutions.

To find the relation between our parameterization and the coordinates used in \cite{HT}, we observe that the action by $H=SU(2)_{diag}$ changes components of $g_l$ and $g_r$ in (\ref{GandGprime}), but three expressions remain invariant:
\bea\label{VecInvar}
\vec{\alpha}^2\equiv \sum_{i=1}^3\alpha_i\alpha_i,\quad 
\vec{\beta}^2\equiv\sum_{i=1}^3\beta_i\beta_i,\quad 
\vec{\alpha}\cdot\vec{\beta}\equiv\sum_{i=1}^3\alpha_i\beta_i\,.
\eea
Although the gauge used in \cite{HT} was different from ours, we can find the map between two sets of coordinates by matching the expressions (\ref{VecInvar}) in two descriptions. The authors of \cite{HT} used parameterization in terms of the Euler's angles:
\bea\label{TrigParam}
g^{trig}=\exp[i\varphi \sigma_3\oplus (-\sigma_3)]\exp[i\zeta \sigma_1\oplus\sigma_1]\exp[i \phi \sigma_3\oplus\sigma_3].
\eea
Evaluating the invariants (\ref{VecInvar}) for parameterizations (\ref{GaugeFixedS3})--(\ref{OurVariablesAdS3}) and (\ref{TrigParam}), and comparing the results, we arrive at the map\footnote{Recall that to simplify notation we introduced $\alpha=\alpha_0$ and $\beta=\beta_0$, and all our results were written in these variables.}
\bea
\alpha=\cos(\varphi+\phi)\cos\zeta,\ \beta=\cos(\varphi-\phi)\cos\zeta,\
\gamma=\cos2\varphi-\frac{\cos2\varphi+\cos2\phi}{2}\cos^2\zeta.
\eea

Another interesting coordinate system comes from parameterizing the coset $SO(4)/SO(3)$ in terms of a three--dimensional vector $X$ and an anti--symmetric $3\times 3$ matrix $A$ \cite{Bars,SfetsosAdS5}. Such parameterization of $SO(n+1)/SO(n)$ will be used in the next section for studying the deformed AdS$_5\times$S$^5$, so it is important to introduce similar coordinates in the present case to make comparisons. The detailed discussion of parameterization and the gauge fixing is presented in section \ref{SectAdS5metr}, here we just write the result\footnote{We use variables $Y_1$ and $Y_2$ in (\ref{SO4gauge}) to make comparison with AdS$_5\times$S$^5$ case easier: the variable $Y_1$ is a counterpart of $X_1$, and $Y_2$ is a counterpart of $X_5$ in (\ref{SO6gauge}).}:
\bea\label{SO4gauge}
g&=&
\left[\begin{array}{cc}
1&0\\
0&(1+A)(1-A)^{-1}
\end{array}\right]
\left[\begin{array}{cccc}
b-1&bX_i\\
-bX_i&\delta_i^j-bX_iX^j
\end{array}\right]\,,\\
A&=&\left[\begin{array}{ccc}
0&a&0\\
-a&0&0\\
0&0&0
\end{array}\right],\quad 
b=\frac{2}{1+(Y_1)^2+(Y_2)^2},\quad {\vec X}=\{Y_1,0,Y_2\}\,.\nonumber
\eea 
The parameterizations (\ref{SO4gauge}) and (\ref{GandGprime}), (\ref{CosetElementS3}) correspond to different representations of $SO(4)$, so to relate them we should compare quantities which don't depend on the representation. We have already encountered such an object before:
\bea\label{DabMap}
D_{AB}=\mbox{Tr}(T_A g^{-1} T_B g).
\eea
To establish the map between generators, we recall that the subgroup $H=SU(2)_{diag}$ corresponds to
\bea
T_{H}^{SU(2)\times SU(2)}=
\frac{1}{2}\left[\begin{array}{cc}
x_a\sigma_a&0\\
0&x_a\sigma_a
\end{array}\right],\quad
T_{H}^{SO(4)}=\frac{i}{\sqrt{2}}\left[\begin{array}{cccc}
0&0&0&0\\
0&0&x_3&-x_2\\
0&-x_3&0&x_1\\
0&x_2&-x_1&0
\end{array}\right]
\eea
and the coset generators correspond to
\bea
T_{coset}^{SU(2)\times SU(2)}=
\frac{1}{2}\left[\begin{array}{cc}
y_a\sigma_a&0\\
0&-y_a\sigma_a
\end{array}\right],\quad
T_{coset}^{SO(4)}=\frac{i}{\sqrt{2}}\left[\begin{array}{cccc}
0&y_1&y_2&y_3\\
-y_1&0&0&0\\
-y_2&0&0&0\\
-y_3&0&0&0
\end{array}\right]
\eea
Evaluating (\ref{DabMap}) for (\ref{SO4gauge}) and $\{$(\ref{GandGprime}), (\ref{CosetElementS3})$\}$, using appropriate generators, and matching the results, we arrive at the map
\bea
&&\alpha=\frac{1-a Y_2}{\sqrt{1+a^2}Y},\quad \beta=\frac{1+a Y_2}{\sqrt{1+a^2}\,Y},\quad \gamma=-\frac{Y_1^2+a^2(Y_1^2-1)+Y_2^2}{(1+a^2)\,Y^2},\\
&&Y^2=1+(Y_1)^2+(Y_2)^2.\nonumber
\eea
and its inverse
\bea
&&a=-\frac{\sqrt{2(1+\gamma^2)-\alpha^2-\beta^2}}{\alpha+\beta},\quad Y_2=\frac{\alpha-\beta}{\sqrt{2(1+\gamma)-\alpha^2-\beta^2}},\nn
&&Y_1=-\sqrt{\frac{2[(1-\alpha^2)(1-\beta^2)-\gamma^2]}{[1+\alpha\beta+\gamma][2(1+\gamma)-(\alpha^2+\beta^2)]}}.
\eea
The AdS coordinates are obtained by the replacement
\bea
Y_1\to i {\tilde Y}_1,\quad Y_2\rightarrow Y_2,\quad a\rightarrow{\tilde a}.
\eea
In coordinates $(Y_1,Y_2,a,{\tilde Y}_1,{\tilde Y}_2,{\tilde a})$ the dilaton becomes
\bea
e^{\Phi}&=&Q e^{\Phi_B},\quad e^{\Phi_B}=\frac{\sqrt{1+a^2}Y \sqrt{1+\tilde{a}^2}\tilde{Y}}{16 a\tilde{a}Y_1\tilde{Y}_1},\quad Y^2=1+Y_1^2+Y_2^2,\quad \tilde{Y}^2=1-\tilde{Y}_1^2+\tilde{Y}_2^2,\nn
Q
&=&(1-\lambda^2)^4\Big[-\frac{Y_1^2+a^2(Y_1^2-1)+Y_2^2}{(1+a^2)Y^2}+\frac{\tilde{Y}_1^2+\tilde{a}^2(\tilde{Y}_1^2+1)+\tilde{Y}_2^2}{(1+\tilde{a}^2)\tilde{Y}^2}\\
&&-\frac{8\la(1+\la^2)}{(1-\la^2)^2}\frac{1-a\tilde{a}Y_2\tilde{Y}_2}{\sqrt{1+a^2}\sqrt{1+\tilde{a}^2}Y\tilde{Y}} 
+\frac{\la^4+6\la^2+1}{(\la^2-1)^2}\left(\frac{1-a^2 Y_2^2}{(1+a^2)Y^2}+\frac{1-\tilde{a}^2\tilde{Y}^2_2}{(1+\tilde{a}^2)\tilde{Y}^2} \right)\Big]^2.\nonumber
\eea
In particular, for the gauged WZW model ($\la=0$) we find
\bea\label{QforWZW3}
Q=4\left[\frac{X^2+{\tilde X}^2-X^2{\tilde X}^2}{X^2{\tilde X}^2}\right]^2\,.
\eea
Notice that this expression does not depend on coordinates $a$ and ${\tilde a}$, and the same phenomenon is encountered in the AdS$_5\times$S$^5$ case, see the last factor in (\ref{WZWforAdS5}).


\section{Towards the deformation of AdS$_5\times $S$^5$}

\label{SectAdS5}

In this section we apply the procedure described in section \ref{SecReview} to construct the $\lambda$--deformed AdS$_5\times $S$^5$ supercoset. Our final result includes the metric and the dilaton, but since the latter looks rather complicated, we were not able to solve the equations for the Ramond--Ramond fluxes. 

Superstrings on AdS$_5\times $S$^5$ are described by a  sigma model on the supercoset \cite{AdS5PSU}
\bea\label{AdS5suCoset}
\frac{PSU(2,2|4)}{SO(4,1)\times SO(5)}.
\eea
The corresponding superalgebra is represented by $4\times 4$ matrices, and an explicit parameterization is presented in the appendix \ref{AppPSU}. The bosonic part of the supercoset (\ref{AdS5suCoset}) is given by
\bea\label{AdS5BosCoset}
 \frac{SU(2,2)}{SO(4,1)}\times \frac{SU(4)}{SO(5)}= 
 \frac{SO(4,2)}{SO(4,1)}\times \frac{SO(6)}{SO(5)},
\eea 
and, as in the AdS$_3\times$S$^3$ case, the two subgroups decouple in the metric (\ref{GeneralMetric}) and in the bosonic contribution to the dilaton (\ref{BosDilaton1}). While these objects have been computed in \cite{SfetsosAdS5}, to evaluate the fermionic contribution to the dilaton we will have to use a different parameterization, so we begin with specifying our coordinates, finding the metric and the bosonic dilaton 
for them, and comparing the results with \cite{SfetsosAdS5}. The fermionic contribution to the dilaton will be evaluated in section \ref{SecFerDilAdS5}.

\subsection{Metric and the bosonic dilaton}
\label{SectAdS5metr}

To apply the procedure outlined in section \ref{SecReview}, we need an explicit form of the coset (\ref{AdS5BosCoset}). The most natural way to parameterize the sphere $S^5=SO(6)/SO(5)$ is to use the Euler angles, and such description has been used in \cite{SfetsosAdS5}, but unfortunately these coordinates make the evaluation of the ferminonic contribution to the dilaton nearly impossible. Thus we use the alternative coordinates introduced in \cite{Bars,SfetsosAdS5}, in which all expressions remain algebraic\footnote{It appears that the authors of \cite{SfetsosAdS5} used the same coordinates while computing the metric and rewrote the final answers in terms of the Euler's angles. We find the algebraic coordinates more convenient.}.

Specifically, we write the element of $SO(6)$ as
\bea\label{SO6gauge}
g=
\left[\begin{array}{cc}
1&0\\
0&{h_m}^n
\end{array}\right]
\left[\begin{array}{cc}
b-1&bX^j\\
-bX_i&\delta_i^j-bX_iX^j
\end{array}\right]\,,\qquad b=\frac{2}{{1+X^m X_m}},
\eea 
where $X^i$ is a five--dimensional vector and ${h_m}^n$ is an element of $SO(5)$. The defining condition for $SO(5)$, $h^T h=I$, can be solved by writing $h$ in terms of an anti-symmetric matrix $A$ as
\bea
{h_m}^n={[(1+A)(1-A)^{-1}]_m}^n\,.
\eea
The $SO(5)$ rotations act on $A$ and $X$ as
\bea
A\rightarrow \Lambda A\Lambda^{-1},\qquad X\rightarrow \Lambda X.
\eea
To fix this gauge freedom, we follow the procedure discussed in \cite{Bars}: first we rotate $A$ to a block form\footnote{Notice that there is a slight difference in gauge fixing between $SO(n)/SO(n-1)$ for odd and even $n$: matrix $A$ has $[(n-1)/2]$ independent components, and there are $[n/2]$ independent $X$.}:
\bea\label{SO6alphaGg}
A=\left(\begin{array}{ccccc}
0&a&0&0&0\\
-a&0&0&0&0\\
0&0&0&b&0\\
0&0&-b&0&0\\
0&0&0&0&0\\
\end{array}\right)\,,
\eea
and then we use the remaining $[SO(2)]^2$ rotations to set $X_2=X_4=0$. 

The $\mathfrak{so}(6)$ algebra has $15$ generators, first ten of them form $\mathfrak{so}(5)$, while the last five correspond to the coset. Specifically, in our parameterization, the coset generators are\footnote{Recall that throughout this chapter we use hermitian generator, so the element of a group is constructed as $g=\exp[iT x]$.} 
\bea\label{CosetGenerSO6}
(T_\alpha)_{mn}=-\frac{i}{\sqrt{2}}\left[\delta_{m1}\delta_{n(\alpha-9)}-\delta_{n1}\delta_{m(\alpha-9)}
\right]\qquad \alpha=11,\dots 15.
\eea
Application of the procedure (\ref{MainCoset}) leads to the bosonic contribution to the dilaton (\ref{BosDilaton1}) 
\bea\label{AdS5bosDil}
e^{-2\Phi_B}&=&\frac{1024a^2b^2(a^2-b^2)^2X_1^2X_3^2}{(1+a^2)^3(1+b^2)^3 X^2}\, 
\frac{(1-\la^2)^3(1+\la^2)^2}{\la^{10}},
\eea
where we defined
\bea
X^2&\equiv&1+X_1^2+X_3^2+X_5^2.\nonumber
\eea
Note that the lambda dependence factorizes in (\ref{AdS5bosDil}), and this is a general feature of the bosonic dilaton, as discussed in the end of section \ref{SecReview}. Specifically, in the present case, matrix ${\mathbf P}$ defined in (\ref{TriangM}) has the form
\bea
{\mathbf P}&=&\left[\begin{array}{ccc}
W(a)&0&0\\
0&W(b)&0\\
0&0&1
 \end{array}\right],\quad \mbox{where}\quad W(x)\equiv (1+x^2)^{-1}\left[\begin{array}{cc}1&-x\\-x& -1 \end{array}\right].
 \eea
This matrix has eigenvalues $\la^{-2}\pm 1$ and
\bea
\mbox{det}\, {\mathbf P}=\frac{(1-\la^2)^3(1+\la^2)^2}{\la^{10}}\,.
\eea

The metric for $\la$--deformation is constructed using (\ref{MainCoset}), and the result reads
\bea
ds^2_{(\lambda)}&=&\sum_\alpha (e^\alpha_{(\lambda)})^2,\quad 
e^\alpha_{(\lambda)}=\frac{\sqrt{k(1-\la^4)}}{2\la^2}{[{\mathbf P}^{-1}]^\alpha}_\beta e^\beta_{(0)},
\eea
where $e^\beta_{(0)}$ refer to the frames describing the gauged WZW model ($\la=0$):
\bea\label{WZWframesAdS5}
e_{(0)}^{6}&=&\frac{a^2(1+b^2)X_3^2+(a^2-b^2)X_5^2}{a(1+a^2)(a^2-b^2)X_1}da+\frac{(1+a^2)bX_1}{(a^2-b^2)(1+b^2)}db\nn
&&\qquad+\frac{1}{X^2}\left[-(X^2-X_1^2)dX_1+X_1X_3 dX_3+X_1X_5 dX_5\right],\nn
e_{(0)}^7&=&\frac{da}{X_1(1+a^2)},\quad e_{(0)}^9=\frac{db}{X_3(1+b^2)},\\
e_{(0)}^{10}&=&-\frac{X_5 da}{a(1+a^2)}-\frac{X_5 db}{b(1+b^2)}+\frac{1}{X^2}\left[X_1X_5 dX_1+X_3X_5 dX_3-(X^2-X_5^2)dX_5\right],\nn
e_{(0)}^8&=&-\frac{a(1+b^2)X_3 da}{(1+a^2)(a^2-b^2)}-\frac{(1+a^2)b^2X_1^2-(a^2-b^2)X_5^2}{b(1+b^2)(a^2-b^2)X_3}db\nn
&&\qquad+\frac{1}{X^2}\left[X_1X_3 dX_1-(X^2-X_3^2)dX_3+X_3X_5 dX_5 \right].\nonumber
\eea
The $AdS_5$ counterparts of the metric and the dilaton are obtained by an analytic continuation
\bea\label{AnalContAdS5}
X_1\to i X_1,\quad X_3\to i X_3,\quad k\to -k,
\eea
and the corresponding frames are denoted by $e^1_{(0)}$,\dots,$e^5_{(0)}$.

\subsection{Fermionic dilaton: general discussion}
\label{SecFerDilAdS5}

Although the $SO(6)/SO(5)$ representation (\ref{SO6gauge}) of the five--dimensional sphere is very intuitive, the construction of the supercoset (\ref{AdS5suCoset}) requires embedding of $SO(6)$ into $SU(4)$ and identifying the fermionic degrees of freedom corresponding to the supercoset. We begin with finding the $SU(4)$ matrices in parameterization (\ref{SO6gauge}). 

The $SU(4)$ matrices $g$ describe a representation of $SO(6)$, which acts on anti--symmetric $4\times 4$
matrices $A$ as 
\bea
A\rightarrow gAg^T\,.
\eea 
Specifically, starting with the fundamental representation of $SO(6)$ acting on six--dimensional vectors $(x_1,x_2,x_3,y_1,y_2,y_3)$, one can construct matrix $A$ as
\bea\label{AmatrSU4}
A=\left[\begin{array}{cccc}
0&x_3-iy_3&-x_2+iy_2&x_1+iy_1\\
-x_3+iy_3&0&x_1-iy_1&x_2+iy_2\\
x_2-iy_2&-x_1+iy_1&0&x_3+iy_3\\
-x_1-iy_1&-x_2-iy_2&-x_3-iy_3&0
\end{array}\right]\,.
\eea
The generators of $SU(4)$ are hermitian $4\times 4$ matrices, and to proceed with the coset construction, we need to identify the elements $t_\alpha$ corresponding to the generators (\ref{CosetGenerSO6}). Comparing the action $T_\alpha$ on $(x_1,x_2,x_3,y_1,y_2,y_3)$ and the action of $g\in \mathfrak{su}(4)$  on (\ref{AmatrSU4}), we find
\bea\label{CosetGenerSU4}
T^\alpha c_\alpha=\frac{1}{2}\left[\begin{array}{cccc}
c_{13}&c_{14}+ic_{11}&c_{15}+ic_{12}&0\\
c_{14}-ic_{11}&-c_{13}&0&-c_{15}-ic_{12}\\
c_{15}-ic_{12}&0&-c_{13}&c_{14}+ic_{11}\\
0&ic_{12}-c_{15}&c_{14}-ic_{11}&c_{13}
\end{array}\right]\,.
\eea
All generators of $SU(4)$, including (\ref{CosetGenerSU4}), are hermitian, while generators of $SU(2,2)$ satisfy the modified hermiticity relation
\bea
(T_A)^\dagger=\Sigma T_A\Sigma,\qquad 
\Sigma=\left[\begin{array}{cccc}
0&\sigma_3\\
\sigma_3&0
\end{array}\right].
\eea
For example, the counterparts of the coset generators (\ref{CosetGenerSU4}) are obtained by an analytic continuation
\bea
c_{11}\rightarrow i{\tilde c}_{11},\quad c_{12}\rightarrow i{\tilde c}_{12},\quad c_{13}\rightarrow i{\tilde c}_{13},\quad c_{14}\rightarrow i{\tilde c}_{14},\quad c_{15}\rightarrow {\tilde c}_{15}.
\eea

To proceed we need to construct an automorphism $J_1$ which satisfies (\ref{JprojBos}) for all generators $g\in \mathfrak{su}(4)$ with the exception of (\ref{CosetGenerSU4}). While it is easy to find this $J_1$ for $6\times 6$ matrices and coset generators (\ref{CosetGenerSO6}) (specifically, $J_1=\pm\mbox{diag}(-1,1,1,1,1)$), such matrix does not exist in the four--dimensional representation of $\mathfrak{so}(6)$, and the closest analog of (\ref{JprojBos}) is
\bea\label{JprojBosT}
J_1^{-1} {g}J_1={g^T},\quad 
J_1=\left[\begin{array}{cccc}
0&0&0&1\\
0&0&1&0\\
0&-1&0&0\\
-1&0&0&0
\end{array}\right]\,.
\eea
This means that condition (\ref{SuProj}) will be modified as  
\bea\label{SuProjT}
{\mathcal P}^{-1}{\mathcal M}{\mathcal P}={\mathcal M}^T,
\eea
and such grading is a familiar feature of $PSU(2,2|4)$ (see, for example, \cite{ArutFrolov} for a detailed discussion). In our parameterization,
\bea\label{PforSU4}
{\mathcal P}=\left[\begin{array}{cc}
J_1&0\\
0&J_1
\end{array}\right]\,,
\eea
and the detailed discussion of fermions projected out by (\ref{PforSU4}) and relation to other conventions used in the literature is presented in the Appendix \ref{AppPSU}. Here we only mention that if $8\times 8$ supercoset matrix is written as
\bea
{\mathcal M}=\left[\begin{array}{cc}
A&B\\C&D
\end{array}\right]\,,\quad 
B\equiv\left[\begin{array}{cccc}
b_1&b_2\\
b_3&b_4
\end{array}\right],\quad 
C\equiv\left[\begin{array}{cccc}
c_1&c_2\\
c_3&c_4
\end{array}\right]=-i\left[\begin{array}{cccc}
b^\dagger_3\sigma_3&b^\dagger_1\sigma_3\\
b^\dagger_4\sigma_3&b^\dagger_2\sigma_3
\end{array}\right],
\eea
then projector $P_3$ entering $P_\la$ (\ref{PlaDef}) selects the components satisfying an additional relation (\ref{qqq2}):
\bea
C=\left[\begin{array}{rr}
-[\sigma_1 b_4\sigma_1]^T&[\sigma_1 b_2\sigma_1]^T\\
\ [\sigma_1 b_3\sigma_1]^T&[\sigma_1 b_1\sigma_1]^T
\end{array}\right]\,.
\eea

The last ingredient for constructing the fermionic contribution to the dilaton is the explicit expression for the element of $SU(4)/SO(5)$ in the gauge (\ref{SO6gauge}), (\ref{SO6alphaGg}):
\bea\label{gSAdS5}
g_S&=&\frac{1}{\Delta_S}
\left[\begin{array}{cccc}
1&b&ab&a\\
-b&1&a&-ab\\
a b &-a&1&-b\\
-a &-a b&b&1
\end{array}\right]
\left[\begin{array}{cccc}
1-iX_3&X_1&-iX_5&0\\
-X_1&1+iX_3&0&iX_5\\
-iX_5&0&1+iX_3&X_1\\
0 &iX_5&-X_1&1-iX_3
\end{array}\right]\nn
\phantom{\frac{a}{b}} \\
\Delta_S&=&\sqrt{1+a^2}\sqrt{1+b^2}\sqrt{1+(X_1)^2+(X_3)^2+(X_5)^2}\,.\nonumber
\eea
The element of $SU(2,2)/SO(4,1)$ is obtained by making the analytic continuation (\ref{AnalContAdS5}) in the last expression. Notice that the symmetry
\bea
X_1\leftrightarrow X_3,\quad a\leftrightarrow b,
\eea
which was obvious in the $SO(6)$ parameterization (\ref{SO6gauge}), (\ref{SO6alphaGg}), is less explicit in (\ref{gSAdS5}). 

Evaluation of the fermionic contribution to the dilaton involves a straightforward but tedious calculation of the determinant
\bea
{\mbox{det}[(Ad_f-1-(\lambda^{-2}-1)P_\lambda)|_{\hat{f}_1\oplus \hat{f}_3}]},
\eea
and the results are rather complicated. We collect them and discuss some of their features in the next two subsections. 

\subsection{Dilaton for the gauged WZW model}
\label{SecAdS5WZW}

Geometry with $\la=0$ describes the gauged WZW model, and the solution in this case is given by the frames (\ref{WZWframesAdS5}), along with their AdS$_5$ counterpart and the dilaton
\bea\label{WZWforAdS5}
e^{-2\Phi}&=&2^{20}\frac{a^2b^2(a^2-b^2)^2X_1^2X_3^2}{(1+a^2)^3(1+b^2)^3 X^2}
\frac{{\tilde a}^2{\tilde b}^2({\tilde a}^2-{\tilde b}^2)^2{\tilde X}_1^2{\tilde X}_3^2}{
(1+{\tilde a}^2)^3(1+{\tilde b}^2)^3 {\tilde X}^2}
\left[\frac{X^{2}{\tilde X}^{2}}{X^2+{\tilde X}^2-X^2{\tilde X}^2}\right]^8\\
&&X^2=1+X_1^2+X_3^2+X_5^2,\quad {\tilde X}^2=1-{\tilde X}_1^2-{\tilde X}_3^2+{\tilde X}_5^2\,.\nonumber
\eea
The bosonic contribution to the dilaton is obtained by dropping the expression in the brackets, and the bosonic coset does not require any Ramond--Ramond fluxes. The situation for the supercoset is different, as we have already seen in the 
AdS$_3\times$S$^3$ case: the Ramond--Ramond fluxes are turned on even at $\la=0$. In the present case we were not able to construct the fluxes explicitly, but we verified that the solution (\ref{WZWframesAdS5})--(\ref{WZWforAdS5}) can be supported by $F_5$. 

Recall that the stress--energy tensor for the self--dual five--form,
\bea
T_{mn}=\frac{1}{96}F_{mabcd}{F_n}^{abcd}
\eea
satisfies the Rainich conditions \cite{Rainich}\footnote{For a recent discussion of the original Rainich conditions for electromagnetism and their generalizations to higher dimensions see, for example, \cite{Rainich1}.} :
\bea\label{Rainich}
{T_m}^m\equiv \mbox{Tr}\, T=0,\quad \mbox{Tr}\, T^3=0,\quad \mbox{Tr}\, T^5=0,\quad \mbox{Tr}\, T^7=0,\quad 
\mbox{Tr}\,  T^9=0,\quad 
\eea
and for geometry supported only by the dilaton and the metric the $T_{mn}$ can be expressed as\footnote{In this chapter we are working in the string frame.}
\bea
T_{mn}=R_{mn}+2\nabla_m\nabla_n \Phi.
\eea
The right--hand side vanishes for the ``bosonic'' dilaton, while for the full solution (\ref{WZWforAdS5}) it gives a nontrivial result which satisfies the constraints (\ref{Rainich}). It would be very interesting to find the corresponding flux $F_5$.

\subsection{Special cases for $\la\ne 0$}
\label{SecAdS5SpecCases}

Although the dilaton for arbitrary values of $\la$ can be computed by evaluating the appropriate determinants, unfortunately the results are not very illuminating. In this subsection we will collect the answers for some special cases which give manageable expressions. Since the general expression for the bosonic dilaton is already given by (\ref{AdS5bosDil}), we will focus only on the fermionic contribution to (\ref{SuDilaton1}):
\bea\label{SuDilatonFerm}
e^{2\Phi_F}&=&{\mbox{det}[(Ad_f-1-(\lambda^{-2}-1)P_\lambda)|_{\hat{f}_1\oplus \hat{f}_3}]}\,.
\eea

First we observe that at $\la=0$ the expression for $e^{2\Phi_F}$ depends only on $X_k$ and ${\tilde X}_k$. While this property does not hold for general values of $\la$, setting $a={\tilde a}=b={\tilde b}$ we still find an interesting result:
\bea
e^{2\Phi_F}\Big|_{a={\tilde a}=b={\tilde b}=0}=\left[\frac{(1-\mu X{\tilde X})^2-(1-X^2)(1-{\tilde X}^2)}{X^2{\tilde X}^2}\right]^8,\qquad \mu\equiv \frac{2\la}{\la^2+1}\,.
\eea
In the opposite case, where all $X$ are switched off, the expression is much more complicated, for example at $\la=1$ it has the form
\bea
&&e^{2\Phi_F}\Big|_{X_m={\tilde X}_m=0,\la=1}=F^{12}\left[8F-2+(FP_{-1}-2)^2-2(P_1-1)^2+2P_2\right]^2\,,\\
&&F\equiv(a^2+1)(b^2+1)({\tilde a}^2+1)({\tilde b}^2+1),\nn
&&P_k\equiv (a^2+1)^k+(b^2+1)^k+({\tilde a}^2+1)^k+({\tilde b}^2+1)^k.\nonumber
\eea
In particular, we observe that the last expression is fully symmetric under interchanging the elements of the list $(a,b,{\tilde a},{\tilde b})$. This property persists for all values of $\la$, as long as $X_m={\tilde X}_m=0$, but the general expression is not very illuminating, so we will not write it here. 

The last two interesting cases corresponds to looking only at the sphere or only at the AdS space:
\bea
e^{2\Phi_F}\Big|_S&=&\left[\frac{(AB-\mu X)^2+\mu^2[(AbX_3)^2+(aBX_1)^2-a^2b^2 X^2]}{A^2B^2X^2} \right]^8\,,\nn
e^{2\Phi_F}\Big|_{AdS}&=&\left[\frac{({\tilde A}{\tilde B}-\mu {\tilde X})^2-\mu^2[
({\tilde A}{\tilde b}{\tilde X}_3)^2+({\tilde a}{\tilde B}{\tilde X}_1)^2+{\tilde a}^2{\tilde b}^2 {\tilde X}^2]}{{\tilde A}^2{\tilde B}^2{\tilde X}^2} \right]^8\,,\\
A&=&\sqrt{1+a^2},\quad B=\sqrt{1+b^2},\quad {\tilde A}=\sqrt{1+{\tilde a}^2},\quad {\tilde B}=\sqrt{1+{\tilde b}^2}\,.\nonumber
\eea 
The complexity of our results beyond $\la=0$ suggests that the full solution for the $\la$--deformation of AdS$_5\times$S$^5$ cannot be constructed unless one finds better coordinates, and we leave this problem for future investigation.

\section{Discussion}

In this chapter we have constructed the supergravity background describing the $\la$--deformation of AdS$_3\times$S$^3$ supercoset and reported some progress towards the analogous result for AdS$_5\times$S$^5$. Our main result is summarized by equations (\ref{AdS3metr}), (\ref{S3metr}),
(\ref{SuDilatonAdS3}), (\ref{C2forAdS3}). In the AdS$_5\times$S$^5$ case we have constructed the metric and the dilaton describing the supercoset, and while the results presented in section \ref{SectAdS5} are rather complicated, there are striking similarities with lower--dimensional cases. For example, at the WZW point, where the expression (\ref{WZWforAdS5}) for the  ten--dimensional dilaton is rather simple, one finds a very close analogy with the six--dimensional case (\ref{QforWZW3}), and we hope that a further exploration of such analogies will lead to construction of full gravity solution for the deformed AdS$_5\times$S$^5$.

		\chapter{Generalized $\lambda$--deformations of $AdS_p\times S^p$}\label{ChapterGenLambda}

		In the previous chapter we explored $\lambda$-deformation and constructed new families of integrable string theories each depending on one parameter, now we extend our analysis to a deformation depending on several parameters, the generalized $\lambda$-deformation. In particular, in this chapter we study analytical properties of the generalized $\lambda$-deformation and construct the explicit backgrounds corresponding to $AdS_p \times S^p$, including the Ramond-Ramond fluxes. For an arbitrary coset, we find the general form of the R-matrix underlying the deformation, and prove that the dilaton is not modified by the deformation, while the frames are multiplied by a constant matrix. Our explicit solutions describe families of integrable string theories depending on several continuous parameters.

\bigskip

This chapter is based on the results published in \cite{Chervonyi:2016bfl} and has the following organization. In section \ref{SectReview} we review the procedure for finding the generalized $\la$--deformation introduced in \cite{GenLambda}. This construction is based on solutions of the classical modified Yang--Baxter equation, and in section \ref{Rmatrix} we find large classes of such solutions for general cosets $G/F$, as well as the most general solutions that can be used to deform string theory on  AdS$_p\times$S$^p$ ($p=2,3,5$). We also construct very large classes of graded R--matrices, which can be used for extending the procedure of \cite{GenLambda} to supercosets, along the lines of the analysis presented in \cite{HMS}. In section \ref{SubsGenMetr} we uncover some analytical properties of the deformed metric and the dilaton, which are applicable to all cosets. The remainder of section \ref{SUGRAemb} is devoted to constructing the supergravity backgrounds supporting the generalized $\la$--deformations of  AdS$_p\times$S$^p$. Appendix \ref{AppDmtr} is devoted to exploration of analytical properties of a matrix that plays a pivotal role in constructing the generalized $\la$--deformations.

\section{Introduction}

The last few years witnessed an impressive progress in finding new families of integrable string theories. Initially integrability was discovered in isolated models, such as strings on AdS$_p\times$S$^q$ \cite{bethe,RoibPolch,AdS3Int,IntAdS2}, and in their extensions called beta deformations \cite{margin}. Recent developments, stimulated by the mathematical literature \cite{Cherednyk}, led to construction of very large classes of integrable string theories. One of the approaches originated from studies of the Yang--Baxter sigma models \cite{BosonicYangBaxter,Qdeform,Yoshida}, and it culminated in construction of new integrable string theories, which became known as $\eta$--deformations \cite{Delduc1,OtherEta,MoreEta}.
A different approach originated from the desire to relate two classes of solvable systems, the Wess--Zumino--Witten \cite{WZW} and the Principal Chiral \cite{PCM} sigma models, and it culminated in the discovery of a one--parameter family of integrable conformal field theories, which has WZW and PCM as its endpoints \cite{SftsGr,ST}\footnote{See \cite{PreST} for earlier work in this direction.}. Such line of conformal field theories becomes especially interesting when the PCM point represents a string theory on AdS$_p\times$S$^q$ space, and the corresponding families, which became known as $\la$--deformations, have been subjects of intensive investigations \cite{SfetsosAdS5,HT,HMS,BTW,Holl2}. Recently the powers of the two approaches were combined to construct the generalized $\la$--deformations \cite{GenLambda}\footnote{See 
\cite{HT} for the earlier exploration of the connection between the $\eta$ and $\la$ deformations.}, the largest class on integrable string theories known to date, which encompasses all earlier examples.  In this chapter we study the generalized $\la$--deformations of cosets with a special emphasis on describing integrable extensions of strings on AdS$_2\times$S$^2$, AdS$_3\times$S$^3$, and AdS$_5\times$S$^5$.
 
\bigskip

While the procedure for constructing the generalized $\la$--deformation has been outlined in \cite{GenLambda}, its practical implementation presents some technical challenges. Moreover, just as in the case of the standard $\la$-- and $\eta$--deformations, the CFT construction gives only the NS--NS fields, and evaluation of the Ramond--Ramond fluxes relies on supergravity computations. On the CFT side one encounters two types of challenges: construction of the classical R--matrix, which is the central element of the generalized $\la$--deformation, and evaluation of the modified metric. R--matrices are solutions of the modified classical Yang--Baxter equation (mCYB), and while many examples have been studied in the literature \cite{CYB, BosonicYangBaxter}, the full classification of R--matrices is still missing. In section \ref{Rmatrix} we find a rather general class of solutions of the mCYB equation for arbitrary cosets $G/F$, and for specific examples arising in the description of strings on AdS$_p\times$S$^p$ we construct \textit{all} solutions. Keeping in mind that the prescription of \cite{GenLambda} might have a counterpart involving supercosets (as it happened in the case of the ordinary $\la$--deformation \cite{HMS,BTW,ChLLambda}), we also find a large class of R--matrices solving the graded mCYB equation, which governs the deformations of supercosets. Deforming various supercosets using such matrices would be an interesting topic for future work.

Finding the R--matrices is not the only technical challenge associated with the generalized $\la$--deformation. 
While the procedure for finding the metric is algorithmic, and in principle it can be applied to any coset\footnote{In practice, the difficulty of such `brute force' calculation grows exponentially with the size of the coset and the number of deformation parameters. This presents an additional motivation for understanding the hidden symmetries of the problem and for simplifying the calculations.}, the calculations can be tedious, and one finds a lot of `accidental cancellations' in the final results. Such surprises have been encountered in the past \cite{ST, SfetsosAdS5}, and in some instances they have been explained on a case-by-case basis \cite{SfetsosAdS5}. In section \ref{SUGRAemb} we demonstrate that the `accidental cancellations' are guaranteed by the symmetries of the underlying problem, thus they must be present for all deformations, and they can be used to drastically simplify the calculations. 
Even apart from this practical usefulness, our study of hidden symmetries contributes to the general 
analytical understanding of integrable deformations. 

Application of the algebraic procedure outlined in \cite{GenLambda} yields the metric and the dilaton for the deformed backgrounds, but recovery of the Ramond--Ramond fluxes from the sigma model is a very complicated task \cite{BTW}. In practice, it is much easier to find such fluxes by solving the supergravity equations of motion, and in the past this technique has been successfully implemented for several families of integrable string theories \cite{OtherEta,ST, SfetsosAdS5, ChLLambda}. Following the same path in section \ref{SUGRAemb}, we recover the fluxes supporting the generalized $\la$--deformation of  AdS$_2\times$S$^2$ and AdS$_3\times$S$^3$. Interestingly, the construction of \cite{GenLambda} does not allow one to deform AdS$_5\times$S$^5$ unless a trivial R--matrix is chosen.

\section{Review of the generalized $\lambda$-deformation}
\label{SectReview}

Lambda deformations of the Principal Chiral Models (PCM) were introduced in \cite{SftsGr} and further studied in \cite{HMS, ST, SfetsosAdS5, HT, BTW, ChLLambda}. Application of such deformation to any PCM leads to a one--parameter family of integrable conformal field theories. This deformation was generalized to a larger family in \cite{GenLambda}, and we begin with reviewing this construction following section 5 of \cite{GenLambda}.

\bigskip

The $\lambda$ deformation  interpolated between Conformal Field Theories described by a Principal Chiral Model (PCM) and a Wess--Zumino--Witten model (WZW), and we begin with looking at the WZW side:
\bea\label{gWZW}
S_{WZW,k}(g)=\frac{k}{4\pi}\int_\Sigma d^2\sigma R_+^a R_-^a-\frac{k}{24\pi}\int_B f_{abc} R^a\wedge R^b\wedge R^c,\quad \d B=\Sigma.
\eea
Here $g\in G$ is an element of some group $G$ with generators $T_a$, $k$ is the level of the WZW model, 
$R_\pm$ are the right-invariant Maurer-Cartan forms,
\bea
R^a_\pm=-i \mbox{Tr}(T^a \d_\pm g  g^{-1})\,,
\eea
and $f_{abc}$ are the structure constants:
\bea
[T_a,T_b]=i f_{ab}{}^c T_c\,.
\eea
To construct the $\lambda$ deformation one adds the action (\ref{gWZW}) to a generalized PCM on a group manifold\footnote{In comparison with \cite{GenLambda} we have rescaled the constant coefficients ${E}_{ab}$ by $k$ so the level of the WZW appears as an overall factor in the sum of (\ref{gWZW}) and (\ref{genPCM}). Such rescaling simplifies the formulas associated with $\la$--deformation.},
\bea\label{genPCM}
S_{gPCM}(\hat{g})=\frac{k}{2\pi}\int d^2\sigma {E}_{ab} R^a_+(\hat{g}) R^b_-(\hat{g}),\qquad
\hat{g}\in G\,,
\eea
and gauges away half of the degrees of freedom in the resulting 
sum\footnote{See \cite{GenLambda} for more details.}.
Parameters ${E}_{ab}$ in (\ref{genPCM}) represent an arbitrary constant matrix, and later its form will be restricted by the requirements of conformal invariance and integrability. The gauging procedure in the sum of (\ref{gWZW}) and (\ref{genPCM}) leads to the action \cite{PreST,GenLambda}
\bea\label{GenLambdaAction}
S_{k,\lambda}(g)=S_{WZW,k}(g)+\frac{k}{2\pi}\int d^2\sigma L_+^a(\hat{\lambda}^{-1}-D)^{-1}R_-^b,
\eea
where\footnote{Following \cite{GenLambda}, we denote the \textit{matrix} appearing in (\ref{GenLambdaAction}), (\ref{lambdaMtrDef}) by $\hat{\lambda}$ to distinguish it from the \textit{scalar} deformation parameter $\la$.}  
\bea\label{lambdaMtrDef}
\hat{\lambda}^{-1}={E}+ I,
\quad D_{ab}=\mbox{Tr}(T_a g T_b g^{-1}),\quad
L_\pm^a
=i\mbox{Tr}(T_ag^{-1}\d_\pm g),\quad R_\mu^a=D_{ab}L_\mu^b.
\eea
Application of this prescription to the standard PCM,
\bea\label{StandLambda}
{E}_{ab}=\frac{\kappa^2}{k}\delta_{ab},\quad \hat{\lambda}^{-1}=\frac{k+\kappa^2}{k}I,
\eea
leads to a one-parameter $\lambda$--deformation, and integrability of the corresponding conformal field theory (\ref{GenLambdaAction}) was demonstrated in \cite{SftsGr}. It is clear that the sigma model \eqref{GenLambdaAction} would not be integrable for a generic matrix ${E}$, but the authors of \cite{GenLambda} found a large class of integrable models extending (\ref{StandLambda}). We begin with reviewing this construction for groups, and then discuss the cosets, which will be the main objects of our study.

\bigskip
\noindent
\textbf{Generalized $\lambda$-deformation for groups.}

To arrive at an integrable deformation (\ref{GenLambdaAction}), one should start with an integrable generalized PCM (\ref{genPCM}), and this already imposes severe restrictions on the constant matrix 
${ E}_{ab}$. Extending the standard choice (\ref{StandLambda}), one can start with the action of the $\eta$--deformed PCM \cite{BosonicYangBaxter}:
\bea\label{YangBaxterAction}
S_{gPCM}=\frac{1}{2\pi {\tilde t}}\int d^2\sigma R_+^T(I-\tilde{\eta} \mathcal{R})^{-1}R_-,\quad \eta>0.
\eea
As demonstrated in \cite{BosonicYangBaxter}, this model is integrable, as long as the constant matrix $\mathcal{R}$ satisfies the modified classical Yang-Baxter (mCYB) equation\footnote{The constant matrix $\mathcal{R}$ satisfying the Yang-Baxter equation is called the Yang-Baxter operator or the R--matrix. In this work we use both names.}
\bea\label{modYBeq}
[\mathcal{R} A,\mathcal{R} B]-\mathcal{R}([\mathcal{R} A,B]+[A,\mathcal{R} B])=-c^2 [A,B],\quad A,B\in \mathfrak{g},\quad c\in\mathbb{C}.
\eea
Then the interpolating model (\ref{GenLambdaAction}) with 
\bea
{E}_{YB}=\frac{1}{\tilde{t}}\left( I-\tilde{\eta}\mathcal{R} \right)^{-1}
\eea
is integrable as well, and it is called the generalized $\lambda$-deformation of (\ref{YangBaxterAction}) \cite{GenLambda}.

\bigskip
\noindent
\textbf{Generalized $\lambda$-deformation for cosets.}

The authors of \cite{GenLambda} also extended the construction of the generalized $\lambda$-deformation to cosets $G/F$ by defining 
\bea\label{EmtrCosetA}
{E}={E}_H\oplus{E}_{G/F},\quad {E}_F=0,\quad {E}_{G/F}=\frac{1}{\tilde{t}}(I-\tilde{\eta}\mathcal{R})^{-1},\quad \mathfrak{g}=\mathfrak{f}+\mathfrak{l},
\eea
This ansatz for ${E}$ leads to inconsistent equations of motion for (\ref{GenLambdaAction}) unless all elements of the coset satisfy the constraint \cite{GenLambda}\footnote{This constraint is multiplied by $\tilde\eta$, but since we are interested in the deformed theory, $\tilde\eta\ne 0$}:
\bea\label{YBcosetConstr}
([\mathcal{R}X,Y]+[X,\mathcal{R}Y])|_\mathfrak{f}=0,\quad X,Y\in\mathfrak{l}.
\eea
Assuming that this constraint is satisfied, the equations of motion for the action \eqref{GenLambdaAction} with the matrix ${E}$ from \eqref{EmtrCosetA} can be written as the integrability condition of a Lax pair (see \cite{GenLambda} for details).

\bigskip

To summarize, the generalized $\lambda$ deformation can be defined on cosets, but integrability puts a severe restriction (\ref{YBcosetConstr}) on the Yang-Baxter operator $\mathcal{R}$. In the next section we will consider several cosets arising in the type II string theory and discuss the corresponding Yang-Baxter operators $\mathcal{R}$ solving the modified classical Yang-Baxter (mCYB) equation \eqref{modYBeq} and the coset constraint \eqref{YBcosetConstr}. Then in section \ref{SUGRAemb} we will use these solutions to embed the generalized $\lambda$ deformations of the corresponding cosets into supergravity.


\section{R-matrices for Lie algebras and cosets}\label{Rmatrix}

In string theory integrability was discovered by studying strings on AdS$_p\times$S$^q$ \cite{RoibPolch,AdS3Int,IntAdS2} and the corresponding CFTs are the Principal Chiral models on various cosets. In this chapter we are interested in the generalized $\lambda$ deformations of such backgrounds, so as outlined in the last section, we should find the Yang--Baxter operators $\mathcal{R}$ satisfying the mCYB equation (\ref{modYBeq}) and the constraint (\ref{YBcosetConstr}) on the relevant coset. In subsection \ref{SubsRgen} we will discuss some general features of such operators, and in the remaining part of this section we will apply this construction to the specific cosets arising in string theory.

\subsection{General construction}
\label{SubsRgen}

The generalized $\la$ deformation reviewed in section \ref{SectReview} is based on the Yang-Baxter operator satisfying the mCYB equation (\ref{modYBeq})\footnote{We set $c=i$ in \eqref{modYBeq}.},
\bea\label{mYBeq}
\label{YBmatrix}
[\mathcal{R}X,\mathcal{R}Y]-\mathcal{R}([\mathcal{R}X,Y]+[X,\mathcal{R}Y])=[X,Y],\quad X,Y\in \mathfrak{g},
\eea
and the constraint (\ref{YBcosetConstr})
\bea\label{YBcosetConstr1}
\label{CosetConstrMtr}
([\mathcal{R}\tilde{X},\tilde{Y}]+[\tilde{X},\mathcal{R}\tilde{Y}])|_\mathfrak{f}=0,\quad \mathfrak{g}=\mathfrak{f}+\mathfrak{l},\quad \tilde{X},\tilde{Y}\in\mathfrak{l}\,.
\eea
We further impose the skew-symmetry condition
\bea\label{skewSym}
(\mathcal{R}X,Y)_{\mathfrak{g}}+(X,\mathcal{R}Y)_{\mathfrak{g}}=0,
\eea
where $(.,.)_\mathfrak{g}$ is the Killing-Cartan form on the Lie algebra. While acting on generators $T_a$, the operator $\mathcal{R}$ can be viewed as a tensor with one lower and one upper index (${\mathcal{R}_b}^a$) and the skew-symmetry condition \eqref{skewSym} means that
\bea\label{skewSymMatr}
\mathcal{R}_{ab}=-\mathcal{R}_{ba}\,.
\eea

Finding the most general solution of (\ref{mYBeq}) for an arbitrary group is an open problem, but one solution is well-known \cite{BosonicYangBaxter}, and now we will introduce its generalization. We will also find the most general solution of (\ref{mYBeq})--(\ref{skewSym}) for specific cosets arising in string theory. 

\bigskip

Equations (\ref{mYBeq}), (\ref{skewSymMatr}) in the adjoint representation imply that $\mathcal{R}_{ab}$ is a real antisymmetric matrix, so it can be diagonalized using a \textit{unitary} rotation, and all its eigenvalues are imaginary. In particular, some of these eigenvalues might vanish, then equation (\ref{mYBeq}) implies that the corresponding eigenvectors (which are generators of $\mathfrak{g}$) must commute. Thus we conclude that the kernel of operator ${\mathcal{R}_b}^a$ is a subset of the Cartan subalgebra $\mathfrak{h}$ and
\bea\label{ineqR}
\mbox{rank}\,\mathcal{R}\ge \mbox{dim}\,\mathfrak{g}-\mbox{rank}\,\mathfrak{g}\,.
\eea
The standard solution of the classical Yang--Baxter equation \cite{BosonicYangBaxter} corresponds to the case where the last inequality saturates, so the kernel of ${\mathcal{R}_b}^a$ coincides with the Cartan subalgebra:
\bea\label{CartanR}
\mathcal{R}H_i=0\quad\mbox{for all}\quad H_i\in \mathfrak{h}.
\eea 
Looking at an arbitrary $X=H$ from this subalgebra, and representing this generator as an operator ${\hat H}$ acting in the adjoint representation, we can rewrite (\ref{mYBeq}) as
\bea\label{RHRCartan}
-\mathcal{R}{\hat H}\mathcal{R}Y={\hat H}Y.
\eea
If $Y$ is an eigenvector of $\mathcal{R}$ with an eigenvalue $\la_Y$, then ${\hat H}Y$ is an eigenvector with an eigenvalue $-\frac{1}{\la_Y}$ for any ${\hat H}$. 

To proceed, we expand the eigenvector $Y$ in the Weyl--Cartan basis,
\bea\label{YexpandCW}
Y=\sum c_k |\alpha^{(k)}\rangle,
\eea
where each $|\alpha^{(k)}\rangle$ is an eigenvector of all Cartan generators\footnote{Equation (\ref{CartWeyl}) gives a more explicit expression, but it is not needed here.}. Focusing on a particular Cartan generator 
${\hat H}_i$, we conclude that $[{\hat H}_i]^NY$ is an eigenvector of $\mathcal{R}$, which is dominated by $ |\alpha^{(k)}\rangle$ with the largest eigenvalue of ${\hat H}_i$. Removing this vector and repeating the argument for the second largest eigenvalue and so on, one can demonstrate that all 
$|\alpha^{(k)}\rangle$ are eigenvectors of $\mathcal{R}$. In other words, we have shown that matrix $\mathcal{R}$ must be diagonal in the Cartan--Weyl basis. 

Let us now specify the  Cartan--Weyl basis in more detail. Any semisimple Lie algebra admits a decomposition into the Cartan generators $H_i$ and ladder operators $E_\alpha$ so that the full commutation relations have the form
\bea\label{CartWeyl}
[H_i,H_j]=0,\quad [H_i,E_\alpha]=\alpha_i E_\alpha,\quad [E_\alpha,E_\beta]=e_{\alpha,\beta}E_{\alpha+\beta},\quad [E_\alpha,E_{-\alpha}]=\sum_i {\tilde\alpha}^i H_i\,.
\eea
In the expansion (\ref{YexpandCW}) the generator $E_\alpha$ was denoted as $|\alpha^{(k)}\rangle$. 
By an appropriate rescaling of the ladder operators one can go to a more restrictive Chevalley basis, but such specification will not play any role in our discussion. As we have demonstrated, relation (\ref{CartanR}) implies that the R--matrix must be diagonal in the basis (\ref{CartWeyl}), this leads to the explicit form of the Yang--Baxter operator:
\bea
\mathcal{R}H_i=0,\quad \mathcal{R}E_\alpha=\la_\alpha E_\alpha
\eea
Substitution into (\ref{RHRCartan}) leads to $\la_\alpha=\pm i$, and application of the Yang--Baxter equation (\ref{mYBeq}) to $(X,Y)=(E_\alpha,E_\beta)$ gives a constraint on the eigenvalues
\bea
\la_\alpha\la_\beta-\la_{\alpha+\beta}(\la_\alpha+\la_{\beta})=1.
\eea
In particular, $\la_\alpha\la_{-\alpha}=1$, so the Yang--Baxter operator becomes:
\bea\label{CanonYBsln}
\mathcal{R} H_i=0,\quad \mathcal{R} E_\alpha=-i E_\alpha,\quad \mathcal{R} E_{-\alpha}=i E_{-\alpha},
\eea
where $\alpha$ are positive roots. This construction is known as the canonical R--matrix, and we have derived it from (\ref{CartanR}), which in turn follows from the assumption that the inequality (\ref{ineqR}) saturates. 

\bigskip

The canonical R--matrix \eqref{CanonYBsln} can be easily generalized by modifying the first relation in (\ref{CanonYBsln}), and such extension will play an important role in the analysis presented in the rest of this section. Specifically, it is clear that equation (\ref{mYBeq}) is solved by
\bea\label{CanonYBslnG}
\mathcal{R}H_i={R_i}^j H_j,\quad \mathcal{R}E_\alpha=-i E_\alpha,\quad 
\mathcal{R}E_{-\alpha}=i E_{-\alpha}
\eea
for an \textit{arbitrary} matrix ${R_i}^j$. In other words, the R--matrix can be modified in the Cartan subalgebra\footnote{A similar construction has been discussed in the mathematical literature \cite{Skprypnik}.}. Notice that for the deformation (\ref{CanonYBsln}) the inequality (\ref{ineqR}) is replaced by
\bea
\mbox{rank}\,\mathcal{R}= \mbox{dim}\,\mathfrak{g}-\mbox{rank}\,\mathfrak{g}+\mbox{rank}\,{R}\,.
\eea
For future reference we also give the real form of (\ref{CanonYBslnG}):
\bea
&&B_\alpha=\frac{i}{\sqrt{2}}(E_\alpha+E_{-\alpha}),\quad C_\alpha=\frac{1}{\sqrt{2}}(E_\alpha-E_{-\alpha}),\nn
\label{RoperCanon}
&&\mathcal{R}H_i={R_i}^j H_j,\quad \mathcal{R}B_\alpha=C_\alpha,\quad 
\mathcal{R}C_\alpha=-B_\alpha,
\eea
The undeformed version of this solution (i.e., the one with $R=0$) has been widely discussed in the literature \cite{BosonicYangBaxter, Skprypnik}, and the general form of (\ref{RoperCanon}) will be used later in this section.

While (\ref{CanonYBsln}) was the most general solution with saturated inequality (\ref{ineqR}), the construction (\ref{CanonYBslnG}) is just one possible option for non--saturating (\ref{ineqR}), and later we will present explicit examples of R--matrices which do not fit into (\ref{CanonYBslnG}). However, we will now demonstrate that any solution that can be obtained as a continuous perturbation of (\ref{CanonYBsln}) must have the form (\ref{CanonYBslnG}). 

\bigskip

Let us start with the canonical solution (\ref{CanonYBsln}), which will be called $\mathcal{R}_0$, and perturb it  by $\eps \mathcal{R}_1$ with a small parameter $\eps$. Applying (\ref{mYBeq}) to two elements of the Cartan subalgebra ($(X,Y)\in {\mathfrak{h}}$) and expanding the result to the first order in $\eps$, we find a system of linear constraints on $\mathcal{R}_1$:
\bea\label{ConstrR1}
-\mathcal{R}_0([\mathcal{R}_1X,Y]+[X,\mathcal{R}_1Y])=0,\quad X,Y\in \mathfrak{h},
\eea
Clearly, our ansatz (\ref{CanonYBslnG}) solves these constraints with
\bea
\mathcal{R}_1H_i={R_i}^j H_j,\quad \mathcal{R}_1E_\alpha=0,\quad 
\mathcal{R}_1E_{-\alpha}=0,\nonumber
\eea 
and since equations (\ref{ConstrR1}) are linear in $\mathcal{R}_1$, one can always subtract an appropriate solution (\ref{CanonYBslnG}) to ensure that $\mathcal{R}_1X$ has a trivial projection on the Cartan subalgebra. In other words, without the loss of generality, we can write
\bea\label{R1onX}
\mathcal{R}_1X=\sum_\alpha c_X(\alpha)E_\alpha\,,
\eea
where sum is extended over all roots of the Lie algebra, and $c_X(\alpha)$ are some numerical coefficients.
Substitution into (\ref{ConstrR1}) gives
\bea\label{ConstrR1Prime}
-\sum_{\alpha}[-c_X(\alpha)Y(\alpha)+c_Y(\alpha)X(\alpha)\Big]\Big[\mathcal{R}_0 E_\alpha\Big]=0,
\eea
where coefficients $X(\alpha)$ are defined using the commutation relations (\ref{CartWeyl}):
\bea
[X,E_\alpha]=\Big[\sum_i x^i H_i,E_{\alpha}\Big]=E_\alpha \sum_i x^i\alpha_i\quad
\Rightarrow\quad [X,E_\alpha]\equiv X(\alpha)E_\alpha.
\eea
Since the roots $E_\alpha$ are eigenvectors of $\mathcal{R}_0$ (recall (\ref{CanonYBsln})), and they are linearly independent, equation (\ref{ConstrR1Prime}) implies that\footnote{For every root $\alpha$ we can always start with $Y\in {\mathfrak{g}}$, such that $Y(\alpha)\ne 0$, so the right hand side of (\ref{ConstrR1Prime2}) is well-defined.}
\bea\label{ConstrR1Prime2}
c_X(\alpha)=X(\alpha)\frac{c_Y(\alpha)}{Y(\alpha)}\equiv c(\alpha)X(\alpha)\,.
\eea
Substitution into (\ref{R1onX}) leads to
\bea\label{ConstrR1Prime4}
\mathcal{R}_1X=\sum_\alpha X(\alpha) c(\alpha) E_\alpha\,,
\eea
where $c(\alpha)$ depends on the root, but not on the element $X$ of the Cartan subalgebra. To complete the argument, we define
\bea\label{ConstrR1Prime3}
{\tilde X}\equiv X-\eps \sum_\alpha X(\alpha) c(\alpha) \left[\frac{E_\alpha}{\mathcal{R}_0 E_\alpha}\right]E_\alpha\,.
\eea
Notice that relations (\ref{CanonYBsln}) for $\mathcal{R}_0$ imply that expressions in the square brackets are $c$--numbers equal to $\pm i$. Using (\ref{CanonYBsln}), we conclude that 
\bea\label{ConstrR1Prime5}
(\mathcal{R}_0+\eps \mathcal{R}_1){\tilde X}=O(\eps^2),
\eea
so in the leading order in $\eps$ operator $\mathcal{R}$ has the same number of zero modes as $\mathcal{R}_0$, so the solution is still given by (\ref{CanonYBsln}), but the Cartan subalgebra is rotated 
by (\ref{ConstrR1Prime3}). To simplify the discussion we started with equation (\ref{R1onX}) by subtracting the part of $\mathcal{R}_1$ that acts on the Cartan subalgebra, and in general equations (\ref{ConstrR1Prime4}) and (\ref{ConstrR1Prime5}) are replaced by
\bea\label{ConstrR1Prime6}
&&\mathcal{R}_1X=R X+\sum_\alpha X(\alpha) c(\alpha) E_\alpha\,,\nn
&&(\mathcal{R}_0+\eps \mathcal{R}_1){\tilde X}=\eps R {\tilde X}+ O(\eps^2),
\eea
while equation (\ref{ConstrR1Prime3}) remains the same. Here $R$ is an operator mapping the Cartan subalgebra on itself, so equation (\ref{ConstrR1Prime6}) is a perturbative expansion of (\ref{CanonYBslnG}).  

\bigskip

To summarize, we have demonstrated that the most general solution of the mCYB equation (\ref{mYBeq}) with $\mbox{rank}\,\mathcal{R}= \mbox{dim}\,\mathfrak{g}-\mbox{rank}\,\mathfrak{g}$ is given by (\ref{CanonYBsln}), and its most general perturbation fits the ansatz (\ref{CanonYBslnG}). It would be interesting to find the most general solution of the mCYB equation without relying on perturbative argument, but such investigation is beyond the scope of this work.

\bigskip

So far we have focused on the Yang--Baxter equation (\ref{mYBeq}) and have ignored the coset constraint 
(\ref{YBcosetConstr1}). This leads to the expression (\ref{CanonYBslnG}), which is not sensitive to the choice of the coset, but condition (\ref{YBcosetConstr1}) projects out some solutions. If fact, as we will see in subsection \ref{SubsAdS5R}, in the case of the SO(6)/SO(5) coset the constraint  (\ref{YBcosetConstr1}) eliminates all solutions preventing the construction of the generalized $\la$--deformation for AdS$_5\times$S$^5$. Note that while the construction (\ref{CanonYBslnG}) can be applied to any 
Cartan subalgebra and all resulting R--matrices would be related by a group rotation, a specific embedding of the subgroup $F$ removes equivalence between different choices of the Cartan subalgebra. Thus the 
constraint (\ref{YBcosetConstr1}) should be imposed on the R--matrices which have the form (\ref{CanonYBslnG}) for \textit{at least one} Cartan subalgebra.  Starting with one Cartan subalgebra, applying the prescription (\ref{CanonYBslnG}), and rotating the result by an arbitrary element of the group, one constructs the most general R--matrix in the class (\ref{CanonYBslnG}), which depends on $N$ parameters with 
\bea
N=\frac{r(r-1)}{2}+(d-r),\qquad r=\mbox{rank}\,\mathfrak{g},\quad d=\mbox{dim}\,\mathfrak{g}.
\eea
The constraint (\ref{YBcosetConstr1}) should be imposed in the end. 

We conclude this subsection by presenting an explicit example of the construction (\ref{CanonYBslnG}), (\ref{YBcosetConstr1}) for the simplest coset SU(2)/U(1). Since SU(2) has a one--dimensional Cartan subalgebra, the antisymmetric matrix $R_{ij}$ entering (\ref{CanonYBslnG}) must be trivial, so in the real basis the $R$--matrix has only two non--zero elements:
\bea
\mathcal{R}_{12}=-\mathcal{R}_{21}=1.
\eea
Rotation by a group element leads to a more general matrix in terms of the Euler angles
\bea\label{RmtrSU2}
\mathcal{R}=\left[\begin{array}{ccc}
0 &\cos\theta & \sin\theta\cos\phi\\
-\cos\theta & 0 & \sin\theta\sin\phi\\
-\sin\theta\cos\phi & -\sin\theta\sin\phi & 0
 \end{array} \right].
\eea
Direct calculation shows that this is the most general solution of the Yang--Baxter equation (\ref{mYBeq}).
The coset constraint \eqref{CosetConstrMtr} is satisfied trivially.

In the next few subsections we will discuss some examples of cosets arising in string theory.

\subsection{Solution for SO(3)/SO(2)}

Let us discuss the most general solutions of the modified Yang-Baxter equation for the cosets SO(3)/SO(2) and SO(2,1)/SO(1,1), which arise in the deformation of AdS$_2\times$S$^2$. Strings on this background are described by the supercoset $\mathfrak{psu(1,1|2)}$ \cite{AdS2PSU}, whose bosonic sector is represented by two $2\times 2$ matrices $\mathfrak{g_{u(2)}}$, $\mathfrak{g_{u(1,1)}}$:
\bea
\mathfrak{g_{psu(1,1)}}=\left[
\begin{array}{cc}
\mathfrak{g_{u(1,1)}}& 0\\
0& \mathfrak{g_{u(2)}}
\end{array}
\right],\quad \mathfrak{g^\dagger_{u(1,1)}} \Sigma\,\mathfrak{g_{u(1,1)}}=\Sigma, \quad 
\mathfrak{g^\dagger_{u(2)}}\mathfrak{g_{u(2)}}=I,\quad
\Sigma=\left[
\begin{array}{cc}
1&0\\
0&-1
\end{array}
 \right]\,.\nonumber
\eea
We will use the following explicit parameterization of generators\footnote{Labels 6-12 are usually reserved for the fermionic generators.}:
\bea
\mathfrak{g_{u(1,1)}}&=&
\left[\begin{array}{cc}
F_1+F_4& F_2+i F_3\\
-F_2+i F_3& -F_1+F_4
 \end{array} \right],\quad
\mathfrak{g_{u(2)}}=
\left[\begin{array}{cc}
F_{13}+F_{16}& F_{14}+iF_{15}\\
F_{14}-iF_{15}& -F_{13}+F_{16}
 \end{array} \right].
\eea
U(2) subgroup has two--dimensional Cartan subalgebra spanned by $(F_{13},F_{16})$, and the construction (\ref{CanonYBsln}) gives
\bea\label{SimpleRU2}
\mathcal{R}_{U(2)}&=&\left[
\begin{array}{cccc} 
0&0&0&a \\
0 & 0 & 1& 0\\
0& -1& 0 & 0\\
-a&0& 0& 0
\end{array} 
\right]\,.
\eea
Rotation by a general group element gives
\bea\label{RmatrU2}
\mathcal{R}_{U(2)}&=&\left[
\begin{array}{cccc} 
0& \cos\gamma\sin\theta& \sin\gamma\sin\theta &a \\
-\cos\gamma\sin\theta & 0 & \cos\theta& -a \sin\gamma\tan\theta\\
-\sin\gamma\sin\theta& -\cos\theta& 0 & a\cos\gamma\tan\theta\\
-a& a \sin\gamma\tan\theta& -a\cos\gamma\tan\theta& 0
\end{array} 
\right]\,,
\eea
and direct calculation shows that this is the most general R--matrix for U(2). Choosing the subgroup $F$ spanned by $(F_{13},F_{16})$, one can check that the constraint (\ref{YBcosetConstr1}) is satisfied.

The R--matrix for U(1,1) is obtained by rotating the counterpart of (\ref{SimpleRU2}) by an appropriate group element, and the result is
\bea\label{RmatrU11}
\mathcal{R}_{U(1,1)}&=&\left[
\begin{array}{cccc} 
0& \cos\gamma\sinh\xi& \sin\gamma\sinh\xi &a \\
-\cos\gamma\sinh\xi & 0 & \cosh\xi& a \sin\gamma\tanh\xi\\
-\sin\gamma\sinh\xi& -\cosh\xi& 0 & -a\cos\gamma\tanh\xi\\
-a& -a \sin\gamma\tanh\xi& a\cos\gamma\tanh\xi& 0
\end{array} 
\right].
\eea
While constructing the integrable deformations of strings on AdS$_2\times$S$^2$, one can obtain the fields for U(1,1)/U(1) by analytic continuation of the result for U(2)/U(1). This is slightly easier than performing a separate calculations using (\ref{RmatrU11}), but the answers are the same.

\subsection{Solution for SO(4)/SO(3)}

Next, we consider the coset
\bea
\frac{SO(4)}{SO(3)}=\frac{SU(2)_L\times SU(2)_R}{SU(2)_{diag}}.
\eea
This coset, along with its counterpart $SO(2,2)/SO(1,1)$, arises in description of strings on AdS$_3\times$S$^3$.

To simplify the evaluation of the R--matrix we pick the following generators of SU(2)$\times$SU(2)
\bea\label{GeneratorsSO4}
T^{[SU(2)]^2}=\{T^L, T^R \},\quad 
T^L_i=\left[
\begin{array}{cc}
\sigma_i& 0\\
0& 0
\end{array}
 \right],\quad 
 T^R_i=\left[
\begin{array}{cc}
0& 0\\
0& \sigma_i
\end{array}
 \right],
\eea
where $\sigma_i$ are the Pauli matrices. The subgroup SU(2)$_{diag}$ is generated by
\bea
T_i^{diag}=\frac{1}{2}\left[\begin{array}{cc}
\sigma_i &0\\ 0&\sigma_i
\end{array}\right].
\eea
Starting with  the most general antisymmetric $R$ matrix 
\bea\label{RmtrSO4ansatz}
\mathcal{R}=\left[\begin{array}{cc}
A&B\\
-B^T&C
\end{array}\right]
\eea
and performing an SU(2)$_{diag}$ rotation,
we can put the antisymmetric matrix $A$ in the form
\bea
A= \left[\begin{array}{ccc}
0&0&0\\
0&0&1\\
0&-1&0
\end{array}\right]
\eea
An additional rotation in the 2--3 plane can be used to set $B_{31}=0$. 

Direct substitution of \eqref{RmtrSO4ansatz} into the modified Yang-Baxter equation \eqref{YBmatrix} and the coset constraint \eqref{CosetConstrMtr} leads to three families of the R matrices and one special solution $\mathcal{R}_4$:

\bea\label{RmatricesSO4}
\mathcal{R}_1&=\left[\begin{array}{cccccc}
0& 0& 0& a& 0& 0\\
0& 0& 1& 0& 0& 0\\
0& -1& 0& 0& 0& 0\\
-a& 0& 0& 0& 0& 0\\
0& 0& 0& 0& 0& -1\\
0& 0& 0& 0& 1& 0
 \end{array}\right],\quad&
\mathcal{R}_2=\left[\begin{array}{cccccc}
0& 0& 0& i& b& -ib\\
0& 0& 1& 0& ic& c\\
0& -1& 0& 0& c& -ic\\
-i& 0& 0& 0& b& -ib\\
-b& -ic& -c& -b& 0& -1\\
ib& -c& ic& ib& 1& 0
 \end{array}\right]\nn
\mathcal{R}_3&=\left[\begin{array}{cccccc}
0& 0& 0& -i& b& ib\\
0& 0& 1& 0& 0& 0\\
0& -1& 0& 0& 0& 0\\
i& 0& 0& 0& b& ib\\
-b& 0& 0& -b& 0& -1\\
-ib& 0& 0& -ib& 1& 0
 \end{array}\right],\quad&
\mathcal{R}_4=\left[\begin{array}{cccccc}
0& 0& 0& i& 0 & 0\\
0& 0& 1& 0& i& 1\\
0& -1& 0& 0& -1& i\\
-i& 0& 0& 0& 0& 0\\
0& -i& 1& 0& 0& 1\\
0& -1& -i& 0& -1& 0
 \end{array}\right],
\eea
As expected from the general analysis of subsection \ref{SubsRgen}, only $\mathcal{R}_1$, which fits the ansatz (\ref{CanonYBslnG}), can be continuously connected to the canonical solution (\ref{CanonYBsln}). All other matrices are complex, and they cannot be transformed into $\mathcal{R}_1$ or into each other by any 
action of SU(2)$\times$SU(2) (recall that $g\in$ SU(2)$\times$SU(2) acts as a rotation $\mathcal{R}\rightarrow g \mathcal{R} g^{-1}$). Since matrices $\mathcal{R}_{2,3,4}$ are complex, they are not acting in a proper real section of the SU(2)$\times$SU(2) algebra, so they will not play any role in our construction. Interestingly, the generalized canonical solution (\ref{CanonYBslnG}) exhausts all real R matrices. While this result was proven in subsection \ref{SubsRgen} using perturbative techniques, the current example suggests that it might hold in general. On the other hand, example (\ref{RmatricesSO4}) illustrates that in complexified algebras solution (\ref{CanonYBslnG}) is not unique beyond perturbation theory. It would be interesting to study the counterparts of $\mathcal{R}_{2,3,4}$ for other complexified algebras.

\subsection{Absence of solution for SO(6)/SO(5)}
\label{SubsAdS5R}

Finally let us apply the construction (\ref{CanonYBslnG}) to the coset 
\bea
\frac{SO(6)}{SO(5)}\,,
\eea
which arises in description of strings on AdS$_5\times$S$^5$.

The generators of $SO(6)$ are defined as 
\bea
(T_{mn})_{ab}=\delta_{ma} \delta_{nb}-\delta_{mb}\delta_{na}, \quad m,n,a,b=1,...,6,
\eea
the Cartan subgroup is three--dimensional, and it can be represented by 
\bea
H=\{T_{23}, T_{45}, T_{61}\}.
\eea
The standard diagonalization procedure leads to twelve roots:
\bea
\alpha_\alpha=\{(0,a,b), (a, 0, b), (a,b, 0)\},\qquad a,b=\pm 1.
\eea
A root will be considered positive if the first non-zero entry is positive, and for such roots prescription 
(\ref{CanonYBslnG}) gives $\mathcal{R}E_\alpha=-i E_\alpha$. For negative roots we have $\mathcal{R}E_{-\alpha}=i E_{-\alpha}$. Since $SO(6)$ has rank three, the antisymmetric matrix ${R_{ij}}$ appearing in (\ref{CanonYBslnG}) has only one non--zero element.

Next we should specify the subgroup and check the coset constraint \eqref{CosetConstrMtr}. Instead of choosing a particular subgroup, we parametrize the entire family of SO(5) embeddings, which are in the one--to--one correspondence with the unit vectors in $\mathbb{R}^6$. In the simplest case of the unit vector with only one nontrivial component $v_1=1$, the coset generators are given by 
\bea
(T^{(cos,0)}_i)_{ab}=\delta_{ia} \delta_{1b}-\delta_{ib}\delta_{1a},\quad i=2,...,6,
\eea
and in general we find
\bea\label{CosetRotate}
T^{(cos)}=(g^{SO(6)})^{-1}T^{(cos,0)}g^{SO(6)}
\eea
The SO(6) group element is parameterized in terms of the Euler angles as \cite{Fradkin}
\bea
g^{SO(n)}=\prod_{i=1}^n \prod_{j=i}^1 g_j(\theta_j^i),\quad g_k(x)=\exp\left[ x T_{n+1-k,n+1-(k+1)} \right].
\eea
and the independent choices of the cosets (\ref{CosetRotate}) correspond to $\theta_{i,5}$.
Plugging the extended canonical R--matrix \eqref{CanonYBslnG} into the coset constraint \eqref{CosetConstrMtr} we find that there are no solutions, which means that the coset SO(6)/SO(5) does not satisfy the coset constraint, and it is impossible to construct the generalized $\lambda$ deformation of AdS$_5\times$S$^5$.

\subsection{Graded Yang-Baxter equation}
\label{SecGraded}

Although in this chapter we are focusing on deformations of bosonic cosets, in the future it might be interesting to extend the generalized lambda deformation to supercosets describing string theories on AdS$_p\times$S$^p$ \cite{AdS2PSU,AdS3PSU,AdS5PSU}. For the ordinary lambda deformation this has been done in \cite{HMS}, but the generalized deformation is more involved. However, preliminary analysis indicates that an extension to supercoset would involve the graded Yang-Baxter equation, and in this subsection we will briefly discuss its properties and some solutions.

To define the Yang-Baxter equation on superalgebras and supercosets, one replaces the commutators in \eqref{modYBeq} by the graded commutators
\bea\label{GradedModYBeq}
[\mathcal{R} X,\mathcal{R} Y\}-\mathcal{R}([\mathcal{R} X,Y\}+[X,\mathcal{R} Y\})=-c^2 [X,Y\},\quad A,B\in \mathfrak{g},\quad c\in\mathbb{C}.
\eea
To define the graded commutator we start with supermatrices $X,Y$ written in the block form
\bea\label{supermtrs}
X=\left[
\begin{array}{c|c}
A& B\\
\hline
C&D
\end{array}
 \right],\quad 
 Y=\left[
\begin{array}{c|c}
E& F\\
\hline
G&H
\end{array}
 \right],
\eea
where the blocks in the left upper and right bottom corners are called even (bosonic), and the blocks in the right upper and left bottom corners - odd (fermionic). If terms of supermatrices \eqref{supermtrs} the graded commutator is \cite{Beisert}
\bea
[X,Y\}=
\left[
\begin{array}{c|c}
AE+BG-EA+FC& AF+BH-EB-FD\\
\hline
CE+DG-GA-HC& CF+DH+GB-HD
\end{array}
\right].
\eea
The generalized canonical R--matrix for the supercoset can be constructed by a simple extension of (\ref{CanonYBslnG}). After choosing bosonic Cartan subalgebras for blocks $A$ and $B$ in (\ref{supermtrs}), we find the roots and the counterparts of the ladder operators $E_\alpha$ in (\ref{CartWeyl}), 
\bea\label{GradedCartWeyl}
[H_i,E_\alpha\}=\alpha_i E_\alpha,
\eea
but now some of $E_\alpha$ are fermionic. Direct calculation shows that the R--matrix
\bea\label{CanonYBsuMatr}
\mathcal{R}H_i={R_i}^j H_j,\quad \mathcal{R}E_\alpha=-i E_\alpha,\quad 
\mathcal{R}E_{-\alpha}=i E_{-\alpha}
\eea
solves the graded Yang-Baxter equation (\ref{GradedModYBeq}). Let us present an explicit solution for the 
 superalgebra $\mathfrak{psu(1,1|2)}$, which arises in description of strings on AdS$_2\times$S$^2$ \cite{AdS2PSU}. 

\bigskip 

The superalgebra $\mathfrak{psu(1,1|2)}$ is defined in terms of the $4\times 4$ supermatrices
\bea
\mathcal{M}=\left[
\begin{array}{c|c}
A& B\\
\hline
C&D
\end{array}
 \right]
\eea
subject to constraint
\bea
\left[\begin{array}{cc}
A&B\\C&D
\end{array}\right]=\left[\begin{array}{cc}
\Sigma A^\dagger \Sigma^{-1}&-i\Sigma C^\dagger\\-iB^\dagger \Sigma^{-1}&D^\dagger
\end{array}\right]\,,\qquad \Sigma=\mbox{diag}(1,-1).
\eea
Parameterizing such matrix as
\bea
\mathcal{M}=
\left[
\begin{array}{cc|cc}
F_1+F_4& F_2+i F_3& F_5+i F_6& F_7+i F_8\\
-F_2+i F_3& -F_1+F_4& F_9+i F_{10}& F_{11}+i F_{12}\\
\hline
-i F_5-F_6& i F_9+F_{10}& F_{13}+F_{16}& F_{14}+i F_{15}\\
-i F_7-F_8& iF_{11}+F_{12}& F_{14}-iF_{15}& -F_{13}+F_{16}
\end{array}
\right]
\eea
and choosing the canonical solution (\ref{CanonYBsuMatr}) with $R=0$, we find $6\times 2$ nonzero elements
\bea
\mathcal{R}_{23}=1,\quad \mathcal{R}_{14,15}=1,\quad
\mathcal{R}_{5,6}=\mathcal{R}_{7,8}=-\mathcal{R}_{9,10}=-\mathcal{R}_{11,12}=-i,\quad \mathcal{R}_{ab}=-\mathcal{R}_{ba}\,.
\eea
In the alternative parametrization of the $\mathfrak{psu(1,1|2)}$ matrix in terms of the holomorphic variables, which is often used in the literature \cite{HT},
\bea\label{HolomSuMatr}
\mathcal{M}=
\left[
\begin{array}{cc|cc}
F_1+F_4& F_2+i F_3& i F_8& F_5\\
-F_2+i F_3& -F_1+F_4& i F_{6}& F_{7}\\
\hline
i F_9& -i F_{11}& F_{13}+F_{16}& F_{14}+i F_{15}\\
F_{10}& -F_{12}& F_{14}-iF_{15}& -F_{13}+F_{16}
\end{array}
\right].
\eea
the R--matrix is
\bea\label{SuYangBaxt}
\mathcal{R}_{23}=1,\quad \mathcal{R}_{14,15}=1,\quad \mathcal{R}_{9,8}=\mathcal{R}_{5,10}=-\mathcal{R}_{11,6}=-\mathcal{R}_{7,12}=\frac{i}{2},\quad
\mathcal{R}_{ab}=-\mathcal{R}_{ba}\,.
\eea
Supercoset (\ref{HolomSuMatr}) has been used to construct the standard $\la$--deformation of strings on AdS$_2\times$S$^2$, and the generalized $\la$--deformation would be based on the solution  (\ref{SuYangBaxt}) of the modified Classical Yang--Baxter equation (\ref{GradedModYBeq}). However, before constructing such solutions one should prove that the resulting deformed supercoset leads to integrable theories, as was done for the standard $\la$ deformation in \cite{HMS}, and such analysis is beyond the scope of this work. In the remaining part of this chapter we will focus on bosonic cosets.

\section{SUGRA embeddings of the generalized $\lambda$--deformations}\label{SUGRAemb}

The general construction reviewed in section \ref{SectReview} gives the bosonic part of the string action  (\ref{gWZW}), (\ref{GenLambdaAction}) for the integrable $\la$--deformation, and in this section we will extract metric and the dilaton from these expressions. After introducing the general procedure in subsection 
\ref{SubsGenMetr}, we use it to derive the deformations of AdS$_2\times$S$^2$ and AdS$_3\times$S$^3$ in subsections \ref{SubsAdS2} and \ref{SubsAdS3}. As in the case of integrable deformations encountered earlier \cite{OtherEta,ST,SfetsosAdS5,HT,BTW, ChLLambda}, the Ramond--Ramond fluxes are recovered from solving the equations of motion of supergravity rather than from the fermionic part of the sigma model\footnote{It has been shown in \cite{BTW} that the extraction of the RR fluxes from the fermionic part of the sigma model is notoriously complicated.}.

\subsection{General construction}
\label{SubsGenMetr}

We begin with constructing the metric and the dilaton for deformations of arbitrary cosets $G/F$. To do so, we need three ingredients from section \ref{SectReview}:  the matrix $D_{ab}$, the left--invariant form $L$ parameterizing the coset, and the matrix $\hat{\lambda}^{-1}$ specifying the deformation. These ingredients are given by (\ref{lambdaMtrDef}) and (\ref{EmtrCosetA})\footnote{Most results of this subsection would apply to any matrix ${E}_G$, not only the one given in by (\ref{SmryDmatr}).}:
\bea\label{SmryDmatr}
&&D_{ab}=\mbox{Tr}(T_a g T_b g^{-1}), \quad
L_a
=i\mbox{Tr}(T_ag^{-1}d g),\nn
&&\hat{\lambda}^{-1}=(I-P){E}_G(I-P)+I,\quad
{E}_G=\frac{1}{\tilde{t}}(I-\tilde{\eta}\mathcal{R})^{-1}\,.
\eea
Here $P$ is the projector on the subgroup $F$, $\mathcal{R}$ is a solution of the modified Classical Yang--Baxter equation (\ref{YBmatrix}) satisfying the constraint (\ref{YBcosetConstr}), and $({\tilde t},{\tilde\eta})$ are free parameters. The authors of \cite{GenLambda} introduced two convenient parameters $(\la,\zeta)$ instead of $({\tilde t},{\tilde\eta})$,
\bea\label{EmtrYB}
\tilde{t}=\frac{\lambda}{(1-\lambda)},\quad \tilde{\eta}=-\frac{\zeta (2 \tilde{t}+1)}{2\tilde{t}},
\eea
and to compare with the existing literature, our final solution will be expressed in terms of $(\la,\zeta)$. Note, however, that the deformation depends on $(\la,\zeta)$ \textit{and} all free parameters appearing in the R--matrix, so the generalized $\la$--deformation can produce \textit{very large families} of integrable string theories.

The metric can be extracted from the symmetric part of the action (\ref{gWZW}), (\ref{GenLambdaAction})\footnote{Here we expressed everything in terms of $L$ using $R=DL$ and the orthogonality relation $D^TD=1$.}:
\bea\label{MetricA}
ds^2=\frac{k}{4\pi}L^T[I-\mathfrak{D}D-(\mathfrak{D}D)^T]L,\quad 
\mathfrak{D}\equiv[D-\hat{\lambda}^{-1}]^{-1}\,.
\eea
To rewrite this in terms of frames, we perform some algebraic manipulations which lead to
\bea\label{MetrPreFrame}
ds^2=\frac{k}{4\pi}L^T(\hat{\lambda}^{-1}-D)^{-1}[\hat{\lambda}^{-1}\hat{\lambda}^{-T}-I](\hat{\lambda}^{-1}-D)^{-T}L.
\eea
In the case of the isotropic deformation, where $\hat{\lambda}$ is proportional to the identity matrix, the expression in the square brackets is a constant, so the frames are given by
\bea
e=\sqrt{\frac{k(\lambda^{-2}-1)}{4\pi}}[\hat{\lambda}^{-1}-D]^{-T}L.
\eea
In general we begin with diagonalizing the symmetric matrix $\hat{\lambda}^{-1}\hat{\lambda}^{-T}$ using an orthogonal transformation $A$:
\bea\label{DefAmatr}
\hat{\lambda}^{-1}\hat{\lambda}^{-T}=A\Lambda^{-2}A^T,\quad AA^T=I,
\eea
then the metric (\ref{MetrPreFrame}) can be recovered from the frames
\bea\label{frames}
e=\sqrt{\frac{k}{4\pi}}\sqrt{\Lambda^{-2}-I}A^T[\hat{\lambda}^{-1}-D]^{-T}L.
\eea
Note that a general $n\times n$ matrix $\hat{\lambda}^{-1}$ can be parameterized in terms of a diagonal matrix $\Lambda$ and two orthogonal matrices $A$, $B$:
\bea\label{DgnlzLam}
\hat{\lambda}^{-1}=A{\Lambda}^{-1}B,\qquad AA^T=I,\quad BB^T=I\,,
\eea
and for computational purposes we will use a slightly different but equivalent expression for the frames:
\bea\label{simpleFrames}
e
&=&\sqrt{\frac{k}{4\pi}}\sqrt{I-\Lambda^2}[(I-D^T \hat{\lambda}^T) B^{-1} ]^{-1}L.
\eea

The dilaton is defined analogously to the regular $\lambda$-deformation \cite{ST}
\bea\label{Dilaton}
e^{-2\Phi}=e^{-2\Phi_0}\mbox{det}[\hat{\lambda}^{-1}-D].
\eea
One can also extract the Kalb--Ramond field by taking an antisymmetric part of the action (\ref{GenLambdaAction}), but such $B$ field vanishes in all our examples, so it will not be discussed further.

\bigskip
 
Expressions  (\ref{frames}) and (\ref{Dilaton}) have some remarkable properties which follow from the structure of matrices $D$ and ${\hat\la}$. As shown in the appendix,
\begin{quote}
For any coset $G/F$ there exists a canonical gauge, where matrix $\mathfrak{D}=[D-\hat{\lambda}^{-1}]^{-1}$ has three properties:
\begin{enumerate}[(i)]
\item matrix $(I-P)\mathfrak{D}(I-P)$ has constant entries;
\item matrix $\mathfrak{D}(I-P)$ factorizes as $\mathfrak{D}(I-P)=ST$, where $S$ does not depend on the deformation, and $T$ is a constant matrix;
\item the dependences upon coordinates and constant deformation parameters factorizes in 
$[\mbox{det}\, \mathfrak{D}]$.
\end{enumerate}
\end{quote}
The canonical gauge is defined by the commutation relations (\ref{CanonGauge}), and such gauge will be imposed throughout this chapter. We will now demonstrate that properties (i)--(iii) lead to drastic simplifications in the frames (\ref{frames}) and in the dilaton (\ref{Dilaton}). 

\bigskip

The implication for the dilaton is obvious: property (iii) ensures that the deformation parameters appear in (\ref{Dilaton}) only in a constant prefactor, and thus they can be absorbed into a shift of $\Phi_0$. For specific examples this property has been seen in \cite{SfetsosAdS5}, but the analysis presented in the appendix establishes the factorization in full generality. It is worth mentioning that in the case of the ordinary $\la$--deformation (i.e., for $\zeta=0$), the metric (\ref{MetricA}) can support two integrable string theories: one is based on the coset construction, and its dilaton is given by (\ref{Dilaton}) \cite{ST,SfetsosAdS5}, while the alternative is based on super--coset, and the resulting dilaton does not factorize between the coordinates and the deformation parameters \cite{HMS,HT,BTW,ChLLambda}. It would be very interesting to find the supercoset counterpart of  (\ref{Dilaton}) for nonzero $\zeta$, but such investigation is beyond the scope of this work. 

To find the implications of the properties (ii)--(iii) for the frames, we rewrite equation  (\ref{frames}) as
\bea\label{framesCurlD}
e=-\sqrt{\frac{k}{4\pi}}\sqrt{\Lambda^{-2}-I}A^T\mathfrak{D}^{T}L.
\eea
Recalling that $P\hat{\lambda}^{-1}=\hat{\lambda}^{-1}P=P$ (see (\ref{SmryDmatr})), we conclude that 
matrices $(A,B,\Lambda)$ in (\ref{DgnlzLam}) \textit{can be chosen} in such a way that\footnote{Since matrix  $\hat{\lambda}^{-1}$ has degenerate eigenvalues, relation (\ref{DefAmatr}) does not define $A$ uniquely. In addition, one has a freedom of permuting eigenvalues, and equation (\ref{ChooseABLa}) would be satisfied only for a particular ordering.} 
\bea\label{ChooseABLa}
PA=AP=P,\quad PB=BP=B,\quad \Rightarrow\quad P\Lambda=\Lambda P=P.
\eea
Introducing an explicit split between the generators of the subgroup $F$ and the coset $G/F$, one can rewrite (\ref{ChooseABLa}) more explicitly:
\bea
A=\left[\begin{array}{cc}
I&0\\
0&{\tilde A}
\end{array}\right],\quad
B=\left[\begin{array}{cc}
I&0\\
0&{\tilde B}
\end{array}\right],\quad
\Lambda=\left[\begin{array}{cc}
I&0\\
0&{\tilde \Lambda}
\end{array}\right].
\eea
Relations (\ref{ChooseABLa}) imply that 
\bea
\sqrt{\Lambda^{-2}-I}=(I-P)\sqrt{\Lambda^{-2}-I}(I-P),
\eea
then, using the property $P^T=P$, the frames (\ref{framesCurlD}) can be rewritten as
\bea
e=-\sqrt{\frac{k}{4\pi}}[I-P]\sqrt{\Lambda^{-2}-I}A^T\Big[\mathfrak{D}[I-P]\Big]^{T}L.
\eea
Application of the property (iii) leads to the final result:
\bea\label{CosetFrmFnl}
e=-\sqrt{\frac{k}{4\pi}}[I-P]\Big(\sqrt{\Lambda^{-2}-I}[TA]^T\Big)\Big(S^{T}L\Big)\,.
\eea
Equation (\ref{CosetFrmFnl}) has three distinct matrix factors. The first one ensures that frames point only along the coset directions. The second factor depends on the deformation, but not on the spacetime. The last factor gives the frames of the undeformed background, and it is not modified by the deformation. Thus application of the generalized $\la$--deformation (\ref{SmryDmatr}) simply rotates the frames by 
\textit{constant} matrices. This feature has been observed for several explicit examples \cite{ST,SfetsosAdS5}, but it is proven in full generality by the analysis presented here and in the Appendix.

\subsection{Deformation of AdS$_2\times$S$^2$} 
\label{sec:supergravity_background_for_ads2}
\label{SubsAdS2}

In this subsection we embed the generalized $\lambda$-deformation of $\frac{SU(2)}{U(1)}\times \frac{SU(1,1)}{U(1)}$ into the type IIB supergravity. First we discuss the coset 
$G/F\equiv SU(2)/U(1)$ corresponding to the sphere, and the AdS part of the geometry will be obtained by an analytic continuation. 

The embedding of $F=U(1)$ into $G=SU(2)$ is unique up to an SU(2) rotation, so without loss of generality we choose the generators of $F$ and $G/F$ as 
\bea\label{S2generat}
F:\{\sigma_3\}\,,\qquad G/F:\{\sigma_1,\sigma_2 \}\,.
\eea
A general element of SU(2) can be written as 
\bea
g=e^{i(\phi_1-\phi_2)\sigma_3/2}e^{i\omega\sigma_1}e^{i(\phi_1+\phi_2)\sigma_3/2}\,,
\eea
and the gauge freedom corresponding to U(1) is fixed by setting $\phi_2=0$. As discussed in the end of subsection \ref{SubsRgen}, the R--matrix for SU(2) is unique up to a global rotations parameterized by two Euler angles (see (\ref{RmtrSU2})), but since we have already chosen the embedding of $F$ into $G$, the deformations related by global rotations may not be equivalent. Since the rotation in $(\sigma_1,\sigma_2)$ plane does not distort the embedding (\ref{S2generat}), R--matrices  (\ref{RmtrSU2}) with different angles $\phi$ lead to equivalent deformations, but dependence on the parameter $\theta$ is nontrivial. Thus the most general deformation of the SU(2)/U(1) coset is parameterized by the R--matrix
\bea\label{RmatrAdS2}
\mathcal{R}=\left[ 
\begin{array}{ccc}
0& \cos\theta& \sin\theta\\
-\cos\theta & 0 & 0\\
-\sin\theta & 0 & 0
\end{array}
\right]\, .
\eea
We begin with discussion of the simplest deformation with $\theta=0$, and we will comment on the general case in the end of this subsection. The deformation matrix $\hat{\lambda}$ is evaluated using equations \eqref{SmryDmatr}, \eqref{EmtrYB} and the projector
\bea
P=\left[\begin{array}{ccc}
0& 0& 0\\
0& 0& 0\\
0& 0& 1
 \end{array}\right].
\eea
Then equation \eqref{simpleFrames} gives the explicit expression for the frames, and to simplify them, we introduce new coordinates $(p,q)$ following \cite{HT}:
\bea
\omega=\arccos\sqrt{p^2+q^2},\quad \phi_1=\arccos\frac{p}{\sqrt{p^2+q^2}}.
\eea
The frames become
\bea\label{FramesS2}
e^i&=&U^{i}{}_je^j_{(0)},\quad e^1_{(0)}=\sqrt{\frac{k}{2\pi(1-p^2-q^2)}}dp,\quad e^2_{(0)}=\sqrt{\frac{k}{2\pi(1-p^2-q^2)}}dq,\\
U^i{}_j&=&\frac{1}{\sqrt{(1-\lambda^2)(4\lambda^2+(1+\lambda)^2\zeta^2)}}\left[
\begin{array}{cc}
-(1+\lambda)(\zeta^2+\lambda(2+\zeta^2)) & \zeta(1-\lambda^2)\\
-(1-\lambda^2)\zeta & -2(1-\lambda)\lambda
\end{array}
 \right],\nonumber
\eea
where $i,j=1,2$. The metric and the $SU(2)$ contribution to the dilaton (see \eqref{Dilaton}) are
\bea\label{defS2metr}
2\pi k^{-1}ds_S^2&=&\frac{(1+\lambda)^2(1+\zeta^2)dp^2+2(1-\lambda^2)\zeta dp dq+(1-\lambda)^2dq^2}{(1-p^2-q^2)(1-\lambda^2)}\,,\\
\label{lambdaS2dil}
e^{-2\Phi_S}&=&1-p^2-q^2.
\eea
The AdS$_2$ counterparts of the metric and the dilaton are found by performing the analytic continuation which has been used in the case of the regular $\la$ deformation \cite{ST},
\bea
q\to i y,\quad p\to x,\quad k\to -k\,,
\eea
and the result is
\bea\label{defAdS2metr}
2\pi k^{-1}ds_{AdS}^2&=&-\frac{(1+\lambda)^2(1+\zeta^2)dx^2+2i(1-\lambda^2)\zeta dx dy-(1-\lambda)^2dy^2}{
(1-x^2+y^2)(1-\lambda^2)}\, .\nn
e^{-2\Phi_{AdS}}&=&-(1-x^2+y^2).
\eea
Note that the dilaton is real since we are working in the domain where $1-x^2+y^2<0$. 

The Ramond--Ramond fluxes can be found by solving the equations of motion for type IIB supergravity
\bea
&&\nabla^2 e^{-2\Phi}=0,\nn
&&\d_m \left(\sqrt{-g} F^{mn} \right)=0,\nn
&&R_{mn}+2\nabla_m\nabla_n \Phi=\frac{e^{2\Phi}}{2}\left( F_{mk}F_n{}^k-\frac{1}{4}g_{mn}F_{ij}F^{ij} \right),
\eea
and the result is\footnote{For example, one can start  for the $\lambda$-deformation, which corresponds to $\zeta=0$, and develop the perturbation theory in $\zeta$.} 
\bea\label{defAdS2flux}
F^{(2)}&=&c_1[S\zeta (dxdp-i dydq)-S^{-1}dxdq]+c_2[S\zeta(idxdp+dydq)+S^{-1}dydp],\nn
S&=&\sqrt{\frac{1-\lambda^2}{4\lambda+(1+\lambda)^2\zeta^2}},\quad c_1^2+c_2^2=\frac{2k}{\pi}.
\eea

Notice that the metric (\ref{defAdS2metr}) and the flux (\ref{defAdS2flux}) are complex unless $\zeta=0$. This is a peculiar feature of the generalized lambda deformation of AdS$_2\times$S$^2$, which does no persist for AdS$_3\times$S$^3$ (the metric and the fluxed are real there). Although the metric (\ref{defAdS2metr})  can be made real by an additional continuation of $y$ ($y\rightarrow i y$), this procedure is not very appealing since even the undeformed metric ($\la=\zeta=0$) has a wrong signature (2,2) and a wrong isometry SO(3)$\times$SO(3). Moreover, the fluxes remain complex. 

To compare the geometry (\ref{defS2metr}), (\ref{defAdS2metr}) with the standard lambda deformation constructed in \cite{ST}, we rescale coordinates by a convenient quantity \cite{HT}
\bea
\kappa=\frac{1-\la}{1+\la}
\eea
This leads to the solution
\bea\label{RescaledAdS2sln}
\frac{2\pi}{k}ds^2&=&\frac{dp^2+(dq+\zeta dp)^2}{1-\k p^2-\k^{-1} q^2}-
\frac{dx^2-(dy-i\zeta dx)^2}{1-\k x^2+\k^{-1}y^2}\\
F^{(2)}&=&c_1[S\zeta (\k dxdp-i\k^{-1} dydq)-S^{-1}dxdq]+
c_2[S\zeta(i\k dxdp+\k^{-1}dydq)+S^{-1}dydp]\nn
e^{2\Phi}&=&-\frac{1}{(1-\k p^2-\k^{-1} q^2)(1-\k x^2+\k^{-1}y^2)}\,,\nonumber
\eea
which generalizes the geometry ({2.7}) of \cite{BTW}.

For the standard $\la$ deformation (i.e., for $\zeta=0$), the AdS$_2\times$S$^2$ geometry is recovered in the limit of small $\k$ \cite{HT}, and application of such limit to (\ref{RescaledAdS2sln}) leads to a very simple $\zeta$--dependence after some shifts and rescaling of coordinates. Indeed, the leading order in $\k$ is
\bea
\frac{2\pi}{k\k}ds^2&=&-\frac{dp^2+(dq+\zeta dp)^2}{q^2}-
\frac{dx^2-(dy-i\zeta dx)^2}{y^2}\\
F^{(2)}&=&\frac{c_1}{\sqrt{\k}}[-i{\tilde S}\zeta dydq-{\tilde S}^{-1}dxdq]+
\frac{c_2}{\sqrt{\k}}[{\tilde S}\zeta dydq+{\tilde S}^{-1}dydp]\nn
e^{2\Phi}&=&\frac{\k^2}{q^2y^2}\,,\quad {\tilde S}=\frac{1}{\sqrt{1+\zeta^2}}\nonumber
\eea
In the new coordinates defined as 
\bea
{\tilde x}= \frac{1}{1+\zeta^2}\left[x+\frac{i\zeta y}{{1+\zeta^2}}\right],\quad 
{\tilde p}=\frac{1}{1+\zeta^2}\left[p+\frac{\zeta q}{{1-\zeta^2}}\right],
\eea
the metric and fluxes become real, and $\zeta$ appears only in the radius of the AdS$_2\times$S$^2$ and in the overall normalization of the fluxes:
\bea\label{LimitAdS2def}
\frac{2\pi}{k\k}ds^2&=&\frac{1}{1+\zeta^2}\left[-\frac{d{\tilde p}^2+dq^2}{q^2}-
\frac{d{\tilde x}^2-dy^2}{y^2}\right],\quad e^{2\Phi}=\frac{\k^2}{q^2y^2}\,,\nn
F^{(2)}&=&\frac{1+\zeta^2}{\sqrt{\k}}[-c_1d{\tilde x}dq+c_2 dyd{\tilde p}],\quad c_1^2+c_2^2=\frac{2k}{\pi}\,.
\eea
To summarize, the generalized $\la$--deformation of AdS$_2\times$S$^2$ is given by (\ref{RescaledAdS2sln}). For generic values of $\la$ and nonzero $\zeta$ the fluxes and metric are complex, if one insists on the correct signature. In the $\la=1$ limit one finds the 
real solution (\ref{LimitAdS2def}), and apart from a very simple $\zeta$ dependence, it coincides with analytic continuation of AdS$_2\times$S$^2$  discussed in \cite{BTW}. 

\bigskip

We conclude this subsection by writing the solution corresponding to the general R--matrix (\ref{RmatrAdS2}). To simplify the result, it is convenient to redefine the deformation parameters as
\bea
a&=&\frac{4\lambda^2+(1-\cos^2\theta(1-\lambda))(1+\lambda)^2\zeta^2}{4\lambda+(1-\cos^2\theta(1-\lambda))(1+\lambda)^2\zeta^2},\quad 
b=-\frac{2\cos\theta\lambda(1-\lambda^2)\zeta}{4\lambda+(1-\cos^2\theta(1-\lambda))(1+\lambda)^2\zeta^2},\nn
c&=&\frac{\lambda(4\lambda+(1+\lambda)^2\zeta^2)}{4\lambda+(1-\cos^2\theta(1-\lambda))(1+\lambda)^2\zeta^2},
\eea
This brings matrix $\hat{\lambda}$ into a simple form, 
\bea
\hat{\lambda}=\left[ 
\begin{array}{ccc}
a& -b& 0\\
b& c& 0\\
0&0& 1
\end{array}
\right].
\eea
and the deformed metric becomes
\bea
2\pi k^{-1} ds_S^2&=&\frac{(1+b^2+ac+a+c)dp^2+4b dp dq+(1+b^2+ac-a-c)dq^2}{(1-b^2-ac-a+c)(1-p^2-q^2)}.
\eea
The expressions for the fluxes are not very illuminating.

\subsection{Deformation of AdS$_3\times$S$^3$} 
\label{sec:supergravity_background_for_ads3}
\label{SubsAdS3}

In this subsection we construct SUGRA embedding of the generalized lambda-deformation based on the  coset
\bea
\frac{SU(2)\times SU(2)}{SU(2)_{diag}} \times \frac{SU(1,1)\times SU(1,1)}{SU(1,1)_{diag}}.
\eea
The element of the first coset can be conveniently parameterized as \eqref{CosetElementS3}
\bea\label{CosetElementS3GenLa}
g=\left(\begin{array}{cc}
g_l&0\\ 0&g_r
\end{array}\right),\qquad g^\dagger g=I
\eea
with
\bea\label{GandGprimeGenLa}
g_l=\left[\begin{array}{cc}
\alpha_0+i\alpha_3&\alpha_2+i\alpha_1\\
-\alpha_2+i\alpha_1&\alpha_0-i\alpha_3
\end{array}\right],\qquad
g_r=\left[\begin{array}{cc}
\beta_0+i\beta_3&\beta_2+i\beta_1\\
-\beta_2+i\beta_1&\beta_0-i\beta_3
\end{array}\right]\,.
\eea
The variables $\alpha_k, \beta_k$ introduced in \cite{ST} are subject to two constraints
\bea\label{DetConstS3}
\sum (\alpha_k)^2=1,\qquad \sum (\beta_k)^2=1.
\eea
Following \cite{ST}, we fix the gauge for $SU(2)_{diag}$ by setting
\bea
\alpha_2=\alpha_3=\beta_3=0,
\eea
and solve the constraints \eqref{DetConstS3} by introducing a convenient variable $\gamma$:
\bea\label{OurVariablesdS3}
 \beta_1\equiv \frac{\gamma}{\sqrt{1-\alpha_0^2}},\qquad\alpha_1=\sqrt{1-\alpha_0^2},\quad \beta_2=\sqrt{1-\beta_0^2-\frac{\gamma^2}{{1-\alpha_0^2}}}\,.
\eea
Note that the three remaining coordinates $\alpha\equiv \alpha_0$, $\beta\equiv \beta_0$ and $\gamma$ have the following ranges \eqref{Range1}:
\bea\label{Range1GenLa}
0<\alpha^2<1,\quad 0<\beta^2<1,\quad \gamma^2<(1-\alpha^2)(1-\beta^2)\,.
\eea
The generators corresponding to the subgroup and the coset are related to (\ref{GeneratorsSO4}) by a linear transformation:
\bea\label{SO4SO3gens}
F:&&T_a=\frac{1}{2}\left[\begin{array}{cc}
\sigma_a &0\\ 0&\sigma_a
\end{array}\right]=\frac{1}{2}[T^L_a+T^R_a],\quad a=1,2,3;\nn
G/F:&&T_\alpha=\frac{1}{2}\left[\begin{array}{cc}
\sigma_{\alpha-3} &0\\ 0&-\sigma_{\alpha-3}
\end{array}\right]=\frac{1}{2}[T^L_{\alpha-3}+T^R_{\alpha-3}],\quad \alpha=4,5,6.
\eea
In this basis the matrix $\mathcal{R}_1$ from (\ref{RmatricesSO4}) becomes
\bea
\mathcal{R}=\left[\begin{array}{cccccc} 
0& 0& 0& 0& -1 & 0\\
0& 0& 0& 1& 0& 0\\
0& 0& 0& 0& 0& a\\
0& -1& 0& 0& 0& 0\\
1& 0& 0& 0& 0& 0\\
0& 0& -a& 0& 0& 0\\
 \end{array} \right].
\eea
The deformation matrix $\hat{\lambda}$ is obtained from \eqref{SmryDmatr}, \eqref{EmtrYB}, where the projector on the subgroup is
\bea
P=\left[
\begin{array}{cc}
I_{3\times 3}&0\\
0&0
\end{array}
\right].
\eea
Evaluation of frames using \eqref{simpleFrames} gives
\bea\label{FramesS3}
e^4_{(0)}&=&-\frac{d\alpha}{\sqrt{1-\alpha^2}},\quad
e^5_{(0)}=\left[\frac{\gamma d\alpha+(1-\alpha^2)d\beta}{\gamma'\sqrt{1-\alpha^2}}\right],\quad
e^6_{(0)}=-\frac{\beta d\alpha+\alpha d\beta-d\gamma}{\gamma'},\nn
e^4&=&c_1e^4_{(0)},\quad e^5=c_1 e^5_{(0)},\quad e^6=c_2 e^6_{(0)},\\
c_1&=&\sqrt{\frac{k}{2\pi}}\sqrt{\frac{(1+\lambda)(\zeta^2+\lambda(2+\zeta^2))}{\lambda(1-\lambda)}},\quad c_2=\sqrt{\frac{k}{2\pi}}\sqrt{\frac{\lambda(1-\lambda)}{(1+\lambda)(2\lambda+a^2\zeta^2(1+\lambda))}}.\nonumber
\eea
where we defined
\bea
\gamma'&=&\sqrt{(1-\alpha^2)(1-\beta^2)-\gamma^2}.
\eea
Interestingly, the frames (\ref{FramesS3}) depend on $\la$ and $\zeta$ only through constant prefactors, exactly as it happened for the standard $\la$--deformation \cite{ST,SfetsosAdS5}. This feature is guaranteed by the general discussion presented in subsection \ref{SubsGenMetr}. Frames (\ref{FramesS3}) exhibit one more interesting feature\footnote{We thank Ben Hoare for making this observation.}: four parameters 
$(k,\la,a,\zeta)$ appear only through two independent combinations $(c_1,c_2)$. This implies that the generalized lambda deformation describes the same set of geometies as its standard counterpart \cite{ST,SfetsosAdS5}. It would be very interesting to see whether the same feature persists for other cosets. 

\bigskip

The AdS counterpart of (\ref{FramesS3}) is obtained by performing an analytic continuation \eqref{AnalContAdS3}
\bea\label{AnalContAdS3GenLa}
\alpha\to \tilde{\alpha},\quad \beta\to \tilde{\beta},\quad \gamma\to \tilde{\gamma},\quad k\rightarrow -k,
\eea
and changing the the range of coordinates from (\ref{Range1GenLa}) to \eqref{Range2}
\bea\label{Range2GenLa}
1<\tilde{\alpha}^2,\quad 1<\tilde{\beta}^2,\quad \tilde{\gamma}^2<(\tilde{\alpha}^2-1)(\tilde{\beta}^2-1).
\eea
Relation \eqref{Dilaton} gives the dilaton 
\bea\label{DilAdS3}
e^{-2\Phi}=e^{-2\Phi_0} \gamma'{\tilde\gamma}'\,,
\eea
and for the Ramond--Ramond fluxes, we take a simple ansatz inspired by the regular $\lambda$--deformation \cite{ST}:
\bea\label{F3ansatz}
F^{(3)}&=&C\gamma'\tilde{\gamma}'\left[e^3_{(0)}\wedge e^4_{(0)}\wedge e^5_{(0)}+e^1_{(0)}\wedge e^2_{(0)}\wedge e^6_{(0)} \right]\,.
\eea
Here $C$ is an unknown constant, which is determined by solving the equations of type IIB supergravity reduced to six dimensions:
\bea
&&\nabla^2 e^{-2\Phi}=0,\nn
&&\d_m \left(\sqrt{-g} F^{mnp} \right)=0,\nn
&&R_{mn}+2\nabla_m\nabla_n \Phi=\frac{e^{2\Phi}}{4}\left( F_{mkl}F_n{}^{kl}-\frac{1}{6}g_{mn}F_{ijk}F^{ijk} \right)\,.
\eea
The final answer is
\bea\label{FinalCAdS3}
C=\frac{k\sqrt{16\lambda^3+2(1+a^2)\lambda(1+\lambda)^3+a^2(1+\lambda)^4\zeta^4}\sqrt{\zeta^2+\lambda(2+\zeta^2)}}{4\pi(1-\lambda)\lambda\sqrt{2\lambda+a^2\zeta^2(1+\lambda)}}.
\eea
and in contrast to the deformation of $AdS_2\times S^2$, the solution (\ref{FramesS3}), (\ref{AnalContAdS3GenLa}), (\ref{F3ansatz}), (\ref{FinalCAdS3}) is real.


\section{Discussion}

In this chapter we have elaborated on the general procedure of constructing generalized $\la$--deformations of coset CFTs, and we have found several explicit solutions relevant for string theory. The main results of this chapter can be separated into three categories. 

In section \ref{Rmatrix} we found rather general solutions of the modified classical Yang--Baxter (mCYB) equation for arbitrary cosets and supercosets, and we also constructed the \textit{most general} R--matrices for the cosets arising in string theory. It would be very interesting to find the most general solutions of the mCYB for any (super)coset and to apply the results of our section \ref{SecGraded} toward generalizing the $\la$--deformation of supercosets discussed in \cite{HMS}.

The second category of our results concerns insights into the analytical structure of the generalized $\la$--deformations. In section \ref{SubsGenMetr} we demonstrated that under and \textit{arbitrary} deformation of an \textit{arbitrary} coset, the frames are rotated by a \textit{constant} matrix and the dilaton is multiplied by a \textit{constant} factor. These properties have been observed a-posteriori in several specific examples \cite{ST,SfetsosAdS5}, but our general proof allows one to drastically simplify calculations by focusing on the relevant constant matrices rather than evaluating coordinate--dependent frames.  

Finally, in sections \ref{SubsAdS2}, \ref{SubsAdS3} we constructed the generalized $\la$--deformations of AdS$_2\times$S$^2$ and AdS$_3\times$S$^3$, including the relevant Ramond--Ramond fluxes. Interestingly, while the solution corresponding to AdS$_3\times$S$^3$ is real, the deformation of AdS$_2\times$S$^2$ leads to complex metric and fluxes. It would be interesting to get a better analytical understanding of this phenomenon. In 
the AdS$_5\times$S$^5$ case we demonstrated that the construction introduced in \cite{GenLambda} does not lead to new solutions beyond the standard $\la$--deformation.

		\chapter{Summary of the results}\label{ChapterResults}

		Gauge/gravity duality relates two very different physical theories, in its most studied example the AdS/CFT correspondence, it relates string theory on five dimensional anti-de Sitter space to $\mathcal{N}$=4 supersymmetric Yang-Mills theory. This duality provides a powerful toolkit for studying two biggest puzzles of the modern theoretical physics: strongly coupled systems and quantum gravity, and connects seemingly unconnected objects such as black holes and quark gluon plasma. Many applications of string theory to strongly coupled systems comes from the fact that it is a strong-weak duality: when quantum fields are strongly interacting, their gravitational duals are weakly interacting, which are more tractable with the current mathematical tools. The duality also allows to apply string theory to many aspects of nuclear and condensed matter physics. Moreover, it works the other way around and provides a non-perturbative formulation of string theory on some particular backgrounds, which allows to use the well-developed tools such as conformal field theory to study quantum gravity, in particular black holes.

During the past 15 years rapid progress has been made in exploring the duality, and the key to this progress has been integrability of gravity/gauge pair of models. Integrability is a wonderful (and very rare in particle physics and gravity) property of some theories to be exactly solvable, which allows to determine spectrum and other observables without relying on the traditional perturbative expansions. It not only leads to drastic simplifications, but also allows to compute observables which were previously inaccessible by all practical means.

In this thesis we study integrability on the gravity side and it would be very interesting to understand what happens on the gauge theory side.

Chapters \ref{ChapterGeodesics}, \ref{ChapterKillingYano} and \ref{ChapterBlackRings} are mostly devoted to studying integrability (hidden symmetries) of the point--like particles, geodesics; and the dynamics of strings is investigated in Chapters \ref{ChapterLambda}, \ref{ChapterGenLambda}.

\section{Geodesics in D--brane backgrounds and Killing tensors}

Integrability is directly related to separability of equations of motion, so in Chapter \ref{ChapterGeodesics} we used this fact and studied separability of equations of motion governing the dynamics of massless (the Hamilton--Jacobi equation) and massive (the Klein--Gordon equation) geodesics on backgrounds produced by various configurations of (supersymmetric) D--branes. In particular we considered geometries produced by:
\begin{itemize}
\item a single type of D branes in Section \ref{SecGnrGds};
\item static configurations of D$p$ and D$p+4$ branes in Section \ref{DpDp4};
\item stationary D1--D5 branes in Section \ref{Rotations};
\item bubbling geometries in Section \ref{Bubbles}.
\end{itemize}
Surprisingly, it turned out that in all cases separability happened only in the elliptic coordinates. Moreover, in Section \ref{StdPar} we showed that the standard parametrization of $AdS_p\times S^q$ can be identified with the elliptic coordinates on the flat base. Our analysis concluded that there are no new integrable geometries produced by D branes and the search should be extended beyond supersymmetric cases. The results have been published in \cite{ChL}.

\bigskip

A logical extension of our analysis was to investigate non--supersymmetric backgrounds, the largest class of which is governed by the black holes solutions. Furthermore, it is known that the most convenient way of studying hidden symmetries of such geometries is to use so--called Killing tensors. In Chapter \ref{ChapterKillingYano} we followed this approach.

Killing tensors encode symmetries beyond isometries and provide a set of conserved quantities along with the coordinates in which the corresponding equations of motion (the Hamilton--Jacobi and Klein-Gordon equations) separate. Killing--Yano tensors, on the other hand, are connected to separability of the Dirac equation. In Section \ref{SectionKTKYTGeneral} we first analyzed general properties of Killing--(Yano) tensors and provided prescription for constructing conserved quantities starting from a single Killing tensor based on analysis of its eigenvalues. Next, we showed that Killing tensors in higher dimensions are always associated with ellipsoidal coordinates. Using the developed framework in Section \ref{SecMyersPerry} we wrote in a compact form the Killing--(Yano) tensors for rotating black holes in any dimensions. 

As a next step in Section \ref{SecKillingsDualities} we extended this analysis to charged black holes in string theory, which required analysis of Killing--(Yano) tensors (and Killing vectors) under string dualities. In particular we showed that:
\begin{itemize}
\item Killing vectors depending on the direction of T duality are usually destroyed by the duality;
\item Killing vectors with trivial dependence of the duality direction survive through the T duality;
\item Killing--(Yano) tensors are usually destroyed by T duality, moreover the Killing--Yano equation gets modified.
\end{itemize}
In Section \ref{SecExamplesF1NS5} we used this analysis to construct Killing--(Yano) tensors for the most general singly charged rotating black hole in any number of dimensions and several cases of charged rotating black holes with two charges. The results of Chapter \ref{ChapterKillingYano} were published in \cite{KYTinSt}.

Finally in Chapter \ref{ChapterBlackRings} we applied the framework of Killing tensors to study symmetries of black holes with non--spherical topology, the black rings. We showed that a peculiar symmetry of the five–dimensional black ring - separability of the base - cannot occur in dimensions higher than five. We also constructed supersymmetric solutions that have symmetries of 5D supersymmetric black ring and showed that they do not have event horizons. These results were published in \cite{Chervonyi:2015uua}.

\section{Integrable $\lambda$--deformations}

Another direction in constructing integrable string theories originated from the idea of deforming already known integrable backgrounds while preserving the desired hidden symmetries, the method of integrable deformations. This method has been very successful and resulted in constructing several very non--trivial integrable string theories, some of which are not only non--supersymmetric, but not possessing any isometries at all!

One of the recent developments, the $\lambda$--deformation interpolates between two integrable theories -- the Principal Chiral model and the Wess-Zumino-Witten model. Moreover there are two different ways of deforming the models: based on cosets and supercosets. Interestingly they give the same metric, but different configuration of other fields. While the coset-based deformations have been constructed some time ago, constructing the supercoset-based ones was technically challenging, and only one such deformation based on $AdS_2\times S^2$ supercoset has been constructed until not a long time ago. So in Chapter \ref{ChapterLambda} we constructed $\lambda$-deformation based on the $AdS_3\times S^3$ supercoset and made some progress in constructing the deformation based on the $AdS_5\times S^5$ supercoset. Out results were published in \cite{ChLLambda}.

Next, in Chapter \ref{ChapterGenLambda} we extended our analysis to so--called generalized $\lambda$-deformation, which deforms already deformed integrable backgrounds. In this way one obtains deformations depending on several continuous parameters. The first deformation is performed using so--called R-matrix and the second one, $\lambda$--deformation, imposes severe constraints on the R--matrix, so in Section \ref{Rmatrix} we investigated such constraints and extended the analysis to supersymmetric (graded) R--matrices, which might be used in constructing generalized $\lambda$--deformations based on supercosets.

Furthermore, in Section \ref{SubsGenMetr} we proved several interesting analytical properties of (generalized) $\lambda$--deformations, which have been conjectured in the literature before. We demonstrated that under arbitrary deformation of an arbitrary coset:
\begin{itemize}
\item frames composing the metric are rotated by a constant matrix,
\item the dilaton is multiplied by a constant factor.
\end{itemize}
These properties have been observed in several specific cases, but our proof allows to significantly simplify calculations by focusing on constant matrices rather than computing coordinate-dependent frames.

Finally, in Section \ref{SubsAdS2} and Section \ref{SubsAdS3} we constructed the generalized $\lambda$--deformations of $AdS_2 \times S^2$ and $AdS_3 \times S^3$ cosets, including the relevant Ramond–Ramond fluxes. Interestingly, while the solution corresponding to $AdS_3 \times S^3$ is real, the deformation of $AdS_2 \times S^2$ leads to complex metric and fluxes. It would be very interesting to get a better analytical understanding of this phenomenon.

		\appendix

		\chapter{Appendix for (Non)-Integrability of Geodesics}

\section{Examples of separable Hamilton--Jacobi equations}

In this appendix we provide some technical details pertaining to derivation of the results presented in sections \ref{Examples} and \ref{SecGnrGds}. In particular, we write down the explicit expressions for $S_L$ on the sphere (\ref{SLSphere}), the complete integral for $R$ in the elliptic coordinates (\ref{RActEll}) and make the connection between the holomorphic function introduced in (\ref{holomZ}) and the standard elliptic coordinates (\ref{DefEllCoord}).

\subsection{Motion on a sphere}\label{AppSphere} 

While discussing separation of variables in the Hamilton--Jacobi equation, on several occasions we have encountered equation
\bea\label{HJSphAp}
{h^{ij}}\frac{\d S^{(k)}_{L_k}}{\d y^i}
\frac{\d S^{(k)}_{L_k}}{\d y^j}=L_k^2,\qquad
\eea
on a $k$--dimensional sphere $S^k$ (see, for example, (\ref{AngHJ}), (\ref{HJSphere})). In this appendix we will write the complete integral of (\ref{HJSphere}) using induction. 

Writing the metric on $S^k$ as
\bea
h_{ij}dy^idy^j=dy_k^2+\sin^2 y_k d\Omega_{k-1}^2,
\eea
and splitting $S^{(k)}_{L}$ as
\bea
S^{(k)}_{L_k}(y_1,\dots,y_k)=F_k(y_k)+
S^{(k-1)}_{L_{k-1}}(y_1,\dots,y_{k-1}),
\eea
we can rewrite equation (\ref{HJSphAp}) as
\bea
&&(F'_k)^2+\frac{L^2_{k-1}}{\sin^2y_k}=L_k^2,\\
&&{h^{ij}}\frac{\d S^{(k-1)}_{L_{k-1}}}{\d y^i}
\frac{\d S^{(k-1)}_{L_{k-1}}}{\d y^j}=L_{k-1}^2,\qquad
\eea
Solving the first equation\footnote{Although the integral in (\ref{AnsFk}) can be performed, the result is not very illuminating.},
\bea\label{AnsFk}
F_k=\int \frac{dy_k}{\sin y_k}\sqrt{L_k^2 \sin^2y_k-L^2_{k-1}},
\eea
and applying induction, we arrive at the complete integral of (\ref{HJSphAp}) that depends on $k$ parameters $L_j$:
\bea\label{SLSphere}
S^{(k)}_{L_k}(y_1,\dots,y_k)=
\sum_{j=2}^{k}\int \frac{dy_j}{\sin y_j}\sqrt{L_j^2 \sin^2y_k-L^2_{j-1}}+
L_1y_1.
\eea
This explicit solution should be substituted into (\ref{HJact}).

\subsection{Two--center potential and elliptic coordinates}
\label{SprElliptic}

In section \ref{Examples} we reviewed separation of variables in elliptic coordinates, and 
here we present some details of that construction. 

The motion of a particle in the geometry produced by two stacks of Dp branes is governed by the Hamilton--Jacobi equation (\ref{EllHJ}) 
\bea\label{EllHJAp}
(\d_r R)^2+\frac{1}{r^2}(\d_\theta R)^2+\frac{L^2}{r^2\sin^2\theta}+ H p_\mu p^\mu=0,
\eea
where $H$ is given by (\ref{EllHarm})
\bea\label{EllHarmAp}
H=a+\frac{Q}{\rho_+^{6-p}}+\frac{Q}{\rho_-^{6-p}},\qquad
\rho_\pm=\sqrt{r^2+d^2\pm 2rd\cos\theta}
\eea
To rewrite (\ref{EllHJAp}) in terms the elliptic coordinates $(\xi,\eta)$ defined by (\ref{DefEllCoord}), we notice that 
(\ref{EllHJAp}) can be viewed as a Hamilton--Jacobi equation corresponding to an effective Lagrangian for $r$ and $\theta$:
\bea
{L}_{eff}={\dot r}^2+r^2{\dot\theta}^2-
\frac{L^2}{r^2\sin^2\theta}-H p_\mu p^\mu.
\eea
Rewriting the last expression in terms of $\xi$, $\eta$,
\bea
{L}_{eff}=d^2(\xi^2-\eta^2)\left[\frac{\dot\xi^2}{\xi^2-1}+\frac{\dot\eta^2}{1-\eta^2}\right]-
\frac{L^2}{d^2(\xi^2-1)(1-\eta^2)}-H p_\mu p^\mu
\eea
and going back to the Hamilton--Jacobi equation for $S(\xi,\eta)$, we find
\bea\label{HJell}
\frac{1}{d^2(\xi^2-\eta^2)}\left[(\xi^2-1)(\d_\xi R)^2+
(1-\eta^2)(\d_\eta R)^2+\left(\frac{1}{\xi^2-1}+\frac{1}{1-\eta^2}\right)L^2\right]+H p_\mu p^\mu=0.\nonumber\\
\eea
This equation separates in variables $(\xi,\eta)$ is and only if
\bea
(\xi^2-\eta^2)H=U_1(\xi)+U_2(\eta).
\eea
Rewriting the harmonic function (\ref{EllHarmAp}) in terms of elliptic coordinates, 
we find that the left hand side of the last expression,
\bea
(\xi^2-\eta^2)H=(\xi^2-\eta^2)\left[a+
\frac{Q}{[d(\xi+\eta)]^{6-p}}+\frac{Q}{[d(\xi-\eta)]^{6-p}}\right],
\eea
separates only for $p=5$. In this case equation (\ref{HJell}) becomes
\bea\label{SepHJell}
&&\left[(\xi^2-1)(\d_\xi R)^2+\frac{L^2}{\xi^2-1}+
p_\mu p^\mu(a\xi^2+2Q\xi)\right]
\nonumber\\
&&\qquad+
\left[(1-\eta^2)(\d_\eta R)^2+\frac{L^2}{1-\eta^2}
-p_\mu p^\mu a\eta^2\right]=0,
\eea
and its complete integral is
\bea\label{RActEll}
R&=&\int \frac{d\xi}{\sqrt{\xi^2-1}}\left[-\la-\frac{L^2}{\xi^2-1}+M^2(a\xi^2+2Q\xi)\right]^{1/2}\nonumber\\
&&+\int \frac{d\eta}{\sqrt{1-\eta^2}}\left[
\la-\frac{L^2}{1-\eta^2}
-M^2 a\eta^2
\right]^{1/2}
\eea
Here $\lambda$ is a separation constant and 
$M^2=-p_\mu p^\mu$. 

To embed the elliptic coordinates in the general framework presented in section \ref{SectPrcdr}, we have to find the holomorphic function $h(w)$ defined by (\ref{holomZ}). Comparing the kinetic terms in (\ref{HJell}) with equation (\ref{XYdiff}), we 
find the relation
\bea
\frac{1}{d^2(\xi^2-\eta^2)}\left[(\xi^2-1)(\d_\xi R)^2+
(1-\eta^2)(\d_\eta R)^2\right]=\frac{1}{A(x,y)}\left[(\d_x {\tilde R})^2+(\d_y {\tilde R})^2\right],\nonumber
\eea
which leads to expressions for $(\xi,\eta)$ in terms of $(x,y)$:
\bea\label{XiEta}
\xi=\cosh x,\qquad \eta=\cos y.
\eea
The relation between coordinates $(r,\theta)$ and $(x,y)$ looks rather complicated (see (\ref{DefEllCoord})),
\bea\label{coshOne}
&&\cosh x= \xi\equiv\frac{\rho_++\rho_-}{2d},\qquad \cos y=\eta\equiv\frac{\rho_+-\rho_-}{2d}\\
\label{coshTwo}
&&\rho_\pm=\sqrt{r^2+d^2\pm 2rd\cos\theta},
\eea
but it can be simplified by making use of the complex variables (\ref{defCmplVar}). 

First we rewrite equations (\ref{coshOne}) 
as an expression for $z$ in terms of $\xi$ and $\eta$:
\bea\label{zInterm}
&&e^x=\xi+\sqrt{\xi^2-1},\qquad e^{iy}=\eta+i\sqrt{1-\eta^2}\nonumber\\
&&z=x+iy=\ln\left[(\xi+\sqrt{\xi^2-1})(\eta+i\sqrt{1-\eta^2})\right]
\eea
Next we recall the definition (\ref{defCmplVar}) of the complex variable $w$ and use (\ref{coshTwo}) to write $\rho_\pm$ in terms of it:
\bea
\rho_+=\sqrt{(r+de^{i\theta})
(r+de^{-i\theta})}=d\sqrt{\left(\frac{l}{d}e^{\bar w}+1\right)
\left(\frac{l}{d}e^w+1\right)},\quad
\rho_-=d\sqrt{\left(\frac{l}{d}e^{\bar w}-1\right)\left(\frac{l}{d}e^w-1\right)}.\nonumber
\eea
To simplify the expressions for various ingredients appearing in (\ref{zInterm}) it is convenient to define holomorphic functions $W_\pm$:
\bea
W_+\equiv \sqrt{\frac{l}{d}e^w+1},\qquad W_-\equiv \sqrt{\frac{l}{d}e^w-1}.
\eea
Then we find
\bea
&&\xi=\frac{W_+{\overline{W}_+}+W_-{\overline{W}_-}}{2},
\qquad
\eta=\frac{W_+{\overline{W}_+}-W_-{\overline{W}_-}}{2}.\nonumber\\
&&\sqrt{\xi^2-1}=
\frac{W_-{\overline{W}_+}+W_+{\overline{W}_-}}{2},\qquad
\sqrt{1-\eta^2}=
\frac{W_-{\overline{W}_+}-W_+{\overline{W}_-}}{2i}.
\nonumber
\eea
Substitution of these results into (\ref{zInterm}) leads to the desired relation between $z$ and $w$:
\bea\label{EllEqnZ}
z&=&\ln\left[\frac{1}{4}(W_++W_-)^2((\overline{W}_+)^2-(\overline{W}_-)^2)
\right]\nonumber\\
&=&\ln\left[\frac{l}{d}e^w+
\sqrt{\frac{l^2}{d^2}e^{2w}-1}\right]
\eea
Finally, the asymptotic behavior (\ref{BoundCond}) determines $l$ in terms of $d$: $l=d/2$. 

As expected from the discussion in section \ref{SectPrcdr}, $z$ turns out to be a holomorphic function: 
\bea\label{EllEqnZpr}
z=h(w)=\ln\left[\frac{1}{2}\left\{e^w+
\sqrt{e^{2w}-4}\right\}\right].
\eea
Finally, by comparing (\ref{SepHJell}) with (\ref{XYdiff}), (\ref{XYalg}), we extract the expressions for the potentials $U_1(x)$ and $U_2(y)$:
\bea\label{EllPot}
U_1(x)&=&\frac{L^2}{\sinh^2 x}-
M^2(a\cosh^2 x+2Q\cosh x)\nonumber\\
U_2(y)&=&\frac{L^2}{\sin^2 y}+M^2 a\cos^2y.
\eea 

\bigskip

To summarize, in this section we demonstrated that the HJ equation (\ref{EllHJAp}) with $H$ given by (\ref{EllHarmAp}) separates in the elliptic coordinates (\ref{DefEllCoord}), (\ref{zInterm}), and the relevant potentials are given by (\ref{EllPot}). We also embedded the elliptic coordinates into the general discussion presented in section \ref{SectPrcdr} by deriving equation (\ref{EllEqnZpr}).

\section{Ellipsoidal coordinates}\label{AppEllips}

As demonstrated in section \ref{RedTo2D}, if the HJ equation (\ref{HJaa}) separates for $k>2$, such separation must occur in ellipsoidal coordinates, including degenerate cases. In this appendix we will demonstrate that a combination of the Laplace equation (\ref{LaplRk}) and the requirement (\ref{NoRotation}) rules out such separation. To avoid unnecessary complications, we will first give the detailed discussion of the $k=3$ case, and in section \ref{AppElpKG3} we will comment on minor changes which emerge from generalization to $k>3$.

\subsection{Ellipsoidal coordinates for $k=3$}

The ellipsoidal coordinates have been introduced by Jacobi \cite{Jacobi}, and there are several equivalent definitions. 
We will follow the notation of \cite{LLv8}. 

Ellipsoidal coordinates, $(x_1,x_2,x_3)$ are defined as three solutions of a cubic equation for $x$:
\bea
\frac{r_1^2}{x+a^2}+\frac{r_2^2}{x+b^2}+
\frac{r_3^2}{x+c^2}=1,
\eea
where $a>b>c>0$ and $r_1, r_2, r_3$ correspond to our $r_k$ from (\ref{BaseMetr}). The roots are arranged in the following order:
\bea\label{ElpsdRng}
x_1\ge-c^2\ge x_2\ge -b^2\ge x_3\ge -a^2
\eea
Coordinates $(r_1,r_2,r_3)$ can be expressed through $(x_1,x_2,x_3)$ by\footnote{We recall that $r_j$ must be non-negative if $d_j>0$ in (\ref{BaseMetr}), otherwise $r_j$ varies from minus infinity to infinity.}
\bea
r_1=\pm \left[\frac{(x_1+a^2)(x_2+a^2)(x_3+a^2)}{
(b^2-a^2)(c^2-a^2)}\right]^{1/2},\quad
r_2=\pm \left[\frac{(x_1+b^2)(x_2+b^2)(x_3+b^2)}{
(c^2-b^2)(a^2-b^2)}\right]^{1/2},\nonumber
\eea
\bea
r_3=\pm \left[\frac{(x_1+c^2)(x_2+c^2)(x_3+c^2)}{
(a^2-c^2)(b^2-c^2)}\right]^{1/2}.\nonumber
\eea
In terms of the elliptic coordinates, the metric of the flat three dimensional space becomes
\bea\label{IntroEpsd}
ds^2&=&dx_1^2+dr_2^2+dr_3^2=h_1^2 dx_1^2+h_2^2 dx_2^2+h_3^2 dx_3^2,\\
h_1^2&=&\frac{(x_1-x_2)(x_1-x_3)}{4R_1},\quad 
R_1=(x_1+a^2)(x_1+b^2)(x_1+c^2)\nonumber
\eea
Expressions for $h_2,h_3,R_2,R_3$ are obtained by making a cyclic permutation of indices. To simplify the formulas appearing below, it is convenient to introduce $d_{ij}\equiv x_i-x_j$. 

In ellipsoidal coordinates equation (\ref{HJaa}) becomes
\bea\label{EllisoidHJ}
\frac{4}{d_{12}d_{13}d_{23}}
\left[
d_{23}R_1(\d_1 S)^2+d_{31}R_2(\d_2 S)^2+
d_{12}R_3(\d_3 S)^2\right]+\sum_{j=1}^3 \frac{L_j^2}{r_j^2}+Hp_\mu p^\mu=0.
\eea
Before imposing the Laplace equation (\ref{LaplRk}) we will demonstrate that separation requires that $d_1=d_2=d_3=0$ in (\ref{BaseMetr}). Indeed, the separation implies that
\bea\label{AsmpSpr}
S(x_1,x_2,x_3)=S_1(x_1)+S_2(x_2)+S_3(x_3).
\eea
Then multiplying equation (\ref{EllisoidHJ}) by $d_{12}d_{13}d_{23}$ and applying 
$\d_1^2\d_2^2$ to the result, we find a relation which does not involve $S$:
\bea\label{IntCndH}
\d_1^2\d_2^2\left[d_{12}d_{13}d_{23}\left\{
\sum_{j=1}^3 \frac{L_j^2}{r_j^2}+Hp_\mu p^\mu
\right\}\right]=0.
\eea
Let us assume that $d_1>0$ in (\ref{BaseMetr}), then the last relation must hold for all values of $L_1$, thus 
\bea
\d_1^2\d_2^2\left[\frac{d_{12}d_{13}d_{23}}{r_1^2}
\right]=\d^2_1\d_2^2\left[\frac{(b^2-a^2)(c^2-a^2)d_{12}d_{13}d_{23}}{(x_1+a^2)(x_2+a^2)(x_3+a^2)}\right]=0.
\eea
This condition can only be satisfied if $a=b$, this degenerate case corresponds to oblate spheroidal coordinates:
\bea\label{ProlSpher}
r_1+ir_2=\left[\frac{(x_1+a^2)(x_2+a^2)}{
a^2-c^2}\right]^{1/2}e^{i\phi},\qquad
r_3=\pm \left[\frac{(x_1+c^2)(x_2+c^2)}{
c^2-a^2}\right]^{1/2}
\eea 
We will now demonstrate that separation of the HJ equation in spheroidal coordinates implies that $\d_\phi H=0$, i.e., violation of (\ref{NoRotation}) for $i=1,j=2$. This will falsify our assumption $d_1>0$, and similar arguments will show that separability of the HJ equation requires $d_2=d_3=0$.

To simplify notation and to connect to our discussion in section \ref{SectPrcdr} it is convenient to use an alternative form of the oblate spheroidal coordinates (\ref{ProlSpher}):
\bea
x_1=(a^2-c^2)\cosh^2 x-a^2,\quad 
x_2=(a^2-c^2)\cos^2 y-a^2
\eea 
This gives expressions for the radii:
\bea
r_1+ir_2=A\cosh x \cos y e^{i\phi},\quad r_3=\pm A\sinh x\sin y,\quad A=\sqrt{a^2-c^2},
\eea
and the metric becomes
\bea
ds^2=A^2(\sinh^2 x+\sin^2 y)(dx^2+dy^2)+A^2\cosh^2 x\cos^2 y d\phi^2
\eea
Notice that these are precisely the elliptic coordinates found in section \ref{SectPrcdr}. The HJ equation (\ref{HJaa}) in coordinates $(x,y)$ has the form
\bea\label{HJaaxy}
\frac{(\d_x S)^2+
(\d_y S)^2}{A^2(\sinh^2 x+\sin^2 y)}+
\frac{1}{A^2\cosh^2 x\cos^2 y}(\d_\phi S)^2
+\sum_j^3\frac{L_j^2}{r_j^2}+ H p_\mu p^\mu=0,
\eea
and it does not separate unless 
\bea
H\cosh^2 x\cos^2 y=X(x,y)+\Phi(\phi)
\eea
Recalling that at large values of $x$ function $H$ does not depend on $\phi$ (see (\ref{Hasymt})), we conclude that $\Phi'(\phi)=0$, then $\d_\phi H=0$ everywhere. As already mentioned, this relation violates the condition (\ref{NoRotation}), so our assumption $d_1>0$ was false. Repeating this arguments for the remaining two spheres, we conclude that $d_1=d_2=d_3=0$ in (\ref{BaseMetr}) and (\ref{LaplRk}).

\bigskip

We will now combine the integrability condition (\ref{IntCndH}),
\bea\label{IntCndHa}
\d_i^2\d_j^2\left[d_{12}d_{13}d_{23}H\right]=0,
\eea
with Laplace equation (\ref{LaplRk}) to rule out separability of the HJ equation in ellipsoidal coordinates,
\bea\label{ElpsdHJ}
\frac{4}{d_{12}d_{13}d_{23}}
\left[
R_1d_{23}(\d_1 S)^2+d_{31}R_2(\d_2 S)^2+
d_{12}R_3(\d_3 S)^2\right]+Hp_\mu p^\mu=0.
\eea
Since we have already established that $d_1=d_2=d_3=0$, equation (\ref{LaplRk}) reduces to the Laplace equation on a flat three--dimensional space formed by $(r_1,r_2,r_3)$, and it can be rewritten in terms of ellipsoidal coordinates using (\ref{IntroEpsd}):
\bea\label{ElpsdLapl}
\frac{4}{d_{12}d_{31}d_{23}}\left[
d_{23}\sqrt{R_1}\d_1(\sqrt{R_1}\d_1 H)+d_{31}\sqrt{R_2}\d_2(\sqrt{R_2}\d_2 H)+
d_{12}\sqrt{R_3}\d_3(\sqrt{R_3}\d_3 H)\right]=0.
\nonumber\\
\eea
In the remaining part of this subsection we will demonstrate that equations (\ref{IntCndHa}), (\ref{ElpsdLapl}), (\ref{NoRotation}), (\ref{Hasymt}) are inconsistent with separability of (\ref{ElpsdHJ}).

Let us assume that the HJ equation (\ref{ElpsdHJ}) separates. Although the ellipsoidal coordinates are restricted by (\ref{ElpsdRng}) the formal separation (\ref{AsmpSpr}) must persist beyond this range. In particular, it is convenient to look at the limit $x_1\rightarrow x_2$, where $x_2$ is kept as a free parameter. The Laplace equation (\ref{ElpsdLapl}) guarantees that $H$ remains finite in this limit, then equation (\ref{ElpsdHJ}) reduces to
\bea\label{x1ISx2}
\left[R_1(S_1')^2-R_2(S_2')^2\right]_{x_1=x_2}=0
\eea
This implies that
\bea
S_1(x_1)=F(x_1),\qquad S_2(x_2)=F(x_2)
\eea
with the same function $F$. Repeating this argument for $x_1=x_3$, we conclude that 
\bea
S_3(x_3)=F(x_3).
\eea
Rewriting (\ref{ElpsdHJ}) in terms of $F$, we conclude that $H$ must be symmetric under interchange of its arguments\footnote{Of course, such formal interchange takes us outside of the physical range (\ref{ElpsdRng}).}, and combination of this symmetry with integrability conditions (\ref{IntCndHa}) leads to severe restrictions on the form of $H$.

Taking $(i,j)=(1,2)$ in (\ref{IntCndHa}), and solving the resulting equation, we find
\bea
H=\frac{1}{d_{12}d_{13}d_{23}}\left[G_1(x_1,x_3)x_2+G_2(x_2,x_3)x_1
+G_{01}(x_1,x_3)+G_{02}(x_2,x_3)\right],
\eea
where $G_1,G_2,G_{01},G_{02}$ are some undetermined functions of their arguments. Applying (\ref{IntCndHa}) with different values of $(i,j)$, we find further restrictions on the form of $H$:
\bea
H=\frac{1}{d_{12}d_{13}d_{23}}\left[
\sum_{i\ne j}G_{ij}(x_i)x_j+\sum_i G_{0i}(x_i)\right]
\eea
Symmetry of $H$ under interchange of any pair of coordinates implies that $G_{0i}=0$, matrix $G_{ij}$ is anti-symmetric, 
and all $G_{ij}$ can be reduced to a single function\footnote{Specifically, $G_{12}=-G_{21}=G_{23}=-G_{32}=G_{31}=-G_{13}$ so it is convenient to introduce $G\equiv G_{12}$.} $G$:
\bea\label{HarmElpsd}
H&=&\frac{1}{d_{12}d_{13}d_{23}}\left[
G(x_1)(x_2-x_3)+G(x_2)(x_3-x_1)+G(x_3)(x_1-x_2)
\right]\nonumber\\
&=&\frac{G(x_1)}{d_{12}d_{13}}+
\frac{G(x_2)}{d_{21}d_{23}}+\frac{G(x_3)}{d_{31}d_{32}}.
\eea
Notice that constant and linear functions $G$ lead to $H=0$, and quadratic $G$ gives constant $H$. 

Substitution of (\ref{HarmElpsd}) into (\ref{ElpsdLapl}) leads to a complicated equation for function $G$:
\bea\label{ElpsdLaplA}
&&\frac{R_1 d_{23}}{d_{12}d_{13}}G_1''-2\frac{d_{23}}{d^2_{12}d^2_{13}}(d_{12}+d_{13})R_1G_1'+
\frac{d_{23}}{2d_{12}d_{13}}R_1'G_1'\nonumber\\
&&+\frac{d_{23}}{d_{12}^2d_{13}^2}\left[3(R_1-R_0)-(x_1x_2+x_1x_3+x_2x_3)\frac{R_0''}{2}-R'_0(x_1+x_2+x_3)-
3x_1x_2x_3\right]G_1\nonumber\\
&&+perm=0
\eea
Here we used a shorthand notation: $R_0=R(0)$, $R'_0=R'(0)$, $R''_0=R''(0)$, $R_1=R(x_1)$, $G_1=G(x_1)$. Equation 
(\ref{ElpsdLaplA}) should work around $x_2=x_3$ as long as $x_1$ is sufficiently large\footnote{All sources of the harmonic function are localized in some finite region of space, which cannot protrude to large $x_1$, which is analogous to radial coordinate.},  so it can be expanded in powers of $d_{23}$. Taking the leading piece proportional to $d_{23}$, multiplying by $d_{12}^3$, differentiating the result four times with respect to $x_1$, we find a closed--form equation for $F(x_1)\equiv G'''_1$:
\bea\label{FFFeqn}
10R'''F+15 R_1''F'+9R_1'F''+2R_1 F'''=0.
\eea
To analyze this equation, it is convenient to set $c=0$ by shifting $x_i$, $a^2$, $b^2$ by $-c^2$ (see definition (\ref{IntroEpsd})), and
set $a=1$ by rescaling $x_i$. The resulting equation (\ref{FFFeqn}) has a general solution 
\bea\label{ElpsdFb}
F(x)=\frac{(b^2+2x+x^2)(b^2-x^2)(b^2+2xb^2+x^2)}{x^{5/2}(b^2+x)^{5/2}(1+x)^{5/2}}
\left(C_1+C_2 I_2+C_3 I_3\right),
\eea
where
\bea
I_2&=&\int_{q}^x \frac{x^{3/2}(1+x)^{3/2}(b^2+x)^{3/2}}{(x^2+b^2+2xb^2)^2(b^2+2x+x^2)^2}dx,
\nonumber\\
I_3&=&\int_{q}^x
\frac{x^{3/2}(1+x)^{3/2}(b^2+x)^{3/2}}{(x^2+b^2+2xb^2)^2(b^2+2x+x^2)^2}\frac{x(1+x)(b^2+x)}{(b^2-x^2)^2}dx
\eea
The low limit of integration, $q$, will be defined in a moment. 
Function $H$ should remain finite and smooth as long as $x_1$ is sufficiently large, this implies that functions $G(x_2)$ and $G(x_3)$ must be finite for all 
$0\ge x_2\ge-b^2\ge x_3\ge -1$ (recall the region (\ref{ElpsdRng}) and our normalization $c=0$, $a=1$), so function $G(x)$ must be well--defined for all $0>x>-1$. This gives restrictions on $C_2$ and $C_3$. 

We will now demonstrate that function $G(x)$ cannot remain finite for $x<-b$ and $x>-b$ unless $C_3=0$. 
Let us choose $q=-b-\eps$ and assume that function $G(p)$ is
finite. Recalling the definition of $F$ ($F(x)=G'''(x)$), we find
\bea
G(x)=\int_q^x dy_1\int_q^{y_1} dy_2
\int_q^{y_2} dy_3 F(y_3)+G(q)+(x-q)G'(q)+
\frac{1}{2}(x-q)^2G''(q)
\eea
As $x$ changes from $-b-\eps$ to $-b+\delta$, the last three terms as well as contributions proportional to $C_1$ and $C_2$ remain finite, so we focus on the term containing $C_3$:
\bea\label{G3Elpsd}
G_3(x)&=&C_3\int_q^x dy_1\int_q^{y_1} dy_2
\int_q^{y_2} dy_3
(b+y_3)f(y_3)I_3(y_3)\nonumber\\
&=&C_3\int_q^x dy_1\int_q^{y_1} dy_2
\int_q^{y_2} dy_3
(b+y_3)f(y_3)\int_q^{y_3}\frac{g(y)}{(b+y)^2}
\eea
Here we introduced two functions, 
\bea
f(x)&=&\frac{(b^2+2x+x^2)(b-x)(b^2+2xb^2+x^2)}{x^{5/2}(b^2+x)^{5/2}(1+x)^{5/2}},\nonumber\\
g(x)&=&\frac{x^{3/2}(1+x)^{3/2}(b^2+x)^{3/2}}{(x^2+b^2+2xb^2)^2(b^2+2x+x^2)^2}\frac{x(1+x)(b^2+x)}{(b+x)^2},
\eea
which remain finite and non--zero in some vicinity of $x=-b$. Changing the order of integration in (\ref{G3Elpsd}), we find 
\bea
G_3(x)&=&C_3\int_q^{x}\frac{g(y)}{(b+y)^2}dy
\int_y^{x} dx_3
(b+y_3)f(y_3)\int_{y_3}^{x} dy_2\int_{y_2}^{x} dy_1
\nonumber\\
&=&C_3\int_q^{x}\frac{g(y)}{(b+y)^2}dy
\int_y^{x} dy_3
(b+y_3)f(y_3)\frac{(x-y_3)^2}{2}
\eea
The integral in the right--hand side does not make sense at $x>-b$ unless 
\bea\label{IntEqlZero}
\int_{-b}^{x} dy_3
(b+y_3)f(y_3)\frac{(x-y_3)^2}{2}=0,
\eea
which is clearly not the case. We conclude that $G_3(x)$ (and thus $G(x)$) is ill--defined at $x>-b$ unless $C_3=0.$\footnote{ 
Although a simpler analysis shows that $I_3$ is ill--defined for $x>-b$, this condition by itself is not sufficient to rule out  $C_3$: in particular, had function $f$ satisfied (\ref{IntEqlZero}), $G_3(x)$ would have existed for $x>-b$, even though $I_3$ would have not.} Once $C_3$ is set to zero, a similar analysis in a vicinity of $x=\sqrt{1-b^2}-1$ leads to $C_2=0$, and integrating the resulting $F(x)$, we find 
\bea
G(x)=C_1\sqrt{x(x+b^2)(x+1)}+C_3x^2+C_4 x+C_5
\eea
Putting back $a$ and $c$, we  can rewrite the last expression as
\bea\label{ElpsdSlnG}
G(x)=C_1\sqrt{(x+a^2)(x+b^2)(x+c^2)}+C_3x^2+C_4 x+C_5,
\eea
This function gives the most general solution of (\ref{FFFeqn}) consistent with physical requirements imposed on $G$,
and direct substitution of (\ref{ElpsdSlnG}) into (\ref{ElpsdLaplA}) shows that the corresponding function $H$ (see (\ref{HarmElpsd})) is harmonic. 

Although solution (\ref{ElpsdSlnG}) is well-defined everywhere (unlike contributions proportional to $C_2$ and $C_3$), the corresponding harmonic function (\ref{HarmElpsd}) has singular points at arbitrarily large $x_1$ unless $C_1=0$. Indeed, consider (\ref{HarmElpsd}) near $x_2=-b^2,x_3=-b^2$ for large $x_1$:
\bea
H
&=&\frac{G(x_1)}{d_{12}d_{13}}+
\frac{G(x_2)}{d_{21}d_{23}}+\frac{G(x_3)}{d_{31}d_{32}}
\nonumber\\
&=&
C_1\frac{\sqrt{(b^2-c^2)(b^2-a^2)}}{d_{21}d_{23}}\left[\sqrt{x_2+b^2}-\sqrt{x_3+b^2}\right]+
finite\nonumber\\
&=&C_1\frac{\sqrt{(b^2-c^2)(b^2-a^2)}}{d_{21}d_{23}}
\frac{d_{32}}{\sqrt{x_2+b^2}+\sqrt{x_3+b^2}}+finite\\
&=&C_1\frac{\sqrt{(b^2-c^2)(b^2-a^2)}}{d_{21}(\sqrt{x_2+b^2}+\sqrt{x_3+b^2})}+finite\nonumber
\eea
This harmonic function diverges as $x_2$ and $x_3$ approach $-b^2$ unless $C_1=0$. 

\bigskip

To summarize, we have demonstrated that separation of the HJ equation in ellipsoidal coordinates with $k=3$ implies that the transverse space is three--dimensional, and the harmonic function is given by (\ref{HarmElpsd}) with 
\bea
G(x)=C_3x^2+C_4 x+C_5. 
\eea
Such harmonic function, $H=C_3$, does not satisfy the required boundary condition (\ref{Hasymt}), so ellipsoidal coordinates for $k=3$ do not no separate the HJ equation for geodesics in D--brane backgrounds. In the next subsection we will outline the extension of this result to $k>3$. 

\subsection{Ellipsoidal coordinates for $k>3$}
\label{AppElpKG3}

After giving a detailed description of ellipsoidal coordinates for $k=3$, we will briefly comment on the extension of the results obtained in the last subsection to $k>3$. 

The ellipsoidal coordinates in higher dimensions were introduced in \cite{Jacobi}\footnote{We use a slightly modified notation to connect with discussion in the last subsection.}:
\bea
r_i^2=-\left[\prod_j(a_i^2+x_j)\right]\left[\prod_{j\ne i}\frac{1}{(a_j^2-a_i^2)}\right].
\eea
Here $r_i$ are the $k$ radii introduced in (\ref{BaseMetr}), and $x_i$ are ellipsoidal coordinates corresponding to $k$ root of an algebraic equation
\bea\label{EqnElpsd}
\sum_{i=1}^k\frac{r_k^2}{x+a_k^2}=1,
\eea
The ranges of ellipsoidal coordinates are analogous to (\ref{ElpsdRng}) in the three--dimensional case:
\bea
x_1\ge -a_1^2\ge x_2\ge -a_2^2\ge \dots x_k\ge -a_k^2.
\eea 
In terms of the ellipsoidal coordinates, the radial part of the metric (\ref{BaseMetr}),
\bea
ds^2_r=dr_1^2+\dots+dr_k^2,
\eea
becomes
\bea
ds_r^2=\sum_{i=1}^k h_i^2 (dx_i)^2,\qquad h_i^2\equiv\frac{1}{4}\left[\prod_{j\ne i}(x_i-x_j)\right]\left[\prod_j\frac{1}{a_j^2+x_i}\right]
\eea
To simplify some formulas appearing below, it is convenient to 
define
\bea
D_i=\prod_{j\ne i}(x_i-x_j),\qquad 
R_i=\prod_j(a_j^2+x_i),\qquad D=\prod_{i< j}(x_i-x_j).
\eea

A counterpart of the HJ equation (\ref{EllisoidHJ}) for $k>3$ is 
\bea\label{ElpsNHJ}
4\sum_{i=1}^k\frac{R_i}{D_i}(\d_i S)^2+\sum_{j=1}^k \frac{L_j^2}{r_j^2}+Hp_\mu p^\mu=0.
\eea
Assuming that this equation separates in ellipsoidal coordinates,
\bea
S(x_1,\dots,x_k)=S_1(x_1)+\dots+S_k(x_k),
\eea
and applying derivatives to (\ref{ElpsNHJ}), we find one of the integrability conditions (cf. (\ref{IntCndH})):
\bea\label{ElpsNInt}
\d_i^{n-1}\d_j^{n-1}\left[D\left\{\sum_{l=1}^k \frac{L_l^2}{r_l^2}+Hp_\mu p^\mu\right\}\right]=0
\eea

As before, this relation can be used to show that $d_1=d_2=\dots=d_k=0$ in (\ref{BaseMetr}). Indeed, for  $d_1>0$, equation (\ref{ElpsNInt}) must be satisfied for all values of $L_1$, then
\bea
\d_1^{n-1}\d_2^{n-1}\left[\frac{D}{r_1^2}\right]=
-\left[\prod_{j>1}(a_j^2-a_1^2)\right]
\d_1^{n-1}\d_2^{n-1}\left[\frac{D}{R_1}\right]=0
\eea
The last relation is false, and the easiest way to see this is to notice that the leading contribution in the vicinity of 
$x_1=-a_1^2$ comes when all $\d_1$ derivatives hit $(x_1+a_1^2)^2$, giving $n$--order pole, however, the remaining $\d_2$ derivatives do not kill $D$. As before the degenerate case (e.g., $a_2=a_1$) requires a separate consideration, and it can be eliminated using the arguments that followed equation (\ref{ProlSpher}). 

Next, we demonstrate that functions $S$ and $H$ must have a formal symmetry under interchange of their arguments. Indeed, taking the limit $x_1\rightarrow x_2$ in (\ref{ElpsNHJ}), we find a counterpart of (\ref{x1ISx2}):
\bea
\left[\frac{R_1d_{12}}{D_1}(S_1')^2-\frac{R_2d_{21}}{D_2}(S_2')^2\right]_{x_1=x_2}=0
\eea
This and other similar limits imply that
\bea
S_1(x_1)=F(x_1),\quad S_2(x_2)=F(x_2),\quad\dots\quad S_k(x_k)=F(x_k),
\eea
then equation (\ref{ElpsNHJ}) ensures the symmetry of $H$.\footnote{Recall that we already established that $d_1=d_2=\dots=d_k=0$, so angular momenta disappear from 
(\ref{ElpsNHJ}).}

Modifying the arguments that led to (\ref{HarmElpsd}), we conclude that
\bea\label{ElpsNHarm}
H=\sum \frac{G(x_i)}{D_i}.
\eea
Indeed, integrability condition (\ref{ElpsNInt}) for $(i,j)=(1,2)$ gives
\bea
H=\frac{1}{D}\left[
\sum_{l=0}^{k-2}C_l(x_2,x_3\dots x_k)x_1^l
-\sum_{l=0}^{k-2}{\tilde C}_l(x_1,x_3\dots x_k)x_2^l
\right].
\eea
Symmetry of $H$ implies that $C_l$ and ${\tilde C}_l$ have the same functional form. Repeating this argument for other pairs $(i,j)$, we find
\bea
H=\frac{1}{D}\sum_j\sum_{l=0}^{(k-2)(k-1)} 
{G}_l(x_j)P_l\left[x_1\dots x_{j-1},x_{j+1}\dots x_k\right],
\eea 
where $P_l$ is a polynomial of $l$--th degree, which is anti-symmetric under interchange of any pair of its arguments. Any such polynomial is proportional to
\bea
{\tilde P}\left[y_1\dots \dots y_{k-1}\right]\equiv\prod_{i<j}
(y_i-y_j),
\eea
which already has degree $(k-1)(k-2)$, so 
$P_{(k-1)(k-2)}={\tilde P}$ and using the relation 
\bea
{\tilde P}\left[x_1\dots x_{j-1},x_{j+1}\dots x_k\right]D_j=
(-1)^{j+1}D,\nonumber
\eea
we find
\bea
H=\frac{1}{D}\sum_j {G}_{(k-2)(k-1)}(x_j)\left\{c{\tilde P}\left[x_1\dots x_{j-1},x_{j+1}\dots x_n\right]\right\}=
\sum_j \frac{G(x_j)}{D_j}
\eea
This proves (\ref{ElpsNHarm}).

Since we have already established that $d_1=d_2=\dots=d_k=0$, it becomes easy to write the Laplace equation (\ref{LaplRk}) in ellipsoidal coordinates (cf. (\ref{ElpsdLapl})):
\bea\label{ElpsNLpl}
\sum_{i=1}^k\frac{\sqrt{R_i}}{D_i}\d_i\left[
\sqrt{R}_i\d_i H\right]=0.
\eea
Substitution of (\ref{ElpsNHarm}) into (\ref{ElpsNLpl}) leads to a counterpart of equation (\ref{ElpsdLaplA}):
\bea\label{ElpsdLaplAN}
&&\frac{R_1}{D_1^2}G_1''-2\frac{R_1}{D_1^2}\left[
\frac{1}{d_{12}}+\frac{1}{d_{13}}+\dots+
\frac{1}{d_{1k}}\right]G_1'+
\frac{1}{2D_1^2}R_1'G_1'\nonumber\\
&&+\left[\frac{R_1}{D_1}\d_1^2\left(\frac{1}{D_1}\right)+
\frac{R_1'}{2D_1}\d_1\left(\frac{1}{D_1}\right)+
\sum_{j=2}^k\left\{\frac{R_j}{D_j}~\frac{2}{D_1d_{1j}^2}+
\frac{R_j'}{2D_j}~\frac{1}{D_1d_{1j}}\right\}
\right]G_1\nonumber\\
&&+perm=0,
\eea
Repeating the analysis which led from equation (\ref{ElpsdLaplA}) to (\ref{ElpsdSlnG}) for $k=3$, we find the most general solution 
of (\ref{ElpsdLaplAN}) that exists for all $x\in (-a_j^2,-a_{j+1}^2)$:
\bea
G(x)=C \sqrt{\prod_j(a_j^2+x)}+\sum_{j=0}^{k-1}C_j x^j.
\eea
Requiring $H$ to remain finite at sufficiently large $x_1$, we conclude that most integration constants in the last relation must vanish, and $H$ must be a constant, just as in the $k=3$ case. 

\section{Equation for geodesics in D--brane backgrounds}
\label{AppPrcdr}

In this appendix we implement the program outlined in section \ref{SectPrcdr}. Specifically, we introduce the perturbative expansions (\ref{AppPert}) for functions $h(w)$, $U_1(x)$ and $U_2(y)$, substitute the results into (\ref{HarmRTheta}), and find the restrictions imposed by the Laplace equation (\ref{LaplEqn}). 

\subsection{Particles without angular momentum}

First we analyze the harmonic function (\ref{LaplEqn}) for $L_1=L_2=0$, and angular momenta will be added in the next subsection. To stress the fact that we are dealing with this special case, the potentials will be denoted as ${\tilde U}_1(x)$, ${\tilde U}_2(y)$.
Far away from the sources, the harmonic function is given by (\ref{LeadHarm}), and Laplace equation separates in spherical coordinates, so we find the leading asymptotics\footnote{Due to linearity of the Laplace equation (\ref{LaplEqn}), a constant term can always be added to $H$, so the boundary condition (\ref{AsympApp}) can be easily extended to the asymptotically--flat case $H\approx a+\frac{Q}{r^{7-p}}$.}:
\bea\label{AsympApp}
x+iy\approx\ln\frac{r}{l}+i\theta,\qquad H\approx \frac{Q}{r^{7-p}}.
\eea  
In this appendix we will construct the most general function $H$ and coordinates $(x,y)$ which satisfy four conditions:
\label{PageCondAD}
\begin{enumerate}[(a)]
\item{The Hamilton--Jacobi equation (\ref{HJone}) separates in variables $(x,y)$.}
\item{Away from the sources, function $H(r,\theta)$ satisfies the Laplace equation (\ref{LaplEqn}).}
\item{At large values of $r$, function $H$ and coordinates $(x,y)$ approach the asymptotic expressions given by (\ref{AsympApp}).}
\item{The sources are localized at finite values of $r$, in particular, $H$ is regular for $r>R$.}
\end{enumerate}

To derive the expressions for $(x,y)$ and $H$ consistent with (\ref{AsympApp}), it is convenient to introduce expansions in powers of 
$l/r\sim e^{-w}\sim e^{-x}$:
\bea\label{AppPert}
h(w)=w+\sum_{k>0}a_k e^{-kw},\qquad
{\tilde U}_1(x)=e^{(2-n-m)x}
\left[1+\sum_{k>0}b_k e^{-kx}\right].
\eea
Substituting this into the Laplace equation (\ref{LaplEqn}), matching the results for all powers of $l/r$, and resumming the series using Mathematica, we find that the most general solution for $h(w)$ is parameterized by one constant $a_2$:
\bea\label{TheSoln}
h(w)&=&\ln\left[\frac{1}{2}\left\{e^w+\sqrt{4a_2+e^{2w}}\right\}\right],
\eea
and corresponding potentials are given by\footnote{For $p\ge 5$ we find additional solutions for ${\tilde U}_1$ and ${\tilde U}_2$, which will be discussed below.}
\bea\label{PertPotential}
{\tilde U}_1(x)=\frac{C_1}{(e^x+a_2 e^{-x})^{n-1}
(e^x-a_2 e^{-x})^{m-1}},\quad {\tilde U}_2(y)=0.
\eea
We now start with coordinates defined by (\ref{TheSoln}) and find the most general separable solution of the Laplace equation (\ref{LaplEqn}).

First we observe that starting with an arbitrary solution (\ref{TheSoln}) and adjusting $l$, we can set $a_2$ to one of three values $(-1,0,1)$. Indeed, solution (\ref{TheSoln}), definitions (\ref{defCmplVar}), and  boundary conditions (\ref{AsympApp}) remain invariant under the transformation
\bea\label{GaugeTr}
l\rightarrow e^{\la} l,\quad w\rightarrow w-\la,\quad
x\rightarrow x-\la,\quad a_2\rightarrow a_2 e^{-2\la}
\eea
for any real $\la$. Since $a_2$ is a real number\footnote{This follows from reality of potential  ${\tilde U}_1(x)$ in (\ref{PertPotential}).}, it can be set to zero or to $\pm 1$ by choosing an appropriate $\la$ in (\ref{GaugeTr}). It is convenient to analyze these cases separately.
~\\
~\\
\noindent
\textbf{Case I:} $a_2=0$, $z=w$.\\
Variables $(x,y)$ correspond to spherical coordinates, and equation (\ref{HarmRTheta}) becomes
\bea\label{HUtilde}
H(r,\theta)=\frac{1}{M^2}\left[-\frac{1}{r^2}\left\{
{\tilde U}_1\left(\ln\frac{r}{l}\right)+
{\tilde U}_2(\theta)\right\}\right]
\eea
Definig a new function
\bea
{\hat U}_1(r)\equiv {\tilde U}_1\left(\ln\frac{r}{l}\right),
\eea
we find that the Laplace equation (\ref{LaplEqn}) reduces to two relations:
\bea
&&\frac{1}{r^{8-p}}\frac{d}{dr}\left[r^{8-p}\frac{d}{dr}\frac{{\hat U}_1}{r^2}\right]+\frac{\la}{r^4}=0,
\nonumber\\
&&\frac{1}{r^{8-p}}\frac{d}{dr}\left[
r^{8-p}\frac{d}{dr}\frac{1}{r^2}\right]{\tilde U}_2+
\frac{1}{r^4\sin^{n}\theta\cos^{m}\theta}
\d_\theta(\sin^{n}\theta\cos^{m}\theta\d_\theta 
{\tilde U}_2)=\frac{\la}{r^4}\nonumber
\eea
with separation constant $\la$. Solving these equations, we find the harmonic function 
\bea
H(r,\theta)&=&-\frac{1}{M^2r^2}
[{\tilde U}_1+{\tilde U}_2]=-\frac{1}{M^2r^2}\left[\frac{C_1}{r^{5-p}}+C_2r^2\right.\nonumber\\
&&\left.+\frac{C_3}{\sin^{n-1}\theta\cos^{m-1}\theta}
+\frac{C_4}{\sin^{n-1}\theta} F\left(\frac{m-1}{2},
\frac{3-n}{2},\frac{m+1}{2};\cos^2\theta\right)\right]\nonumber
\eea
Notice that $\la$ gives constant contributions to ${\tilde U}_1$ and ${\tilde U}_2$, which cancel in the sum. Condition (c) implies that $C_2=0$, and condition (d) gives $C_3=C_4=0$. We conclude that in this case the harmonic function is sourced by a single stack of Dp branes:
\bea
H(r,\theta)=\frac{Q}{r^{7-p}}.
\eea
\noindent
\textbf{Case II:} $a_2=-1$, 
$
\displaystyle z=\ln\left[\frac{1}{2}(e^w+\sqrt{e^{2w}-4})\right]
$.\\
This change of variables has been analyzed in Appendix \ref{SprElliptic} (see equation (\ref{EllEqnZ})). Reversing the steps which led from (\ref{coshOne})--(\ref{coshTwo}) to (\ref{EllEqnZ}), we can rewrite (\ref{TheSoln}) as
\bea\label{CaseTwoRep}
\cosh x=\frac{\rho_++\rho_-}{2d},\quad \cos y=\frac{\rho_+-\rho_-}{2d},\quad
\rho_\pm=\sqrt{r^2+d^2\pm 2rd\cos\theta},\quad
d=2l.
\eea

~\\
\noindent
Let us now implement the requirements (a)--(d) listed in page \pageref{PageCondAD}.

(a) As discussed in section \ref{SectPrcdr}, the Hamilton--Jacobi equation is separable if $z=h(w)$ is a holomorphic function, and $H$ has the form (\ref{HarmRTheta}):
\bea\label{SeparHS}
H&=&-
\frac{1}{M^2}\left[
\frac{|h'|^2}{r^2}\left[
{\tilde U}_1\left(x\right)+
{\tilde U}_2\left(y\right)\right]
\right].
\eea
We recall that in this section we are focusing on the special case 
$L_1=L_2=0$.

(b) We will now find the most general function  
${\tilde U}(r,\theta)={\tilde U}_1(x)+{\tilde U}_2(y)$ which  
satisfies the Laplace equation (\ref{LaplEqn}) in the metric
\bea\label{metrBB}
ds^2_{9-p}=
dr^2+r^2d\theta^2+
r^2\sin^{2}\theta d\Omega_{n}^2+
r^2\cos^{2}\theta d\Omega_{m}^2.
\eea
Using a relation
\bea
\left|\frac{h'}{r}\right|^2=\left|\frac{e^w}{\sqrt{e^{2w}-4}}
\frac{1}{le^{(w+{\bar w})/2}}\right|^2=
\frac{1}{l^2|e^{2w}-4|}=\frac{1}{d^2}
\left|\frac{4e^{2z}}{(e^{2z}-1)^2}\right|=
\frac{1}{d^2(\cosh^2x-\cos^2y)},\nonumber
\eea
we can rewrite (\ref{metrBB}) in terms of $x$ and $y$:
\bea\label{metrAA}
ds^2_{9-p}=
d^2\left[(\cosh^2x-\cos^2y)(dx^2+dy^2)+
\sinh^2 x\sin^2 y d\Omega_{n}^2+
\cosh^2 x\cos^2 y d\Omega_{m}^2\right]
\eea
and to rewrite the Laplace equation (\ref{LaplEqn}) in these coordinates:
\bea\label{LaplInXY}
\frac{1}{\sinh^{n}x\cosh^{m}x}\frac{\d}{\d x}\left[
\sinh^{n}x\cosh^{m}x\frac{\d H}{\d x}
\right]+
\frac{1}{\sin^{n}y\cos^{m}y}\frac{\d}{\d y}\left[
\sin^{n}y\cos^{m}y\frac{\d H}{\d y}
\right]=0
\eea
We are looking for a separable solution of equation (\ref{LaplInXY}) which has the form (\ref{SeparHS}):
\bea\label{SeparH}
H
&=&-\frac{1}{(Md)^2(\cosh^2 x-\cos^2 y)}
\left[
{\tilde U}_1\left(x\right)+
{\tilde U}_2\left(y\right)\right].
\eea
Straightforward algebraic manipulations with (\ref{LaplInXY}) lead to equation for 
${\tilde U}={\tilde U}_1\left(x\right)+{\tilde U}_2\left(y\right)$
\bea\label{EqnForTilde}
0&=&\frac{1}{\sinh^{n}x\cosh^{m}x}\frac{\d}{\d x}\left[
\sinh^{n}x\cosh^{m}x\frac{\d \tilde U}{\d x}
\right]+
\frac{1}{\sin^{n}y\cos^{m}y}\frac{\d}{\d y}\left[
\sin^{n}y\cos^{m}y\frac{\d \tilde U}{\d y}
\right]\nonumber\\
&&-2\frac{\left[\{(m+n-2)(\cosh 2x+\cos 2y)-2(m-n)\}+2(\sinh 2x\d_x+\sin 2y\d_y)\right]{\tilde U}}{\cosh 2x-\cos2y}.\nonumber\\
\eea
Since the first line of this equation is the sum of $x$-- and $y$--dependent terms, the second line must have the same structure, so it can be rewritten as 
\bea\label{SplitEqn}
F(x)+G(y)&=&2(m+n-2)({\tilde U}_1(x)-{\tilde U}_2(y))-4\frac{K(x)+L(y)}{\cosh 2x-\cos2y}\\
\label{SplitEqnK}
K(x)&\equiv&\left[(m+n-2)\cosh 2x-(m-n)+\sinh 2x\d_x\right]{\tilde U}_1\\
\label{SplitEqnL}
L(y)&\equiv&\left[(m+n-2)\cos 2y-(m-n)+\sin 2y\d_y\right]{\tilde U}_2
\eea
Consistency of equation (\ref{SplitEqn}) requires that
\bea\label{KandL}
K(x)=\la_1\cosh 2x+\la_2\cosh^2 2x+\la_0,\quad L(y)=-\la_1\cos 2y-\la_2\cos^2 2y-\la_0
\eea
with constant $\la_0$, $\la_1$ and $\la_2$. Substituting this $K(x)$, $L(y)$ in (\ref{SplitEqnK})--(\ref{SplitEqnL}) and solving the resulting equations for 
${\tilde U}_1(x)$ and ${\tilde U}_2(y)$, we find
\bea\label{SolnUtilde}
{\tilde U}_1&=&\frac{1}{\cosh^{m-1}x\sinh^{n-1}x}\left[C_1+\frac{1}{2}\int dx K(x)\cosh^{m-2}x\sinh^{n-2}x\right]\nonumber\\
{\tilde U}_2&=&\frac{1}{\cos^{m-1}y\sin^{n-1}y}\left[C_2+\frac{1}{2}\int dy L(y)\cos^{m-2}y\sin^{n-2}y\right]
\eea
The last two equations give necessary, but not sufficient condition for ${\tilde U}={\tilde U}_1\left(x\right)+{\tilde U}_2\left(y\right)$ to solve 
(\ref{EqnForTilde}). Going back to equation (\ref{EqnForTilde}), we can now rewrite it as
\bea
&&\frac{1}{\sinh^{n}x\cosh^{m}x}\frac{\d}{\d x}\left[
\sinh^{n}x\cosh^{m}x\frac{\d \tilde U_1}{\d x}
\right]+
\frac{1}{\sin^{n}y\cos^{m}y}\frac{\d}{\d y}\left[
\sin^{n}y\cos^{m}y\frac{\d \tilde U_2}{\d y}
\right]\nonumber\\
&&\qquad\qquad+2(m+n-2)({\tilde U}_1(x)-{\tilde U}_2(y))-4\la_1-4\la_2(\cosh 2x+\cos 2y)=0.\nonumber
\eea
This relation is satisfied identically for all $\la_0$, $\la_1$ and $\la_2$.

(c) Although expressions (\ref{SolnUtilde}) produce solutions 
of the Laplace equation (\ref{EqnForTilde}) for all $K(x)$ and $L(x)$ given by 
(\ref{SplitEqnK})--(\ref{SplitEqnL}), the resulting harmonic function (\ref{SeparH}) may not satisfy the boundary conditions (\ref{AsympApp}). 
To analyze these boundary conditions, we find the leading behavior of $H$ at large values of $r$ by substituting 
the asymptotic expressions for various coordinates,
\bea\label{BBeqn}
h(w)\approx w,\quad x\approx \ln\frac{r}{l},\quad y\approx\theta,
\eea
in equations (\ref{SeparH}), (\ref{SplitEqnK})--(\ref{SplitEqnL}), (\ref{SolnUtilde}). 
We conclude that the leading contribution to $H$ satisfies the boundary conditions (\ref{AsympApp})\footnote{We recall that $7-p=m+n$ due 
to (\ref{MNasPQ}).} for the following values of $(C_2,\la_1,\la_2)$:
\bea\label{FromBC}
m=n=0:&&\la_2=0;\nonumber\\
m+n=1:&&\la_2=0;\nonumber\\
m=n=1:&&\la_0=\la_1=\la_2=0;\\
(m,n)=(2,0)\mbox{ or }(0,2):&&\la_1=\la_2=0,\quad C_2=0;\nonumber\\\
m+n>2:&&\la_0=\la_1=\la_2=0,\quad C_2=0.\nonumber
\eea

(d) Function $H$ given by (\ref{SeparH}), (\ref{SolnUtilde}), (\ref{KandL}) satisfies the Laplace equation and the boundary conditions (\ref{AsympApp}) for all values of $(C_2,\la_1,\la_2)$ listed in (\ref{FromBC}). However, regularity of $H$ at sufficiently large values of $r$ imposes some additional requirements. First we notice that regularity in the asymptotic region (\ref{BBeqn}) implies that ${\tilde U}_2(y)$ must remain finite for all values of $y$. This gives additional restrictions for the first three cases listed in (\ref{FromBC})
\bea\label{FromReg}
m=n=0\mbox{ or }m+n=1\mbox{ or }m=n=1:&&\la_1=0,\quad\la_2=-\la_0.
\eea
and to additional restriction $\la_0=0$ for $(m,n)=(2,0)\mbox{ or }(0,2)$. Since for $m=n=1$ the constant $C_2$ can be absorbed into $C_1$, restrictions (\ref{FromBC}), (\ref{FromReg}) can be summarized as
\bea\label{FromAll}
m+n\le1:&&\la_0=\la_1=\la_2=0;\nonumber\\
m+n>1:&&\la_0=\la_1=\la_2=0,\quad C_2=0.
\eea
For $m+n>1$ we recover the solution (\ref{PertPotential}):
\bea\label{SolnUtildeGen}
p<6:\quad {\tilde U}_1&=&\frac{C_1}{\cosh^{m-1}x\sinh^{n-1}x},\qquad
{\tilde U}_2=0,
\eea
and for $m+n\le1$ we find three special cases:
\bea\label{SolnUSpec}
(p,m,n)=(7,0,0):&&
{\tilde U}_1=\frac{C_1}{2}\sinh 2x,\quad
{\tilde U}_2=\frac{C_2}{2}\sin 2y\nonumber\\
(p,m,n)=(6,1,0):&&
{\tilde U}_1=C_1\sinh x,\quad
{\tilde U}_2=C_2\sin y\\
(p,m,n)=(6,0,1):&&
{\tilde U}_1=C_1\cosh x,\quad
{\tilde U}_2=C_2\cos y\nonumber
\eea
The corresponding harmonic function is given by (\ref{HarmTilde}). 
~\\
~\\
~\\
\noindent
\textbf{Case III:} $a_2=1$, 
$z=\ln\left[e^w+\sqrt{e^{2w}+4}\right]
$.\\
This change of variables can be obtained from case 2 by making replacements\footnote{Notice that this replacement also interchanges $S^n$ and $S^m$ in (\ref{SSmetr}).}
\bea\label{ShiftTheta}
w\rightarrow w-\frac{i\pi}{2},\quad 
z\rightarrow z-\frac{i\pi}{2}\quad\Rightarrow\quad
\theta\rightarrow \theta-\frac{\pi}{2},\quad
y\rightarrow y-\frac{\pi}{2},
\eea
which lead to
\bea\label{RepCase3}
\cosh x=\frac{\rho_++\rho_-}{2d},\quad 
\sin y=\frac{\rho_+-\rho_-}{2d},\quad
\rho_\pm=\sqrt{r^2+d^2\pm 2rd\sin\theta},
\eea
Repeating the steps which led to (\ref{SolnUtildeGen}), we arrive at 
\bea\label{SolnUtildeGenII}
p<6:\quad {\tilde U}_1&=&\frac{C_1}{\cosh^{n-1}x\sinh^{m-1}x},\qquad
{\tilde U}_2=0.
\eea
while the counterpart of (\ref{SolnUSpec}) becomes
\bea\label{SolnUSpecII}
(p,m,n)=(7,0,0):&&
{\tilde U}_1=\frac{C_1}{2}\sinh 2x,\quad
{\tilde U}_2=\frac{C_2}{2}\sin 2y\nonumber\\
(p,m,n)=(6,0,1):&&
{\tilde U}_1=C_1\sinh x,\quad
{\tilde U}_2=C_2\cos y\\
(p,m,n)=(6,1,0):&&
{\tilde U}_1=C_1\cosh x,\quad
{\tilde U}_2=C_2\sin y\nonumber
\eea
The corresponding harmonic function is given by (\ref{HarmTildeAlt}).

\subsection{General case}\label{Momenta}

As shown in the last subsection, the harmonic function (\ref{HarmRTheta}),
\bea\label{Harmbb}
H(r,\theta)=
\frac{1}{M^2}\left[
\frac{L_2^2}{r^2\sin^2\theta}+
\frac{L_1^2}{r^2\cos^2\theta}-
\frac{|h'|^2}{r^2}\left[
U_1\left(x\right)+
U_2\left(y\right)\right]
\right].
\eea
solves the Laplace equation (\ref{LaplEqn}) as long as $L_1=L_2=0$ and $U_1(x)$, $U_2(y)$ are given by (\ref{SolnUtildeGen}) or (\ref{SolnUSpec}). We will now find the potentials $U_1(x)$ and $U_2(y)$ for non-zero values of $L_1$ and $L_2$. 

Recalling the change of coordinates (\ref{CaseTwoRep}), we can write the first two terms in (\ref{Harmbb}) as
\bea
&&\frac{L_2^2}{r^2\sin^2\theta}+
\frac{L_1^2}{r^2\cos^2\theta}=
\frac{L_2^2}{d^2\sinh^2 x\sin^2 y}+
\frac{L_1^2}{d^2\cosh^2 x\cos^2 y}\nonumber\\
&&\qquad=\frac{1}{d^2(\cosh^2 x-\cos^2 y)}
\left[\left(\frac{L_2^2}{\sinh^2 x}-
\frac{L_1^2}{\cosh^2 x}\right)+\left(
\frac{L_2^2}{\sin^2 y}+\frac{L_1^2}{\cos^2 y}\right)\right]
\eea
Then the harmonic function (\ref{Harmbb}) can be rewritten as
\bea\label{HarmTilde}
H(r,\theta)&=&-\frac{|h'|^2}{M^2r^2}\left[{\tilde U}_1(x)+
{\tilde U}_2(y)\right]\nonumber\\
&=&-\frac{1}{(Md)^2(\cosh^2 x-\cos^2 y)}\left[{\tilde U}_1(x)+
{\tilde U}_2(y)\right],\\
{U}_1(x)&=&{\tilde U}_1(x)+\frac{L_2^2}{\sinh^2 x}-\frac{L_1^2}{\cosh^2 x},\qquad
{U}_2(y)={\tilde U}_2(y)+
\frac{L_1^2}{\cos^2 y}+\frac{L_2^2}{\sin^2 y},\nonumber
\eea
where ${\tilde U}_1$ and ${\tilde U}_2$ are given by (\ref{SolnUtildeGen}) or (\ref{SolnUSpec}).

Reparametrization (\ref{RepCase3}) can be obtained from (\ref{CaseTwoRep}) by making the replacement (\ref{ShiftTheta}), then instead of (\ref{HarmTilde}) we 
find\footnote{Notice that the interchange $\sin\theta\leftrightarrow\cos\theta$ is accompanied by the interchange of angular momenta $L_1\leftrightarrow L_2$}
\bea\label{HarmTildeAlt}
H(r,\theta)&=&-\frac{|h'|^2}{M^2r^2}\left[{\tilde U}_1(x)+
{\tilde U}_2(y)\right]\nonumber\\
&=&-\frac{1}{(Md)^2(\cosh^2 x-\sin^2 y)}\left[{\tilde U}_1(x)+
{\tilde U}_2(y)\right],\\
{U}_1(x)&=&{\tilde U}_1(x)+\frac{L_1^2}{\sinh^2 x}-\frac{L_2^2}{\cosh^2 x},\qquad
{U}_2(y)={\tilde U}_2(y)+\frac{L_1^2}{\cos^2 y}+\frac{L_2^2}{\sin^2 y}.\nonumber
\eea
In this case, functions ${\tilde U}_1$ and ${\tilde U}_2$ are given by (\ref{SolnUtildeGenII}) or (\ref{SolnUSpecII}).

\section{Equations for Killing tensors}
\label{AppKill}

In section \ref{SecKill} we discussed the Killing tensors associated with separation of the HJ equation in elliptic coordinates. The detailed calculations are presented in this appendix.

\subsection{Killing tensors for the metric produced by D--branes}

In this appendix we will solve the equations for Killing tensors in the geometry (\ref{EinstFrame})--(\ref{WWbase}) and derive the expression (\ref{Jul25}). To simplify some formulas appearing below, we modify the parameterization of 
(\ref{EinstFrame})--(\ref{WWbase}):
\bea\label{SmplPrmMtr}
ds^2=FG dx^\mu dx_\mu+F\left[A(dx^2+dy^2)+e^B d\Omega_m^2+e^C d\Omega_n^2\right]
\eea
and introduce four types of indices:
\bea
R^{1,p}:\, \mu,\nu,\dots;\qquad
(x,y):\, i,j,\dots;\qquad S^m:\, a,b,\dots;\qquad 
S^n:\, {\dot a},{\dot b},\dots
\eea
Separation of the wave equation implies an existence of rank-two conformal Killing tensors which satisfy equation 
(\ref{CKTdef})
\bea\label{CnfKillEqn}
\nabla_{(M}K_{NL)}=\frac{1}{2}W_{(M}g_{NL)}.
\eea
We begin with analyzing $(xxx)$ and $(yyy)$ components of this relation:
\bea\label{Jul30f}
&&\nabla_xK_{xx}=\frac{1}{2}W_x g_{xx}\,\Rightarrow\, W^x=2\nabla_x K^{xx},\nonumber\\
&&\nabla_yK_{yy}=\frac{1}{2}W_y g_{yy}\,\Rightarrow\, W^y=2\nabla_y K^{yy}.
\eea
Substituting this into $(xxy)$ component and recalling that $g_{xx}=g_{yy}$, we find
\bea\label{Jul30}
\nabla_yK_{xx}+2\nabla_x K_{xy}=\frac{1}{2}W_y g_{xx}=\nabla_y K_{yy}\,\Rightarrow\,
\nabla_y(K^{yy}-K^{xx})=2\nabla_x K^{xy}.
\eea
Using the explicit expressions for the Christoffel symbols,
\bea
\Gamma_{xx}^x=\Gamma_{xy}^y=-\Gamma_{yy}^x=\frac{1}{2}\d_x\ln g_{xx},\quad
\Gamma_{yy}^y=\Gamma_{xy}^x=-\Gamma_{xx}^y=\frac{1}{2}\d_y\ln g_{yy},
\eea
equation (\ref{Jul30}) can be rewritten as
\bea
\d_y(K^{yy}-K^{xx})=2\d_x K^{xy}.
\eea
Combining this with a similar relation coming from $(xyy)$ component of (\ref{CnfKillEqn}),
\bea
\d_x(K^{xx}-K^{yy})=2\d_y K^{xy},
\eea
we conclude that
\bea\label{KillXY}
K^{yy}=K^{xx}+N,\qquad 2d K^{xy}=\star_2 dN.
\eea
so $K^{xy}$ and $N\equiv K^{xx}-K^{yy}$ are dual harmonic functions. We also quote the expressions for $W_i$, which can be obtained by evaluating the covariant derivatives in  (\ref{Jul30f}):
\bea\label{KillVecXY}
W_x=2\left(\d_x\left[g_{xx}K^{xx}\right]+K^{xy}\d_y g_{yy}
\right),
\quad
W_y=2\left(\d_y\left[g_{yy}K^{yy}\right]+K^{xy}\d_x g_{xx}\right).
\eea

Next we look at the Killing equation (\ref{CnfKillEqn}) with three legs on $R^{1,p}$:
\bea\label{Jul30a}
{\d}_{(\mu} K_{\nu\la)}-2\Gamma_{(\mu\nu}^i K_{\la)i}=\frac{1}{2}FGW_{(\la}\eta_{\mu\nu)}.
\eea
Using the expression for the relevant Christoffel symbol,
\bea
\Gamma_{\mu\nu}^i=-\frac{1}{2}g^{ij}\d_jg_{\mu\nu}=-\frac{1}{2FA}\d_i(FG)\eta_{\mu\nu},
\eea
we can rewrite (\ref{Jul30a}) as an equation for a conformal Killing tensor on $R^{1,p}$:
\bea\label{FlatCnfKill}
\d_{(\la} K_{\mu\nu)}=\frac{1}{2}\eta_{(\mu\nu}Z_{\la)},\qquad
Z_\la\equiv\left[GFW_{\la}-
\frac{2}{FA}\d_i(FG)K_{\la i}
\right].
\eea
Tensors satisfying this equation have already been encountered in section \ref{RedTo2D}, and the general solution of 
(\ref{FlatCnfKill}) has been found in \cite{Weir}:
\bea\label{WeirKill}
K_{\mu\nu}&=&A\eta_{\mu\nu}+B_{ab}K^{(a)}_\mu K^{(b)}_\nu,\\
Z_\mu&=&4B_{ab}f^{(a)}K^{(b)}_\mu.\nonumber
\eea
Here $K^{(a)}_\mu$ are conformal Killing vectors on $R^{1,p}$,
\bea\label{FlatKillEqn}
\nabla_{(\mu}K^{(a)}_{\nu)}=f^{(a)}\eta_{\mu\nu},
\eea
and coefficients $A$, $B_{ab}$ can depend on the directions transverse to $R^{1,p}$. Solutions of equation (\ref{SmplPrmMtr}) are reviewed in Appendix \ref{AppFltKil}. The conformal Killing vectors $K^{(a)}_\mu$ are responsible for separation on $R^{1,p}$ (which is already taken into account in the ansatz 
(\ref{GenAction})), and to study the separation in $(x,y)$ coordinates\footnote{In contrast to the explicit symmetries related to Killing vectors, the origin of separation in $(x,y)$ is usually called a ``hidden symmetry".}, it is sufficient to focus on the first term in (\ref{WeirKill}), which is invariant under $ISO(p,1)$ transformations. Repeating this arguments for equation (\ref{CnfKillEqn}) with three legs on the spheres, we conclude that the Killing tensor responsible for separation of $x$ and $y$ is invariant under $SO(m+1)\times SO(n+1)\times ISO(p,1)$, in particular, this implies that 
\bea\label{SymmKT}
K^{MN}p_Mp_N=K_1(x,y)\eta^{\mu\nu}p_\mu p_\nu+
K_2(x,y)h^{ab}p_ap_b+
K_3h^{\dot a\dot b}p_{\dot a}p_{\dot b}+K^{ij}p_ip_j,
\eea 
\bea
W_\mu=0,\qquad W_a=0,\qquad W_{\dot a}=0.\nonumber
\eea
Here $h_{ab}$ and $h_{\dot a\dot b}$ are metrics on $S^m$ and $S^n$. Ansatz (\ref{SymmKT}) ensures that all components of equation (\ref{CnfKillEqn}) which do not contain legs along $x$ or $y$ are satisfied.

Next we look at $(x,\mu,\nu)$ and $(y,\mu,\nu)$ components:
\bea\label{Jul30b}
\d_i K_{\mu\nu}-4\Gamma^\la_{i(\mu}K_{\nu)\la}-
2\Gamma^j_{\mu\nu}K_{ij}=\frac{1}{2}W_i g_{\mu\nu}
\eea
Taking into account the expressions for the Christoffel symbols,
\bea
\Gamma^\la_{i\mu}=\frac{1}{2}\delta^\la_\mu \d_i \ln(FG),
\quad \Gamma^j_{\mu\nu}=
-\frac{1}{2FA}\eta_{\mu\nu}\d_j(FG),
\eea
we can rewrite equation (\ref{Jul30b}) as
\bea\label{Jul30mnk}
\d_i(K^{\mu\nu})-
K_{ij}\eta^{\mu\nu}\frac{1}{FA}\d_j\frac{1}{FG}=
\frac{1}{2}W_i g^{\mu\nu}\,\Rightarrow\,
\d_i K_1+\frac{K_{ij}}{FA}\d_j\frac{1}{FG}=
\frac{1}{2FG}W_i
\eea
Similar analysis of equations along sphere directions gives:
\bea\label{Jul30sph}
\d_i K_2-\frac{K_{ij}}{FA}\d_j\frac{e^{-B}}{F}=
\frac{e^{-B}}{2F}W_i ,\qquad
\d_i K_3-\frac{K_{ij}}{FA}\d_j\frac{e^{-C}}{F}=\frac{e^{-C}}{2F}W_i 
\eea
Equations (\ref{KillXY}), (\ref{KillVecXY}), (\ref{Jul30mnk}), 
(\ref{Jul30sph}) give a complete system which is equivalent to 
(\ref{CnfKillEqn}) for the ansatz (\ref{SymmKT}). 

In the special case $K^{xx}=K^{xy}=0$ (which corresponds to the Killing tensor (\ref{Jul25})), we find
\bea\label{Oct23a}
K^{yy}=c,\quad W_x=0,\quad W_y=2c\d_y(AF),\quad
\d_x K_1=\d_x K_2=\d_x K_3=0
\eea
\bea\label{Oct23b}
\d_y K_1=c\d_y\frac{A}{G},\quad
\d_y K_2=c\d_y\frac{A}{e^B},\quad
\d_y K_2=c\d_y\frac{A}{e^C}
\eea
Integrability conditions require that
\bea\label{Oct23}
\d_x\d_y\frac{A}{G}=\d_x\d_y\frac{A}{e^B}=\d_x\d_y\frac{A}{e^C}
\eea
and they are satisfied by our solution (\ref{EinstFrame}), (\ref{WWbase}), (\ref{KGharm}). As expected, we did not get any restrictions on function $F$ from (\ref{SmplPrmMtr}) since conformal rescaling of the metric does not affect equations for the null geodesics. However, as demonstrated in section \ref{SecWave}, separability of the HJ equation does not persist for the wave equation unless function $F$ has a special form. In particular, we found that the wave equation on the D--brane background is only separable in the Einstein frame. 

To summarize, we have demonstrated that metric (\ref{SmplPrmMtr}) possesses a conformal Killing tensor of the form 
(\ref{SymmKT}) if and only if equations (\ref{KillXY}), (\ref{KillVecXY}), (\ref{Jul30mnk}), (\ref{Jul30sph}) are satisfied. In the special case $K^{xx}=K^{xy}=0$, the system reduces to equations (\ref{Oct23a})--(\ref{Oct23b}), and their integrability conditions are given by (\ref{Oct23}). Application of this construction to the geometry (\ref{EinstFrame})--(\ref{WWbase}) 
gives (\ref{Jul25}).

\subsection{Review of Killing tensors for flat space}
\label{AppFltKil}

In the last subsection we have encountered Killing equations on $R^{1,p}$ space and on a sphere,
\bea\label{KillFlat}
\d_{(\la} K_{\mu\nu)}=g_{(\mu\nu}W_{\la)}
\eea
and we chose the simplest solution,
\bea
K_{\mu\nu}=K g_{\mu\nu},\qquad W_\mu=0,\quad
\d_\mu K=0.
\eea
In this section we review the most general solution of (\ref{KillFlat}).

Killing tensors on symmetric spaces were studied in \cite{Weir}, where it was shown that the most general solution of (\ref{KillFlat}) on a sphere can be written as 
\bea\label{WeirKillA}
K_{\mu\nu}&=&A\eta_{\mu\nu}+B_{ab}K^{(a)}_\mu K^{(b)}_\nu,\\
W_\mu&=&4B_{ab}f^{(a)}K^{(b)}_\mu,\nonumber
\eea
where $K^{(a)}_\mu$ are conformal Killing vectors satisfying equation
\bea\label{FlatKillEqnA}
\nabla_{(\mu}K^{(a)}_{\nu)}=f^{(a)}\eta_{\mu\nu}. 
\eea
Although an explicit solution of equation (\ref{FlatKillEqnA}) was not given in \cite{Weir}, it can be easily derived, and in this section we will present such derivation for flat space since this case played an important role in the construction presented in section \ref{RedTo2D}\footnote{Specifically, an explicit construction of Killing tensors (\ref{WeirKillA}), (\ref{FltCKV}) can be used to show that any coordinate system leading to separation of the Laplace equation in flat space with $d>2$ is a special case of the ellipsoidal coordinates (\ref{ref317}).}. Out result can be easily extended to the conformal Killing tensors on a sphere.

Consider the equation for the Conformal Killing Vector (CKV) on flat space:
\bea\label{CKVeqn}
\d_i K_j+\d_j K_i=2f \delta_{ij}
\eea
From equations with $i=j$ we conclude that
\bea\label{CKVfK}
\d_1 K_1=\d_2 K_2=\dots=f
\eea
Then applying $\d_1\d_2$ to the $(1,2)$ component of equation (\ref{CKVeqn}),
\bea\label{KV12}
\d_1 K_2+\d_2 K_1=0,
\eea
we find a restriction on $f$:
\bea\label{Aug17}
(\d_1^2+\d_2^2)f=0\quad \Rightarrow\quad
f=g(x_1+ix_2,x_3,x_4\dots)+\overline{g(x_1+ix_2,x_3,x_4\dots)}.
\eea
In two dimensions this exhausts all equations for the CKV, thus the most general solution is parameterized by one holomorphic function $F$:
\bea
d=2:\quad K_1=F(x_1+ix_2)+cc,\quad K_2=iF(x_1+ix_2)+cc,\quad
f=\d_1 K_1.
\eea

For $d\ge 3$ we can combine (\ref{KV12}) with its counterparts for the $(1,3)$ and $(2,3)$ component of (\ref{CKVeqn}) to further restrict the form of $f$. Differentiating 
(\ref{KV12}) with respect to $x_3$ and combining the result with
\bea
\d_2\d_1 K_3+\d_2\d_3 K_1=0,\quad 
\d_1\d_2 K_3+\d_1\d_3 K_2=0,
\eea
we conclude that
\bea\label{Aub14b}
\d_2\d_3 K_1=0\quad\Rightarrow\quad \d_2\d_3 f=0.
\eea
Repetition of this argument for all $i\ne j$ gives
\bea
\d_i\d_j f=0\quad\Rightarrow\quad f=\sum h_j(x_j).
\eea
Substituting this into (\ref{Aug17}), we find a relation between $D_1$ and $D_2$:
\bea
h_1''(x_1)+h_2''(x_2)=0\quad\Rightarrow\quad
\begin{array}{ll}
h_1=D_1 x_1^2+A_1 x_1+B_1\\
h_2=D_2 x_2^2+A_2 x_2+B_2
\end{array},
\quad D_2=-D_1.
\eea
Similar arguments show that 
\bea
D_3=-D_2,\quad D_3=-D_1 \quad\Rightarrow\quad
D_i=0.
\eea
To summarize, we have demonstrated that $f$ is a linear function,
\bea\label{Aug14a}
f=A_i x_i+B,
\eea
and (\ref{CKVfK}) leads to the following expressions for $K_1$, $K_2$:
\bea
K_1&=&
x_1A_j x_j-\frac{A_1x_1^2}{2}+Bx_1+h_1(x_2,x_3,\dots),
\nonumber\\
K_2&=&
x_2B_j x_j-\frac{B_2x_2^2}{2}+Bx_2+h_2(x_1,x_3,\dots).
\nonumber
\eea
Substitution of these relations into (\ref{KV12}) gives an equation for $h_1$ and $h_2$: 
\bea
&&A_2x_1+\d_2h_1(x_2,x_3,\dots)
+A_1 x_2+\d_1 h_2(x_1,x_3,\dots)=0,
\eea
and the solution reads
\bea
h_1(x_2,x_3,\dots)&=&-\frac{A_1 x_2^2}{2}+x_2h_{12}(x_3,\dots)+
{\tilde h}_1(x_3,\dots)\nonumber\\
h_2(x_1,x_3,\dots)&=&-\frac{A_2 x_1^2}{2}-x_1h_{12}(x_3,\dots)+
{\tilde h}_2(x_3,\dots)\nonumber
\eea
Equation (\ref{Aub14b}) implies that $\d_3 h_{12}=0$.  Analogs of (\ref{Aub14b}) with different indices ensure that $h_{12}$ does not depend on $(x_4,\dots,x_d)$, this leads to the complete determination of $h_1$ as a function of $x_2$:
\bea
h_1(x_2,x_3,\dots)&=&-\frac{A_1 x_2^2}{2}+C_{12}x_2+
{\tilde h}_1(x_3,\dots)\nonumber\\
\eea
Repeating this argument for other pairs, we find the most general expression 
for $K_i$:
\bea\label{FltCKV}
K_i&=&
(A_j x_j)x_i-\frac{A_i r^2}{2}+Bx_i+C_{ij}x_j+D_i,\qquad
C_{ij}=-C_{ji}
\eea
To summarize, we have demonstrated that any conformal Killing vector in flat space has form (\ref{FltCKV}) with constant parameters $A_i$, $B$, $C_{ij}$. The arguments of \cite{Weir} imply that the general Killing tensor on such space can be written as combination of such Killing vectors (\ref{WeirKillA})
with additional constants $A$, $B_{ab}$.

\section{Non--integrability of strings on D--brane backgrounds}
\label{AppNVE}
In section \ref{SecGnrGds} we have classified Dp--brane backgrounds leading to integrable geodesics, and in section 
\ref{NonintStr} we outlined the procedure for studying dynamics of strings on such geometries. This appendix contains 
technical details supporting the analysis of section \ref{NonintStr}. Appendix \ref{NVEReview} reviews one of the 
approaches to integrable systems, and section \ref{NonIntStrsDBr} applies this general construction to specific configurations of strings on Dp--brane backgrounds. In particular, in appendix \ref{AppNintGen} we focus on the geometries leading to integrable geodesics\footnote{As discussed in the Introduction, integrability of geodesics is a necessary condition for integrability of strings.} and rule out integrability of strings for the majority of such backgrounds.
 
\subsection{Review of analytical non--integrability}
\label{NVEReview}

A dynamical system is called integrable if the number of integrals of motion is equal to the number of degrees of freedom. To demonstrate that a system is \textit{not integrable}, it is sufficient to look at linear perturbations around a particular solution and to demonstrate that the resulting \textit{linear equations} do not have a sufficient number of integrals of motion. The equation for linear perturbations is known as the Normal Variational Equation (NVE), and the Kovacic algorithm \cite{Kovacic_alg} gives a powerful analytical tool for studying this equation \cite{ziglinMor}. Let us outline the procedure developed in \cite{ziglinMor}.

Consider a Hamiltonian system parameterized by canonical variables as $(x_i, p_i)$, ($i$ ranges from $1$ to $N$). To rule 
out integrability using NVE, one has to perform the following steps.
\begin{enumerate}
\item Start with a Hamiltonian and write down the equations of motion
\bea
H=\frac{1}{2}\sum_{i=1}^N p_i+V(x_1,..., x_N),
\eea
\bea\label{eom}
\dot{x}_i&=&\frac{\partial H}{\partial p_i}=f_i(x_1,...,x_N,p_1,...,p_N),\nonumber\\
\dot{p}_i&=&-\frac{\partial H}{\partial x_i}=g_i(x_1,...,x_N,p_1,...,p_N),\\ i&=&1,...,N.\nonumber
\eea
\item Write the full variational equations of (\ref{eom}):
\bea\label{var_eqs}
\delta\dot{x}_i&=&\sum_{j=1}^N\frac{\partial f_j}{\partial x_j} \delta x_j+\sum_{i=j}^N\frac{\partial f_j}{\partial p_j} \delta p_j,\nonumber\\
\delta\dot{p}_i&=&\sum_{j=1}^N\frac{\partial g_j}{\partial x_j} \delta x_j+\sum_{j=1}^N\frac{\partial g_j}{\partial p_j} \delta p_j,\\ i&=&1,...,N.\nonumber
\eea
\item Choose a particular solution $x_i=\hat{x}_i,p_i=\hat{p}_i, i=1,...,N$ and one pair of the canonical coordinates, 
$(x_k, p_k)$. The Normal Variational Equation (NVE) is a subsystem of (\ref{var_eqs}) with $i=k$:
\bea\label{Nnv01}
\delta\dot{x}_k=\sum_{j=1}^N\frac{\partial f_j}{\partial x_j} \delta x_j\Big|_{\hat{x}_i,\hat{p}_i}+\sum_{i=j}^N\frac{\partial f_j}{\partial p_j} \delta p_j\Big|_{\hat{x}_i,\hat{p}_i},\\
\delta\dot{p}_k=\sum_{j=1}^N\frac{\partial g_j}{\partial x_j} \delta x_j\Big|_{\hat{x}_i,\hat{p}_i}+\sum_{j=1}^N\frac{\partial g_j}{\partial p_j} \delta p_j\Big|_{\hat{x}_i,\hat{p}_i}.\nonumber
\eea
\item Rewrite the system (\ref{Nnv01}) as a second-order differential equation for $\delta x_k$:
\bea
\delta \ddot{x}_k+q(t)\delta \dot{x}_k+r(t)\delta x_k=0,
\eea
or in the standard notation
\bea\label{nve}
\ddot{\eta}+q(t)\dot{\eta}+r(t)\eta=0,\quad \delta x_k=\eta.
\eea
Equation (\ref{nve}) is known as NVE \cite{ziglinMor}.
\item Make the NVE suitable for using the Kovacic algorithm, namely, algebrize the NVE (i.e. rewrite it as differential equation with rational coefficients) by using a change of variables $t\to x(t)$:
\bea\label{alg_nve}
\eta^{''}+\left(\frac{\ddot{x}}{\dot{x}^2}+\frac{q(t(x))}{\dot{x}}\right)\eta'+\frac{r(t(x))}{\dot{x}^2}\eta=0, \quad F'\equiv\frac{dF}{dx}.
\eea
The Kovacic algorithm requires the coefficients in this equation to be rational functions of $x$. The next step is to convert (\ref{alg_nve}) to the normal form by redefining function $\eta$:
\begin{eqnarray}\label{norm_nve}
{\tilde \eta}{''}(x)+U(x){\tilde\eta}(x)=0.
\end{eqnarray}

\item Apply the Kovacic algorithm to (\ref{norm_nve}), and if it fails then the system is not integrable. 
\end{enumerate}
A more detailed explanation of these steps, containing several examples, can be found in \cite{StepTs}. 

\subsection{Application to strings in Dp--brane backgrounds}\label{NonIntStrsDBr}

Let us now apply the NVE method described in Appendix \ref{NVEReview} to equations for strings in the D$p$--brane background given by (\ref{SSmetr}):
\begin{eqnarray}\label{p-brane_ansatz_1}
ds^2&=&\frac{1}{\sqrt{H}} \eta_{\mu\nu} dx^{\mu} dx^{\nu}+\sqrt{H}(dr^2+r^2d\theta^2+r^2\cos^2\theta d\Omega_{m}^2+r^2\sin^2\theta d\Omega_{n}^2).
\end{eqnarray}
To compare with \cite{StepTs} we introduce a notation $f=H^{1/4}$. The Polyakov action for the string,
\begin{equation}\label{Polyakov}
S=-\frac{1}{4 \pi \alpha^{'}} \int d\sigma d\tau G_{MN}(X) \partial_a X^{M} \partial^{a} X^{N},
\end{equation}
leads to the equations of motion and Virasoro constraints,
\bea\label{Virasoro}\label{Virasoro1}
G_{MN}\dot{X^M} X^{'N}=0,\\
\label{Virasoro2}
G_{MN}(\dot{X^M} \dot{X^N}+X^{'M} X^{'N})=0.
\eea

To demonstrate that system is not integrable, it is sufficient to look only at a specific sector and show that the number of conserved quantities does not match the number of variables. Specifically, we consider the following ansatz:
\begin{equation}\label{string_ansatz}
x^{0}=t(\tau),\quad r=r(\tau), \quad \phi=\phi(\sigma), \quad \theta=\theta(\tau).
\end{equation}
All other coordinates are considered to be constants. Substitution (\ref{string_ansatz}) into (\ref{Polyakov}) leads to the Lagrangian
\begin{equation}\label{OrLag}
L=-f^{-2}\dot{t}^2+f^2\dot{r}^2+f^2r^2(-\sin^2\theta\phi^{'2}+\dot{\theta}^2).
\end{equation}
Solving the equations of motion for cyclic variables $t,\phi$ 
\begin{eqnarray}
\dot{t}&=&Ef^2,\nonumber\\
\phi'&=&\nu=const,
\end{eqnarray}
and substituting the Virasoro constraint (\ref{Virasoro2}), 
\bea
E^2&=&\dot{r}^2+r^2\dot{\theta}^2+\nu^2r^2\sin\theta^2,
\eea
back to the original Lagrangian (\ref{OrLag}) we obtain the effective Lagrangian
\begin{equation}
L=f^2(\dot{r}^2+r^2\dot{\theta}^2-r^2\sin^2\theta\nu^2-E^2),
\end{equation}
and the corresponding Hamiltonian
\begin{equation}
H=\frac{p_r^2}{4f^2}+\frac{p_{\theta}^2}{4f^2r^2}+f^2\nu^2r^2\sin^2\theta+E^2.
\end{equation}
Hamiltonian equations for $\theta$ and $p_{\theta}$ are
\bea\label{Heqptheta}
\dot{\theta}&=&\frac{p_{\theta}}{2f^2r^2},\nonumber\\
\\
\dot{p_{\theta}}&=&\frac{f'_{\theta}}{2f^3}\left(p_r^2+\frac{p_{\theta}^2}{r^2}\right)-2ff'_{\theta}\nu^2r^2\sin^2\theta-2f^2\nu^2r^2\sin\theta\cos\theta.\nonumber
\eea

Let us choose a particular solution
\bea\label{Oct23s}
\theta=\frac{\pi}{2},\qquad p_{\theta}=0
\eea
corresponding to a string wrapped on the equator of $S^2$ and moving only in $r$. Then by shifting coordinate $\tau$ we can set $r(0)=0$, and the Virasoro constraint (\ref{Virasoro2}) gives
\begin{equation}\label{Oct23s1}
r=\hat{r}(\tau)=\frac{E}{\nu}\sin{\nu\tau}.
\end{equation}
Expanding (\ref{Heqptheta}) around solution (\ref{Oct23s})--(\ref{Oct23s1}), we get equations for variations:
\bea\label{Jj23}
\delta\dot{\theta}&=&\frac{\delta p_{\theta}}{2f^2r^2},\nonumber\\
\\
\delta\dot{p_{\theta}}&=&
\frac{f^{''}_{\theta\theta}\delta\theta}{2f^3}p_r^2-\frac{3}{2}\frac{f'_{\theta}}{f^4}f'_{\theta}\delta\theta p_r^2-2f'_{\theta}\delta\theta-2ff^{''}_{\theta\theta}\delta\theta\nu^2r^2+2f^2\nu^2r^2\delta\theta.\nonumber
\nonumber
\eea
Substituting $p_r=2f^2\dot{r}$ into the last equation, we find equation for $\delta\dot{p_{\theta}}$,
\bea\label{Jj23a}
\delta\dot{p_{\theta}}&=&
\left[ 2f^{''}_{\theta\theta}\dot{r}^2 - 6(f'_{\theta})^2\dot{r}^2-2(f'_{\theta})^2\nu^2r^2 -2ff^{''}_{\theta\theta}\nu^2r^2+2f^2\nu^2r^2 \right]\delta\theta,\nonumber
\eea
and substitution of this result into (\ref{Jj23}) leads to the NVE for $\delta\theta\equiv\eta$:
\bea
\ddot\eta+2\dot{r}\left( \frac{f'_r}{f}+\frac{1}{r} \right) \dot\eta - \left[ \frac{f^{''}_{\theta\theta}}{f} \left(\frac{\dot{r}}{r}\right)^2 -3\left(\frac{f'_{\theta}}{f}\right)^2\left(\frac{\dot{r}}{r}\right)^2-\left(\frac{f'_{\theta}}{f}\right)^2\nu^2-\frac{f^{''}_{\theta\theta}}{f}\nu^2+\nu^2 \right]\eta=0.\nonumber
\eea
Finally by changing variable $\displaystyle r=\frac{E}{\nu}\sin{\nu\tau}$ we obtain
\begin{equation}\label{NVE_frtheta}
\eta^{''}+\left[ \frac{\ddot{r}}{\dot{r}^2}+2\left(\frac{f'_r}{f}+\frac{1}{r}\right) \right] \eta' - \left[ \frac{f^{''}_{\theta\theta}}{f} \frac{1}{r^2} -3\left(\frac{f'_{\theta}}{f}\right)^2\frac{1}{r^2}-\left(\frac{f'_{\theta}}{f}\right)^2\frac{\nu^2}{\dot{r}^2}-\frac{f^{''}_{\theta\theta}}{f}\frac{\nu^2}{\dot{r}^2}+\frac{\nu^2}{\dot{r}^2} \right]\eta=0.
\end{equation}

To summarize, in this Appendix we have derived the NVE for the metric (\ref{p-brane_ansatz_1}) and the ``pulsating'' string ansatz (\ref{string_ansatz}). To analyze the resulting equation (\ref{NVE_frtheta}) the function $f$ should be specified. In the next Appendix \ref{AppNintGen} we consider a particular function $f$ corresponding to the most general geodesics-integrable harmonic function derived in section \ref{SecGnrGds}.

\subsection{Application to geometries with integrable geodesics}
\label{AppNintGen}

In this Appendix we consider the most general harmonic function $H=f^4$ leading to integrable geodesics (see (\ref{TheSolnMone}) and (\ref{TheSoln67})). First we express this particular $f$ in terms of $(r,\theta)$, then we use NVE (\ref{NVE_frtheta})  derived in the previous Appendix to check its integrability. The first step leads to
\bea
f^4&=&\frac{d^2}{\rho_+\rho_-}\Bigg\{ \tilde{Q} \left[ \left(\frac{\rho_++\rho_-}{2d}\right)^2-1 \right]^{\frac{1-n}{2}}\left[ \frac{\rho_++\rho_-}{2d} \right]^{1-m}  \nonumber\\ && + P\left[  1-\left( \frac{\rho_+-\rho_-}{2d} \right)^2 \right]^{\frac{1-m}{2}} \left[  \frac{\rho_+-\rho_-}{2d} \right]^{1-n} \Bigg\}, \quad n+m<2 \label{fnmll2}\\
f^4&=&\frac{d^2\tilde{Q}}{\rho_+\rho_-}\left[ \frac{\rho_++\rho_-}{2d} \right]^{1-m}\left[ \left(\frac{\rho_++\rho_-}{2d}\right)^2-1\right]^{\frac{1-n}{2}},\quad n+m\ge2\label{fnmge2}
\eea
where we used the mapping between $(x,y)$ and $(\rho_+,\rho_-)$, namely 
\bea
\displaystyle \cosh x=\frac{\rho_++\rho_-}{2d}, \qquad
\cos y=\frac{\rho_+-\rho_-}{2d}.
\eea
The NVEs for $P=0$ are the same in both cases (\ref{fnmll2}), (\ref{fnmge2}):
\bea\label{NVEmnge2}
\eta^{''}&+&\frac{U}{D}\eta=0,\nonumber\\
U&=&E^4 \left[-d^4(n-1)^2-2 d^2 r^2 \left((m-3) n-5 m+n^2+2\right)-r^4
   (m+n-4) (m+n)\right] \nonumber\\
   &&+2 E^2 r^2 \nu^2 \Big[d^4 ((n-2) n+5)+2 d^2 r^2 (m (n-5)+(n-4) n+9)\nonumber\\
   &&+r^4 \left(m^2+2 m (n-3)+(n-6) n+4\right)\Big]-r^4 \nu^4 \Big[d^4 (n-3) (n+1)\\&&+2 d^2
   r^2 \left((m-5) n-5 m+n^2+4\right)+r^4 \left(m^2+2 m (n-4)+(n-8) n-4\right)\Big],\nonumber\\
D&=&16 r^2 \left(d^2+r^2\right)^2 [E^2-(r \nu)^2]^2.\nonumber
\eea
Application of the Kovacic algorithm shows non--integrability of the system unless $d=0$ and $m+n=4$. The integrable case corresponds to strings on $AdS_5\times S^5$, then  
equation (\ref{NVEmnge2}),
\begin{equation}\label{NVEAdS5S5}
\eta^{''}+\frac{5x^2-2}{4(x^2-1)^2}\eta=0,\qquad x\equiv \frac{r\nu}{E},
\end{equation}
coincides with (3.26) from \cite{StepTs}. 

Non--vanishing $P$--term in (\ref{fnmll2}) for $n=0,m=0$ and $n=0,m=1$ gives rise to divergences at $\theta=\pi/2$ ($f(r,\pi/2)=0$), so the NVE around solution (\ref{Oct23s}) is not well defined. The last remaining case of (\ref{fnmll2}) is $n=1,m=0$. 
Setting $\tilde Q=0$ gives the following NVE
\bea\label{Nov5}
\eta''+\frac{E^4(3d^2+2r^2)+E^2\nu^2(2d^4-d^2r^2-5r^4)+r^2\nu^4(d^4+4d^2r^2+6r^4)}{4(d^2+r^2)^2(E^2-r^2\nu^2)^2}\eta=0.
\eea
Using the Kovacic algorithm we see that this term corresponds to an integrable system. Unfortunately the corresponding geometry is unphysical since it has singularities at arbitrarily large values of $r$. To see this we recall that metric 
(\ref{p-brane_ansatz_1}) becomes singular when $f=0$, and function $f$ vanishes at $\rho_+-\rho_-=2d$, which corresponds to 
$\theta=0$ and arbitrary value of $r$ (recall (\ref{EllHarm})). 

To summarize, in this appendix we have demonstrated that supergravity backgrounds with integrable geodesics do not lead to integrable string theories, with the exception of $AdS_5\times S^5$.

\section{Geodesics in bubbling geometries.}\label{AppBbl}

In this appendix we will present the analysis of geodesics in the 1/2--BPS geometries constructed in \cite{LLM}. Our conclusions are summarized in section \ref{Bubbles}. 

The BPS geometries constructed in \cite{LLM} are supported by the five--form field strength, and the metric is given by 
\bea\label{GenBubble}
ds^2&=&-h^{-2}(dt+V_i dx^i)^2+h^2(dY^2+dx_1^2+dx_2^2)+Ye^G d\Omega_3^2+Ye^{-G} d{\tilde\Omega}_3^2,\\
h^{-2}&=&2Y\cosh G,\qquad 
YdV=\star_3 dz,\qquad
z=\frac{1}{2}\tanh G,\nonumber
\eea
where function $z(x_1,x_2,Y)$, which satisfies the Laplace equation,
\bea\label{bbLpl}
\d_i\d_i z+Y\d_Y\left(Y^{-1}\d_Y z\right)=0,
\eea
obeys the boundary conditions 
\bea\label{Droplet}
z(Y=0)=\pm \frac{1}{2}.
\eea
As discussed in section \ref{Bubbles}, only three classes of configurations can potentially lead to integrable geodesics, and we will discuss these classes in three separate subsections.  

\subsection{Geometries with AdS$_5\times$S$^5$ asymptotics}
\label{BubblesCircles}

The boundary conditions depicted in figure \ref{Fig:Bubbles}(a) lead to geometries (\ref{GenBubble}) which are invariant under rotations in $(x_1,x_2)$ plane, and such solutions are conveniently formulated in terms of coordinates introduced in 
(\ref{RingRed}):
\bea\label{RngBubble}
ds^2=-h^{-2}(dt+V_\phi d\phi)^2+h^2(dr^2+r^2d\theta^2+r^2\cos^2\theta d\phi^2)+
Ye^G d\Omega_3^2+Ye^{-G} d{\tilde\Omega}_3^2.
\nonumber\\
\eea
The complete solution of the Laplace equation (\ref{bbLpl}) and expression for $V_\phi$ for this case were found in \cite{LLM}:
\bea\label{zV}
{z}=\frac{1}{2}+\frac{1}{2}\sum_{i=1}^n(-1)^{i+1}\left[\frac{r^2-R_i^2}{\sqrt{(r^2+R_i^2)^2-4R_i^2r^2\cos^2\theta}}-1\right],
\nonumber \\
V_\phi=-\frac{1}{2}\sum_{i=1}^n(-1)^{i+1}\left[ \frac{r^2+r_i^2}{\sqrt{(r^2+R_i^2)^2-4R_i^2r^2\cos^2\theta}} - 1 \right],
\eea 
Summation in (\ref{zV}) is performed over $n$ circles with radii $R_i$, and following conventions of \cite{LLM}, we take $R_1$ to be the radius of the largest circle. For example, a disk corresponds to one circle, a ring to two circles, and so on. 

To write the HJ equation (\ref{HJone}) in the metric 
(\ref{RngBubble}), we notice that coordinates $(t,\phi)$, 
as well as two spheres, separate in a trivial way, this gives
\bea\label{HJbblSep}
S=-Et+J\phi+S^{(3)}_{L_1}(y)+
{\tilde S}^{(3)}_{L_2}({\tilde y})+T(r,\theta),
\eea
where $S^{(3)}_{L_1}(y)$ and 
${\tilde S}^{(3)}_{L_2}({\tilde y})$ satisfy equations (\ref{HJSphere}). Then equation (\ref{HJone}) becomes
\bea\label{HJ_rings}
0&=&-\left(h^4-\frac{V_{\phi}^2}{r^2\cos^2\theta}
\right)E^2+
\left(\frac{\partial T}{\partial r}\right)^2+
\frac{1}{r^2}\left(
\frac{\partial T}{\partial \theta}\right)^2\nonumber\\
&&+2\frac{V_{\phi}}{r^2\cos^2\theta}JE+
\frac{J^2}{r^2\cos^2\theta}+
\frac{L_1^2(\frac{1}{2}-z)}{Y^2}+\frac{L_2^2(z+\frac{1}{2})}{Y^2}.
\eea
We begin with analyzing equations for geodesics with $L_1=L_2=J=0$:
\bea\label{RingsA}
(\d_r T)^2+\frac{1}{r^2}(\d_\theta T)^2
-\left(h^4-\frac{V_{\phi}^2}{r^2\cos^2\theta}\right)E^2=0,
\eea
which is similar to (\ref{TheHJ}). To evaluate the last term in (\ref{RingsA}), we need expressions (\ref{zV}) as well as relation
\bea
h^4=-\frac{(z+\frac{1}{2})(z-\frac{1}{2})}{Y^2}=
-\frac{(z+\frac{1}{2})(z-\frac{1}{2})}{r^2\sin^2\theta}.\nonumber
\eea

Applying the techniques developed in section \ref{SecGnrGds} to (\ref{RingsA}), we conclude that this equation is separable if and 
only if there exists a holomorphic function $g(w)=x+iy$, such that
\bea\label{RingsCompl}
\frac{r^2}{|g'(w)|^2}\left(h^4-\frac{V_{\phi}^2}{r^2\cos^2\theta}\right)=U_1(x)+U_2(y).
\eea
Complex variable $w$ is defined in terms of $(r,\theta)$ by (\ref{defCmplVar}). To find $g(w)$, we employ the same perturbative technique that was used to derive (\ref{TheSolnMthree})--(\ref{TheSolnMtwo}): starting with a counterpart of (\ref{AppPert}), 
\bea\label{AppPertRing}
g(w)=w+\sum_{k>0}a_k e^{-kw},\qquad
{U}_1(x)=e^{-2x}
\left[1+\sum_{k>0}b_k e^{-kx}\right],
\eea
and solving (\ref{RingsCompl}) order--by--order in $e^{-x}$, we find the necessary condition for the existence of a solution:
\bea\label{ReqnRing}
(D_2)^{k-1}D_{2(k+1)}=(D_4)^k.
\eea
Here we defined combinations of $R_j$ appearing in (\ref{zV}):
\bea\label{RiRing}
D_{p}\equiv\sum_{j=1}^n (-1)^{j+1} (R_{j})^p.
\eea
Relations (\ref{ReqnRing}) are clearly satisfied for $n=0$ (flat space) and for $n=1$ (AdS$_5\times$S$^5$), and we will now demonstrate that they fail for $n\ge 2$.

If $n\ge 2$ in (\ref{RiRing}), we can define 
$\eps\equiv R_2/R_1< 1$ and 
\bea\label{defDtld}
{\tilde D}_{p}\equiv\sum_{j=2}^n (-1)^{j+1} (R_{j})^p.
\eea
Notice that
\bea\label{IneqDtld}
|{\tilde D}_{p}|\le \sum_{j=2}^n (R_{j})^p\le
(n-1)(R_{2})^p<(n-1)(\eps R_1)^p.
\eea
Rewriting an infinite set of relations (\ref{ReqnRing}) as
\bea\label{RdfnDD}
D_2D_{2(k+1)}=D_4 D_{2k},
\eea
and extracting explicit powers of $R_1$ in the last relation we find an equation
\bea
{\tilde D}_2-R_1^{-2}{\tilde D}_4=
R_1^{2(1-k)}{\tilde D}_{2k}-R_1^{-2k}{\tilde D}_{2k+2}.
\eea
Then inequality (\ref{IneqDtld}) implies that
\bea\label{DDineq}
|{\tilde D}_2-R_1^{-2}{\tilde D}_4|\le R_1^{2(1-k)}|{\tilde D}_{2k}|+R_1^{-2k}|{\tilde D}_{2k+2}|<
(n-1)\eps^{2k}R_1^2(1+\eps^2).
\eea
The left--hand side of this inequality does not depend on $k$, and the right--hand side goes to zero as $k$ goes to infinity ($n$ is an arbitrary but fixed number), then we conclude that
\bea
{\tilde D}_2=R_1^{-2}{\tilde D}_4\quad\Rightarrow\quad
D_2=R_1^{-2}D_4.
\eea
Substituting this into (\ref{RdfnDD}), we express all 
${\tilde D}_k$ through ${\tilde D}_2$ and $R_1$:
\bea
{\tilde D}_{2k+2}=R_1^2{\tilde D}_{2k}=\dots=R_1^{2k}{\tilde D}_2,
\eea
then inequality (\ref{IneqDtld}) ensures that ${\tilde D}_2=0$:
\bea
|{\tilde D}_2|=R_1^{-2k}|{\tilde D}_{2k+2}|=
(n-1)R_1^2\eps^{2k+2}\rightarrow 0 \quad \mbox{as}\ k\rightarrow\infty.
\eea
On the other hand, the definition (\ref{defDtld}) implies that ${\tilde D}_2$ is strictly negative\footnote{Recall that $(R_2)^2>(R_3)^2>\dots (R_n)^2$.}, as long as $n\ge 2$. We conclude that relation (\ref{ReqnRing}) holds only for $n=0,1$. To show this for all values of $n$ we used an infinite set of relations (\ref{ReqnRing}), however, for any given $n$ is it sufficient to use (\ref{ReqnRing}) with $k=0,\dots,n$. Note that the equations with $k=0,1$ are trivial. 

Going back to the solution with $n=1$, we can extract the relevant holomorphic function and potential $U_1(x)$ (see (\ref{RingsCompl}))
\bea\label{hDisk}
g(w)&=&\ln\left[ \frac{1}{2}\left( e^w+\sqrt{e^{2w}-(R_1/l)^2} \right) \right],\nonumber\\
U_1(x)&=&\frac{e^{2x}E^2R_1^2}{l^2(e^{2x}+(R_1/2l)^2)^2},\quad U_2(y)=0.
\eea
We recall the $w$ is defined by (\ref{defCmplVar}), and this relation has a free dimensionful parameter $l$, which can be chosen in a convenient way. Setting $l=R_1/2$, we recover the standard elliptic coordinates (\ref{TheSolnMone}), which can also be rewritten as (\ref{ElliptCase}):
\bea\label{BblElptCrdA}
\cosh x=\frac{\rho_++\rho_-}{2d},\quad \cos y=\frac{\rho_+-\rho_-}{2d},\quad
\rho_\pm=\sqrt{r^2+R_1^2\pm 2rR_1\cos\theta}.
\eea
The relation between $(x,y)$ and standard coordinates on AdS$_5\times$S$^5$ will be discussed in section \ref{StdPar}.

We now go back to the general HJ equation (\ref{HJ_rings}) and show that presence of (angular) momenta does not spoil separation of the HJ equation for $n=1$ (\ref{hDisk}). To prove this we express the contribution from momenta as sums of two functions $\tilde{U}_1(x)+\tilde{U}_2(y)$ (analogously to (\ref{RingsCompl})). A series of transformations is resulted in
\bea
&&\frac{r^2}{|g'(w)|^2}\left(2\frac{V_{\phi}}{r^2\cos^2\theta}JE+\frac{J^2}{r^2\cos^2\theta}\right)\nonumber\\
&&\qquad=-\frac{32e^{2x}(R_1/l)^2JE}{(4e^{2x}+(R/l)^2)^2}-\left[\frac{16e^{2x}(R_1/l)^2}{(4e^{2x}+(R_1/l)^2)^2} +\frac{1}{\cos^2y} \right]J^2,\nonumber\\
&&\frac{r^2}{|g'(w)|^2}\frac{L_1^2(\frac{1}{2}-z)}{Y^2}=\frac{16L_1^2e^{2x}(R_1/l)^2}{(-4e^{2x}+(R_1/l)^2)^2},\\
&&\frac{r^2}{|g'(w)|^2}\frac{L_2^2(z+\frac{1}{2})}{Y^2}=\frac{L_2^2}{\sin^2y}.\nonumber
\eea
Clearly the right hand sides of these relations are separable.

\subsection{Geometries with pp--wave asymptotics}
\label{StripsPlane}

We will now discuss the geometries with translational $U(1)$ symmetry (\ref{StripRed}), which correspond to parallel strips in the $(x_1,x_2)$ plane (see figure \ref{Fig:Bubbles}(b,c)). It is convenient to distinguish two possibilities: $z$ can either approach different values $x_2\rightarrow \pm\infty$ (as in figure \ref{Fig:Bubbles}(b)) or approach the same value on both sides (as in figure \ref{Fig:Bubbles}(c)). In this subsection we will focus on the first option, which corresponds to geometries with plane wave asymptotics, and the second case will be discussed in subsection \ref{SecStrips}.

Pp--wave can be obtained as a limit of $AdS_5\times S^5$ geometry by taking the five--form flux to infinity \cite{BMN}. This limit has a clear representation in terms of boundary conditions in $(x_1,x_2)$ plane: taking the radius of a disk to infinity, we recover a half-filled plane corresponding to the pp--wave (see figure \ref{Fig:BubblesInt}(c)). Taking a similar limit for a system of concentric circles, we find excitations of pp--wave geometry by 
a system of parallel strips (see figure \ref{Fig:Bubbles}(b)). Since strings are integrable on the pp--wave geometry \cite{BMN}, it is natural to ask whether such integrability persists for the deformations represented in figure \ref{Fig:Bubbles}(b). In this subsection we will rule out integrability on the deformed backgrounds by demonstrating that even equations for massless geodesics are not integrable. 

Solutions of the Laplace equation (\ref{bbLpl}) corresponding to the boundary conditions depicted in figure \ref{Fig:Bubbles}(b) are 
given by \cite{LLM}
\bea\label{zVpwave}
z&=&-\frac{x_2-d_0}{2\sqrt{(x_2-d_0)^2+Y^2}}-\frac{1}{2} \sum_{i=1}^n \left[ \frac{x_2-d_{2i}}{\sqrt{(x_2-d_{2i})^2+Y^2}}-\frac{x_2-d_{2i-1}}{\sqrt{(x_2-d_{2i-1})^2+Y^2}} \right],\nonumber \\
&&\\
V_1&=&-\frac{1}{2\sqrt{(x_2-d_0)^2+Y^2}}-\frac{1}{2} \sum_{i=1}^n \left[ \frac{1}{\sqrt{(x_2-d_{2i})^2+Y^2}}-\frac{1}{\sqrt{(x_2-d_{2i-1})^2+Y^2}} \right],\nonumber\\
V_2&=&0.\nonumber
\eea
Here $n$ is the number of black strips (strip number $i$ is located at $d_{2i-1}<x_2<d_{2i}$), and $x_2=d_0$ is the boundary of the half--filled plane. We will assume that the set of $d_j$ is ordered:
\bea
d_{2N}>d_{2N-1}>\dots>d_1>d_0.
\eea
Although one can shift $x_2$ to set $d_0=0$, we will keep this value free and use the shift symmetry later to simplify some equations. 

Repeating the steps performed in the last subsection, we write the counterpart of (\ref{HJbblSep})
\bea\label{F26}
S=-Et+px_1+S^{(3)}_{L_1}(y)+
{\tilde S}^{(3)}_{L_2}({\tilde y})+T(r,\theta),
\eea
where coordinates $(r,\theta)$ are defined by (\ref{StripRed}). The HJ equation is given by the counterpart of (\ref{HJ_rings}):
\bea\label{HJppWave}
0&=&-\left(h^4-V_1^2\right)E^2+
\left(\frac{\partial T}{\partial r}\right)^2+
\frac{1}{r^2}
\left(\frac{\partial T}{\d\theta}\right)^2\nonumber\\
&&+2pV_1E+p^2+
\frac{L_1^2(\frac{1}{2}-z)}{Y^2}+\frac{L_2^2(z+\frac{1}{2})}{Y^2}.
\eea
As before, we begin with analyzing this equation for  $L_1=L_2=p=0$:
\bea\label{PwaveA}
(\d_r T)^2+\frac{1}{r^2}(\d_\theta T)^2
-\left(h^4-V_1^2\right)E^2=0.
\eea
Results of section \ref{SecGnrGds} ensure that equation (\ref{PwaveA}) is separable if and only if there exists a holomorphic function $g(w)$, such that
\bea\label{PwaveCompl}
\frac{r^2}{|g'(w)|^2}\left(h^4-V_1^2\right)=U_1\left(x\right)+
U_2\left(y\right),
\eea
where complex variable $w$ is defined by (\ref{defCmplVar}) and $x+iy=g(w)$.
Solving equation (\ref{PwaveCompl}) in perturbation theory, we find relations for $d_i$, which are much more complicated that (\ref{ReqnRing}). Introducing the convenient notation $D_{p}=\displaystyle\sum_{j=1}^{2n} (-1)^{j} (d_{j})^p$ as in the previous subsection we find two options for the relations between $D_i$:
\bea\label{CondStripsPlane1}
&&D_4=\frac{D_2^3+D_3^2}{D_2},\quad
D_5=2D_2D_3+\frac{D_3^3}{D_2^2},\quad
D_6=D_2^3+3D_3^3+\frac{D_3^4}{D_2^3}, \\
&&D_7=\frac{D_3(D_2^3+D_3^2)(3D_2^3+D_3^2)}{D_2^4},\quad
D_8=D_2^4+6D_2D_3^2+\frac{5D_3^4}{D_2^2}+\frac{D_3^6}{D_2^5},\quad\dots\nonumber 
\eea
or
\bea\label{CondStripsPlane2}
&&D_4=\frac{D_2^2}{2}+\frac{8D_3^2}{9D_2},\quad
D_5=\frac{5D_2D_3}{6}+\frac{20D_3^3}{27D_2^2},\quad
D_6=\frac{D_2^3}{4}+D_3^2+\frac{16D_3^4}{27D_2^3},\\
&&D_7=\frac{7D_3(9D_2^3+8D_3^2)^2}{972D_2^4},\quad
D_8=\frac{D_2^4}{8}+\frac{8D_2D_3^2}{9}+\frac{80D_3^4}{81D_2^2}+\frac{256D_3^6}{729D_2^5},\quad\dots\nonumber
\eea
The systems (\ref{CondStripsPlane1}) and (\ref{CondStripsPlane2}) are complicated, but they can be analyzed using Mathematica, and 
here we just quote the result: equation (\ref{PwaveCompl}) has no solutions if $n>1$ in (\ref{zVpwave}). 

One special case can be studied analytically: setting $D_1=D_3=0$ in (\ref{CondStripsPlane1}) and (\ref{CondStripsPlane2}) we find simple sets of relations:
\bea\label{Oct14}
D_{2i+1}=0,\quad D_{2i}=D_2^i,\quad\mbox{or}\quad
D_{2i+1}=0,\quad D_{2i}=\frac{D_2^i}{2^{i-1}}, \quad i\ge 1.
\eea
Interestingly, the same system of equations (\ref{CondStrips}) will be encountered in the next subsection\footnote{Solutions (\ref{Oct14}) correspond to $D_1=1$ and to $D_1=2$ in (\ref{CondStrips}).}, where we will prove that 
$n\le 1$.

For $n=0$ equation (\ref{PwaveCompl}) gives no restrictions on $g(w)$ since $h^4-V_1^2=0$ for the pp--wave. Going back to (\ref{HJppWave}) and requiring it to separate for all $(p,L_1,L_2)$, we find the standard pp--wave coordinates:
\bea
ds^2&=&-2dt dx_1-(r_1^2+r_2^2)dt^2+dr_1^2+r_1^2d\Omega^2+dr_2^2+r_2^2d{\tilde\Omega}^2,\\
&&r_1=x,\quad r_2=y.\nonumber
\eea
To write the solution for $n=1$, it is convenient to choose $d_1=0$ by shifting $x_2$. This gives for (\ref{PwaveCompl})
\bea\label{hOneStripHalfPlane}
g(w)&=&w+\ln\left[\frac{1}{2}\left(\sqrt{1-\frac{d_0}{l}e^{-w}}+1\right)\right] +\ln\left[\frac{1}{2}\left(\sqrt{1-\frac{d_2}{l}e^{-w}}+1\right)\right]\\
U_1(x)&=&-\frac{128d_0 d_2 l^2 e^{2x} E^2}{(d_0 d_2-16l^2e^{2x})^2},\quad U_2(y)=0.\nonumber
\eea 
Analysis of the HJ equation (\ref{HJppWave}) with non-vanishing momenta $p,L_1,L_2$ is much more complicated than in case of geometries with $AdS_5\times S^5$ asymptotics, so we performed perturbative analysis. Specifically we compared the function ${\tilde g}(w)$ separating momenta terms with $g(w)$ from (\ref{hOneStripHalfPlane}) separating the rest of the HJ equation (\ref{HJppWave}) and saw that $g(w)$ and ${\tilde g}(w)$ are not compatible for $n=1$.

We conclude that solution with $n=0$ (the standard pp--wave) always gives separable geodesics, solution with $n=1$ leads to integrable geodesics only for vanishing momenta, and solutions with $n\ge 2$ never gives integrable geodesics. 

\subsection{Geometries dual to SYM on a circle}
\label{SecStrips}

Finally we consider configuration depicted in figure \ref{Fig:Bubbles}(c). As discussed in \cite{LLM}, these configurations are dual to Yang--Mills theory on $S^3\times S^1\times R$, and since we are only keeping zero modes on the sphere, the solutions (\ref{GenBubble}) correspond to BPS states in two--dimensional gauge theory on a circle.   

Following the discussion presented in the last subsection, we arrive at equations (\ref{HJppWave}) and (\ref{PwaveA}), however, instead of (\ref{zVpwave}) we find
\bea\label{zVstrip}
z&=&-\frac{1}{2} \sum_{i=1}^n \left[ \frac{x_2-d_{2i}}{\sqrt{(x_2-d_{2i})^2+Y^2}}-\frac{x_2-d_{2i-1}}{\sqrt{(x_2-d_{2i-1})^2+Y^2}} \right],\nonumber \\
V_1&=&-\frac{1}{2} \sum_{i=1}^n \left[ \frac{1}{\sqrt{(x_2-d_{2i})^2+Y^2}}-\frac{1}{\sqrt{(x_2-d_{2i-1})^2+Y^2}} \right],\\
V_2&=&0.\nonumber
\eea
where $n$ is the number of black strips (strip number $i$ is located at $d_{2i-1}<x_2<d_{2i}$), and we assume that the set 
of $d_j$ is ordered:
\bea\label{OrderStrip}
d_{2n}>d_{2n-1}>\dots>d_1.
\eea
Solving equation (\ref{PwaveCompl}) in perturbation theory (here we again start with (\ref{HJppWave}) in absence of momenta), we find a sequence of restrictions on the set of $d_i$. Introducing a convenient notation 
\bea\label{IneqD}
D_{p}=\sum_{j=1}^{2n} (-1)^{j} (d_{j})^p,
\eea
we can write the first three equations as
\bea\label{EqnForDst}
D_1^2D_4-2D_1D_3D_2+D_2^3=0,\nonumber\\
D_1^3D_5-D_1^2D_3^2-D_1D_3D_2^2+D_2^4=0,\\
D_1^4 D_6 - 3 D_1^2 D_2 D_3^2 + D_3 D_2^3 D_1^2=0.
\nonumber
\eea
Notice that inequality (\ref{OrderStrip}) ensures that $D_1>0$, but we can set $D_2=0$ by introducing a shift
\bea
d_n\rightarrow d_n+\alpha.
\eea 
Indeed, $D_1$ remains invariant under this shift, and $D_2$ transforms as 
\bea
D_2\rightarrow D_2+2\alpha D_1,
\eea
so by choosing an appropriate $\alpha$, we can set $D_2=0$. This choice leads to great simplifications in (\ref{EqnForDst}), in particular, we find that $D_4=0$. We will now show that $D_{2p}=0$ for all values of $p$.

First we observe that the structure of our perturbative expansion guarantees that restriction at order $p-1$ has the form
\bea\label{Dpolyn}
D_1^{p-2}D_{p}+P[D_p,D_{p-1},\dots,D_1]=0.
\eea
where $P$ is some polynomial of $p$ arguments. Next we notice that if any configuration of strips leads to a separable Hamilton--Jacobi equation, the configuration which is obtained by the reflection about $x_2$ axis must have the same property. 
The reflection corresponds to the transformation
\bea\label{MirrorD}
d'_{i}= -d_{2N+1-i}\ \Rightarrow\ 
D'_{2k}= -D_{2k},\quad D'_{2k+1}= D_{2k+1},
\eea
so if equations (\ref{Dpolyn}) are solved by the set 
$\{D_k\}$, they should also be solved by the set $\{D'_k\}$. 
We can now use induction to demonstrate that any solution of  (\ref{Dpolyn}) with $D_2=0$ must have
\bea\label{DkZero}
D_{2k}=0\quad\mbox{for all}\quad k.
\eea
The statement is trivial for $k=1$, and we assume that it holds for all $k<K$. Then equation (\ref{Dpolyn}) for $p=2K$ becomes
\bea
D_1^{2K-2}D_{2K}+P[D_{2K-1},0,D_{2K-3},0\dots,D_1]=0
\eea
This equation is symmetric under (\ref{MirrorD}) if and only if $D_{2K}$ is equal to zero\footnote{We also used that $D_1$ is always positive.}. This completes the proof of (\ref{DkZero}) by induction.

For configurations satisfying (\ref{DkZero}) perturbative expansion becomes very simple, and restrictions of $d_i$ can be formulated as
\bea\label{CondStrips}
(D_1)^{p-1}D_{2p+1}=(D_3)^p, \quad p\ge 1.
\eea
Moreover, restrictions (\ref{DkZero}) can be used to simplify the expressions for $D_{2p+1}$. Let us demonstrate that two sets,
\bea\label{StripBBa}
\{d_{2n-1}^2,d_{2n-3}^2,\dots,d_1^2\}\quad\mbox{and}\quad
\{d_{2n}^2,d_{2n-4}^2,\dots,d_2^2\},
\eea 
contain the same elements, although these elements may appear in different order. Indeed, consider 
\bea
d_+=\max\{d_{2n-1}^2,d_{2n-3}^2,\dots,d_1^2\},\quad
d_-=\max\{d_{2n}^2,d_{2n-4}^2,\dots,d_2^2\}
\eea
and assume that $d_+>d_-$. Then
\bea
D_{2p}=\sum_{j=1}^{n}
\left[(d_{2j-1})^{2p}-(d_{2j})^{2p}\right]\ge
(d_+)^{2p}-(n-1)(d_-)^{2p}
\eea
becomes positive for sufficiently large $p$, thus relations (\ref{DkZero}) imply that $d_+\le d_-$. Similar argument 
shows that $d_+\ge d_-$, so $d_+=d_-$ and using (\ref{OrderStrip}) we conclude that $d_{2n}=-d_1$. Repeating this argument for the sets 
\bea\label{StripBB}
\{d_{2n-3}^2,\dots,d_1^2\}\quad\mbox{and}\quad
\{d_{2n-2}^2,\dots,d_2^2\},
\eea 
we find $d_{2n-2}=-d_3$, and continuing this procedure we conclude that\footnote{Notice that symmetry (\ref{MirrorD}) and conditions (\ref{DkZero}) are not sufficient for this conclusion: equation (\ref{OrderStrip}) played a crucial role in or derivation.}
\bea\label{ddOne}
d_{j}= -d_{2n+1-j}\,.
\eea
For such distribution we find an alternative expression for $D_{2p+1}$:
\bea\label{IneqDprm}
D_{2p+1}=2\sum_{j=n+1}^{2n} (-1)^{j+1} (d_{j})^{2p+1}
\eea
Combining (\ref{IneqD}) and (\ref{ddOne}), we conclude that 
\bea\label{OrderDDsr}
d_{2n}>d_{2n-1}>\dots>d_{n+1}>0,
\eea
so equations (\ref{CondStrips}), (\ref{IneqDprm}) are very similar to equations (\ref{RiRing}), (\ref{ReqnRing}), and we can used the same logic\footnote{The proof would not work without inequality (\ref{OrderDDsr}).}  to conclude that $n=1$.  

To separate equation (\ref{PwaveA}) for one strip, it is convenient to set $d_1=0$ by shifting the origin of $x_2$. Then we find
\bea\label{h_one_strip}
g(w)&=&2\ln\left[ \frac{1}{2}
\left(\sqrt{e^w-(d_2/l)}+e^{w/2}\right)\right],\nonumber\\
U_1(x)&=&\frac{8d_2le^xE^2}{(d_2+4le^x)^2},\quad U_2(y)=0.
\eea

Now we go back to the original HJ equation (\ref{HJppWave}) containing non-vanishing momenta $L_1,L_2,p$ and demonstrate that momenta do not spoil separability. Here we consider the separable case of one strip.

Based on the logic used through the entire chapter separability of each momentum $p,L_1,L_2$ requires 
\bea\label{ppWaveMomenta}
\frac{r^2}{|g'(w)|^2}\left[
2pV_1E+p^2+\frac{L_1^2(\frac{1}{2}-z)}{Y^2}+\frac{L_2^2(z+\frac{1}{2})}{Y^2}
\right] =\tilde{U}_1(x)+\tilde{U}_2(y),
\eea
Expressing the left--hand side of (\ref{ppWaveMomenta}) in terms of $(x,y)$ and using the holomorphic function $g(w)$ from (\ref{h_one_strip}), we find
\bea
{\tilde U}_1(x)&=&p^2\left(\frac{d_2^4}{256l^4}e^{-2x}+e^{2x}\right)+\frac{8L_1^2d_2e^x[(d_2/l)^2+16e^{2x}]}{l[(d_2/l)^2-16e^{2x}]^2}+
\frac{8d_2L_2^2e^x}{l(d_2/l+4e^x)^2},\nonumber\\
{\tilde U}_2(y)&=&-2pE\frac{d_2}{l}\cos y-\frac{p^2}{8l^2}d_2^2\cos2y+\frac{L_2^2}{\sin^2y}.
\eea
Thus addition of momenta does not spoil separability.

\bigskip

All results obtained in this appendix can be summarized by listing all bubbling geometries leading to separable geodesics (here we use the language introduced in \cite{LLM} to describe the solutions):
\begin{enumerate}
\item $AdS_5\times S^5$: the boundary conditions are given by the disk depicted in figure \ref{Fig:BubblesInt}(a).
\item Pp--wave: the boundary conditions are depicted in figure \ref{Fig:BubblesInt}(c).
\item Single M2 brane: the boundary conditions (one strip) are depicted in figure \ref{Fig:BubblesInt}(b).
\item Pp--wave with an additional strip (see figure \ref{Fig:BubblesInt}(d)): geodesics are only separable in all momenta and angular momenta vanish.
\end{enumerate}

		\chapter{Appendix for Killing--Yano tensors in string theory}

\section{Conformal transformations of Killing tensors}
\label{AppConfResc}

In this appendix we analyze the behavior of Killing vectors and tensors under conformal rescaling of the metric. In the context of string theory such rescalings appear when one goes from the string to the Einstein frame or when one compares the string frames before and after S duality. In this appendix we will find the restrictions on the dilaton which guarantee that Killing vectors and tensors survive after S duality. We study general conformal Killing vectors and tensors, and reduction to the standard objects is obtained by setting the conformal factors to zero.

\subsection{Killing vectors}\label{AppConfRescKV}

We begin with considering an equation for the conformal Killing vector (CKV):
\bea\label{AppConfRescCKV}
\nabla_M V_{N}+\nabla_N V_{M}=2g_{M N}v
\eea
and writing its counterpart in the rescaled metric:
\bea\label{AppConfRescConfResc}
g'_{M N}=e^C g_{M N}:\qquad \nabla'_M V'_{N}+\nabla'_N V'_{M}=2g'_{MN}v'.
\eea
Recalling the transformation of the connections,
\bea\label{AppConfRescConnResc}
(\Gamma^M_{NP})'=\Gamma^M_{NP}+\frac{1}{2}\left[\delta^M_P\d_N C+\delta^M_N\d_P C-
g_{N P}g^{M A}\d_A C\right].
\eea
we can rewrite the equation for $V'$ in terms of the original covariant derivatives:
\bea
\nabla_M(e^{-C} V'_{N})+\nabla_N (e^{-C}V'_{M})=
2g_{MN}(v'+\frac{1}{2}V'^{A}\d_A e^{-C}).
\eea
Comparing this to (\ref{AppConfRescCKV}), we find the transformation law for the CKV:
\bea\label{AppSdualKVa}
V'_M&=&e^C V_M\quad \Rightarrow\quad V'^M=g'^{MN}V'_N=V^M,\nn
v'&=&v-\frac{1}{2}V^A \d_A e^{-C}.
\eea
This implies that CKV always survives the conformal rescaling, but the KV (which must have $v=0$) disappears unless
\bea
V^A \d_A e^{-C}=0.
\eea
In the context of S duality and transition between string and Einstein frames, the last condition implies that Lie derivatives of the dilaton along the Killing vector must vanish, which is a very natural requirement.

\subsection{Killing(--Yano) tensors}\label{AppCKYT}

Next we look at transformation properties of the conformal Killing--Yano tensor, which satisfies equation
\bea\label{AppConfRescCKYT}
\nabla_M Y_{NP}+\nabla_N Y_{MP}=
2g_{MN}W_P-g_{MP}W_N-g_{NP}W_M.
\eea
Using \eqref{AppConfRescConnResc} we can rewrite the left hand side of (\ref{AppConfRescCKYT}) in the rescaled frame as
\bea
&&\nabla'_M Y'_{N P}+\nabla'_N Y'_{M P}=
\nabla_M Y'_{N P}-\frac{1}{2}\left[
\d_M C Y'_{N P}+\d_N CY'_{M P}-g_{M N}g^{AB}\d_B CY'_{AP}
\right]\nonumber\\
&&\qquad-\frac{1}{2}\left[
\d_M C Y'_{NP}+\d_P CY'_{NM}-g_{M P}g^{AB}\d_B CY'_{NA}
\right]+(M \leftrightarrow N)\nonumber\\
&&=\nabla_M Y'_{NP}-\frac{3}{2}\d_MCY'_{NP}+\frac{1}{2}g_{MN}g^{AB}\d_B C Y'_{AP}+\frac{1}{2}g_{MP}g^{AB}\d_B C Y'_{NA}+(M \leftrightarrow N)\nonumber
\eea
and the full equation becomes
\bea
&&e^{3C/2}\nabla_M (e^{-3C/2}Y'_{NP})+
e^{3C/2}\nabla_N (e^{-3/2C}Y'_{MP})\nonumber\\
&&\qquad=2g_{MN}(W'_P e^C-\frac{1}{2}g^{AB}\d_B C Y'_{AP})-
\left[g_{MP}(W'_N e^C-\frac{1}{2}g^{AB}\d_B C Y'_{AN})+(M\leftrightarrow N)\right]\nonumber.
\eea
To recover the original equation \eqref{AppConfRescCKYT}, we must set
\bea\label{eqnA8}
Y'_{NP}=e^{3C/2}Y_{NP}, \quad W'_M=e^{C/2}(W_M+\frac{1}{2}g^{AB}\d_B CY_{AM}).
\eea
The conformal Killing--Yano tensors of higher rank can be analyzed in a similar fashion, and for the rank $k$ tensor we find
\bea\label{May15c}
Y'_{M_1\dots M_k}&=&e^{(k+1)C/2}Y_{M_1\dots M_k}	,\\
W'_{M_2\dots M_k}&=&e^{(k-1)C/2}W_{M_2\dots M_k}+\frac{e^{(3k-5)C/2}}{2}g^{AB}\d_B CY_{A{M_2\dots M_k}}\nonumber
\eea
The same calculations show that for Killing tensors we have
\be\label{May15d}
K'_{MN}=e^{2C}K_{MN}, \quad W'_M=e^C(W_M+g^{AB}\d_B C K_{AM}).
\ee
Equations (\ref{May15c}) and (\ref{May15d}) summarize the behavior of Killing(--Yano) tensors under conformal rescalings.


\section{Killing tensors and ellipsoidal coordinates}

In this appendix we will justify the procedure for extracting separation of variables from a nontrivial Killing tensor and review an example of ellipsoidal coordinates and their degeneration.
 
\subsection{Ellipsoidal coordinates from Killing tensors}
\label{AppEllipsKYT}

As discussed in section \ref{SecSeparKT}, existence of a non--trivial Killing tensor leads to separation of variables, and in this appendix we will provide some details of the procedure for extracting the relevant coordinates and the separation function. 

We will focus on studying the reduced metric (\ref{KillLaMain}), and  to simplify the notation we will drop the subscript $red$. Assuming that non--cyclic coordinates separate, we find
\bea\label{KTApr22}
ds^2=\sum g_k dx_k^2,\quad 
K=\sum \Lambda_k g_k dx_k^2
\eea
where $g_k$ and $\Lambda_k$ are functions of all coordinates. 
Equations for the Killing tensor give
\bea\label{Apr22a}
\d_i\Lambda_i=0,\quad \d_j\ln g_i=\d_j\ln(\Lambda_i-\Lambda_j),\quad 
j\ne i
\eea
and there are no summations in these relations. We will now make an additional assumption of separability:
\bea\label{g1ElptSepar}
\d_j\d_k\ln g_m=0\quad\mbox{for different}\quad (i,j,k).
\eea
and determine the form of $g_k$ and $\Lambda_k$. The procedure involves several steps:
\begin{enumerate}[1.]
\item Equation (\ref{g1ElptSepar}) leads to factorization of $g_1$ 
\bea\label{g1Elpt}
g_1=\prod f_{1j}(x_1,x_j),\quad \d_j \ln f_{1j}(x_1,x_j)=\d_j \ln(\Lambda_1-\Lambda_j)
\eea
which implies factorization of
\bea\label{temp12}
\Lambda_1-\Lambda_2=f_{12}(x_1,x_2)g_{12}(x_1,x_3\dots).
\eea
The same expression can also be obtained by starting with $g_2$, but this leads to a different factorization:
\bea\label{temp12a}
\Lambda_1-\Lambda_2=f_{21}(x_2,x_1)g_{21}(x_2,x_3\dots).
\eea
Applying $\d_1\d_3$ to the logs of (\ref{temp12}), (\ref{temp12a}), we conclude that $x_1$ dependence factorizes in $g_{12}$. Absorbing the $x_1$--dependent factor in $f_{12}(x_1,x_2)$, we find
\bea
&&\Lambda_1-\Lambda_2=f_{12}(x_1,x_2)g_{12}(x_3\dots).\nonumber
\eea
The left--hand side of the last relation is killed by $\d_1\d_2$ (recall the first relation in (\ref{Apr22a})), so
\bea
f_{12}(x_1,x_2)=f_{12}^{(1)}(x_1)-f_{12}^{(2)}(x_2).
\eea
Repeating the same steps for $x_3,\dots, x_n$, we conclude that 
\bea\label{eqn23}
&&g_1=h_1(x_1)\prod_j [f_{1j}^{(1)}(x_1)-f_{1j}^{(j)}(x_j)],\nonumber\\
&&\Lambda_1-\Lambda_j=[f_{1j}^{(1)}(x_1)-f_{1j}^{(j)}(x_j)]g_{1j}(x_3\dots x_n),\quad
\d_j g_{1j}=0.
\eea
Since coordinate $x_1$ is not special, the last equation can be generalized:
\bea\label{eqn23a}
\Lambda_k-\Lambda_j=[f_{kj}^{(k)}(x_k)-f_{kj}^{(j)}(x_j)]g_{kj}(x_1\dots x_n),\quad
\d_j g_{kj}=\d_k g_{kj}=0.
\eea

\item
Assuming that $f_{1j}^{(j)}(x_j)$ are nontrivial functions of their arguments\footnote{This assumption of generality eventually leads to ellipsoidal coordinates for curved spaces. Relaxing this assumption, one arrives at degenerate cases, and some examples are presented in Appendix \ref{AllFlatEllips}. We conjecture that any degenerate case can be obtained by a singular limit of ellipsoidal coordinates, but we will not prove this statement. The proof for flat three dimensional space is implicitly contained in \cite{MorseFesh}.}, we can define new coordinates by setting
\bea
{\tilde x}_j\equiv f_{1j}^{(j)}(x_j),\quad j>1.
\eea
and dropping the tildes. We still have the freedom of making a linear transformation of $x_k$, which will be fixed later. Taking a second derivative of (\ref{eqn23}) with respect 
to $x_j$,
\bea
\d^2_j\Lambda_1=\d^2_j\Lambda_j+\d^2_j\Big([f_{1j}^{(1)}(x_1)-x_j]g_{1j}(x_2\dots x_n)\Big)=0.\nonumber
\eea
we conclude that $\Lambda_1$ is a linear polynomial in every coordinate  
$(x_2,\dots x_n)$.
Furthermore, since $\d_2\Lambda_2=0$ we find
\bea
\Lambda_2=\Lambda_1-[f_{12}^{(1)}(x_1)-x_2]g_{12}(x_3,\dots, x_n)=
\Lambda_1+[f_{12}^{(1)}(x_1)-x_2]\d_2 \Lambda_1\nonumber
\eea
and similarly
\bea
\Lambda_j=\Lambda_1+[f_{1j}^{(1)}(x_1)-x_j]\d_j \Lambda_1.
\eea
\item
Next we look at 
\bea
&&\Lambda_2-\Lambda_3=(f_{12}^{(1)}(x_1)-x_2)\d_2 \Lambda_1-
(f_{13}^{(1)}(x_1)-x_3)\d_3 \Lambda_1\nn
&&\quad=[f_{12}^{(1)}\d_2\Lambda_1-f_{13}^{(1)}\d_3\Lambda_1]_0+
x_3[f_{12}^{(1)}\d_2\d_3\Lambda_1+\d_3\Lambda_1]_0-
x_2[f_{13}^{(1)}\d_2\d_3\Lambda_1+\d_2\Lambda_1]_0\nonumber
\eea
Expressions in the square brackets are evaluated at $x_2=x_3=0$.  
Equation (\ref{eqn23a}) implies that $(x_2,x_3)$ dependence in the last equation must factorize, and this is possible only if
\bea\label{eqn23c}
f_{13}^{(1)}(x_1)=c_{32}f_{12}^{(1)}(x_1)+d_{32},\quad 
[f_{12}^{(1)}\d_2\Lambda_1-f_{13}^{(1)}\d_3\Lambda_1]_0=e_{32}
\eea
with \textit{constant} $(c_{32},d_{32},e_{32})$. Similar arguments demonstrate that all 
$f_{1j}(x_1)$ are linear polynomials in $f_{12}^{(1)}(x_1)$, so by re-defining this coordinate,
\bea
x_1\rightarrow f_{12}^{(1)}(x_1),\nonumber
\eea
we conclude that all $f_{1j}(x_1)$ are linear functions of their arguments. For example,
\bea
f_{13}^{(1)}(x_1)-x_3=c_{32}x_1+d_{32}-x_3,\nonumber
\eea
so by making a linear transformation of $x_3$, we can simplify the last expression:
\bea
f_{13}^{(1)}(x_1)-x_3\rightarrow c_{32}(x_1-x_3).\nonumber
\eea
Repeating this for $(x_4\dots x_n)$, we find
\bea\label{eqn23b}
g_1=h_1(x_1)\prod_j [x_1-x_j],\qquad
\Lambda_j=\Lambda_1+[x_1-x_j]\d_j \Lambda_1.
\eea
\item
We will now demonstrate that polynomial $\Lambda_1(x_2,\dots,x_n)$ must be symmetric under interchange of any pair of its arguments. Without the loss of generality, we focus on $x_2$ and $x_3$ and write $\Lambda_1$ as
\bea
\Lambda_1=P_{1}x_2x_3+P_{2}x_2+P_{3}x_3+P_{4},
\eea
where $P_k$ are polynomials in $(x_4\dots x_n)$. The second equation in 
(\ref{eqn23c}) gives
\bea\label{eqn23e} 
e_{32}=x_1[\d_2\Lambda_1-\d_3\Lambda_1]_0=x_1[P_2-P_3].
\eea
Consistency of this relation requires $P_2=P_3$, i.e., symmetry of $\Lambda_1$ under the interchange of $x_2$ and $x_3$.
\item
Once we established that $\Lambda_1(x_2\dots x_n)$ is symmetric, it is convenient to introduce a ``generating" linear polynomial $\Lambda(x_1\dots x_n)$ symmetric in its arguments and define
\bea
\Lambda_1=\d_1 \Lambda.
\eea
Then the second relation in (\ref{eqn23b}) implies
\bea
\Lambda_j=\d_1\Lambda+(x_1-x_j)\d_1\d_j\Lambda=
\d_1\Lambda|_{x_j=0}+x_1\d_1\d_j\Lambda=
\d_j\Lambda|_{x_1=0}+x_1\d_1\d_j\Lambda=\d_j\Lambda.\nonumber
\eea
\end{enumerate}

To summarize, we have demonstrated that in the generic case existence of the Killing tensor in the non--cyclic part of the metric (\ref{KTApr22}) implies that 
\bea
g_k=h_1(x_k)\prod_{j\ne k} [x_k-x_j],\qquad
\Lambda_j=\d_j\Lambda,
\eea
where $\Lambda(x_1\dots x_n)$ is a linear polynomial in every $(x_1\dots x_n)$ symmetric under interchange of every pair of arguments. This completes the justification of (\ref{KillLaMain})--(\ref{SeparHJmain}), which summarize the extraction of the separable coordinates from a Killing tensor.

\subsection{Ellipsoidal coordinates in flat space}
\label{AllFlatEllips}

In section \ref{SecSeparKT} we demonstrated that separation of non--cyclic coordinates generically leads to ellipsoidal coordinates. Our derivation was based on the assumption of generality: we postulated that metric components have non--trivial dependence on all non--cyclic coordinates. If this assumption is dropped, one recovers degenerate cases of ellipsoidal coordinates, and in this appendix we will illustrate this using a well-known example of flat three--dimensional space. Degeneration in higher dimensions is very similar, but its detailed discussion is beyond the scope of this work.

Consider a flat three--dimensional space with a metric
\bea\label{EllAppFlat}
ds^2=dr_1^2+dr_2^2+dr_3^2.
\eea
The ellipsoidal coordinates $(x_0,x_1,x_2)$ are defined as three solutions of a cubic equation for $x$ \cite{Jacobi}:
\bea
\frac{r_1^2}{x-a}+\frac{r_2^2}{x-b}+
\frac{r_3^2}{x-c}=1,
\eea
Without the loss of generality we assume that non--degenerate coordinates have $a> b> c$ and the roots are arranged in the following order:
\bea\label{ElpsdRngKYT}
x_0>a>x_1>b>x_2>c.
\eea
Cartesian coordinates $(r_1,r_2,r_3)$ can be expressed in terms of $(x_0,x_1,x_2)$ as
\bea\label{EllipsoidalCoords}
r_1=\left[\frac{(x_0-a)(x_1-a)(x_2-a)}{
(a-b)(a-c)}\right]^{1/2},\quad
r_2=\left[\frac{(x_0-b)(x_1-b)(x_2-b)}{
(b-a)(b-c)}\right]^{1/2},\quad
\eea
\bea
r_3=\left[\frac{(x_0-c)(x_1-c)(x_2-c)}{
(c-a)(c-b)}\right]^{1/2}.\nonumber
\eea
This transformation turns the metric (\ref{EllAppFlat}) into
\bea\label{EllAppElps}
ds^2&=&\frac{(x_0-x_1)(x_0-x_2)dx_0^2}{4(x_0-a)(x_0-b)(x_0-c)}+
\frac{(x_1-x_0)(x_1-x_2)dx_1^2}{4(x_1-a)(x_1-b)(x_1-c)}\nn
&&+\frac{(x_2-x_0)(x_2-x_1)dx_2^2}{4(x_2-a)(x_2-b)(x_2-c)}.
\eea
Shifting six quantities $(x_i,a,b,c)$ by $c$, one usually sets $c=0$, and we will follow this convention\footnote{In section \ref{SecMyersPerry} we use a different convention: $a=0$, $b=-a_1^2$, $c=-a_2^2$.}. 

The degenerate cases of the ellipsoidal coordinates are discussed in great detail in \cite{MorseFesh}\footnote{There are ten of them:
rectangular, oblate/prolate spheroidal, circular/elliptic/parabolic cylinder, spherical, conical, paraboloidal, and parabolic.}, and we will focus only on oblate spheroidal and spherical coordinates. Oblate spheroidal coordinates are obtained from \eqref{EllipsoidalCoords} by writing
\bea
x_0=a+\xi_0,\quad x_1=a-a\xi_1,\quad x_2=b \xi_2
\eea
and sending $b$ to zero. Then metric (\ref{EllAppElps}) becomes
\bea
ds^2&=&\frac{(\xi_0+a\xi_1)d\xi_0^2}{4\xi_0(\xi_0+a)}+
\frac{(\xi_0+a\xi_1)d\xi_1^2}{4\xi_1(1-\xi_1)}
+(\xi_0+a)(1-\xi_1)\frac{d\xi_2^2}{4\xi_2(1-\xi_2)}.
\eea
This expression has a very simple interpretation: $\xi_2$ gives rise to a new cyclic coordinate $\zeta$ ($\xi_2=\cos^2\zeta$), while $(\xi_0,\xi_1)$ form two--dimensional elliptic coordinates. This is in a perfect agreement with general analysis of non--cyclic directions presented in section \ref{SecSeparKT}.   

As a next example we consider spherical coordinates, which can be obtained by writing
\bea
b=a-\epsilon,\quad x_0=\xi_0,\quad x_1=a-\epsilon \xi_1,\quad x_2=a\xi_2,
\eea
sending $\epsilon$ to zero, and setting $a=0$ in the resulting expression. This gives
\bea
ds^2&=&\frac{d\xi_0^2}{4\xi_0}+
\frac{\xi_0(1-\xi_2)d\xi_1^2}{4\xi_1(1-\xi_1)}
+\frac{\xi_0d\xi_2^2}{4(1-\xi_2)\xi_2}=dr^2+r^2\sin^2\theta d\phi^2+d\theta^2.
\eea
We see that although $\xi_2$ (which is related to the polar angle $\theta$) remains a non--cyclic coordinate, it does not appear in $g_{11}$, so spherical coordinates violate one of the assumptions made in section \ref{SecSeparKT}. Nevertheless such parameterization can be obtained as a degenerate case of ellipsoidal coordinates, and we conjecture that any separable frame in the non--cyclic coordinates can be obtained as a similar singular limit from the systems derived in section \ref{SecSeparKTsub}. The proof of this conjecture is beyond the scope of this work.



\section{Principal CKYT for the Myers--Perry black hole}
\label{AppPCKYT}

In section \ref{SecMyersPerry} we found a family of the Killing--Yano tensors (\ref{Kub2}) for the Myers--Perry black hole, and the construction was based on three statements:
\begin{enumerate}
\item The anti--symmetric tensor $h$ defined by (\ref{Kub1}) is a Conformal Killing--Yano tensor and the form (\ref{Kub1}) is closed. Such tensors are called Principal Conformal Killing--Yano tensors (PCKYT) \cite{Kub1}.
\item A wedge product of two PCKY tensors is again a PCKYT\footnote{Note that this is not true for KY tensors.}, so the expression $\wedge h^n$ is a PCKYT for any value of $n$.
\item If ${\mathcal Y}$ is a PCKYT then $Y=\star {\mathcal Y}$ is a Killing--Yano tensor.
\end{enumerate}
The proofs of these statements are scattered throughout the literature 
\cite{Carig,Kub1,KubThes}, and the goal of this appendix is to present a simpler derivation of properties 1-3. We will begin with properties 2 and 3 since they are not specific to the Myers--Perry black hole.

\bigskip

We begin with writing the condition $d{\mathcal Y}=0$ for a Principal Conformal Killing--Yano tensor ${\mathcal Y}$ of rank $p$:
\bea\label{Jun10}
\nabla_a {\mathcal Y}_{bcd\dots}-\nabla_b {\mathcal Y}_{acd\dots}-\nabla_c {\mathcal Y}_{bad\dots}+\dots=0.
\eea
There are $p$ terms in this equation. Using the defining relation (\ref{CKTdefKYT}) for the CKYT,
\bea\label{Jun10a}
\nabla_b {\mathcal Y}_{acd\dots}=-\nabla_a {\mathcal Y}_{bcd\dots}+2 g_{ab}Z_{cd\dots}-
[g_{ca}Z_{bd\dots}+g_{cb}Z_{ad\dots}]+\dots,
\eea
equation (\ref{Jun10}) can be rewritten as
\bea\label{Jun10b}
\nabla_a {\mathcal Y}_{bcd\dots}=g_{ab}Z_{cd\dots}-g_{ac}Z_{bd\dots}+\dots=
p g_{a[b}Z_{cd\dots]}.
\eea
The PCKYT is defined as an object satisfying relations (\ref{Jun10}), (\ref{Jun10a}), but one can use the equivalent set of defining relation (\ref{Jun10}) and (\ref{Jun10b}) instead. In particular, we observe that any Killing--Yano tensor which is also closed must be covariantly constant. Such objects are closely related to complex structures on K\"ahler manifolds, which are discussed in the Appendix \ref{AppComplexStructure}. 

To prove property 2, we observe that a product of two PCKYT, ${\mathcal Y}^{(p)}\wedge {\mathcal Y}^{(q)}$ is closed, and it satisfies equation (\ref{Jun10b}) with 
\bea
Z^{(p+q)}=\frac{1}{p+q}\left[p Z^{(p)}\wedge {\mathcal Y}^{(q)}+(-1)^{p+q}q Z^{(q)}\wedge {\mathcal Y}^{(p)}\right].
\eea
To prove property 3, we consider
\bea
\nabla_m\left[\eps_{a_1\dots a_q}{}^{b_1\dots b_p}{\mathcal Y}_{b_1\dots b_p}\right]=
\eps_{a_1\dots a_q}{}^{b_1\dots b_p}p g_{m[b_1}Z_{b_2\dots b_p]}=
p \eps_{a_1\dots a_q m}{}^{b_2\dots b_p}Z_{b_2\dots b_p}.
\eea
Symmetrization over $(m,a_1)$ gives zero, so 
$Y_{a_1\dots a_q}\equiv \eps_{a_1\dots a_q}{}^{b_1\dots b_p}{\mathcal Y}_{b_1\dots b_p}$ is a Killing--Yano tensor. This completes the proof of properties 2 and 3 which hold for all spaces admitting PCKYT.

\bigskip

Next we focus on the Myers--Perry black hole and demonstrate that the closed form
\bea
h&=&\h \sum a_id\mu_i^2\wedge\left[a_i dt+(r^2+a_i^2)d\phi_i\right]+
\h dr^2\wedge\left[dt+a_i\mu_i^2 d\phi_i\right]\nonumber\\
&=&\h d[r^2+\sum a_i^2\mu_i^2]\wedge dt+\h\sum  d[a_i(r^2+a_i^2)\mu_i^2]\wedge d\phi_i
\eea
is a Conformal Killing--Yano tensor. The proof will go in two steps: first we will verify the CKYT equation for $m=0$, and then we will show that $m$ dependence does not affect the result.

For $m=0$ the geometry (\ref{MPodd}) is flat, and it is convenient to rewrite it in the Cartesian coordinates. In odd dimensions such coordinates are defined by
\bea
X_k+iY_k=\sqrt{r^2+a_k^2}\mu_k e^{i\phi_k},\quad ds^2=-dt^2+\sum [(dX_k)^2+(dY_k)^2],
\eea
and the two--form $h$ becomes
\bea
h=\h d[\sum (X_k^2+Y_k^2)]\wedge dt+\sum a_k dX_k\wedge dY_k\,.
\eea
This gives interesting relations for the derivatives of $h_{MN}$,
\bea
&&\nabla_M h_{NP}+\nabla_N h_{MP}=0,\quad\mbox{if}\quad (MNP)\ne t,\nn
&&\nabla_M h_{Nt}+\nabla_N h_{Mt}=2[\delta_{MN}-\delta_{Mt}\delta_{Nt}],\\
&&\nabla_M h_{tP}+\nabla_t h_{MP}=-[\delta_{MP}-\delta_{Mt}\delta_{Pt}],\nonumber
\eea
which can be summarized as an equation for the CKYT (\ref{CKTdefKYT}):
\bea\label{PrincKYT}
\nabla_M h_{NP}+\nabla_N h_{MP}=2g_{MN}Z_P-g_{MP}Z_N-g_{NP}Z_M,\quad 
Z^M\d_M=\d_t.
\eea
The argument for even dimensions works in a similar way. 
This concludes the first part of the proof ($h$ is a CKYT for the flat space), and now we will demonstrate that (\ref{PrincKYT}) holds for $m\ne 0$ as well.

While it is possible to verify (\ref{PrincKYT}) using the explicit form of the Christoffel's symbols\footnote{Such `brute force' calculation is performed in the Appendix B.3 of \cite{KubThes}.}, this calculation is tedious and not very instructive since it does not take advantage of the high degree of symmetry of the Myers--Perry solution. We will use an alternative method based on spin connections, which gives the answer in an easier and more transparent way. First we rewrite (\ref{PrincKYT}) in terms of frame indices:
\bea\label{AddPCKYTframes}
&&\nabla_a h_{bc}+\nabla_b h_{ac}=2\eta_{ab}Z_c-\eta_{ac}Z_b-\eta_{bc}Z_a\\
&&h=re^{\hat r}\wedge e^{\hat t}+\sum_i\sqrt{-x_i}e^{{\hat x^i}}\wedge e^{\hat i},\quad Z_a=e_{at}
\nonumber
\eea
To derive the desired result we should analyze the $m$--dependence of
\bea\label{DefTcky}
T_{abc}\equiv\nabla_a h_{bc}+\nabla_b h_{ac}
\eea
Covariant derivatives of the objects with frame indices are evaluated using the standard relations
\bea
\nabla_a V^b=e_a^M\d_M V^b+\omega_{a,}{}^b{}_c V^c,\quad
\nabla_a W_b=e_a^M\d_M W_b-\omega_{a,}{}^c{}_b W_c,
\eea
and the spin connection $\omega_{a,}{}^b_{e}$ is related to the anholonomy coefficients $\Gamma_{a,be}$ by
\bea\label{SpinConn}
\omega_{c,ab}=\frac{1}{2}\left[\Gamma_{c,ab}+\Gamma_{b,ac}-\Gamma_{a,bc}\right],\quad
de^a=\frac{1}{2}\Gamma^a{}_{,bc}e^b\wedge e^c
\eea
In particular, the explicit expressions for $e^{\hat t}$ in (\ref{AllFramesMP}) and (\ref{AllFramesMPOdd}) imply that 
\bea
\Gamma^{\hat{t}}{}_{,\beta \hat{r}}=0.
\eea
Here and below the greek letters denote the frame indices excluding $(\hat r,\hat t)$. Although it is not obvious from $e^{\hat i}$ and $e^{\hat \phi_i}$, the anholonomy coefficients $\Gamma_{\alpha,\hat t\gamma}$ vanish as well. To see this, we use an 
alternative expression for $\Gamma$:
\bea
\Gamma_{a,bc}&=&(de_a)_{\mu\nu}e^\mu_be^\nu_c=(\d_\mu e_{a\nu}-\d_\nu e_{a\mu})e^\mu_be^\nu_c=
-e^\mu_be_{a\nu}\d_\mu e^\nu_c+e^\mu_ce_{a\nu}\d_\mu e^\nu_b,
\eea
which gives for (\ref{AllFramesMP}) and (\ref{AllFramesMPOdd}):
\bea
\Gamma_{\alpha,\hat t\gamma}&=&e^\mu_\gamma e_{\alpha\nu}\d_\mu e^\nu_{\hat{t}}=-\frac{1}{2}
e^\mu_\gamma e_{\alpha\nu} e^\nu_{\hat{t}} \d_\mu\ln F=-\frac{1}{2}
e^\mu_\gamma \eta_{\alpha\hat{t}}\, \d_\mu\ln F=0.
\eea
Next we use the frames (\ref{AllFramesMP}) and (\ref{AllFramesMPOdd}) to compute the anholonomy $\Gamma_{a,bc}$ coefficients and spin connections $\omega_{a,bc}$ in terms of their counterparts ${\tilde\Gamma}_{a,bc}$ and ${\tilde\omega}_{a,bc}$ for $m=0$. The simplicity of the $m$ dependence in the frames combined with relations $\Gamma_{\alpha,\hat t\gamma}=\Gamma^{\hat{t}}{}_{,\beta \hat{r}}=0$ allows us to write the answers without doing complicated calculations which are normally associated with evaluation of the spin connection. Introducing convenient notation
\bea
S=\left\{
\begin{array}{ll}
\sqrt{\frac{R-mr}{R}},&\mbox{even\ }d \\
\sqrt{\frac{R-mr^2}{R}},&\mbox{odd\ }d
\end{array}
\right.\,,
\eea
we can summarize the anholonomy coefficients as
\bea\label{SpConGam}
&&\Gamma^\alpha{}_{,\beta\gamma}={\tilde\Gamma}^\alpha{}_{,\beta\gamma},\quad
\Gamma^\alpha{}_{,\beta \hat r}=S{\tilde\Gamma}^\alpha{}_{,\beta \hat r},\quad
\Gamma^{\hat{r}}{}_{,\beta \hat{r}}={\tilde\Gamma}^{\hat{r}}{}_{,\beta \hat{r}},\nonumber\\
&&\Gamma^\alpha{}_{,\beta \hat{t}}=0,\quad
\Gamma^{\hat t}{}_{,\beta \hat t}={\tilde\Gamma}^{\hat t}{}_{,\beta t},\quad
\Gamma^{\hat{t}}{}_{,\beta \gamma}=S{\tilde\Gamma}^{\hat{t}}{}_{,\beta \gamma}\\
&&\Gamma^{\hat{t}}{}_{,\beta \hat{r}}=0,
\quad\Gamma^\alpha{}_{,\hat{t}\hat{r}}={\tilde\Gamma}^\alpha{}_{,\hat{t}\hat{r}},\quad
\Gamma^{\hat{t}}{}_{,\hat{t} \hat{r}}=S{\tilde\Gamma}^{\hat{t}}{}_{,\hat{t} \hat{r}}-\frac{1}{F}\d_r S,\quad \Gamma^{\hat{r}}_{\hat{r}\hat{t}}=0,
\nonumber
\eea
and the spin connections as
\bea\label{SpConOm}
&&\omega_{\alpha,\beta\gamma}={\tilde\omega}_{\alpha,\beta\gamma},\quad
\omega_{\hat{r},\alpha\beta}=S {\tilde\omega}_{\hat{r},\alpha\beta},\quad
\omega_{\alpha,\hat{r}\beta}=S{\tilde\omega}_{\alpha,\hat{r}\beta},\quad 
\omega_{\hat{r},\hat{r}\beta}={\tilde\omega}_{\hat{r},\hat{r}\beta},\nn
&&
\omega_{\hat{t},\hat{t}\alpha}={\tilde\omega}_{\hat{t},\hat{t}\alpha},\quad 
\omega_{\hat{t},\alpha\beta}=S{\tilde\omega}_{\hat{t},\alpha\beta},
\quad \omega_{\alpha,\beta \hat{t}}=S{\tilde \omega}_{\alpha,\beta \hat{t}},\\
&&\omega_{\alpha,\hat{r}\hat{t}}=
{\tilde\omega}_{\alpha,\hat{r}\hat{t}},\quad \omega_{\hat{r},\alpha \hat{t}}=
{\tilde\omega}_{\hat{r},\alpha \hat{t}},\quad \omega_{\hat{t},\alpha \hat{r}}=
{\tilde\omega}_{\hat{t},\alpha \hat{r}},\quad \omega_{\hat{r},\hat{r}\hat{t}}=0,\quad \omega_{\hat{t},\hat{t}\hat{r}}=\Gamma_{\hat{t},\hat{t}\hat{r}}.\nonumber
\eea
Substituting the expressions (\ref{SpinConn}) into (\ref{DefTcky}) and introducing ${\hat\d}_a\equiv e_a^M\d_M$, we find
\bea\label{AddJun8}
T_{abc}&=&{\hat\d}_a h_{bc}+{\hat\d}_b h_{ac}+\left[\omega_{a,be}+\omega_{b,ae}\right]h^e{}_c+
\omega_{a,ce}h_b{}^e+\omega_{b,ce}h_a{}^e\nn
&=&{\hat\d}_a h_{bc}+{\hat\d}_b h_{ac}+(\Gamma_{a,be}+\Gamma_{b,ae})h^e{}_c+
\omega_{a,ce}h_b{}^e+\omega_{b,ce}h_a{}^e
\eea
and the explicit expressions (\ref{SpConGam}), (\ref{SpConOm}) give
\bea
T_{\alpha\beta\gamma}={\tilde T}_{\alpha\beta\gamma},\quad
T_{\alpha {\hat m}{\hat n}}={\tilde T}_{\alpha {\hat m}{\hat n}},\quad 
T_{{\hat m}{\hat n}\alpha}={\tilde T}_{{\hat m}{\hat n}\alpha},\quad
T_{{\hat m}\beta\gamma}=S{\tilde T}_{{\hat m}\beta\gamma},\quad
T_{\alpha\beta {\hat m}}=S{\tilde T}_{\alpha\beta {\hat m}},
\eea
where $({\hat m},{\hat n})$ take values ${\hat t}$ or ${\hat r}$. The remaining components are
\bea
T_{\hat{r}\hat{r}\hat{r}}&=&0,\quad T_{\hat{t}\hat{t}\hat{t}}=0,\quad
T_{\hat{r}\hat{t}\hat{t}}=0,\quad T_{\hat{t}\hat{r}\hat{t}}=0,\nn
T_{\hat{r}\hat{r}\hat{t}}
&=&2{\hat\d}_{\hat{r}} h_{\hat{r}\hat{t}}+2\Gamma_{\hat{r},\hat{r}\hat{r}}h^{\hat{r}}{}_{\hat{t}}+
2\omega_{\hat{r},\hat{t}\hat{t}}h_{\hat{r}}{}^{\hat{t}}=2{\hat\d}_{\hat{r}} h_{\hat{r}\hat{t}}=S{\tilde T}_{\hat{r}\hat{r}\hat{t}},\\
T_{\hat{t}\hat{r}\hat{r}}
&=&{\hat\d}_{\hat{r}} h_{\hat{t}\hat{r}}+\Gamma_{\hat{t},\hat{r}\hat{t}}h^{\hat{t}}{}_{\hat{r}}+
2\omega_{\hat{t},\hat{r}\hat{t}}h_{\hat{r}}{}^{\hat{t}}={\hat\d}_{\hat{r}} h_{\hat{t}\hat{r}}=S{\tilde T}_{\hat{t}\hat{r}\hat{r}}\,.\nonumber
\eea
Recalling that 
\bea
Z_\alpha=e_{\alpha t}={\tilde Z}_\alpha,\quad Z_{\hat t}=e_{\hat t t}=S{\tilde Z}_{\hat t},\quad 
Z_{\hat r}=0,
\eea
we conclude that equation (\ref{AddPCKYTframes}),
\bea
T_{abc}=2\eta_{ab}Z_c-\eta_{ac}Z_b-\eta_{bc}Z_a,
\eea
is equivalent to 
\bea
{\tilde T}_{abc}=2\eta_{ab}{\tilde Z}_c-\eta_{ac}{\tilde Z}_b-\eta_{bc}{\tilde Z}_a,
\eea
which has been verified earlier. This completes the proof of the relation (\ref{PrincKYT}) for the Myers--Perry black hole and verification of statements 1-3 made in the beginning of this appendix.



\section{Dimensional reduction and T duality}\label{AppDimRed}

This appendix discusses dimensional reduction of equations for Killing vectors, Killing--(Yano) tensors and their conformal counterparts. Section \ref{AppDRconv} sets up the conventions, section \ref{AppDimRed1} discusses dimensional reduction of arbitrary tensors, and these results are applied to Killing vectors in section \ref{AppDRKV}, to symmetric Killing tensors in section \ref{AppDRKT4}, and to Killing--Yano tensors in section \ref{AppBeqKYT}. Conformal Killing tensors are discussed in section \ref{AppCKT}, conformal Killing vectors are analyzed in section \ref{SubsectionCKV} and some comments about conformal Killing--Yano tensors are made in the end of section \ref{SectionModifiedKYT}.

We demonstrate that equations for the KV and KT are consistent with T duality, but equation for the KYT should be modified, and we find the unique modification. Also we find that consistency between continuous symmetries and T duality leads to constraints on the Kalb--Ramond field if one is present, and such constraints suggest an interesting generalization of a standard Lie derivative along vector field to the derivative along Killing tensors. This construction is discussed in section \ref{AppLieDer}.

\subsection{Conventions}
\label{AppDRconv}

We begin with setting up the conventions. Consider a geometry which admits a Killing vector $\d_z$ and 
write the metric and the Kalb--Ramond field in the form
\bea\label{AppDimRedBdy}
\label{DimRedSetup}
ds^2&=&e^C[dz+A_mdx^m]^2+\hat g_{mn}dx^mdx^n,\nn
B&=&{\tilde A}_n dx^n\wedge [dz+\frac{1}{2}A_m dx^m]+
\frac{1}{2}{\hat B}_{mn}dx^m\wedge dx^m.
\eea
Here $(m,n)$ run over all coordinates excluding $z$, and an unusual notation for $B$ field will be justified below. Ramond--Ramond fields may also be present, but they will not affect our discussion. For future reference we also write the metric and its inverse in matrix form:
\bea\label{SetupBefore}
g_{MN}=\begin{pmatrix}
  e^C & e^C A_i\\
              e^CA_j & e^CA_iA_j+\hat{g}_{ij}\end{pmatrix}, \qquad
g^{MN}=\left(\begin{array}{cc}
	 e^{-C}+A_iA^i & -A^i\\
              -A^j & \hat{g}^{ij}\end{array}\right). 
\eea
Since $z$ is a cyclic coordinate in (\ref{DimRedSetup}), it is possible to perform T duality along this direction using the Buscher's rules
\bea\label{Buscher}\label{TdualityMap}
\tilde{g}_{zz}&=&\frac{1}{g_{zz}},\quad e^{2\tilde{\Phi}}=\frac{e^{2\Phi}}{g_{zz}},\quad
\tilde{g}_{m z}=\frac{B_{m z}}{g_{zz}},\qquad \tilde{B}_{m z}=\frac{g_{m z}}{g_{zz}},
\\
\tilde{g}_{mn}&=&g_{mn}-\frac{G_{m z}G_{n z}-B_{m z}B_{n z}}{g_{zz}},\quad
\tilde{B}_{mn}=B_{mn}-\frac{B_{m z}g_{n z}-g_{m z}B_{n z}}{g_{zz}}.\nonumber
\eea
Application of this procedure to  (\ref{DimRedSetup}) gives
\be\label{SetupAfter}
d\tilde s^2=e^{-C}(dz+{\tilde A}_mdx^m)^2+\hat g_{mn}dx^mdx^n, \quad 
\tilde{B}={A}_n dx^n\wedge [dz+\frac{1}{2}\tilde{A}_m dx^m]+
\frac{1}{2}{\hat B}_{mn}dx^m\wedge dx^m. 
\ee
Notice that $A_m$ and ${\tilde A}_m$ are interchanged by T duality making the notation (\ref{DimRedSetup}) very natural. 

\bigskip
\noindent
In this work we use the following conventions:
\begin{itemize} 
\item capital letters run through all the coordinates, $\{M,N, ... \}= \{ 1,...,d\}$;
\item lower case letters run through all the coordinates except $z$, $\{ m, n, ... \}= \{ 1,...,d-1\}$;
\item objects after T duality are marked with tilde, e.g. $\tilde V_i$, $\tilde K_{mn}$;
\item objects not affected by T duality are marked by hat, e.g. $\hat g_{ij}$, $\hat\nabla_m$.
\end{itemize}


\subsection{Dimensional reduction and covariant derivatives}

\label{AppDimRed1}

In this appendix we will express covariant derivatives in the geometry (\ref{DimRedSetup}) 
in terms of derivatives on the base $d{\hat s}^2$ assuming that all objects are 
$z$--independent. 

We begin with analyzing covariant derivatives of a vector:
\bea
W_{MN}=\nabla_M V_N.
\eea
The connections corresponding to the metric  (\ref{DimRedSetup}) are:
\bea
&&\Gamma^z_{zz}=\frac{1}{2}A^a\d_a e^C,\quad
\Gamma^m_{zz}=-\frac{1}{2}{\hat g}^{ma}\d_a e^C,\quad
\Gamma^z_{mz}=\frac{1}{2}\left[\d_m C-A^a e^C F_{ma}-2A^aA_{[a}\d_{m]}e^C\right],
\nonumber\\
&&\Gamma^m_{nz}=\frac{1}{2}{\hat g}^{ma}(e^C F_{na}-A_{n}\d_{a}e^C),\quad
\Gamma^z_{mn}=-A_a {\hat\Gamma}^a_{mn}+
\frac{1}{2}e^{-C}(\d_m g_{nz}+\d_n g_{mz})\\
&&\Gamma^s_{mn}={\hat\Gamma}^s_{mn}-\frac{1}{2}A^s(\d_m g_{nz}+\d_n g_{mz}).\nonumber
\eea
Indices of the gauge field $A_i$ are raised using ${\hat g}^{ij}$, and ${\hat\Gamma}^s_{mn}$ denotes Christoffel symbols on the base.

Explicit calculations give various components of $W$:
\bea\label{DimRedVectApp}
W_{zz}&=&\frac{1}{2}V^a\d_a e^C,
\nonumber\\
{W_z}^n&=&\frac{1}{2}{\hat g}^{nb}e^C F_{ab}V^a-
\frac{1}{2}{\hat g}^{na}V_z\d_a C,\\
{W^n}_z&=&\hat{g}^{na}\d_a V_z-\frac{1}{2}\hat{g}^{na}V_z \d_a C-\frac{1}{2}\hat{g}^{nb}e^C F_{ba}V^a,\nonumber\\
W^{mn}&=&{\hat \nabla}^m V^n
+\frac{1}{2}\hat{g}^{ma}\hat{g}^{nb} F_{ab}V_z.\nonumber
\eea
All components of $W_{MN}$ can be obtained by taking linear combinations of the expressions written above, for example,
\bea\label{WzmDown}
W_{zm}&=&g_{ma}{W_z}^a+g_{mz}{W_z}^z=g_{ma}{W_z}^a+\frac{g_{mz}}{g_{zz}}\left[W_{zz}-g_{az}{W_z}^a\right]=
A_m W_{zz}+{\hat g}_{ma}{W_z}^a\nonumber\\
&=&-\frac{1}{2}V_z\d_m C-\frac{1}{2}e^C F_{ma}V^a+\frac{1}{2}A_m V^a\d_a e^C.
\eea
The relation (\ref{DimRedVectApp}), (\ref{WzmDown}) are used in section \ref{SecKVDuality}. While discussing conformal Killing vectors in section \ref{SubsectionCKV} we also need generalization of (\ref{DimRedVectApp}) to derivatives of a $z$--dependent vector:
\bea\label{AppzdepKVred}
W_{zz}&=& \d_z V_z+\h  V^a\d_a e^C,\nn
W^m{}_{z}+W_{z}{}^m&=& \hat{g}^{ma}\left[\d_aV_z-\d_a CV_z-e^C F_{ab}V^b+\hat{g}_{ab}\d_z V^b-A_a\d_z V_z\right],\\
W^{mn}+W^{nm}&=& \hat\nabla^m V^n+\hat\nabla^n V^m-A^m\d_z V^n-A^n\d_z V^m.\nonumber
\eea

Once the action of covariant derivatives on various types of indices is specified, their application to a tensor of rank 2 becomes straightforward:
\bea\label{DimRedL}
\nabla_z L_{zz}&=&\frac{1}{2}[{L^a}_z+{L_z}^a]\d_a e^C,\nonumber\\
\nabla_z {L_z}^n&=&\frac{1}{2}{L^{an}}\d_a e^C+\frac{1}{2}{\hat g}^{nb}e^C F_{ab}{L_z}^a-
\frac{1}{2}{\hat g}^{na}L_{zz}\d_a C,\nonumber\\
\nabla_z L^{mn}&=&\frac{1}{2}[{\hat g}^{mb}L^{an}+{\hat g}^{nb}L^{ma}]e^C F_{ab}-
\frac{1}{2}[{\hat g}^{ma}{L_z}^n+{\hat g}^{na}{L^m}_z]\d_a C,\nonumber\\
\nabla^n {L_{zz}}&=&\hat{g}^{na}\d_a L_{zz}-\hat{g}^{na}L_{zz} \d_a C-\frac{1}{2}\hat g^{nb}e^C F_{ba}[{L^a}_z+{L_z}^a],
\\
\nabla^m {L_z}^n&=&{\hat\nabla}^m {L_z}^n-\frac{1}{2}\hat{g}^{ma}{L_z}^n \d_a C-\frac{1}{2}\hat{g}^{mb}e^C F_{ba}L^{an}
+\frac{1}{2}\hat{g}^{ma}\hat{g}^{nb} F_{ab}L_{zz},\nonumber\\
{\nabla}^m L^{np}&=&{\hat \nabla}^m L^{np}
+\frac{1}{2}\hat{g}^{ma}\hat{g}^{nb} F_{ab}{L_z}^p
+\frac{1}{2}\hat{g}^{ma}\hat{g}^{pb} F_{ab}{L^n}_z.\nonumber
\eea
These formulas are used in section \ref{SecKillingsDualities} to study the reduction of Killing--(Yano) tensors.

\subsection{Dimensional reduction for Killing vectors}
\label{AppDRKV}

In this subsection we will consider the behavior of Killing vectors under T duality. We will start with an object which satisfies the Killing equation
\be\label{AppKVeq}
\nabla_M V_N+\nabla_N V_M=0,
\ee
in the geometry (\ref{AppDimRedBdy}) supported by the NS--NS fields.  
T duality along $z$ direction gives the geometry (\ref{SetupAfter}) which has the same form with replacements
\bea\label{TdualityMapBdy}
C\rightarrow -C,\quad A\leftrightarrow {\tilde A},\quad e^{2\phi}\rightarrow e^{2\phi-C},\quad
\mbox{fixed}\quad {\hat g}_{mn},\ {\hat B}_{mn},
\eea 
If present, Ramond--Ramond fields would also transform under such duality, but such fields will not affect our analysis. 

Let us assume that before T duality geometry (\ref{AppDimRedBdy}) admitted a Killing vector that satisfied equation
\bea\label{KVeqnBdy}
Z_{MN}=0,\qquad Z_{MN}\equiv \nabla_M V_N+\nabla_N V_M.
\eea
As demonstrated in section \ref{AppDimRed1}, equation (\ref{KVeqnBdy}) can be written as a system\footnote{This follows from (\ref{DimRedVectApp}) by noticing that $Z_{MN}=W_{MN}+W_{NM}$.}
\bea\label{Oct3}
Z_{zz}&=&V^a\d_a e^C=0,\nonumber\\
Z^{mn}&=&{\hat \nabla}^m V^n+{\hat \nabla}^n V^m=0,\\
{Z_z}^m&=&{\hat g}^{ma}\d_a(e^{-C}V_z)-
{\hat g}^{mb} F_{ba}V^a=0.\nonumber
\eea
T duality (\ref{TdualityMapBdy}) leaves the first two equations invariant as long as we make identification
\bea
{\tilde V}^a=V^a, 
\eea
and it maps the last equation (\ref{Oct3}) into a restriction on the $B$ field:
\bea
{\hat g}^{ma}\d_a {\tilde W}_z+{\hat g}^{mb} 
{\tilde H}_{baz}V^a=0,
\qquad {\tilde W}_z\equiv -e^{-C}V_z.
\eea
Similarly, before the T duality we must have
\bea
{\hat g}^{ma}\d_a {W}_z+{\hat g}^{mb} 
{H}_{baz}V^a=0,
\qquad {W}_z\equiv -e^{C}{\tilde V}_z.
\eea
The last equation is a $(mz)$ component of a covariant relation:
\bea\label{BfieldKV}
H_{MNS}V^S=\nabla_M W_N-\nabla_N W_M,
\eea
as now we will discuss its origin and implications coming from the remaining components. 

To give a geometrical interpretation of (\ref{BfieldKV}) we look at a Lie derivative of the $B$ field along the Killing vector $V$:
\bea
{\mathcal L}_V B_{MN}&=&V^A\nabla_A B_{MN}+B_{AN}\nabla_M V^A+B_{MA}\nabla_N V^A\nonumber\\
&=&
V^A H_{MNA}-\nabla_M(V^AB_{AN})+\nabla_N(V^AB_{AM})\nonumber
\eea
and recall that if $V^A$ is a Killing vector, then this derivative must be a pure gauge, i.e.,
\bea\label{KillDerBfield}
{\mathcal L}_V B_{MN}=\nabla_M W'_N-\nabla _N W'_M
\eea
for some vector $W'_M$. Combining the last two relations, we find 
\bea
V^A H_{MNA}=\nabla_M (W'_N+V^AB_{AN})-\nabla _N (W'_M+V^AB_{AM}),\nonumber
\eea
which coincides with (\ref{BfieldKV}) if we define
\bea\label{WgaugeKV}
W_N=W'_N+V^AB_{AN}.
\eea
At this point we have demonstrated that condition (\ref{BfieldKV}) comes from requiring that the Lie derivative of the $B$ field is a pure gauge, and we found the T duality map for various components of $V$ and $W$:
\bea
{\tilde V}^a=V^a,\quad {\tilde W}_z= -e^{-C}V_z,\quad 
{\tilde V}_z= -e^{-C}W_z.
\eea
To complete the proof that the system
\bea\label{KVArray}
\left\{
\begin{array}{l}
\nabla_M V_N+\nabla_N V_M=0\\
H_{MNS}V^S=\nabla_M W_N-\nabla_N W_M
\end{array}
\right.
\eea
remains invariant, we have to analyze the $(mn)$ components of the last equation and find the map between $W_m$ and ${\tilde W}_m$. 

Let us start with a $B$ field that satisfies the constraint (\ref{BfieldKV}) in the original frame. In particular this implies 
\bea
\nabla_m W_n-\nabla_n W_m=H_{mna}V^a+H_{mnz}V^z=
[d{\hat B}+\frac{1}{2}d({\tilde A}\wedge A)]_{mna}V^a+
{\tilde F}_{mn}V^z.
\eea
Assuming that the counterpart of this relation after T duality is also satisfied, we can subtract it from the last relation to find
\bea
&&\nabla_m (W_n-{\tilde W}_n)-
\nabla_n (W_m-{\tilde W}_m)=[d({\tilde A}\wedge A)]_{mna}V^a+
{\tilde F}_{mn}V^z-F_{mn}{\tilde V}^z\nonumber\\
&&\qquad={\tilde F}_{mn}[e^{-C}V_z-A_aV^a]-
F_{mn}[e^C{\tilde V}_z-{\tilde A}_a V^a]+[d({\tilde A}\wedge A)]_{mna}V^a\nonumber\\
&&\qquad={\tilde F}_{mn}e^{-C}V_z-F_{mn}e^C{\tilde V}_z-
[{\tilde F}_{ma}A_n-F_{ma}{\tilde A}_n-(m\leftrightarrow n)]V^a.
\eea
Using the last equation in (\ref{Oct3}) and its counterpart after T duality, we can simplify the last bracket:
\bea
&&\nabla_m (W_n-{\tilde W}_n)-
\nabla_n (W_m-{\tilde W}_m)\nonumber\\
&&\qquad={\tilde F}_{mn}e^{-C}V_z-F_{mn}e^C{\tilde V}_z-
[\d_m(e^{C}{\tilde V}_z)A_n-\d_m(e^{-C}V_z){\tilde A}_n-(m\leftrightarrow n)]
\nonumber\\
&&\qquad=\d_m[{\tilde A}_n e^{-C}V_z-A_n e^C{\tilde V}_z]-
\d_n[{\tilde A}_m e^{-C}V_z-A_m e^C{\tilde V}_z].
\eea
We conclude that the system (\ref{KVArray}) remains invariant under T duality if the standard rules (\ref{TdualityMapBdy}) are supplemented by 
\bea
&&{\tilde V}^a=V^a,\quad {\tilde W}_z= -e^{-C}V_z,\quad 
{\tilde V}_z= -e^{-C}W_z,\nonumber\\
&&{\tilde W}_n=W_n-{\tilde A}_n e^{-C}V_z-A_n W_z+\d_n f,
\eea
where $f$ is an arbitrary function. The last line can also be written as
\bea
{\tilde W}^n=W^n+{\hat g}^{na}\d_a f,
\eea
and the transformation law can be made symmetric between $V$ and $W$ by setting $f=0$.

\subsection{Dimensional reduction of the Killing tensor equation}
\label{AppDRKT4}

Next we look at the equation for the Killing tensor:
\bea\label{AppKTEq}
M_{MNP}=0,\quad M_{MNP}\equiv\nabla_M K_{NP}+\nabla_N K_{MP}+\nabla_P K_{MN}, 
\quad K_{MN}=K_{NM}.
\eea
Assuming that geometry (\ref{DimRedSetup}) does not have a $B$ field and that all components of $K_{MN}$ are $z$--independent, we can use (\ref{DimRedL}) to perform dimensional reduction along $z$ direction:
\bea\label{DimRedKeqn}
M_{zzz}&=&{K^a}_z\d_a e^C,\nonumber\\
{M_{zz}}^p&=&{K^{an}}\d_a e^C+2{\hat g}^{pb}e^C F_{ab}{K_z}^a-
2{\hat g}^{pa}K_{zz}\d_a C+{\hat g}^{pa}\d_a K_{zz},\\
M^{mn}{}_z&=&{\hat\nabla}^m {K_z}^n+{\hat\nabla}^n {K_z}^m
+[{\hat g}^{mb}K^{an}+{\hat g}^{nb}K^{ma}]e^C F_{ab}-
[{\hat g}^{ma}{K_z}^n+{\hat g}^{na}{K_z}^m]\d_a C,\nonumber\\
M^{mnp}&=&{\hat \nabla}^m K^{np}+{\hat \nabla}^n K^{mp}+{\hat \nabla}^p K^{mn}.\nonumber
\eea
and match equations for the Killing tensor before and after the duality:
\bea\label{AppAllKTEqs}
\begin{array}{|r|l|l|}
\hline
zzz&K_z^{\ t}\d_t e^C=0&\tilde K_z{}^t\d_t e^{-C}=0\Big.\\
zzp&2{\hat g}^{pa}F_{ba}K_z^{\ b}e^{-C}= \d_a e^{-C} K^{ap}- 
{\hat g}^{pa}\d_a (e^{-2C}K_{zz})&
\d_a e^C \tilde K^{ap}-{\hat g}^{pa}\d_a(e^{2C}\tilde K_{zz})=0\Big.\\
mnz&{\hat g}^{ma}\left[\hat\nabla_a(e^{-C}K^n{}_{z})+F_{ba}K^{nb}\right]+(m\leftrightarrow n)=0&
\hat\nabla^m(e^{C}\tilde K^n{}_{z})+(m\leftrightarrow n)=0\Big.\\
mnp&{\hat\nabla}^mK^{np}+{\hat\nabla}^nK^{mp}+{\hat\nabla}^p K^{mn}=0&
{\hat\nabla}^m\tilde K^{np}+{\hat\nabla}^n\tilde K^{mp}+{\hat\nabla}^p\tilde K^{mn}=0
\Big.
\\
\hline
\end{array}\nonumber
\eea
From $mnp$ components we obtain 
\bea
\tilde K^{mn}=K^{mn}.
\eea
Next we rewrite the $(mnz)$ components before T duality using the relation 
${\tilde H}_{mnz}=F_{mn}$:
\bea
g^{ma}\left[{\tilde H}_{abz}K^{nb}-{\hat\nabla}_a (e^{-C}K^n_{\ z})\right]+
(m\leftrightarrow n)=0.
\eea
Using the general reduction (\ref{DimRedL}) \textit{after} duality, we find
\bea
{\tilde\nabla}^m {L_z}^n&=&{\hat\nabla}^m {L_z}^n+\frac{1}{2}g^{ma}{L_z}^n \d_a C
\quad\Rightarrow\quad
{\hat\nabla}^m {L_z}^n=e^{C/2}{\tilde\nabla}^m [e^{-C/2}{L_z}^n],\nonumber
\eea
and applying this relation to ${L_z}^n=e^{-C/2}K^n_{\ z}$, we find a constraint on the Kalb--Ramond field after duality.\footnote{Recall that ${\tilde K}^{mn}={K}^{mn}$, 
so we can write the left hand side of (\ref{addGuee}) in terms of dual variables.}
\bea\label{addGuee}
{\tilde g}^{ma}{\tilde H}_{abz}{\tilde K}^{nb}+
{\tilde g}^{na}{\tilde H}_{abz}{\tilde K}^{mb}=e^{C/2}{\tilde\nabla}^m [e^{-C/2}K^n_{\ z}]+
e^{C/2}{\tilde\nabla}^n [e^{-C/2}K^m_{\ \ z}]
\eea
The only covariant extension of this equation for the $B$--field is\footnote{As a consistency check, we note that the trivial 
Killing tensor $\tilde K_{MN}=g_{MN}$ does not give any restriction on the $B$ field.}
\bea\label{AppBEqGuessFinal}
{\tilde H}_{AMP}\tilde K_N{}^A+{\tilde H}_{ANP}\tilde K_M{}^A=e^{C/2}{\tilde\nabla}_M[e^{-C/2}{\tilde W}_{NP}]+e^{C/2}{\tilde\nabla}_N[e^{-C/2}{\tilde W}_{MP}].
\eea
Equation (\ref{addGuee}) recovers the $(mnz)$ component of this constraint, but other components require additional analysis. Here we just mention that the constraint (\ref{AppBEqGuessFinal}) admits a special solution
\be\label{SpecSolnConstr}
\begin{aligned}
&\tilde K^n{}_z=0, \quad W^n{}_z=-e^{-C}K^n{}_z, \quad W_{zz}=0,\quad W_{mn}=0,\\
&F_{pa}K_n{}^a-F_{na}K_p{}^a+2F_{np}\tilde K_z{}^z=\tilde \nabla_n (-e^{-C}K_{zp})-\tilde \nabla_p (-e^{-C}K_{zn}),\\
&\d_a e^C g_{pb}K^{ab}-\d_p(e^{2C}\tilde K_{zz})=0.
\end{aligned}
\ee
To summarize, we found that T duality maps equations for KT to a combination of the same equation and a constraint on the $B$ field:
\bea\label{AppClaimKT}
\nabla_{(M} K_{NP)}=0 
\Longleftrightarrow\left\{
\begin{array}{l}
 \nabla_{(M} \tilde K_{NP)}=0 ,\\
H_{A P (M}\tilde K_{N)}{}^A+e^{C/2}\nabla_{(M}[e^{-C/2}W_{N)P}]=0.
\end{array}\right.
\eea

\subsubsection{Lie derivative along KT}
\label{AppLieDer}

Note that the third equation in (\ref{DimRedKeqn}) has an interesting interpretation in terms of Lie derivatives. To see this, we rewrite the $M^{mn}{}_{z}$ as
\bea
0&=&{\hat g}^{ma}\left[{\hat\nabla}_a(e^{-C}K^n{}_{z})+
({\hat\nabla}_b A_a-{\hat\nabla}_a A_b)K^{nb}\right]+(m\leftrightarrow n)\\
&=&
g^{ma}\left[{\tilde\nabla}_a 
(e^{-C}K^n_{\ z}-A_b K^{nb})+{\hat\nabla}_b A_a K^{nb}+A_b{\hat\nabla}_a K^{nb}\right]+
(m\leftrightarrow n)\nonumber\\
&=&
\left[{{\hat\nabla}}^m
(e^{-C}K^n_{\ z}-A_b K^{nb})+
(m\leftrightarrow n)\right]+{\hat\nabla}_b A^m K^{nb}+
{\hat\nabla}_b A^n K^{mb}-A_b{\hat\nabla}^b K^{mn}.\nonumber
\eea
At the final step we used the equation for the Killing tensor. The last equation implies 
an interesting relation for the Killing tensor
\bea\label{AppLieTenA}
{
\nabla_a A^m K^{na}+
\nabla_a A^n K^{ma}-A_a\nabla^a K^{mn}=\nabla^m W^n+\nabla^n W^m,}
\eea
which generalizes the expression (\ref{KillDerBfield}) involving the Lie derivative of the $B$ field along a Killing vector. Specifically, rewriting (\ref{AppLieTenA}) as
\bea\label{AppLieTenB}
A_a\nabla^a K_{mn}-
K_m^{\ a}\nabla_a A_n-K_n^{\ a}\nabla_a A_m=-\nabla^m W^n-\nabla^n W^m
\eea
we are tempted to interpret the left--hand side of the last equation as a ``Lie derivative of $A_m$ along a Killing tensor". Although the analogy with the usual Lie derivative has limitations (for example, the rank of the lhs is higher than the rank of $A_m$), equation (\ref{AppLieTenB}) does reduce to the combination of Lie derivative if Killing tensor has a form $K^{mn}=\la^m\la^n$:
\bea
lhs&=&\la^n\la^a\nabla_a A^m +
\nabla_a A^n \la^m\la^a-A_r\nabla^a (\la^m\la^n)
\nonumber\\
&=&\la^n\left[\la^a\nabla_a A^m-A_a\nabla^a\la^m\right]+
\la^m\left[\la^a\nabla_a A^n-A_a\nabla^a\la_n\right]
\nonumber\\
&=&
\la^n\left[\la^a\nabla_a A^m+A_a\nabla^m\la^a\right]+
\la^m\left[\la^a\nabla_a A^n+A_a\nabla_n\la^a\right]\\
&=&\la^n{\mathcal L}_\la A^m+\la^m{\mathcal L}_\la A^n.\nonumber
\eea
It would be interesting to  investigate the relation between (\ref{AppLieTenB}) and Lie derivatives further.


\subsection{Extension to CKT}
\label{AppCKT}

In this appendix our results are extended to the conformal Killing tensor assuming that the original geometry has vanishing $B$ field and that there is no mixture between $z$ and other coordinates.  Starting with equation for the CKT,
\be
3\nabla_{(M} {\mathcal K}_{NP)}=g_{MN}W_P+g_{MP}W_N+g_{NP}W_M,
\ee
and performing reduction with $A_m=0$, we find 
\bea
zzz:&&\d_z {\mathcal K}_{zz}+{{\mathcal K}_z}^m\d_m e^C=e^C W_z\nonumber\\
mnz:&&\left[{\tilde\nabla}^m{{\mathcal K}^n}_z-{{\mathcal K}^n}_z\nabla^n C\right]+
(m\leftrightarrow n)+\d_z {\mathcal K}^{mn}=W_z g^{mn}
\nonumber\\
zzp:&&{\mathcal K}^{ap}\d_a e^C-2{\mathcal K}_{zz}\nabla^p C+\nabla^p {\mathcal K}_{zz}+
2\d_z {{\mathcal K}_z}^p=e^C W^p\\
mnp:&&\nabla^m{\mathcal K}^{np}+\nabla^n{\mathcal K}^{mp}+
\nabla^n{\mathcal K}^{mp}=W^m g^{np}+W^n g^{mp}+W^p g^{nm}
\nonumber
\eea
Motivated by the discussion of the CKV in subsection \ref{SubsectionCKV} we allowed the components of CKT to depend on the $z$ coordinate. We will assume that $\d_z=0$ before T duality, but the $z$--dependence appears afterward. 

To satisfy the $(mnp)$ equations before and after duality, we require
\bea
{\tilde W}^p=W^p,\qquad {\tilde {\mathcal K}}^{mn}={\mathcal K}^{mn}. 
\eea
Comparing $(mnz)$ equations before and after duality, and taking into account that 
$\d_z {\mathcal K}^{mn}=0$, we  set
\bea
{\tilde W}_z=e^{-2C}W_z+2ve^{-C},\quad 
{\tilde {\mathcal K}^n}_{\ \ z}=e^{-2C}{{\mathcal K}^n}_z+e^{-C}{\mathcal V}^n,
\eea
where ${\mathcal V}^n$ is a CKV with conformal factor $v$. 
Then $(zzz)$ equation after T duality gives
\bea
\d_z {\tilde {\mathcal K}}_{zz}&=&
2e^{-3C} W_z+e^{-2C}({\mathcal V}^a\d_a C+2v),\nonumber\\
{\tilde {\mathcal K}}_{zz}&=&
z\left[2e^{-3C} W_z+e^{-2C}({\mathcal V}^a\d_a C+2v)\right]+N_{zz},
\eea
where $N_{zz}$ is $z$--independent ``integration constant".

Comparing the $(zzp)$ equations before and after duality,
\bea\label{Jul25KYT}
&&e^{-C}\nabla^p 
(e^{2C}\tilde {\mathcal K}_{zz})+e^C\nabla^p (e^{-2C}{\mathcal K}_{zz})+
2e^{C}\d_z {\tilde{\mathcal K}_z}^{\ p}=2W^p,\nonumber\\
&&e^{-C}\nabla^p 
(e^{2C}\tilde {\mathcal K}_{zz})-e^C\nabla^p (e^{-2C}{\mathcal K}_{zz})+
2e^{C}\d_z {\tilde {\mathcal K}_z}^{\ p}=2{\mathcal K}^{ap}\d_a C,
\eea
and assuming that $\d_z {\mathcal V}^n=0$ (and 
thus $\d_z {\tilde {\mathcal K}_z}^{\ p}=0$), we conclude that $z$--dependence disappears from the last two equations if
\bea
&&\d_p\left[2e^{-C} W_z+({\mathcal V}^a\d_a C+2v)\right]=0,
\nonumber\\
\label{CKTConstr1}
&&\d_p\left[2{\tilde W}^z+({\mathcal V}^a\d_a C-2v)\right]=0.
\eea
The last equation is a counterpart of the homothety condition for the CKV. The remaining equations are (\ref{Jul25KYT}):
\bea\label{CKTConstr2}
&&e^{-C}\nabla^p 
(e^{2C}N_{zz})+e^C\nabla^p (e^{-2C}{\mathcal K}_{zz})=2W^p,\nonumber\\
&&e^{-C}\nabla^p 
(e^{2C}N_{zz})-e^C\nabla^p (e^{-2C}{\mathcal K}_{zz})=2{\mathcal K}^{ap}\d_a C.
\eea
To summarize, we have to satisfy two constraints (\ref{CKTConstr1}) and (\ref{CKTConstr2}) on 
constraints on $W^p$ and ${\mathcal K}^{tp}\d_t C$, then all equations can be solved.


\subsection{mKTY equation and the constraint on the $B$ field}
\label{AppBeqKYT}

This subsection is dedicated to the derivation of our main result: invariance of the modified Killing--Yano (mKYT) equation (\ref{KYTBfield}),
\bea\label{mKYTapp}
\nabla_M Y_{NP}+\h H_{MPA}g^{AB}Y_{NB}+(M \leftrightarrow N)=0,
\eea 
under the  T--duality transformations
. Starting with a geometry (\ref{AppDimRedBdy}) that admits a modified 
Killing--Yano tensor (mKYT) satisfying 
(\ref{mKYTapp}), we will show that the system (\ref{SetupAfter}) related to (\ref{AppDimRedBdy}) by T duality 
admits a mKYT ${\tilde Y}_{MN}$ with components
\bea\label{TransYaa}
{\tilde Y}^{mn}=Y^{mn},\qquad  {\tilde Y}_z{}^s= e^{-C}Y_z{}^s.
\eea

To demonstrate the invariance of the mKYT equation, we perform a dimensional reduction of
\bea
T_{MNP}\equiv \nabla_M Y_{NP}+\h H_{MPA}g^{AB}Y_{NB}+(M \leftrightarrow N)\,.
\eea
As discussed in section \ref{AppDimRed1}, it is sufficient to look only at components with covariant indices $z$ and contravariant indices $(m,n\dots)$, and since tensor $T_{MNP}$ is symmetric in the first two indices, we have to analyze five types of  components\footnote{Notice that $T_{zzz}=0$.}:
\bea\label{TensZnmp}
{T_{zz}}^p,\quad {T^m}_{zz},\quad {T^{mn}}_z,\quad {T_z}^{mp},\quad T^{mnp}.
\eea
and demonstrate that they are invariant under the T duality (\ref{TdualityMap}).

\bigskip

\noindent
\textbf{1. $(zzp)$ component.}

The first component in (\ref{TensZnmp}) is
\bea
{T_{zz}}^p&=&2\nabla_z Y_{z}{}^p+H_z{}^p{}_AY_z{}^{A}=
\d_ae^C Y^{ap}+g^{pa}e^C F_{ba}Y_z{}^b+H_{zsa}g^{sp}Y_z{}^a{}\\
&&=\d_ae^C Y^{ap}+g^{pa}e^C F_{ba}Y_z{}^b+\tilde F_{ab}g^{ap}Y_z{}^b.
\nonumber
\eea
Here we used expression (\ref{DimRedL}) for the covariant derivative $\nabla_z L_{z}{}^p$ of 
an arbitrary rank--2 tensor. Rewriting the last equation  as
\bea\label{TzzpOne}
e^{-C}{T_{zz}}^p=\d_aC Y^{ap}+g^{pa}F_{ba}Y_z{}^b-g^{ap}e^{-C}\tilde F_{ba}Y_z{}^b,
\eea
we observe that is it invariant under the T duality transformation (\ref{TdualityMap}) if we require that 
\bea
Y^{mn}\rightarrow Y^{mn},\qquad Y_z{}^m\rightarrow e^{C}Y_z{}^m.
\eea
To keep track of the last rescaling in the remaining equations, we introduce
\bea\label{hatY}
{\hat Y}_z{}^m\equiv e^{-C/2}Y_z{}^m
\eea
that remains invariant under T duality. Then equation (\ref{TzzpOne}) becomes more symmetric:
\bea\label{TzzpTwo}
{T_{zz}}^p=\d_aC Y^{ap}+g^{pa}F_{ba}e^{C/2}{\hat Y}_z{}^b-g^{ap}e^{-C/2}\tilde F_{ba}{\hat Y}_z{}^b
\eea
and invariance of equation ${T_{zz}}^p=0$ under T duality becomes explicit.
\bigskip 

\noindent
\textbf{2. $(mzz)$ component.}

The second component in (\ref{TensZnmp}),
\bea\label{Tmzz}
{T^m}_{zz}=\nabla_z {Y^m}_z+\frac{1}{2}{H^m}_{zA}{Y^A}_z=-
\frac{1}{2}{T_{zz}}^m,
\eea
is also invariant under T duality.

\bigskip 

\noindent
\textbf{3. $(mnz)$ component.}

The third component of (\ref{TensZnmp}) is
\bea
{T^{mn}}_z&=&\nabla^m {Y^n}_z+\h {H^m}_{zA}Y^{An}+(m \leftrightarrow n)
\nonumber\\
&=&\hat\nabla^m Y^n{}_z-\h g^{ma}\d_a C Y^n{}_z-\h g^{ma} e^C F_{ar} Y^{nr}+\h g^{ma}H_{azb}Y^{nb}+ (m\leftrightarrow n)\nonumber
\eea
Here we used (\ref{DimRedL}) to express $\nabla^m Y^n{}_z$ in terms of the covariant derivative $\hat\nabla^m Y^n{}_z$ in the reduced metric ${\hat g}_{mn}$. 
Rewriting the last equation in terms of the field strengths $(F_{ij},{\tilde F}_{ij})$,
\bea\label{TmnzOne}
{T^{mn}}_z=
{\hat\nabla}^m Y^n{}_z-\h g^{ma}\d_a C Y^n{}_z-\h g^{ma} \left[e^C F_{ar}+
\tilde F_{ab}\right]Y^{nb}+ (m\leftrightarrow n),
\eea
and expressing the result in terms of ${\hat Y}$ defined by (\ref{hatY}), we find
\bea\label{TmnzTwo}
{T^{mn}}_z=
{\hat\nabla}^m [{\hat Y}^n{}_z]-\h g^{ma} \left[e^{C/2} F_{ab}+
e^{-C/2}\tilde F_{ab}\right]Y^{nb}+ (m\leftrightarrow n).
\eea
Clearly this expression is invariant under T duality. 

\bigskip

\noindent
\textbf{4. $(zmp)$ component.}

To simplify the fourth component of (\ref{TensZnmp}) we again use (\ref{DimRedL}):
\bea
{T_z}^{mp}&=&\nabla^m Y_{z}{}^p+\nabla_z Y^{mp}+\h H^{mpA}Y_{Az}+
\h {H_z}^{pA}{Y^m}_A\nonumber\\
 &=&{\hat\nabla}^m Y_{z}{}^p-\h g^{ma}\d_aC Y_{z}{}^p-\h g^{ma}e^CF_{ab}Y^{bp}
\nonumber\\
&&-\h g^{ma}\d_aCY_z{}^p+\h g^{ma}e^CF_{ba} Y^{bp}-\h g^{pa}\d_aCY^m{}_z+\h g^{pa}e^CF_{ba}Y^{mb}\nonumber\\
&&+\h {H^{mp}}_a{Y^a}_{z}+\h {H^{mp}}_z{Y^z}_{z}+\h {{H_z}^{p}}_{a}{Y}^{ma}
\nonumber
\eea
Using expressions
\bea
{H^{mp}}_s{Y^s}_{z}&=&g^{ma}g^{pb}\left[H_{abs}-A_a H_{zbs}-A_b H_{azs}\right]{Y^s}_{z}=
g^{ma}g^{pb}\left[H_{abs}-A_a {\tilde F}_{bs}-A_b {\tilde F}_{sa}\right]{Y^s}_{z}\,,
\nonumber\\
{H^{mp}}_z{Y^z}_{z}&=&g^{ma}g^{pb}{\tilde F}_{ab}~\frac{1}{g_{zz}}\left[Y_{zz}-g_{zs}{Y^s}_z\right]=-g^{ma}g^{pb}A_s{\tilde F}_{ab}{Y^s}_z\,,
\nonumber\\
{{H_z}^{p}}_{s}{Y}^{ms}&=&g^{pa}{\tilde F}_{as}{Y}^{ms}\,,\nonumber
\eea
we find
\bea\label{TzmpOne}
{T_z}^{mp}&=&{\hat\nabla}^m Y_{z}{}^p-g^{ms}\d_sC Y_{z}{}^p+g^{ms}e^CF_{sr}Y^{pr}
-\h g^{ps}\d_sCY^m{}_z+\h g^{ps}e^CF_{rs}Y^{mr}\nonumber\\
&&+\frac{1}{2}g^{ma}g^{pb}\left[H-A\wedge {\tilde F}\right]_{abs}{Y^s}_{z}+
\frac{1}{2}g^{pa}{\tilde F}_{as}{Y}^{ms}
\eea
Recalling the expression for $H$ in terms of duality--invariant ${\hat B}$ (see 
(\ref{DimRedSetup}), (\ref{SetupAfter})),
\bea
H=d{\hat B}+\frac{1}{2}({\tilde A}\wedge A),
\eea
we observe that 
\bea\label{HhatDef}
{\hat H}\equiv H-A\wedge {\tilde F}=d{\hat B}-\frac{1}{2}[A\wedge \tilde F+{\tilde A}\wedge F]
\eea
is invariant under T duality. To demonstrate the invariance of (\ref{TzmpOne}), we rewrite that expression as
\bea\label{TzmpTwo}
{T_z}^{mp}&=&{\hat\nabla}^m {\hat Y}_{z}{}^p
+\frac{1}{2}g^{ms}(e^{C/2}F_{sr}+e^{-C/2}{\tilde F}_{sr})Y^{pr}+\frac{1}{2}g^{ma}g^{pb}{\hat H}_{abs}{{\hat Y}^s}{}_{z}\nonumber\\
&&+\frac{1}{2}\left[g^{ms}\d_sC {\hat Y}^p{}_{z}-g^{ps}\d_sC{\hat Y}^m{}_z+
g^{ps}G_{sr}Y^{rm}-g^{ms}G_{sr}Y^{rp}\right]\\
G_{sr}&\equiv& e^{C/2}F_{sr}-e^{-C/2}{\tilde F}_{sr}.\nonumber
\eea
The first line of this equation is invariant under T--duality, while the second line changes sign. Thus to make ${T_z}^{mp}$ invariant, we must impose a constraint on $F_{mn}$ and 
${\tilde F}_{mn}$:
\bea\label{ConstrBcomp}
{\tilde S}_{}^{mp}&\equiv& 
\left[g^{ms}\d_sC {\hat Y}^p{}_{z}-g^{ps}\d_sC{\hat Y}^m{}_z+
g^{ps}G_{sr}Y^{rm}-g^{ms}G_{sr}Y^{rp}\right]=0,\\
G_{sr}&\equiv& e^{C/2}F_{sr}-e^{-C/2}{\tilde F}_{sr}\nonumber
\eea
The physical meaning of this constraint is discussed in section \ref{SectionModifiedKYT}. 

\bigskip

\noindent
\textbf{5. $(mnp)$ component.}

The final component of (\ref{TensZnmp}) gives\footnote{We used (\ref{DimRedL}) to express 
$\nabla^m Y^{np}$ in terms of $\hat\nabla^m Y^{np}$.}
\bea\label{Sept29}
T^{mnp}=\hat\nabla^m Y^{np} + \h F^{mp} Y^n{}_z + \h H^{mp}{}_{A}Y^{nA}+ (m \leftrightarrow n).
\eea
Simplifying the term that involves flux
\bea
&&H^{mp}{}_{D}Y^{nD}=g^{mA}g^{pB} H_{AB D}Y^{nD}\nonumber\\
&&=g^{mz}g^{pb}H_{zbc}Y^{nc}+g^{ma}g^{pz} H_{azc}Y^{nc}+g^{ma}g^{pb} H_{abz}Y^{nz}+g^{ma}g^{pb} H_{abc}Y^{nc}\nonumber\\
&&=-A^mg^{pb}\tilde F_{bc}Y^{nc}+g^{ma}A^{p} \tilde F_{ac}Y^{nc}+g^{ma}g^{pb} \tilde F_{ab}(e^{-C}Y^n{}_z-A_cY^{nc})+g^{ma}g^{pb} H_{abc}Y^{nc}\nonumber\\
&&=g^{ma}g^{pb}e^{-C} \tilde F_{ab}Y^n{}_z+
g^{ma}g^{pb}Y^{nc}(H_{abc}-A_a\tilde F_{bc}+A_b \tilde F_{ac}-A_c \tilde F_{ab})
\nonumber\\
&&=g^{ma}g^{pb}e^{-C} \tilde F_{ab}Y^n{}_z+
g^{ma}g^{pb}Y^{nc}(H-A\wedge \tilde F)_{abc}
\eea
and recalling expression (\ref{HhatDef}) for the duality--invariant $\hat H$, we find
\bea
H^{mp}{}_{A}Y^{nA}=g^{ma}g^{pb}e^{-C} \tilde F_{ab}Y^n{}_z+
g^{ma}g^{pb}Y^{nc}{\hat H}_{abc}
\eea
Then equation (\ref{Sept29}) becomes
\bea\label{TmnpOne}
T^{mnp}=\tilde\nabla^m Y^{np} +\h g^{ma}g^{pb}\Big[[F_{ab}+e^{-C} \tilde F_{ab}] Y^n{}_z +Y^{nc}{\hat H}_{abc}\Big]
+ 
(m \leftrightarrow n),
\eea
and rewriting it in terms of ${\hat Y}$
\bea\label{TmnpTwo}
T^{mnp}=\tilde\nabla^m Y^{np} +\h g^{ma}g^{pb}\Big[[
e^{C/2}F_{ab}+e^{-C/2} \tilde F_{ab}] {\hat Y}^n{}_z +Y^{nc}{\hat H}_{abc}\Big]
+ 
(m \leftrightarrow n)
\eea
make the invariance under T duality explicit. 

The constraint (\ref{ConstrBcomp}) treat $z$ direction in a special way, but it would be nice to write it in a covariant form. This can be accomplished in an important special case when $F_{mn}=0$, which implies the T--dual configuration has no $B$ field. Then (\ref{ConstrBcomp}) reduces to
\be\label{AppmKYTMismatch}
H_{naz}g^{ab}Y_{bp}+H_{zpa}g^{ab}Y_{nb}+[\d_n C\check{Y}_{zp}-\d_p C\check{Y}_{zn}]=0.
\ee
The unique covariant form of this relation is
\be\label{AppRestBKYTGuess}
\begin{aligned}
H_{AMP}\tilde Y_N{}^A-H_{ANP}\tilde Y_M{}^A-H_{AMN}\tilde Y_P{}^A-(\d_M C \tilde Y_{NP}-\d_N C \tilde Y_{MP}-\d_P C \tilde Y_{N M})\\=\d_M W_{NP}-\d_N W_{MP}-\d_P W_{N M},
\end{aligned}
\ee
where $W$ is auxiliary field introduced to satisfy the $mnp$ components of the last equation, which would be too restrictive otherwise.

\bigskip

To summarize, we have demonstrated that all independent components of $T_{MNP}$ given by 
(\ref{TensZnmp}) can be written in a way that makes invariance under T duality (\ref{TdualityMap}) very explicit (see (\ref{TzzpTwo}), (\ref{Tmzz}), (\ref{TmnzTwo}), (\ref{TzmpTwo}), (\ref{TmnpTwo})), as long as constraint (\ref{ConstrBcomp}) is satisfied. 

\subsubsection{KT from mKYT}

Finally we show that the modified Killing-Yano equation reduces to a standard Killing tensor equation. To do so we begin with the modified equation for KYT 
\bea
\nabla_M Y_{NP}+\nabla_N Y_{MP}+
\frac{1}{2}H_{MPA}{Y_N}^A+
\frac{1}{2}H_{NPA}{Y_M}^A=0
\eea
and construct various combinations:
\bea
&&{Y_B}^P\left[\nabla_M Y_{NP}+\nabla_N Y_{MP}+
\frac{1}{2}H_{MPA}{Y_N}^A+
\frac{1}{2}H_{NPA}{Y_M}^A\right]=0,\nonumber\\
&&{Y_N}^P\left[\nabla_M Y_{BP}+\nabla_B Y_{MP}+
\frac{1}{2}H_{MPA}{Y_B}^A+
\frac{1}{2}H_{BPA}{Y_M}^A\right]=0,\nonumber\\
&&{Y_M}^P\left[\nabla_B Y_{NP}+\nabla_N Y_{BP}+
\frac{1}{2}H_{BPA}{Y_N}^A+
\frac{1}{2}H_{NPA}{Y_B}^A\right]=0.\nonumber
\eea
Adding these equations, we find the standard Killing tensor equation
\bea
&&\nabla_M K_{BN}+\nabla_N K_{MB}+
\nabla_B K_{MN}+
\frac{1}{2}\left[H_{MPA}({Y_N}^A{Y_B}^P+{Y_N}^P{Y_B}^A)+
perm\right]=0,\nonumber\\
&&\nabla_M K_{BN}+\nabla_N K_{MB}+
\nabla_B K_{MN}=0.
\eea
Here 
\bea
K_{MN}\equiv {Y_M}^A Y_{NA}.
\eea
To summarize, we demonstrated that the standard relation ``KT=KYT$^2$'' persists for the modified Killing--Yano tensors as well.  




\section{The restrictions on the $B$ field from the F1 $\to$ NS5 duality chain}
\label{AppCondBNS5}

In Section \ref{KTOdd} we derived the restrictions on the metric and the B-field \eqref{SepCondB}
by requiring separability of the Hamilton--Jacobi equation along all $O(d,d)$ orbits which start with a pure metric. 
In this section we will extend those results to $O(d,d)$ orbits starting with NS5 solutions (thus generating the 
entire F1--NS5--P family) and show that separability leads to additional constraints (\ref{Hcond2f}), (\ref{Hcond2fields}), (\ref{Hcond2sep}) on the $B$ field. 

We start with conditions on the $B$ field \eqref{Hcond1} and \eqref{Hcond}
\bea\label{AppHconds}
&&\d_x\d_y (fg_{ab})-fg^{MN}H_{y aM}H_{xNb}-fg^{MN}H_{xaM}H_{y Nb}=0,\\
&&\d_y(f g^{mM})H_{xMb}+\d_x (f g^{mM}) H_{y Mb}+f g^{mM}\d_xH_{y Mb}=0.\nonumber
\eea
Next we consider the first equation and require this constraint to hold on the entire $O(d,d)$ orbit containing NS5 brane. Comparing  
\eqref{AppHconds} for F1 orbit with its counterpart for NS5, we find
\bea\label{AppHcond21}
&&\d_x\d_y (fg_{ab})-fH_{y aM} H_{x}{}^M{}_b-fH_{xaM} H_{y}{}^M{}_b=0,\\
&&\d_y \d_x [F^2fg_{ab}]-fF (H^{(NS5)}_{y aM} H^{(NS5)}_{x}{}^M{}_{b}+H^{(NS5)}_{xaM} H^{(NS5)}_{y}{}^M{}_{b})=0.\nonumber
\eea
Here we used the transformation law for the metric and defined a convenient function $F$
\bea\label{AppF1NS5trans}
g_{MN}^{NS5}=F g_{MN}^{F1}, \quad f^{NS5}=F f^{F1},\quad F\equiv \sqrt{\mbox{det}G}\,\mbox{det}H.
\eea
Expressions without superscript in (\ref{AppHcond21}) refer to the fundamental string. The field strengths of the Kalb--Ramond fields for NS5 and F1 systems are related by the electric--magnetic duality
\bea
H^{(NS5)}_{y aM}=\frac{1}{7!} e^{2\Phi_{NS5}}e_{y a M}{}^{x NP z_1z_2z_3z_4} G^{(NS5)}_{x NP z_1z_2z_3z_4}=
\frac{1}{7!} e^{2\Phi_{NS5}}e_{y a M}{}^{xNP z_1z_2z_3z_4} H^{(F1)}_{x NP}.
\eea
In particular, the product of the field strengths is
\bea\label{AppEqH2}
H^{(NS5)}_{yaM}H^{(NS5)}_{x}{}^M{}_b&=&
\frac{e^{4\Phi_{NS5}}}{(3!)^2}e_{ya M}{}^{x NP} e_{x}{}^M{}_b{}^{y}{}_A{}^{B} H_{x NP} H_{y}{}^A {}_{B}=\frac{e^{4\Phi_{NS5}}}{(2!)^2}e_{a M}{}^{NP} e^M{}_{bA}{}^{B} H_{x NP} H_{y}{}^A {}_{B} \nonumber\\
&=&e^{4\Phi_{NS5}}\left[-H_{xbM}H_y{}^M{}_a-\frac{1}{2}H_{xMN}H_y{}^{MN}g_{ab}^{(NS5)}\right]_{g^{(NS5)}}.
\eea
In the last line all indices are contracted with $g_{MN}^{(NS5)}$. In terms of the F1 metric we find
\bea
H^{(NS5)}_{y aM} H^{(NS5)}_{x}{}^M{}_{b}=
F\left[-H_{xbM}H_y{}^M{}_a-\frac{1}{2}H_{xMN}H_y{}^{MN}
g_{ab}\right].
\eea
We can now rewrite the conditions \eqref{AppHcond21} in terms of the F1 fields:
\bea\label{AppHcond22}
&&\d_x\d_y(f g_{ab})-fH_{y aM} H_{x}{}^M{}_b-fH_{xaM} H_{y}{}^M{}_b=0,\\
&&\d_x\d_y\Big[f g_{ab}F^2 \Big]+fF^2(H_{y aM}H_x{}^M{}_b+(x\to y)+H_{xMN}H_y{}^{MN}g_{ab})=0.\nonumber
\eea
Subtracting the first equation from the second one we get the relation
\bea
\d_x\d_y\Big[f g_{ab}F^2 \Big]+F^2(\d_x\d_y[f g_{ab}]+fH_{xMN}H_y{}^{MN}g_{ab})=0,\nonumber
\eea 
which can be rewritten as
\bea\label{AppHcond2f}
\d_x\d_y[g_{ab}fF]+\frac{f}{F}g_{ab}\Big[\d_x\ln F \d_y\ln F+ \frac{1}{2}H_{xMN}H_y{}^{MN}\Big]=0.
\eea
Remarkably in all our examples the two terms entering this expression vanish separately, so we conjecture that this will always happen for the systems obtained from fundamental stings via the  duality chain, although we will not attempt to prove this fact. Recalling that  $F=e^{-2\Phi_{F1}}$,  we conclude that vanishing of the first term in (\ref{AppHcond2f}) implies separation of the duality--invariant expression
\bea
g^{(F1)}_{ab}f^{(F1)}e^{-2\Phi_{F1}}=g^{(NS5)}_{ab}f^{(NS5)}e^{-2\Phi_{NS5}}.
\eea
In other words vanishing of the first term in \eqref{AppHcond2f} can be written as 
\bea\label{AppHcond2sep}
\d_x\d_y\left[g_{ab}fe^{-2\Phi}\right]=0
\eea
in \textit{every frame} containing only NS--NS fields. Vanishing of the second term in \eqref{AppHcond2f} gives the relation in the F1 frame
\bea\label{AppHcond2fields}
\d_x\Phi \d_y\Phi+\frac{1}{8} H_{xMN}H_y{}^{MN}=0.
\eea

Now we consider the the second condition in \eqref{AppHconds}
\bea
\d_y(f g^{mM})H_{xMb}+\d_x (f g^{mM}) H_{y Mb}+f g^{mM}\d_xH_{y Mb}=0.
\eea
Writing it for F1 and for NS5, and using \eqref{AppF1NS5trans} we get
\bea\label{AppHcond1f}
&&\d_y(f g^{mM})H_{xMb}+\d_x (f g^{mM}) H_{y Mb}+f g^{mM}\d_xH_{y Mb}=0,\\
&&\d_y(f g^{mM})\tilde{H}_{xMb}+\d_x (f g^{mM}) \tilde{H}_{y Mb}+\frac{f}{F} g^{mM}\d_x(F\tilde{H}_{y Mb})=0.\nonumber
\eea
Here $\tilde{H}=\star_6H^{(F1)}$ is six--dimensional Hodge dual of the field strength for F1. Note that the first equation (and its dual counterpart) can be written in two different ways (using $\d_xH_{y Mb}=\d_y H_{xMb}$). The difference gives equation of motion for the $B$ field
\bea
g^{mM}\Big[\d_x(e^{2\Phi_{NS5}}\tilde{H}_{y Mb}) -\d_y( e^{2\Phi_{NS5}}\tilde{H}_{x Mb}) \Big]=0,\quad e^{2\Phi_{NS5}}=\mbox{det}H\sqrt{\mbox{det}G}.
\eea

To summarize we have found two additional constrains \eqref{AppHcond2f}, \eqref{AppHcond1f} on the $B$ field that guarantee separability of F1--NS5. Remarkably in the studied examples the first condition decouples into two very simple equations - separation condition \eqref{AppHcond2sep} and the field equation \eqref{AppHcond2fields}.




\section{Modified KY tensor for the charged Myers--Perry black hole}
\label{AppChargedMP}

In section \ref{SecMPF1} we presented the modified Killing--Yano tensor for the charged counterpart of the Myers--Perry black hole. In this appendix we will outline the derivation of (\ref{ChMPHYT})--(\ref{ChMPframe}).

We begin with the original Myers--Perry metric and its Killing--Yano tensor written in terms of frames (\ref{AllFramesMP}) and apply the first two steps in the duality chain 
(\ref{DualChain}). The boost leads to replacements
\bea
\begin{array}{l}
dt\rightarrow \ch_\alpha dt+\sh_\alpha dy\\
dy\rightarrow  \ch_\alpha dy+\sh_\alpha dt\\
\end{array},\qquad
\begin{array}{l}
\d_t\rightarrow \ch_\alpha\d_t-\sh_\alpha\d_y\\
\d_y\rightarrow \ch_\alpha\d_y-\sh_\alpha\d_t
\end{array}
\eea
in the frames (\ref{AllFramesMP}), but it does not modify the expressions 
(\ref{Kub2}), (\ref{HInFrame}). T duality along $y$ direction leaves the contravariant components $g^{mn}={\hat g}^{mn}$ and $Y^{mn}$ invariant, so it is reasonable to assume that neither expressions (\ref{Kub2}), (\ref{HInFrame})
nor components of $e_A$ which don't involve $y$ are modified.  In other words, we will assume that after T duality the frames have the form
\bea\label{ChMPframePrm}
e_r&=&\sqrt{\frac{R-mr}{FR}}\d_r,\quad 
e_{x_i}=\sqrt{-\frac{4x_iH_i}{d_i(r^2-x_i)}}\d_{x_i},\quad
e_y=C_y\ch_\alpha\d_y-\sh_\alpha\d_t,
\nn
e_t&=&\sqrt{\frac{R^2}{FR(R-mr)}}\left[\ch_\alpha \d_t-C_t{\sh_\alpha}\d_y
-\sum_k\frac{a_k}{r^2+a_k^2}\d_{\phi_k}\right],\\
e_i&=&\sqrt{\frac{H_i}{d_i(r^2-x_i)}}\left[\ch_\alpha\d_t-C_i\sh_\alpha \d_y-\sum_k\frac{a_k}{x_i+a_k^2}\d_{\phi_k}
\right]\nonumber
\eea
with some functions $(C_y,C_t,C_i)$. This assumption will be justified by the explicit calculation that recovers transformation rules (\ref{DimRedSetup}), (\ref{SetupAfter}) and (\ref{TransYaa}) and determines the functions $(C_y,C_t,C_i)$.

We begin with recovering the relation ${\tilde g}^{ym}=0$, which must hold after T duality. Equations (\ref{ChMPframePrm}) give
\bea\label{TempGyM}
{\tilde g}^{ym}\d_m&=&
-C_y\ch_\alpha\sh_\alpha\d_t+
{\frac{R^2}{FR(R-mr)}}C_t{\sh_\alpha}\left[\ch_\alpha \d_t-
\sum_k\frac{a_k}{r^2+a_k^2}\d_{\phi_k}\right]\nn
&&-
\sum_i\left[{\frac{(-x_i)H_i}{d_i(r^2-x_i)}}C_i\sh_\alpha\left[\ch_\alpha\d_t-\sum_k\frac{a_k}{x_i+a_k^2}\d_{\phi_k}
\right]\right]=0.
\eea
Coefficients in front of $\d_t$ and all $\d_{\phi_k}$ must vanish, so we find $n$ equations for $(n+1)$ variables $(C_y,C_t,C_i)$, which are completely determined up to one overall factor. Thus it is sufficient to guess the solution and check the result. 
To determine the coefficients $(C_y,C_t,C_i)$ we set $m=0$ in the boosted frames before T duality, which can be extracted from (\ref{ChMPframePrm}) by setting $C_y=C_t=C_i=1$. This gives the off--diagonal components before T duality
\bea\label{TempGyM1}
{g}^{yp}\d_p|_{m=0}&=&-\ch_\alpha\sh_\alpha\d_t+
{\frac{R}{F}}{\sh_\alpha}\left[\ch_\alpha \d_t-
\sum_k\frac{a_k}{r^2+a_k^2}\d_{\phi_k}\right]\nn
&&-
\sum_i\left[{\frac{(-x_i)H_i}{d_i(r^2-x_i)}}\sh_\alpha\left[\ch_\alpha\d_t-\sum_k\frac{a_k}{x_i+a_k^2}\d_{\phi_k}
\right]\right]=0.
\eea
The last expression must vanish since for $m=0$ time and $y$ coordinate enter the Myers--Perry metric (\ref{MPeven}) only through the boost--invariant combination $-dt^2+dy^2$. Comparison of (\ref{TempGyM}) with (\ref{TempGyM1}) gives the unique expressions for the unknown functions in terms of $C_y$:
\bea
C_i=C_y, \quad C_t=\frac{R-mr}{R}C_y.
\eea
To determine the last remaining coefficient we compute ${\tilde g}^{yy}$:
\bea
{\tilde g}^{yy}&=&C_y^2\left[\ch^2\alpha-\frac{(R-mr)}{FR}\sh^2\alpha+
\sum_i\left[{\frac{(-x_i)H_i}{d_i(r^2-x_i)}}\sh^2\alpha\right]\right]\nonumber\\
&=&C_y^2\left[1+\frac{mr}{FR}\sh^2\alpha\right].
\eea
To simplify this expression we again used the trick of setting $m$ to zero. For the boosted version of (\ref{MPeven}) we find
\bea
g_{yy}=1+\frac{mr}{FR}\sh^2\alpha.
\eea
Matching this with ${\tilde g}^{yy}$, we conclude that $C_y=1$.

To summarize, we have demonstrated that the frames (\ref{ChMPframePrm}) with 
\bea\label{CframeDual}
C_i=C_y=1, \quad C_t=\frac{R-mr}{R}
\eea
reproduce the metric after T duality and expression (\ref{ChMPHYT}) recovers the correct components $Y^{mn}$, it only remains to check that the correct transformation of $Y_z{}^s$ is also recovered. 

According to our conjecture (\ref{ChMPHYT}), the mKYT in the original and T dual frames are given by
\bea
Y^{(p)}=\sum A_{a_1,\dots a_p}e^{a_1}\wedge \dots\wedge e^{a_p},\quad
{\tilde Y}^{(p)}=\sum A_{a_1,\dots a_p}{\tilde e}^{a_1}\wedge \dots\wedge {\tilde e}^{a_p}
\eea
with the \textit{same} coefficients $A_{a_1,\dots a_p}$. The original frames $e^{a}$ are given by  (\ref{ChMPframePrm}) with $C_i=C_y=C_t=1$, and the dual frames $\tilde e^{a}$ have different values of coefficients  (\ref{CframeDual}). Observing that 
\bea
\tilde e^{a}_y=\frac{1}{h_1}e^{a}_y,\quad \tilde e_{a}^m=e_{a}^m,
\eea
we find the perfect agreement with transformation (\ref{KYTtransYcomp}),
\bea
{\tilde Y}^{m_1\dots m_p}= Y^{m_1\dots m_p},\qquad 
{\tilde Y}_z{}^{m_2\dots m_p}= e^{-C}{Y}_z{}^{m_2\dots m_p},
\eea
since
\bea
e^C\equiv g_{yy}=1+\frac{mr}{FR}\sh^2\alpha=h_1.
\eea
This concludes the derivation of the Killing--Yano tensors (\ref{ChMPHYT}), (\ref{ChMPframe}), (\ref{ChMPframeDwn}) for the charged Myers--Perry black holes in even dimensions. The arguments for the odd dimensions are identical, and the answer is given by (\ref{ChMPframeOdd}), (\ref{ChMPframeDwnOdd}).


\section{Killing tensors for the F1--NS5 system}
\label{App5D6DandKT}

In this appendix we will present some technical details of calculations leading to the Killing tensors for the examples discussed in section \ref{SecExamplesF1NS5}. 

\subsection{F1--NS5 from the four--dimensional Kerr metric}
\label{App4DKerr}

Starting with Kerr metric (\ref{Kerr4D}) and using the duality chain (\ref{DualChain}), we generate the F1--NS5 solution
\bea\label{NonExtr5D}
ds^2&=&\frac{1}{h_{\beta}}dy^2+\frac{4ma\sh_\beta \sh_\alpha \cos\theta}{\rho^2 h_\beta}dzdy-
\left(1-\frac{2mr \ch^2_\beta}{\rho^2 h_{\beta}}\right)dt^2
-\frac{4mra \ch_\alpha \ch_\beta \sin^2\theta}{\rho^2 h_{\beta}}dtd\phi
\nn
&&+\Bigg[
(r^2+a^2)h_\alpha+\frac{2mra^2\sin^2\theta}{\rho^2}-
\frac{(2mar \ch_\alpha \sh_\beta \sin\theta)^2}{\rho^4 h_{\beta}}
 \Bigg]\sin^2\theta d\phi^2\\
&&+\frac{h_\alpha \rho^2}{\Delta}dr^2+h_\alpha \rho^2 d\theta^2+\Big[1+
\frac{2m\sh^2_\alpha(2m \sh^2_\beta+r)}{\rho^2 h_\beta}\Big]dz^2,\nn
B_2&=& \frac{mr \sh_{2\beta}}{h_\beta \rho^2} dy\wedge dt - 
\frac{2amr \ch_\alpha \sh_\beta\sin^2\theta }{h_\beta\rho^2} dy\wedge d\phi 
+ \frac{2am\cos\theta \ch_\beta \sh_\alpha}{h_\beta \rho^2} dt\wedge d z\nn
&& - \frac{m\cos\theta 
\sh_{2\alpha}(a^2+2mr \sh^2_\beta+r^2)}{h_\beta \rho^2} d\phi\wedge dz,\nn
e^{2\Phi}&=&\frac{h_\alpha}{h_\beta},\nn
\Delta&=&r^2+a^2-2mr, \quad \rho^2=r^2+a^2\cos^2\theta, \quad 
h_\alpha=1+\frac{2mr \sh^2_\alpha}{\rho^2}, \quad h_\beta=1+\frac{2mr \sh^2_\beta}{\rho^2}.\nonumber
\eea
The charges associated with NS5 branes and fundamental strings are defined by 
\be\label{ChargesDef}
Q_5=2A^2=2m\sinh^2\alpha, \quad Q_1=2B^2=2{m}\sinh^2\beta
\ee
The nontrivial Killing tensor for (\ref{NonExtr5D}) can be extracted either from solving a system of differential equations (\ref{KTeqnDef}) or by separating variables in the massive Hamilton--Jacobi equation. The second approach is easier and more instructive, so we begin with equation
\be\label{HJMass}
g^{MN}\frac{\d S}{\d x^M}\frac{\d S}{\d x^N}+\mu^2=0,
\ee
multiply it by $\rho^2 h_\alpha$, and rewrite the result as a system of two differential equations
\bea\label{5Dlambda}
\Lambda&=&(2A^2+r)(2B^2+r)(\d_y S)^2-\Big[\frac{(r^2+2A^2r+a^2)(r^2+2B^2r+a^2)}{\Delta}-\frac{a^2}{2}\Big](\d_t S)^2\nn
&&-\frac{4ar\sqrt{(A^2+m)(B^2+m)}}{\Delta}\d_tS\d_\phi S -\frac{a^2}{\Delta}(\d_\phi S)^2+\Delta (\d_r S)^2 +r^2(\d_z S)^2\nn
&&+\mu^2 (2B^2r+r^2),\\
\Lambda&=&-a^2c^2_\theta(\d_y S)^2+4aABc_\theta \d_zS \d_y S+ \frac{a^2c_{2\theta}}{2}(\d_t S)^2- \frac{1}{s^2_\theta}(\d_\phi S)^2- (\d_\theta S)^2\nn
&&-a^2c^2_\theta(\d_z S)^2-\mu^2a^2c^2_\theta.\nonumber
\eea 
In general $\Lambda$ can depend on all coordinates, but for separable solutions,
\bea
S=-Et+J\phi+p_z z+p_y y+S_r(r)+S_\theta(\theta)
\eea
this function must be constant. This constant gives rise to a Killing tensor
\bea\label{KT5dApp}
K^{MN}\d_M\d_N=-a^2c^2_\theta\d^2_y+4aABc_\theta \d_z\d_y+
\frac{a^2c_{2\theta}}{2} \d^2_t-\frac{1}{s^2_\theta}\d^2_\phi-\d^2_\theta-a^2c^2_\theta\d^2_z
+a^2c^2_\theta g^{MN}\d_M \d_N.
\nn
\eea
Here we removed $\mu^2$ from (\ref{5Dlambda}) using the relation
\bea
g^{MN}\d_M S\d_N S+\mu^2=0.\nonumber
\eea
This Killing tensor (\ref{KT5dApp}) is used in section \ref{SecEx5d}. Note that even though we found KT, the square root of \eqref{KT5dApp} does not solve either standard or modified KYT equation for arbitrary charges. The special cases for which modified KYT exists are discussed in subsection \ref{SubsectionExamples5D}.

\subsection{F1--NS5 from the five--dimensional black hole}

The chain of dualities (\ref{DualChain}) can also be applied to a five--dimensional black hole, but fortunately this procedure has been performed in \cite{3charge}\footnote{The metric has been constructed earlier in \cite{cvet} using different methods, and in the full solution (\ref{NonExtr6D}) for the extremal case was found in \cite{MultiStr}.}. Here we will focus on solution with one rotation which can be obtained by setting  $\delta_p=0, a_1=0, a_2=a$ in equation  (3.6) of \cite{3charge} and performing an S duality. The result reads
\bea\label{NonExtr6D}
ds^2 &=& - \left( 1- \frac{M}{f}
\right)\frac{dt^2}{H_{1}}  + \frac{ dy^2}{H_{1}} +   fH_{5}\left(\frac{ dr^2
}{r^2+a^2-M} +  d\theta^2\right) \nonumber \\
  &+& \left[ r^2H_{5} + \frac{a^2
K_{1}K_{5} \cos^2\theta}{H_{1}} \right] \cos^2\theta
d\psi^2 + \left[ (r^2+a^2) H_{5} -
\frac{a^2  K_{1}K_{5} \sin^2\theta}{H_{1}}
\right] \sin^2\theta d\phi^2 \nonumber \\
&+& \frac{M  }{fH_{1}}a^2\sin^4\theta d\phi^{2} + \frac{2\cos^2\theta}{fH_{1} } aAB dyd\psi +  \frac{2\sin^2\theta}{fH_{1} }
a\sqrt{A^2+M}\sqrt{B^2+M} dt d\phi +\sum_{i=1}^{4} dz_{i}^2 \nonumber \\
B_{2} &=& \frac{\cos^2\theta}{fH_{1}} aA \sqrt{B^2+M} dt \wedge d\psi + \frac{\sin^2\theta }{f H_{1}} aB \sqrt{A^2+M}dy\wedge  d\phi \nonumber \\
  &-& \frac{B\sqrt{B^2+M} }{fH_{1}}  dt\wedge dy - \frac{A\sqrt{A^2+M}}{f H_{1}}\left(r^2+a^2+B^2\right)\cos^2\theta d\psi\wedge d\phi,\nonumber\\
e^{2\Phi}&=&\frac{H_5}{H_1},\nn
f&=&r^2+a^2\cos^2\theta,\quad K_1=\frac{B^2}{f}, \quad K_5=\frac{A^2}{f},\quad H_i\equiv 1+K_i, ~ i=1,5.
\eea
Multiplying the Hamilton--Jacobi equation (\ref{HJMass}) for the metric (\ref{NonExtr6D}) by $f H_5$ and separating variables, we find
\bea\label{AppSep6D}
&&-\left(A^2+B^2+M+r^2+\frac{(A^2+M)(B^2+M)}{a^2-M+r^2}\right)(\d_t S)^2+\frac{2a\sqrt{A^2+M}\sqrt{B^2+M}}{a^2-M+r^2}\d_tS\d_\phi S\nonumber\\
&&+\frac{(A^2+r^2)(B^2+r^2)}{r^2}(\d_y S)^2-\frac{2aAB}{r^2}\d_y S\d_\psi S+(a^2-M+r^2)(\d_r S)^2\\
&&+\frac{a^2}{r^2}(\d_\psi S)^2-\frac{a^2}{a^2-M+r^2}(\d_\phi S)^2+(A^2+r^2)\mu^2=\nonumber\\
&&a^2\cos^2\theta(\d_t S)^2-a^2\cos^2\theta(\d_y S)^2
-(\d_\theta S)^2-\frac{1}{\cos^2\theta}(\d_\psi S)^2
-\frac{1}{\sin^2\theta}(\d_\phi S)^2-a^2\cos^2\theta\mu^2.\nonumber
\eea
This equation clearly separates in $\theta,r$ and gives rise to the Killing tensor
\bea\label{KT6dApp}
K^{MN}\d_M\d_N&=&a^2\cos^2\theta\d^2_t-a^2\cos^2\theta\d^2_y-\d^2_\theta-\frac{1}{\cos^2\theta}\d^2_\psi-\frac{1}{\sin^2\theta}\d^2_\phi\nn
&&+a^2\cos^2\theta g^{MN}\d_M S\d_N S.
\eea
In contrast to the F1--NS5 system constructed from the four--dimensional Kerr solution (there was no mKYT) the square root of \eqref{KT6dApp} give rises to a rank-3 modified Killing-Yano tensor discussed in subsection \ref{SubsectionExamples6D}.

\subsection{F1--NS5 from the Plebanski--Demianski solutions}

Our final example is F1--NS5 constructed from the Plebanski--Demianski metric \cite{PlebDem}:
\bea\label{Plebanski}
ds^2&=&\frac{p^2+q^2}{X}dp^2+\frac{p^2+q^2}{Y}dq^2+\frac{X}{p^2+q^2}(d\tau+q^2 d\sigma)^2-\frac{Y}{p^2+q^2}(d\tau-p^2 d\sigma)^2,\nn
X&=&\gamma-g^2-\epsilon p^2-\lambda p^4+2lp, \quad Y=\gamma+e^2+\epsilon q^2-\lambda q^4-2mq.
\eea
Here $\lambda$ is a cosmological constant, $e$ and $g$ are electric and magnetic charges (we will set these quantities to zero). The remaining constants $(\gamma, m , l , \epsilon)$ effectively comprise 3 real continuous parameters and one discrete parameter, since one can always rescale coordinates to set $\eps$ to one of three values $(+1,-1,0)$. The remaining  continuous parameters $(\gamma, m , l)$ are related to the angular momentum, mass, and the NUT charge. The Kerr solution (\ref{Kerr4D}) is recovered by setting 
\bea
\gamma=a^2,~ \epsilon=1-\lambda a^2,~ p=a \cos\theta,~ q=r,~ \tau=t-\frac{a}{1+\lambda a^2}\phi,~ \sigma=-\frac{1}{a(1+\lambda a^2)}\phi.\nonumber
\eea
In string theory applications one usually sets $e=g=0$, and since asymptotic flatness is a crucial part of our solution generating technique, we set $\la=0$ as well. Applying the chain of dualities (\ref{DualChain}) to such truncated version of \eqref{Plebanski} we get an 
F1--NS5 solution
\bea\label{F1NS5Pleb}
ds^2&=&\frac{f_\alpha}{X}dp^2+\frac{f_\alpha}{Y}dq^2+\frac{X-Y}{f_\beta}d\tau^2+2\frac{q^2X+p^2Y}{f_\beta}\ch_\alpha\ch_\beta d\tau d\sigma\nn
&-&\Big[ \frac{p^4Y-q^4X}{p^2+q^2}\ch_\alpha^2+XY\sh_\alpha^2+\frac{(q^2X+p^2Y)^2\ch_\alpha^2 \sh_\beta^2}{f_\beta(p^2+q^2)}\Big]d\sigma^2\nn
&+&\frac{p^2+q^2}{f_\beta}dy^2+\frac{4(mp-lq)\sh_\alpha\sh_\beta}{f_\beta}dy dz+\Big[\frac{f_\alpha}{p^2+q^2}+\frac{4(mp-lq)^2\sh_\alpha^2\sh_\beta^2}{f_\beta(p^2+q^2)} \Big]dz^2,\nn
B&=&\Bigg[ \frac{p^2+q^2+X-Y}{f_\beta}\ch_\beta\sh_\beta d\tau+\frac{q^2X+p^2Y}{f_\beta}\ch_\alpha\sh_\beta d\sigma \Bigg]\wedge dy\nn
&+&\Bigg[ \frac{2(lq-mp)}{f_\beta}\ch_\beta\sh_\alpha d\tau \\
&&- \frac{f_\beta pq(lp+mq)+(lq-mp)(q^2X+p^2Y)\sh_\beta^2}{f_\beta(p^2+q^2)}\sh_{2\alpha}d\sigma \Bigg]\wedge dz,\nn
f_\alpha&=& (p^2+q^2)\left[1+\frac{X-Y+p^2+q^2}{p^2+q^2}\sh_\alpha^2 \right].\nonumber
\eea
Writing the HJ equation for the metric \eqref{F1NS5Pleb} and multiplying it  by $f_\alpha$, we extract the Killing tensor from separation of variables as in the previous subsections
\bea
K^{MN}\d_M\d_N&=&-\bar{p}_\alpha\bar{p}_\beta\d_y^2+4mp\sh_\alpha\sh_\beta \d_y \d_z-p^2\d_z^2-X\d_p^2-\frac{1}{X}\d_\phi^2-\frac{2p^2}{X}\ch_\alpha \ch_\beta \d_\tau\d_\phi\nn
&&-\Big[p^2\sh_\beta^2+\frac{p^4\ch_\beta^2}{X}+\frac{p^4\ch_\beta^2\sh_\alpha^2}{X}+p\sh_\alpha^2(p+(2l+2p-\epsilon p)\sh^2_\beta)\Big]\d_\tau^2,\nn
&&+p\bar{p}_\alpha g^{MN}\d_M\d_N,
\eea
where we defined
\bea
\bar{p}_\alpha=p\ch_\alpha^2+(2l-\epsilon p)\sh_\alpha^2.
\eea
Note that setting the NUT charge to zero and choosing $\epsilon=1$ gives
\bea
\bar{p}|_{l=0,\epsilon=1}=p.
\eea
This example shows that the NUT charge does not spoil separability and consistent with results from Appendix \ref{App4DKerr}.


\section{Double Field Theory}\label{AppDFT}

In this appendix we review the Double Field Theory (DFT) \cite{DFT} and use rewrite the action of T duality on Killing vectors in a more symmetric form.

Double Field Theory is an elegant way of incorporating T duality as a symmetry of field theory. This is accomplished by extending the standard $D$ coordinates $x^m$ into a larder $2D$--dimensional space $x^M=(\tilde x_m, x^m)$. In this appendix we deviate from the notation used throughout this work and denote the spacetime indices by lower--case letters, while reserving the capital ones to label the ``double space" spanning over regular and barred indices $N=(n,\bar{n})$. This notation is standard in the DFT literature. The theory is formulated with full duality group $O(D,D)$. 

Recall that the T duality group is associated to string compactifications on $T^n$ is $O(n,n)$, so we see that DFT gives a geometric interpretation to  the T duality transformation.

The next step in constructing DFT is defining the fields. One is looking for $O(D,D)$ invariant tensors. It turns out that the metric $g_{mn}$ and the $B_{mn}$ field can be unified into such kind of tensor called the generalized metric \cite{Odd, GenMetric}
\bea\label{DefDFT}
\cH_{MN}=\begin{pmatrix}g^{mn}& -g^{mk}B_{kn}\\B_{mk}g^{kn}&g_{mn}-B_{mk}g^{kl}B_{ln} \end{pmatrix}.
\eea 

Note that the generalized metric does not play the same role as the regular metric in general relativity: the indices are raised and lowered with the constant $O(D,D)$ invariant metric $\eta_{MN}$ rather than $\cH_{MN}$, where
\be
\eta_{MN}=\begin{pmatrix} 0& \delta^m{}_n\\ \delta_m{}^n&0\end{pmatrix}.
\ee
To define diffeomorphisms in DFT theory one needs to introduce the generalized Lie derivative \cite{DFTGaugeTransf} of the generalized metric
\be\label{AppGenLie}
{L}_{\xi}\mathcal{H}_{MN}=\xi^P\d_P\cH_{MN}+(\d_M\xi^P-\d^P\xi_M)\cH_{PN}+(\d_N\xi^P-\d^P\xi_N)\cH_{MP}.
\ee
where $\xi^I=(\tilde\lambda_i,\lambda^i), \xi_I=(\lambda^i,\tilde\lambda_i)$ is the generalized gauge parameter. Here $\tilde\lambda_i$ corresponds to the gauge transformation of the Kalb--Ramond field $B_{ij}$ and $\lambda^i$ is a usual diffeomorphism. 

Transformation (\ref{AppGenLie}) differs from the standard diffeomorphisms in $2D$ dimensions since the following condition must be preserved
\bea\label{AppchSquare}
\cH_{MA}\eta^{AB}\cH_{BN}=\eta_{MN}.
\eea
To demonstrate that (\ref{AppGenLie}) accomplishes this task, one begins with observing that 
\bea
{L}_{\xi}\eta_{MN}=(\d_M\xi^P-\d^P\xi_M)\eta_{PN}+(\d_N\xi^P-\d^P\xi_N)\eta_{MP}=0.
\eea
Then
\bea
&&{L}_{\xi}(\cH_{MA}\eta^{AB}\cH_{BN})\nonumber\\
&&=\left[\xi^P\d_P\cH_{MA}+(\d_M\xi^P-\d^P\xi_M)\cH_{PA}+(\d_A\xi^P-\d^P\xi_A)\cH_{MP}\right]\eta^{AB}\cH_{BN}
+(M\leftrightarrow N)
\nonumber\\
&&=\left[(\d_M\xi^P-\d^P\xi_M)\eta_{PN}+(\d_A\xi^P-\d^P\xi_A)\cH_{MP}\eta^{AB}\cH_{BN}\right]
+(M\leftrightarrow N)
\nonumber\\
&&=(\d_A\xi_Q-\d_Q\xi_A)\eta^{PQ}\eta^{AB}(\cH_{MP}\cH_{BN}+\cH_{NP}\cH_{BM})=0
\eea
This leads to the conclusion that the condition (\ref{AppchSquare}) 
is preserved by the modified diffeomorphism (\ref{AppGenLie}).

\addtocontents{toc}{\protect\setcounter{tocdepth}{0}}
\subsection{Killing vectors in DFT}
\addtocontents{toc}{\protect\setcounter{tocdepth}{6}}

\label{AppKVfromDFT}

To incorporate Killing vectors in the DFT framework, we recall that in the Riemannian geometry the Lie derivative of the metric $g_{mn}$ along a Killing vector $\la$ vanishes
\bea\label{DFTstandKV}
{\mathcal L}_\la g_{mn}=\nabla_m \lambda_n+\nabla_n \lambda_n=0.
\eea 
So to define the ``double Killing vector'' $\xi^M=({\tilde\la}_m,\la^m)$ we require vanishing of the generalized Lie derivative (\ref{AppGenLie})
\bea\label{AppDFTGenKV}
L_\xi \mathcal{H}_{MN}=\xi^P\d_P\cH_{MN}+(\d_M\xi^P-\d^P\xi_M)\cH_{PN}+(\d_N\xi^P-\d^P\xi_N)\cH_{MP}=0.
\eea
Next we will demonstrate that this equation incorporates both gauge transformation of $B$ field and usual diffeomorphism of the metric\footnote{Appearance of both ingredients in the generalized Lie derivative has been discussed in \cite{DFTGaugeTransf}.}. 

Let us begin with $\bar m\bar n$ components of equation \eqref{AppDFTGenKV}\footnote{In the following calculations we use the strong constraint $\tilde\d=0$ \cite{StrongConstraint}.} with $\cH_{MN}$ from (\ref{DefDFT})
\bea\label{LieDerMetr}
{L}_{\xi}\mathcal{H}_{\bar m\bar n}&=&\xi^P\d_P\cH_{\bar m\bar n}+\d_{\bar m}\xi^P\cH_{P\bar n}-\d^P\xi_{\bar m}\cH_{P\bar n}+\d_{\bar n}\xi^P\cH_{\bar mP}-\d^P\xi_{\bar n}\cH_{\bar mP}\nonumber\\
&=&\xi^p\d_p\cH_{\bar m\bar n}-\d_p\xi_{\bar m}\cH^{p}{}_{\bar n}-\d_p\xi_{\bar n}\cH_{\bar m}{}^p=\xi^p\d_pg^{mn}-\d_p\xi_{\bar m}g^{pn}-\d_p\xi_{\bar n}g^{mp}\nonumber\\&=&\lambda^p\d_pg^{mn}-\d_p\lambda^{ m}g^{pn}-\d_p\lambda^{ n}g^{mp}={\mathcal L}_{\xi}(g^{mn})=0.
\eea
This recovers the standard equation (\ref{DFTstandKV}) for the Killing vector. 
For the $\bar m n$ components of equation \eqref{AppDFTGenKV} we find
\bea\label{LieDerGau}
{L}_{\xi}\mathcal{H}_{\bar m n}&=&\xi^p\d_p\cH_{\bar mn}-\d_p\xi_{\bar m}\cH^p{}_{n}+\d_n\xi^P\cH_{\bar mP}-\d_p\xi_n\cH_{\bar m}{}^p\nonumber\\
&=&\lambda^p\d_p(-g^{mk}B_{kn})-\d_p\lambda^m(-g^{pk}B_{kn})+\d_n\lambda^p(-g^{mk}B_{kp})+\d_n\tilde\lambda_pg^{mp}-\d_p\tilde\lambda_ng^{mp}\nonumber\\
&=&\lambda^p\d_pB_{n}{}^m-\d_p\lambda^mB_{n}{}^p+\d_n\lambda^pB_{p}{}^m+(\d_n\tilde\lambda_p-\d_p\tilde\lambda_n)g^{mp}=0.
\eea
The first two terms give the regular Lie derivative of $B_n{}^m$ along the Killing vector $\la^m$, but this derivative does bot have to vanish since the Kalb--Ramond is defined only up to a gauge transformation. Equation (\ref{LieDerGau}) states that the Lie derivative of $B$ must be a pure gauge (with gauge parameter ${\tilde\la}_m$), which means that all physical effects from the Kalb--Ramond field are invariant under the diffeomorphisms generated by $\la^m$. The $mn$ components of (\ref{AppGenLie}) give nothing new due to the constraint (\ref{AppchSquare}). 

We conclude that the Lie derivative (\ref{AppGenLie}) can be used to formulate generalized Killing equation
\bea\label{AppGenKil}
\xi^P\d_P\cH_{MN}+(\d_M\xi^P-\d^P\xi_M)\cH_{PN}+(\d_N\xi^P-\d^P\xi_N)\cH_{MP}=0,
\eea
whose components give equation (\ref{LieDerMetr}) for the regular Killing vector and relation (\ref{LieDerGau}) for the Lie derivative of the $B$ field. 

For future reference we rewrite equations (\ref{LieDerMetr}) and (\ref{LieDerGau}) in terms of the covariant derivatives. For the first equation the transition is standard:
\be\label{AppGenLieKillEq1}
\mathcal{L}_\xi\cH_{\bar m\bar n}=0 \quad\Rightarrow\quad 
\nabla_m\lambda_n+\nabla_n\lambda_m=0,
\ee
and equation (\ref{AppchSquare}), 
\be\label{AppGenLieKillEq2}
\mathcal{L}_\xi\cH_{\bar m n}=0\quad \Rightarrow\quad \lambda^p\d_pB_n{}^m-\d_p\lambda^mB_n{}^p+\d_n\lambda^pB_p{}^m+(\d_n\tilde\lambda_p-\d_p\tilde\lambda_n)g^{mp}=0.	
\ee
requires additional work. Straightforward transformations lead to 
\bea\label{AppGenLieBCov}
\lambda^p\nabla_pB_n{}^m-\nabla_p\lambda^mB_n{}^p+\nabla_n\lambda^pB_p{}^m+
\nabla^m\tilde\lambda_n-\nabla_n\tilde\lambda^m=0,
\eea
and using the Killing equation (\ref{AppGenLieKillEq1}) the last relation can be rewritten in terms of the gauge--invariant field strength $H=dB$:
\bea
{H_{mnp}\lambda^p=\nabla_m\tilde\lambda'_n-\nabla_n\tilde\lambda'_m.}
\eea
where we defined
\bea
\tilde\lambda'_m=\tilde\lambda_m+\lambda_pb_m{}^p\,.
\eea
Notice that under the $O(D,D)$ transformations act as a rotation between ${\tilde\la}_m$ and $\lambda^m$, and 
$\tilde\lambda'_m$ transforms in a more complicated way.


\section{Complex structures}\label{AppComplexStructure}

Killing--Yano tensors are closely related to K{\"a}hler forms on complex manifolds, and in this appendix we will apply the reduction used for the KYT to arrive at the modified K{\"a}hler condition on manifolds with torsion to recover the well--known results \cite{Strominger,HullCS}. 
We begin with an arbitrary anti--symmetric tensor $J$ and define
\be\label{AppCSEq}
T_{PMN}=\nabla_P J_{MN}.
\ee
The Killing--Yano equation for $J$ can be written as
\bea
T_{(PM)N}=0,
\eea
and the K{\"a}hler condition, $dJ=0$, is
\bea
T_{[PMN]}=0.
\eea
Combination of the K{\"a}hler condition with integrability of the complex structure is equivalent to a simple constraint \cite{Zumino}
\bea\label{NewKahler}
T_{PMN}=0,
\eea
and we will now analyze its transformation under T duality.

Starting with a pure metric (\ref{DimRedSetup}) with $B=0$ 
and performing the dimensional reduction of (\ref{AppCSEq}) using (\ref{DimRedL}), we find
\bea\label{CStemp}
T_{zz}{}^n&=&\frac{1}{2}{J^{an}}\d_a e^C+\frac{1}{2}{\hat g}^{nb}e^C F_{ab}{J_z}^a,\nonumber\\
T_z{}^{mn}&=&\frac{1}{2}[{\hat g}^{mb}J^{an}-{\hat g}^{nb}J^{am}]e^C F_{ab}-
\frac{1}{2}[{\hat g}^{ma}{J_z}^n-{\hat g}^{na}{J_z}^m]\d_a C\\
T^{p}{}_z{}^n&=&{\hat\nabla}^p {J_z}^n-\frac{1}{2}g^{pa}{J_z}^n \d_a C-\frac{1}{2}g^{pb}e^C F_{ba}J^{an}
\nonumber\\
T^{pmn}&=&{\hat \nabla}^p J^{mn}
+\frac{1}{2}g^{pa}g^{mb} F_{ab}{J_z}^n
-\frac{1}{2}g^{pa}g^{nb} F_{ab}{J_z}^m.\nonumber
\eea
Introducing rescaled quantities 
\bea
{\tilde J}_z^{~m}=e^{-C}  J_z^{~m},\qquad {\tilde J}^{mn}= J^{mn},
\eea 
we can rewrite these relations as
\bea
T_{zz}{}^n&=&\frac{1}{2}{{\tilde J}^{an}}\d_a e^C+\frac{1}{2}{\hat g}^{nb}e^{2C} {\tilde H}_{abz}{{\tilde J}_z}{}^a,\nonumber\\
T_z{}^{mn}&=&\frac{1}{2}[{\hat g}^{mb}J^{an}-{\hat g}^{nb}J^{am}]e^C {\tilde H}_{abz}+e^C
\frac{1}{2}[{\hat g}^{ma}{\tilde J_z}{}^n-{\hat g}^{na}{\tilde J_z}{}^m]\d_a {\tilde C},\nn
T^{p}{}_z{}^n&=&e^C{\hat\nabla}^p {\tilde J_z}{}^n-\frac{1}{2}e^Cg^{pa}{J_z}^n \d_a {\tilde C}-\frac{1}{2}g^{pb}e^C {\tilde H}_{baz}J^{an},\\
T^{pmn}&=&{\hat \nabla}^p J^{mn}
+\frac{e^C}{2}g^{pa}g^{mb} {\tilde H}_{abz}{J_z}^n
-\frac{e^C}{2}g^{pa}g^{nb} {\tilde H}_{abz}{J_z}^m,\nonumber
\eea
where tildes refer to expressions after the T duality. If we define a tensor
\bea\label{ModifComplStr}
{\tilde T}_{PMN}\equiv {\nabla_P {\tilde J}_{MN}+\frac{1}{2}{\tilde H}_{PNA}{\tilde g}^{AB}{\tilde J}_{MB}-
\frac{1}{2}{\tilde H}_{PMA}{\tilde g}^{AB}{\tilde J}_{NB}}
\eea
after duality, then
\bea
{\tilde T}_{zz}{}^n=-e^{-2C}{T}_{zz}{}^n,\quad
{\tilde T}_z{}^{mn}=-e^{-C}T_z{}^{mn},\quad 
{\tilde T}^{p}{}_z{}^n=e^{-C}T^{p}{}_z{}^n,\quad
{\tilde T}^{mnp}=T^{mnp}.
\eea
In particular we observe that the K{\"a}hler condition (\ref{NewKahler}) is preserved by the T duality, as long as one uses the modified expression (\ref{ModifComplStr}) for ${\tilde T}_{PMN}$ in the presence of the $B$ field. Expression (\ref{ModifComplStr}) can be interpreted as a covariant derivative on a manifold with torsion, and equation 
${\tilde T}_{PMN}=0$ coincides with well--known requirement of supersymmetry for geometries supported by the Kalb--Ramond field \cite{Strominger}.

		\chapter{Appendix for Towards higher dimensional black rings}

\section{Myers--Perry black holes and the neutral black ring}
\label{AppMPandBR}

In this appendix we review neutral rotating black holes and the neutral five--dimensional black ring, in particular we are interested in black holes/ring with one rotation. Starting with the Myers--Perry black hole \cite{MyersPerry}, setting all the rotation parameters except one to zero and introducing 
\be
\mu_1=\sin \theta\equiv s_\theta,\quad\cos\theta\equiv c_\theta
\ee
one gets
\bea\label{AppMPmetric}
ds^2&=&-dt^2+\frac{m}{(r^2+a^2c_\theta^2)r^{D-5}}\Big(dt+a s_\theta^2 d\phi\Big)^2+(r^2+a^2c_\theta^2)\left(\frac{dr^2}{r^2+a^2-mr^{5-D}}
+d\theta^2\right)\nn
&&\quad+(r^2+a^2)s_\theta^2 d\phi^2+r^2c_\theta^2d\Omega_{D-4}.
\eea
Note that even though in this chapter we study separability of the bases, we should mention that the whole Myers--Perry metric in any dimensions is also separable \cite{FrolovKerrNUTAdS,Kub1,KYTinSt}.

Next we recall the neutral black ring with one rotation \cite{BRElvang}\footnote{Note that here comparing to \cite{BRElvang} we swapped $\phi$ and $\psi$ to be consistent with black holes.}
\bea\label{AppBRmetric}
ds^2 &=& -\frac{F(x)}{F(y)} \left(dt+ R\sqrt{\lambda\nu} (1 + y) d\phi\right)^2  \\
&&+\frac{R^2F(x)F(y)^2}{(x-y)^2}\left[ -\frac{1}{F(y)^2} \left( G(y) d\phi^2 + \frac{F(y)}{G(y)} dy^2 \right)+ \frac{1}{F(x)} \left( \frac{dx^2}{G(x)} + \frac{G(x)}{F(x)}d\psi^2\right)\right],\nonumber
\eea
where
\bea\label{AppFGdef}
F(\xi) = 1 - \lambda\xi, \quad  G(\xi) = (1 - \xi^2)(1-\nu \xi),\quad -1\le x\le 1,\quad -\infty<y\le-1.
\eea
In order to avoid conical singularity at $x=-1$ and $y=-1\ne x$ one must set \cite{BRwick}
\bea
\Delta\phi=\Delta\psi=
\frac{2\pi\sqrt{1+\lambda}}{1+\nu}.
\eea
Further one needs to avoid a singularity at $x=1$ which can be done in two ways
\bea\label{Applambdadef}
\lambda_r=\frac{2\nu}{1+\nu^2},\quad \lambda_h=1,
\eea
where the first choice corresponds to a black ring and the second one to a black hole.
In particular the metric \eqref{AppBRmetric} with $\lambda=1$ and the five--dimensional neutral black hole with one rotation (\eqref{AppMPmetric} with $D=5$) are related through
\bea\label{AppMP-ringMap}
r^2&=&\frac{2R^2(1-y)(1-\nu x)}{x-y},\quad c_\theta^2=\frac{(1+x)(1-y)}{2(x-y)},\nn
m&=&2R^2-2\nu R^2+a^2, \quad a=2\sqrt{\nu}R,\nn
t_{MP}&=&\frac{t_{BR}}{\sqrt{\alpha}},\quad \phi_{MP}=\frac{\phi_{BR}}{\sqrt{\alpha}}, \quad \psi_{MP}=\frac{\psi_{BR}}{\sqrt{\alpha}},\quad \alpha=\frac{2}{(1+\nu)^2}.
\eea
For the static black hole $\lambda=1,\nu=0$ we get
\bea\label{AppStaticMap}
F(\xi)=1-\xi,\quad G(\xi)=1-\xi^2,\quad x=-1+\frac{4R^2c_\theta^2}{r^2},\quad y=-1-\frac{4R^2s_\theta^2}{r^2-2R^2},\quad m=2R^2.\nn
\eea
Finally the flat space limit is obtained by setting $\nu=0$ followed by writing
\bea
x=-1+4R^2\tilde{x},\quad y=-1+4R^2\tilde{y}
\eea
and sending $R$ to zero
\bea
r^2&=&\frac{1}{\tilde{x}-\tilde{y}},\quad c_\theta^2=\frac{\tilde{x}}{\tilde{x}-\tilde{y}},\quad s_\theta^2=\frac{\tilde{y}}{\tilde{y}-\tilde{x}}.
\eea

		\chapter{Appendix for Background of the $\lambda$--deformed $AdS_3\times S^3$ supercoset}

\section{$\lambda$--deformation for AdS$_2\times $S$^2$}
\label{AppAdS2}

For comparison with the results obtained in this chapter, we review the geometry of $\lambda$--deformed $AdS_2\times S^2$ constructed in \cite{BTW}. We also extend the solution of \cite{BTW} by one free parameter which makes the fluxes symmetric between the sphere and AdS space.  Applying the procedure reviewed in section \ref{SecReview} to a coset $SU(2)/U(1)$, the authors of \cite{BTW} constructed the metric and the supercoset version of the dilaton (\ref{SuDilaton})\footnote{The solution  corresponding to the bosonic dilaton (\ref{BosDilaton1}) had been constructed earlier in \cite{ST}.}:
\bea\label{AdS2HT}
ds^2&=&\frac{-dx^2+dy^2}{1-\kappa x^2+\kappa^{-1} y^2}+\frac{dp^2+dq^2}{1-\kappa p^2-\kappa^{-1} q^2},\nn
e^{\Phi}&=&\frac{\kappa-x^2+y^2-p^2-q^2+2\sqrt{1-\kappa^2}xp}{\sqrt{-(1-\kappa x^2+\kappa^{-1}y^2)}\sqrt{1-\kappa p^2-\kappa^{-1}q^2}},
\eea
where
\bea
\kappa=\frac{1-\lambda^2}{1+\lambda^2}.
\eea
This background is supported by the Ramond--Ramond flux 
\bea\label{FluxForAdS2}
A&=&\frac{c_1}{M} [ydx-(x-\sqrt{1-\kappa^2}p)dy]+ \frac{c_2}{M}  [q dp-(p-\sqrt{1-\kappa^2}x) dq],\\
M&=&\frac{\kappa-x^2+y^2-p^2-q^2+
2\sqrt{1-\kappa^2}xp}{\sqrt{-(1-\kappa x^2+\kappa^{-1}y^2)(1-\kappa p^2-\kappa^{-1}q^2)}},
\quad c_1^2+c_2^2= 4 \kappa^{-1},\nonumber
\eea
which solves the supergravity equations
\bea
R_{mn}+2\nabla_m\nabla_n \Phi=\frac{e^{2\Phi}}{2}(F_{mp}F_n{}^p-\frac{1}{4}g_{mn}F_{kl}F^{kl}),\nn
\d_n(\sqrt{-g}F^{mn})=0,\quad \nabla^2 e^{-2\Phi}=0,
\eea
and article \cite{BTW} presented the answer (\ref{FluxForAdS2}) for $c_2=0$. 

It is interesting that the flux (\ref{FluxForAdS2}) has a free parameter which interpolates between the components on the sphere and on AdS, while the AdS$_3\times$S$^3$ solution (\ref{C2forAdS3}) has no freedom. This difference can already be seen for the undeformed AdS$_p\times$S$^p$, and it can be traced to the different structure of ``electric--magnetic'' duality groups in four and six dimensions ($U(1)$ in 4d vs $Z_2$ in 6d). 

Since in this chapter we use parameterization of cosets in terms of $X,A$ coordinates introduced in (\ref{SO6gauge}), we will conclude this appendix by writing the relations between coordinate systems used in \cite{ST,BTW} and a three--dimensional version of (\ref{SO6gauge})--(\ref{SO6alphaGg}) describing $SO(3)/SO(2)$: 
\bea\label{SO3gauge}
g_{so}&=&
\left[\begin{array}{cc}
1&0\\
0&(1+A)(1-A)^{-1}
\end{array}\right]
\left[\begin{array}{ccc}
b-1&bX&0\\
-bX&1-bX^2&-bX\\
0&-bX&1\\
\end{array}\right]\,,\\
A&=&\left[\begin{array}{cc}
0&a\\
-a&0
\end{array}\right],\quad 
b=\frac{2}{{1+X^2}}.\nonumber
\eea 
To compare this with the parameterization in terms of the Euler's angles used in \cite{ST,HT},
\bea
g^{trig}=\exp[i(\phi_1-\phi_2)\sigma_3/2]\exp(i\omega \sigma_2)\exp[i(\phi_1+\phi_2)\sigma_3/2]\,,
\eea
we follow the procedure outlined in section  \ref{SecAltParamAdS3}. Specifically, computing
the matrix $D$ (\ref{DabMap}) and comparing the result with a general parameterization (\ref{SO6gauge}) applied to $SO(3)$, we find 
\bea
X_1&=&-\frac{4(\cos^2\omega\sin2\phi_1+\sin^2\omega\sin2\phi_2)}{4+\cos[2(\omega-\phi_1)]+\cos[2(\omega+\phi_1)]+2\cos2\phi_1+4\cos2\phi_2 \sin^2\omega},\nn
X_2&=&-\frac{4\sin2\omega\sin(\phi_1-\phi_2)}{4+\cos[2(\omega-\phi_1)]+\cos[2(\omega+\phi_1)]+2\cos2\phi_1+4\cos2\phi_2 \sin^2\omega},\nn
a&=&\frac{\cos\phi_2\tan\omega}{\cos\phi_1}.
\eea
A $U(1)$ gauge transformation relates this to (\ref{SO3gauge}) with
\bea
X&=&-\frac{4\sqrt{\sin^2(2\phi_1)+\sin^2(\phi_1-\phi_2)\sin^2(2\omega)}}{4+\cos[2(\omega-\phi_1)]+\cos[2(\omega+\phi_1)]+2\cos2\phi_1+4\cos2\phi_2 \sin^2\omega},\nn
a&=&\frac{\cos\phi_2\tan\omega}{\cos\phi_1}.
\eea
The authors of \cite{ST} fixed the gauge by setting $\phi_2=0$, while the authors of \cite{BTW} chose $\phi_2=\phi_1$ and changed coordinates as 
\bea
\omega=\arccos\sqrt{\kappa p^2+\kappa^{-1}q^2},\qquad \phi_1=\arccos\frac{\sqrt{\kappa}p}{\sqrt{\kappa p^2+\kappa^{-1}q^2}}
\eea
to arrive at (\ref{AdS2HT}).


\section{Parametrization of $\mathfrak{psu}(1,1|2)$ and $\mathfrak{psu}(2,2|4)$}
\label{AppPSU}

In this appendix we briefly summarize the parameterization of $\mathfrak{psu}(1,1|2)$, $\mathfrak{psu}(2,2|4)$, and their cosets used in sections \ref{SecAdS3} and \ref{SectAdS5}. We will mostly follow the notation of \cite{ArutFrolov,Beisert}, although our parameterization of fermions differs from the one in \cite{ArutFrolov}, and we will comment on the difference.

The Lie superalgebras $\mathfrak{psu}(n,n|2n)$ can be defined in terms of $(4n)\times (4n)$ supermatrices
\bea\label{GenSuAlg}
{\mathcal M}=\left[\begin{array}{cc}
A&B\\C&D
\end{array}\right]\,,
\eea
with even $(2n)\times (2n)$ blocks $A$, $D$ and odd $(2n)\times (2n)$ blocks $B$, $C$. The graded Lie bracket is defined as
\bea
[{\mathcal M},{\mathcal M}'\}=\left[\begin{array}{cc}
AA'+BC'-A'A+B'C&AB'+BD'-A'B-B'D\\CA'+DC'-C'A-D'C&CB'+DD'+C'B-D'D
\end{array}\right]\,,
\eea
Matrix ${\mathcal M}$ is subject to the hermiticity condition
\bea\label{HermitPSU}
\left[\begin{array}{cc}
A&B\\C&D
\end{array}\right]=\left[\begin{array}{cc}
\Sigma A^\dagger \Sigma^{-1}&-i\Sigma C^\dagger\\-iB^\dagger \Sigma^{-1}&D^\dagger
\end{array}\right]\,,
\eea
where $\Sigma$ is a hermitian matrix of signature $(n,n)$. Convention for $\mathfrak{su}(n,n)$ represented by $A$ fixes the matrix $\Sigma$ and the parameterization of fermions $B$, $C$.

\bigskip

For $\mathfrak{psu}(1,1|2)$ we choose $\Sigma=\mbox{diag}(1,-1)$. This leads to the relation
\bea
C=-iB^\dagger\left[\begin{array}{rr}
1&0\\
0&-1
\end{array}\right],
\eea
or more explicitly
\bea\label{BCcomponents}
B=\left[\begin{array}{rr}
b_{11}&b_{12}\\
b_{21}&b_{22}
\end{array}\right],\quad C=\left[\begin{array}{rr}
-ib^\dagger_{11}&ib^\dagger_{21}\\
-ib^\dagger_{12}&ib^\dagger_{22}
\end{array}\right]\,.
\eea
To construct the algebra for the coset 
\bea
\frac{PSU(1,1|2)_l\times PSU(1,1|2)_r}{SU(1,1)_{diag}\times SU(2)_{diag}},
\eea
we take two copies of $\mathfrak{psu}(1,1|2)$,
\bea\label{CalMwrngBlcks}
{\mathcal M}'=\left[\begin{array}{cc}
{\mathcal M}_1&0\\
0&{\mathcal M}_2
\end{array}\right]\,,
\eea
and project to the subgroup $H$ by imposing the relation (\ref{SuProj})
\bea\label{SuProjApp}
{\mathcal P}^{-1}{\mathcal M}'{\mathcal P}={\mathcal M}'\,,
\eea
as discussed in section \ref{SecAdS3ferm}. Notice that AdS$_3$ and S$_3$ blocks are mixed in the matrix (\ref{CalMwrngBlcks}), and to make the separation more explicit we rearrange the components of the matrix ${\mathcal M}'$ using the parameterization (\ref{GenSuAlg}) for ${\mathcal M}_1$ and ${\mathcal M}_2$. Specifically we define
\bea\label{MAdS3Split}
{\mathcal M}=\left[\begin{array}{cc|cc}
A_1&0&B_1&0\\
0&A_2&0&B_2\\
\hline
C_1&0&D_1&0\\
0&C_2&0&D_2
\end{array}\right]\,.
\eea
The top left block of this matrix describes AdS space, the bottom right block describes the sphere, and the matrix ${\mathcal P}$ corresponding to this supercoset is given by (\ref{CalPforAdS3}):
\bea\label{AdS5Papp}
{\mathcal P}=\left[\begin{array}{cccc}
0&1_{2\times 2}&0&0\\ 
1_{2\times 2}&0&0&0\\
0&0&0&1_{2\times 2}\\ 
0&0&1_{2\times 2}&0
\end{array}\right]\,.
\eea
In particular, this matrix does not mix the $B_i$ and $C_i$ components, so in section \ref{SecAdS3ferm} we computed the fermionic contribution to the dilaton by treating the holomorphic and anti--holomorphic components ($b_{ij}$ and $b^\dagger_{ij}$) as independent variables.

\bigskip

Let us now discuss the $\mathfrak{psu}(2,2|4)$ superalgebra, which emerges in the description of strings on AdS$_5\times$S$^5$ \cite{AdS5PSU}. In this case equation (\ref{HermitPSU}) involves $4\times 4$ blocks, and we choose the matrix $\Sigma$ involved in the hermiticity condition (\ref{HermitPSU}) to be 
\bea
\Sigma=\left[\begin{array}{cccc}
0&\sigma_3\\
\sigma_3&0
\end{array}\right]
\eea
This choice leads to a relation between $2\times 2$ blocks of $B$ and $C$ in (\ref{GenSuAlg}):
\bea\label{qqq0}
B\equiv\left[\begin{array}{cccc}
b_1&b_2\\
b_3&b_4
\end{array}\right],\quad 
C\equiv\left[\begin{array}{cccc}
c_1&c_2\\
c_3&c_4
\end{array}\right]=-i\left[\begin{array}{cccc}
b^\dagger_3\sigma_3&b^\dagger_1\sigma_3\\
b^\dagger_4\sigma_3&b^\dagger_2\sigma_3
\end{array}\right],
\eea
A choice of holomporhic and anti--holomorphic fermions is no longer convenient since the coset projection mixes them. As discussed in section \ref{SecFerDilAdS5}, for the $\mathfrak{psu}(2,2|4)$ supercoset, the condition (\ref{SuProj}) is replaced by  (\ref{SuProjT})
\bea\label{qqq1}
{\mathcal P}^{-1}{\mathcal M}{\mathcal P}={\mathcal M}^T,
\eea
with ${\mathcal P}$ given by (\ref{PforSU4}). An explicit calculation shows that projection (\ref{qqq1}) chooses the elements which satisfy 
\bea\label{qqq2}
B=\left[\begin{array}{cccc}
b_1&b_2\\
b_3&b_4
\end{array}\right],\quad 
C=\left[\begin{array}{rr}
-[\sigma_1 b_4\sigma_1]^T&[\sigma_1 b_2\sigma_1]^T\\
\ [\sigma_1 b_3\sigma_1]^T&[\sigma_1 b_1\sigma_1]^T
\end{array}\right]
\eea
in addition to (\ref{qqq0}). The coset corresponds to the generators included in (\ref{qqq0}), but not in (\ref{qqq2}). In other words, generators satisfying both (\ref{qqq2}) and (\ref{qqq0}) survive under projection $P_3$, and $P_1$ is defined as $P_1=1-P_3$. 

We conclude this appendix by relating our conventions with notation used in \cite{ArutFrolov}. We chose a different embedding of the coset into $SU(4)\times SU(2,2)$, and this led to a following relation between our generators and the ones used by Arutyunov and Frolov (AF) \cite{ArutFrolov}:
\bea
T_{su(4)}&=&R T_{su(4)}^{AF} R^{-1},\quad 
R=\left[\begin{array}{cccc} 
i&0&0&0\\
0&0&0&i\\
0&0&1&0\\
0&1&0&0
\end{array}\right],\\
T_{su(2,2)}&=&\tilde{R} T^{AF}_{su(2,2)} \tilde{R}^{-1},\quad 
\tilde{R}=\frac{1}{\sqrt{2}}\left[\begin{array}{cccc} 
1&0&-1&0\\
0&1&0&1\\
1&0&1&0\\
0&-1&0&1
\end{array}\right].
\eea
While our generators are convenient for evaluating the $\la$--deformation, the generators of Arutyunov and Frolov are better suited for imposing kappa symmetry. Specifically, elimination of this freedom in the notation of  \cite{ArutFrolov} gives
\bea
B^{AF}=\left[\begin{array}{cccc}
0&b_2\\
b_3&0
\end{array}\right],\quad 
C^{AF}=\left[\begin{array}{rr}
0&c_2\\
c_3&0
\end{array}\right]
\eea
while in our notation
\bea
B=\left[\begin{array}{cccc}
b_1&b_2\\
\sigma_3b_1\sigma_3&-\sigma_3 b_2\sigma_3
\end{array}\right],\quad 
C=\left[\begin{array}{rr}
i\sigma_3 b_1^\dagger&ib_1^\dagger\sigma_3\\
-i\sigma_3b^\dagger_2&ib_2^\dagger\sigma_3
\end{array}\right]\,.
\eea
The expressions for kappa symmetry are not used in this work.

		\chapter{Appendix for Generalized $\lambda$--deformations of $AdS_p\times S^p$}

\section{Properties of the matrix $D$}\label{AppDmtr}

In this appendix we study some properties of the matrix\footnote{For the reason which will become clear below, in this appendix we use capital letters $(A,B)$ to denote indices on the algebra $\mathfrak{g}$. This is a minor change of notation in comparison with (\ref{lambdaMtrDef}), which was more convenient in the main text.}
\bea\label{DabApp}
D_{AB}=\mbox{Tr}(T_A g T_B g^{-1}),
\eea
which plays the central role in constructing the generalized $\la$--deformation. While some empirical evidence for these properties has been accumulated from the impressive explicit calculations performed on a case--by--case basis \cite{ST,SfetsosAdS5}, to our knowledge, a general study of matrix $D_{AB}$ has not been carried out.  Using group theory, we derive several important features of this matrix which significantly simplify the construction of integrable deformations for arbitrary cosets in comparison with the explicit calculations performed in \cite{ST, SfetsosAdS5} and explain the nice `surprising relations' observed in these articles.

\bigskip

We begin with recalling the context in which matrix $D_{AB}$ arises in the $\la$--deformation of cosets. The metric is constructed using the frames (\ref{simpleFrames}), the dilaton is given by (\ref{Dilaton}), and both relations contain the expression 
\bea\label{AppEqn1}
\mathfrak{D}=[D-\hat{\lambda}^{-1}]^{-1}\,.
\eea
To construct the deformation of a coset $G/F$, one takes $g\in G/F$ and a constant matrix $\hat{\lambda}^{-1}$ given by (\ref{SmryDmatr})
\bea
\hat{\lambda}^{-1}=I+(I-P){E}_{G}(I-P)\,.
\eea
Here $P$ is a projection on a subgroup $F$, and the explicit form of matrix ${E}_{G}$, given by (\ref{SmryDmatr}), will not be important for our group theoretic discussion here. The results of this appendix can be summarized in the following statement:
\begin{quote}
For any coset $G/F$ there exists a canonical gauge (\ref{CanonGauge}), where matrix $\mathfrak{D}$ has three properties:
\begin{enumerate}[(i)]
\item matrix $(I-P)\mathfrak{D}(I-P)$ has constant entries;
\item matrix $\mathfrak{D}(I-P)$ factorizes as $\mathfrak{D}(I-P)=ST$, where $S$ does not depend on the deformation, and $T$ is a constant matrix;
\item the dependences upon coordinates and constant deformation parameters factorizes in 
$[\mbox{det}\, \mathfrak{D}]$.
\end{enumerate}
\end{quote}
By choosing the canonical gauge in sections \ref{sec:supergravity_background_for_ads2} and \ref{sec:supergravity_background_for_ads3}, we found a very simple deformation dependence in the dilatons  (\ref{lambdaS2dil}), (\ref{DilAdS3}) and frames (\ref{FramesS2}), (\ref{FramesS3}), in agreement with the general statements above.  The specific examples discussed in \cite{ST, SfetsosAdS5} provide additional illustrations of these statements.

\bigskip

We begin with specifying the convenient canonical gauge. The coset $G/F$ introduces a decomposition of the Lie algebra into a subalgebra $\mathfrak{f}$ and the remaining space $\mathfrak{l}$, and in this appendix the generators of $\mathfrak{f}$ and $\mathfrak{l}$ will be denotes using different 
labels\footnote{This decomposition shows the convenience of denoting indices in (\ref{DabApp}) by capital letters.}:
\bea
T_A\in \mathfrak{g}=\mathfrak{f}+\mathfrak{l},\qquad T_a\in \mathfrak{f}, \quad T_\alpha\in \mathfrak{l}\,.
\eea
Algebra $\mathfrak{f}$ closes under commutations, while the commutators of $T_\alpha$ are gauge--dependent, and we will choose a convenient gauge where the structure constants have only three nontrivial blocks:
\bea\label{CanonGauge}
[T_a,T_b]=\sum_c i{f_{ab}}^c T_c\,,\quad
[T_a,T_\beta]=\sum_\gamma i{f_{a\beta}}^\gamma T_\gamma\,\quad
[T_\alpha,T_\beta]=\sum_\gamma i{f_{\alpha\beta}}^c T_c\,.
\eea
In this gauge the Killing metric $\eta_{AB}\propto {f_{AM}}^N {f_{BN}}^M$ splits into two blocks $(\eta_{ab},\eta_{\alpha\beta})$ with vanishing off--diagonal elements $\eta_{a\alpha}=0$.

\bigskip

Our statement (i) reduces to coordinate independence of $\mathfrak{D}_{\alpha\beta}$, and to prove this, as well as the properties (ii) and (iii),  we begin with writing matrices $D$ and ${\hat\la}^{-1}$ in the canonical basis:
\bea\label{DfromH}
\mathfrak{D}^{-1}=D-\hat{\lambda}^{-1}=\left[\begin{array}{cc}
D_{ab}-\delta_{ab}&D_{a\beta}\\
D_{\alpha b}&D_{\alpha\beta}-H_{\alpha\beta}
\end{array}\right]\,,\quad H_{\alpha\beta}=(I+{E}_G)_{\alpha\beta}\,.
\eea
Notice that the all information about the deformation is contained in the \textit{constant matrix} $H_{\alpha\beta}$, which has indices only on the coset. To proceed it is convenient to label various components of (\ref{DfromH}) by different letters:
\bea
\mathfrak{D}^{-1}\equiv \left[\begin{array}{cc}
A&B\\
C&F-H
\end{array}\right]\,.
\eea
To invert the matrix $\mathfrak{D}^{-1}$ and to compute its determinant, we introduce a triangular decomposition:\footnote{In a special case an analogous decomposition was used in \cite{SfetsosAdS5}.}
\bea\label{TriangDecomp}
\mathfrak{D}^{-1}=\left[\begin{array}{cc}
A&0\\
C&M
\end{array}\right]\left[\begin{array}{cc}
I&A^{-1}B\\
0&I
\end{array}\right]\,, \qquad M\equiv F-H-CA^{-1}B.
\eea
Then matrix $\mathfrak{D}$ is given by 
\bea
\mathfrak{D}=\left[\begin{array}{cc}
I&-A^{-1}B\\
0&I
\end{array}\right]\left[\begin{array}{cc}
A^{-1}&0\\
-M^{-1}CA^{-1}&M^{-1}
\end{array}\right],
\eea
in particular,
\bea
\mathfrak{D}_{a\beta}=-[A^{-1}BM^{-1}]_{a\beta}, \quad 
\mathfrak{D}_{\alpha\beta}=[M^{-1}]_{\alpha\beta},\quad 
\mbox{det}\, \mathfrak{D}=[\mbox{det}\, A^{-1}][\mbox{det}\, M^{-1}].
\eea
Recalling that matrices $(A,B,C)$ do not depend on the deformation, we conclude that proving the properties (i)--(iii) amounts to demonstrating than the matrix $M$ does not depend on the coordinates. For example,
equation (\ref{TriangDecomp}) implies that
\bea
\mathfrak{D}(1-P)=S \left[\begin{array}{cc}0&0\\0&M^{-1}\end{array}\right],
\eea
where $S$ does not depend on the deformation and $S_{\alpha\beta}=-\delta_{\alpha\beta}$, so the trivial coordinate dependence of $M$ implies (i) and (ii). 

\bigskip

To summarize, the properties (i)--(iii) would be proven if we demonstrate that $M$ does not depend on coordinates, and this is equivalent to showing that 
\bea
M_0=F-CA^{-1}B
\eea
is a constant matrix. Since the deformation does not enter the last expression, we have arrived at a purely group--theoretic statement, and the rest of this appendix will be dedicated to proving it. 

Let us define $\mathfrak{D}_0$ as the inverse of $(D-\hat{\lambda}^{-1})$ for $H=0$:
\bea\label{DefineD0}
\mathfrak{D}_0=\left[\begin{array}{cc}
D_{ab}-\delta_{ab}&D_{a\beta}\\
D_{\alpha b}&D_{\alpha\beta}
\end{array}\right]^{-1}= \left[\begin{array}{cc}
A&B\\
C&F
\end{array}\right]^{-1}\,.
\eea
Note that $[\mathfrak{D}_0]_{\alpha\beta}=[M_0]_{\alpha\beta}$, and we will show that  these matrix elements do not depend on the coordinates (i.e., on $g$ in (\ref{DabApp})) by demonstrating that they remain constant along \textit{any} one--parametric trajectory on a coset. Let us consider such a trajectory:
\bea
g=\exp\left[ixc^\alpha T_\alpha\right]
\eea
Evaluating the derivative of the matrix $D_{AB}$, we find
\bea\label{DerDab}
\frac{d}{dx}D_{AB}=ic^\alpha {f_{B\alpha}}^C D_{AC}
\eea
Introducing a matrix 
\bea
{f_B}^C\equiv c^\alpha {f_{B\alpha}}^C,
\eea
we can solve the differential equation (\ref{DerDab}):
\bea\label{ee1}
D_{AB}(x)={\exp[ix f]_B}^C D_{AC}(0).
\eea
In the canonical gauge (\ref{CanonGauge}) matrix $f$ has only two types of components, ${f_a}^\beta$ and ${f_\alpha}^b$, so we can write\footnote{Due to antisymmetry of the structure constants, matrices $M$ and $N$ are related by $(M\eta)^T=-N\eta$, where $\eta$ is the Killing form. To avoid unnecessary complications, we use canonical generators with $\eta_{AB}=\delta_{AB}$, but obviously the final results (i)--(iii) hold for any normalization, as long as conditions (\ref{CanonGauge}) are satisfied.}
\bea
f=\left[\begin{array}{cc}
0&N^T\\
M^T&0
\end{array}\right]\,,\quad N=-M^T
\eea
and evaluate the exponent
\bea\label{ee2}
\exp[ix f]^T=\left[\begin{array}{cc}
\cos\left[x\sqrt{MN}\right]&ixM\frac{\sin\left[x\sqrt{NM}\right]}{x\sqrt{NM}}\\
ixN\frac{\sin\left[x\sqrt{MN}\right]}{x\sqrt{MN}}&\cos\left[x\sqrt{NM}\right]
\end{array}\right]\,.
\eea
Here we defined two formal functions of matrix variables using series expansions:
\bea\label{MatrSin}
\cos[\sqrt{A}]\equiv\sum_{n=0}^\infty \frac{(-1)^n}{(2n)!}A^n,\qquad
\frac{\sin[\sqrt{A}]}{\sqrt{A}}\equiv\sum_{n=0}^\infty \frac{(-1)^n}{(2n+1)!}A^n\,.
\eea
Matrix $\mathfrak{D}_0$ is determined by substituting (\ref{ee1}) and (\ref{ee2}) into (\ref{DefineD0}).

We begin with analyzing the generic case with $\mbox{det}[MN]\ne 0$. It is natural to identify the starting point $D_{AB}(0)$ of the  trajectory (\ref{ee1}) with the unit element of the group (i.e., with $g=I$ in (\ref{DabApp})), and in our normalization this choice gives\footnote{In general, $D_{AB}$ in the origin is proportional to the Killing form $\eta_{AB}$. To avoid unnecessary complications, we normalized the generators to have $\eta_{AB}=\delta_{AB}$.}
\bea\label{DatZero}
D_{AB}(0)=\delta_{AB}\,.
\eea
Substitution of (\ref{ee1}) and (\ref{ee2}) into (\ref{DefineD0}) with the initial condition (\ref{DatZero}) gives
\bea\label{D0Trig}
\mathfrak{D}_0=\left[\begin{array}{cc}
\cos\left[x\sqrt{MN}\right]-I&ixM\frac{\sin\left[x\sqrt{NM}\right]}{x\sqrt{NM}}\\
ixN\frac{\sin\left[x\sqrt{MN}\right]}{x\sqrt{MN}}&\cos\left[x\sqrt{NM}\right]
\end{array}\right]^{-1}\,.
\eea
Direct calculation shows that, as long as matrices $(MN)$ and $(NM)$ are non--degenerate,  
\bea\label{InvertDnonsing}
\mathfrak{D}_0=\left[\begin{array}{cc}
\cos\left[x\sqrt{MN}\right]&-ixM\frac{\sin\left[x\sqrt{NM}\right]}{x\sqrt{NM}}\\
-ixN\frac{\sin\left[x\sqrt{MN}\right]}{x\sqrt{MN}}&\cos\left[x\sqrt{NM}\right]-I
\end{array}\right]\left[\begin{array}{c}
I-\cos\left[x\sqrt{MN}\right]\qquad   0\quad \\
\qquad 0\qquad  I-\cos\left[x\sqrt{NM}\right]
\end{array}\right]^{-1}\,.
\eea
In particular, it is clear that
\bea\label{D0isOne}
[\mathfrak{D}_0]_{\alpha\beta}=-I
\eea
does not depend on the coordinate $x$. This completes our proof of the statements (i)--(iii) for the trajectories with $\mbox{det}[MN]\ne 0$,  $\mbox{det}[NM]\ne 0$. The rest of this appendix is devoted to the study of degenerate cases. 

First we assume $\mbox{det}[NM]=0$ while still keeping the condition $\mbox{det}[MN]\ne 0$. Then a symmetric matrix $NM$ can be diagonalized by a \textit{constant} orthogonal transformation $A$, and after such diagonalization, matrix $M$ can be written in a block form:
\bea
M=\left[\begin{array}{cc}
{\tilde M}&0
\end{array}
\right]A^T\,,\quad \mbox{det}{\tilde M}\ne 0.
\eea
Note that
\bea
N=-A\left[\begin{array}{c}
{\tilde M}^T\\ 0
\end{array}
\right]\,,\quad MN=-{\tilde M}{\tilde M}^T,\quad 
NM=-A\left[\begin{array}{cc}
{\tilde M}^T{\tilde M}&0\\
0&0
\end{array}
\right]A^T\,.
\eea
Substitution into (\ref{D0Trig}) gives
\bea\label{D0TrigA}
\mathfrak{D}_0=
\left[\begin{array}{c|c}
I&0\\
\hline
0&A^T
\end{array}\right]^{-1}
\left[\begin{array}{c|cc}
\cosh\left[x\sqrt{{\tilde M}{\tilde M}^T}\right]-I&ix{\tilde M}
\frac{\sinh\left[x\sqrt{{\tilde M}^T{\tilde M}}\right]}{x\sqrt{{\tilde M}^T{\tilde M}}}&0\\
\hline\\
-ix{\tilde M}^T\frac{\sinh\left[x\sqrt{{\tilde M}{\tilde M}^T}\right]}{x\sqrt{{\tilde M}{\tilde M}^T}}&\cosh\left[x\sqrt{{\tilde M}^T{\tilde M}}\right]&0\\
0&0&I
\end{array}\right]^{-1}
\left[\begin{array}{c|c}
I&0\\
\hline
0&A
\end{array}\right]^{-1}
\,.\nonumber
\eea
Performing the inversion as in (\ref{InvertDnonsing}), we conclude that $(\mathfrak{D}_0)_{\alpha\beta}$ is a constant matrix:
\bea
(\mathfrak{D}_0)_{\alpha\beta}=A\left[\begin{array}{cc}
-I&0\\
0&I
\end{array}\right]A^T\,.
\eea
This completes the proof of the statements (i)--(iii) for all trajectories with $\mbox{det}[MN]\ne 0$.

Finally, we look at the most general case. Diagonalzing symmetric matrices $[MN]$ and $[NM]$ with constant orthogonal rotations $A$ and $B$, we can bring $M$ to a canonical form
\bea
M=B\left[\begin{array}{cc}
{\tilde M}&0\\
0&0
\end{array}
\right]A^T\,,\quad \mbox{det}{\tilde M}\ne 0.
\eea
This gives
\bea
N=-A\left[\begin{array}{cc}
{\tilde M}^T&0\\ 0&0
\end{array}
\right]B^T\,,\ 
MN=-B\left[\begin{array}{cc}
{\tilde M}{\tilde M}^T&0\\
0&0
\end{array}
\right]B^T,\ 
NM=-A\left[\begin{array}{cc}
{\tilde M}^T{\tilde M}&0\\
0&0
\end{array}
\right]A^T\,\nonumber
\eea
and
\bea\label{GeneralGrpElem}
\exp[ix f]^T&=&R\left[\begin{array}{cc|cc}
\cosh\left[x\sqrt{{\tilde M}{\tilde M}^T}\right]&0&ix{\tilde M}
\frac{\sinh\left[x\sqrt{{\tilde M}^T{\tilde M}}\right]}{x\sqrt{{\tilde M}^T{\tilde M}}}&0\\
0&I_{d_1}&0&0\\
\hline
-ix{\tilde M}^T\frac{\sinh\left[x\sqrt{{\tilde M}{\tilde M}^T}\right]}{x\sqrt{{\tilde M}{\tilde M}^T}}&0&\cosh\left[x\sqrt{{\tilde M}^T{\tilde M}}\right]&0\\
0&0&0&I_{d_2}
\end{array}\right]R^{-1},\\
R&=&\left[\begin{array}{c|c}
B&0\\
\hline
0&A\end{array}\right],\quad A^T=A^{-1},\quad B^T=B^{-1}\,\quad \mbox{det}{\tilde M}\ne0.\nonumber
\eea
Substitution of (\ref{GeneralGrpElem}) and (\ref{DatZero}) into (\ref{ee1}) leads to a non--invertible matrix in the right--hand side of (\ref{DefineD0}) unless $d_1=0$. To cure this problem, we observe that under a gauge transformation
\bea
g\rightarrow gh,\quad h\in F,
\eea
matrix (\ref{DabApp}) transforms as
\bea
D_{AB}\rightarrow {{\hat h}_B}~^C D_{AC},
\eea
where ${{\hat h}_B}~^C$ is the image of $h$ in the adjoint representation:
\bea
h T_B h^{-1}\equiv {{\hat h}_B}~^C T_C
\eea
In the basis (\ref{CanonGauge}) matrix ${{\hat h}_B}~^C$ has a block--diagonal form:
\bea
 {{\hat h}_B}~^C=\left[\begin{array}{c|c}
\bullet&0\\
\hline
0&\bullet\end{array}\right]
\eea
To regularize the expression for $\mathfrak{D}_0$ corresponding to (\ref{GeneralGrpElem}), we replace the condition (\ref{DatZero}) by its gauge-transformed version:
\bea
D_{AB}(0)={\hat h}_{BA}.
\eea
Then definition (\ref{DefineD0}) gives
\bea
[\mathfrak{D}_0]^{-1}={\hat h}^TR\left[\begin{array}{cc|cc}
\cosh\left[x\sqrt{{\tilde M}{\tilde M}^T}\right]&0&ix{\tilde M}
\frac{\sinh\left[x\sqrt{{\tilde M}^T{\tilde M}}\right]}{x\sqrt{{\tilde M}^T{\tilde M}}}&0\\
0&I_{d_1}&0&0\\
\hline
-ix{\tilde M}^T\frac{\sinh\left[x\sqrt{{\tilde M}{\tilde M}^T}\right]}{x\sqrt{{\tilde M}{\tilde M}^T}}&0&\cosh\left[x\sqrt{{\tilde M}^T{\tilde M}}\right]&0\\
0&0&0&I_{d_2}
\end{array}\right]R^{-1}-
\left[\begin{array}{c|c}
I&0\\
\hline
0&0\end{array}\right]\nonumber
\eea
Note that the last term in the right--hand side can be written as
\bea
\left[\begin{array}{c|c}
I&0\\
\hline
0&0\end{array}\right]=
{\hat h}^TR\left[\begin{array}{c|c}
{\tilde h}&0\\
\hline
0&0\end{array}\right]R^{-1},
\eea
where ${\tilde h}$ is some matrix. It is convenient to parameterize its components as
\bea
{\tilde h}\equiv \left[\begin{array}{cc}
{\tilde h}_1&{\tilde h}_2\\
{\tilde h}_3&{\tilde h}_4+I_{d_1}\end{array}\right]\,.
\eea
If $d_1$ is even, the we can choose a gauge where ${\tilde h}_2={\tilde h}^T_3=0$, ${\tilde h}_1=I$, and 
\bea
{\tilde h}_4=\exp\left[\begin{array}{cc}0&iq\\ -iq^T&0\end{array}\right]-I_{d_1}
\eea
is a non--degenerate matrix. For odd $d_1$ a similar gauge can be used to reduce the problem to $d_1=1$. Furthermore, by choosing appropriate matrices $A$ and $B$ in (\ref{GeneralGrpElem}), we can make ${\tilde M}$ diagonal, then for $d_1=1$ we can further specify the gauge\footnote{To make the next expression compact, we introduced shortcuts: $\sh=\sinh$, $\ch=\cosh$.}:
\bea
[\mathfrak{D}_0]^{-1}={\hat h}^TR\left[\begin{array}{ccc|ccc}
\ch[x\hat{M}]-I&0&0&i\,\sh[x{\hat M}]&0&0\\
0&\ch [x m]-\ch\, y&i\,\sh\, y&0&i\,\sh[x m]&0\\
0&-i\,\sh\, y&1-\ch\, y&0&0&0\\
\hline
-i\,\sh[x{\hat M}]&0&0&\ch[x{\hat M}]&0&0\\
0&-i\,\sh[x{m}]&0&0&\ch[x m]&0\\
0&0&0&0&0&I_{d_2}
\end{array}\right]R^{-1}\nonumber
\eea
Here ${\hat M}$ is a non--degenerate diagonal matrix, and $m\ne 0$ is a number. 
The inverse of the last matrix is
\bea
\mathfrak{D}_0=R\left[\begin{array}{ccc|ccc}
\frac{\cosh[x\hat{M}]}{\cosh[x\hat{M}]-I}&0&0&i\coth[\frac{x}{2}{\hat M}]&0&0\\
0&\bullet&\bullet&0&\bullet&0\\
0&\bullet&\bullet&0&\bullet&0\\
\hline
-i\coth[\frac{x}{2}{\hat M}]&0&0&-I&0&0\\
0&\bullet&\bullet&0&1&0\\
0&0&0&0&0&I_{d_2}
\end{array}\right][{\hat h}^TR]^{-1}\nonumber
\eea
Bullets denote some complicated expressions which are irrelevant for our analysis. 

To summarize, we have demonstrated that even in the degenerate case when $\mbox{det}[MN]= 0$, there exists a gauge where $[\mathfrak{D}_0]_{\alpha\beta}$ remains constant along any one--parametric trajectory. This completes the proof of the statements (i)--(iii).


	\backmatter

		\bibliographystyle{apsrev4-1long}

\end{document}